\newif\ifpaper \papertrue
\newif\iffull \fullfalse
\newif\ifdraft \draftfalse
\newif\ifpreprint \preprintfalse

\preprinttrue

\ifpreprint
\documentclass[acmsmall]{acmart}
\else
\documentclass[acmsmall,review,anonymous]{acmart}
\fi

\ifpreprint

\makeatletter
\@ACM@screentrue
\makeatother

\else

\acmJournal{PACMPL}
\acmVolume{1}
\acmNumber{CONF} 
\acmArticle{1}
\acmYear{2018}
\acmMonth{1}
\acmDOI{} 
\startPage{1}

\fi

\setcopyright{none}

\usepackage{booktabs}   
\usepackage{subcaption} 

\usepackage{multirow}

\usepackage{ifthen}
\newboolean{show-common} \setboolean{show-common}{true}
\newboolean{shade-common} \setboolean{shade-common}{false}
\newboolean{show-variant} \setboolean{show-variant}{true}
\newboolean{show-ectx} \setboolean{show-ectx}{true}
\newboolean{show-grad} \setboolean{show-grad}{true}
\newboolean{show-grad-term} \setboolean{show-grad-term}{true}
\newboolean{show-typename} \setboolean{show-typename}{true}
\newboolean{shade-dyn} \setboolean{shade-dyn}{false}
\newboolean{shade-variant} \setboolean{shade-variant}{false}

\newcommand{\ifthen}[2]{\ifthenelse{#1}{#2}{}}
\newcommand{\ifelse}[2]{\ifthenelse{#1}{}{#2}}

\usepackage{xcolor}
\newcommand{\shadeif}[2]{%
\begingroup \ifthen{#1}{\color{gray}}#2 \endgroup}
\newcommand{\mayshadecommon}[1]{\shadeif{\boolean{shade-common}}{#1}}
\newcommand{\mayshadevariant}[1]{\shadeif{\boolean{shade-variant}}{#1}}

\newtheorem{defn}{Definition}

\ifpaper
\else
\usepackage{amsthm}

\newtheorem{lemm}{Lemma}

\newenvironment{lemma}[1]
{\begin{lemm} \label{lem:#1} \noindent}
{\end{lemm}}

\newtheorem{thm}{Theorem}
\newenvironment{theorem}[2][]
{\begin{thm}[#1] \label{thm:#2} \noindent}
 {\end{thm}}
 
\fi

\usepackage{enumitem}

\def\case#1:{\item[\textmd{Case {#1}:}]}

\newcommand{\surfacelang}{\ensuremath{\text{F}^\rho_\text{G}}}
\newcommand{\interlang}{\ensuremath{\text{F}^\rho_\text{C}}}
\newcommand{\staticlang}{\ensuremath{\text{F}^\rho}}
\newcommand{\defeq}{\stackrel{\rm \tiny def}{=}}

\newcommand{\reffig}[1]{Figure~\ref{fig:#1}}
\newcommand{\refdef}[1]{Definition~\ref{def:#1}}
\newcommand{\sect}[1]{Section~\ref{sec:#1}}

\newcommand{\Rule}[2]{\ensuremath{\text{({\sc{{#1}\_{#2}}})}}}
\newcommand{\WF}[1]{\Rule{WF}{#1}}
\newcommand{\WFs}[1]{\Rule{WFs}{#1}}

\newcommand{\Ts}[1]{\Rule{Ts}{#1}}
\newcommand{\R}[1]{\Rule{R}{#1}}
\newcommand{\Rs}[1]{\Rule{Rs}{#1}}
\newcommand{\E}[1]{\Rule{E}{#1}}
\newcommand{\T}[1]{\Rule{T}{#1}}

\newcommand{\Cns}[1]{\Rule{C}{#1}}
\newcommand{\CE}[1]{\Rule{CE}{#1}}

\newcommand{\Cv}[1]{\Rule{Cv}{#1}}

\newcommand{\RuleWoP}[2]{\ensuremath{\text{{\sc{{#1}\_{#2}}}}}}
\newcommand{\RWoP}[1]{\RuleWoP{R}{#1}}
\newcommand{\RsWoP}[1]{\RuleWoP{Rs}{#1}}
\newcommand{\EWoP}[1]{\RuleWoP{E}{#1}}
\newcommand{\EsWoP}[1]{\RuleWoP{Es}{#1}}

\newcommand{\TS}[1]{\ifdraft\textcolor{red}{TS: #1}\fi}
\newcommand{\AI}[1]{\ifdraft\textcolor{blue}{#1 -- AI}\fi}

\usepackage{multirow}

\newcommand{\ottdrule}[4][]{{\displaystyle\frac{\begin{array}{l}#2\end{array}}{#3}\quad\ottdrulename{#4}}}

\newcommand{\ottpremise}[1]{ #1 \\}
\newenvironment{ottdefnblock}[3][]{ \framebox{\mbox{#2}} \quad #3 \\[0pt]}{}

\newcommand{\ottnt}[1]{\mathit{#1}}
\newcommand{\ottmv}[1]{\mathit{#1}}

\newcommand{\ottsym}[1]{#1}

\newcommand{\ottdrulename}[1]{\textsc{#1}}

\usepackage{amssymb}
\usepackage{amsmath,bm}
\usepackage{centernot}
\usepackage{xcolor}

\newif\ifvector \vectortrue

\newcommand{\variantliftrow}[2]{\uparrow\! {#1} \, {#2} }
\newcommand{\variantlift}[3]{\variantliftrow{\langle {#1} : {#2} \rangle}{#3} }

\newcommand{\variantliftdown}[4]{\,\downarrow\!^{#1}_{\langle {#2} : {#3} \rangle} {#4} }

\newcommand{\makestatic}[1]{ {#1}^\mathit{s} }

\definecolor{lightgray}{rgb}{0.83, 0.83, 0.83}
\def\shadeback:#1:{\colorbox{lightgray}{$#1$} }

\def\undline<#1>{\underline{\smash{#1} } }
\def\undlinesq[#1]{\underline{#1} }

\newcommand{\ottdruleCEXXConsL}[1]{\ottdrule[#1]{%
\ottpremise{ \rho_{{\mathrm{2}}} \,  \triangleright _{ \ell }  \, \ottnt{B}  \ottsym{,}  \rho'_{{\mathrm{2}}}  \quad   \ottnt{A}  \simeq  \ottnt{B}  \quad  \rho_{{\mathrm{1}}}  \simeq  \rho'_{{\mathrm{2}}}  }%
}{
  \ell  \mathbin{:}  \ottnt{A}   ;  \rho_{{\mathrm{1}}}   \simeq  \rho_{{\mathrm{2}}}}{%
{\ottdrulename{CE\_ConsL}}{}%
}}

\newcommand{\ottdruleCXXDynR}[1]{\ottdrule[#1]{%
}{
\ottnt{A}  \sim  \star}{%
{\ottdrulename{C\_DynR}}{}%
}}

\newcommand{\ottdruleCXXDynL}[1]{\ottdrule[#1]{%
}{
\star  \sim  \ottnt{A}}{%
{\ottdrulename{C\_DynL}}{}%
}}

\newcommand{\ottdruleCXXRefl}[1]{\ottdrule[#1]{%
}{
\ottnt{A}  \sim  \ottnt{A}}{%
{\ottdrulename{C\_Refl}}{}%
}}

\newcommand{\ottdruleCXXFun}[1]{\ottdrule[#1]{%
\ottpremise{ \ottnt{A_{{\mathrm{1}}}}  \sim  \ottnt{A_{{\mathrm{2}}}}  \quad  \ottnt{B_{{\mathrm{1}}}}  \sim  \ottnt{B_{{\mathrm{2}}}} }%
}{
\ottnt{A_{{\mathrm{1}}}}  \rightarrow  \ottnt{B_{{\mathrm{1}}}}  \sim  \ottnt{A_{{\mathrm{2}}}}  \rightarrow  \ottnt{B_{{\mathrm{2}}}}}{%
{\ottdrulename{C\_Fun}}{}%
}}

\newcommand{\ottdruleCXXPoly}[1]{\ottdrule[#1]{%
\ottpremise{\ottnt{A_{{\mathrm{1}}}}  \sim  \ottnt{A_{{\mathrm{2}}}}}%
}{
 \text{\unboldmath$\forall\!$}  \,  \mathit{X}  \mathord{:}  \ottnt{K}   \ottsym{.} \, \ottnt{A_{{\mathrm{1}}}}  \sim   \text{\unboldmath$\forall\!$}  \,  \mathit{X}  \mathord{:}  \ottnt{K}   \ottsym{.} \, \ottnt{A_{{\mathrm{2}}}}}{%
{\ottdrulename{C\_Poly}}{}%
}}

\newcommand{\ottdruleCXXPolyL}[1]{\ottdrule[#1]{%
\ottpremise{ \mathbf{QPoly} \, \ottsym{(}  \ottnt{A_{{\mathrm{2}}}}  \ottsym{)}  \quad   \mathit{X} \,  \not\in  \,  \mathit{ftv}  (  \ottnt{A_{{\mathrm{2}}}}  )   \quad  \ottnt{A_{{\mathrm{1}}}}  \sim  \ottnt{A_{{\mathrm{2}}}}  }%
}{
 \text{\unboldmath$\forall\!$}  \,  \mathit{X}  \mathord{:}  \ottnt{K}   \ottsym{.} \, \ottnt{A_{{\mathrm{1}}}}  \sim  \ottnt{A_{{\mathrm{2}}}}}{%
{\ottdrulename{C\_PolyL}}{}%
}}

\newcommand{\ottdruleCXXPolyR}[1]{\ottdrule[#1]{%
\ottpremise{ \mathbf{QPoly} \, \ottsym{(}  \ottnt{A_{{\mathrm{1}}}}  \ottsym{)}  \quad   \mathit{X} \,  \not\in  \,  \mathit{ftv}  (  \ottnt{A_{{\mathrm{1}}}}  )   \quad  \ottnt{A_{{\mathrm{1}}}}  \sim  \ottnt{A_{{\mathrm{2}}}}  }%
}{
\ottnt{A_{{\mathrm{1}}}}  \sim   \text{\unboldmath$\forall\!$}  \,  \mathit{X}  \mathord{:}  \ottnt{K}   \ottsym{.} \, \ottnt{A_{{\mathrm{2}}}}}{%
{\ottdrulename{C\_PolyR}}{}%
}}

\newcommand{\ottdruleCXXRecord}[1]{\ottdrule[#1]{%
\ottpremise{\rho_{{\mathrm{1}}}  \sim  \rho_{{\mathrm{2}}}}%
}{
 [  \rho_{{\mathrm{1}}}  ]   \sim   [  \rho_{{\mathrm{2}}}  ] }{%
{\ottdrulename{C\_Record}}{}%
}}

\newcommand{\ottdruleCXXVariant}[1]{\ottdrule[#1]{%
\ottpremise{\rho_{{\mathrm{1}}}  \sim  \rho_{{\mathrm{2}}}}%
}{
 \langle  \rho_{{\mathrm{1}}}  \rangle   \sim   \langle  \rho_{{\mathrm{2}}}  \rangle }{%
{\ottdrulename{C\_Variant}}{}%
}}

\newcommand{\ottdruleCXXCons}[1]{\ottdrule[#1]{%
\ottpremise{ \ottnt{A_{{\mathrm{1}}}}  \sim  \ottnt{A_{{\mathrm{2}}}}  \quad  \rho_{{\mathrm{1}}}  \sim  \rho_{{\mathrm{2}}} }%
}{
  \ell  \mathbin{:}  \ottnt{A_{{\mathrm{1}}}}   ;  \rho_{{\mathrm{1}}}   \sim    \ell  \mathbin{:}  \ottnt{A_{{\mathrm{2}}}}   ;  \rho_{{\mathrm{2}}} }{%
{\ottdrulename{C\_Cons}}{}%
}}

\newcommand{\ottdruleCXXConsL}[1]{\ottdrule[#1]{%
\ottpremise{ \ell \,  \not\in  \, \mathit{dom} \, \ottsym{(}  \rho_{{\mathrm{2}}}  \ottsym{)}  \quad    \rho_{{\mathrm{2}}}  \text{ ends with }  \star   \quad  \rho_{{\mathrm{1}}}  \sim  \rho_{{\mathrm{2}}}  }%
}{
  \ell  \mathbin{:}  \ottnt{A}   ;  \rho_{{\mathrm{1}}}   \sim  \rho_{{\mathrm{2}}}}{%
{\ottdrulename{C\_ConsL}}{}%
}}

\newcommand{\ottdruleCXXConsR}[1]{\ottdrule[#1]{%
\ottpremise{ \ell \,  \not\in  \, \mathit{dom} \, \ottsym{(}  \rho_{{\mathrm{1}}}  \ottsym{)}  \quad    \rho_{{\mathrm{1}}}  \text{ ends with }  \star   \quad  \rho_{{\mathrm{1}}}  \sim  \rho_{{\mathrm{2}}}  }%
}{
\rho_{{\mathrm{1}}}  \sim    \ell  \mathbin{:}  \ottnt{A}   ;  \rho_{{\mathrm{2}}} }{%
{\ottdrulename{C\_ConsR}}{}%
}}

\newcommand{\ottdruleCvXXDyn}[1]{\ottdrule[#1]{%
}{
 \Sigma   \vdash   \star  \prec^{ \Phi }  \star }{%
{\ottdrulename{Cv\_Dyn}}{}%
}}

\newcommand{\ottdruleCvXXTyVar}[1]{\ottdrule[#1]{%
}{
 \Sigma   \vdash   \mathit{X}  \prec^{ \Phi }  \mathit{X} }{%
{\ottdrulename{Cv\_TyVar}}{}%
}}

\newcommand{\ottdruleCvXXTyName}[1]{\ottdrule[#1]{%
\ottpremise{ \mathit{name} (  \Phi  )  \,  \not=  \, \alpha}%
}{
 \Sigma   \vdash   \alpha  \prec^{ \Phi }  \alpha }{%
{\ottdrulename{Cv\_TyName}}{}%
}}

\newcommand{\ottdruleCvXXReveal}[1]{\ottdrule[#1]{%
\ottpremise{\Sigma  \ottsym{(}  \alpha  \ottsym{)} \,  =  \, \ottnt{A}}%
}{
 \Sigma   \vdash   \alpha  \prec^{ \ottsym{+}  \alpha }  \ottnt{A} }{%
{\ottdrulename{Cv\_Reveal}}{}%
}}

\newcommand{\ottdruleCvXXConceal}[1]{\ottdrule[#1]{%
\ottpremise{\Sigma  \ottsym{(}  \alpha  \ottsym{)} \,  =  \, \ottnt{A}}%
}{
 \Sigma   \vdash   \ottnt{A}  \prec^{ \ottsym{-}  \alpha }  \alpha }{%
{\ottdrulename{Cv\_Conceal}}{}%
}}

\newcommand{\ottdruleCvXXBase}[1]{\ottdrule[#1]{%
}{
 \Sigma   \vdash   \iota  \prec^{ \Phi }  \iota }{%
{\ottdrulename{Cv\_Base}}{}%
}}

\newcommand{\ottdruleCvXXFun}[1]{\ottdrule[#1]{%
\ottpremise{  \Sigma   \vdash   \ottnt{A_{{\mathrm{2}}}}  \prec^{  \overline{ \Phi }  }  \ottnt{A_{{\mathrm{1}}}}   \quad   \Sigma   \vdash   \ottnt{B_{{\mathrm{1}}}}  \prec^{ \Phi }  \ottnt{B_{{\mathrm{2}}}}  }%
}{
 \Sigma   \vdash   \ottnt{A_{{\mathrm{1}}}}  \rightarrow  \ottnt{B_{{\mathrm{1}}}}  \prec^{ \Phi }  \ottnt{A_{{\mathrm{2}}}}  \rightarrow  \ottnt{B_{{\mathrm{2}}}} }{%
{\ottdrulename{Cv\_Fun}}{}%
}}

\newcommand{\ottdruleCvXXPoly}[1]{\ottdrule[#1]{%
\ottpremise{ \Sigma   \vdash   \ottnt{A_{{\mathrm{1}}}}  \prec^{ \Phi }  \ottnt{A_{{\mathrm{2}}}} }%
}{
 \Sigma   \vdash    \text{\unboldmath$\forall\!$}  \,  \mathit{X}  \mathord{:}  \ottnt{K}   \ottsym{.} \, \ottnt{A_{{\mathrm{1}}}}  \prec^{ \Phi }   \text{\unboldmath$\forall\!$}  \,  \mathit{X}  \mathord{:}  \ottnt{K}   \ottsym{.} \, \ottnt{A_{{\mathrm{2}}}} }{%
{\ottdrulename{Cv\_Poly}}{}%
}}

\newcommand{\ottdruleCvXXRecord}[1]{\ottdrule[#1]{%
\ottpremise{ \Sigma   \vdash   \rho_{{\mathrm{1}}}  \prec^{ \Phi }  \rho_{{\mathrm{2}}} }%
}{
 \Sigma   \vdash    [  \rho_{{\mathrm{1}}}  ]   \prec^{ \Phi }   [  \rho_{{\mathrm{2}}}  ]  }{%
{\ottdrulename{Cv\_Record}}{}%
}}

\newcommand{\ottdruleCvXXVariant}[1]{\ottdrule[#1]{%
\ottpremise{ \Sigma   \vdash   \rho_{{\mathrm{1}}}  \prec^{ \Phi }  \rho_{{\mathrm{2}}} }%
}{
 \Sigma   \vdash    \langle  \rho_{{\mathrm{1}}}  \rangle   \prec^{ \Phi }   \langle  \rho_{{\mathrm{2}}}  \rangle  }{%
{\ottdrulename{Cv\_Variant}}{}%
}}

\newcommand{\ottdruleCvXXREmp}[1]{\ottdrule[#1]{%
}{
 \Sigma   \vdash    \cdot   \prec^{ \Phi }   \cdot  }{%
{\ottdrulename{Cv\_REmp}}{}%
}}

\newcommand{\ottdruleCvXXCons}[1]{\ottdrule[#1]{%
\ottpremise{  \Sigma   \vdash   \ottnt{A_{{\mathrm{1}}}}  \prec^{ \Phi }  \ottnt{A_{{\mathrm{2}}}}   \quad   \Sigma   \vdash   \rho_{{\mathrm{1}}}  \prec^{ \Phi }  \rho_{{\mathrm{2}}}  }%
}{
 \Sigma   \vdash     \ell  \mathbin{:}  \ottnt{A_{{\mathrm{1}}}}   ;  \rho_{{\mathrm{1}}}   \prec^{ \Phi }    \ell  \mathbin{:}  \ottnt{A_{{\mathrm{2}}}}   ;  \rho_{{\mathrm{2}}}  }{%
{\ottdrulename{Cv\_Cons}}{}%
}}

\newcommand{\ottdruleEqXXSwap}[1]{\ottdrule[#1]{%
\ottpremise{\ell \,  \not=  \, \ell'}%
}{
  \ell  \mathbin{:}  \ottnt{A}   ;    \ell'  \mathbin{:}  \ottnt{B}   ;  \rho    \equiv    \ell'  \mathbin{:}  \ottnt{B}   ;    \ell  \mathbin{:}  \ottnt{A}   ;  \rho  }{%
{\ottdrulename{Eq\_Swap}}{}%
}}

\newcommand{\ottdruleWFsXXTyVar}[1]{\ottdrule[#1]{%
\ottpremise{  \mathrel{ \makestatic{\vdash} }  \Gamma   \quad   \mathit{X}  \mathord{:}  \ottnt{K}  \,  \in  \, \Gamma }%
}{
 \Gamma  \mathrel{ \makestatic{\vdash} }  \mathit{X}  :  \ottnt{K} }{%
{\ottdrulename{WFs\_TyVar}}{}%
}}

\newcommand{\ottdruleWFsXXBase}[1]{\ottdrule[#1]{%
\ottpremise{ \mathrel{ \makestatic{\vdash} }  \Gamma }%
}{
 \Gamma  \mathrel{ \makestatic{\vdash} }  \iota  :   \mathsf{T}  }{%
{\ottdrulename{WFs\_Base}}{}%
}}

\newcommand{\ottdruleWFsXXFun}[1]{\ottdrule[#1]{%
\ottpremise{  \Gamma  \mathrel{ \makestatic{\vdash} }  \ottnt{A}  :   \mathsf{T}    \quad   \Gamma  \mathrel{ \makestatic{\vdash} }  \ottnt{B}  :   \mathsf{T}   }%
}{
 \Gamma  \mathrel{ \makestatic{\vdash} }  \ottnt{A}  \rightarrow  \ottnt{B}  :   \mathsf{T}  }{%
{\ottdrulename{WFs\_Fun}}{}%
}}

\newcommand{\ottdruleWFsXXPoly}[1]{\ottdrule[#1]{%
\ottpremise{ \Gamma  \ottsym{,}   \mathit{X}  \mathord{:}  \ottnt{K}   \mathrel{ \makestatic{\vdash} }  \ottnt{A}  :   \mathsf{T}  }%
}{
 \Gamma  \mathrel{ \makestatic{\vdash} }   \text{\unboldmath$\forall\!$}  \,  \mathit{X}  \mathord{:}  \ottnt{K}   \ottsym{.} \, \ottnt{A}  :   \mathsf{T}  }{%
{\ottdrulename{WFs\_Poly}}{}%
}}

\newcommand{\ottdruleWFsXXRecord}[1]{\ottdrule[#1]{%
\ottpremise{ \Gamma  \mathrel{ \makestatic{\vdash} }  \rho  :   \mathsf{R}  }%
}{
 \Gamma  \mathrel{ \makestatic{\vdash} }   [  \rho  ]   :   \mathsf{T}  }{%
{\ottdrulename{WFs\_Record}}{}%
}}

\newcommand{\ottdruleWFsXXVariant}[1]{\ottdrule[#1]{%
\ottpremise{ \Gamma  \mathrel{ \makestatic{\vdash} }  \rho  :   \mathsf{R}  }%
}{
 \Gamma  \mathrel{ \makestatic{\vdash} }   \langle  \rho  \rangle   :   \mathsf{T}  }{%
{\ottdrulename{WFs\_Variant}}{}%
}}

\newcommand{\ottdruleWFsXXREmp}[1]{\ottdrule[#1]{%
\ottpremise{ \mathrel{ \makestatic{\vdash} }  \Gamma }%
}{
 \Gamma  \mathrel{ \makestatic{\vdash} }   \cdot   :   \mathsf{R}  }{%
{\ottdrulename{WFs\_REmp}}{}%
}}

\newcommand{\ottdruleWFsXXCons}[1]{\ottdrule[#1]{%
\ottpremise{  \Gamma  \mathrel{ \makestatic{\vdash} }  \ottnt{A}  :   \mathsf{T}    \quad   \Gamma  \mathrel{ \makestatic{\vdash} }  \rho  :   \mathsf{R}   }%
}{
 \Gamma  \mathrel{ \makestatic{\vdash} }    \ell  \mathbin{:}  \ottnt{A}   ;  \rho   :   \mathsf{R}  }{%
{\ottdrulename{WFs\_Cons}}{}%
}}

\newcommand{\ottdruleTsXXVar}[1]{\ottdrule[#1]{%
\ottpremise{  \mathrel{ \makestatic{\vdash} }  \Gamma   \quad   \mathit{x}  \mathord{:}  \ottnt{A}  \,  \in  \, \Gamma }%
}{
 \Gamma  \mathrel{ \makestatic{\vdash} }  \mathit{x}  :  \ottnt{A} }{%
{\ottdrulename{Ts\_Var}}{}%
}}

\newcommand{\ottdruleTsXXConst}[1]{\ottdrule[#1]{%
\ottpremise{ \mathrel{ \makestatic{\vdash} }  \Gamma }%
}{
 \Gamma  \mathrel{ \makestatic{\vdash} }  \kappa  :   \mathit{ty}  (  \kappa  )  }{%
{\ottdrulename{Ts\_Const}}{}%
}}

\newcommand{\ottdruleTsXXLam}[1]{\ottdrule[#1]{%
\ottpremise{ \Gamma  \ottsym{,}   \mathit{x}  \mathord{:}  \ottnt{A}   \mathrel{ \makestatic{\vdash} }  \ottnt{M}  :  \ottnt{B} }%
}{
 \Gamma  \mathrel{ \makestatic{\vdash} }   \lambda\!  \,  \mathit{x}  \mathord{:}  \ottnt{A}   \ottsym{.}  \ottnt{M}  :  \ottnt{A}  \rightarrow  \ottnt{B} }{%
{\ottdrulename{Ts\_Lam}}{}%
}}

\newcommand{\ottdruleTsXXApp}[1]{\ottdrule[#1]{%
\ottpremise{  \Gamma  \mathrel{ \makestatic{\vdash} }  \ottnt{M_{{\mathrm{1}}}}  :  \ottnt{A}  \rightarrow  \ottnt{B}   \quad   \Gamma  \mathrel{ \makestatic{\vdash} }  \ottnt{M_{{\mathrm{2}}}}  :  \ottnt{A}  }%
}{
 \Gamma  \mathrel{ \makestatic{\vdash} }  \ottnt{M_{{\mathrm{1}}}} \, \ottnt{M_{{\mathrm{2}}}}  :  \ottnt{B} }{%
{\ottdrulename{Ts\_App}}{}%
}}

\newcommand{\ottdruleTsXXTLam}[1]{\ottdrule[#1]{%
\ottpremise{ \Gamma  \ottsym{,}   \mathit{X}  \mathord{:}  \ottnt{K}   \mathrel{ \makestatic{\vdash} }  \ottnt{M}  :  \ottnt{A} }%
}{
 \Gamma  \mathrel{ \makestatic{\vdash} }   \Lambda\!  \,  \mathit{X}  \mathord{:}  \ottnt{K}   \ottsym{.} \, \ottnt{M}  :   \text{\unboldmath$\forall\!$}  \,  \mathit{X}  \mathord{:}  \ottnt{K}   \ottsym{.} \, \ottnt{A} }{%
{\ottdrulename{Ts\_TLam}}{}%
}}

\newcommand{\ottdruleTsXXTApp}[1]{\ottdrule[#1]{%
\ottpremise{  \Gamma  \mathrel{ \makestatic{\vdash} }  \ottnt{M}  :   \text{\unboldmath$\forall\!$}  \,  \mathit{X}  \mathord{:}  \ottnt{K}   \ottsym{.} \, \ottnt{A}   \quad   \Gamma  \mathrel{ \makestatic{\vdash} }  \ottnt{B}  :  \ottnt{K}  }%
}{
 \Gamma  \mathrel{ \makestatic{\vdash} }  \ottnt{M} \, \ottnt{B}  :   \ottnt{A}    [  \ottnt{B}  /  \mathit{X}  ]   }{%
{\ottdrulename{Ts\_TApp}}{}%
}}

\newcommand{\ottdruleTsXXREmp}[1]{\ottdrule[#1]{%
\ottpremise{ \mathrel{ \makestatic{\vdash} }  \Gamma }%
}{
 \Gamma  \mathrel{ \makestatic{\vdash} }  \ottsym{\{}  \ottsym{\}}  :   [   \cdot   ]  }{%
{\ottdrulename{Ts\_REmp}}{}%
}}

\newcommand{\ottdruleTsXXRExt}[1]{\ottdrule[#1]{%
\ottpremise{  \Gamma  \mathrel{ \makestatic{\vdash} }  \ottnt{M_{{\mathrm{1}}}}  :  \ottnt{A}   \quad   \Gamma  \mathrel{ \makestatic{\vdash} }  \ottnt{M_{{\mathrm{2}}}}  :   [  \rho  ]   }%
}{
 \Gamma  \mathrel{ \makestatic{\vdash} }  \ottsym{\{}  \ell  \ottsym{=}  \ottnt{M_{{\mathrm{1}}}}  \ottsym{;}  \ottnt{M_{{\mathrm{2}}}}  \ottsym{\}}  :   [    \ell  \mathbin{:}  \ottnt{A}   ;  \rho   ]  }{%
{\ottdrulename{Ts\_RExt}}{}%
}}

\newcommand{\ottdruleTsXXRLet}[1]{\ottdrule[#1]{%
\ottpremise{  \Gamma  \mathrel{ \makestatic{\vdash} }  \ottnt{M_{{\mathrm{1}}}}  :   [    \ell  \mathbin{:}  \ottnt{A}   ;  \rho   ]    \quad   \Gamma  \ottsym{,}   \mathit{x}  \mathord{:}  \ottnt{A}   \ottsym{,}   \mathit{y}  \mathord{:}   [  \rho  ]    \mathrel{ \makestatic{\vdash} }  \ottnt{M_{{\mathrm{2}}}}  :  \ottnt{B}  }%
}{
 \Gamma  \mathrel{ \makestatic{\vdash} }  \mathsf{let} \, \ottsym{\{}  \ell  \ottsym{=}  \mathit{x}  \ottsym{;}  \mathit{y}  \ottsym{\}}  \ottsym{=}  \ottnt{M_{{\mathrm{1}}}} \, \mathsf{in} \, \ottnt{M_{{\mathrm{2}}}}  :  \ottnt{B} }{%
{\ottdrulename{Ts\_RLet}}{}%
}}

\newcommand{\ottdruleTsXXVInj}[1]{\ottdrule[#1]{%
\ottpremise{  \Gamma  \mathrel{ \makestatic{\vdash} }  \ottnt{M}  :  \ottnt{A}   \quad   \Gamma  \mathrel{ \makestatic{\vdash} }  \rho  :   \mathsf{R}   }%
}{
 \Gamma  \mathrel{ \makestatic{\vdash} }  \ell \, \ottnt{M}  :   \langle    \ell  \mathbin{:}  \ottnt{A}   ;  \rho   \rangle  }{%
{\ottdrulename{Ts\_VInj}}{}%
}}

\newcommand{\ottdruleTsXXVLift}[1]{\ottdrule[#1]{%
\ottpremise{  \Gamma  \mathrel{ \makestatic{\vdash} }  \ottnt{M}  :   \langle  \rho  \rangle    \quad   \Gamma  \mathrel{ \makestatic{\vdash} }  \ottnt{A}  :   \mathsf{T}   }%
}{
 \Gamma  \mathrel{ \makestatic{\vdash} }   \variantlift{ \ell }{ \ottnt{A} }{ \ottnt{M} }   :   \langle    \ell  \mathbin{:}  \ottnt{A}   ;  \rho   \rangle  }{%
{\ottdrulename{Ts\_VLift}}{}%
}}

\newcommand{\ottdruleTsXXVCase}[1]{\ottdrule[#1]{%
\ottpremise{  \Gamma  \mathrel{ \makestatic{\vdash} }  \ottnt{M}  :   \langle    \ell  \mathbin{:}  \ottnt{A}   ;  \rho   \rangle    \quad    \Gamma  \ottsym{,}   \mathit{x}  \mathord{:}  \ottnt{A}   \mathrel{ \makestatic{\vdash} }  \ottnt{M_{{\mathrm{1}}}}  :  \ottnt{B}   \quad   \Gamma  \ottsym{,}   \mathit{y}  \mathord{:}   \langle  \rho  \rangle    \mathrel{ \makestatic{\vdash} }  \ottnt{M_{{\mathrm{2}}}}  :  \ottnt{B}   }%
}{
 \Gamma  \mathrel{ \makestatic{\vdash} }   \mathsf{case} \,  \ottnt{M}  \,\mathsf{with}\, \langle  \ell \,  \mathit{x}   \rightarrow   \ottnt{M_{{\mathrm{1}}}}   \ottsym{;}   \mathit{y}   \rightarrow   \ottnt{M_{{\mathrm{2}}}}  \rangle   :  \ottnt{B} }{%
{\ottdrulename{Ts\_VCase}}{}%
}}

\newcommand{\ottdruleTsXXEquiv}[1]{\ottdrule[#1]{%
\ottpremise{  \Gamma  \mathrel{ \makestatic{\vdash} }  \ottnt{M}  :  \ottnt{A}   \quad   \ottnt{A}  \equiv  \ottnt{B}  \quad   \Gamma  \mathrel{ \makestatic{\vdash} }  \ottnt{B}  :   \mathsf{T}    }%
}{
 \Gamma  \mathrel{ \makestatic{\vdash} }  \ottnt{M}  :  \ottnt{B} }{%
{\ottdrulename{Ts\_Equiv}}{}%
}}

\newcommand{\ottdruleTgXXRLet}[1]{\ottdrule[#1]{%
\ottpremise{ \Gamma  \vdash  \ottnt{M_{{\mathrm{1}}}}  \ottsym{:}  \ottnt{A}  \quad   \ottnt{A}  \triangleright   [  \rho  ]   \quad   \rho \,  \triangleright _{ \ell }  \, \ottnt{B}  \ottsym{,}  \rho'  \quad  \Gamma  \ottsym{,}   \mathit{x}  \mathord{:}  \ottnt{B}   \ottsym{,}   \mathit{y}  \mathord{:}   [  \rho'  ]    \vdash  \ottnt{M_{{\mathrm{2}}}}  \ottsym{:}  \ottnt{C}   }%
}{
\Gamma  \vdash  \mathsf{let} \, \ottsym{\{}  \ell  \ottsym{=}  \mathit{x}  \ottsym{;}  \mathit{y}  \ottsym{\}}  \ottsym{=}  \ottnt{M_{{\mathrm{1}}}} \, \mathsf{in} \, \ottnt{M_{{\mathrm{2}}}}  \ottsym{:}  \ottnt{C}}{%
{\ottdrulename{Tg\_RLet}}{}%
}}

\newcommand{\ottdruleWFXXTyVar}[1]{\ottdrule[#1]{%
\ottpremise{ \Sigma  \vdash  \Gamma  \quad   \mathit{X}  \mathord{:}  \ottnt{K}  \,  \in  \, \Gamma }%
}{
\Sigma  \ottsym{;}  \Gamma  \vdash  \mathit{X}  \ottsym{:}  \ottnt{K}}{%
{\ottdrulename{WF\_TyVar}}{}%
}}

\newcommand{\ottdruleWFXXTyName}[1]{\ottdrule[#1]{%
\ottpremise{ \Sigma  \vdash  \Gamma  \quad   \alpha  \mathord{:}  \ottnt{K}   \ottsym{:=}  \ottnt{A} \,  \in  \, \Sigma }%
}{
\Sigma  \ottsym{;}  \Gamma  \vdash  \alpha  \ottsym{:}  \ottnt{K}}{%
{\ottdrulename{WF\_TyName}}{}%
}}

\newcommand{\ottdruleWFXXDyn}[1]{\ottdrule[#1]{%
\ottpremise{\Sigma  \vdash  \Gamma}%
}{
\Sigma  \ottsym{;}  \Gamma  \vdash  \star  \ottsym{:}  \ottnt{K}}{%
{\ottdrulename{WF\_Dyn}}{}%
}}

\newcommand{\ottdruleWFXXBase}[1]{\ottdrule[#1]{%
\ottpremise{\Sigma  \vdash  \Gamma}%
}{
\Sigma  \ottsym{;}  \Gamma  \vdash  \iota  \ottsym{:}   \mathsf{T} }{%
{\ottdrulename{WF\_Base}}{}%
}}

\newcommand{\ottdruleWFXXFun}[1]{\ottdrule[#1]{%
\ottpremise{ \Sigma  \ottsym{;}  \Gamma  \vdash  \ottnt{A}  \ottsym{:}   \mathsf{T}   \quad  \Sigma  \ottsym{;}  \Gamma  \vdash  \ottnt{B}  \ottsym{:}   \mathsf{T}  }%
}{
\Sigma  \ottsym{;}  \Gamma  \vdash  \ottnt{A}  \rightarrow  \ottnt{B}  \ottsym{:}   \mathsf{T} }{%
{\ottdrulename{WF\_Fun}}{}%
}}

\newcommand{\ottdruleWFXXPoly}[1]{\ottdrule[#1]{%
\ottpremise{\Sigma  \ottsym{;}  \Gamma  \ottsym{,}   \mathit{X}  \mathord{:}  \ottnt{K}   \vdash  \ottnt{A}  \ottsym{:}   \mathsf{T} }%
}{
\Sigma  \ottsym{;}  \Gamma  \vdash   \text{\unboldmath$\forall\!$}  \,  \mathit{X}  \mathord{:}  \ottnt{K}   \ottsym{.} \, \ottnt{A}  \ottsym{:}   \mathsf{T} }{%
{\ottdrulename{WF\_Poly}}{}%
}}

\newcommand{\ottdruleWFXXRecord}[1]{\ottdrule[#1]{%
\ottpremise{\Sigma  \ottsym{;}  \Gamma  \vdash  \rho  \ottsym{:}   \mathsf{R} }%
}{
\Sigma  \ottsym{;}  \Gamma  \vdash   [  \rho  ]   \ottsym{:}   \mathsf{T} }{%
{\ottdrulename{WF\_Record}}{}%
}}

\newcommand{\ottdruleWFXXVariant}[1]{\ottdrule[#1]{%
\ottpremise{\Sigma  \ottsym{;}  \Gamma  \vdash  \rho  \ottsym{:}   \mathsf{R} }%
}{
\Sigma  \ottsym{;}  \Gamma  \vdash   \langle  \rho  \rangle   \ottsym{:}   \mathsf{T} }{%
{\ottdrulename{WF\_Variant}}{}%
}}

\newcommand{\ottdruleWFXXREmp}[1]{\ottdrule[#1]{%
\ottpremise{\Sigma  \vdash  \Gamma}%
}{
\Sigma  \ottsym{;}  \Gamma  \vdash   \cdot   \ottsym{:}   \mathsf{R} }{%
{\ottdrulename{WF\_REmp}}{}%
}}

\newcommand{\ottdruleWFXXCons}[1]{\ottdrule[#1]{%
\ottpremise{ \Sigma  \ottsym{;}  \Gamma  \vdash  \ottnt{A}  \ottsym{:}   \mathsf{T}   \quad  \Sigma  \ottsym{;}  \Gamma  \vdash  \rho  \ottsym{:}   \mathsf{R}  }%
}{
\Sigma  \ottsym{;}  \Gamma  \vdash    \ell  \mathbin{:}  \ottnt{A}   ;  \rho   \ottsym{:}   \mathsf{R} }{%
{\ottdrulename{WF\_Cons}}{}%
}}

\newcommand{\ottdruleTXXVar}[1]{\ottdrule[#1]{%
\ottpremise{ \Sigma  \vdash  \Gamma  \quad   \mathit{x}  \mathord{:}  \ottnt{A}  \,  \in  \, \Gamma }%
}{
\Sigma  \ottsym{;}  \Gamma  \vdash  \mathit{x}  \ottsym{:}  \ottnt{A}}{%
{\ottdrulename{T\_Var}}{}%
}}

\newcommand{\ottdruleTXXConst}[1]{\ottdrule[#1]{%
\ottpremise{\Sigma  \vdash  \Gamma}%
}{
\Sigma  \ottsym{;}  \Gamma  \vdash  \kappa  \ottsym{:}   \mathit{ty}  (  \kappa  ) }{%
{\ottdrulename{T\_Const}}{}%
}}

\newcommand{\ottdruleTXXLam}[1]{\ottdrule[#1]{%
\ottpremise{\Sigma  \ottsym{;}  \Gamma  \ottsym{,}   \mathit{x}  \mathord{:}  \ottnt{A}   \vdash  \ottnt{e}  \ottsym{:}  \ottnt{B}}%
}{
\Sigma  \ottsym{;}  \Gamma  \vdash   \lambda\!  \,  \mathit{x}  \mathord{:}  \ottnt{A}   \ottsym{.}  \ottnt{e}  \ottsym{:}  \ottnt{A}  \rightarrow  \ottnt{B}}{%
{\ottdrulename{T\_Lam}}{}%
}}

\newcommand{\ottdruleTXXApp}[1]{\ottdrule[#1]{%
\ottpremise{ \Sigma  \ottsym{;}  \Gamma  \vdash  \ottnt{e_{{\mathrm{1}}}}  \ottsym{:}  \ottnt{A}  \rightarrow  \ottnt{B}  \quad  \Sigma  \ottsym{;}  \Gamma  \vdash  \ottnt{e_{{\mathrm{2}}}}  \ottsym{:}  \ottnt{A} }%
}{
\Sigma  \ottsym{;}  \Gamma  \vdash  \ottnt{e_{{\mathrm{1}}}} \, \ottnt{e_{{\mathrm{2}}}}  \ottsym{:}  \ottnt{B}}{%
{\ottdrulename{T\_App}}{}%
}}

\newcommand{\ottdruleTXXTLam}[1]{\ottdrule[#1]{%
\ottpremise{\Sigma  \ottsym{;}  \Gamma  \ottsym{,}   \mathit{X}  \mathord{:}  \ottnt{K}   \vdash  \ottnt{e}  \ottsym{:}  \ottnt{A}}%
}{
\Sigma  \ottsym{;}  \Gamma  \vdash    \Lambda\!  \,  \mathit{X}  \mathord{:}  \ottnt{K}   \ottsym{.}   \ottnt{e}  ::  \ottnt{A}   \ottsym{:}   \text{\unboldmath$\forall\!$}  \,  \mathit{X}  \mathord{:}  \ottnt{K}   \ottsym{.} \, \ottnt{A}}{%
{\ottdrulename{T\_TLam}}{}%
}}

\newcommand{\ottdruleTXXTApp}[1]{\ottdrule[#1]{%
\ottpremise{ \Sigma  \ottsym{;}  \Gamma  \vdash  \ottnt{e}  \ottsym{:}   \text{\unboldmath$\forall\!$}  \,  \mathit{X}  \mathord{:}  \ottnt{K}   \ottsym{.} \, \ottnt{A}  \quad  \Sigma  \ottsym{;}  \Gamma  \vdash  \ottnt{B}  \ottsym{:}  \ottnt{K} }%
}{
\Sigma  \ottsym{;}  \Gamma  \vdash  \ottnt{e} \, \ottnt{B}  \ottsym{:}   \ottnt{A}    [  \ottnt{B}  /  \mathit{X}  ]  }{%
{\ottdrulename{T\_TApp}}{}%
}}

\newcommand{\ottdruleTXXREmp}[1]{\ottdrule[#1]{%
\ottpremise{\Sigma  \vdash  \Gamma}%
}{
\Sigma  \ottsym{;}  \Gamma  \vdash  \ottsym{\{}  \ottsym{\}}  \ottsym{:}   [   \cdot   ] }{%
{\ottdrulename{T\_REmp}}{}%
}}

\newcommand{\ottdruleTXXRExt}[1]{\ottdrule[#1]{%
\ottpremise{ \Sigma  \ottsym{;}  \Gamma  \vdash  \ottnt{e_{{\mathrm{1}}}}  \ottsym{:}  \ottnt{A}  \quad  \Sigma  \ottsym{;}  \Gamma  \vdash  \ottnt{e_{{\mathrm{2}}}}  \ottsym{:}   [  \rho  ]  }%
}{
\Sigma  \ottsym{;}  \Gamma  \vdash  \ottsym{\{}  \ell  \ottsym{=}  \ottnt{e_{{\mathrm{1}}}}  \ottsym{;}  \ottnt{e_{{\mathrm{2}}}}  \ottsym{\}}  \ottsym{:}   [    \ell  \mathbin{:}  \ottnt{A}   ;  \rho   ] }{%
{\ottdrulename{T\_RExt}}{}%
}}

\newcommand{\ottdruleTXXRLet}[1]{\ottdrule[#1]{%
\ottpremise{ \Sigma  \ottsym{;}  \Gamma  \vdash  \ottnt{e_{{\mathrm{1}}}}  \ottsym{:}   [    \ell  \mathbin{:}  \ottnt{A}   ;  \rho   ]   \quad  \Sigma  \ottsym{;}  \Gamma  \ottsym{,}   \mathit{x}  \mathord{:}  \ottnt{A}   \ottsym{,}   \mathit{y}  \mathord{:}   [  \rho  ]    \vdash  \ottnt{e_{{\mathrm{2}}}}  \ottsym{:}  \ottnt{B} }%
}{
\Sigma  \ottsym{;}  \Gamma  \vdash  \mathsf{let} \, \ottsym{\{}  \ell  \ottsym{=}  \mathit{x}  \ottsym{;}  \mathit{y}  \ottsym{\}}  \ottsym{=}  \ottnt{e_{{\mathrm{1}}}} \, \mathsf{in} \, \ottnt{e_{{\mathrm{2}}}}  \ottsym{:}  \ottnt{B}}{%
{\ottdrulename{T\_RLet}}{}%
}}

\newcommand{\ottdruleTXXVInj}[1]{\ottdrule[#1]{%
\ottpremise{ \Sigma  \ottsym{;}  \Gamma  \vdash  \ottnt{e}  \ottsym{:}  \ottnt{A}  \quad  \Sigma  \ottsym{;}  \Gamma  \vdash  \rho  \ottsym{:}   \mathsf{R}  }%
}{
\Sigma  \ottsym{;}  \Gamma  \vdash  \ell \, \ottnt{e}  \ottsym{:}   \langle    \ell  \mathbin{:}  \ottnt{A}   ;  \rho   \rangle }{%
{\ottdrulename{T\_VInj}}{}%
}}

\newcommand{\ottdruleTXXVLift}[1]{\ottdrule[#1]{%
\ottpremise{ \Sigma  \ottsym{;}  \Gamma  \vdash  \ottnt{e}  \ottsym{:}   \langle  \rho  \rangle   \quad  \Sigma  \ottsym{;}  \Gamma  \vdash  \ottnt{A}  \ottsym{:}   \mathsf{T}  }%
}{
\Sigma  \ottsym{;}  \Gamma  \vdash   \variantlift{ \ell }{ \ottnt{A} }{ \ottnt{e} }   \ottsym{:}   \langle    \ell  \mathbin{:}  \ottnt{A}   ;  \rho   \rangle }{%
{\ottdrulename{T\_VLift}}{}%
}}

\newcommand{\ottdruleTXXVCase}[1]{\ottdrule[#1]{%
\ottpremise{ \Sigma  \ottsym{;}  \Gamma  \vdash  \ottnt{e}  \ottsym{:}   \langle    \ell  \mathbin{:}  \ottnt{A}   ;  \rho   \rangle   \quad   \Sigma  \ottsym{;}  \Gamma  \ottsym{,}   \mathit{x}  \mathord{:}  \ottnt{A}   \vdash  \ottnt{e_{{\mathrm{1}}}}  \ottsym{:}  \ottnt{B}  \quad  \Sigma  \ottsym{;}  \Gamma  \ottsym{,}   \mathit{y}  \mathord{:}   \langle  \rho  \rangle    \vdash  \ottnt{e_{{\mathrm{2}}}}  \ottsym{:}  \ottnt{B}  }%
}{
\Sigma  \ottsym{;}  \Gamma  \vdash   \mathsf{case} \,  \ottnt{e}  \,\mathsf{with}\, \langle  \ell \,  \mathit{x}   \rightarrow   \ottnt{e_{{\mathrm{1}}}}   \ottsym{;}   \mathit{y}   \rightarrow   \ottnt{e_{{\mathrm{2}}}}  \rangle   \ottsym{:}  \ottnt{B}}{%
{\ottdrulename{T\_VCase}}{}%
}}

\newcommand{\ottdruleTXXCast}[1]{\ottdrule[#1]{%
\ottpremise{ \Sigma  \ottsym{;}  \Gamma  \vdash  \ottnt{e}  \ottsym{:}  \ottnt{A}  \quad   \Sigma  \ottsym{;}  \Gamma  \vdash  \ottnt{B}  \ottsym{:}   \mathsf{T}   \quad  \ottnt{A}  \simeq  \ottnt{B}  }%
}{
\Sigma  \ottsym{;}  \Gamma  \vdash  \ottnt{e}  \ottsym{:}  \ottnt{A} \,  \stackrel{ \ottnt{p} }{\Rightarrow}  \ottnt{B}   \ottsym{:}  \ottnt{B}}{%
{\ottdrulename{T\_Cast}}{}%
}}

\newcommand{\ottdruleTXXConv}[1]{\ottdrule[#1]{%
\ottpremise{ \Sigma  \vdash  \Gamma  \quad   \Sigma  \ottsym{;}   \emptyset   \vdash  \ottnt{e}  \ottsym{:}  \ottnt{A}  \quad   \Sigma  \ottsym{;}   \emptyset   \vdash  \ottnt{B}  \ottsym{:}   \mathsf{T}   \quad   \Sigma   \vdash   \ottnt{A}  \prec^{ \Phi }  \ottnt{B}    }%
}{
\Sigma  \ottsym{;}  \Gamma  \vdash  \ottnt{e}  \ottsym{:}  \ottnt{A} \,  \stackrel{ \Phi }{\Rightarrow}  \ottnt{B}   \ottsym{:}  \ottnt{B}}{%
{\ottdrulename{T\_Conv}}{}%
}}

\newcommand{\ottdruleTXXBlame}[1]{\ottdrule[#1]{%
\ottpremise{\Sigma  \ottsym{;}  \Gamma  \vdash  \ottnt{A}  \ottsym{:}   \mathsf{T} }%
}{
\Sigma  \ottsym{;}  \Gamma  \vdash  \mathsf{blame} \, \ottnt{p}  \ottsym{:}  \ottnt{A}}{%
{\ottdrulename{T\_Blame}}{}%
}}

\newcommand{\ottdruleTransXXApp}[1]{\ottdrule[#1]{%
\ottpremise{ \Gamma  \vdash  \ottnt{M_{{\mathrm{1}}}}  \ottsym{:}  \ottnt{A_{{\mathrm{1}}}}  \hookrightarrow  \ottnt{e_{{\mathrm{1}}}}  \quad   \Gamma  \vdash  \ottnt{M_{{\mathrm{2}}}}  \ottsym{:}  \ottnt{A_{{\mathrm{2}}}}  \hookrightarrow  \ottnt{e_{{\mathrm{2}}}}  \quad   \ottnt{A_{{\mathrm{1}}}}  \triangleright  \ottnt{A_{{\mathrm{11}}}}  \rightarrow  \ottnt{A_{{\mathrm{12}}}}  \quad  \ottnt{A_{{\mathrm{2}}}}  \simeq  \ottnt{A_{{\mathrm{11}}}}   }%
}{
\Gamma  \vdash  \ottnt{M_{{\mathrm{1}}}} \, \ottnt{M_{{\mathrm{2}}}}  \ottsym{:}  \ottnt{A_{{\mathrm{12}}}}  \hookrightarrow  \ottsym{(}  \ottnt{e_{{\mathrm{1}}}}  \ottsym{:}  \ottnt{A_{{\mathrm{1}}}} \,  \stackrel{ \ottnt{p} }{\Rightarrow}  \ottnt{A_{{\mathrm{11}}}}  \rightarrow  \ottnt{A_{{\mathrm{12}}}}   \ottsym{)} \, \ottsym{(}  \ottnt{e_{{\mathrm{2}}}}  \ottsym{:}  \ottnt{A_{{\mathrm{2}}}} \,  \stackrel{ \ottnt{q} }{\Rightarrow}  \ottnt{A_{{\mathrm{11}}}}   \ottsym{)}}{%
{\ottdrulename{Trans\_App}}{}%
}}

\newcommand{\ottdruleEsXXRed}[1]{\ottdrule[#1]{%
\ottpremise{ \ottnt{M_{{\mathrm{1}}}}  \mathrel{ \makestatic{\rightsquigarrow} }  \ottnt{M_{{\mathrm{2}}}} }%
}{
  F  [  \ottnt{M_{{\mathrm{1}}}}  ]   \mathrel{ \makestatic{\longrightarrow} }   F  [  \ottnt{M_{{\mathrm{2}}}}  ]  }{%
{\ottdrulename{Es\_Red}}{}%
}}

\newcommand{\ottdruleEXXRed}[1]{\ottdrule[#1]{%
\ottpremise{\ottnt{e_{{\mathrm{1}}}}  \rightsquigarrow  \ottnt{e_{{\mathrm{2}}}}}%
}{
 \Sigma  \mid   \ottnt{E}  [  \ottnt{e_{{\mathrm{1}}}}  ]    \longrightarrow   \Sigma  \mid   \ottnt{E}  [  \ottnt{e_{{\mathrm{2}}}}  ]  }{%
{\ottdrulename{E\_Red}}{}%
}}

\newcommand{\ottdruleEXXBlame}[1]{\ottdrule[#1]{%
\ottpremise{\ottnt{E} \,  \not=  \,  [\,] }%
}{
 \Sigma  \mid   \ottnt{E}  [  \mathsf{blame} \, \ottnt{p}  ]    \longrightarrow   \Sigma  \mid  \mathsf{blame} \, \ottnt{p} }{%
{\ottdrulename{E\_Blame}}{}%
}}

\renewcommand{\ottdrule}[4][]{{\displaystyle\frac{\begin{array}{c}#2\end{array}}{#3}\ \ottdrulename{#4}}}

\def\syntax-type-row-name{
 \textbf{Names for types and rows} \quad \alpha
}

\def\syntaxTypeRowDef{
  \ottnt{A}, \ottnt{B}, \ottnt{C}, \ottnt{D}, \rho & ::= &
                      \ifthenelse{\boolean{show-common}}
                      {
                      \mayshadecommon{\mathit{X}}
                      \ifthen{\boolean{show-typename}}{
                      \mayshadecommon{\mid}
                      \alpha
                      }
                      \ifthen{\boolean{show-grad} \OR \boolean{show-typename}}{
                      \shadeif{\( \boolean{shade-common} \AND
                                  \NOT \boolean{show-typename} \) \OR
                               \boolean{shade-dyn}}
                              {\mid}
                      \shadeif{\boolean{shade-dyn}}{
                       \star 
                      }
                      }
                      \mayshadecommon{
                      \mid
                      \iota \mid \ottnt{A}  \rightarrow  \ottnt{B} \mid  \text{\unboldmath$\forall\!$}  \,  \mathit{X}  \mathord{:}  \ottnt{K}   \ottsym{.} \, \ottnt{A} \mid
                       [  \rho  ] 
                      }
                      }
                      {...}
                      \mayshadevariant{
                      \ifthen{\boolean{show-variant}}{
                       \mayshadecommon{\mid}
                        \langle  \rho  \rangle 
                      }
                      }
                      \ifthen{\boolean{show-common}}
                      {
                      \mayshadecommon{
                      \mid
                       \cdot  \mid   \ell  \mathbin{:}  \ottnt{A}   ;  \rho 
                      }
                      }
}

\def\syntax-type-row{
 \textbf{Types and rows} \quad \hfill
 \syntaxTypeRowDef{}
}

\def\syntaxtype{
 \csname ifshow-common\endcsname
  \multicolumn{3}{l}{
   \textbf{Variables for types and rows} \quad \mathit{X}
   \qquad
   \textbf{Kinds} \quad \ottnt{K} ::= \mayshadecommon{ \mathsf{T}  \mid  \mathsf{R} }
  } \\[1ex]
 \fi

 \csname ifshow-typename\endcsname
  \multicolumn{3}{l}{
   \textbf{Type-and-row names} \quad \alpha
  } \\[1ex]
 \fi

 \csname ifshow-common\endcsname
  \multicolumn{3}{l}{
  \textbf{Base types} \quad
  \iota  ::= \mayshadecommon{ \mathsf{bool}  \mid  \mathsf{int}  \mid ...}
  \qquad
  \textbf{Constants} \quad
  \kappa ::= \mayshadecommon{
                \mathsf{true}  \mid  \mathsf{false}  \mid
               \ottsym{0} \mid  \mathsf{+}  \mid ...
               }
  } \\[1ex]
 \fi

 \syntax-type-row{} \\[.5ex]
}

\def\syntaxMterm{
 \textbf{Terms} \hfill
  \ottnt{M}   & ::= & \ifthenelse{\boolean{show-common}}
                  {
                  \mayshadecommon{
                  \mathit{x} \mid \kappa \mid  \lambda\!  \,  \mathit{x}  \mathord{:}  \ottnt{A}   \ottsym{.}  \ottnt{M} \mid \ottnt{M_{{\mathrm{1}}}} \, \ottnt{M_{{\mathrm{2}}}} \mid
                   \Lambda\!  \,  \mathit{X}  \mathord{:}  \ottnt{K}   \ottsym{.} \, \ottnt{M} \mid \ottnt{M} \, \ottnt{A} \mid
                  }
                  \\ &&
                  \mayshadecommon{
                  \ottsym{\{}  \ottsym{\}} \mid \ottsym{\{}  \ell  \ottsym{=}  \ottnt{M_{{\mathrm{1}}}}  \ottsym{;}  \ottnt{M_{{\mathrm{2}}}}  \ottsym{\}} \mid
                  \mathsf{let} \, \ottsym{\{}  \ell  \ottsym{=}  \mathit{x}  \ottsym{;}  \mathit{y}  \ottsym{\}}  \ottsym{=}  \ottnt{M_{{\mathrm{1}}}} \, \mathsf{in} \, \ottnt{M_{{\mathrm{2}}}}
                  }
                  }
                  {...}
                  \ifthen{\boolean{show-variant}}{
                  \mid \ifthen{\boolean{show-common}}{\\ &&}
                  \ell \, \ottnt{M} \mid
                   \mathsf{case} \,  \ottnt{M}  \,\mathsf{with}\, \langle  \ell \,  \mathit{x}   \rightarrow   \ottnt{M_{{\mathrm{1}}}}   \ottsym{;}   \mathit{y}   \rightarrow   \ottnt{M_{{\mathrm{2}}}}  \rangle  \mid
                   \variantlift{ \ell }{ \ottnt{A} }{ \ottnt{M} } 
                  }
                  \\[.5ex]
}

\def\syntaxtctx{
 \ifthen{\boolean{show-common}}{
 \textbf{Typing contexts} \hfill
  \Gamma   & ::= & \mayshadecommon{
                   \emptyset  \mid \Gamma  \ottsym{,}   \mathit{x}  \mathord{:}  \ottnt{A}  \mid \Gamma  \ottsym{,}   \mathit{X}  \mathord{:}  \ottnt{K} 
                  }
 }
}

\def\syntaxWvalue{
 \multicolumn{3}{l}{
 \textbf{Values} \quad
  w       \ ::= \  \ifthenelse{\boolean{show-common}}
                       {
                       \kappa \mid  \lambda\!  \,  \mathit{x}  \mathord{:}  \ottnt{A}   \ottsym{.}  \ottnt{M} \mid  \Lambda\!  \,  \mathit{X}  \mathord{:}  \ottnt{K}   \ottsym{.} \, \ottnt{M} \mid
                       \ottsym{\{}  \ottsym{\}} \mid \ottsym{\{}  \ell  \ottsym{=}  w_{{\mathrm{1}}}  \ottsym{;}  w_{{\mathrm{2}}}  \ottsym{\}}
                       }
                       {...}
                       \ifthen{\boolean{show-variant}}{
                       \mid
                        { w }^{ \ell } 
                       \qquad
  { w }^{ \ell }  \ ::= \ \ell \, w \mid  \variantlift{ \ell }{ \ottnt{A} }{  { w }^{ \ell }  } 
                       }
 } \\[.5ex]
}

\def\syntaxFctx{
 \textbf{Evaluation contexts} \hfill
  F & ::= & \ifthenelse{\boolean{show-common}}
                {
                 [\,]  \mid F \, \ottnt{M_{{\mathrm{2}}}} \mid w_{{\mathrm{1}}} \, F \mid F \, \ottnt{A} \mid \\ &&
                \ottsym{\{}  \ell  \ottsym{=}  F  \ottsym{;}  \ottnt{M_{{\mathrm{2}}}}  \ottsym{\}} \mid \ottsym{\{}  \ell  \ottsym{=}  w_{{\mathrm{1}}}  \ottsym{;}  F  \ottsym{\}} \mid
                \mathsf{let} \, \ottsym{\{}  \ell  \ottsym{=}  \mathit{x}  \ottsym{;}  \mathit{y}  \ottsym{\}}  \ottsym{=}  F \, \mathsf{in} \, \ottnt{M_{{\mathrm{2}}}}
                }
                {...}
                \ifthen{\boolean{show-variant}}{
                \mid \ifthen{\boolean{show-common}}{\\ &&}
                \ell \, F \mid
                 \mathsf{case} \,  F  \,\mathsf{with}\, \langle  \ell \,  \mathit{x}   \rightarrow   \ottnt{M_{{\mathrm{1}}}}   \ottsym{;}   \mathit{y}   \rightarrow   \ottnt{M_{{\mathrm{2}}}}  \rangle  \mid
                 \variantlift{ \ell }{ \ottnt{A} }{ F } 
                }
                \\[.5ex]
}

\def\syntaxGty{
 \textbf{Ground types} \quad
 \mathit{G}, \mathit{H} & ::= & \ifthenelse{\boolean{show-common}}{
                          \alpha \mid \iota \mid \star  \rightarrow  \star \mid
                           [  \star  ] 
                          }{...}
                          \ifthen{\boolean{show-variant}}{
                          \mid  \langle  \star  \rangle 
                          }
}

\def\syntaxEterm{
 \textbf{Terms} \quad
 \ottnt{e}  & ::= & \ifthenelse{\boolean{show-common}}{
                \mayshadecommon{
                \mathit{x} \mid \kappa \mid  \lambda\!  \,  \mathit{x}  \mathord{:}  \ottnt{A}   \ottsym{.}  \ottnt{e} \mid \ottnt{e_{{\mathrm{1}}}} \, \ottnt{e_{{\mathrm{2}}}} \mid
                }
                  \Lambda\!  \,  \mathit{X}  \mathord{:}  \ottnt{K}   \ottsym{.}   \ottnt{e}  ::  \ottnt{A} 
                \mayshadecommon{
                \mid \ottnt{e} \, \ottnt{A} \mid
                }
                \\ &&
                \mayshadecommon{
                \ottsym{\{}  \ottsym{\}} \mid \ottsym{\{}  \ell  \ottsym{=}  \ottnt{e_{{\mathrm{1}}}}  \ottsym{;}  \ottnt{e_{{\mathrm{2}}}}  \ottsym{\}} \mid
                \mathsf{let} \, \ottsym{\{}  \ell  \ottsym{=}  \mathit{x}  \ottsym{;}  \mathit{y}  \ottsym{\}}  \ottsym{=}  \ottnt{e_{{\mathrm{1}}}} \, \mathsf{in} \, \ottnt{e_{{\mathrm{2}}}} \mid
                }
                \\ &&
                }{... \mid}
                \ifthen{\boolean{show-variant}}{
                \csname ifshade-variant\endcsname \color{gray} \fi
                \ell \, \ottnt{e} \mid
                 \mathsf{case} \,  \ottnt{e}  \,\mathsf{with}\, \langle  \ell \,  \mathit{x}   \rightarrow   \ottnt{e_{{\mathrm{1}}}}   \ottsym{;}   \mathit{y}   \rightarrow   \ottnt{e_{{\mathrm{2}}}}  \rangle  \mid
                 \variantlift{ \ell }{ \ottnt{A} }{ \ottnt{e} } 
                \ifthen{\boolean{show-common}}{\mid \\ &&}
                }
                \ifthen{\boolean{show-common}}{
                \ottnt{e}  \ottsym{:}  \ottnt{A} \,  \stackrel{ \ottnt{p} }{\Rightarrow}  \ottnt{B}  \mid \ottnt{e}  \ottsym{:}  \ottnt{A} \,  \stackrel{ \Phi }{\Rightarrow}  \ottnt{B}  \mid \mathsf{blame} \, \ottnt{p}
                }
}

\def\syntaxVvalue{
 \textbf{Values} \quad
 \ottnt{v}  & ::= & \ifthenelse{\boolean{show-common}}{
                \mayshadecommon{
                \kappa \mid  \lambda\!  \,  \mathit{x}  \mathord{:}  \ottnt{A}   \ottsym{.}  \ottnt{e} \mid
                }
                  \Lambda\!  \,  \mathit{X}  \mathord{:}  \ottnt{K}   \ottsym{.}   \ottnt{e}  ::  \ottnt{A} 
                \mayshadecommon{
                \mid
                \ottsym{\{}  \ottsym{\}} \mid \ottsym{\{}  \ell  \ottsym{=}  \ottnt{v_{{\mathrm{1}}}}  \ottsym{;}  \ottnt{v_{{\mathrm{2}}}}  \ottsym{\}} \mid
                }
                }{... \mid}
                \ifthen{\boolean{show-variant}}{
                \ell \, \ottnt{v} \mid  \variantlift{ \ell }{ \ottnt{A} }{ \ottnt{v} }  \mid
                }
                \ifthen{\boolean{show-common}}{
                \\ &&
                \ottnt{v}  \ottsym{:}  \mathit{G} \,  \stackrel{ \ottnt{p} }{\Rightarrow}  \star  \mid
                \ottnt{v}  \ottsym{:}   [  \gamma  ]  \,  \stackrel{ \ottnt{p} }{\Rightarrow}   [  \star  ]   \mid
                }
                \ifthen{\boolean{show-variant}}{
                \ottnt{v}  \ottsym{:}   \langle  \gamma  \rangle  \,  \stackrel{ \ottnt{p} }{\Rightarrow}   \langle  \star  \rangle  
                \ifthen{\boolean{show-common}}{\mid \\ &&}
                }
                \ifthen{\boolean{show-common}}{
                \ottnt{v}  \ottsym{:}  \ottnt{A} \,  \stackrel{ \ottsym{-}  \alpha }{\Rightarrow}  \alpha  \mid
                \ottnt{v}  \ottsym{:}   [  \rho  ]  \,  \stackrel{ \ottsym{-}  \alpha }{\Rightarrow}   [  \alpha  ]  
                }
                \ifthen{\boolean{show-variant}}{
                \mid
                \ottnt{v}  \ottsym{:}   \langle  \rho  \rangle  \,  \stackrel{ \ottsym{-}  \alpha }{\Rightarrow}   \langle  \alpha  \rangle  
                }
}

\def\syntaxEctx{
 \textbf{Evaluation contexts} \quad
 \ottnt{E} & ::= & \ifthenelse{\boolean{show-common}}{
               \mayshadecommon{
                [\,]  \mid \ottnt{E} \, \ottnt{e_{{\mathrm{2}}}} \mid \ottnt{v_{{\mathrm{1}}}} \, \ottnt{E} \mid \ottnt{E} \, \ottnt{A} \mid
               \ottsym{\{}  \ell  \ottsym{=}  \ottnt{E}  \ottsym{;}  \ottnt{e_{{\mathrm{2}}}}  \ottsym{\}} \mid \ottsym{\{}  \ell  \ottsym{=}  \ottnt{v_{{\mathrm{1}}}}  \ottsym{;}  \ottnt{E}  \ottsym{\}} \mid
               }
               \\ &&
               \mayshadecommon{
               \mathsf{let} \, \ottsym{\{}  \ell  \ottsym{=}  \mathit{x}  \ottsym{;}  \mathit{y}  \ottsym{\}}  \ottsym{=}  \ottnt{E} \, \mathsf{in} \, \ottnt{e_{{\mathrm{2}}}} \mid
               }
               }{... \mid}
               \ifthen{\boolean{show-variant}}{
               \ifthen{\boolean{show-common}}{\\ &&}
               \csname ifshade-variant\endcsname \color{gray} \fi
               \ell \, \ottnt{E} \mid
                \mathsf{case} \,  \ottnt{E}  \,\mathsf{with}\, \langle  \ell \,  \mathit{x}   \rightarrow   \ottnt{e_{{\mathrm{1}}}}   \ottsym{;}   \mathit{y}   \rightarrow   \ottnt{e_{{\mathrm{2}}}}  \rangle  \mid
                \variantlift{ \ell }{ \ottnt{A} }{ \ottnt{E} } 
               \ifthen{\boolean{show-common}}{\mid \\ &&}
               }
               \ifthen{\boolean{show-common}}{
               \ottnt{E}  \ottsym{:}  \ottnt{A} \,  \stackrel{ \ottnt{p} }{\Rightarrow}  \ottnt{B}  \mid \ottnt{E}  \ottsym{:}  \ottnt{A} \,  \stackrel{ \Phi }{\Rightarrow}  \ottnt{B} 
               }
}

\newcommand{\lambdab}{$\lambda$B}
\newcommand{\systemfc}{System $\text{F}_\text{C}$}
\newcommand{\valdecl}[2]{\ensuremath{\mathsf{val} \ \ {#1} \ : \ {#2}}}
\newenvironment{myfigure}[1][\unskip]{\begin{figure}[#1]\small}{\end{figure}}

\begin{document}


\title{Gradual Typing for Extensibility by Rows}


\author{Taro Sekiyama}
\orcid{0000-0001-9286-230X}             
\affiliation{
  \department{}              
  \institution{National Institute of Informatics and The Graduate University for Advanced Studies, SOKENDAI}            
  \state{Tokyo}
  \country{Japan}                    
}
\email{tsekiyama@acm.org}          

\author{Atsushi Igarashi}
\orcid{0000-0002-5143-9764}             
\affiliation{
  \institution{Kyoto University}           
  \state{Kyoto}
  \country{Japan}                   
}
\email{igarashi@kuis.kyoto-u.ac.jp}         

\begin{abstract}
 This work studies gradual typing for row types and row polymorphism.
 Key ingredients in
 this work are the \emph{dynamic row type}, which represents a statically
 unknown part of a row, and \emph{consistency for row types}, which allows
 injecting static row types into the dynamic row type and, conversely,
 projecting the dynamic row type to any static row type.  While consistency
 captures the behavior of the dynamic row type statically, it makes the
 semantics of a gradually typed language incoherent when combined with row
 equivalence which identifies row types up to field reordering.  To solve this
 problem, we develop \emph{consistent equivalence}, which characterizes
 composition of consistency and row equivalence.  Using consistent equivalence,
 we propose a polymorphic blame calculus {\interlang} for row types and row
 polymorphism.  In {\interlang}, casts perform not only run-time checking with
 the dynamic row type but also field reordering in row types.  To simplify our
 technical development for row polymorphism, we adopt \emph{scoped} labels,
 which are employed by the language Koka and are also emerging in the context of
 effect systems.  We give the formal definition of {\interlang} with these
 technical developments and prove its type soundness.  We also sketch the gradually
 typed surface language {\surfacelang} and type-preserving translation from
 {\surfacelang} to {\interlang} and discuss conservativity of {\surfacelang} over
 typing of a statically typed language with row types and row polymorphism.
\end{abstract}

\ccsdesc[500]{Software and its engineering~General programming languages}
\ccsdesc[300]{Social and professional topics~History of programming languages}

\keywords{gradual typing, row types, row polymorphism}  

\maketitle

\ifdraft
\section*{TODO}
\begin{itemize}
 \item Rename rules related to lifts.
 \item Remove ascription from {\surfacelang}
 \item WFS --> WFG (well-formedness of {\surfacelang})
 \item Write a note about a difference of definitions between paper and
       supplementary material; the material has the full definition while the
       paper defines them as a closure satisfying some of the full set of the
       inferences rules.
 \item Clarify the problem discussed in \sect{tour:row-parametricity}
       \url{https://lab8.slack.com/archives/D02MDH9MA/p1561430434066000}
 \item Clarify distinction between consistency and consistent equivalence.
 \item Prove that cast between consistently equivalent types works just as
       reordering fields.
 \item Prove conservativity.
\end{itemize}
\fi

\section{Introduction}

\subsection{Background: extensibility by rows and gradual typing}

Extensibility is a measure of how open and how adaptive software is to future
extensions or changes of system requirements.  Extensibility is important not
only for maintenance and adding new features but also for continuous, evolutionary
software engineering practices such as Agile development.  For example, consider a software system using a
database.  We may like to add a new column to a table in the database when
extending the system, and to change the name of an existing column for
refactoring.  For such changes, software should be extensible---i.e., no
modification to the existing code should be necessary except for the parts
directly influenced by the changes.

One type-based approach to software extensibility is \emph{row
types}~\cite{Wand_1987_LICS}, which address extensibility in terms of data
types.  A row type is a finite sequence $  \ell_{{\mathrm{1}}}  \mathbin{:}  \ottnt{A_{{\mathrm{1}}}}   ;  ...  ;   \ell_{\ottmv{n}}  \mathbin{:}  \ottnt{A_{\ottmv{n}}}  $ of pairs of a
label $\ell_{\ottmv{i}}$ and its type $\ottnt{A_{\ottmv{i}}}$; it captures a common form of data
types, such as record types and variant types, ubiquitous among various
programming paradigms.

\sloppy{
Row types are prominent in research on static typing and have been used in
practice in many situations.  From the outset, row types were developed for
extensibility---Wand proposed row types for achieving extensibility originating from inheritance in object-oriented
programming by records~\cite{Wand_1987_LICS,Wand_1991_IC}.  That embedding of
object-oriented features by row types is also adopted by the object system of
OCaml in a more sophisticated way~\cite{Remy/Vouillon_1998_TAPOS}.  Row types
are also able to make variant types extensible.  Extensible variant types, also
called polymorphic variants, are one of the techniques to resolve the Expression
Problem~\cite{Garrigue_2000_FOSE}, which is a litmus test to evaluate the
suitability of a language for modular software development.  Extensible
variant types also provide a theoretical foundation for exceptions open to
extension with user-defined errors.  These extensible record and variant
types with row types are implicitly or explicitly available in many languages such
as OCaml, Haskell, PureScript, Gluon, Koka, etc.  Another, more recent
application of row types is to support effect systems for algebraic effects
and handlers~\cite{Plotkin/Pretnar_2009_ESOP}, and multiple languages with such
an effect system are
emerging~\cite{Leijen_2014_MSFP,Hillerstrom/Lindley_2016_TyDe,Leijen_2017_POPL,Lindley/MacBrid/McLaughlin_2017_POPL}.
}

While the practicality of row types has been demonstrated with many
applications, strict enforcement of this static typing discipline might interfere with
rapid software development, and in such contexts a dynamic typing discipline would be
more suitable.  On the other hand, as development progresses and
software is scaled up, static typing provides more benefits, such as
extensibility in a safe manner as well as early error detection and better
maintainability.



\emph{Gradual typing}, proposed independently by Siek and Taha~\shortcite{Siek/Taha_2006_SFPW} and
Tobin-Hochstadt and Felleisen~\shortcite{Tobin-Hochstadt/Felleisen_2006_OOPSLA}, has been studied
for resolving the conflict between static and dynamic typing and for enabling
gradual, smooth evolution from fully dynamically typed code to fully statically
typed code.  Gradual typing was first proposed for higher-order
functions and later extended with various programming features such as
subtyping~\cite{Siek/Taha_2007_ECOOP,Xie/Bi/Oliveira_2018_ESOP}, parametric
polymorphism~\cite{Ahmed/Findler/Siek/Wadler_2011_POPL,Ahmed/Jamner/Siek/Wadler_2017_ICFP,Igarashi/Sekiyama/Igarashi_2017_ICFP,Xie/Bi/Oliveira_2018_ESOP,Toro/Labrada/Tanter_2019_POPL},
control
operators~\cite{Takikawa/Strickland/Tobin-Hochstadt_2013_ESOP,Sekiyama/Ueda/Igarashi_2015_APLAS},
\iffull
mutable
state~\cite{Siek/Taha_2006_SFPW,Herman/Tomb/Flanaga_2010_HOSC,Siek/Vitousek/Cimini/Tobin-Hochstadt/Garcia_2015_ESOP},
\fi
and type
inference~\cite{Siek/Vachharajani_2008_DLS,Garcia/Cimini_2015_POPL,Miyazaki/Sekiyama/Igarashi_2019_POPL}.
A key ingredient for achieving gradual evolution is the \emph{dynamic type},
denoted by $ \star $, which is the type of dynamically typed code.  The dynamic
type makes it possible to inject any statically typed values into the dynamic type
and, conversely, to project dynamically typed values to any static type with
run-time type conversions, called \emph{casts}.{\iffull\footnote{Semantics for
polymorphic gradual typing involves two kinds of conversion: one performs
run-time check and the other does
not~\cite{Ahmed/Findler/Siek/Wadler_2011_POPL,Ahmed/Jamner/Siek/Wadler_2017_ICFP}.
We call the former a cast as usual and the latter just a conversion, following
\citet{Ahmed/Jamner/Siek/Wadler_2017_ICFP}.}\fi}  Gradual type systems reflect this
semantic ability of the dynamic type to \TS{the use of ''using'' is suggested by a reviewer.} \emph{consistency}.  Consistency plays
the role of type equality in gradual typing and tells where the cast is
necessary.  Consistency is designed to be flexible enough to allow possibly successful
casts---e.g., between $ \mathsf{int} $ and $ \star $ and between $ \mathsf{int}   \rightarrow  \star$ and
$\star  \rightarrow   \mathsf{bool} $---but strict enough not to miss definitely unsafe casts,
e.g., between $ \mathsf{int} $ and $ \mathsf{bool} $ and between $ \mathsf{int}   \rightarrow  \star$ and
$ \mathsf{bool}   \rightarrow  \star$.

\subsection{Our work}

This work aims at gradual evolution between dynamically typed code and
statically typed, safely extensible code, and to this end we study gradual
typing for row types.  Key ingredients in our work are the \emph{dynamic row
type}, denoted by the same notation $ \star $ as the dynamic type, and
consistency for row types.  The dynamic row type has been proposed first by
\citet{Garcia/Clark/Tanter_2016_POPL} for making the effective use of
monomorphic record types in gradual typing, and we extend it to handle variant
types as well.  The dynamic row type intuitively represents a statically unknown
part of a row.  For example, row type $  \ell  \mathbin{:}   \mathsf{int}    ;  \star $ ensures that there is an
$\ell$ field coupled with type $ \mathsf{int} $ but it guarantees nothing about other
fields, neither their presence nor absence.  Thus, a record with that row type
\emph{must} have an $\ell$ field holding an integer value and \emph{may} have
other fields; a variant with that row type \emph{requires} consumers of the
variant to handle the case where the variant is constructed by injecting an
integer value with label $\ell$ and \emph{allows} them to handle other cases.
We define consistency for row types taking into account this intuition.
Interestingly, the dynamic row type not only enables gradual evolution of code
with record and variant types but also provides fine-grained control over
interfaces of program components, as seen in \sect{tour}.

To bring extensibility achieved by static row typing into gradual typing, we
also deal with \emph{row
polymorphism}~\cite{Wand_1987_LICS,Gaster/Jones_1996_TR},\footnote{Another major
form of polymorphism is subtyping possibly with bounded
polymorphism~\cite{Cardelli/Wegner_1985_ACMCS}.} which gives great modularity
and reusability to components with row types by enabling a type signature of an
expression to expose interesting fields and to abstract and take the remaining,
uninteresting row information as a parameter.  Introduction of row polymorphism
to gradual typing, however, gives rise to two technical issues.  The first is on
row parametricity.  To ensure row parametricity, we need to protect
polymorphically typed values from untyped code.  We resolve this issue by
applying the idea in the earlier work on polymorphic gradual
typing~\cite{Ahmed/Findler/Siek/Wadler_2011_POPL,Ahmed/Jamner/Siek/Wadler_2017_ICFP,Igarashi/Sekiyama/Igarashi_2017_ICFP,Toro/Labrada/Tanter_2019_POPL}
to row polymorphism.  The second issue is on row equivalence.  In a monomorphic
setting, we can assume that the label set is totally ordered and consider only
the canonical form of a row.  In a polymorphic setting, however, a row type
obtained by substitution for a row type variable may not be in a canonical form,
and therefore row types may be syntactically different even if they are
semantically equivalent.  A standard approach to this issue is to identify row
types up to field reordering~\cite{Gaster/Jones_1996_TR}.  However, perhaps
surprisingly, the semantics based on the earlier polymorphic gradual typing is
\emph{not} well defined in the sense that the behavior of some program changes
depending on representative row types.  To solve this problem, we develop
\emph{consistent equivalence}, which characterizes composition of consistency
and row equivalence, and incorporate it into our gradually typed language
instead of consistency and row equivalence.  Thanks to consistent equivalence,
the behavior of programs in the language---especially, the order of run-time
checks---is determined by type annotations, not by representative row types.
Thus, the behavior of a program is determined to be unique.

To ease technical development for row polymorphism, we allow duplicate labels in
a single row; such labels are also called \emph{scoped}~\cite{Leijen_2005_TFP}.
An alternative approach to row polymorphism is to assume labels in a row to be
unique and to introduce qualified types~\cite{Gaster/Jones_1996_TR} or a kind
system~\cite{Pottier/Remy_2005_ATTaPL} which assert that a row type variable can
be instantiated only with rows without some labels.  While we could give a
gradually typed language with such a restriction on row variables, it would
make the run-time checking of the language complicated.  By
contrast, row polymorphism with scoped labels does not need restriction on row
variables and simplifies our technical development.  In addition, the fact that
emerging applications of row types---i.e., effect systems for algebraic effect
handlers---adopt scoped
labels~\cite{Leijen_2014_MSFP,Leijen_2017_POPL,Lindley/MacBrid/McLaughlin_2017_POPL,Biernack/Pirog/Polesiuk/Siezkowski_2018_POPL}
motivates us to take this approach.

Employing scoped labels also enables us to use the
\emph{embedding operation}~\cite{Leijen_2005_TFP}, which embeds a variant
expression into a variant type with a wider row statically and wraps a variant
value by a dummy label dynamically.  The embedding operation was originally proposed
to align rows in variant types with a polymorphic row variable, and
its usefulness is also found in effect
systems~\cite{Leijen_2014_MSFP,Biernack/Pirog/Polesiuk/Siezkowski_2018_POPL}.
The embedding operation also plays an important role to make the type system of
our calculus syntax-directed.

Below is a summary of the contributions by this work.
\begin{itemize}
 \item We define consistency for value and row types.  While consistency
       captures the essence of the dynamic type and the dynamic row type, it is
       problematic when used together with row equivalence.  To solve the problem with
       consistency, we also give consistent equivalence.

 \item We define a polymorphic $\lambda$-calculus {\interlang} equipped with
       run-time checking by casts, row types with scoped labels, the dynamic
       row type, record and variant types, row polymorphism, and the embedding
       operation, using consistent equivalence.
       \iffull
       Our
       calculus is an extension of earlier polymorphic blame calculi
       {\lambdab}~\cite{Ahmed/Jamner/Siek/Wadler_2017_ICFP} and
       {\systemfc}~\cite{Igarashi/Sekiyama/Igarashi_2017_ICFP}.
       \fi

 \item We show that consistent equivalence characterizes composition of
       consistency and row equivalence and that {\interlang} satisfies type
       soundness.
       %
       We sketch a surface language {\surfacelang} for {\interlang}
       and type-preserving translation from {\surfacelang} to {\interlang} and
       also state conservativity of {\surfacelang} over typing of a statically
       typed language for row types and row polymorphism.
\end{itemize}

The rest of this paper is organized as follows.
\sect{tour} presents motivating examples of records and variants with the
dynamic row type.  Next, we review a statically typed language {\staticlang}
with row types and row polymorphism in \sect{static_lang}.  \sect{consistency}
defines consistency and consistent equivalence that support the dynamic row type
and \sect{blame} formalizes {\interlang}.  \sect{blame} also sketches
{\surfacelang} and translation from {\surfacelang} to {\interlang} and states
properties of {\interlang} and {\surfacelang}.  After discussing related work in
\sect{relwork}, we conclude in \sect{conclusion}.

This paper omits some parts of definitions and the details of proofs.  The full
definitions, including those of {\surfacelang} and the translation from
{\surfacelang} to {\interlang}, and the complete proofs are found in the
supplementary material.

\section{Programming with gradual typing for row types}
\label{sec:tour}

Records and variants are fundamental building blocks to represent and manipulate
data structures.  Records provide a means to put several pieces of data together
and to access them by specifying labels.  Variants enables us to do case
analysis on labels safely.  This section shows multiple motivating examples of
gradual typing for record and variant types.  The programs presented
in this section are in the surface language {\surfacelang}, but it is easy to
translate them to {\interlang}.

\subsection{Records}

\paragraph{Gradual evolution of data structures}
A trivial application of ``gradualizing'' record types is evolving shapes of
data structures gradually.  For instance, let us consider development of a
window system.  Assume that we have a function $ \mathsf{window} $ that returns the
current window frame.

When development starts with fully dynamic typing, $ \mathsf{window} $ is given the
dynamic type $ \star $.  Since a value of $ \star $ can be supposed to have any
type, we can use $ \mathsf{window} $ as a function of $ \mathsf{unit}   \rightarrow  \star$.  We also
assume that a window frame is represented by a record.  Then, for example, an
expression checking that the current window is valid can be given as follows:
\[
 \ottnt{M} {\ \ \defeq \ \ }
 \mathsf{let} \,  \mathit{w}   \ottsym{=}   \mathsf{window}  \, \ottsym{()} \, \mathsf{in} \,  \mathit{w}   \ottsym{.}   \mathit{width}  \,  \leq  \,  2560  \,  \mathrel{\mathsf{\&} }  \,  \mathit{w}   \ottsym{.}   \mathit{height}  \,  \leq  \,  1440 .
\]
which checks that the width and height of the current window frame are
valid.  Variable $ \mathit{w} $ bound to the current window is assigned type
$ \star $ and used as a record holding $ \mathit{width} $ and $ \mathit{height} $ fields
having integer values.  The static assumptions---whether $ \mathsf{window} $ is a
function and whether $ \mathit{w} $ is such a record---are checked at run time; for
example, if the $ \mathit{width} $ field has a string value, then the run-time check
for the $ \mathit{width} $ field will fail and an exception will be raised.

As development progresses, type specifications would gradually become concrete and stable.
Now, suppose that the type of $ \mathsf{window} $ is refined to be $ \mathsf{unit}   \rightarrow   [     \mathit{width}   \mathbin{:}   \mathsf{int}    ;     \mathit{height}   \mathbin{:}   \mathsf{int}    ;  \star    ] $, where $ \star $ is the \emph{dynamic
row type} and $ [  \rho  ] $ is a record type with row type $\rho$.  Thus, this
function type means that a window frame is represented by a record that holds
$ \mathit{width} $ and $ \mathit{height} $ fields with integer values \emph{surely} and, in
addition, may hold other fields.  Since this refinement is consistent with the
assumptions on $ \mathsf{window} $ and $ \mathit{w} $ in $\ottnt{M}$, the expression $\ottnt{M}$
works still without any change.  If the change is inconsistent with the
assumption---e.g., the type of $ \mathsf{window} $ is changed to $ \mathsf{unit}   \rightarrow   [     \mathit{width}   \mathbin{:}   \mathsf{str}    ;     \mathit{height}   \mathbin{:}   \mathsf{str}    ;  \star    ] $---the type system would detect the type mismatch
statically.

The dynamic row type $ \star $ left in the record type indicates a possibility
that a window frame has other field specifications which are not fixed.
This gives the ability to develop a prototype implementation rapidly.  For
example, let us consider prototype development of window drawing in the stack
order, where a window frame with lower $ \mathit{depth} $ field is drawn in front of
other windows with greater $ \mathit{depth} $ fields.  Since $ \mathsf{window} $ returns the
current window frame, it should be the topmost, i.e., its $ \mathit{depth} $ field should
be $0$.  Thus, the checking expression would be rewritten as:
\[
  \mathsf{let} \,  \mathit{w}   \ottsym{=}   \mathsf{window}  \, \ottsym{()} \, \mathsf{in} \,  \mathit{w}   \ottsym{.}   \mathit{width}  \,  \leq  \,  2560  \,  \mathrel{\mathsf{\&} }  \,  \mathit{w}   \ottsym{.}   \mathit{height}  \,  \leq  \,  1440  \,  \mathrel{\mathsf{\&} }  \,  \mathit{w}   \ottsym{.}   \mathit{depth}  \,  =  \,  0 .
\]
Here, we do not need to change the type of $ \mathsf{window} $ because the dynamic row
type allows us to \emph{suppose} the window frame to have a $ \mathit{depth} $ field.
This flexibility of the dynamic row type lets us concentrate on extending
software and avoid being bothered by type puzzles.  Once it is decided to deploy
this drawing system into production, we could opt to detect typing errors statically
and make the software safer by changing the record type to
$ [     \mathit{width}   \mathbin{:}   \mathsf{int}    ;     \mathit{height}   \mathbin{:}   \mathsf{int}    ;     \mathit{depth}   \mathbin{:}   \mathsf{int}    ;  \star     ] $.

\paragraph{Optional information.}
Record types combined with the dynamic row type are also useful to attach
optional information.  For example, let us consider a function that tests if a
given string matches a given regular expression and returns not only the testing
result of Boolean but also a matching substring if the test succeeds.
We also suppose that users have to give an option in order to make the function
return the matching substring for reducing memory consumption.
We can give such a function $ \mathsf{matching} $ the following type:
\[
 \valdecl{ \mathsf{matching} }{ [     \mathit{re}   \mathbin{:}   \mathsf{str}    ;     \mathit{match}   \mathbin{:}   \mathsf{str}    ;  \star    ]   \rightarrow   [     \mathit{res}   \mathbin{:}   \mathsf{bool}    ;  \star   ] }.
\]
The fields that appear explicitly in the argument type are \emph{mandatory}
arguments: users have to give a regular expression by the $ \mathit{re} $ field and a
string to match by the $ \mathit{match} $ field.  The dynamic row type there
corresponds to \emph{optional} arguments: in order for the function to return a
matching substring, one sets the $ \mathit{return\_sub} $ field to
$ \mathsf{true} $:\footnote{Here we assume that a language supports \emph{dynamic field
testing} on records.  We do not deal with such an operation in this paper, but
it is easy to add, like type testing on dynamically typed
values~\cite{Ahmed/Findler/Siek/Wadler_2011_POPL}.}
\[
 \ottnt{M} \defeq  \mathsf{matching}  \, \ottsym{\{}   \mathit{re}   \ottsym{=}   \texttt{"o}^{\ast}\texttt{"}   \ottsym{;}   \mathit{match}   \ottsym{=}   \texttt{"foo"}   \ottsym{;}   \mathit{return\_sub}   \ottsym{=}   \mathsf{true}   \ottsym{\}}.
\]
The return type of $ \mathsf{matching} $ means that $ \mathsf{matching} $ returns whether the
string matches the pattern by the Boolean $ \mathit{res} $ field.  The dynamic row type
in the return type enables augmenting the Boolean result with the matching
substring, if any, by the $ \mathit{substr} $ field.  Then, we can write a program that
returns the length of the matched substring (if any) or returns -1.
\[
 \mathsf{let} \, \mathit{x}  \ottsym{:}   [     \mathit{res}   \mathbin{:}   \mathsf{bool}    ;  \star   ]   \ottsym{=}  \ottnt{M} \, \mathsf{in} \,  \mathsf{if}  \  \mathit{x}  \ottsym{.}   \mathit{res}   \  \mathsf{then}  \  \ottsym{(}   \mathsf{length}  \, \mathit{x}  \ottsym{.}   \mathit{substr}   \ottsym{)}  \  \mathsf{else}  \   -\!1  
\]
$ \mathsf{matching} $ does not produce the matching substring if the
$ \mathit{return\_sub} $ field is missing or set to $ \mathsf{false} $:
\[
 \mathsf{let} \, \mathit{x}  \ottsym{:}   [     \mathit{res}   \mathbin{:}   \mathsf{bool}    ;  \star   ]   \ottsym{=}   \mathsf{matching}  \, \ottsym{\{}   \mathit{re}   \ottsym{=}   \texttt{"o}^{\ast}\texttt{"}   \ottsym{;}   \mathit{match}   \ottsym{=}   \texttt{"foo"}   \ottsym{\}} \, \mathsf{in} \, \mathit{x}  \ottsym{.}   \mathit{substr}   \longrightarrow^{*}  exception
\]
Thus, the dynamic row type can give natural and flexible type interfaces beyond
gradual evolution.

\paragraph{Dynamic data type definition.}
The dynamic row type in record types is also useful when one deals with values
whose structures are determined by external environments.  For example, loading
JSON files and constructing object-relational mappings by analyzing SQL queries
at run time are such practical applications.

\subsection{Variants}
A key operation on variants is injection, which injects values of different
types into a single type representation, a variant type $ \langle  \rho  \rangle $, by
tagging the values with labels that occur in row type $\rho$.  The injected
values can be projected to the field types of $\rho$ safely.  Variants are seen
throughout programming--their applications include enumerated types,
heterogeneous collections, and algebraic data types, sometimes together with
recursive types.


Variant types combined with the dynamic row type not only allow gradual
evolution of code with variant types but also can represent cases with
uncertainty.
%
%
For example, let us consider a function $ \mathsf{input\_event } $ that returns an input
event from users.  There are several kinds of events, such as key down, key up,
mouse move, mouse click, etc.  We can use variant types to represent what event happens
with additional information of the event (e.g., key codes if key events happen).
\[
 \valdecl{ \mathsf{input\_event } }{ \mathsf{unit}   \rightarrow   \langle     \mathit{key\_down}   \mathbin{:}   \mathsf{int}    ;     \mathit{key\_up}   \mathbin{:}   \mathsf{int}    ;  ...  ;   \cdot     \rangle }
\]
where $ \cdot $ is the empty row.
Suppose that we have to handle all key and mouse events but do not have to
handle events from other input devices such as touchscreens and gamepads.  We
could naturally imagine that $ \mathsf{input\_event } $ is changed to take optional
arguments to specify what additional events we are interested in:
\[
 \valdecl{ \mathsf{input\_event } }{ [  \star  ]   \rightarrow   \langle     \mathit{key\_down}   \mathbin{:}   \mathsf{int}    ;     \mathit{key\_up}   \mathbin{:}   \mathsf{int}    ;  ...  ;   \cdot     \rangle }.
\]
For example, if we are interested in touchscreen events as well, we would call
$ \mathsf{input\_event } $ with an additional argument to enable monitoring touchscreen
events, like:
\[
  \mathsf{input\_event }  \, \ottsym{\{}   \mathit{touch}   \ottsym{=}   \mathsf{true}   \ottsym{\}}.
\]
For the return type of $ \mathsf{input\_event } $, enumerating all possible events in the variant
type would be inconvenient from the viewpoints of both efficiency and
engineering because it seems that we have to handle even uninteresting, not
happening events.  Variant types with the dynamic row type allow us to take care
of only mandatory and interesting events by changing the type
signature of $ \mathsf{input\_event } $ as:
\[
 \valdecl{ \mathsf{input\_event } }{ [  \star  ]   \rightarrow   \langle     \mathit{key\_down}   \mathbin{:}   \mathsf{int}    ;     \mathit{key\_up}   \mathbin{:}   \mathsf{int}    ;  ...  ;  \star    \rangle }
\]
where $ \mathit{key\_down} ,  \mathit{key\_up} , ...$ are mandatory events that must be handled and
$ \star $ in the return type is for events handled only when interesting.  If we
do not have additional interesting events, we can convert $ \star $ to the empty
row $ \cdot $:
\[
  \mathsf{input\_event }  \, \ottsym{\{}  \ottsym{\}}  \ottsym{:}   \langle     \mathit{key\_down}   \mathbin{:}   \mathsf{int}    ;     \mathit{key\_up}   \mathbin{:}   \mathsf{int}    ;  ...  ;   \cdot     \rangle .
\]
If interested in touchscreen devices, we can convert $ \star $ to fields for
touchscreen events:
\[
  \mathsf{input\_event }  \, \ottsym{\{}   \mathit{touch}   \ottsym{=}   \mathsf{true}   \ottsym{\}}  \ottsym{:}   \langle     \mathit{key\_up}   \mathbin{:}   \mathsf{int}    ;  ...  ;     \mathit{touch\_start}   \mathbin{:}   \mathsf{pos}    ;     \mathit{touch\_end}   \mathbin{:}   \mathsf{pos}    ;   \cdot      \rangle 
\]
where $ \mathsf{pos} $ is the type of positions.  While we can choose
optional events by passing an optional argument and converting $ \star $, we
\emph{cannot} drop mandatory events, such as $ \mathit{key\_up} $ and
$ \mathit{key\_down} $.

Furthermore, the flexibility of the dynamic row type makes it possible to
monitor events even from devices unknown to the provider of $ \mathsf{input\_event } $.
Let us suppose that $ \mathsf{input\_event } $ supports dynamic loading of device driver
libraries to monitor events from unknown devices.  Such events could not appear
in a type signature of $ \mathsf{input\_event } $ because $ \mathsf{input\_event } $ does not know
at compile time what events will be triggered by an unknown device, though it
can know at run time by dynamic library loading.  The dynamic row type enables
users of $ \mathsf{input\_event } $ to assert what events are monitored by
$ \mathsf{input\_event } $ when a device driver is loaded.  For example, if a barcode
reader is not supported by $ \mathsf{input\_event } $ but it provides a device driver
library, we can assert that an event from the barcode reader may happen by converting
$ \star $:
\[
  \mathsf{input\_event }  \, \ottsym{\{}   \mathit{load}   \ottsym{=}   \texttt{"barcode\_lib"}   \ottsym{\}}  \ottsym{:}   \langle     \mathit{key\_up}   \mathbin{:}   \mathsf{int}    ;  ...  ;     \mathit{barcode}   \mathbin{:}   \mathsf{str}    ;   \cdot     \rangle .
\]
It would be difficult to give this flexibility only by static typing.

\setboolean{show-common}{true}
\setboolean{shade-common}{false}
\setboolean{show-variant}{true}
\setboolean{show-ectx}{true}
\setboolean{show-grad}{false}
\setboolean{show-grad-term}{false}
\setboolean{show-typename}{false}

\section{A polymorphically typed language for row types}
\label{sec:static_lang}

We start with reviewing a statically typed language {\staticlang} with row types
and row polymorphism.  Our language {\staticlang} is a variant of the language
given by \citet{Hillerstrom/Lindley/Atkey/Sivaramakrishnan_2017_FSCD}, from
which {\staticlang} differs in that it adopts scoped labels and incorporates row
equivalence as a typing rule.

\subsection{Syntax}

\begin{myfigure}[t]
 \[
\begin{array}{lrl}
 \syntaxtype{}
 \syntaxMterm{} \\[-3ex]
 \syntaxWvalue{}
 \syntaxFctx{}
 \syntaxtctx{}
\end{array}\]

 \caption{Syntax of {\staticlang}.}
 \label{fig:static_lang:syntax}
\end{myfigure}

\reffig{static_lang:syntax} defines the syntax of {\staticlang}, a statically
typed $\lambda$-calculus equipped with polymorphism, records, variants, and a
kind system to classify value types and row types.  Metavariable $\mathit{X}$ ranges
over type and row variables and $\ottnt{K}$ over kinds.  Kind $ \mathsf{T} $ is the kind
of value types, and $ \mathsf{R} $ is that of row types.  We often just say ``types''
for value types and ``rows'' for row types.  Evaluation contexts $F$ and
typing contexts $\Gamma$ are defined in a standard manner.

\paragraph{Types and rows.}
We use $\ottnt{A}$, $\ottnt{B}$, $\ottnt{C}$, and $\ottnt{D}$ to mean types and $\rho$ to mean
rows.  Types are: variables $\mathit{X}$; base types $\iota$; function types
$\ottnt{A}  \rightarrow  \ottnt{B}$; universal types $ \text{\unboldmath$\forall\!$}  \,  \mathit{X}  \mathord{:}  \ottnt{K}   \ottsym{.} \, \ottnt{A}$, where $\mathit{X}$ is bound in $\ottnt{A}$ and
it will be instantiated with inhabitants of $\ottnt{K}$; record types $ [  \rho  ] $; or variant types $ \langle  \rho  \rangle $.  Rows are variables, the empty row
$ \cdot $, or extension $\ottsym{(}    \ell  \mathbin{:}  \ottnt{A}   ;  \rho   \ottsym{)}$ of row $\rho$ with label $\ell$ and
$\ottnt{A}$.  For example, $\ottsym{(}    \ell_{{\mathrm{1}}}  \mathbin{:}   \mathsf{int}    ;    \ell_{{\mathrm{2}}}  \mathbin{:}   \mathsf{bool}    ;   \cdot     \ottsym{)}$ is a row type having two
fields, $\ell_{{\mathrm{1}}}$ with $ \mathsf{int} $ and $\ell_{{\mathrm{2}}}$ with $ \mathsf{bool} $.  Record type
$ [    \ell_{{\mathrm{1}}}  \mathbin{:}   \mathsf{int}    ;    \ell_{{\mathrm{2}}}  \mathbin{:}   \mathsf{bool}    ;   \cdot     ] $ is given to records that hold an integer
value accessed by $\ell_{{\mathrm{1}}}$ and a Boolean value accessed by $\ell_{{\mathrm{2}}}$.
Variant type $ \langle    \ell_{{\mathrm{1}}}  \mathbin{:}   \mathsf{int}    ;    \ell_{{\mathrm{2}}}  \mathbin{:}   \mathsf{bool}    ;   \cdot     \rangle $ is given to an integer value
tagged with $\ell_{{\mathrm{1}}}$ or a Boolean value tagged with $\ell_{{\mathrm{2}}}$.  Types and rows
are not distinguished by the syntax; they are by the kind system given in
\sect{static_lang:typing}.


We make a remark on scoped (i.e., duplicate) labels.  For example, scoped labels
allow row type $\ottsym{(}    \ell  \mathbin{:}   \mathsf{int}    ;    \ell  \mathbin{:}   \mathsf{bool}    ;   \cdot     \ottsym{)}$ though the same label $\ell$
occurs twice there.  Scoped labels make row polymorphism easy to use.  For
example, let us consider a function that removes field $ \ell_{{\mathrm{1}}}  \mathbin{:}  \ottnt{A} $ from a given
record and instead appends field $ \ell_{{\mathrm{2}}}  \mathbin{:}  \ottnt{B} $ to it.  A promising type of that
function would be $ \text{\unboldmath$\forall\!$}  \,  \mathit{X}  \mathord{:}   \mathsf{R}    \ottsym{.} \,  [    \ell_{{\mathrm{1}}}  \mathbin{:}  \ottnt{A}   ;  \mathit{X}   ]   \rightarrow   [    \ell_{{\mathrm{2}}}  \mathbin{:}  \ottnt{B}   ;  \mathit{X}   ] $, and, indeed,
{\staticlang} would allow it to have that type.  Similarly, a function that
handles only the case that a given variant is tagged with label $\ell_{{\mathrm{1}}}$ would
be able to have type $ \text{\unboldmath$\forall\!$}  \,  \mathit{X}  \mathord{:}   \mathsf{R}    \ottsym{.} \,  \langle    \ell_{{\mathrm{1}}}  \mathbin{:}  \ottnt{A}   ;  \mathit{X}   \rangle   \rightarrow  \ottnt{B}$ for some type $\ottnt{B}$.
These type representations are acceptable thanks to scoped labels.  In other
words, if labels in {\staticlang} were not scoped (i.e., had to be unique in a
row), {\staticlang} would not allow such type representations because row
variable $\mathit{X}$ may be instantiated with a row including a field with label
$\ell_{{\mathrm{1}}}$ or $\ell_{{\mathrm{2}}}$.

\iffull
One might consider an alternative approach that takes care about \AI{of?}
substituted row types so that they do not contain fields with $\ell_{{\mathrm{1}}}$ nor
$\ell_{{\mathrm{2}}}$.  \AI{The following sentence not easy to understand.} But this is also
unsound because the function $\mathit{f}$ may augment the record value of $ [  \mathit{X}  ] $ with any fields in the computation.  A sound alternative---and more
traditional than scoped labels---is to introduce the ``lacks''
predicate~\cite{Gaster/Jones_1996_TR}, which can state lack of some labels in a
row type, and to allow only record extension with the lacked labels.  While it
is reasonable and the literature has demonstrated its practicality~\cite{}, it
makes the metatheory of a language complicated.  Row types with scoped labels
can support row polymorphism naturally without such a complicated mechanism and
they let us concentrate on challenges in gradual typing for row polymorphism.
Nevertheless, it is an interesting direction for future work to deal with a
general treatment of row types and row
polymorphism~\cite{Morris/McKinna_2019_POPL}.
\fi

\paragraph{Terms and values.}
Terms are ranged over by $\ottnt{M}$.  In \reffig{static_lang:syntax}, the first
line for terms---i.e., variables $\mathit{x}$; constants $\kappa$; functions
$ \lambda\!  \,  \mathit{x}  \mathord{:}  \ottnt{A}   \ottsym{.}  \ottnt{M}$, where $\mathit{x}$ is bound in $\ottnt{M}$; function applications $\ottnt{M_{{\mathrm{1}}}} \, \ottnt{M_{{\mathrm{2}}}}$; type abstractions $ \Lambda\!  \,  \mathit{X}  \mathord{:}  \ottnt{K}   \ottsym{.} \, \ottnt{M}$, where $\mathit{X}$ is bound in $\ottnt{M}$; and
type applications $\ottnt{M} \, \ottnt{A}$---comes from System F~\cite{Reynolds_1974}.  The
only difference from System F is that type abstractions abstract not only over
value types but also over row types.

The second line shows operations on records: $\ottsym{\{}  \ottsym{\}}$ is the empty record; $\ottsym{\{}  \ell  \ottsym{=}  \ottnt{M_{{\mathrm{1}}}}  \ottsym{;}  \ottnt{M_{{\mathrm{2}}}}  \ottsym{\}}$ is the extension of record $\ottnt{M_{{\mathrm{2}}}}$ with $\ottnt{M}$ using label
$\ell$; and record decomposition $\mathsf{let} \, \ottsym{\{}  \ell  \ottsym{=}  \mathit{x}  \ottsym{;}  \mathit{y}  \ottsym{\}}  \ottsym{=}  \ottnt{M_{{\mathrm{1}}}} \, \mathsf{in} \, \ottnt{M_{{\mathrm{2}}}}$ decomposes the
record of $\ottnt{M_{{\mathrm{1}}}}$ into a value held by the outermost $\ell$ field and the rest
of the record and binds $\mathit{x}$ to the value and $\mathit{y}$ to the remaining record
in $\ottnt{M_{{\mathrm{2}}}}$.
These operations are fundamental enough to implement the basic operations on
records~\cite{Cardelli/Mitchell_1991_MSCS,Leijen_2005_TFP}: Extension just
corresponds to the record extension $\ottsym{\{}  \ell  \ottsym{=}  \ottnt{M_{{\mathrm{1}}}}  \ottsym{;}  \ottnt{M_{{\mathrm{2}}}}  \ottsym{\}}$; Restriction, which
removes an $\ell$ field from a record $\ottnt{M}$, is implemented by $\mathsf{let} \, \ottsym{\{}  \ell  \ottsym{=}  \mathit{x}  \ottsym{;}  \mathit{y}  \ottsym{\}}  \ottsym{=}  \ottnt{M} \, \mathsf{in} \, \mathit{y}$; Extraction (written $\ottnt{M}  \ottsym{.}  \ell$ in \sect{tour}), which extracts
the value of an $\ell$ field from a record $\ottnt{M}$, is by $\mathsf{let} \, \ottsym{\{}  \ell  \ottsym{=}  \mathit{x}  \ottsym{;}  \mathit{y}  \ottsym{\}}  \ottsym{=}  \ottnt{M} \, \mathsf{in} \, \mathit{x}$.  The notation $\ottsym{\{}  \ell_{{\mathrm{1}}}  \ottsym{=}  \ottnt{M_{{\mathrm{1}}}}  \ottsym{;}  ... \, \ottsym{;}  \ell_{\ottmv{n}}  \ottsym{=}  \ottnt{M_{\ottmv{n}}}  \ottsym{\}}$ used in \sect{tour} is an
abbreviation of $\ottsym{\{}  \ell_{{\mathrm{1}}}  \ottsym{=}  \ottnt{M_{{\mathrm{1}}}}  \ottsym{;}  \ottsym{\{} \, ... \, \ottsym{;}  \ottsym{\{}  \ell_{\ottmv{n}}  \ottsym{=}  \ottnt{M_{\ottmv{n}}}  \ottsym{;}  \ottsym{\{}  \ottsym{\}}  \ottsym{\}} \, ... \, \ottsym{\}}  \ottsym{\}}$.

Terms in the third line are for variants.  Injection $\ottsym{(}  \ell \, \ottnt{M}  \ottsym{)}$ tags the value
of $\ottnt{M}$ with $\ell$.  Case expression $ \mathsf{case} \,  \ottnt{M}  \,\mathsf{with}\, \langle  \ell \,  \mathit{x}   \rightarrow   \ottnt{M_{{\mathrm{1}}}}   \ottsym{;}   \mathit{y}   \rightarrow   \ottnt{M_{{\mathrm{2}}}}  \rangle $
(where $\mathit{x}$ and $\mathit{y}$ are bound in $\ottnt{M_{{\mathrm{1}}}}$ and $\ottnt{M_{{\mathrm{2}}}}$ respectively)
tests the variant value of $\ottnt{M}$ on $\ell$; if it is tagged with $\ell$,
$\ottnt{M_{{\mathrm{1}}}}$ will be evaluated with binding of $\mathit{x}$ to the injected value;
otherwise, $\ottnt{M_{{\mathrm{2}}}}$ will be evaluated with binding of $\mathit{y}$ to the variant.
The last is the so-called \emph{embedding operation}~\cite{Leijen_2005_TFP},
tailored to variant types with scoped labels.  Embedding $ \variantlift{ \ell }{ \ottnt{A} }{ \ottnt{M} } $
embeds the variant value of $\ottnt{M}$ into a variant type extended with field
$ \ell  \mathbin{:}  \ottnt{A} $. That is, if $\ottnt{M}$ has type $ \langle  \rho  \rangle $, the type of embedding
term $ \variantlift{ \ell }{ \ottnt{A} }{ \ottnt{M} } $ is $ \langle    \ell  \mathbin{:}  \ottnt{A}   ;  \rho   \rangle $.  The embedding
operation may seem to be just an operation to enable width subtyping.  This is the
case if it is sure that $\rho$ never contains any $\ell$ field.  However, if
$\rho$ could contain an $\ell$ field---this includes the case that $\rho$
ends with a row variable because it may be instantiated with a row holding an
$\ell$ field---it is not the case.  In such a case, the embedding operation
works as inserting a dummy field with label $\ell$, and the label $\ell$
attached by the embedding operation does not match with the label $\ell$ in a
case expression.  Instead, the case expression peels off the label given by the
embedding operation.  For instance, case expression $ \mathsf{case} \,   \variantlift{ \ell }{ \ottnt{A} }{ \ottsym{(}  \ell \, \ottnt{M}  \ottsym{)} }   \,\mathsf{with}\, \langle  \ell \,  \mathit{x}   \rightarrow   \ottnt{M_{{\mathrm{1}}}}   \ottsym{;}   \mathit{y}   \rightarrow   \ottnt{M_{{\mathrm{2}}}}  \rangle $ will be reduced to $\ottnt{M_{{\mathrm{2}}}}$ with binding of $\mathit{y}$ to the
value of $\ottsym{(}  \ell \, \ottnt{M}  \ottsym{)}$.  The embedding operation is useful especially to align
variant types containing row variables.  For example, suppose that an expression
$\ottnt{M}$ has type $ \langle  \mathit{X}  \rangle $ and consider writing a program that returns
$\ottnt{M}$ if some condition $\ottnt{M'}$ holds and, otherwise, returns $\ell \,  0 $.
We can make such a program acceptable using the embedding operation:
\[
  \mathsf{if}  \  \ottnt{M'}  \  \mathsf{then}  \   \variantlift{ \ell }{  \mathsf{int}  }{ \ottnt{M} }   \  \mathsf{else}  \  \ottsym{(}  \ell \,  0   \ottsym{)} .
\]
Without the embedding operation, the program would be rejected because
terms of $ \langle  \mathit{X}  \rangle $ could not have any variant type including an $\ell$
field.  More practical applications of the embedding operation can be found in
the literature on
effects~\cite{Leijen_2014_MSFP,Biernack/Pirog/Polesiuk/Siezkowski_2018_POPL}.

Values, ranged over $w$, are constants, functions, type abstractions, the
empty row, records holding only values, or variant values.  Variant values,
ranged over by $ { w }^{ \ell } $, are values injected with label $\ell$ or
application of the embedding operation to a variant value with $\ell$.  Note
that injection and embedding in a variant value $ { w }^{ \ell } $ shares the same
label $\ell$.

\paragraph{Notation.}
We introduce standard notions and notation.
The set of type and row variables that occur free in $\ottnt{A}$ is written
$ \mathit{ftv}  (  \ottnt{A}  ) $.  We define capture-avoiding substitution $\ottnt{M} \,  [  w  \ottsym{/}  \mathit{x}  ] $ (resp.\ %
$\ottnt{M} \,  [  \ottnt{A}  /  \mathit{X}  ] $) of $w$ (resp.\ $\ottnt{A}$) for $\mathit{x}$ (resp.\ $\mathit{X}$) in
$\ottnt{M}$ as usual.  We also write $ \ottnt{A}    [  \ottnt{B}  /  \mathit{X}  ]  $ for the capture avoiding
substitution of $\ottnt{B}$ for $\mathit{X}$ in $\ottnt{A}$.  Filling the hole of evaluation
context $F$ with term $\ottnt{M}$ is denoted by $ F  [  \ottnt{M}  ] $.  We write
$\mathit{dom} \, \ottsym{(}  \Gamma  \ottsym{)}$ for the set of variables (both $\mathit{x}$ and $\mathit{X}$) bound by
$\Gamma$.  We use similar notation for other syntax classes throughout the paper.

\subsection{Semantics}

\begin{myfigure}[t]
 \begin{flushleft}
  \textbf{Reduction rules} \quad \framebox{$ \ottnt{M_{{\mathrm{1}}}}  \mathrel{ \makestatic{\rightsquigarrow} }  \ottnt{M_{{\mathrm{2}}}} $}
\qquad
\textbf{Evaluation rule} \quad \framebox{$ \ottnt{M_{{\mathrm{1}}}}  \mathrel{ \makestatic{\longrightarrow} }  \ottnt{M_{{\mathrm{2}}}} $}
\[\begin{array}{r@{\ }c@{\ }lll}
 %
 %
 \multicolumn{5}{@{}c@{}}{
   \kappa_{{\mathrm{1}}} \, \kappa_{{\mathrm{2}}}  \mathrel{ \makestatic{\rightsquigarrow} }   \zeta  (  \kappa_{{\mathrm{1}}}  ,  \kappa_{{\mathrm{2}}}  )   \ \ \RsWoP{Const} \quad
   \ottsym{(}   \lambda\!  \,  \mathit{x}  \mathord{:}  \ottnt{A}   \ottsym{.}  \ottnt{M}  \ottsym{)} \, w  \mathrel{ \makestatic{\rightsquigarrow} }  \ottnt{M} \,  [  w  \ottsym{/}  \mathit{x}  ]   \ \ \RsWoP{Beta} \quad
    \ottsym{(}   \Lambda\!  \,  \mathit{X}  \mathord{:}  \ottnt{K}   \ottsym{.} \, \ottnt{M}  \ottsym{)} \, \ottnt{A}  \mathrel{ \makestatic{\rightsquigarrow} }  \ottnt{M} \,  [  \ottnt{A}  /  \mathit{X}  ]   \ \ \RsWoP{TyBeta}
 } \\
 \ifthen{\boolean{show-common}}{
 \mathsf{let} \, \ottsym{\{}  \ell  \ottsym{=}  \mathit{x}  \ottsym{;}  \mathit{y}  \ottsym{\}}  \ottsym{=}  w \, \mathsf{in} \, \ottnt{M_{{\mathrm{2}}}} &  \makestatic{\rightsquigarrow} 
 & \ottnt{M_{{\mathrm{2}}}} \,  [  w_{{\mathrm{1}}}  \ottsym{/}  \mathit{x}  \ottsym{,}  w_{{\mathrm{2}}}  \ottsym{/}  \mathit{y}  ]  & \text{(if $w \,  \triangleright _{ \ell }  \, w_{{\mathrm{1}}}  \ottsym{,}  w_{{\mathrm{2}}}$)}
 & \RsWoP{Record} \\
 }
 %
 %
 \ifthen{\boolean{show-variant}}{
  \variantlift{ \ell }{ \ottnt{A} }{ \ottsym{(}   { w }^{ \ell' }   \ottsym{)} }  &  \makestatic{\rightsquigarrow} 
 &  { w }^{ \ell' }  & \text{(if $\ell \,  \not=  \, \ell'$)}
 & \RsWoP{Embed} \\
  \mathsf{case} \,  \ottsym{(}  \ell \, w  \ottsym{)}  \,\mathsf{with}\, \langle  \ell \,  \mathit{x}   \rightarrow   \ottnt{M_{{\mathrm{1}}}}   \ottsym{;}   \mathit{y}   \rightarrow   \ottnt{M_{{\mathrm{2}}}}  \rangle  &  \makestatic{\rightsquigarrow} 
 & \ottnt{M_{{\mathrm{1}}}} \,  [  w  \ottsym{/}  \mathit{x}  ] 
 && \RsWoP{CaseL} \\
  \mathsf{case} \,   \variantlift{ \ell }{ \ottnt{A} }{ \ottsym{(}   { w }^{ \ell }   \ottsym{)} }   \,\mathsf{with}\, \langle  \ell \,  \mathit{x}   \rightarrow   \ottnt{M_{{\mathrm{1}}}}   \ottsym{;}   \mathit{y}   \rightarrow   \ottnt{M_{{\mathrm{2}}}}  \rangle  &  \makestatic{\rightsquigarrow} 
 & \ottnt{M_{{\mathrm{2}}}} \,  [   { w }^{ \ell }   \ottsym{/}  \mathit{y}  ] 
 && \RsWoP{CaseR1} \\
  \mathsf{case} \,   { w }^{ \ell' }   \,\mathsf{with}\, \langle  \ell \,  \mathit{x}   \rightarrow   \ottnt{M_{{\mathrm{1}}}}   \ottsym{;}   \mathit{y}   \rightarrow   \ottnt{M_{{\mathrm{2}}}}  \rangle  &  \makestatic{\rightsquigarrow} 
 & \ottnt{M_{{\mathrm{2}}}} \,  [   { w }^{ \ell' }   \ottsym{/}  \mathit{y}  ]  & \text{(if $\ell \,  \not=  \, \ell'$)}
 & \RsWoP{CaseR2} \\[1ex]
 }
  F  [  \ottnt{M_{{\mathrm{1}}}}  ]  &  \makestatic{\longrightarrow}  &  F  [  \ottnt{M_{{\mathrm{2}}}}  ]  & \text{(if $ \ottnt{M_{{\mathrm{1}}}}  \mathrel{ \makestatic{\rightsquigarrow} }  \ottnt{M_{{\mathrm{2}}}} $)} & \EsWoP{Red}
\end{array}\]

  \iffull
  \textbf{Evaluation rule} \quad \framebox{$ \ottnt{M_{{\mathrm{1}}}}  \mathrel{ \makestatic{\longrightarrow} }  \ottnt{M_{{\mathrm{2}}}} $}
\begin{center}
 $\ottdruleEsXXRed{}$
\end{center}

  \fi
 \end{flushleft}
 \caption{Semantics of {\staticlang}.}
 \label{fig:static_lang:semantics}
\end{myfigure}

The semantics of {\staticlang} is given by two relations between terms: the
reduction rule $ \makestatic{\rightsquigarrow} $ and the evaluation relation $ \makestatic{\longrightarrow} $, which are
defined by the rules in \reffig{static_lang:semantics}.

The reduction rules are shown at the top of \reffig{static_lang:semantics}.
The first three rules are standard.  Reduction of constant application $\kappa_{{\mathrm{1}}} \, \kappa_{{\mathrm{2}}}$ depends on the denotation mapping $ \zeta $, which maps a pair of
constants to the constant corresponding to their denotation.  Function and type
applications reduce to the bodies of the abstractions with substitution of
the arguments.
The rule \Rs{Record} splits a record value $w$ into $w_{{\mathrm{1}}}$, which is
associated to $\ell$, and $w_{{\mathrm{2}}}$, which is the result of removing $w_{{\mathrm{1}}}$
from record $w$.  The values $w_{{\mathrm{1}}}$ and $w_{{\mathrm{2}}}$ are obtained by splitting
function $ \triangleright _{ \ell } $ defined as follows.
\begin{defn}[Record splitting]
 \label{def:static_lang:record_split}
 $w \,  \triangleright _{ \ell }  \, w_{{\mathrm{1}}}  \ottsym{,}  w_{{\mathrm{2}}}$ is defined as follows:
 \ifpaper
 \[
  \ottsym{\{}  \ell  \ottsym{=}  w_{{\mathrm{1}}}  \ottsym{;}  w_{{\mathrm{2}}}  \ottsym{\}} \,  \triangleright _{ \ell }  \, w_{{\mathrm{1}}}  \ottsym{,}  w_{{\mathrm{2}}} \qquad
  \ottsym{\{}  \ell'  \ottsym{=}  w_{{\mathrm{1}}}  \ottsym{;}  w_{{\mathrm{2}}}  \ottsym{\}} \,  \triangleright _{ \ell }  \, w_{{\mathrm{21}}}  \ottsym{,}  \ottsym{\{}  \ell'  \ottsym{=}  w_{{\mathrm{1}}}  \ottsym{;}  w_{{\mathrm{22}}}  \ottsym{\}} {\ }
   \text{(if $\ell \,  \not=  \, \ell'$ and $w_{{\mathrm{2}}} \,  \triangleright _{ \ell }  \, w_{{\mathrm{21}}}  \ottsym{,}  w_{{\mathrm{22}}}$)}
 \]
 \else
 \[\begin{array}{lll}
  \ottsym{\{}  \ell  \ottsym{=}  w_{{\mathrm{1}}}  \ottsym{;}  w_{{\mathrm{2}}}  \ottsym{\}} \,  \triangleright _{ \ell }  \, w_{{\mathrm{1}}}  \ottsym{,}  w_{{\mathrm{2}}} \\
  \ottsym{\{}  \ell'  \ottsym{=}  w_{{\mathrm{1}}}  \ottsym{;}  w_{{\mathrm{2}}}  \ottsym{\}} \,  \triangleright _{ \ell }  \, w_{{\mathrm{21}}}  \ottsym{,}  \ottsym{\{}  \ell'  \ottsym{=}  w_{{\mathrm{1}}}  \ottsym{;}  w_{{\mathrm{22}}}  \ottsym{\}} &
   \text{(if $\ell \,  \not=  \, \ell'$ and $w_{{\mathrm{2}}} \,  \triangleright _{ \ell }  \, w_{{\mathrm{21}}}  \ottsym{,}  w_{{\mathrm{22}}}$)}
   \end{array}\]
 \fi
\end{defn}

\noindent
Then, the subsequent term
$\ottnt{M_{{\mathrm{2}}}}$ will be executed after substituting $w_{{\mathrm{1}}}$ and $w_{{\mathrm{2}}}$.

The last four reduction rules are for variants.
The first is for embedding terms, and it means that embedding is discarded if a
label of an embedding term is different from the one of the variant value
$ { w }^{ \ell } $.  This is justified by the fact that a variant type of
$ { w }^{ \ell } $ can contain any field other than $\ell$ fields and, therefore,
only retaining applications of the embedding operation with the same label
$\ell$ is important.  The other rules are for case expressions $ \mathsf{case} \,   { w }^{ \ell' }   \,\mathsf{with}\, \langle  \ell \,  \mathit{x}   \rightarrow   \ottnt{M_{{\mathrm{1}}}}   \ottsym{;}   \mathit{y}   \rightarrow   \ottnt{M_{{\mathrm{2}}}}  \rangle $.  If $ { w }^{ \ell' } $ is an injection with $\ell$,
the branch $\ottnt{M_{{\mathrm{1}}}}$ will be evaluated with substitution of the injected value
for $\mathit{x}$ \Rs{CaseL}.  If $ { w }^{ \ell' } $ is an embedding term with $\ell$,
as explained above, $\ottnt{M_{{\mathrm{2}}}}$ will be evaluated with substitution of the
underlying variant value for $\mathit{y}$ \Rs{CaseR1}.  If $\ell \,  \not=  \, \ell'$, $\ottnt{M_{{\mathrm{2}}}}$
will be evaluated with substitution of the same variant value for $\mathit{y}$
\Rs{CaseR2}.

\begin{myfigure}[t]
 \begin{flushleft}
  \textbf{Well-formedness rules (selected)} \quad
\framebox{$ \mathrel{ \makestatic{\vdash} }  \Gamma $} \quad \framebox{$ \Gamma  \mathrel{ \makestatic{\vdash} }  \ottnt{A}  :  \ottnt{K} $}
\begin{center}
 \ifthen{\boolean{show-common}}{
$\ottdruleWFsXXTyVar{}$ \hfil
\ifpaper\else
$\ottdruleWFsXXBase{}$ \hfil
$\ottdruleWFsXXFun{}$ \\[1ex]
\fi
$\ottdruleWFsXXPoly{}$ \hfil
$\ottdruleWFsXXRecord{}$
}
\ifthen{\boolean{show-variant}}{
\ifpaper \\[1ex] \else \hfil \fi
$\ottdruleWFsXXVariant{}$
}
\ifthen{\boolean{show-common}}{
\ifpaper \hfil \else \\[1ex] \fi
$\ottdruleWFsXXREmp{}$ \hfil
$\ottdruleWFsXXCons{}$
}

\end{center}

  \textbf{Typing rules} \quad
\framebox{$ \Gamma  \mathrel{ \makestatic{\vdash} }  \ottnt{M}  :  \ottnt{A} $}
\begin{center}
 \ifthen{\boolean{show-common}}{
$\ottdruleTsXXVar{}$ \hfil
$\ottdruleTsXXConst{}$ \hfil
$\ottdruleTsXXLam{}$ \\[1ex]
$\ottdruleTsXXApp{}$ \hfil
$\ottdruleTsXXTLam{}$ \\[1ex]
$\ottdruleTsXXTApp{}$ \hfil
$\ottdruleTsXXREmp{}$ \\[1ex]
$\ottdruleTsXXRExt{}$ \hfil
$\ottdruleTsXXRLet{}$ \\[1ex]
}
\ifthen{\boolean{show-variant}}{
$\ottdruleTsXXVInj{}$ \hfil
$\ottdruleTsXXVLift{}$ \\[1ex]
$\ottdruleTsXXVCase{}$ \\[1ex]
}
\ifthen{\boolean{show-common}}{
$\ottdruleTsXXEquiv{}$
}

\end{center}

 \end{flushleft}
 \caption{The type system of {\staticlang}.}
 \label{fig:static_lang:typing}
\end{myfigure}

\subsection{Type system}
\label{sec:static_lang:typing}

As other type systems for row types, the type system of {\staticlang} also
identifies row types up to reordering of fields only with \emph{distinct}
labels~\cite{Berthomieu/Sagazan_1995_TPA,Leijen_2005_TFP}.
\begin{defn}[Type-and-row equivalence]
 Type-and-row equivalence $\ottnt{A}  \equiv  \ottnt{B}$ is the smallest congruence relation
 satisfying the following rule:
 \[ \ottdruleEqXXSwap{} \]
\end{defn}
\noindent
For example, this definition deems row type $\ottsym{(}    \ell_{{\mathrm{1}}}  \mathbin{:}   \mathsf{int}    ;    \ell_{{\mathrm{2}}}  \mathbin{:}   \mathsf{bool}    ;  \rho    \ottsym{)}$
equivalent to $\ottsym{(}    \ell_{{\mathrm{2}}}  \mathbin{:}   \mathsf{bool}    ;    \ell_{{\mathrm{1}}}  \mathbin{:}   \mathsf{int}    ;  \rho    \ottsym{)}$ if and only if $\ell_{{\mathrm{1}}} \,  \not=  \, \ell_{{\mathrm{2}}}$.  The
restriction on inequality of labels is necessary for type soundness.  For
example, a record $\ottsym{\{}  \ell  \ottsym{=}   0   \ottsym{;}  \ottsym{\{}  \ell  \ottsym{=}   \mathsf{true}   \ottsym{;}  \ottsym{\{}  \ottsym{\}}  \ottsym{\}}  \ottsym{\}}$ should not be typed at $ [    \ell  \mathbin{:}   \mathsf{bool}    ;    \ell  \mathbin{:}   \mathsf{int}    ;   \cdot     ] $ because record decomposition for $\ell$ extracts the
value of the outermost $\ell$ field.

The type system of {\staticlang} is given by three judgments: well-formedness of
typing contexts $ \mathrel{ \makestatic{\vdash} }  \Gamma $, well-formedness of types $ \Gamma  \mathrel{ \makestatic{\vdash} }  \ottnt{A}  :  \ottnt{K} $, and
typing judgment $ \Gamma  \mathrel{ \makestatic{\vdash} }  \ottnt{M}  :  \ottnt{A} $.  The inference rules of these judgments are
in \reffig{static_lang:typing} (where trivial well-formedness rules are
omitted), and most of them are standard or easy to understand.  We explain only
the key rules in what follows.
The rules for well-formedness of types assign kind $ \mathsf{T} $ to value types and
$ \mathsf{R} $ to row types; the kind of a type variable is given by a typing context
\WFs{TyVar}.
The type of a constant $\kappa$ is assigned by function $ \mathit{ty} $ \Ts{Const};
we assume that the type respects the denotation of $\kappa$.
\iffull
The empty record is given the empty record type \Ts{REmp}.  The type of record
extension $\ottsym{\{}  \ell  \ottsym{=}  \ottnt{M_{{\mathrm{1}}}}  \ottsym{;}  \ottnt{M_{{\mathrm{2}}}}  \ottsym{\}}$ is composed of the type of $\ottnt{M_{{\mathrm{1}}}}$ and the record
type of $\ottnt{M_{{\mathrm{2}}}}$ \Ts{RExt}.  The typing rule \Ts{RLet} for record decomposition
$\mathsf{let} \, \ottsym{\{}  \ell  \ottsym{=}  \mathit{x}  \ottsym{;}  \mathit{y}  \ottsym{\}}  \ottsym{=}  \ottnt{M_{{\mathrm{1}}}} \, \mathsf{in} \, \ottnt{M_{{\mathrm{2}}}}$ assumes that $\mathit{x}$ is bound to a value of the
outermost $\ell$ field in record $\ottnt{M_{{\mathrm{1}}}}$ and $\mathit{y}$ is bound to the record
value obtained by removing the $\ell$ field from $\ottnt{M_{{\mathrm{1}}}}$.
\fi
Injection $\ell \, \ottnt{M}$ can be given any variant type where the first $\ell$ field
has the same type as $\ottnt{M}$.  Embedding $ \variantlift{ \ell }{ \ottnt{A} }{ \ottnt{M} } $ extends the variant
type of $\ottnt{M}$ with field $ \ell  \mathbin{:}  \ottnt{A} $.  For case expression $ \mathsf{case} \,  \ottnt{M}  \,\mathsf{with}\, \langle  \ell \,  \mathit{x}   \rightarrow   \ottnt{M_{{\mathrm{1}}}}   \ottsym{;}   \mathit{y}   \rightarrow   \ottnt{M_{{\mathrm{2}}}}  \rangle $, matched expression $\ottnt{M}$ must have a variant type holding an
$\ell$ field and branches $\ottnt{M_{{\mathrm{1}}}}$ and $\ottnt{M_{{\mathrm{2}}}}$ are typechecked under the
assumptions that $\mathit{x}$ and $\mathit{y}$ are bound to a value injected with $\ell$
and a variant value discarding the first $\ell$ field, respectively.
The last rule \Ts{Equiv} allows reordering of fields with distinct labels by
employing type-and-row equivalence.  Thanks to \Ts{Equiv}, the type system can
accept terms like:
\[
  \lambda\!  \,  \mathit{f}  \mathord{:}   \text{\unboldmath$\forall\!$}  \,  \mathit{X}  \mathord{:}   \mathsf{R}    \ottsym{.} \,  [    \ell_{{\mathrm{1}}}  \mathbin{:}   \mathsf{int}    ;  \mathit{X}   ]   \rightarrow  \ottnt{A}   \ottsym{.}  \mathit{f} \, \ottsym{(}    \ell_{{\mathrm{2}}}  \mathbin{:}   \mathsf{bool}    ;   \cdot    \ottsym{)} \, \ottsym{\{}  \ell_{{\mathrm{2}}}  \ottsym{=}   \mathsf{true}   \ottsym{;}  \ottsym{\{}  \ell_{{\mathrm{1}}}  \ottsym{=}   0   \ottsym{;}  \ottsym{\{}  \ottsym{\}}  \ottsym{\}}  \ottsym{\}} .
\]
This term would be rejected without \Ts{Equiv}, because $\mathit{f} \, \ottsym{(}    \ell_{{\mathrm{2}}}  \mathbin{:}   \mathsf{bool}    ;   \cdot    \ottsym{)}$
requires arguments of $ [    \ell_{{\mathrm{1}}}  \mathbin{:}   \mathsf{int}    ;    \ell_{{\mathrm{2}}}  \mathbin{:}   \mathsf{bool}    ;   \cdot     ] $ but the type of the
actual argument $\ottsym{\{}  \ell_{{\mathrm{2}}}  \ottsym{=}   \mathsf{true}   \ottsym{;}  \ottsym{\{}  \ell_{{\mathrm{1}}}  \ottsym{=}   0   \ottsym{;}  \ottsym{\{}  \ottsym{\}}  \ottsym{\}}  \ottsym{\}}$ is $ [    \ell_{{\mathrm{2}}}  \mathbin{:}   \mathsf{bool}    ;    \ell_{{\mathrm{1}}}  \mathbin{:}   \mathsf{int}    ;   \cdot     ] $, which is syntactically different from the type
required by $\mathit{f} \, \ottsym{(}    \ell_{{\mathrm{2}}}  \mathbin{:}   \mathsf{bool}    ;   \cdot    \ottsym{)}$.  Type-and-row equivalence makes these
two record types interchangeable and, therefore, the above function application is accepted by
giving $ [    \ell_{{\mathrm{1}}}  \mathbin{:}   \mathsf{int}    ;    \ell_{{\mathrm{2}}}  \mathbin{:}   \mathsf{bool}    ;   \cdot     ] $ to the argument record.


\setboolean{show-common}{true}
\setboolean{shade-common}{true}
\setboolean{show-variant}{true}
\setboolean{show-grad}{true}
\setboolean{show-grad-term}{true}
\setboolean{show-typename}{false}
\setbool{shade-dyn}{false}
\setbool{shade-variant}{true}

\section{Consistency and consistent equivalence}
\label{sec:consistency}

This section presents \emph{consistency}.  Consistency for row types allows the
dynamic row type to be interpreted as any row.  However, consistency does not
consider type-and-row equivalence, which is problematic when we derive a
gradually typed language from {\staticlang} with implicit type conversion by
type-and-row equivalence.  To resolve the issue on consistency, we introduce
\emph{consistent equivalence}, which characterizes composition of type-and-row
equivalence and consistency.

In this section, we consider \emph{gradual} types and rows, which are obtained
by extending static types given in \reffig{static_lang:syntax} with $ \star $,
which denote the dynamic type or the dynamic row type depending on contexts.
\[\begin{array}{lll}
 \syntaxTypeRowDef{}
  \end{array}\]
We show the kind system for the extended types in \sect{blame}.

\subsection{Consistency}

Consistency $ \sim $ is fundamental to the static aspect of gradual typing
and decides possible interaction between statically typed and dynamically typed
code.  Usually, it is defined as a binary relation between types and,
intuitively, types are consistent if casts between them could
be successful.  For example, statically typed values can be injected to the
dynamic type and, conversely, dynamically typed values could be projected to any
type (whether a cast succeeds depends on whether run-time values can
behave as the target type of the cast, though).  This is axiomatized by the
following rules.
\[
 \ottnt{A}  \sim  \star \qquad \star  \sim  \ottnt{A}
\]
Consistency is also defined so that type constructors are compatible with it.
For example, a consistency rule for function types is:
\[
 \frac{\ottnt{A_{{\mathrm{1}}}}  \sim  \ottnt{B_{{\mathrm{1}}}} \quad \ottnt{A_{{\mathrm{2}}}}  \sim  \ottnt{B_{{\mathrm{2}}}}}{\ottnt{A_{{\mathrm{1}}}}  \rightarrow  \ottnt{A_{{\mathrm{2}}}}  \sim  \ottnt{B_{{\mathrm{1}}}}  \rightarrow  \ottnt{B_{{\mathrm{2}}}}}
\]

In what follows, we discuss how to extend consistency to deal with row types and
universal types and then give its formal definition.  After that, we show issues
with consistency in designing a gradually typed language with it.  These issues
motivate us to introduce consistent equivalence.

\subsubsection{Consistency for row types}
\label{sec:consistency:consistency-row}
A trivial extension of consistency to row types is to allow relating the dynamic
row type to any row type ($\rho  \sim  \star$ and $\star  \sim  \rho$) and to add the
following compatible rules for the empty row and row extension.
\[
   \cdot   \sim   \cdot 
  \qquad\qquad
  \frac{\ottnt{A}  \sim  \ottnt{B} \quad \rho_{{\mathrm{1}}}  \sim  \rho_{{\mathrm{2}}}}{  \ell  \mathbin{:}  \ottnt{A}   ;  \rho_{{\mathrm{1}}}   \sim    \ell  \mathbin{:}  \ottnt{B}   ;  \rho_{{\mathrm{2}}} }
\]
These rules make, e.g., $\ottsym{(}    \ell  \mathbin{:}   \mathsf{int}    ;   \cdot    \ottsym{)}$ and $\ottsym{(}    \ell  \mathbin{:}  \star   ;   \cdot    \ottsym{)}$
consistent.

While necessary and reasonable, these compatibility rules are not sufficient to
contain all pairs of row types such that casts between them could
be successful.  The problem is in a case that row types to be related end with
$ \star $ (i.e., they take the form $  \ell_{{\mathrm{1}}}  \mathbin{:}  \ottnt{A_{{\mathrm{1}}}}   ;  ...  ;    \ell_{\ottmv{n}}  \mathbin{:}  \ottnt{A_{\ottmv{n}}}   ;  \star  $)
and they hold field labels distinct from those of each other.  For example, let
us consider row types $  \ell_{{\mathrm{1}}}  \mathbin{:}   \mathsf{int}    ;  \star $ and $  \ell_{{\mathrm{2}}}  \mathbin{:}   \mathsf{str}    ;  \star $ where $\ell_{{\mathrm{1}}} \,  \not=  \, \ell_{{\mathrm{2}}}$.  While these row types are not consistent only with the above extension
of consistency, it is desirable that they are consistent because casts
between record types and between variant types with these rows could be
successful.  Casts between record types $ [    \ell_{{\mathrm{1}}}  \mathbin{:}   \mathsf{int}    ;  \star   ] $ and
$ [    \ell_{{\mathrm{2}}}  \mathbin{:}   \mathsf{str}    ;  \star   ] $ could be successful because a record value of either of
them could hold both $\ell_{{\mathrm{1}}}$ and $\ell_{{\mathrm{2}}}$ fields.  Similarly, casts
between variant types $ \langle    \ell_{{\mathrm{1}}}  \mathbin{:}   \mathsf{int}    ;  \star   \rangle $ and $ \langle    \ell_{{\mathrm{2}}}  \mathbin{:}   \mathsf{str}    ;  \star   \rangle $
could be successful because they accommodate both of values injected with
$\ell_{{\mathrm{1}}}$ and $\ell_{{\mathrm{2}}}$.  It is notable that the assumption that $\ell_{{\mathrm{1}}}$ and
$\ell_{{\mathrm{2}}}$ are distinct labels is critical here.
For example, $ [    \ell_{{\mathrm{1}}}  \mathbin{:}   \mathsf{int}    ;  \star   ] $ and $ [    \ell_{{\mathrm{1}}}  \mathbin{:}   \mathsf{str}    ;  \star   ] $ should not
be consistent since the types $ \mathsf{int} $ and $ \mathsf{str} $ of their $\ell$ fields
are inconsistent.

The example showing the insufficiency of the simple extension above guides
the design of a new consistency rule for row extension: given consistent
row types $\rho_{{\mathrm{1}}}$ and $\rho_{{\mathrm{2}}}$, extension of $\rho_{{\mathrm{1}}}$ with label $\ell$
preserves consistency with $\rho_{{\mathrm{2}}}$ if $\rho_{{\mathrm{2}}}$ ends with $ \star $ and $\ell$
does not appear in $\rho_{{\mathrm{2}}}$ .  Formally:
\[ \ottdruleCXXConsL{} \]
where $\mathit{dom} \, \ottsym{(}  \rho_{{\mathrm{2}}}  \ottsym{)}$ is the set of the field labels of $\rho_{{\mathrm{2}}}$.  This rule is
justified by the intuition that: first, the occurrence of $ \star $ in $\rho_{{\mathrm{2}}}$
allows assuming that $\rho_{{\mathrm{2}}}$ could contain a field of $ \ell  \mathbin{:}  \star $; and
then the $\ell$ field can move to the head of $\rho_{{\mathrm{2}}}$ by type-and-row
equivalence since $\rho_{{\mathrm{2}}}$ is assumed not to have other $\ell$ fields.
We can apply the same discussion for extension of $\rho_{{\mathrm{2}}}$ and indeed require
consistency to satisfy the symmetric version of \Cns{ConsL}.
\iffull
\[
 \frac{\quad  \rho_{{\mathrm{1}}}  \text{ ends with }  \star  \quad \ell \,  \not\in  \, \mathit{dom} \, \ottsym{(}  \rho_{{\mathrm{1}}}  \ottsym{)} \quad \rho_{{\mathrm{1}}}  \sim  \rho_{{\mathrm{2}}} \quad}{\rho_{{\mathrm{1}}}  \sim    \ell  \mathbin{:}  \ottnt{A}   ;  \rho_{{\mathrm{2}}} }
\]
\fi
Then, row types $\ottsym{(}    \ell_{{\mathrm{1}}}  \mathbin{:}   \mathsf{int}    ;  \star   \ottsym{)}$ and
$\ottsym{(}    \ell_{{\mathrm{2}}}  \mathbin{:}   \mathsf{str}    ;  \star   \ottsym{)}$ are consistent.

\subsubsection{Consistency for universal types}
Consistency for universal types in this work follows the earlier work on
polymorphic gradual typing by \citet{Igarashi/Sekiyama/Igarashi_2017_ICFP}.
Their consistency relates a universal type not only to another universal type but also
to what they call a non-$\forall$ type (i.e., a type such that
its top type constructor is not $\forall$).  The flexibility of their
consistency enables interaction between statically typed code with polymorphism
and dynamically typed code without polymorphism.  For example, in their work,
universal type $ \text{\unboldmath$\forall\!$}  \,  \mathit{X}  \mathord{:}   \mathsf{T}    \ottsym{.} \, \mathit{X}  \rightarrow  \mathit{X}$ is consistent with non-$\forall$ type
$\star  \rightarrow  \star$.  Igarashi et al.\ present a few conditions on non-$\forall$
types to be consistent with universal types; non-$\forall$ types satisfying the
conditions are called \emph{quasi}-universal types\footnote{{Igarashi et al.}
call such types quasi-polymorphic types, but we use that term for consistent use
of terminology.} because they are not actual universal types but could behave as
such by casts.  We adjust their notion of quasi-universal
types to our setting with row types.
\begin{defn}[Quasi-universal types]
 The predicate $\mathbf{QPoly} \, \ottsym{(}  \ottnt{A}  \ottsym{)}$ is defined by: $\mathbf{QPoly} \, \ottsym{(}  \ottnt{A}  \ottsym{)}$ if and only if
 \ifpaper
 (1) $\ottnt{A}$ is none of $ \text{\unboldmath$\forall\!$}  \,  \mathit{X}  \mathord{:}  \ottnt{K}   \ottsym{.} \, \ottnt{B}$, $ \cdot $ (the empty row), and $  \ell  \mathbin{:}  \ottnt{B}   ;  \rho $
 for any $\mathit{X}$, $\ottnt{K}$, $\ottnt{B}$, $\ell$, and $\rho$; and
 (2) $ \star $ occurs somewhere in $\ottnt{A}$.
 \else
 \begin{itemize}
  \item $\ottnt{A} \,  \not=  \,  \text{\unboldmath$\forall\!$}  \,  \mathit{X}  \mathord{:}  \ottnt{K}   \ottsym{.} \, \ottnt{B}$ for any $\mathit{X}$, $\ottnt{K}$, and $\ottnt{B}$,
  \item $\ottnt{A} \,  \not=  \,  \cdot $,
  \item $\ottnt{A} \,  \not=  \,   \ell  \mathbin{:}  \ottnt{B}   ;  \rho $ for any $\ell$, $\ottnt{B}$, and $\rho$, and
  \item $ \star $ occurs somewhere in $\ottnt{A}$.
 \end{itemize}
 \fi
 Type $\ottnt{A}$ is a quasi-universal type if and only if $\mathbf{QPoly} \, \ottsym{(}  \ottnt{A}  \ottsym{)}$.
\end{defn}

Then, we introduce a consistency rule
\[ \ottdruleCXXPolyL{} \]
and its symmetric version.

We make a remark on other choices of consistency for universal types.
\citet{Ahmed/Findler/Siek/Wadler_2011_POPL,Ahmed/Jamner/Siek/Wadler_2017_ICFP}
give \emph{compatibility} instead of consistency.  Their compatibility is
designed to capture as many possibly successful casts as
possible, and, as a result, it deems even perhaps apparently incompatible
types---e.g., $ \text{\unboldmath$\forall\!$}  \,  \mathit{X}  \mathord{:}   \mathsf{T}    \ottsym{.} \, \mathit{X}  \rightarrow  \mathit{X}$ and $ \mathsf{int}   \rightarrow   \mathsf{str} $---compatible.  In return for
this great flexibility, their calculus lacks conservativity over typing of
System F, the underlying calculus of their gradually typed language (i.e., a
static typing error found by System F may not be found by their gradual type
system).  Another definition of consistency is given by
\citet{Toro/Labrada/Tanter_2019_POPL}.  Their consistency relates a universal
type only to another universal type and not to any non-$\forall$ type.  Their
gradually typed language achieves conservativity over typing of System F, but
the strict distinction between universal types and non-$\forall$ types
prevents dynamically typed code, where no type information appears, from using
polymorphic values.
\iffull
For example, let us consider a program $\ottnt{M} \defeq
 \lambda\!  \,  \mathit{f}  \mathord{:}  \star   \ottsym{.}  \mathit{f} \,  0 $, which calls function $\mathit{f}$ of the dynamic type $ \star $ with
integer $\ottsym{0}$.  In Toro et al.'s language, any application of $\ottnt{M}$ to a
polymorphic function would raise an exception:
\[
 \ottnt{M} \, \ottsym{(}   \Lambda\!  \,  \mathit{X}  \mathord{:}   \mathsf{T}    \ottsym{.} \,  \lambda\!  \,  \mathit{x}  \mathord{:}  \mathit{X}   \ottsym{.}  \mathit{x}  \ottsym{)}  \longrightarrow^{*}  \mathsf{blame} \, \ottnt{p}
\]
because polymorphic values can be used successfully only if it is applied to
types.  This means that polymorphic values cannot be used in dynamically typed
code because it should have no type information including type application.
\fi

We follow \citet{Igarashi/Sekiyama/Igarashi_2017_ICFP} because of its balance
between flexibility---it allows dynamically typed code to use polymorphic
values---and strictness---it makes a gradually typed language conservative over typing of System
F.  However, we believe that how to deal with universal types in consistency is
orthogonal to consistency for row types and that we can choose a suitable
treatment depending on cases.

\subsubsection{Formal definition}
Now, we present a formal definition of consistency.
\iffull We start with formalizing the notions used in the consistency rules for
row extension.
\begin{defn}[Labels in row]
 We define $\mathit{dom} \, \ottsym{(}  \rho  \ottsym{)}$, the set of the field labels in $\rho$, as follows.
 \[\begin{array}{lll}
  \mathit{dom} \, \ottsym{(}   \cdot   \ottsym{)}     &\defeq&  \emptyset  \\
  \mathit{dom} \, \ottsym{(}  \star  \ottsym{)}      &\defeq&  \emptyset  \\
  \mathit{dom} \, \ottsym{(}  \mathit{X}  \ottsym{)}        &\defeq&  \emptyset  \\
  \ifthen{\boolean{show-typename}}{
  \mathit{dom} \, \ottsym{(}  \alpha  \ottsym{)}        &\defeq&  \emptyset  \\
  }
  \mathit{dom} \, \ottsym{(}    \ell  \mathbin{:}  \ottnt{A}   ;  \rho   \ottsym{)} &\defeq& \mathit{dom} \, \ottsym{(}  \rho  \ottsym{)} \,  \mathbin{\cup}  \, \ottsym{\{}  \ell  \ottsym{\}} \\
   \end{array}\]
\end{defn}

\begin{defn}[Row concatenation]
 Row concatenation $\rho_{{\mathrm{1}}}  \odot  \rho_{{\mathrm{2}}}$ is defined as follows:
 \[\begin{array}{lll}
   \cdot   \odot  \rho_{{\mathrm{2}}} &\defeq& \rho_{{\mathrm{2}}} \\
  \ottsym{(}    \ell  \mathbin{:}  \ottnt{A}   ;  \rho_{{\mathrm{1}}}   \ottsym{)}  \odot  \rho_{{\mathrm{2}}} &\defeq&   \ell  \mathbin{:}  \ottnt{A}   ;  \ottsym{(}  \rho_{{\mathrm{1}}}  \odot  \rho_{{\mathrm{2}}}  \ottsym{)}  \\
   \end{array}\]
\end{defn}

\begin{defn}[Rows ending with $ \star $]
 Row type $\rho$ ends with $ \star $ if and only if
 $\rho \,  =  \, \rho'  \odot  \star$ for some $\rho'$.
\end{defn}

\AI{Do you have to define ``ending with $ \star $'' in terms of concatenation?
I thought you could define as a simple recursively defined predicate.}
\fi
We say that a relation between types is compatible if and only if it is closed
under type and row constructors.

\iffull
\begin{myfigure}[t]
 \textbf{Consistency rules} \quad
\framebox{$\ottnt{A}  \sim  \ottnt{B}$}
\begin{center}
 \ifpaper\else
 $\ottdruleCXXRefl{}$ \hfil
 \fi
 $\ottdruleCXXDynL{}$ \hfil
 \ifpaper\else
 $\ottdruleCXXDynR{}$ \\[3ex]
 $\ottdruleCXXFun{}$ \hfil
 $\ottdruleCXXPoly{}$ \\[3ex]
 \fi
 $\ottdruleCXXPolyL{}$ \hfil
 \ifpaper\else
 $\ottdruleCXXPolyR{}$ \\[3ex]
 $\ottdruleCXXRecord{}$ \hfil
 $\ottdruleCXXVariant{}$ \hfil
 $\ottdruleCXXCons{}$
 \fi
 \\[3ex]
 $\ottdruleCXXConsL{}$ \hfil
 \ifpaper\else
 $\ottdruleCXXConsR{}$ \hfil
 \fi
\end{center}

 \caption{Consistency.}
 \label{fig:grad_lang:consistency}
\end{myfigure}
\fi

\begin{defn}[Consistency]
 Consistency $\ottnt{A}  \sim  \ottnt{B}$ is the smallest compatible symmetric relation
 satisfying
 \iffull
 the rules given by \reffig{grad_lang:consistency}.
 \else
 (1) $\star  \sim  \ottnt{A}$ for any $\ottnt{A}$, (2) \Cns{ConsL}, and (3) \Cns{PolyL}.
 \fi
\end{defn}

\subsubsection{Consistency issues}
\label{sec:consistency-issue}
Consistency does not subsume type-and-row equivalence.  Thus, if a gradually
typed language employed consistency directly, it would be combined with
type-and-row equivalence, particularly in the form of composition $ \equiv  \circ
 \sim $.  However, use of that composition gives rise to two issues, which were
first found in the work on gradual typing for
subtyping~\cite{Siek/Taha_2007_ECOOP}.

The first issue is on typechecking.  A typechecking algorithm for a type system
using $ \equiv  \circ  \sim $ for type comparison would have to decide whether
given two types $\ottnt{A}$ and $\ottnt{B}$ are in $ \equiv  \circ  \sim $. Thus, it would
need to find an intermediate type $\ottnt{C}$ such that $\ottnt{A}  \equiv  \ottnt{C}$ and $\ottnt{C}  \sim  \ottnt{B}$.
But, how?  This issue may not be as serious as the case of
subtyping~\cite{Siek/Taha_2007_ECOOP} because $ \equiv $ just reorders fields in a
row, but it should be still resolved.

The second issue is more serious: incoherent semantics.  For example, let us
consider the following gradually typed term:
\[
\ottnt{M} \defeq \ottsym{\{}  \ell_{{\mathrm{1}}}  \ottsym{=}   \Lambda\!  \,  \mathit{X}  \mathord{:}  \ottnt{K}   \ottsym{.} \, \ottnt{M_{{\mathrm{1}}}}  \ottsym{;}  \ottsym{\{}  \ell_{{\mathrm{2}}}  \ottsym{=}   \Lambda\!  \,  \mathit{X}  \mathord{:}  \ottnt{K}   \ottsym{.} \, \ottnt{M_{{\mathrm{2}}}}  \ottsym{;}  \ottsym{\{}  \ottsym{\}}  \ottsym{\}}  \ottsym{\}}  \ottsym{:}  \star
\]
where we suppose that $\ell_{{\mathrm{1}}}$ and $\ell_{{\mathrm{2}}}$ are distinct, $\ottnt{M_{{\mathrm{1}}}}$ is a
divergent term, and $\ottnt{M_{{\mathrm{2}}}}$ is a term involving run-time checking that always
fails; ascription $\ottnt{M'}  \ottsym{:}  \ottnt{A}$ is a shorthand of $\ottsym{(}   \lambda\!  \,  \mathit{x}  \mathord{:}  \ottnt{A}   \ottsym{.}  \mathit{x}  \ottsym{)} \, \ottnt{M'}$.
In this example, the record value is injected into the dynamic type $ \star $.
In the course of the injection, each field value would be also injected into
$ \star $ so that it can be used in dynamically typed code.  The problem here is
that (1) under the semantics of earlier polymorphic gradually typed
languages~\cite{Ahmed/Findler/Siek/Wadler_2011_POPL,Ahmed/Jamner/Siek/Wadler_2017_ICFP,Igarashi/Sekiyama/Igarashi_2017_ICFP},
the evaluation result of $\ottnt{M}$ changes depending on which field value is
injected into $ \star $ first and (2) the use of $ \equiv  \circ  \sim $ prevents
determining the order of the injections to be unique.
Let us start with seeing the first observation.  In the semantics of earlier
work on polymorphic gradual
typing~\cite{Ahmed/Findler/Siek/Wadler_2011_POPL,Ahmed/Jamner/Siek/Wadler_2017_ICFP,Igarashi/Sekiyama/Igarashi_2017_ICFP},
injection of the type abstractions $ \Lambda\!  \,  \mathit{X}  \mathord{:}  \ottnt{K}   \ottsym{.} \, \ottnt{M_{{\mathrm{1}}}}$ and $ \Lambda\!  \,  \mathit{X}  \mathord{:}  \ottnt{K}   \ottsym{.} \, \ottnt{M_{{\mathrm{2}}}}$ into
$ \star $ reduces to terms containing $\ottnt{M_{{\mathrm{1}}}} \,  [  \star  /  \mathit{X}  ] $ and $\ottnt{M_{{\mathrm{2}}}} \,  [  \star  /  \mathit{X}  ] $ as redexes,
respectively.  Thus, if $ \Lambda\!  \,  \mathit{X}  \mathord{:}  \ottnt{K}   \ottsym{.} \, \ottnt{M_{{\mathrm{1}}}}$ is injected first, the evaluation result
would be divergence since $\ottnt{M_{{\mathrm{1}}}}$ is a divergent term; otherwise, if
$ \Lambda\!  \,  \mathit{X}  \mathord{:}  \ottnt{K}   \ottsym{.} \, \ottnt{M_{{\mathrm{2}}}}$ is first, the result would be a failure of run-time checking since
$\ottnt{M_{{\mathrm{2}}}}$ contains a failing check.
Therefore, in order for the semantics to be coherent, the order of injections of
$ \Lambda\!  \,  \mathit{X}  \mathord{:}  \ottnt{K}   \ottsym{.} \, \ottnt{M_{{\mathrm{1}}}}$ and $ \Lambda\!  \,  \mathit{X}  \mathord{:}  \ottnt{K}   \ottsym{.} \, \ottnt{M_{{\mathrm{2}}}}$ into $ \star $ has to be unique.  However, a
gradual type system employing $ \equiv  \circ  \sim $ could not determine the
order to be unique.  Why not?  In gradual typing, how to inject values into
$ \star $ is decided by instances of consistency appearing in a typing
derivation.  In the example term $\ottnt{M}$, composition $ \equiv  \circ  \sim $
would be used to compare $ [    \ell_{{\mathrm{1}}}  \mathbin{:}   \text{\unboldmath$\forall\!$}  \,  \mathit{X}  \mathord{:}  \ottnt{K}   \ottsym{.} \, \ottnt{A}   ;    \ell_{{\mathrm{2}}}  \mathbin{:}   \text{\unboldmath$\forall\!$}  \,  \mathit{X}  \mathord{:}  \ottnt{K}   \ottsym{.} \, \ottnt{B}   ;   \cdot     ] $ and $ \star $
(where $\ottnt{A}$ and $\ottnt{B}$ are types of $\ottnt{M_{{\mathrm{1}}}}$ and $\ottnt{M_{{\mathrm{2}}}}$, respectively),
and there are two possible instances of consistency to derive
$ [    \ell_{{\mathrm{1}}}  \mathbin{:}   \text{\unboldmath$\forall\!$}  \,  \mathit{X}  \mathord{:}  \ottnt{K}   \ottsym{.} \, \ottnt{A}   ;    \ell_{{\mathrm{2}}}  \mathbin{:}   \text{\unboldmath$\forall\!$}  \,  \mathit{X}  \mathord{:}  \ottnt{K}   \ottsym{.} \, \ottnt{B}   ;   \cdot     ] \mathrel{( \equiv \circ \sim )} \star $:
one is $ [    \ell_{{\mathrm{1}}}  \mathbin{:}   \text{\unboldmath$\forall\!$}  \,  \mathit{X}  \mathord{:}  \ottnt{K}   \ottsym{.} \, \ottnt{A}   ;    \ell_{{\mathrm{2}}}  \mathbin{:}   \text{\unboldmath$\forall\!$}  \,  \mathit{X}  \mathord{:}  \ottnt{K}   \ottsym{.} \, \ottnt{B}   ;   \cdot     ]   \sim  \star$ and the other is
$ [    \ell_{{\mathrm{2}}}  \mathbin{:}   \text{\unboldmath$\forall\!$}  \,  \mathit{X}  \mathord{:}  \ottnt{K}   \ottsym{.} \, \ottnt{B}   ;    \ell_{{\mathrm{1}}}  \mathbin{:}   \text{\unboldmath$\forall\!$}  \,  \mathit{X}  \mathord{:}  \ottnt{K}   \ottsym{.} \, \ottnt{A}   ;   \cdot     ]   \sim  \star$.  If a typing derivation with the
former instance is given, $ \Lambda\!  \,  \mathit{X}  \mathord{:}  \ottnt{K}   \ottsym{.} \, \ottnt{M_{{\mathrm{1}}}}$ would be injected into $ \star $ first;
otherwise, if one with the latter instance is given, $ \Lambda\!  \,  \mathit{X}  \mathord{:}  \ottnt{K}   \ottsym{.} \, \ottnt{M_{{\mathrm{2}}}}$ would be
first---thus, the evaluation result of $\ottnt{M}$ depends on which consistency
instance a given typing derivation has. Although we might be able to design
coherent semantics with respect to choice of consistency instances, we take another
approach, consistent equivalence, which seems more standard in gradual
typing~\cite{Siek/Taha_2007_ECOOP,Xie/Bi/Oliveira_2018_ESOP}.

\subsection{Consistent equivalence}

To resolve the issues on consistency, we give \emph{consistent equivalence}
$ \simeq $, which characterizes composition of consistency and type-and-row
equivalence.  Our idea is to extend the consistency rule \Cns{ConsL} for row
extension in such a way as to take into account when a label used for
extension on the left-hand side does and does not appear in a row on the
right-hand side.  (The rule \Cns{ConsL} in \sect{consistency:consistency-row} handles only the latter case.)  A promising rule
that handles only the former case is:
\[
 \frac{\rho_{{\mathrm{2}}}  \equiv    \ell  \mathbin{:}  \ottnt{B}   ;  \rho'_{{\mathrm{2}}}  \quad \ottnt{A}  \simeq  \ottnt{B} \quad \rho_{{\mathrm{1}}}  \sim  \rho'_{{\mathrm{2}}}}{  \ell  \mathbin{:}  \ottnt{A}   ;  \rho_{{\mathrm{1}}}   \simeq  \rho_{{\mathrm{2}}}}
\]
We merge this rule and \Cns{ConsL} into a single rule as follows.  First, we
split $\rho_{{\mathrm{2}}}$ into the first field labeled with $\ell$ and the remaining row;
these field and row correspond to $ \ell  \mathbin{:}  \ottnt{B} $ and $\rho'_{{\mathrm{2}}}$ in the rule above,
respectively.  Even in the case that $\rho_{{\mathrm{2}}}$ includes no field labeled with
$\ell$, \emph{if $\rho_{{\mathrm{2}}}$ ends with $ \star $}, then we can \emph{suppose} that
$\rho_{{\mathrm{2}}}$ includes a $\ell$ field because the dynamic row type can be supposed
to be any row.  Since we cannot know what type such a missing $\ell$ field has, we
regard the type as $ \star $ conservatively.  Finally, we check consistency
between $\ottnt{A}$ and the type of the $\ell$ field extracted from $\rho_{{\mathrm{2}}}$ and
between $\rho_{{\mathrm{1}}}$ and the remaining row.

The idea above is formalized by the following consistent equivalence
rule, which subsumes even the compatibility rule for row extension (shown in the beginning of \sect{consistency:consistency-row}):
\[ \ottdruleCEXXConsL{} \]
where $\rho_{{\mathrm{1}}} \,  \triangleright _{ \ell }  \, \ottnt{A}  \ottsym{,}  \rho_{{\mathrm{2}}}$ is a formalization of the ``split'' operation on row
types, defined as follows.
\begin{defn}[Row splitting]
 Row splitting $\rho_{{\mathrm{1}}} \,  \triangleright _{ \ell }  \, \ottnt{A}  \ottsym{,}  \rho_{{\mathrm{2}}}$ is defined as follows.
 \ifpaper
 \[\begin{array}{ll}
  \star \,  \triangleright _{ \ell }  \, \star  \ottsym{,}  \star \qquad
    \ell  \mathbin{:}  \ottnt{A}   ;  \rho  \,  \triangleright _{ \ell }  \, \ottnt{A}  \ottsym{,}  \rho \qquad
    \ell'  \mathbin{:}  \ottnt{B}   ;  \rho_{{\mathrm{1}}}  \,  \triangleright _{ \ell }  \, \ottnt{A}  \ottsym{,}  \ottsym{(}    \ell'  \mathbin{:}  \ottnt{B}   ;  \rho_{{\mathrm{2}}}   \ottsym{)} \quad
   \text{(if $\ell \,  \not=  \, \ell'$ and $\rho_{{\mathrm{1}}} \,  \triangleright _{ \ell }  \, \ottnt{A}  \ottsym{,}  \rho_{{\mathrm{2}}}$)}
   \end{array}\]
 \else
 \[\begin{array}{lcll}
    \ell  \mathbin{:}  \ottnt{A}   ;  \rho    &  \triangleright _{ \ell }  & \ottnt{A}  \ottsym{,}  \rho \\
    \ell'  \mathbin{:}  \ottnt{B}   ;  \rho_{{\mathrm{1}}}  &  \triangleright _{ \ell }  & \ottnt{A}  \ottsym{,}  \ottsym{(}    \ell'  \mathbin{:}  \ottnt{B}   ;  \rho_{{\mathrm{2}}}   \ottsym{)} &
   \text{(if $\ell \,  \not=  \, \ell'$ and $\rho_{{\mathrm{1}}} \,  \triangleright _{ \ell }  \, \ottnt{A}  \ottsym{,}  \rho_{{\mathrm{2}}}$)} \\
   \star        &  \triangleright _{ \ell }  & \star  \ottsym{,}  \star
   \end{array}\]
 \fi
\end{defn}

\begin{defn}[Consistent equivalence]
 Consistent equivalence $\ottnt{A}  \simeq  \ottnt{B}$ is the smallest compatible symmetric
 relation satisfying
 \iffull
 \CE{ConsL} and the rules in \reffig{grad_lang:consistency}
 except \Cns{ConsL}.
 \else
 (1) $\star  \simeq  \ottnt{A}$ for any $\ottnt{A}$, (2) \CE{ConsL}, and (3) the rule of the same form as \Cns{PolyL}.
 \fi
\end{defn}

We can confirm that consistent equivalence subsumes both consistency and
type-and-row equivalence by examples.  For example,
$  \ell_{{\mathrm{1}}}  \mathbin{:}   \mathsf{int}    ;  \star   \simeq    \ell_{{\mathrm{2}}}  \mathbin{:}   \mathsf{str}    ;  \star $ and
$\ottsym{(}    \ell_{{\mathrm{1}}}  \mathbin{:}  \ottnt{A}   ;    \ell_{{\mathrm{2}}}  \mathbin{:}  \ottnt{B}   ;   \cdot     \ottsym{)}  \simeq  \ottsym{(}    \ell_{{\mathrm{2}}}  \mathbin{:}  \ottnt{B}   ;    \ell_{{\mathrm{1}}}  \mathbin{:}  \ottnt{A}   ;   \cdot     \ottsym{)}$ are derivable if $\ell_{{\mathrm{1}}} \,  \not=  \, \ell_{{\mathrm{2}}}$.
More generally, it subsumes the composition of consistency and type-and-row
equivalence.  We can show (and indeed have shown) that $ \simeq $ coincides with
$ \equiv  \circ  \sim $, but, following \citet{Xie/Bi/Oliveira_2018_ESOP}, we prove
another form of equivalence between $ \simeq $ and combination of $ \equiv $ and
$ \sim $; the statement in this form expects us to incorporate implicit
higher-order polymorphism easily.
\begin{theorem}
 \label{thm:consistent-eq-consistent-equiv}
 $\ottnt{A}  \simeq  \ottnt{B}$ if and only if $\ottnt{A}  \equiv  \ottnt{A'}$ and $\ottnt{A'}  \sim  \ottnt{B'}$ and $\ottnt{B'}  \equiv  \ottnt{B}$
 for some $\ottnt{A'}$ and $\ottnt{B'}$.
\end{theorem}

We can develop a row-polymorphic gradually typed language easily by using consistent
equivalence (we give it in the supplementary material).  The language
does not rest on consistency and, therefore, does not cause the issues on
typechecking nor semantics raised by consistency.
A typechecking algorithm for that language does not need to infer an
intermediate type because it is enough to check if given two types are in a
single relation, consistent equivalence.  At first glance, one might consider
that it is problematic that consistent equivalence is not syntax-directed.  For
example, when we would like to show
$\ottsym{(}    \ell_{{\mathrm{1}}}  \mathbin{:}  \ottnt{A}   ;    \ell_{{\mathrm{2}}}  \mathbin{:}  \ottnt{B}   ;   \cdot     \ottsym{)}  \simeq  \ottsym{(}    \ell_{{\mathrm{2}}}  \mathbin{:}  \ottnt{B}   ;    \ell_{{\mathrm{1}}}  \mathbin{:}  \ottnt{A}   ;   \cdot     \ottsym{)}$, it may appear unclear which
rule of \CE{ConsL} and its symmetric version should be applied first.
Fortunately, either is fine, which is shown by the following inversion lemma
together with symmetry of consistent equivalence.
\begin{lemma}
 If $  \ell  \mathbin{:}  \ottnt{A}   ;  \rho_{{\mathrm{1}}}   \simeq  \rho_{{\mathrm{2}}}$,
 then $\rho_{{\mathrm{2}}} \,  \triangleright _{ \ell }  \, \ottnt{B}  \ottsym{,}  \rho'_{{\mathrm{2}}}$ and $\ottnt{A}  \simeq  \ottnt{B}$ and $\rho_{{\mathrm{1}}}  \simeq  \rho'_{{\mathrm{2}}}$.
\end{lemma}
\noindent
For semantics, use of consistent equivalence makes the typing rules
syntax-directed and, therefore, derivations for a typing judgment and instances
of consistent equivalence appearing there are determined uniquely.

\setboolean{show-common}{true}
\setboolean{shade-common}{true}
\setboolean{show-variant}{true}
\setboolean{show-grad}{true}
\setboolean{show-grad-term}{true}
\setboolean{show-typename}{true}
\setboolean{shade-dyn}{false}
\setboolean{shade-variant}{true}

\section{Blame calculus {\interlang}}
\label{sec:blame}

This section defines a polymorphic blame calculus {\interlang} equipped with row
types, record and variant types, and row polymorphism.  As earlier polymorphic
blame
calculi~\cite{Ahmed/Findler/Siek/Wadler_2011_POPL,Ahmed/Jamner/Siek/Wadler_2017_ICFP,Igarashi/Sekiyama/Igarashi_2017_ICFP,Toro/Labrada/Tanter_2019_POPL},
our calculus is designed so that parametricity holds.  In fact, our calculus is a
variant of {\lambdab} by \citet{Ahmed/Jamner/Siek/Wadler_2017_ICFP}, but it
differs from {\lambdab} in two points.  First, the behavior of casts for
universal types follow \citet{Igarashi/Sekiyama/Igarashi_2017_ICFP}.  Second,
more importantly, {\interlang} deals with casts for record and variant types.
In what follows, after defining the syntax, we show the type system of
{\interlang} and then present the semantics.

\subsection{Syntax}
\label{sec:blame:syntax}

\begin{myfigure}[t]
 \[\begin{array}{l@{\ }r@{\ }l}
 \multicolumn{3}{l}{
  \textbf{Blame labels} \quad \ottnt{p}, \ottnt{q}
  \hfill
  \textbf{Type-and-row names} \quad \alpha
  \hfill
  \textbf{Conversion labels} \quad
  \Phi \ ::= \  \mathsf{+}  \, \alpha \mid \ottsym{-}  \alpha
 } \\
 \syntax-type-row{} \\
 \syntaxGty{} \\
 \textbf{Ground row types} \quad
  \gamma & ::= & \alpha \mid  \cdot  \mid   \ell  \mathbin{:}  \star   ;  \star  \\
 \syntaxEterm{} \\
 \syntaxVvalue{} \\
 \ifpaper
 \multicolumn{3}{l}{
 \textbf{Evaluation contexts} \quad
 \ottnt{E}  ::= ... \mid \ottnt{E}  \ottsym{:}  \ottnt{A} \,  \stackrel{ \ottnt{p} }{\Rightarrow}  \ottnt{B}  \mid \ottnt{E}  \ottsym{:}  \ottnt{A} \,  \stackrel{ \Phi }{\Rightarrow}  \ottnt{B}  \qquad
 \textbf{Name stores} \quad
 \Sigma ::=  \emptyset  \mid \Sigma  \ottsym{,}   \alpha  \mathord{:}  \ottnt{K}   \ottsym{:=}  \ottnt{A}
 } \\
 \else
 \syntaxEctx{} \\
 \textbf{Name stores} \quad
 \Sigma & ::= &  \emptyset  \mid \Sigma  \ottsym{,}   \alpha  \mathord{:}  \ottnt{K}   \ottsym{:=}  \ottnt{A}
 \fi
\end{array}\]

 \caption{Syntax of {\interlang}.}
 \label{fig:blame:syntax}
\end{myfigure}

The syntax of {\interlang} is presented in \reffig{blame:syntax}, where the
parts overlapping with that of {\staticlang} are displayed in gray.  To explain
some extended parts, we first review run-time enforcement of parametricity by
\citet{Ahmed/Findler/Siek/Wadler_2011_POPL,Ahmed/Jamner/Siek/Wadler_2017_ICFP}.
After that, we detail the extended syntax of {\interlang}.

\subsubsection{Run-time enforcement of parametricity.}
\citet{Ahmed/Findler/Siek/Wadler_2011_POPL} found that type application with
normal substitution-based semantics breaks parametricity.
\iffull
For example, let us consider
$ \mathsf{Id_{int} }  \defeq  \Lambda\!  \,  \mathit{X}  \mathord{:}   \mathsf{T}    \ottsym{.} \,  \lambda\!  \,  \mathit{x}  \mathord{:}  \mathit{X}   \ottsym{.}  \ottsym{(}  \mathit{x}  \ottsym{:}  \star  \ottsym{)}  \ottsym{:}   \mathsf{int} $ of $ \text{\unboldmath$\forall\!$}  \,  \mathit{X}  \mathord{:}   \mathsf{T}    \ottsym{.} \, \mathit{X}  \rightarrow   \mathsf{int} $.
Parametricity indicates that $ \mathsf{Id_{int} } $ behaves uniformly whatever type is
substituted for $\mathit{X}$.  However, if type application reduces as usual (i.e.,
$\ottsym{(}   \Lambda\!  \,  \mathit{X}  \mathord{:}   \mathsf{T}    \ottsym{.} \, \ottnt{e}  \ottsym{)} \, \ottnt{A}  \longrightarrow   \ottnt{e}    [  \ottnt{A}  /  \mathit{X}  ]  $), the behavior of $ \mathsf{Id_{int} } $ would change
depending on whether $\mathit{X}$ is instantiated with $ \mathsf{int} $ or other types:
\[\begin{array}{lllllll}
  \mathsf{Id_{int} }  \,  \mathsf{int}  \,  0      & \longrightarrow & \ottsym{(}    \lambda\!  \,  \mathit{x}  \mathord{:}   \mathsf{int}    \ottsym{.}  \ottsym{(}   \mathit{x}  :  \star   \ottsym{)}  :   \mathsf{int}    \ottsym{)} \,  0 
  & \longrightarrow &  \ottsym{(}    0   :  \star   \ottsym{)}  :   \mathsf{int}      &  \longrightarrow^{*}  & \ottsym{0} \\
  \mathsf{Id_{int} }  \,  \mathsf{bool}  \,  \mathsf{true}  & \longrightarrow & \ottsym{(}    \lambda\!  \,  \mathit{x}  \mathord{:}   \mathsf{bool}    \ottsym{.}  \ottsym{(}   \mathit{x}  :  \star   \ottsym{)}  :   \mathsf{int}    \ottsym{)} \,  \mathsf{true} 
  & \longrightarrow &  \ottsym{(}    \mathsf{true}   :  \star   \ottsym{)}  :   \mathsf{int}   &  \longrightarrow^{*}  & \mathsf{blame} \, \ottnt{p} \\
  \end{array}
\]
The latter expression raises an exception because the Boolean value $ \mathsf{true} $
cannot behave as type $ \mathsf{int} $.  This example shows that polymorphic blame
calculi with normal substitution-based semantics lacks parametricity.
\fi
To recover parametricity in gradual typing,
\citet{Ahmed/Jamner/Siek/Wadler_2017_ICFP} give a semantics that type
application $\ottsym{(}   \Lambda\!  \,  \mathit{X}  \mathord{:}   \mathsf{T}    \ottsym{.} \, \ottnt{e}  \ottsym{)} \, \ottnt{A}$ generates a fresh type name $\alpha$ and
substitutes $\alpha$ for $\mathit{X}$ in $\ottnt{e}$, where type name $\alpha$ works like
an abstract, ``fresh base type'': if a value of type $\alpha$ is injected to the
dynamic type, the resulting value can be projected successfully only to $\alpha$ and
projection to other types always fails.  While abstract inside $\ottnt{e}$, $\alpha$
should be visible as $\ottnt{A}$ outside $\ottnt{e}$.
\citet{Ahmed/Jamner/Siek/Wadler_2017_ICFP} control such revelation and
concealment of actual type information $\ottnt{A}$ of $\alpha$ by \emph{explicit type
conversion}.  With a global store mapping $\alpha$ to $\ottnt{A}$, conversion $\ottnt{e}  \ottsym{:}  \ottnt{B} \,  \stackrel{ \ottsym{+}  \alpha }{\Rightarrow}  \ottnt{C} $ reveals actual type $\ottnt{A}$ of $\alpha$ in type $\ottnt{B}$ of term
$\ottnt{e}$.  By contrast, conversion $\ottnt{e}  \ottsym{:}  \ottnt{B} \,  \stackrel{ \ottsym{-}  \alpha }{\Rightarrow}  \ottnt{C} $ conceals $\ottnt{A}$ in $\ottnt{B}$
by $\alpha$. Type $\ottnt{C}$ is the result of the revelation or concealment.  For
example, let us
\iffull revisit \else consider \fi
type application of $ \mathsf{Id_{int} } 
\iffull\else \defeq  \Lambda\!  \,  \mathit{X}  \mathord{:}   \mathsf{T}    \ottsym{.} \,  \lambda\!  \,  \mathit{x}  \mathord{:}  \mathit{X}   \ottsym{.}  \ottsym{(}  \mathit{x}  \ottsym{:}  \star  \ottsym{)}  \ottsym{:}   \mathsf{int}  \fi
$ which would otherwise break parametricity.  In Ahmed et al.'s semantics, application $ \mathsf{Id_{int} }  \, \ottnt{A} \, \ottnt{v}$ (where
$\ottnt{v}$ is a value of $\ottnt{A}$) is evaluated as follows:
\[\begin{array}{rcl}
  \mathsf{Id_{int} }  \, \ottnt{A} \, \ottnt{v} & \longrightarrow   & \ottsym{(}   \ottsym{(}    \lambda\!  \,  \mathit{x}  \mathord{:}  \mathit{X}   \ottsym{.}  \ottsym{(}   \mathit{x}  :  \star   \ottsym{)}  :   \mathsf{int}    \ottsym{)}    [  \alpha  /  \mathit{X}  ]    \ottsym{:}  \alpha  \rightarrow   \mathsf{int}  \,  \stackrel{ \ottsym{+}  \alpha }{\Rightarrow}  \ottnt{A}  \rightarrow   \mathsf{int}    \ottsym{)} \, \ottnt{v} \\
   \iffull
                &=        & \ottsym{(}  \ottsym{(}    \lambda\!  \,  \mathit{x}  \mathord{:}  \alpha   \ottsym{.}  \ottsym{(}   \mathit{x}  :  \star   \ottsym{)}  :   \mathsf{int}    \ottsym{)}  \ottsym{:}  \alpha  \rightarrow   \mathsf{int}  \,  \stackrel{ \ottsym{+}  \alpha }{\Rightarrow}  \ottnt{A}  \rightarrow   \mathsf{int}    \ottsym{)} \, \ottnt{v} \\
   \fi
                & \longrightarrow^{*}  &  \ottsym{(}   \ottsym{(}   \mathit{x}  :  \star   \ottsym{)}  :   \mathsf{int}    \ottsym{)}    [    \ottnt{v}  \ottsym{:}  \ottnt{A} \,  \stackrel{ \ottsym{-}  \alpha }{\Rightarrow}  \alpha     \ottsym{/}  \mathit{x}  ]    \ottsym{:}   \mathsf{int}  \,  \stackrel{ \ottsym{+}  \alpha }{\Rightarrow}   \mathsf{int}   \\
                &=        & \ottsym{(}   \ottsym{(}   \ottsym{(}  \ottnt{v}  \ottsym{:}  \ottnt{A} \,  \stackrel{ \ottsym{-}  \alpha }{\Rightarrow}  \alpha   \ottsym{)}  :  \star   \ottsym{)}  :   \mathsf{int}    \ottsym{)}  \ottsym{:}   \mathsf{int}  \,  \stackrel{ \ottsym{+}  \alpha }{\Rightarrow}   \mathsf{int}  .
  \end{array}
\]
The type application generates a fresh type name $\alpha$, substitutes it for
bound type variable $\mathit{X}$, and reveals $\ottnt{A}$ to the outside (here, function
application to $\ottnt{v}$) by conversion $\alpha  \rightarrow   \mathsf{int}  \,  \stackrel{ \ottsym{+}  \alpha }{\Rightarrow}  \ottnt{A}  \rightarrow   \mathsf{int}  $.  Applied to
argument $\ottnt{v}$, the conversion conceals the type $\ottnt{A}$ of $\ottnt{v}$ by
$\alpha$, as $\ottnt{v}  \ottsym{:}  \ottnt{A} \,  \stackrel{ \ottsym{-}  \alpha }{\Rightarrow}  \alpha $, and passes the abstracted value to the original
function $  \lambda\!  \,  \mathit{x}  \mathord{:}  \alpha   \ottsym{.}  \ottsym{(}   \mathit{x}  :  \star   \ottsym{)}  :   \mathsf{int}  $ (reduction from the first to the second line).
From the result in the third line, we can find that it will be tested if
$\ottnt{v}  \ottsym{:}  \ottnt{A} \,  \stackrel{ \ottsym{-}  \alpha }{\Rightarrow}  \alpha $ is an integer value.  Since type name $\alpha$ works like a
fresh base type and matches only with $\alpha$ itself, that test will fail
whatever $\ottnt{A}$ is--even if $\ottnt{A} \,  =  \,  \mathsf{int} $.  Therefore, $ \mathsf{Id_{int} } $ behaves
uniformly---raises an exception---whatever type is substituted for $\mathit{X}$.
\iffull
Interestingly, this semantics is permissive enough to allow $ \mathsf{Id }  \defeq
  \Lambda\!  \,  \mathit{X}  \mathord{:}   \mathsf{T}    \ottsym{.} \,  \lambda\!  \,  \mathit{x}  \mathord{:}  \mathit{X}   \ottsym{.}  \ottsym{(}   \mathit{x}  :  \star   \ottsym{)}  :  \mathit{X} $ of type $ \text{\unboldmath$\forall\!$}  \,  \mathit{X}  \mathord{:}   \mathsf{T}    \ottsym{.} \, \mathit{X}  \rightarrow  \mathit{X}$ to behave as an identity
function:
\[\begin{array}{rcl}
  \mathsf{Id }  \, \ottnt{A} \, \ottnt{v}     & \longrightarrow   & \ottsym{(}  \ottsym{(}    \lambda\!  \,  \mathit{x}  \mathord{:}  \alpha   \ottsym{.}  \ottsym{(}   \mathit{x}  :  \star   \ottsym{)}  :  \alpha   \ottsym{)}  \ottsym{:}  \alpha  \rightarrow  \alpha \,  \stackrel{ \ottsym{+}  \alpha }{\Rightarrow}  \ottnt{A}  \rightarrow  \ottnt{A}   \ottsym{)} \, \ottnt{v} \\
                & \longrightarrow^{*}  & \ottsym{(}   \ottsym{(}   \ottsym{(}  \ottnt{v}  \ottsym{:}  \ottnt{A} \,  \stackrel{ \ottsym{-}  \alpha }{\Rightarrow}  \alpha   \ottsym{)}  :  \star   \ottsym{)}  :  \alpha   \ottsym{)}  \ottsym{:}  \alpha \,  \stackrel{ \ottsym{+}  \alpha }{\Rightarrow}  \ottnt{A}  \\
                & \longrightarrow^{*}  & \ottsym{(}  \ottnt{v}  \ottsym{:}  \ottnt{A} \,  \stackrel{ \ottsym{-}  \alpha }{\Rightarrow}  \alpha   \ottsym{)}  \ottsym{:}  \alpha \,  \stackrel{ \ottsym{+}  \alpha }{\Rightarrow}  \ottnt{A}  \\
                & \longrightarrow   & \ottnt{v}
  \end{array}
\]
The reduction in the first and second lines is similar to that of application of
$ \mathsf{Id_{int} } $.  In the second line, injection of the value $\ottnt{v}  \ottsym{:}  \ottnt{A} \,  \stackrel{ \ottsym{-}  \alpha }{\Rightarrow}  \alpha $ to
the dynamic type is projected back to $\alpha$.  Since $\alpha$ matches with
itself successfully, this projection returns the injected value
$\ottnt{v}  \ottsym{:}  \ottnt{A} \,  \stackrel{ \ottsym{-}  \alpha }{\Rightarrow}  \alpha $ (the third line).  Finally, the term in the third line
reveals the type $\ottnt{A}$ of the value $\ottnt{v}  \ottsym{:}  \ottnt{A} \,  \stackrel{ \ottsym{-}  \alpha }{\Rightarrow}  \alpha $ which conceals $\ottnt{A}$;
thus, it evaluates to the original value $\ottnt{v}$.
\fi
Our blame calculus {\interlang} applies this idea for row parametricity as well.

\subsubsection{The extended syntax of {\interlang}.}

Types and rows are augmented with type-and-row names, ranged over by $\alpha$.
Ground types, ranged over by $\mathit{G}$ and $\mathit{H}$, are type tags given to a
value injected to the dynamic type.  Similarly, ground row types, ranged over by
$\gamma$, are row tags given to a row injected to the dynamic row type, being
a row name, the empty row, or a row extension of the form $  \ell  \mathbin{:}  \star   ;  \star $.

Terms, ranged over by $\ottnt{e}$, have three additional constructors.  A cast $\ottnt{e}  \ottsym{:}  \ottnt{A} \,  \stackrel{ \ottnt{p} }{\Rightarrow}  \ottnt{B} $ between consistently equivalent types $\ottnt{A}$ and $\ottnt{B}$ checks if
the value of $\ottnt{e}$ can behave as $\ottnt{B}$ at run time.  Blame label $\ottnt{p}$
represents the location of the cast.  A conversion $\ottnt{e}  \ottsym{:}  \ottnt{A} \,  \stackrel{ \Phi }{\Rightarrow}  \ottnt{B} $ with
conversion label $\Phi$ conceals or reveals type information by the type name
of $\Phi$.  Blame ``$\mathsf{blame} \, \ottnt{p}$'' is an (uncatchable) exception indicating
failure of a cast with $\ottnt{p}$.  We write $\ottnt{e}  \ottsym{:}  \ottnt{A} \,  \stackrel{ \ottnt{p} }{\Rightarrow}  \ottnt{B}  \,  \stackrel{ \ottnt{q} }{\Rightarrow}  \ottnt{C} $ for $\ottsym{(}  \ottnt{e}  \ottsym{:}  \ottnt{A} \,  \stackrel{ \ottnt{p} }{\Rightarrow}  \ottnt{B}   \ottsym{)}  \ottsym{:}  \ottnt{B} \,  \stackrel{ \ottnt{q} }{\Rightarrow}  \ottnt{C} $ and $\ottnt{e}  \ottsym{:}  \ottnt{A} \,  \stackrel{ \Phi_{{\mathrm{1}}} }{\Rightarrow}  \ottnt{B}  \,  \stackrel{ \Phi_{{\mathrm{2}}} }{\Rightarrow}  \ottnt{C} $ for $\ottsym{(}  \ottnt{e}  \ottsym{:}  \ottnt{A} \,  \stackrel{ \Phi_{{\mathrm{1}}} }{\Rightarrow}  \ottnt{B}   \ottsym{)}  \ottsym{:}  \ottnt{B} \,  \stackrel{ \Phi_{{\mathrm{2}}} }{\Rightarrow}  \ottnt{C} $.
Evaluation contexts, ranged over by $\ottnt{E}$, are also extended with casts and
conversions.  Type abstraction $  \Lambda\!  \,  \mathit{X}  \mathord{:}  \ottnt{K}   \ottsym{.}   \ottnt{e}  ::  \ottnt{A} $ is augmented with the type
$\ottnt{A}$ of $\ottnt{e}$.

Values, ranged over by $\ottnt{v}$, have six additional constructors:
the first three values are injections into $ \star $, $ [  \star  ] $, and
$ \langle  \star  \rangle $ with tag $\mathit{G}$, $ [  \gamma  ] $, and $ \langle  \gamma  \rangle $,
respectively.
The next three values are conversions that conceal $\ottnt{A}$ or $\rho$
by $\alpha$.

It is notable that embedding $ \variantlift{ \ell }{ \ottnt{A} }{ \ottnt{v} } $ is a value even if embedded value
$\ottnt{v}$ is injection $\ell' \, \ottnt{v'}$ where $\ell' \,  \not=  \, \ell$, while in {\staticlang}
$\ell$ and $\ell'$ have to be the same in order for the embedding term to be a
value.  This is because we would like to make the type system of {\interlang}
syntax-directed and, for that, we drop the implicit type conversion rule
\Ts{Equiv} from {\interlang}.  Thus, for example, injection $\ell \, \ottnt{v}$ can be
given type $ \langle    \ell  \mathbin{:}  \ottnt{A}   ;    \ell'  \mathbin{:}  \ottnt{B}   ;   \cdot     \rangle $ but cannot be given
$ \langle    \ell'  \mathbin{:}  \ottnt{B}   ;    \ell  \mathbin{:}  \ottnt{A}   ;   \cdot     \rangle $ in {\interlang}.  In order to embed $\ell \, \ottnt{v}$
into $ \langle    \ell'  \mathbin{:}  \ottnt{B}   ;    \ell  \mathbin{:}  \ottnt{A}   ;   \cdot     \rangle $, we use embedding: embedding value
$ \variantlift{ \ell' }{ \ottnt{B} }{ \ottsym{(}  \ell \, \ottnt{v}  \ottsym{)} } $ can have type $ \langle    \ell'  \mathbin{:}  \ottnt{B}   ;    \ell  \mathbin{:}  \ottnt{A}   ;   \cdot     \rangle $.
Conversely, if the type of value $\ottnt{v}$ is a variant type $ \langle    \ell  \mathbin{:}  \ottnt{A}   ;  \rho   \rangle $,
then $\ottnt{v}$ must be either an
injection value $\ell \, \ottnt{v'}$ or an embedding value $ \variantlift{ \ell }{ \ottnt{A} }{ \ottnt{v'} } $ for some
$\ottnt{v'}$.  Thus, the embedding operation is not only useful to make variant
types easy to use in the setting with row polymorphism---this motivates
\citet{Leijen_2005_TFP} to introduce the embedding operation---but also crucial
to make a type system for variant types syntax-directed.

Name stores, ranged over by $\Sigma$, bind names generated during evaluation to
their actual types or rows.  We suppose that names bound by $\Sigma$ are unique.
We write $\Sigma  \ottsym{(}  \alpha  \ottsym{)} \,  =  \, \ottnt{A}$ if and only if $ \alpha  \mathord{:}  \ottnt{K}   \ottsym{:=}  \ottnt{A} \,  \in  \, \Sigma$.

\subsection{Type system}

\begin{myfigure}[t]
 \begin{flushleft}
  \textbf{Convertible rules} \quad \framebox{$ \Sigma   \vdash   \ottnt{A}  \prec^{ \Phi }  \ottnt{B} $} \\[1ex]
\begin{center}
 \ifpaper\else
 $\ottdruleCvXXDyn{}$ \hfil
 $\ottdruleCvXXTyVar{}$ \\[1ex]
 \fi
 $\ottdruleCvXXTyName{}$ \hfil
 $\ottdruleCvXXReveal{}$ \hfil
 $\ottdruleCvXXConceal{}$ \\[1ex]
 \ifpaper\else
 $\ottdruleCvXXBase{}$ \hfil
 \fi
 $\ottdruleCvXXFun{}$ \hfil
 \ifpaper\else
 $\ottdruleCvXXPoly{}$ \\[1ex]
 $\ottdruleCvXXRecord{}$ \hfil
 $\ottdruleCvXXVariant{}$ \\[1ex]
 $\ottdruleCvXXREmp{}$ \hfil
 $\ottdruleCvXXCons{}$
 \fi
 \\[1ex]
\end{center}

  \textbf{Well-formedness rules for types and rows} \quad
\framebox{$\Sigma  \ottsym{;}  \Gamma  \vdash  \ottnt{A}  \ottsym{:}  \ottnt{K}$}
\begin{center}
 \ifpaper\else
 $\ottdruleWFXXTyVar{}$ \hfil
 \fi
 $\ottdruleWFXXTyName{}$ \hfil
 $\ottdruleWFXXDyn{}$ \\[1ex]
 \ifpaper\else
 $\ottdruleWFXXBase{}$ \hfil
 $\ottdruleWFXXFun{}$ \hfil
 $\ottdruleWFXXPoly{}$ \\[1ex]
 $\ottdruleWFXXRecord{}$ \hfil
 $\ottdruleWFXXVariant{}$ \\[1ex]
 $\ottdruleWFXXREmp{}$ \hfil
 $\ottdruleWFXXCons{}$ \hfil
 \fi
\end{center}

  \textbf{Typing rules} \quad
\framebox{$\Sigma  \ottsym{;}  \Gamma  \vdash  \ottnt{e}  \ottsym{:}  \ottnt{A}$}
\begin{center}
 \ifpaper\else
 $\ottdruleTXXVar{}$ \hfil
 $\ottdruleTXXConst{}$ \\[1ex]
 $\ottdruleTXXLam{}$ \hfil
 $\ottdruleTXXApp{}$ \\[1ex]
 \fi
 $\ottdruleTXXTLam{}$ \hfil
 \ifpaper\else
 $\ottdruleTXXTApp{}$ \\[1ex]
 $\ottdruleTXXREmp{}$ \hfil
 $\ottdruleTXXRExt{}$ \\[1ex]
 $\ottdruleTXXRLet{}$ \hfil
 \ifthen{\boolean{show-variant}}{
 $\ottdruleTXXVInj{}$ \\[1ex]
 $\ottdruleTXXVLift{}$ \\[1ex]
 $\ottdruleTXXVCase{}$ \\[1ex]
 }
 \fi
 $\ottdruleTXXBlame{}$ \\[1ex]
 $\ottdruleTXXCast{}$ \hfil
 $\ottdruleTXXConv{}$
\end{center}

 \end{flushleft}
 \caption{The type system of {\interlang} (selected rules).}
 \label{fig:blame:typing}
\end{myfigure}

The type system of {\interlang} also has three judgments taking forms augmented
with $\Sigma$: well-formedness judgments for typing contexts $\Sigma  \vdash  \Gamma$ and
for types $\Sigma  \ottsym{;}  \Gamma  \vdash  \ottnt{A}  \ottsym{:}  \ottnt{K}$, and typing judgment $\Sigma  \ottsym{;}  \Gamma  \vdash  \ottnt{e}  \ottsym{:}  \ottnt{A}$.  Most
of the inference rules of these judgments are similar to those of {\staticlang}
except for three points.  First, the inference rules are also augmented with
$\Sigma$.  Second, new rules for the dynamic type, type-and-row names, casts,
conversions, and blame are added and the typing rule for type abstractions is
adapted for change of syntax; these rules are shown in \reffig{blame:typing}.
Third, the implicit type conversion rule \Ts{Equiv} with type-and-row equivalence is dropped and field
reordering is covered by casts.  Hence, the inference rules of {\interlang} are
syntax-directed.  \reffig{blame:typing} shows only key rules, and the other
rules have the same forms as those of {\staticlang}; interested readers can find
the complete definition of the type system in the supplementary material.

There are two additional well-formedness rules for names and the dynamic type.
The dynamic type $ \star $ can be used as both the dynamic value type and the
dynamic row type \WF{Dyn}.  A type-and-row name is given kind $\ottnt{K}$ assigned
by $\Sigma$ \WF{TyName}.

New typing rules are added for new constructors.  Types in a cast have to be
consistently equivalent.  A conversion $\ottnt{e}  \ottsym{:}  \ottnt{A} \,  \stackrel{ \Phi }{\Rightarrow}  \ottnt{B} $ converts type $\ottnt{A}$
of $\ottnt{e}$ to type $\ottnt{B}$ by revealing type information $\Sigma  \ottsym{(}  \alpha  \ottsym{)}$ of $\alpha$
in $\ottnt{A}$ if $\Phi \,  =  \, \ottsym{+}  \alpha$, or concealing it if $\Phi \,  =  \, \ottsym{-}  \alpha$.  This idea is
formalized by \emph{convertibility} $ \Sigma   \vdash   \ottnt{A}  \prec^{ \Phi }  \ottnt{B} $, which means that, if
$\Phi \,  =  \, \ottsym{+}  \alpha$, $\ottnt{B}$ is obtained by substituting $\Sigma  \ottsym{(}  \alpha  \ottsym{)}$ for $\alpha$ in
$\ottnt{A}$ and that, if $\Phi \,  =  \, \ottsym{-}  \alpha$, $\ottnt{A}$ is obtained by substituting
$\Sigma  \ottsym{(}  \alpha  \ottsym{)}$ for $\alpha$ in $\ottnt{B}$.
Convertibility is the smallest relation such that (1) it satisfies the rules
given at the top of \reffig{blame:typing} and (2) it is closed under type and
row constructors other than names and function types.  The convertibility rules
use two operations on $\Phi$: $ \mathit{name} (  \Phi  ) $ returns the name of $\Phi$, i.e.,
$ \mathit{name} (  \ottsym{+}  \alpha  )  \defeq  \mathit{name} (  \ottsym{-}  \alpha  )  \defeq \alpha$; $ \overline{ \Phi } $ is the negation of
$\Phi$, i.e., $ \overline{ \ottsym{+}  \alpha }  \defeq \ottsym{-}  \alpha$ and $ \overline{ \ottsym{-}  \alpha }  \defeq  \mathsf{+}  \, \alpha$.
The rules \Cv{Reveal} and \Cv{Conceal} reflect the above intuition of
convertibility.  The rule \Cv{TyName} means that type information of
$ \mathit{name} (  \Phi  ) $ must be revealed or concealed.  The rule \Cv{Fun} means that
convertibility is contravariant on argument types with the negated $\Phi$ and
covariant on return types with $\Phi$.
\TS{Check!}

\iffull
\subsection{Translation}
As usual, a term in {\surfacelang} is translated to {\interlang} by inserting
casts between types which are related by type matching and consistent
equivalence in a typing derivation.  For example, the rule of translation
$ \hookrightarrow $ for function applications is:
\[ \ottdruleTransXXApp{} \]
The full definition of the translation is found in the supplementary material.
\fi

\subsection{Semantics}
The semantics of {\interlang} consists of two relations: the reduction relation
$\ottnt{e_{{\mathrm{1}}}}  \rightsquigarrow  \ottnt{e_{{\mathrm{2}}}}$, which handles basic computation irrelevant to name stores,
and the evaluation relation $ \Sigma_{{\mathrm{1}}}  \mid  \ottnt{e_{{\mathrm{1}}}}   \longrightarrow   \Sigma_{{\mathrm{2}}}  \mid  \ottnt{e_{{\mathrm{2}}}} $, which reduces a subterm,
lifts blame, or handles type application with name generation.

\begin{myfigure}[t]
 \begin{flushleft}
  \textbf{Reduction rules} \quad \framebox{$\ottnt{e_{{\mathrm{1}}}}  \rightsquigarrow  \ottnt{e_{{\mathrm{2}}}}$}
\[\begin{array}{rcl@{\qquad}l}
 \kappa_{{\mathrm{1}}} \, \kappa_{{\mathrm{2}}}  \rightsquigarrow   \zeta  (  \kappa_{{\mathrm{1}}}  ,  \kappa_{{\mathrm{2}}}  )  \quad \RWoP{Const} &&
 \ottsym{(}   \lambda\!  \,  \mathit{x}  \mathord{:}  \ottnt{A}   \ottsym{.}  \ottnt{e}  \ottsym{)} \, \ottnt{v}  \rightsquigarrow   \ottnt{e}    [  \ottnt{v}  \ottsym{/}  \mathit{x}  ]               & \RWoP{Beta} \\
 \mathsf{let} \, \ottsym{\{}  \ell  \ottsym{=}  \mathit{x}  \ottsym{;}  \mathit{y}  \ottsym{\}}  \ottsym{=}  \ottsym{\{}  \ell  \ottsym{=}  \ottnt{v_{{\mathrm{1}}}}  \ottsym{;}  \ottnt{v_{{\mathrm{2}}}}  \ottsym{\}} \, \mathsf{in} \, \ottnt{e_{{\mathrm{2}}}} &  \rightsquigarrow  &  \ottnt{e}    [  \ottnt{v_{{\mathrm{1}}}}  \ottsym{/}  \mathit{x}  \ottsym{,}  \ottnt{v_{{\mathrm{2}}}}  \ottsym{/}  \mathit{y}  ]   & \RWoP{Record} \\
 \csname ifshow-variant\endcsname
  \mathsf{case} \,  \ottsym{(}  \ell \, \ottnt{v}  \ottsym{)}  \,\mathsf{with}\, \langle  \ell \,  \mathit{x}   \rightarrow   \ottnt{e_{{\mathrm{1}}}}   \ottsym{;}   \mathit{y}   \rightarrow   \ottnt{e_{{\mathrm{2}}}}  \rangle        &  \rightsquigarrow  &  \ottnt{e_{{\mathrm{1}}}}    [  \ottnt{v}  \ottsym{/}  \mathit{x}  ]   & \RWoP{CaseL} \\
  \mathsf{case} \,   \variantlift{ \ell }{ \ottnt{A} }{ \ottnt{v} }   \,\mathsf{with}\, \langle  \ell \,  \mathit{x}   \rightarrow   \ottnt{e_{{\mathrm{1}}}}   \ottsym{;}   \mathit{y}   \rightarrow   \ottnt{e_{{\mathrm{2}}}}  \rangle  &  \rightsquigarrow  &  \ottnt{e_{{\mathrm{2}}}}    [  \ottnt{v}  \ottsym{/}  \mathit{y}  ]   & \RWoP{CaseR} \\
 \fi
 %
 \ottnt{v}  \ottsym{:}  \ottnt{A} \,  \stackrel{ \ottnt{p} }{\Rightarrow}  \ottnt{A}                      &  \rightsquigarrow  & \ottnt{v} \quad \text{(if $\ottnt{A} =  \star $, $\iota$, or $\alpha$)}& \RWoP{Id} \\
 \ottnt{v}  \ottsym{:}  \ottnt{A} \,  \stackrel{ \ottnt{p} }{\Rightarrow}  \star                    &  \rightsquigarrow  & \ottnt{v}  \ottsym{:}  \ottnt{A} \,  \stackrel{ \ottnt{p} }{\Rightarrow}  \mathit{G}  \,  \stackrel{ \ottnt{p} }{\Rightarrow}  \star  & \RWoP{ToDyn} \\
  \multicolumn{3}{r}{
   \text{(if $\ottnt{A}  \simeq  \mathit{G}$ and $\ottnt{A} \,  \not=  \, \mathit{G}$ and $\ottnt{A} \,  \not=  \, \star$ and $\ottnt{A} \,  \not=  \,  \text{\unboldmath$\forall\!$}  \,  \mathit{X}  \mathord{:}  \ottnt{K}   \ottsym{.} \, \ottnt{B}$)}
  } \\
 \ottnt{v}  \ottsym{:}  \star \,  \stackrel{ \ottnt{p} }{\Rightarrow}  \ottnt{A}                    &  \rightsquigarrow  & \ottnt{v}  \ottsym{:}  \star \,  \stackrel{ \ottnt{p} }{\Rightarrow}  \mathit{G}  \,  \stackrel{ \ottnt{p} }{\Rightarrow}  \ottnt{A}  & \RWoP{FromDyn} \\
  \multicolumn{3}{r}{
   \text{(if $\ottnt{A}  \simeq  \mathit{G}$ and $\ottnt{A} \,  \not=  \, \mathit{G}$ and $\ottnt{A} \,  \not=  \, \star$ and $\ottnt{A} \,  \not=  \,  \text{\unboldmath$\forall\!$}  \,  \mathit{X}  \mathord{:}  \ottnt{K}   \ottsym{.} \, \ottnt{B}$)}
  } \\
 \ottnt{v}  \ottsym{:}  \mathit{G} \,  \stackrel{ \ottnt{p} }{\Rightarrow}  \star  \,  \stackrel{ \ottnt{q} }{\Rightarrow}  \mathit{G}   \rightsquigarrow  \ottnt{v}       \quad \RWoP{Ground} &&
 \ottnt{v}  \ottsym{:}  \mathit{G} \,  \stackrel{ \ottnt{p} }{\Rightarrow}  \star  \,  \stackrel{ \ottnt{q} }{\Rightarrow}  \mathit{H}   \rightsquigarrow  \mathsf{blame} \, \ottnt{q} \quad \text{(if $\mathit{G} \,  \not=  \, \mathit{H}$)} & \RWoP{Blame}
 \\
 \ottnt{v}  \ottsym{:}  \ottnt{A_{{\mathrm{1}}}}  \rightarrow  \ottnt{B_{{\mathrm{1}}}} \,  \stackrel{ \ottnt{p} }{\Rightarrow}  \ottnt{A_{{\mathrm{2}}}}  \rightarrow  \ottnt{B_{{\mathrm{2}}}}        &  \rightsquigarrow  &  \lambda\!  \,  \mathit{x}  \mathord{:}  \ottnt{A_{{\mathrm{2}}}}   \ottsym{.}  \ottnt{v} \, \ottsym{(}  \mathit{x}  \ottsym{:}  \ottnt{A_{{\mathrm{2}}}} \,  \stackrel{  \overline{ \ottnt{p} }  }{\Rightarrow}  \ottnt{A_{{\mathrm{1}}}}   \ottsym{)}  \ottsym{:}  \ottnt{B_{{\mathrm{1}}}} \,  \stackrel{ \ottnt{p} }{\Rightarrow}  \ottnt{B_{{\mathrm{2}}}}  & \RWoP{Wrap} \\
 \ottnt{v}  \ottsym{:}   \text{\unboldmath$\forall\!$}  \,  \mathit{X}  \mathord{:}  \ottnt{K}   \ottsym{.} \, \ottnt{A_{{\mathrm{1}}}} \,  \stackrel{ \ottnt{p} }{\Rightarrow}   \text{\unboldmath$\forall\!$}  \,  \mathit{X}  \mathord{:}  \ottnt{K}   \ottsym{.} \, \ottnt{A_{{\mathrm{2}}}}          &  \rightsquigarrow  &   \Lambda\!  \,  \mathit{X}  \mathord{:}  \ottnt{K}   \ottsym{.}   \ottsym{(}  \ottnt{v} \, \mathit{X}  \ottsym{:}  \ottnt{A_{{\mathrm{1}}}} \,  \stackrel{ \ottnt{p} }{\Rightarrow}  \ottnt{A_{{\mathrm{2}}}}   \ottsym{)}  ::  \ottnt{A_{{\mathrm{2}}}}  & \RWoP{Content} \\
 \ottnt{v}  \ottsym{:}   \text{\unboldmath$\forall\!$}  \,  \mathit{X}  \mathord{:}  \ottnt{K}   \ottsym{.} \, \ottnt{A} \,  \stackrel{ \ottnt{p} }{\Rightarrow}  \ottnt{B}                 &  \rightsquigarrow  & \ottsym{(}  \ottnt{v} \, \star  \ottsym{)}  \ottsym{:}   \ottnt{A}    [  \star  /  \mathit{X}  ]   \,  \stackrel{ \ottnt{p} }{\Rightarrow}  \ottnt{B}  \quad \text{(if $\mathbf{QPoly} \, \ottsym{(}  \ottnt{B}  \ottsym{)}$)} & \RWoP{Inst} \\
 \ottnt{v}  \ottsym{:}  \ottnt{A} \,  \stackrel{ \ottnt{p} }{\Rightarrow}   \text{\unboldmath$\forall\!$}  \,  \mathit{X}  \mathord{:}  \ottnt{K}   \ottsym{.} \, \ottnt{B}                 &  \rightsquigarrow  &   \Lambda\!  \,  \mathit{X}  \mathord{:}  \ottnt{K}   \ottsym{.}   \ottsym{(}  \ottnt{v}  \ottsym{:}  \ottnt{A} \,  \stackrel{ \ottnt{p} }{\Rightarrow}  \ottnt{B}   \ottsym{)}  ::  \ottnt{B}  \quad \text{(if $\mathbf{QPoly} \, \ottsym{(}  \ottnt{A}  \ottsym{)}$)} & \RWoP{Gen} \\[1ex]
 %
 \multicolumn{3}{l}{
 \ottnt{v}  \ottsym{:}  \ottnt{A} \,  \stackrel{ \ottsym{-}  \alpha }{\Rightarrow}  \ottnt{B}  \,  \stackrel{ \ottsym{+}  \alpha }{\Rightarrow}  \ottnt{A}   \rightsquigarrow  \ottnt{v} \quad
 \text{(if (1) $\ottnt{B} \,  =  \, \alpha$; (2) $\ottnt{B} \,  =  \,  [  \alpha  ] $ and $\ottnt{A} \,  =  \,  [  \rho  ] $; or (3) $\ottnt{B} \,  =  \,  \langle  \alpha  \rangle $ and $\ottnt{A} \,  =  \,  \langle  \rho  \rangle $)}
 } & \multicolumn{1}{@{}l}{ \RWoP{CName} } \\
 \multicolumn{3}{l}{
 \ottnt{v}  \ottsym{:}  \ottnt{A} \,  \stackrel{ \Phi }{\Rightarrow}  \ottnt{A}   \rightsquigarrow  \ottnt{v} \ \ \qquad\quad
 \text{(if $\ottnt{A} \,  =  \, \star$, $\alpha$, $\iota$, $ [  \star  ] $, $ [  \alpha  ] $, $ [   \cdot   ] $, $ \langle  \star  \rangle $, or $ \langle  \alpha  \rangle $ for $\alpha \,  \not=  \,  \mathit{name} (  \Phi  ) $)}
 } & \RWoP{CId} \\
 \ottnt{v}  \ottsym{:}  \ottnt{A_{{\mathrm{1}}}}  \rightarrow  \ottnt{B_{{\mathrm{1}}}} \,  \stackrel{ \Phi }{\Rightarrow}  \ottnt{A_{{\mathrm{2}}}}  \rightarrow  \ottnt{B_{{\mathrm{2}}}}  &  \rightsquigarrow  &  \lambda\!  \,  \mathit{x}  \mathord{:}  \ottnt{A_{{\mathrm{2}}}}   \ottsym{.}  \ottnt{v} \, \ottsym{(}  \mathit{x}  \ottsym{:}  \ottnt{A_{{\mathrm{2}}}} \,  \stackrel{  \overline{ \Phi }  }{\Rightarrow}  \ottnt{A_{{\mathrm{1}}}}   \ottsym{)}  \ottsym{:}  \ottnt{B_{{\mathrm{1}}}} \,  \stackrel{ \Phi }{\Rightarrow}  \ottnt{B_{{\mathrm{2}}}}  & \RWoP{CFun} \\
 \ottnt{v}  \ottsym{:}   \text{\unboldmath$\forall\!$}  \,  \mathit{X}  \mathord{:}  \ottnt{K}   \ottsym{.} \, \ottnt{A_{{\mathrm{1}}}} \,  \stackrel{ \Phi }{\Rightarrow}   \text{\unboldmath$\forall\!$}  \,  \mathit{X}  \mathord{:}  \ottnt{K}   \ottsym{.} \, \ottnt{A_{{\mathrm{2}}}}    &  \rightsquigarrow  &   \Lambda\!  \,  \mathit{X}  \mathord{:}  \ottnt{K}   \ottsym{.}   \ottsym{(}  \ottnt{v} \, \mathit{X}  \ottsym{:}  \ottnt{A_{{\mathrm{1}}}} \,  \stackrel{ \Phi }{\Rightarrow}  \ottnt{A_{{\mathrm{2}}}}   \ottsym{)}  ::  \ottnt{A_{{\mathrm{2}}}}  & \RWoP{CForall} \\
 \multicolumn{3}{l}{
 \ottnt{v}  \ottsym{:}   [    \ell  \mathbin{:}  \ottnt{A}   ;  \rho_{{\mathrm{1}}}   ]  \,  \stackrel{ \Phi }{\Rightarrow}   [    \ell  \mathbin{:}  \ottnt{B}   ;  \rho_{{\mathrm{2}}}   ]    \rightsquigarrow  \mathsf{let} \, \ottsym{\{}  \ell  \ottsym{=}  \mathit{x}  \ottsym{;}  \mathit{y}  \ottsym{\}}  \ottsym{=}  \ottnt{v} \, \mathsf{in} \, \ottsym{\{}  \ell  \ottsym{=}  \mathit{x}  \ottsym{:}  \ottnt{A} \,  \stackrel{ \Phi }{\Rightarrow}  \ottnt{B}   \ottsym{;}  \mathit{y}  \ottsym{:}   [  \rho_{{\mathrm{1}}}  ]  \,  \stackrel{ \Phi }{\Rightarrow}   [  \rho_{{\mathrm{2}}}  ]    \ottsym{\}}
 } & \RWoP{CRExt} \\
 \multicolumn{4}{l}{
 \ottnt{v}  \ottsym{:}   \langle    \ell  \mathbin{:}  \ottnt{A}   ;  \rho_{{\mathrm{1}}}   \rangle  \,  \stackrel{ \Phi }{\Rightarrow}   \langle    \ell  \mathbin{:}  \ottnt{B}   ;  \rho_{{\mathrm{2}}}   \rangle    \rightsquigarrow   \mathsf{case} \,  \ottnt{v}  \,\mathsf{with}\, \langle  \ell \,  \mathit{x}   \rightarrow   \ell \, \ottsym{(}  \mathit{x}  \ottsym{:}  \ottnt{A} \,  \stackrel{ \Phi }{\Rightarrow}  \ottnt{B}   \ottsym{)}   \ottsym{;}   \mathit{y}   \rightarrow    \variantlift{ \ell }{ \ottnt{B} }{ \ottsym{(}  \mathit{y}  \ottsym{:}   \langle  \rho_{{\mathrm{1}}}  \rangle  \,  \stackrel{ \Phi }{\Rightarrow}   \langle  \rho_{{\mathrm{2}}}  \rangle    \ottsym{)} }   \rangle 
  \hfil \RWoP{CVar}
 } \\
\end{array}\]

 \end{flushleft}
 \caption{Reduction rules of {\interlang} except casts for record and variant types.}
 \label{fig:blame:red}
\end{myfigure}

\subsubsection{Reduction except cast for records and variants}
The reduction rules except cast for record and variant types are shown in
\reffig{blame:red}.  Most of the reduction rules for casts and conversions there
come from \citet{Ahmed/Jamner/Siek/Wadler_2017_ICFP}.  Cast semantics for
universal types follows \citet{Igarashi/Sekiyama/Igarashi_2017_ICFP}.

The first five rules are for function application, record decomposition, and
case matching.  The rule \R{Record} for record decomposition assumes that the
first field label of a record matches with the pattern label, while the
reduction rule \Rs{Record} of {\staticlang} does not assume that and looks for
the $\ell$ field from a record.  This assumption is valid in {\interlang}
because {\interlang} reorders the record fields by casts so that the static
assumption of \T{Record}---the first field of a decomposed record has the same
label as the pattern---is ensured even at run time.  Similarly, the rules
\R{CaseL} and \R{CaseR} for case matching also assume that a matched term has
the same label as the pattern.  For variants, instead of field reordering,
applications of the embedding operation are inserted.

Casts (except for record and variant types) behave as follows.  Casts where both
sides are the dynamic type, a base type, or a type name behave as identity
functions.  If a value of $\ottnt{A}$ is injected to the dynamic type, it is tagged
with ground type $\mathit{G}$ consistently equivalent to $\ottnt{A}$ \R{ToDyn}.
Conversely, if a value of $ \star $ is projected to $\ottnt{A}$, it will be checked
if the injected value is tagged with $\mathit{G}$ consistently equivalent to
$\ottnt{A}$ \R{FromDyn}.  If the check succeeds, the projection returns the injected
value \R{Ground}; otherwise, it raises an exception \R{Blame}.  Casts between
function types and between universal types produce a wrapper of a given value by
decomposing the types.  Casts from a quasi-universal type to a universal type
also produces a wrapper \R{Gen}.  Casts from a universal type to a
quasi-universal type apply a given type abstraction to $ \star $ \R{Inst}.

The last six rules are for conversions.  Revealing the concealed type $\ottnt{A}$
(or $\rho$) of a value reduces to the value itself \R{CName}.  If types in a
conversion take the same ``atomic'' form, it is just like an identity
function \R{CId}.  If types in a conversion are not atomic, a new term is constructed by
decomposing a given value and applying conversion with the type subcomponents to
the result.

\subsubsection{Cast reduction for records}

\begin{myfigure}[t]
 \begin{flushleft}
  \textbf{Cast rules for records} \quad
\framebox{$\ottnt{e_{{\mathrm{1}}}}  \rightsquigarrow  \ottnt{e_{{\mathrm{2}}}}$}
\[\begin{array}{r@{\ \ }c@{\ \ }ll}
 %
 %
 \ottnt{v}  \ottsym{:}   [  \rho  ]  \,  \stackrel{ \ottnt{p} }{\Rightarrow}   [  \rho  ]   &  \rightsquigarrow  & \ottnt{v} \quad
  \text{(if $\rho \,  =  \,  \cdot $ or $\alpha$)} & \RWoP{RId} \\
 \ottnt{v}  \ottsym{:}   [  \rho  ]  \,  \stackrel{ \ottnt{p} }{\Rightarrow}   [  \star  ]       &  \rightsquigarrow  & \ottnt{v}  \ottsym{:}   [  \rho  ]  \,  \stackrel{ \ottnt{p} }{\Rightarrow}   [   \mathit{grow}  (  \rho  )   ]   \,  \stackrel{ \ottnt{p} }{\Rightarrow}   [  \star  ]   \quad
  \text{(if $\rho \,  \not=  \,  \mathit{grow}  (  \rho  ) $)} \qquad\qquad\qquad\ \  & \RWoP{RToDyn} \\
 \ottnt{v}  \ottsym{:}   [  \gamma  ]  \,  \stackrel{ \ottnt{p} }{\Rightarrow}   [  \star  ]   \,  \stackrel{ \ottnt{q} }{\Rightarrow}   [  \rho  ]   &  \rightsquigarrow  & \ottnt{v}  \ottsym{:}   [  \gamma  ]  \,  \stackrel{ \ottnt{q} }{\Rightarrow}   [  \rho  ]   \quad \text{(if $\gamma  \simeq  \rho$)} & \RWoP{RFromDyn} \\
 \ottnt{v}  \ottsym{:}   [  \gamma  ]  \,  \stackrel{ \ottnt{p} }{\Rightarrow}   [  \star  ]   \,  \stackrel{ \ottnt{q} }{\Rightarrow}   [  \rho  ]   &  \rightsquigarrow  & \mathsf{blame} \, \ottnt{q} \quad \text{(if $\gamma  \not\simeq  \rho$)}& \RWoP{RBlame} \\
 \ottnt{v}  \ottsym{:}   [  \rho_{{\mathrm{1}}}  ]  \,  \stackrel{ \ottnt{p} }{\Rightarrow}   [    \ell  \mathbin{:}  \ottnt{B}   ;  \rho_{{\mathrm{2}}}   ]   &  \rightsquigarrow  &
  \multicolumn{2}{@{}l@{}}{
  \ottsym{\{}  \ell  \ottsym{=}  \ottsym{(}  \ottnt{v_{{\mathrm{1}}}}  \ottsym{:}  \ottnt{A} \,  \stackrel{ \ottnt{p} }{\Rightarrow}  \ottnt{B}   \ottsym{)}  \ottsym{;}  \ottnt{v_{{\mathrm{2}}}}  \ottsym{:}   [  \rho'_{{\mathrm{1}}}  ]  \,  \stackrel{ \ottnt{p} }{\Rightarrow}   [  \rho_{{\mathrm{2}}}  ]    \ottsym{\}} \quad
  \text{(if $\ottnt{v} \,  \triangleright _{ \ell }  \, \ottnt{v_{{\mathrm{1}}}}  \ottsym{,}  \ottnt{v_{{\mathrm{2}}}}$ and $\rho_{{\mathrm{1}}} \,  \triangleright _{ \ell }  \, \ottnt{A}  \ottsym{,}  \rho'_{{\mathrm{1}}}$)} \ \, \RWoP{RRev}
  } \\
 \ottnt{v}  \ottsym{:}   [  \rho_{{\mathrm{1}}}  ]  \,  \stackrel{ \ottnt{p} }{\Rightarrow}   [    \ell  \mathbin{:}  \ottnt{B}   ;  \rho_{{\mathrm{2}}}   ]   &  \rightsquigarrow  &
  \multicolumn{2}{@{}l@{}}{
  \ottnt{v}  \ottsym{:}   [  \rho_{{\mathrm{1}}}  ]  \,  \stackrel{ \ottnt{p} }{\Rightarrow}   [     \rho_{{\mathrm{1}}}  \mathrel{@}   \ell  \mathbin{:}  \ottnt{B}      ]   \,  \stackrel{ \ottnt{p} }{\Rightarrow}   [    \ell  \mathbin{:}  \ottnt{B}   ;  \rho_{{\mathrm{2}}}   ]   \quad
  \text{(if $\ell \,  \not\in  \, \mathit{dom} \, \ottsym{(}  \rho_{{\mathrm{1}}}  \ottsym{)}$ and $\rho_{{\mathrm{1}}} \,  \not=  \, \star$)} \quad \RWoP{RCon}
  }
\end{array}\]

 \end{flushleft}
 \caption{Reduction rules for casts between record types.}
 \label{fig:blame:red_record}
\end{myfigure}

The reduction rules for record casts are given in
\reffig{blame:red_record}.

If record types in a cast are the same and their rows are the empty row or a row
name, the cast behaves as an identity function \R{RId}.

If a record of type $ [  \rho  ] $ is injected into the record type $ [  \star  ] $, it is tagged with a ground row type consistently equivalent to $\rho$
\R{RToDyn}.  However, such a ground row type is not always determined uniquely
especially if $\rho$ is a row extension.  For example, row extension
$\ottsym{(}    \ell  \mathbin{:}  \ottnt{A}   ;  \star   \ottsym{)}$ is consistently equivalent to any ground row type of the form
$\ottsym{(}    \ell'  \mathbin{:}  \star   ;  \star   \ottsym{)}$.  We find a ground row type from $\rho$ by using function
$ \mathit{grow} $, which is defined as follows:
\iffull
\begin{defn}[Ground row types of rows]
 \[\begin{array}{lll}
   \mathit{grow}  (   \cdot   )  &\defeq&  \cdot  \\
     \mathit{grow}  (  \alpha  )  &\defeq& \alpha \\
     \mathit{grow}  (    \ell  \mathbin{:}  \ottnt{A}   ;  \rho   )  &\defeq&   \ell  \mathbin{:}  \star   ;  \star  \\
   \end{array}\]
\end{defn}

\noindent
\else
$ \mathit{grow}  (   \cdot   )  \defeq  \cdot $; $ \mathit{grow}  (  \alpha  )  \defeq \alpha$; and
$ \mathit{grow}  (    \ell  \mathbin{:}  \ottnt{A}   ;  \rho   )  \defeq   \ell  \mathbin{:}  \star   ;  \star $.
\fi
If $\rho \,  =  \,  \mathit{grow}  (  \rho  ) $, term $\ottnt{v}  \ottsym{:}   [  \rho  ]  \,  \stackrel{ \ottnt{p} }{\Rightarrow}   [  \star  ]  $ is a value, not
needed to reduce.

If the type of a record to be cast is $ [  \star  ] $, then the record can be
supposed to be tagged with a ground row type $\gamma$.  If $\gamma$ is
consistently equivalent to the target type $\rho$ of a cast, the cast reduces
to another cast from $\gamma$ to $\rho$ \R{RFromDyn}; otherwise, if $\gamma$ is
not consistently equivalent to $\rho$, an exception is raised \R{RBlame}.  Note
that a cast $\ottnt{v}  \ottsym{:}   [  \star  ]  \,  \stackrel{ \ottnt{p} }{\Rightarrow}   [  \star  ]  $ is handled by \R{RFromDyn}.  One
might consider why reduction of cast $\ottnt{v}  \ottsym{:}   [  \star  ]  \,  \stackrel{ \ottnt{p} }{\Rightarrow}   [  \rho  ]  $ is not
defined as \R{FromDyn} in \reffig{blame:red}, that is, the reduction does not
proceed as the cast first reduces to $\ottnt{v}  \ottsym{:}   [  \star  ]  \,  \stackrel{ \ottnt{p} }{\Rightarrow}   [   \mathit{grow}  (  \rho  )   ]   \,  \stackrel{ \ottnt{p} }{\Rightarrow}   [  \rho  ]  $ and then tests equality of $ \mathit{grow}  (  \rho  ) $ and the ground row type
$\gamma$ attached to $\ottnt{v}$.  We do not give such reduction because a ground
row type of $\rho$ may not be determined to be unique and, therefore, equality
test of $ \mathit{grow}  (  \rho  ) $ and $\gamma$ may fail even if the record $\ottnt{v}$ can
behave as $ \mathit{grow}  (  \rho  ) $.  For example, if $\rho \,  =  \, \ottsym{(}    \ell_{{\mathrm{1}}}  \mathbin{:}  \ottnt{A}   ;    \ell_{{\mathrm{2}}}  \mathbin{:}  \ottnt{B}   ;   \cdot     \ottsym{)}$ and the
ground row type $\gamma$ attached to $\ottnt{v}$ is $  \ell_{{\mathrm{2}}}  \mathbin{:}  \star   ;  \star $, then
$ \mathit{grow}  (  \rho  )  \,  =  \,   \ell_{{\mathrm{1}}}  \mathbin{:}  \star   ;  \star $ is different from $\gamma$ (if $\ell_{{\mathrm{1}}} \,  \not=  \, \ell_{{\mathrm{2}}}$), but
the record $\ottnt{v}$ may hold both of fields labeled with $\ell_{{\mathrm{1}}}$ and $\ell_{{\mathrm{2}}}$.
Instead of syntactic equality, use of consistent equivalence for comparison of
$ \mathit{grow}  (  \rho  ) $ and $\gamma$ might work better; indeed, the cast semantics given by
\R{RFromDyn} uses this approach.

The other cast reduction rules \R{RRev} and \R{RCon} are applied to a cast $\ottnt{v}  \ottsym{:}   [  \rho_{{\mathrm{1}}}  ]  \,  \stackrel{ \ottnt{p} }{\Rightarrow}   [    \ell  \mathbin{:}  \ottnt{B}   ;  \rho_{{\mathrm{2}}}   ]  $ which tests whether record $\ottnt{v}$ of type
$ [  \rho_{{\mathrm{1}}}  ] $ has an $\ell$ field and, if so, whether the value of the
$\ell$ field and the other fields can behave as $\ottnt{B}$ and $\rho_{{\mathrm{2}}}$,
respectively.  The rule \R{RRev} handles the case that the source row type
$\rho_{{\mathrm{1}}}$ holds an $\ell$ field.  In this case, and only in this case, we can
find the value $\ottnt{v_{{\mathrm{1}}}}$ of the $\ell$ field from record $\ottnt{v}$ by record
splitting $\ottnt{v} \,  \triangleright _{ \ell }  \, \ottnt{v_{{\mathrm{1}}}}  \ottsym{,}  \ottnt{v_{{\mathrm{2}}}}$, where $\ottnt{v_{{\mathrm{2}}}}$ is the result of removing $\ottnt{v_{{\mathrm{1}}}}$
from $\ottnt{v}$.  The record splitting on $\ottnt{v}$ is defined as
\refdef{static_lang:record_split}.  Row splitting $\rho_{{\mathrm{1}}} \,  \triangleright _{ \ell }  \, \ottnt{A}  \ottsym{,}  \rho'_{{\mathrm{1}}}$ returns the
type $\ottnt{A}$ of $\ottnt{v_{{\mathrm{1}}}}$ and the row type $\rho'_{{\mathrm{1}}}$ for the fields of $\ottnt{v_{{\mathrm{2}}}}$.
As a result, the cast reduces to a record value composed of an $\ell$ field
holding $\ottnt{v_{{\mathrm{1}}}}  \ottsym{:}  \ottnt{A} \,  \stackrel{ \ottnt{p} }{\Rightarrow}  \ottnt{B} $ and record $\ottnt{v_{{\mathrm{2}}}}  \ottsym{:}   [  \rho'_{{\mathrm{1}}}  ]  \,  \stackrel{ \ottnt{p} }{\Rightarrow}   [  \rho_{{\mathrm{2}}}  ]  $.  If
an $\ell$ field is not found in $\rho_{{\mathrm{1}}}$ (i.e., $\ell \,  \not\in  \, \mathit{dom} \, \ottsym{(}  \rho_{{\mathrm{1}}}  \ottsym{)}$), the
rule \R{RCon} is applied.  In this case, we can find that $\rho_{{\mathrm{1}}}$ ends with
$ \star $ since $\ell \,  \not\in  \, \mathit{dom} \, \ottsym{(}  \rho_{{\mathrm{1}}}  \ottsym{)}$ but $\rho_{{\mathrm{1}}}$ should be consistently
equivalent with $  \ell  \mathbin{:}  \ottnt{B}   ;  \rho_{{\mathrm{2}}} $.  Thus, $\ottnt{v}$ may hold an $\ell$ field in the
part hidden by $ \star $.  The reduction result tests it by the cast from
$ [  \rho_{{\mathrm{1}}}  ] $ to $ [     \rho_{{\mathrm{1}}}  \mathrel{@}   \ell  \mathbin{:}  \ottnt{B}      ] $.  The row type $ \rho_{{\mathrm{1}}}  \mathrel{@}   \ell  \mathbin{:}  \ottnt{B}  $ is the
same as $\rho_{{\mathrm{1}}}$ except that $ \ell  \mathbin{:}  \ottnt{B} $ is added as the last field.  Formally,
$ \rho  \mathrel{@}   \ell  \mathbin{:}  \ottnt{A}  $ is defined as follows.
\begin{defn}[Field postpending]
 Field postpending $ \rho  \mathrel{@}   \ell  \mathbin{:}  \ottnt{A}  $ is defined as follows:
 \ifpaper
 \[
  \ottsym{(}    \ell'  \mathbin{:}  \ottnt{B}   ;  \rho   \ottsym{)}  \mathrel{@}   \ell  \mathbin{:}  \ottnt{A}   \defeq   \ell'  \mathbin{:}  \ottnt{B}   ;  \ottsym{(}   \rho  \mathrel{@}   \ell  \mathbin{:}  \ottnt{A}    \ottsym{)}  \qquad
  \star  \mathrel{@}   \ell  \mathbin{:}  \ottnt{A}   \defeq   \ell  \mathbin{:}  \ottnt{A}   ;  \star 
 \]
 \else
 \[\begin{array}{lll}
   \ottsym{(}    \ell'  \mathbin{:}  \ottnt{B}   ;  \rho   \ottsym{)}  \mathrel{@}   \ell  \mathbin{:}  \ottnt{A}   &\defeq&   \ell'  \mathbin{:}  \ottnt{B}   ;  \ottsym{(}   \rho  \mathrel{@}   \ell  \mathbin{:}  \ottnt{A}    \ottsym{)}  \\
     \star  \mathrel{@}   \ell  \mathbin{:}  \ottnt{A}   &\defeq&   \ell  \mathbin{:}  \ottnt{A}   ;  \star 
   \end{array}\]
 \fi
\end{defn}

\noindent
Note that we can assume that $\rho_{{\mathrm{1}}}$ ends with $ \star $ and, therefore,
$ \rho_{{\mathrm{1}}}  \mathrel{@}   \ell  \mathbin{:}  \ottnt{B}  $ is well defined if the reduced term is well typed.  If record
$\ottnt{v}$ holds an $\ell$ field and its value can behave as type $\ottnt{B}$, then
the subsequent cast from $ [     \rho_{{\mathrm{1}}}  \mathrel{@}   \ell  \mathbin{:}  \ottnt{B}      ] $ to $ [    \ell  \mathbin{:}  \ottnt{B}   ;  \rho_{{\mathrm{2}}}   ] $ will
test if the other fields of $\ottnt{v}$ can behave as $\rho_{{\mathrm{2}}}$.

\paragraph{Examples.}
Let us consider a few examples of reduction.  In what follows, we shade subterms
to be reduced and underline their reduction results.

First, cast $\ottsym{\{}  \ell_{{\mathrm{1}}}  \ottsym{=}   0   \ottsym{;}  \ottsym{\{}  \ell_{{\mathrm{2}}}  \ottsym{=}   \mathsf{true}   \ottsym{;}  \ottsym{\{}  \ottsym{\}}  \ottsym{\}}  \ottsym{\}}  \ottsym{:}   [    \ell_{{\mathrm{1}}}  \mathbin{:}   \mathsf{int}    ;    \ell_{{\mathrm{2}}}  \mathbin{:}   \mathsf{bool}    ;   \cdot     ]  \,  \stackrel{ \ottnt{p} }{\Rightarrow}   [  \star  ]  $ reduces as follows:
{\small
\[\fboxsep=0em
\begin{array}{ll} &
   \shadeback:   \ottsym{\{}  \ell_{{\mathrm{1}}}  \ottsym{=}   0   \ottsym{;}  \ottsym{\{}  \ell_{{\mathrm{2}}}  \ottsym{=}   \mathsf{true}   \ottsym{;}  \ottsym{\{}  \ottsym{\}}  \ottsym{\}}  \ottsym{\}}  \ottsym{:}   [    \ell_{{\mathrm{1}}}  \mathbin{:}   \mathsf{int}    ;    \ell_{{\mathrm{2}}}  \mathbin{:}   \mathsf{bool}    ;   \cdot     ]  \,  \stackrel{ \ottnt{p} }{\Rightarrow}   [  \star  ]      :  \\[.5ex]
  \longrightarrow  &
   \undline<    \shadeback:   \ottsym{\{}  \ell_{{\mathrm{1}}}  \ottsym{=}   0   \ottsym{;}  \ottsym{\{}  \ell_{{\mathrm{2}}}  \ottsym{=}   \mathsf{true}   \ottsym{;}  \ottsym{\{}  \ottsym{\}}  \ottsym{\}}  \ottsym{\}}   \ottsym{:}   [    \ell_{{\mathrm{1}}}  \mathbin{:}   \mathsf{int}    ;    \ell_{{\mathrm{2}}}  \mathbin{:}   \mathsf{bool}    ;   \cdot     ]  \,  \stackrel{ \ottnt{p} }{\Rightarrow}    [    \ell_{{\mathrm{1}}}  \mathbin{:}  \star   ;  \star   ]    :   \,  \stackrel{ \ottnt{p} }{\Rightarrow}   [  \star  ]      >  \\[.5ex]
  \longrightarrow  &
  \undline<    \ottsym{\{}  \ell_{{\mathrm{1}}}  \ottsym{=}   0   \ottsym{:}   \mathsf{int}  \,  \stackrel{ \ottnt{p} }{\Rightarrow}  \star   \ottsym{;}    \shadeback:   \ottsym{\{}  \ell_{{\mathrm{2}}}  \ottsym{=}   \mathsf{true}   \ottsym{;}  \ottsym{\{}  \ottsym{\}}  \ottsym{\}}   \ottsym{:}   [    \ell_{{\mathrm{2}}}  \mathbin{:}   \mathsf{bool}    ;   \cdot    ]  \,  \stackrel{ \ottnt{p} }{\Rightarrow}   [  \star  ]     :   \ottsym{\}}   >   \ottsym{:}   [    \ell_{{\mathrm{1}}}  \mathbin{:}  \star   ;  \star   ]  \,  \stackrel{ \ottnt{p} }{\Rightarrow}   [  \star  ]    \\[.5ex]
  \longrightarrow  &
 \ottsym{\{}  \ell_{{\mathrm{1}}}  \ottsym{=}   0   \ottsym{:}   \mathsf{int}  \,  \stackrel{ \ottnt{p} }{\Rightarrow}  \star   \ottsym{;}    \undline<    \shadeback:   \ottsym{\{}  \ell_{{\mathrm{2}}}  \ottsym{=}   \mathsf{true}   \ottsym{;}  \ottsym{\{}  \ottsym{\}}  \ottsym{\}}    \ottsym{:}   [    \ell_{{\mathrm{2}}}  \mathbin{:}   \mathsf{bool}    ;   \cdot    ]  \,  \stackrel{ \ottnt{p} }{\Rightarrow}    [    \ell_{{\mathrm{2}}}  \mathbin{:}  \star   ;  \star   ]    :   \,  \stackrel{ \ottnt{p} }{\Rightarrow}   [  \star  ]     >   \ottsym{\}}  \ottsym{:}   [    \ell_{{\mathrm{1}}}  \mathbin{:}  \star   ;  \star   ]  \,  \stackrel{ \ottnt{p} }{\Rightarrow}   [  \star  ]   \\[.5ex]
  \longrightarrow  &
 \ottsym{\{}  \ell_{{\mathrm{1}}}  \ottsym{=}   0   \ottsym{:}   \mathsf{int}  \,  \stackrel{ \ottnt{p} }{\Rightarrow}  \star   \ottsym{;}    \undline<   \ottsym{\{}  \ell_{{\mathrm{2}}}  \ottsym{=}   \mathsf{true}   \ottsym{:}   \mathsf{bool}  \,  \stackrel{ \ottnt{p} }{\Rightarrow}  \star   \ottsym{;}  \ottsym{\{}  \ottsym{\}}  \ottsym{:}   [   \cdot   ]  \,  \stackrel{ \ottnt{p} }{\Rightarrow}   [  \star  ]    \ottsym{\}}    >   \ottsym{:}   [    \ell_{{\mathrm{2}}}  \mathbin{:}  \star   ;  \star   ]  \,  \stackrel{ \ottnt{p} }{\Rightarrow}   [  \star  ]    \ottsym{\}}  \ottsym{:}   [    \ell_{{\mathrm{1}}}  \mathbin{:}  \star   ;  \star   ]  \,  \stackrel{ \ottnt{p} }{\Rightarrow}   [  \star  ]  
\end{array}
\]
}
where a term in an odd-numbered line reduces by \R{RToDyn} and one in an
even-numbered line by \R{RRev}.

In order to access to an $\ell$ field of the above reduction result, we have to
project it to, e.g., record type $ [    \ell  \mathbin{:}  \ottnt{A}   ;  \star   ] $.  The result can be written
$\ottnt{v}  \ottsym{:}   [    \ell_{{\mathrm{1}}}  \mathbin{:}  \star   ;  \star   ]  \,  \stackrel{ \ottnt{p} }{\Rightarrow}   [  \star  ]  $ where
\[
 \ottnt{v'} \defeq \ottsym{\{}  \ell_{{\mathrm{2}}}  \ottsym{=}   \mathsf{true}   \ottsym{:}   \mathsf{bool}  \,  \stackrel{ \ottnt{p} }{\Rightarrow}  \star   \ottsym{;}  \ottsym{\{}  \ottsym{\}}  \ottsym{:}   [   \cdot   ]  \,  \stackrel{ \ottnt{p} }{\Rightarrow}   [  \star  ]    \ottsym{\}} \qquad
 \ottnt{v}  \defeq \ottsym{\{}  \ell_{{\mathrm{1}}}  \ottsym{=}   0   \ottsym{:}   \mathsf{int}  \,  \stackrel{ \ottnt{p} }{\Rightarrow}  \star   \ottsym{;}  \ottnt{v'}  \ottsym{:}   [    \ell_{{\mathrm{2}}}  \mathbin{:}  \star   ;  \star   ]  \,  \stackrel{ \ottnt{p} }{\Rightarrow}   [  \star  ]    \ottsym{\}}.
\]
Then,
$
 \ottnt{v}  \ottsym{:}   [    \ell_{{\mathrm{1}}}  \mathbin{:}  \star   ;  \star   ]  \,  \stackrel{ \ottnt{p} }{\Rightarrow}   [  \star  ]   \,  \stackrel{ \ottnt{q} }{\Rightarrow}   [    \ell  \mathbin{:}  \ottnt{A}   ;  \star   ]  
  \longrightarrow 
 \ottnt{v}  \ottsym{:}   [    \ell_{{\mathrm{1}}}  \mathbin{:}  \star   ;  \star   ]  \,  \stackrel{ \ottnt{q} }{\Rightarrow}   [    \ell  \mathbin{:}  \ottnt{A}   ;  \star   ]  
$
by \R{RFromDyn}.

If $\ell \,  =  \, \ell_{{\mathrm{1}}}$, then the result reduces to:
\[
 \ottsym{\{}  \ell_{{\mathrm{1}}}  \ottsym{=}   0   \ottsym{:}   \mathsf{int}  \,  \stackrel{ \ottnt{p} }{\Rightarrow}  \star  \,  \stackrel{ \ottnt{q} }{\Rightarrow}  \ottnt{A}   \ottsym{;}  \ottnt{v'}  \ottsym{:}   [    \ell_{{\mathrm{2}}}  \mathbin{:}  \star   ;  \star   ]  \,  \stackrel{ \ottnt{p} }{\Rightarrow}   [  \star  ]   \,  \stackrel{ \ottnt{q} }{\Rightarrow}   [  \star  ]    \ottsym{\}}
\]
by \R{RRev}.  Thus, if $\ottnt{A} \,  =  \,  \mathsf{int} $, we can extract the integer value held by
the $\ell_{{\mathrm{1}}}$ field in $\ottnt{v}$.  Otherwise, if $\ottnt{A} \,  \not=  \,  \mathsf{int} $, an exception
$\mathsf{blame} \, \ottnt{q}$ will be raised.

Let us return to reduction of $\ottnt{v}  \ottsym{:}   [    \ell_{{\mathrm{1}}}  \mathbin{:}  \star   ;  \star   ]  \,  \stackrel{ \ottnt{q} }{\Rightarrow}   [    \ell  \mathbin{:}  \ottnt{A}   ;  \star   ]  $.
If $\ell \,  \not=  \, \ell_{{\mathrm{1}}}$, then:
\[\fboxsep=0em
\begin{array}{@{}l@{}l} &
   \shadeback:   \ottnt{v}  \ottsym{:}   [    \ell_{{\mathrm{1}}}  \mathbin{:}  \star   ;  \star   ]  \,  \stackrel{ \ottnt{q} }{\Rightarrow}   [    \ell  \mathbin{:}  \ottnt{A}   ;  \star   ]      :  \\[.5ex]
  \longrightarrow  &
   \undline<    \shadeback:   \ottnt{v}   \ottsym{:}   [    \ell_{{\mathrm{1}}}  \mathbin{:}  \star   ;  \star   ]  \,  \stackrel{ \ottnt{q} }{\Rightarrow}    [    \ell_{{\mathrm{1}}}  \mathbin{:}  \star   ;    \ell  \mathbin{:}  \ottnt{A}   ;  \star    ]    :   \,  \stackrel{ \ottnt{q} }{\Rightarrow}   [    \ell  \mathbin{:}  \ottnt{A}   ;  \star   ]      >  \hfill
 \text{\R{RCon}} \\[.3ex]
  \longrightarrow^{*}  &
  \undline<    \ottsym{\{}  \ell_{{\mathrm{1}}}  \ottsym{=}   0   \ottsym{:}   \mathsf{int}  \,  \stackrel{ \ottnt{p} }{\Rightarrow}  \star   \ottsym{;}    \shadeback:   \ottnt{v'}   \ottsym{:}   [    \ell_{{\mathrm{2}}}  \mathbin{:}  \star   ;  \star   ]  \,  \stackrel{ \ottnt{p} }{\Rightarrow}   [  \star  ]   \,  \stackrel{ \ottnt{q} }{\Rightarrow}   [    \ell  \mathbin{:}  \ottnt{A}   ;  \star   ]     :   \ottsym{\}}   >   \ottsym{:}   [    \ell_{{\mathrm{1}}}  \mathbin{:}  \star   ;    \ell  \mathbin{:}  \ottnt{A}   ;  \star    ]  \,  \stackrel{ \ottnt{q} }{\Rightarrow}   [    \ell  \mathbin{:}  \ottnt{A}   ;  \star   ]    \\[.3ex]
  \longrightarrow  &
 \ottsym{\{}  \ell_{{\mathrm{1}}}  \ottsym{=}   0   \ottsym{:}   \mathsf{int}  \,  \stackrel{ \ottnt{p} }{\Rightarrow}  \star   \ottsym{;}     \undline<    \shadeback:   \ottnt{v'}    \ottsym{:}   [    \ell_{{\mathrm{2}}}  \mathbin{:}  \star   ;  \star   ]  \,  \stackrel{ \ottnt{q} }{\Rightarrow}   [    \ell  \mathbin{:}  \ottnt{A}   ;  \star   ]     :    >   \ottsym{\}}  \ottsym{:}   [    \ell_{{\mathrm{1}}}  \mathbin{:}  \star   ;    \ell  \mathbin{:}  \ottnt{A}   ;  \star    ]  \,  \stackrel{ \ottnt{q} }{\Rightarrow}   [    \ell  \mathbin{:}  \ottnt{A}   ;  \star   ]   {\ \ }
 \text{\R{RFromDyn}}.
\end{array}
\]
As in the case of $\ell \,  =  \, \ell_{{\mathrm{1}}}$, if $\ell \,  =  \, \ell_{{\mathrm{2}}}$ and $\ottnt{A} \,  =  \,  \mathsf{bool} $, we can
extract the Boolean value held by the $\ell_{{\mathrm{2}}}$ field in $\ottnt{v}$; if $\ell \,  =  \, \ell_{{\mathrm{2}}}$
but $\ottnt{A} \,  \not=  \,  \mathsf{bool} $, an exception $\mathsf{blame} \, \ottnt{q}$ will be raised.
If $\ell \,  \not=  \, \ell_{{\mathrm{2}}}$, the last shaded part in turn evaluates to:
\[\fboxsep=0em
\begin{array}{lll} &
   \undline<    \shadeback:   \ottnt{v'}   \ottsym{:}   [    \ell_{{\mathrm{2}}}  \mathbin{:}  \star   ;  \star   ]  \,  \stackrel{ \ottnt{q} }{\Rightarrow}    [    \ell_{{\mathrm{2}}}  \mathbin{:}  \star   ;    \ell  \mathbin{:}  \ottnt{A}   ;  \star    ]    :   \,  \stackrel{ \ottnt{q} }{\Rightarrow}   [    \ell  \mathbin{:}  \ottnt{A}   ;  \star   ]      >  \hfill
 \text{\R{RCon}} \\
  \longrightarrow^{*}  &
  \undline<    \ottsym{\{}  \ell_{{\mathrm{2}}}  \ottsym{=}   \mathsf{true}   \ottsym{:}   \mathsf{bool}  \,  \stackrel{ \ottnt{p} }{\Rightarrow}  \star   \ottsym{;}    \shadeback:   \ottsym{\{}  \ottsym{\}}   \ottsym{:}   [   \cdot   ]  \,  \stackrel{ \ottnt{p} }{\Rightarrow}   [  \star  ]   \,  \stackrel{ \ottnt{q} }{\Rightarrow}   [    \ell  \mathbin{:}  \ottnt{A}   ;  \star   ]     :   \ottsym{\}}   >   \ottsym{:}   [    \ell_{{\mathrm{2}}}  \mathbin{:}  \star   ;    \ell  \mathbin{:}  \ottnt{A}   ;  \star    ]  \,  \stackrel{ \ottnt{q} }{\Rightarrow}   [    \ell  \mathbin{:}  \ottnt{A}   ;  \star   ]    \\
  \longrightarrow^{*}  &
   \undline<   \mathsf{blame} \, \ottnt{q}    >  \hfill
 \text{\R{RBlame}}
\end{array}\]
This behavior is expected because $\ottnt{v}$ does not hold any field with label
$\ell$ other than $\ell_{{\mathrm{1}}}$ and $\ell_{{\mathrm{2}}}$.

\subsubsection{Cast reduction for variants}

\begin{myfigure}[t]
 \begin{flushleft}
  \textbf{Cast{\ifthen{\boolean{show-common}}{ and conversion}{ }}reduction rules for variants} \quad
\framebox{$\ottnt{e_{{\mathrm{1}}}}  \rightsquigarrow  \ottnt{e_{{\mathrm{2}}}}$}
\[\begin{array}{rcll}
 %
 \ottnt{v}  \ottsym{:}   \langle  \alpha  \rangle  \,  \stackrel{ \ottnt{p} }{\Rightarrow}   \langle  \alpha  \rangle         &  \rightsquigarrow  & \ottnt{v} & \RWoP{VIdName} \\
 \ottnt{v}  \ottsym{:}   \langle  \rho  \rangle  \,  \stackrel{ \ottnt{p} }{\Rightarrow}   \langle  \star  \rangle       &  \rightsquigarrow  & \ottnt{v}  \ottsym{:}   \langle  \rho  \rangle  \,  \stackrel{ \ottnt{p} }{\Rightarrow}   \langle   \mathit{grow}  (  \rho  )   \rangle   \,  \stackrel{ \ottnt{p} }{\Rightarrow}   \langle  \star  \rangle   \quad
  \text{(if $\rho \,  \not=  \,  \mathit{grow}  (  \rho  ) $)}& \RWoP{VToDyn} \\
 \ottnt{v}  \ottsym{:}   \langle  \gamma  \rangle  \,  \stackrel{ \ottnt{p} }{\Rightarrow}   \langle  \star  \rangle   \,  \stackrel{ \ottnt{q} }{\Rightarrow}   \langle  \rho  \rangle   &  \rightsquigarrow  & \ottnt{v}  \ottsym{:}   \langle  \gamma  \rangle  \,  \stackrel{ \ottnt{q} }{\Rightarrow}   \langle  \rho  \rangle   \quad \text{(if $\gamma  \simeq  \rho$)}& \RWoP{VFromDyn} \\
 \ottnt{v}  \ottsym{:}   \langle  \gamma  \rangle  \,  \stackrel{ \ottnt{p} }{\Rightarrow}   \langle  \star  \rangle   \,  \stackrel{ \ottnt{q} }{\Rightarrow}   \langle  \rho  \rangle   &  \rightsquigarrow  & \mathsf{blame} \, \ottnt{q} \quad \text{(if $\gamma  \not\simeq  \rho$)}& \RWoP{VBlame} \\
 \ottsym{(}  \ell \, \ottnt{v}  \ottsym{)}  \ottsym{:}   \langle    \ell  \mathbin{:}  \ottnt{A}   ;  \rho_{{\mathrm{1}}}   \rangle  \,  \stackrel{ \ottnt{p} }{\Rightarrow}   \langle  \rho_{{\mathrm{2}}}  \rangle   &  \rightsquigarrow  &  \variantliftrow{ \rho_{{\mathrm{21}}} }{ \ottsym{(}  \ell \, \ottsym{(}  \ottnt{v}  \ottsym{:}  \ottnt{A} \,  \stackrel{ \ottnt{p} }{\Rightarrow}  \ottnt{B}   \ottsym{)}  \ottsym{)} }  & \multirow{2}{*}{\RWoP{VRevInj}} \\
  \multicolumn{3}{r}{
   \text{(if $\rho_{{\mathrm{2}}} \,  =  \, \rho_{{\mathrm{21}}}  \odot  \ottsym{(}    \ell  \mathbin{:}  \ottnt{B}   ;   \cdot    \ottsym{)}  \odot  \rho_{{\mathrm{22}}}$ and $\ell \,  \not\in  \, \mathit{dom} \, \ottsym{(}  \rho_{{\mathrm{21}}}  \ottsym{)}$)}
  }
 \\
 \multicolumn{3}{l}{
  \ottsym{(}   \variantlift{ \ell }{ \ottnt{A} }{ \ottnt{v} }   \ottsym{)}  \ottsym{:}   \langle    \ell  \mathbin{:}  \ottnt{A}   ;  \rho_{{\mathrm{1}}}   \rangle  \,  \stackrel{ \ottnt{p} }{\Rightarrow}   \langle  \rho_{{\mathrm{2}}}  \rangle   \  \rightsquigarrow \  \variantliftdown{ \rho_{{\mathrm{21}}} }{ \ell }{ \ottnt{B} }{ \ottsym{(}  \ottnt{v}  \ottsym{:}   \langle  \rho_{{\mathrm{1}}}  \rangle  \,  \stackrel{ \ottnt{p} }{\Rightarrow}   \langle    \rho_{{\mathrm{21}}}  \odot  \rho_{{\mathrm{22}}}    \rangle    \ottsym{)} } 
 } & \multirow{2}{*}{\RWoP{VRevLift}} \\
  \multicolumn{3}{r}{
   \text{(if $\rho_{{\mathrm{2}}} \,  =  \, \rho_{{\mathrm{21}}}  \odot  \ottsym{(}    \ell  \mathbin{:}  \ottnt{B}   ;   \cdot    \ottsym{)}  \odot  \rho_{{\mathrm{22}}}$ and $\ell \,  \not\in  \, \mathit{dom} \, \ottsym{(}  \rho_{{\mathrm{21}}}  \ottsym{)}$)}
  }
 \\
 \ottsym{(}  \ell \, \ottnt{v}  \ottsym{)}  \ottsym{:}   \langle    \ell  \mathbin{:}  \ottnt{A}   ;  \rho_{{\mathrm{1}}}   \rangle  \,  \stackrel{ \ottnt{p} }{\Rightarrow}   \langle  \rho_{{\mathrm{2}}}  \rangle   &  \rightsquigarrow  &  \variantliftrow{ \rho_{{\mathrm{2}}} }{ \ottsym{(}  \ell \, \ottnt{v}  \ottsym{:}   \langle    \ell  \mathbin{:}  \ottnt{A}   ;  \star   \rangle  \,  \stackrel{ \ottnt{p} }{\Rightarrow}   \langle  \star  \rangle    \ottsym{)} }  & \multirow{2}{*}{\RWoP{VConInj}} \\
  \multicolumn{3}{r}{
    \text{(if $\ell \,  \not\in  \, \mathit{dom} \, \ottsym{(}  \rho_{{\mathrm{2}}}  \ottsym{)}$ and $\rho_{{\mathrm{2}}} \,  \not=  \, \star$)}
  }
 \\
 \multicolumn{3}{l}{
  \ottsym{(}   \variantlift{ \ell }{ \ottnt{A} }{ \ottnt{v} }   \ottsym{)}  \ottsym{:}   \langle    \ell  \mathbin{:}  \ottnt{A}   ;  \rho_{{\mathrm{1}}}   \rangle  \,  \stackrel{ \ottnt{p} }{\Rightarrow}   \langle  \rho_{{\mathrm{2}}}  \rangle   \  \rightsquigarrow 
 } & \RWoP{VConLift} \\
  \multicolumn{4}{r}{
   \ottsym{(}   \variantliftdown{ \rho_{{\mathrm{2}}} }{ \ell }{ \ottnt{A} }{ \ottsym{(}  \ottnt{v}  \ottsym{:}   \langle  \rho_{{\mathrm{1}}}  \rangle  \,  \stackrel{ \ottnt{p} }{\Rightarrow}   \langle  \rho_{{\mathrm{2}}}  \rangle    \ottsym{)} }   \ottsym{)}  \ottsym{:}   \langle     \rho_{{\mathrm{2}}}  \mathrel{@}   \ell  \mathbin{:}  \ottnt{A}      \rangle  \,  \stackrel{ \ottnt{p} }{\Rightarrow}   \langle  \rho_{{\mathrm{2}}}  \rangle   \hfil
   \text{(if $\ell \,  \not\in  \, \mathit{dom} \, \ottsym{(}  \rho_{{\mathrm{2}}}  \ottsym{)}$ and $\rho_{{\mathrm{2}}} \,  \not=  \, \star$)}
  }
\end{array}\]

 \end{flushleft}
 \caption{Reduction rules for casts between variant types.}
 \label{fig:blame:red_variant}
\end{myfigure}

The reduction rules for casts between variant types are given in
\reffig{blame:red_variant}.  The first four rules are similar to ones for
records.  The other four rules are for a cast from $ \langle    \ell  \mathbin{:}  \ottnt{A}   ;  \rho_{{\mathrm{1}}}   \rangle $ to
$ \langle  \rho_{{\mathrm{2}}}  \rangle $ where $\rho_{{\mathrm{2}}} \,  \not=  \, \star$.  We can suppose that the cast variant
value is an injection tagged with $\ell$ or an embedding value with $ \ell  \mathbin{:}  \ottnt{A} $
under the assumption that it is typed at $ \langle    \ell  \mathbin{:}  \ottnt{A}   ;  \rho_{{\mathrm{1}}}   \rangle $.

The rules \R{VRevInj} and \R{VRevLift} are applied if $\rho_{{\mathrm{2}}}$ holds an $\ell$
field.  We use \emph{row concatenation} to split $\rho_{{\mathrm{2}}}$ into the preceding fields
$\rho_{{\mathrm{21}}}$ such that $\ell \,  \not\in  \, \mathit{dom} \, \ottsym{(}  \rho_{{\mathrm{21}}}  \ottsym{)}$, the first $\ell$ field with type
$\ottnt{B}$, and the following fields $\rho_{{\mathrm{22}}}$ after the $\ell$ field.
\iffull
\begin{defn}[Row concatenation]
 Row concatenation $\rho_{{\mathrm{1}}}  \odot  \rho_{{\mathrm{2}}}$ is defined as follows:
 \[\begin{array}{lll}
   \cdot   \odot  \rho_{{\mathrm{2}}} &\defeq& \rho_{{\mathrm{2}}} \\
  \ottsym{(}    \ell  \mathbin{:}  \ottnt{A}   ;  \rho_{{\mathrm{1}}}   \ottsym{)}  \odot  \rho_{{\mathrm{2}}} &\defeq&   \ell  \mathbin{:}  \ottnt{A}   ;  \ottsym{(}  \rho_{{\mathrm{1}}}  \odot  \rho_{{\mathrm{2}}}  \ottsym{)}  \\
   \end{array}\]
\end{defn}

\else
Row concatenation $ \odot $ is defined by:
$\ottsym{(}    \ell_{{\mathrm{1}}}  \mathbin{:}  \ottnt{A_{{\mathrm{1}}}}   ;  ...  ;    \ell_{\ottmv{n}}  \mathbin{:}  \ottnt{A_{\ottmv{n}}}   ;   \cdot     \ottsym{)}  \odot  \rho_{{\mathrm{2}}} \,  =  \,   \ell_{{\mathrm{1}}}  \mathbin{:}  \ottnt{A_{{\mathrm{1}}}}   ;  ...  ;    \ell_{\ottmv{n}}  \mathbin{:}  \ottnt{A_{\ottmv{n}}}   ;  \rho_{{\mathrm{2}}}  $.
\fi

If the cast variant value is an injection $\ell \, \ottnt{v}$, the cast reduces by
\R{VRevInj}.  Since the target variant type $ \langle  \rho_{{\mathrm{2}}}  \rangle $ requires a value
injected with $\ell$ to be typed at $\ottnt{B}$, the reduction result injects the
result of casting $\ottnt{v}$ to $\ottnt{B}$ with $\ell$.  Furthermore, the injection
$\ell \, \ottsym{(}  \ottnt{v}  \ottsym{:}  \ottnt{A} \,  \stackrel{ \ottnt{p} }{\Rightarrow}  \ottnt{B}   \ottsym{)}$, which can be typed at $ \langle    \ell  \mathbin{:}  \ottnt{B}   ;  \rho_{{\mathrm{22}}}   \rangle $, is
embedded into $ \langle    \rho_{{\mathrm{21}}}  \odot  \ottsym{(}    \ell  \mathbin{:}  \ottnt{B}   ;   \cdot    \ottsym{)}  \odot  \rho_{{\mathrm{22}}}    \rangle  =  \langle  \rho_{{\mathrm{2}}}  \rangle $
by a sequence of applications of the embedding operation with fields in
$\rho_{{\mathrm{21}}}$, which is defined as follows.
\begin{defn}[Row embedding]
 Row embedding $ \variantliftrow{ \rho }{ \ottnt{e} } $ is defined as follows:
 \ifpaper
 \[
   \variantliftrow{ \ottsym{(}    \ell  \mathbin{:}  \ottnt{A}   ;  \rho   \ottsym{)} }{ \ottnt{e} }  \ \defeq \  \variantlift{ \ell }{ \ottnt{A} }{ \ottsym{(}   \variantliftrow{ \rho }{ \ottnt{e} }   \ottsym{)} }  \qquad
   \variantliftrow{ \rho }{ \ottnt{e} }          \ \defeq \ \ottnt{e} \quad \text{(if $\rho$ is not a row extension)}
 \]
 \else
 \[\begin{array}{lll}
   \variantliftrow{ \ottsym{(}    \ell  \mathbin{:}  \ottnt{A}   ;  \rho   \ottsym{)} }{ \ottnt{e} }   &\defeq&  \variantlift{ \ell }{ \ottnt{A} }{ \ottsym{(}   \variantliftrow{ \rho }{ \ottnt{e} }   \ottsym{)} }  \\
   \variantliftrow{ \rho }{ \ottnt{e} }           &\defeq& \ottnt{e} \quad \text{(if $\rho \,  \not=  \, \ottsym{(}    \ell  \mathbin{:}  \ottnt{A}   ;  \rho'   \ottsym{)}$)}
   \end{array}\]
 \fi
\end{defn}

The rule \R{VRevLift} is applied if the cast variant value is an embedding value
$ \variantlift{ \ell }{ \ottnt{A} }{ \ottnt{v} } $.  In this case, the field $ \ell  \mathbin{:}  \ottnt{B} $ in $\rho_{{\mathrm{2}}}$ is inserted by
applying the embedding operation to the result of casting $\ottnt{v}$ to the variant
type $ \langle    \rho_{{\mathrm{21}}}  \odot  \rho_{{\mathrm{22}}}    \rangle $ with the other fields.  The insertion of
field $ \ell  \mathbin{:}  \ottnt{B} $ is performed by the following operation.
\begin{defn}[Field insertion]
 Function $ \variantliftdown{ \rho }{ \ell }{ \ottnt{A} }{ \ottnt{e} } $ embeds a term $\ottnt{e}$ of type $ \langle    \rho  \odot  \rho'    \rangle $ into $ \langle    \rho  \odot  \ottsym{(}    \ell  \mathbin{:}  \ottnt{A}   ;   \cdot    \ottsym{)}  \odot  \rho'    \rangle $.  Formally, it is
 defined as follows:
 \[\begin{array}{lll}
   \variantliftdown{ \ottsym{(}    \ell'  \mathbin{:}  \ottnt{B'}   ;  \rho   \ottsym{)} }{ \ell }{ \ottnt{A} }{ \ottnt{e} }   &\defeq&  \mathsf{case} \,  \ottnt{e}  \,\mathsf{with}\, \langle  \ell' \,  \mathit{x}   \rightarrow   \ell' \, \mathit{x}   \ottsym{;}   \mathit{y}   \rightarrow    \variantlift{ \ell' }{ \ottnt{B'} }{ \ottsym{(}   \variantliftdown{ \rho }{ \ell }{ \ottnt{A} }{ \mathit{y} }   \ottsym{)} }   \rangle  \\
   \variantliftdown{ \rho }{ \ell }{ \ottnt{A} }{ \ottnt{e} }             &\defeq&  \variantlift{ \ell }{ \ottnt{A} }{ \ottnt{e} }  \quad
   \text{(if \ifpaper $\rho$ is not a row extension\else $\rho \,  \not=  \, \ottsym{(}    \ell'  \mathbin{:}  \ottnt{B'}   ;  \rho'   \ottsym{)}$ for any $\ell'$, $\ottnt{B'}$, and $\rho'$\fi)}
   \end{array}\]
\end{defn}

\noindent
Row embedding $ \variantliftrow{ \rho }{ \ottnt{e} } $ is justified as follows. If $\rho$ is not a row
extension (i.e., it is the empty row), then $\ottnt{e}$ is typed at
$ \langle    \rho  \odot  \rho'    \rangle  \,  =  \,  \langle  \rho'  \rangle $ and, therefore, $ \variantlift{ \ell }{ \ottnt{A} }{ \ottnt{e} } $ has
type $ \langle    \ell  \mathbin{:}  \ottnt{A}   ;  \rho'   \rangle  =  \langle    \rho  \odot  \ottsym{(}    \ell  \mathbin{:}  \ottnt{A}   ;   \cdot    \ottsym{)}  \odot  \rho'    \rangle $.  If
$\rho$ is row extension $  \ell'  \mathbin{:}  \ottnt{B'}   ;  \rho'' $, then it is checked whether
$\ottnt{e}$ is an injection or an embedding term with $\ell'$ by case matching.  If
$\ottnt{e}$ is an injection, it can have type $ \langle    \ottsym{(}    \ell'  \mathbin{:}  \ottnt{B'}   ;  \rho''   \ottsym{)}  \odot  \ottsym{(}    \ell  \mathbin{:}  \ottnt{A}   ;   \cdot    \ottsym{)}  \odot  \rho'    \rangle $ because in general an injection with $\ell$ can be
typed at $ \langle    \ell  \mathbin{:}  \ottnt{A}   ;  \rho   \rangle $ for any row $\rho$.  Otherwise, if $\ottnt{e}$ is an
embedding term, the embedded value $\mathit{y}$ is typed at $ \langle    \rho''  \odot  \rho'    \rangle $.
Thus, row embedding is recursively applied to embed $\mathit{y}$ into
$ \langle    \rho''  \odot  \ottsym{(}    \ell  \mathbin{:}  \ottnt{A}   ;   \cdot    \ottsym{)}  \odot  \rho'    \rangle $, and then, the embedding
operation with $ \ell'  \mathbin{:}  \ottnt{B'} $ is applied to the result in order to embed it into
$ \langle    \ottsym{(}    \ell'  \mathbin{:}  \ottnt{B'}   ;  \rho''   \ottsym{)}  \odot  \ottsym{(}    \ell  \mathbin{:}  \ottnt{A}   ;   \cdot    \ottsym{)}  \odot  \rho'    \rangle $.

The last two rules \R{VConInj} and \R{VConLift} are for the case that $\rho_{{\mathrm{2}}}$
does not hold an $\ell$ field.  In this case, we can suppose that $\rho_{{\mathrm{2}}}$ ends
with $ \star $ under the assumption that the reduced term is well typed.  If the
cast variant value is an injection $\ell \, \ottnt{v}$, it is cast to the variant type
$ \langle  \star  \rangle $ and then embedded into type $ \langle  \rho_{{\mathrm{2}}}  \rangle $ \R{VConInj}.  If
the cast value is an embedding value $ \variantlift{ \ell }{ \ottnt{A} }{ \ottnt{v} } $, the embedded value
$\ottnt{v}$ is cast to $ \langle  \rho_{{\mathrm{2}}}  \rangle $ and field $ \ell  \mathbin{:}  \ottnt{A} $ is inserted, and then
the result is cast to $ \langle  \rho_{{\mathrm{2}}}  \rangle $ again \R{VConLift}.  The field insertion
is necessary for dynamic gradual
guarantee~\cite{Siek/Vitousek/Cimini/Boyland_2015_SNAPL}; in this paper we do
not prove that property, but we will show the need of the field insertion by
examples in the following.

\paragraph{Examples.}
First, let us consider cast
$\ottsym{(}   \variantlift{ \ell_{{\mathrm{2}}} }{  \mathsf{bool}  }{ \ottsym{(}  \ell_{{\mathrm{1}}} \,  0   \ottsym{)} }   \ottsym{)}  \ottsym{:}   \langle    \ell_{{\mathrm{2}}}  \mathbin{:}   \mathsf{bool}    ;    \ell_{{\mathrm{1}}}  \mathbin{:}   \mathsf{int}    ;   \cdot     \rangle  \,  \stackrel{ \ottnt{p} }{\Rightarrow}   \langle  \star  \rangle  $,
which reduces as follows.
{
\[\fboxsep=0em
\begin{array}{@{}lll} &
   \shadeback:   \ottsym{(}   \variantlift{ \ell_{{\mathrm{2}}} }{  \mathsf{bool}  }{ \ottsym{(}  \ell_{{\mathrm{1}}} \,  0   \ottsym{)} }   \ottsym{)}  \ottsym{:}   \langle    \ell_{{\mathrm{2}}}  \mathbin{:}   \mathsf{bool}    ;    \ell_{{\mathrm{1}}}  \mathbin{:}   \mathsf{int}    ;   \cdot     \rangle  \,  \stackrel{ \ottnt{p} }{\Rightarrow}   \langle  \star  \rangle      :  \\[.5ex]
  \longrightarrow  &
   \undline<    \shadeback:    \variantlift{ \ell_{{\mathrm{2}}} }{  \mathsf{bool}  }{ \ottsym{(}  \ell_{{\mathrm{1}}} \,  0   \ottsym{)} }    \ottsym{:}   \langle    \ell_{{\mathrm{2}}}  \mathbin{:}   \mathsf{bool}    ;    \ell_{{\mathrm{1}}}  \mathbin{:}   \mathsf{int}    ;   \cdot     \rangle  \,  \stackrel{ \ottnt{p} }{\Rightarrow}    \langle    \ell_{{\mathrm{2}}}  \mathbin{:}  \star   ;  \star   \rangle    :   \,  \stackrel{ \ottnt{p} }{\Rightarrow}   \langle  \star  \rangle      > 
  & \R{VToDyn} \\[.5ex]
  \longrightarrow  &
  \undline<     \variantlift{ \ell_{{\mathrm{2}}} }{ \star }{     \shadeback:   \ottsym{(}  \ottsym{(}  \ell_{{\mathrm{1}}} \,  0   \ottsym{)}  \ottsym{:}   \langle    \ell_{{\mathrm{1}}}  \mathbin{:}   \mathsf{int}    ;   \cdot    \rangle  \,  \stackrel{ \ottnt{p} }{\Rightarrow}   \langle  \star  \rangle    \ottsym{)}    :    }    >   \ottsym{:}   \langle    \ell_{{\mathrm{2}}}  \mathbin{:}  \star   ;  \star   \rangle  \,  \stackrel{ \ottnt{p} }{\Rightarrow}   \langle  \star  \rangle   
  & \R{VRevLift} \\[.5ex]
  \longrightarrow  &
  \variantlift{ \ell_{{\mathrm{2}}} }{ \star }{     \undline<   \ottsym{(}   \shadeback:   \ottsym{(}  \ell_{{\mathrm{1}}} \,  0   \ottsym{)}   \ottsym{:}   \langle    \ell_{{\mathrm{1}}}  \mathbin{:}   \mathsf{int}    ;   \cdot    \rangle  \,  \stackrel{ \ottnt{p} }{\Rightarrow}    \langle    \ell_{{\mathrm{1}}}  \mathbin{:}  \star   ;  \star   \rangle    :   \,  \stackrel{ \ottnt{p} }{\Rightarrow}   \langle  \star  \rangle    \ottsym{)}    >     \ottsym{:}   \langle    \ell_{{\mathrm{2}}}  \mathbin{:}  \star   ;  \star   \rangle  \,  \stackrel{ \ottnt{p} }{\Rightarrow}   \langle  \star  \rangle   } 
  & \R{VToDyn} \\[.5ex]
  \longrightarrow  &
  \variantlift{ \ell_{{\mathrm{2}}} }{ \star }{ \ottsym{(}    \undline<   \ottsym{(}  \ell_{{\mathrm{1}}} \, \ottsym{(}   0   \ottsym{:}   \mathsf{int}  \,  \stackrel{ \ottnt{p} }{\Rightarrow}  \star   \ottsym{)}  \ottsym{)}    >   \ottsym{:}   \langle    \ell_{{\mathrm{1}}}  \mathbin{:}  \star   ;  \star   \rangle  \,  \stackrel{ \ottnt{p} }{\Rightarrow}   \langle  \star  \rangle    \ottsym{)}  \ottsym{:}   \langle    \ell_{{\mathrm{2}}}  \mathbin{:}  \star   ;  \star   \rangle  \,  \stackrel{ \ottnt{p} }{\Rightarrow}   \langle  \star  \rangle   } 
  & \R{VRevInj}
\end{array}\]
}

Next, let us cast the above result to $ \langle    \ell  \mathbin{:}  \ottnt{A}   ;  \star   \rangle $;
let $\ottnt{v} \defeq \ottsym{(}  \ell_{{\mathrm{1}}} \, \ottsym{(}   0   \ottsym{:}   \mathsf{int}  \,  \stackrel{ \ottnt{p} }{\Rightarrow}  \star   \ottsym{)}  \ottsym{)}  \ottsym{:}   \langle    \ell_{{\mathrm{1}}}  \mathbin{:}  \star   ;  \star   \rangle  \,  \stackrel{ \ottnt{p} }{\Rightarrow}   \langle  \star  \rangle  $.
Then,
\begin{equation}
 \ottsym{(}   \variantlift{ \ell_{{\mathrm{2}}} }{ \star }{ \ottnt{v} }   \ottsym{)}  \ottsym{:}   \langle    \ell_{{\mathrm{2}}}  \mathbin{:}  \star   ;  \star   \rangle  \,  \stackrel{ \ottnt{p} }{\Rightarrow}   \langle  \star  \rangle   \,  \stackrel{ \ottnt{q} }{\Rightarrow}   \langle    \ell  \mathbin{:}  \ottnt{A}   ;  \star   \rangle  
  \longrightarrow 
 \ottsym{(}   \variantlift{ \ell_{{\mathrm{2}}} }{ \star }{ \ottnt{v} }   \ottsym{)}  \ottsym{:}   \langle    \ell_{{\mathrm{2}}}  \mathbin{:}  \star   ;  \star   \rangle  \,  \stackrel{ \ottnt{q} }{\Rightarrow}   \langle    \ell  \mathbin{:}  \ottnt{A}   ;  \star   \rangle  
 \label{eqn:variant:example1}
\end{equation}
by \R{VFromDyn}.

If $\ell \,  =  \, \ell_{{\mathrm{2}}}$, then (\ref{eqn:variant:example1}) reduces to
$ \variantlift{ \ell_{{\mathrm{2}}} }{ \ottnt{A} }{ \ottsym{(}  \ottnt{v}  \ottsym{:}   \langle  \star  \rangle  \,  \stackrel{ \ottnt{q} }{\Rightarrow}   \langle  \star  \rangle    \ottsym{)} } $ by \R{VRevLift}.
Thus, the cast just changes the type given to the embedding operation.

If $\ell \,  \not=  \, \ell_{{\mathrm{2}}}$, then, by \R{VConLift}, (\ref{eqn:variant:example1}) reduces to:
\begin{equation}
 \ottsym{(}   \variantliftdown{   \ell  \mathbin{:}  \ottnt{A}   ;  \star  }{ \ell_{{\mathrm{2}}} }{ \star }{ \ottsym{(}    \shadeback:   \ottnt{v}   \ottsym{:}   \langle  \star  \rangle  \,  \stackrel{ \ottnt{q} }{\Rightarrow}   \langle    \ell  \mathbin{:}  \ottnt{A}   ;  \star   \rangle     :   \ottsym{)} }   \ottsym{)}  \ottsym{:}   \langle    \ell  \mathbin{:}  \ottnt{A}   ;    \ell_{{\mathrm{2}}}  \mathbin{:}  \star   ;  \star    \rangle  \,  \stackrel{ \ottnt{p} }{\Rightarrow}   \langle    \ell  \mathbin{:}  \ottnt{A}   ;  \star   \rangle  .
 \label{eqn:variant:example2}
\end{equation}

If $\ell \,  =  \, \ell_{{\mathrm{1}}}$, then the shaded part in (\ref{eqn:variant:example2}) reduces to
$\ell_{{\mathrm{1}}} \, \ottsym{(}   0   \ottsym{:}   \mathsf{int}  \,  \stackrel{ \ottnt{p} }{\Rightarrow}  \star  \,  \stackrel{ \ottnt{q} }{\Rightarrow}  \ottnt{A}   \ottsym{)}$ by \R{VFromDyn} and \R{VRevInj}.  Thus, if $\ottnt{A} \,  \not=  \,  \mathsf{int} $, $\mathsf{blame} \, \ottnt{q}$ is raised; otherwise, (\ref{eqn:variant:example2})
reduces to
\[
 \smash{\ottsym{(}   \variantliftdown{   \ell  \mathbin{:}  \ottnt{A}   ;  \star  }{ \ell_{{\mathrm{2}}} }{ \star }{ \ottsym{(}    \undline<   \ell_{{\mathrm{1}}} \,  0     >   \ottsym{)} }   \ottsym{)}  \ottsym{:}   \langle    \ell  \mathbin{:}  \ottnt{A}   ;    \ell_{{\mathrm{2}}}  \mathbin{:}  \star   ;  \star    \rangle  \,  \stackrel{ \ottnt{p} }{\Rightarrow}   \langle    \ell  \mathbin{:}  \ottnt{A}   ;  \star   \rangle  }
\]

If $\ell \,  \not=  \, \ell_{{\mathrm{1}}}$, then the shaded part in (\ref{eqn:variant:example2}) reduces
to $ \variantlift{ \ell }{ \ottnt{A} }{ \ottsym{(}  \ell_{{\mathrm{1}}} \, \ottsym{(}   0   \ottsym{:}   \mathsf{int}  \,  \stackrel{ \ottnt{p} }{\Rightarrow}  \star   \ottsym{)}  \ottsym{:}   \langle    \ell_{{\mathrm{1}}}  \mathbin{:}  \star   ;  \star   \rangle  \,  \stackrel{ \ottnt{q} }{\Rightarrow}   \langle  \star  \rangle    \ottsym{)} } $
by \R{VFromDyn} and \R{VConInj}.  Note that if $  \ell  \mathbin{:}  \ottnt{A}   ;  \star $ in the shaded part
were $  \ell  \mathbin{:}  \ottnt{A}   ;   \cdot  $, exception $\mathsf{blame} \, \ottnt{q}$ would be raised by
using \R{VBlame}.

Finally, we show that the field insertion in \R{VConLift} is crucial to
prove dynamic gradual guarantee.  To confirm that, let us suppose that
\R{VConLift} does not perform field insertion and instead takes the following
form.
\[
 \ottsym{(}   \variantlift{ \ell }{ \ottnt{A} }{ \ottnt{v} }   \ottsym{)}  \ottsym{:}   \langle    \ell  \mathbin{:}  \ottnt{A}   ;  \rho_{{\mathrm{1}}}   \rangle  \,  \stackrel{ \ottnt{p} }{\Rightarrow}   \langle  \rho_{{\mathrm{2}}}  \rangle  
 {\ \   \rightsquigarrow  \ \ }
 \ottnt{v}  \ottsym{:}   \langle  \rho_{{\mathrm{1}}}  \rangle  \,  \stackrel{ \ottnt{p} }{\Rightarrow}   \langle  \rho_{{\mathrm{2}}}  \rangle  
 \quad \R{VConLift'}
\]
As an example, consider reduction of term $\ottnt{e}$ given as follows:
\[\fboxsep=0em
\begin{array}{lll}
 \ottnt{v} & \defeq & \ell \, \ottsym{(}   0   \ottsym{:}   \mathsf{int}  \,  \stackrel{ \ottnt{p_{{\mathrm{1}}}} }{\Rightarrow}  \star   \ottsym{)}  \ottsym{:}   \langle    \ell  \mathbin{:}  \star   ;  \star   \rangle  \,  \stackrel{ \ottnt{p_{{\mathrm{2}}}} }{\Rightarrow}   \langle  \star  \rangle   \\
 \ottnt{e} & \defeq &  \shadeback:   \ottsym{(}   \variantlift{ \ell }{  \mathsf{bool}  }{ \ottnt{v} }   \ottsym{)}  \ottsym{:}   \langle    \ell  \mathbin{:}   \mathsf{bool}    ;  \star   \rangle  \,  \stackrel{ \ottnt{p_{{\mathrm{3}}}} }{\Rightarrow}    \langle    \ell'  \mathbin{:}   \mathsf{str}    ;  \mathit{X}   \rangle    :   \,  \stackrel{ \ottnt{p_{{\mathrm{4}}}} }{\Rightarrow}   \langle    \ell  \mathbin{:}   \mathsf{bool}    ;  \star   \rangle   
\end{array}
 \]
Dynamic gradual guarantee~\cite{Siek/Vitousek/Cimini/Boyland_2015_SNAPL} states
that changing types in a program to $ \star $ does not change its behavior.  In
the case of $\ottnt{e}$, it means that, if $ \ottnt{e}    [  \ottsym{(}    \ell  \mathbin{:}   \mathsf{bool}    ;  \star   \ottsym{)}  /  \mathit{X}  ]  $ does not raise
blame, $ \ottnt{e}    [  \star  /  \mathit{X}  ]  $ does not either.
First, let us reduce $ \ottnt{e}    [  \ottsym{(}    \ell  \mathbin{:}   \mathsf{bool}    ;  \star   \ottsym{)}  /  \mathit{X}  ]  $.
{
\[
 \fboxsep=0em
\begin{array}{lll} &
  \undlinesq[    \ottsym{(}   \variantliftdown{   \ell'  \mathbin{:}   \mathsf{str}    ;   \cdot   }{ \ell }{  \mathsf{bool}  }{ \ottsym{(}    \shadeback:   \ottnt{v}   \ottsym{:}   \langle  \star  \rangle  \,  \stackrel{ \ottnt{p_{{\mathrm{3}}}} }{\Rightarrow}   \langle    \ell'  \mathbin{:}   \mathsf{str}    ;  \star   \rangle     :   \ottsym{)} }   \ottsym{)}   ]   \ottsym{:}   \langle    \ell'  \mathbin{:}   \mathsf{str}    ;    \ell  \mathbin{:}   \mathsf{bool}    ;  \star    \rangle  \,  \stackrel{ \ottnt{p_{{\mathrm{4}}}} }{\Rightarrow}   \langle    \ell  \mathbin{:}   \mathsf{bool}    ;  \star   \rangle    & \R{VRevLift} \\[.5ex]
  \longrightarrow^{*} &
  \shadeback:    \ottsym{(}   \variantliftdown{   \ell'  \mathbin{:}   \mathsf{str}    ;   \cdot   }{ \ell }{  \mathsf{bool}  }{ \ottsym{(}    \undline<    \variantlift{ \ell' }{  \mathsf{str}  }{ \ottnt{v} }     >   \ottsym{)} }   \ottsym{)}   :   \ottsym{:}   \langle    \ell'  \mathbin{:}   \mathsf{str}    ;    \ell  \mathbin{:}   \mathsf{bool}    ;  \star    \rangle  \,  \stackrel{ \ottnt{p_{{\mathrm{4}}}} }{\Rightarrow}   \langle    \ell  \mathbin{:}   \mathsf{bool}    ;  \star   \rangle    \\
  \longrightarrow^{*} &
   \shadeback:   \ottsym{(}    \undline<    \variantlift{ \ell' }{  \mathsf{str}  }{ \ottsym{(}   \variantlift{ \ell }{  \mathsf{bool}  }{ \ottnt{v} }   \ottsym{)} }     >   \ottsym{)}  \ottsym{:}   \langle    \ell'  \mathbin{:}   \mathsf{str}    ;    \ell  \mathbin{:}   \mathsf{bool}    ;  \star    \rangle  \,  \stackrel{ \ottnt{p_{{\mathrm{4}}}} }{\Rightarrow}   \langle    \ell  \mathbin{:}   \mathsf{bool}    ;  \star   \rangle      :  & \R{CaseR} \\[.5ex]
  \longrightarrow^{*} &
   \undline<   \ottsym{(}   \variantlift{ \ell }{  \mathsf{bool}  }{ \ottnt{v} }   \ottsym{)}  \ottsym{:}   \langle    \ell  \mathbin{:}   \mathsf{bool}    ;  \star   \rangle  \,  \stackrel{ \ottnt{p_{{\mathrm{4}}}} }{\Rightarrow}   \langle    \ell  \mathbin{:}   \mathsf{bool}    ;  \star   \rangle      >  & \R{VConLift'} \\[.5ex]
  \longrightarrow^{*} &
  \variantlift{ \ell }{  \mathsf{bool}  }{ \ottnt{v} } 
\end{array}\]
}
Thus, $ \ottnt{e}    [  \ottsym{(}    \ell  \mathbin{:}   \mathsf{bool}    ;  \star   \ottsym{)}  /  \mathit{X}  ]  $ evaluates to a value under use of \R{VConLift'}.
If dynamic gradual guarantee holds, so should $ \ottnt{e}    [  \star  /  \mathit{X}  ]  $.  However, it does
not:
{
\[\fboxsep=0em
\begin{array}{llll}
  \ottnt{e}    [  \star  /  \mathit{X}  ]   & \longrightarrow &   \undline<    \shadeback:   \ottnt{v}   \ottsym{:}   \langle  \star  \rangle  \,   \stackrel{ \ottnt{p_{{\mathrm{3}}}} }{\Rightarrow}    \langle    \ell'  \mathbin{:}   \mathsf{str}    ;  \star   \rangle    :     >  \,  \stackrel{ \ottnt{p_{{\mathrm{4}}}} }{\Rightarrow}   \langle    \ell  \mathbin{:}   \mathsf{bool}    ;  \star   \rangle    & \R{VConLift'} \\[.5ex]
              & \longrightarrow^{*} &   \shadeback:     \undline<    \variantlift{ \ell' }{  \mathsf{str}  }{ \ottnt{v} }     >   \ottsym{:}   \langle    \ell'  \mathbin{:}   \mathsf{str}    ;  \star   \rangle  \,  \stackrel{ \ottnt{p_{{\mathrm{4}}}} }{\Rightarrow}   \langle    \ell  \mathbin{:}   \mathsf{bool}    ;  \star   \rangle      :      & \text{\R{VFromDyn} and \R{VConInj}} \\[1ex]
              & \longrightarrow &     \undline<    \shadeback:   \ottnt{v}   \ottsym{:}   \langle  \star  \rangle  \,  \stackrel{ \ottnt{p_{{\mathrm{4}}}} }{\Rightarrow}   \langle    \ell  \mathbin{:}   \mathsf{bool}    ;  \star   \rangle      :    >                            & \R{VConLift'} \\[1ex]
              & \longrightarrow^{*} &    \undline<   \ell \,  \shadeback:   \ottsym{(}   0   \ottsym{:}   \mathsf{int}  \,  \stackrel{ \ottnt{p_{{\mathrm{1}}}} }{\Rightarrow}  \star  \,  \stackrel{ \ottnt{p_{{\mathrm{4}}}} }{\Rightarrow}   \mathsf{bool}    \ottsym{)}     :    >                                       & \text{\R{VFromDyn} and \R{VRevInj}} \\[.5ex]
              & \longrightarrow^{*} &   \undline<   \mathsf{blame} \, \ottnt{p_{{\mathrm{4}}}}    >                                                              & \R{Blame}
\end{array}\]
}

We can confirm that both $ \ottnt{e}    [  \ottsym{(}    \ell  \mathbin{:}   \mathsf{bool}    ;  \star   \ottsym{)}  /  \mathit{X}  ]  $ and $ \ottnt{e}    [  \star  /  \mathit{X}  ]  $ evaluate
to values if we use \R{VConLift}.  We show only the reduction of $ \ottnt{e}    [  \star  /  \mathit{X}  ]  $;
the reduction of $ \ottnt{e}    [  \ottsym{(}    \ell  \mathbin{:}   \mathsf{bool}    ;  \star   \ottsym{)}  /  \mathit{X}  ]  $ is similar to the case of using \R{VConLift'}.
\TS{Add the appendix?}
\[\fboxsep=0em \small
\begin{array}{lll}
 \multicolumn{2}{l}{
  \undlinesq[   \ottsym{(}   \variantliftdown{   \ell'  \mathbin{:}   \mathsf{str}    ;  \star  }{ \ell }{  \mathsf{bool}  }{ \ottsym{(}    \shadeback:   \ottnt{v}   \ottsym{:}   \langle  \star  \rangle  \,  \stackrel{ \ottnt{p_{{\mathrm{3}}}} }{\Rightarrow}   \langle    \ell'  \mathbin{:}   \mathsf{str}    ;  \star   \rangle     :   \ottsym{)} }   \ottsym{)}  \ottsym{:}   \langle    \ell'  \mathbin{:}   \mathsf{str}    ;    \ell  \mathbin{:}   \mathsf{bool}    ;  \star    \rangle  \,  \stackrel{ \ottnt{p_{{\mathrm{3}}}} }{\Rightarrow}    \langle    \ell'  \mathbin{:}   \mathsf{str}    ;  \star   \rangle    ]   \,  \stackrel{ \ottnt{p_{{\mathrm{4}}}} }{\Rightarrow}   \langle    \ell  \mathbin{:}   \mathsf{bool}    ;  \star   \rangle   
 } & \R{VConLift} \\[.5ex]
   \longrightarrow^{*} &
   \shadeback:    \ottsym{(}   \variantliftdown{   \ell'  \mathbin{:}   \mathsf{str}    ;  \star  }{ \ell }{  \mathsf{bool}  }{ \ottsym{(}    \undline<    \variantlift{ \ell' }{  \mathsf{str}  }{ \ottnt{v} }     >   \ottsym{)} }   \ottsym{)}   :   \ottsym{:}   \langle    \ell'  \mathbin{:}   \mathsf{str}    ;    \ell  \mathbin{:}   \mathsf{bool}    ;  \star    \rangle  \,  \stackrel{ \ottnt{p_{{\mathrm{3}}}} }{\Rightarrow}   \langle    \ell'  \mathbin{:}   \mathsf{str}    ;  \star   \rangle   \,  \stackrel{ \ottnt{p_{{\mathrm{4}}}} }{\Rightarrow}   \langle    \ell  \mathbin{:}   \mathsf{bool}    ;  \star   \rangle    \\
   \longrightarrow &
   \shadeback:   \ottsym{(}    \undline<    \variantlift{ \ell' }{  \mathsf{str}  }{ \ottsym{(}   \variantlift{ \ell }{  \mathsf{bool}  }{ \ottnt{v} }   \ottsym{)} }     >   \ottsym{)}  \ottsym{:}   \langle    \ell'  \mathbin{:}   \mathsf{str}    ;    \ell  \mathbin{:}   \mathsf{bool}    ;  \star    \rangle  \,  \stackrel{ \ottnt{p_{{\mathrm{3}}}} }{\Rightarrow}    \langle    \ell'  \mathbin{:}   \mathsf{str}    ;  \star   \rangle    :   \,  \stackrel{ \ottnt{p_{{\mathrm{4}}}} }{\Rightarrow}   \langle    \ell  \mathbin{:}   \mathsf{bool}    ;  \star   \rangle    & \R{CaseR} \\
   \longrightarrow &
   \undline<    \ottsym{(}   \variantlift{ \ell' }{  \mathsf{str}  }{     \shadeback:   \ottsym{(}  \ottsym{(}   \variantlift{ \ell }{  \mathsf{bool}  }{ \ottnt{v} }   \ottsym{)}  \ottsym{:}   \langle    \ell  \mathbin{:}   \mathsf{bool}    ;  \star   \rangle  \,  \stackrel{ \ottnt{p_{{\mathrm{3}}}} }{\Rightarrow}   \langle  \star  \rangle    \ottsym{)}    :    }   \ottsym{)}   >   \ottsym{:}   \langle    \ell'  \mathbin{:}   \mathsf{str}    ;  \star   \rangle  \,  \stackrel{ \ottnt{p_{{\mathrm{4}}}} }{\Rightarrow}   \langle    \ell  \mathbin{:}   \mathsf{bool}    ;  \star   \rangle    & \R{VRevLift} \\
   \longrightarrow &
  \multicolumn{2}{l}{
     \shadeback:   \ottsym{(}   \variantlift{ \ell' }{  \mathsf{str}  }{     \undline<   \ottnt{v'}    >    }   \ottsym{)}  \ottsym{:}   \langle    \ell'  \mathbin{:}   \mathsf{str}    ;  \star   \rangle  \,  \stackrel{ \ottnt{p_{{\mathrm{4}}}} }{\Rightarrow}   \langle    \ell  \mathbin{:}   \mathsf{bool}    ;  \star   \rangle      :  \hfill \text{\R{VToDyn} and \R{VRevLift}}
  } \\
   \multicolumn{2}{r}{\text{(where $\ottnt{v'} \,  =  \, \ottsym{(}   \variantlift{ \ell }{ \star }{ \ottnt{v} }   \ottsym{)}  \ottsym{:}   \langle    \ell  \mathbin{:}  \star   ;  \star   \rangle  \,  \stackrel{ \ottnt{p_{{\mathrm{3}}}} }{\Rightarrow}   \langle  \star  \rangle  $)}} \\
   \longrightarrow &
    \undlinesq[     \shadeback:   \ottsym{(}   \variantliftdown{   \ell  \mathbin{:}   \mathsf{bool}    ;  \star  }{ \ell' }{  \mathsf{str}  }{ \ottsym{(}  \ottnt{v'}  \ottsym{:}   \langle  \star  \rangle  \,  \stackrel{ \ottnt{p_{{\mathrm{4}}}} }{\Rightarrow}   \langle    \ell  \mathbin{:}   \mathsf{bool}    ;  \star   \rangle    \ottsym{)} }   \ottsym{)}    :   \ottsym{:}   \langle    \ell  \mathbin{:}   \mathsf{bool}    ;    \ell'  \mathbin{:}   \mathsf{str}    ;  \star    \rangle  \,  \stackrel{ \ottnt{p_{{\mathrm{4}}}} }{\Rightarrow}   \langle    \ell  \mathbin{:}   \mathsf{bool}    ;  \star   \rangle      ]  & \R{VConLift} \\
   \longrightarrow^{*} &
    \shadeback:     \undline<   \ottsym{(}   \variantlift{ \ell }{  \mathsf{bool}  }{ \ottsym{(}   \variantlift{ \ell' }{  \mathsf{str}  }{ \ottnt{v} }   \ottsym{)} }   \ottsym{)}    >   \ottsym{:}   \langle    \ell  \mathbin{:}   \mathsf{bool}    ;    \ell'  \mathbin{:}   \mathsf{str}    ;  \star    \rangle  \,  \stackrel{ \ottnt{p_{{\mathrm{4}}}} }{\Rightarrow}   \langle    \ell  \mathbin{:}   \mathsf{bool}    ;  \star   \rangle      :  \\
   \longrightarrow^{*} &
  \ottsym{(}   \variantlift{ \ell }{  \mathsf{bool}  }{ \ottsym{(}  \ottsym{(}   \variantlift{ \ell' }{ \star }{ \ottnt{v} }   \ottsym{)}  \ottsym{:}   \langle    \ell'  \mathbin{:}  \star   ;  \star   \rangle  \,  \stackrel{ \ottnt{p} }{\Rightarrow}   \langle  \star  \rangle    \ottsym{)} }   \ottsym{)}
\end{array}\]
Thus, we believe that the field insertion is key to show dynamic gradual
guarantee, though it is left as future work.

\subsubsection{Evaluation}
\begin{myfigure}[t]
 \begin{flushleft}
  \iffull
  \textbf{Evaluation rules} \quad \framebox{$ \Sigma_{{\mathrm{1}}}  \mid  \ottnt{e_{{\mathrm{1}}}}   \longrightarrow   \Sigma_{{\mathrm{2}}}  \mid  \ottnt{e_{{\mathrm{2}}}} $}
\begin{center}
 $\ottdruleEXXRed{}$ \hfil
 $\ottdruleEXXBlame{}$ \\[3ex]
 \[
   \Sigma  \mid   \ottnt{E}  [  \ottsym{(}    \Lambda\!  \,  \mathit{X}  \mathord{:}  \ottnt{K}   \ottsym{.}   \ottnt{e}  ::  \ottnt{A}   \ottsym{)} \, \ottnt{B}  ]    \longrightarrow   \Sigma  \ottsym{,}   \alpha  \mathord{:}  \ottnt{K}   \ottsym{:=}  \ottnt{B}  \mid   \ottnt{E}  [   \ottnt{e}    [  \alpha  /  \mathit{X}  ]    \ottsym{:}   \ottnt{A}    [  \alpha  /  \mathit{X}  ]   \,  \stackrel{ \ottsym{+}  \alpha }{\Rightarrow}     \ottnt{A}    [  \ottnt{B}  /  \mathit{X}  ]       ]   \quad \EWoP{TyBeta}
 \]
\end{center}

  \else
  \textbf{Evaluation rules} \quad \framebox{$ \Sigma_{{\mathrm{1}}}  \mid  \ottnt{e_{{\mathrm{1}}}}   \longrightarrow   \Sigma_{{\mathrm{2}}}  \mid  \ottnt{e_{{\mathrm{2}}}} $}
  \[\begin{array}{l}
    \Sigma  \mid   \ottnt{E}  [  \ottnt{e_{{\mathrm{1}}}}  ]    \longrightarrow   \Sigma  \mid   \ottnt{E}  [  \ottnt{e_{{\mathrm{2}}}}  ]   \ \text{(if $\ottnt{e_{{\mathrm{1}}}}  \rightsquigarrow  \ottnt{e_{{\mathrm{2}}}}$)} \quad \EWoP{Red} \qquad \hfill
    \Sigma  \mid   \ottnt{E}  [  \mathsf{blame} \, \ottnt{p}  ]    \longrightarrow   \Sigma  \mid  \mathsf{blame} \, \ottnt{p}  \ \text{(if $\ottnt{E} \,  \not=  \,  [\,] $)} \quad \EWoP{Blame} \\
    \Sigma  \mid   \ottnt{E}  [  \ottsym{(}    \Lambda\!  \,  \mathit{X}  \mathord{:}  \ottnt{K}   \ottsym{.}   \ottnt{e}  ::  \ottnt{A}   \ottsym{)} \, \ottnt{B}  ]    \longrightarrow   \Sigma  \ottsym{,}   \alpha  \mathord{:}  \ottnt{K}   \ottsym{:=}  \ottnt{B}  \mid   \ottnt{E}  [   \ottnt{e}    [  \alpha  /  \mathit{X}  ]    \ottsym{:}   \ottnt{A}    [  \alpha  /  \mathit{X}  ]   \,  \stackrel{ \ottsym{+}  \alpha }{\Rightarrow}     \ottnt{A}    [  \ottnt{B}  /  \mathit{X}  ]       ]   \quad \EWoP{TyBeta}
    \end{array}\]
  \fi
 \end{flushleft}
 \caption{Evaluation rules of {\interlang}.}
 \label{fig:blame:eval}
\end{myfigure}
The evaluation rules are shown in \reffig{blame:eval}.  A term evaluates if its
subterm under an evaluation context reduces \E{Red}, triggers an exception
\E{Blame}, or involves type application \E{TyBeta}.  As discussed in
\sect{blame:syntax}, type application $\ottsym{(}    \Lambda\!  \,  \mathit{X}  \mathord{:}  \ottnt{K}   \ottsym{.}   \ottnt{e}  ::  \ottnt{A}   \ottsym{)} \, \ottnt{B}$ generates a fresh
name $\alpha$, substitutes $\alpha$ for $\mathit{X}$ in $\ottnt{e}$, stores the actual type
(or row) $\ottnt{B}$ of $\alpha$ in name store $\Sigma$, and reveals $\ottnt{B}$ to
evaluation context $\ottnt{E}$.

\subsection{Properties}
We show type soundness of {\interlang} via progress and subject reduction.
\iffull
\begin{lemma}[Progress] \label{ref:progress}
 If $\Sigma  \ottsym{;}   \emptyset   \vdash  \ottnt{e}  \ottsym{:}  \ottnt{A}$, then one of the followings holds:
 (1) $\ottnt{e}$ is a value; (2) $\ottnt{e} \,  =  \, \mathsf{blame} \, \ottnt{p}$ for some $\ell$; or
 (3) $ \Sigma  \mid  \ottnt{e}   \longrightarrow   \Sigma'  \mid  \ottnt{e'} $ for some $\Sigma'$ and $\ottnt{e'}$.
\end{lemma}
\begin{lemma}[Subject reduction] \label{lem:subject-red}
 If $\Sigma  \ottsym{;}   \emptyset   \vdash  \ottnt{e}  \ottsym{:}  \ottnt{A}$ and $ \Sigma  \mid  \ottnt{e}   \longrightarrow   \Sigma'  \mid  \ottnt{e'} $,
 then $\Sigma'  \ottsym{;}   \emptyset   \vdash  \ottnt{e'}  \ottsym{:}  \ottnt{A}$.
\end{lemma}
\fi
\begin{theorem}[Type soundness]
 If $ \emptyset   \ottsym{;}   \emptyset   \vdash  \ottnt{e}  \ottsym{:}  \ottnt{A}$ and $  \emptyset   \mid  \ottnt{e}   \longrightarrow^{*}   \Sigma'  \mid  \ottnt{e'} $ and
 $\ottnt{e'}$ cannot be evaluated under $\Sigma'$, then
 either $\ottnt{e'}$ is a value or $\ottnt{e'} \,  =  \, \mathsf{blame} \, \ottnt{p}$ for some $\ottnt{p}$.
\end{theorem}

We also show that our surface language {\surfacelang} is conservative over
typing of {\interlang}.  We omit the full presentation of {\surfacelang}, but,
as usual~\cite{Siek/Taha_2006_SFPW}, it is obtained by changing {\staticlang} so that (1) types are
extended with $ \star $ and (2) the typing rules use consistent equivalence
instead of type equality.  We write $\Gamma  \vdash  \ottnt{M}  \ottsym{:}  \ottnt{A}$ if $\ottnt{M}$ has type $\ottnt{A}$
under $\Gamma$ in {\surfacelang}.  For example, the typing rule for record
decomposition in {\surfacelang} is
{\small
\[
 \ottdruleTgXXRLet{}
\]
}
where type matching $\ottnt{A}  \triangleright   [  \rho  ] $ is defined as: $\star  \triangleright   [  \star  ] $
and $ [  \rho  ]   \triangleright   [  \rho  ] $.  Type-preserving translation $\Gamma  \vdash  \ottnt{M}  \ottsym{:}  \ottnt{A}  \hookrightarrow  \ottnt{e}$ from $\ottnt{M}$ of $\ottnt{A}$ under $\Gamma$ in {\surfacelang} to $\ottnt{e}$
in {\interlang} is given by inserting casts where type matching and consistent
equivalence are used.  The full definitions are found in the supplementary
material.
\begin{theorem}
 If $\Gamma  \vdash  \ottnt{M}  \ottsym{:}  \ottnt{A}$, then there exists some $\ottnt{e}$ such that
 $\Gamma  \vdash  \ottnt{M}  \ottsym{:}  \ottnt{A}  \hookrightarrow  \ottnt{e}$ and $ \emptyset   \ottsym{;}  \Gamma  \vdash  \ottnt{e}  \ottsym{:}  \ottnt{A}$.
\end{theorem}
We state that the language {\surfacelang} is a conservative extension of
{\staticlang} in terms of typing.
\begin{theorem}[Conservativity over typing]
 Suppose that $ \star $ does not appear in $\Gamma$, $\ottnt{A}$, and $\ottnt{M}$.
 (1) If $\Gamma  \vdash  \ottnt{M}  \ottsym{:}  \ottnt{A}$, then $ \Gamma  \mathrel{ \makestatic{\vdash} }  \ottnt{M}  :  \ottnt{A} $.
 (2) If $ \Gamma  \mathrel{ \makestatic{\vdash} }  \ottnt{M}  :  \ottnt{A} $, then $\Gamma  \vdash  \ottnt{M}  \ottsym{:}  \ottnt{B}$ for some $\ottnt{B}$
 such that $\ottnt{A}  \equiv  \ottnt{B}$.
\end{theorem}

\section{Related work}
\label{sec:relwork}

\subsection{Row types, row polymorphism, and their applications}

Row types were introduced by \citet{Wand_1987_LICS}, who has studied type
inference for objected-oriented languages and modeled objects in a variant of
$\lambda$-calculus equipped with record types and variant types with rows.  Wand
also introduced row type variables for row type inference and discussed row
polymorphism informally.  Although that work supposed labels in a row type to be
unique, it allowed record extension $\ottsym{\{}  \ell  \ottsym{=}  \ottnt{M_{{\mathrm{1}}}}  \ottsym{;}  \ottnt{M_{{\mathrm{2}}}}  \ottsym{\}}$ even for record $\ottnt{M_{{\mathrm{2}}}}$
holding an $\ell$ field; if $\ottnt{M_{{\mathrm{2}}}}$ contains an $\ell$ field, its value will
be overwritten by $\ottnt{M_{{\mathrm{1}}}}$.  However, this overwriting semantics causes an issue
that some programs do not have principal types~\cite{Wand_1991_IC}.
\citet{Gaster/Jones_1996_TR} resolved this issue by allowing record extension
only when record $\ottnt{M_{{\mathrm{2}}}}$ does not contain an $\ell$ field.
With help of presence and absence types~\cite{Remy_1989_POPL},  they
gave a type inference algorithm that produces a principal type (if any).  In
order for row type substitution to preserve uniqueness of labels, they
employed \emph{qualified types} called ``lacks'' predicates, which
constrain quantified row type variables to be instantiated only with row types
that lack some fields.  The use of presence and absence types also enabled them
to deal with record restriction, which was not handled by
\citet{Wand_1987_LICS,Wand_1991_IC}.

Another approach to principal typing for rows is to lift the uniqueness
restriction and to allow scoped labels.  Scoped labels were first discussed by
\citet{Berthomieu/Sagazan_1995_TPA} in the context of process calculi and later
applied to functional programming by \citet{Leijen_2005_TFP}, who also developed
a sound and complete unification algorithm for inference of row types with
scoped labels.  In this work we adopt scoped labels, which enable us not only to
simplify the metatheory of our calculus but also to use the embedding
operation~\cite{Leijen_2005_TFP}.  The embedding operation is helpful to align
variant types with different row types in a polymorphic setting.  In our work,
it is also important to make the type system of {\interlang} syntax-directed.

Row types have been applied, e.g., to model
objects~\cite{Wand_1987_LICS,Wand_1991_IC,Remy/Vouillon_1998_TAPOS} and
polymorphic variants~\cite{Garrigue_1998_ML} and have been found in many programming
languages.  A more recent application of row types is an effect system for
effect handlers~\cite{Plotkin/Pretnar_2009_ESOP}
with~\cite{Leijen_2014_MSFP,Leijen_2017_POPL,Lindley/MacBrid/McLaughlin_2017_POPL}
or without~\cite{Hillerstrom/Lindley_2016_TyDe} scoped labels.  Actually, our
formalization of scoped labels and the embedding operation is influenced by
\citet{Hillerstrom/Lindley/Atkey/Sivaramakrishnan_2017_FSCD} and
\citet{Biernack/Pirog/Polesiuk/Siezkowski_2018_POPL}, respectively.

\subsection{Gradual typing for records and variants}

\citet{Takikawa/Strickland/Dimoulas/Tobin-Hochstadt/Felleisen_2012_OOPSLA}
studied gradual typing for first-class classes.  They employed row types and row
polymorphism for expressing presence and absence of interesting methods.  Thus,
they did not handle variant types and considered row polymorphism together with
lacks predicates.  Their work dealt with specifications written in the form of
contracts and supposed contracts to play a role of interface for module
components.  This style of gradual typing is called ``macro''-level gradual
typing. i.e., typed and untyped modules are mixed, while we focus on
``micro''-level gradual typing, where typed and untyped expressions are mixed.
Technically, this difference appears, e.g., in the need of consistency.  In
addition, as our work, they also protected polymorphically typed values from
untyped code.  Their development, \emph{sealing contracts}, has finer-grained
control than row names in our work in that sealing contracts can expose absence
of fields, while row names cannot.

\citet{Garcia/Clark/Tanter_2016_POPL} proposed a general framework to derive a
gradually typed language from a statically typed one.  They also developed
\emph{gradual rows}, which are rows possibly ending with the dynamic row type,
for record types via application of their framework to a calculus with width and
depth record subtyping.  Thus, a clear difference between their and our work is
support for variant types and row polymorphism.  The consistency relation in
their work involves row equivalence, and, therefore, it seems to be equivalent
to consistent equivalence given by the present work (modulo support for variant
types and row polymorphism).

\citet{Jafery/Dunfield_2017_POPL} introduced dynamic sums for gradual datasort
refinement.  A dynamic sum $\ottnt{A} \mathrel{+^?} \ottnt{B}$ can be interpreted as
both of a single type $\ottnt{A}$ and $\ottnt{B}$, and its value can be deconstructed by
a case expression having a single branch for $\ottnt{A}$ or $\ottnt{B}$.  In our
calculus dynamic sums can be encoded by two-fold variant type
$ \langle    \ell_{{\mathrm{1}}}  \mathbin{:}  \ottnt{A}   ;    \ell_{{\mathrm{2}}}  \mathbin{:}  \ottnt{B}   ;   \cdot     \rangle $ which are coerced to $ \langle    \ell_{{\mathrm{1}}}  \mathbin{:}  \ottnt{A}   ;   \cdot    \rangle $ or
$ \langle    \ell_{{\mathrm{2}}}  \mathbin{:}  \ottnt{B}   ;   \cdot    \rangle $ in case matching via injection to $ \langle  \star  \rangle $.
Unlike our work, they did not deal with labeled fields and row polymorphism.

\section{Conclusion}
\label{sec:conclusion}

\sloppy{
We have introduced the dynamic row type and consistency for gradual typing with
row types and row polymorphism.  While consistency captures the static aspect of
the dynamic row type, we have found that it is problematic if combined with row
equivalence.  To solve the problem with consistency, we have developed
consistent equivalence and shown that it characterizes composition of
consistency and row equivalence.  We also have given a polymorphic blame
calculus {\interlang} with scoped labels, row types, row polymorphism, and
consistent equivalence and proven its type soundness as well as
type-preservation of translation from surface language {\surfacelang} to
{\interlang} and conservativity of {\surfacelang} over typing of {\staticlang}.
The cast semantics of {\interlang} is designed carefully to take into account
criteria of gradual typing~\cite{Siek/Vitousek/Cimini/Boyland_2015_SNAPL}, but
proving them is left as future work.  Another direction of future work is to
extend our calculus to effect systems for effect
handlers~\cite{Plotkin/Pretnar_2009_ESOP}.  It is also interesting to
``gradualize'' other formalisms, such as presence and absence
types~\cite{Pottier/Remy_2005_ATTaPL} and a general framework for row
types~\cite{Morris/McKinna_2019_POPL}.
}

\begin{acks}                            
 We would like to thank John Toman for proofreading.
 This work was supported in part by:
 JSPS KAKENHI Grant Number
 \grantnum{https://kaken.nii.ac.jp/grant/KAKENHI-PROJECT-17H01723/}{JP17H01723}
 (Igarashi); and
 JSPS KAKENHI Grant Number
 \grantnum{https://kaken.nii.ac.jp/grant/KAKENHI-PROJECT-19K20247/}{JP19K20247}
 and ERATO HASUO Metamathematics for Systems Design Project
 (\grantnum{http://dx.doi.org/10.13039/501100009024}{JPMJER1603}), JST
 (Sekiyama).
\end{acks}

\bibliography{main}



\end{document}


\ifdraft

\section{Notes}

\begin{itemize}
 \item Requirements on $ \oplus $.
       \begin{itemize}
        \item $\rho_{{\mathrm{1}}}  \ottsym{<:}  \rho_{{\mathrm{1}}}  \oplus  \rho_{{\mathrm{2}}}$ and $\rho_{{\mathrm{2}}}  \ottsym{<:}  \rho_{{\mathrm{1}}}  \oplus  \rho_{{\mathrm{2}}}$ (if defined).
        \item If $\rho$ is the least upper bound of $\rho_{{\mathrm{1}}}$ and $\rho_{{\mathrm{2}}}$
              w.r.t.\ subtyping, then $\rho_{{\mathrm{1}}}  \oplus  \rho_{{\mathrm{2}}} = \rho$.
        \item If $\rho_{{\mathrm{1}}}$ and $\rho_{{\mathrm{2}}}$ does not have upper bounds,
              then $\rho_{{\mathrm{1}}}  \oplus  \rho_{{\mathrm{2}}}$ is undefined.
        \item the both casts from $\rho_{{\mathrm{1}}}$ and $\rho_{{\mathrm{2}}}$ to $\rho_{{\mathrm{1}}}  \oplus  \rho_{{\mathrm{2}}}$ never
              fail.  This means $\rho_{{\mathrm{1}}}  \oplus  \rho_{{\mathrm{2}}}$ is weaker than both $\rho_{{\mathrm{1}}}$ and
              $\rho_{{\mathrm{2}}}$.
        \item for any $\rho$ such that the both casts from $\rho_{{\mathrm{1}}}$ and
              $\rho_{{\mathrm{2}}}$ never fail, the cast from $\rho_{{\mathrm{1}}}  \oplus  \rho_{{\mathrm{2}}}$ to $\rho$ also
              never fail.  This means that $\rho_{{\mathrm{1}}}  \oplus  \rho_{{\mathrm{2}}}$ is the strongest among
              rows weaker than both $\rho_{{\mathrm{1}}}$ and $\rho_{{\mathrm{2}}}$.
       \end{itemize}
 \item The form of a ground row for label $\ell$ should be $\ottsym{(}    \ell  \mathbin{:}  \star   ;  \star   \ottsym{)}$, not $\ottsym{(}    \ell  \mathbin{:}  \star   ;   \cdot    \ottsym{)}$, because the latter is not consistent
       with, e.g., $\ottsym{(}    \ell'  \mathbin{:}  \ottnt{A}   ;    \ell  \mathbin{:}  \star   ;  \star    \ottsym{)}$ while the former is.  The value
       $\ell \, \ottnt{v}$ should be able to be typed at $ \langle    \ell  \mathbin{:}  \star   ;    \ell'  \mathbin{:}  \ottnt{A}   ;  \star    \rangle $ which is equivalent to $ \langle    \ell'  \mathbin{:}  \ottnt{A}   ;    \ell  \mathbin{:}  \star   ;  \star    \rangle $.
 \item In \T{Conv} (applied to $\ottnt{e}  \ottsym{:}  \ottnt{A} \,  \stackrel{ \Phi }{\Rightarrow}  \ottnt{B} $), the typing contexts used to
       type $\ottnt{e}$ and $\ottnt{B}$ are empty because type substitution does not
       preserve type convertibility.  For example, $ \Sigma   \vdash   \mathit{X}  \prec^{ \ottsym{+}  \alpha }  \mathit{X} $ holds
       but $ \Sigma   \vdash   \alpha  \prec^{ \ottsym{+}  \alpha }  \alpha $ does not.
\end{itemize}
\clearpage

\fi

\section{Definition}

\setboolean{show-grad}{false}
\setboolean{show-grad-term}{false}
\setboolean{show-typename}{false}

\subsection{Statically typed language {\staticlang}}

\subsubsection{Syntax}
{\def\arraystretch{1.2}
\[
\begin{array}{lrl}
 \syntaxtype{}
 \syntaxMterm{} \\[-3ex]
 \syntaxWvalue{}
 \syntaxFctx{}
 \syntaxtctx{}
\end{array}\]

}

\begin{defn}[Free type variables and type substitution]
 The set $ \mathit{ftv}  (  \ottnt{A}  ) $ of free variables for types and rows in $\ottnt{A}$ is defined
 as usual.
 Substitution $ \ottnt{A}    [  \ottnt{B}  /  \mathit{X}  ]  $ of $\ottnt{B}$ for $\mathit{X}$ in $\ottnt{A}$ is defined in a
 capture-avoiding manner.
\end{defn}

\begin{defn}[Domain of typing contexts]
 We define $\mathit{dom} \, \ottsym{(}  \Gamma  \ottsym{)}$ as follows.
 \[\begin{array}{lll}
  \mathit{dom} \, \ottsym{(}   \emptyset   \ottsym{)}  &\defeq&  \emptyset  \\
  \mathit{dom} \, \ottsym{(}  \Gamma  \ottsym{,}   \mathit{x}  \mathord{:}  \ottnt{A}   \ottsym{)} &\defeq& \mathit{dom} \, \ottsym{(}  \Gamma  \ottsym{)} \,  \mathbin{\cup}  \, \ottsym{\{}  \mathit{x}  \ottsym{\}} \\
  \mathit{dom} \, \ottsym{(}  \Gamma  \ottsym{,}   \mathit{X}  \mathord{:}  \ottnt{K}   \ottsym{)} &\defeq& \mathit{dom} \, \ottsym{(}  \Gamma  \ottsym{)} \,  \mathbin{\cup}  \, \ottsym{\{}  \mathit{X}  \ottsym{\}} \\
   \end{array}\]
\end{defn}

\begin{assum}
 We suppose that each constant $\kappa$ is assigned a first-order type
 $ \mathit{ty}  (  \kappa  ) $ of the form $\iota_{{\mathrm{1}}}  \rightarrow  \cdots  \rightarrow  \iota_{\ottmv{n}}$.

 Suppose that, for any $\iota$, there is a set $ \mathbb{K}_{ \iota } $ of constants of
 $\iota$.
 For any constant $\kappa$, $ \mathit{ty}  (  \kappa  )  \,  =  \, \iota$ if and only if
 $\kappa \,  \in  \,  \mathbb{K}_{ \iota } $.
 The function $ \zeta $ gives a denotation to pairs of constants.
 In particular, for any constants $\kappa_{{\mathrm{1}}}$ and $\kappa_{{\mathrm{2}}}$:
 (1) $ \zeta  (  \kappa_{{\mathrm{1}}}  ,  \kappa_{{\mathrm{2}}}  ) $ is defined if and only if
 $ \mathit{ty}  (  \kappa_{{\mathrm{1}}}  )  \,  =  \, \iota  \rightarrow  \ottnt{A}$ and $ \mathit{ty}  (  \kappa_{{\mathrm{2}}}  )  \,  =  \, \iota$ for some $\ottnt{A}$; and
 (2) if $ \zeta  (  \kappa_{{\mathrm{1}}}  ,  \kappa_{{\mathrm{2}}}  ) $ is defined, $ \zeta  (  \kappa_{{\mathrm{1}}}  ,  \kappa_{{\mathrm{2}}}  ) $ is a
 constant and $ \mathit{ty}  (   \zeta  (  \kappa_{{\mathrm{1}}}  ,  \kappa_{{\mathrm{2}}}  )   )  \,  =  \, \ottnt{A}$ where
 $ \mathit{ty}  (  \kappa_{{\mathrm{1}}}  )  \,  =  \, \iota  \rightarrow  \ottnt{A}$.
\end{assum}

We use the notation and the assumption above also in {\surfacelang} and
{\interlang}.

\subsubsection{Semantics}

\begin{defn}[Record splitting]
 \label{def:static_lang:record_split}
 $w \,  \triangleright _{ \ell }  \, w_{{\mathrm{1}}}  \ottsym{,}  w_{{\mathrm{2}}}$ is defined as follows:
 \ifpaper
 \[
  \ottsym{\{}  \ell  \ottsym{=}  w_{{\mathrm{1}}}  \ottsym{;}  w_{{\mathrm{2}}}  \ottsym{\}} \,  \triangleright _{ \ell }  \, w_{{\mathrm{1}}}  \ottsym{,}  w_{{\mathrm{2}}} \qquad
  \ottsym{\{}  \ell'  \ottsym{=}  w_{{\mathrm{1}}}  \ottsym{;}  w_{{\mathrm{2}}}  \ottsym{\}} \,  \triangleright _{ \ell }  \, w_{{\mathrm{21}}}  \ottsym{,}  \ottsym{\{}  \ell'  \ottsym{=}  w_{{\mathrm{1}}}  \ottsym{;}  w_{{\mathrm{22}}}  \ottsym{\}} {\ }
   \text{(if $\ell \,  \not=  \, \ell'$ and $w_{{\mathrm{2}}} \,  \triangleright _{ \ell }  \, w_{{\mathrm{21}}}  \ottsym{,}  w_{{\mathrm{22}}}$)}
 \]
 \else
 \[\begin{array}{lll}
  \ottsym{\{}  \ell  \ottsym{=}  w_{{\mathrm{1}}}  \ottsym{;}  w_{{\mathrm{2}}}  \ottsym{\}} \,  \triangleright _{ \ell }  \, w_{{\mathrm{1}}}  \ottsym{,}  w_{{\mathrm{2}}} \\
  \ottsym{\{}  \ell'  \ottsym{=}  w_{{\mathrm{1}}}  \ottsym{;}  w_{{\mathrm{2}}}  \ottsym{\}} \,  \triangleright _{ \ell }  \, w_{{\mathrm{21}}}  \ottsym{,}  \ottsym{\{}  \ell'  \ottsym{=}  w_{{\mathrm{1}}}  \ottsym{;}  w_{{\mathrm{22}}}  \ottsym{\}} &
   \text{(if $\ell \,  \not=  \, \ell'$ and $w_{{\mathrm{2}}} \,  \triangleright _{ \ell }  \, w_{{\mathrm{21}}}  \ottsym{,}  w_{{\mathrm{22}}}$)}
   \end{array}\]
 \fi
\end{defn}

\begin{figure}[h]
 \input{defns/static_lang/reduction}
 \input{defns/static_lang/eval}
 \caption{Semantics of {\staticlang}.}
 \label{fig:sup-static-semantics}
\end{figure}

\begin{defn}[Semantics]
 The reduction relation $ \makestatic{\rightsquigarrow} $ and the evaluation relation $ \makestatic{\longrightarrow} $ of
 {\staticlang} are defined by the rules given in
 \reffig{sup-static-semantics}.
\end{defn}

\subsubsection{Type system}

\begin{figure}[h]
 \input{defns/type_equivalence}
 \caption{Type-and-row equivalence of {\staticlang}.}
 \label{fig:sup-static-type-equiv}
\end{figure}

\begin{defn}[Type-and-row equivalence]
 Type-and-row equivalence $ \equiv $ is the smallest relation satisfying the rules
 given by \reffig{sup-static-type-equiv}.
\end{defn}

{\def\arraystretch{1.5}
\begin{figure}[h]
 \input{defns/static_lang/wf_tctx}
 \input{defns/static_lang/wf_type}
 \input{defns/static_lang/typing}
 \caption{Typing of {\staticlang}.}
 \label{fig:sup-static-typing}
\end{figure}
}

\begin{defn}[Typing]
 The well-formedness judgments $ \mathrel{ \makestatic{\vdash} }  \Gamma $ and $ \Gamma  \mathrel{ \makestatic{\vdash} }  \ottnt{A}  :  \ottnt{K} $, and the typing
 judgment $ \Gamma  \mathrel{ \makestatic{\vdash} }  \ottnt{M}  :  \ottnt{A} $ of {\staticlang} are the smallest relations
 satisfying the rules given by \reffig{sup-static-typing}.
\end{defn}

\clearpage

\setboolean{show-grad}{true}
\setboolean{show-grad-term}{true}
\setboolean{show-typename}{true}

\subsection{Gradually typed language {\surfacelang}}

\subsubsection{Syntax}
{\def\arraystretch{1.2}
\[\begin{array}{lrl}
 \syntaxtype{}
 \csname ifshow-grad-term\endcsname \syntaxMterm{} \fi
 \syntaxtctx{}
\end{array}\]

}

\begin{assum}
 We assume that operation $\ottnt{A}  \oplus  \ottnt{B}$ that produces a type is available.
 Assumptions for $ \oplus $ are stated in the beginnings of subsections of proving
 properties (\sect{proof:trans} and \sect{proof:conservativity:typing}).
\end{assum}

\subsubsection{Typing}

\begin{defn}[Type-and-row equivalence]
 Type-and-row equivalence $ \equiv $ is the smallest relation satisfying the rules
 given by \reffig{sup-static-type-equiv}.
\end{defn}

\begin{figure}[h]
 \input{defns/grad_lang/consistency}
 \caption{Consistency.}
 \label{fig:sup-grad-consistency}
\end{figure}

\begin{defn}[Quasi-universal types]
 The predicate $\mathbf{QPoly} \, \ottsym{(}  \ottnt{A}  \ottsym{)}$ is defined by: $\mathbf{QPoly} \, \ottsym{(}  \ottnt{A}  \ottsym{)}$ if and only if
 \ifpaper
 (1) $\ottnt{A}$ is none of $ \text{\unboldmath$\forall\!$}  \,  \mathit{X}  \mathord{:}  \ottnt{K}   \ottsym{.} \, \ottnt{B}$, $ \cdot $ (the empty row), and $  \ell  \mathbin{:}  \ottnt{B}   ;  \rho $
 for any $\mathit{X}$, $\ottnt{K}$, $\ottnt{B}$, $\ell$, and $\rho$; and
 (2) $ \star $ occurs somewhere in $\ottnt{A}$.
 \else
 \begin{itemize}
  \item $\ottnt{A} \,  \not=  \,  \text{\unboldmath$\forall\!$}  \,  \mathit{X}  \mathord{:}  \ottnt{K}   \ottsym{.} \, \ottnt{B}$ for any $\mathit{X}$, $\ottnt{K}$, and $\ottnt{B}$,
  \item $\ottnt{A} \,  \not=  \,  \cdot $,
  \item $\ottnt{A} \,  \not=  \,   \ell  \mathbin{:}  \ottnt{B}   ;  \rho $ for any $\ell$, $\ottnt{B}$, and $\rho$, and
  \item $ \star $ occurs somewhere in $\ottnt{A}$.
 \end{itemize}
 \fi
 Type $\ottnt{A}$ is a quasi-universal type if and only if $\mathbf{QPoly} \, \ottsym{(}  \ottnt{A}  \ottsym{)}$.
\end{defn}

\begin{defn}[Labels in row]
 We define $\mathit{dom} \, \ottsym{(}  \rho  \ottsym{)}$, the set of the field labels in $\rho$, as follows.
 \[\begin{array}{lll}
  \mathit{dom} \, \ottsym{(}   \cdot   \ottsym{)}     &\defeq&  \emptyset  \\
  \mathit{dom} \, \ottsym{(}  \star  \ottsym{)}      &\defeq&  \emptyset  \\
  \mathit{dom} \, \ottsym{(}  \mathit{X}  \ottsym{)}        &\defeq&  \emptyset  \\
  \ifthen{\boolean{show-typename}}{
  \mathit{dom} \, \ottsym{(}  \alpha  \ottsym{)}        &\defeq&  \emptyset  \\
  }
  \mathit{dom} \, \ottsym{(}    \ell  \mathbin{:}  \ottnt{A}   ;  \rho   \ottsym{)} &\defeq& \mathit{dom} \, \ottsym{(}  \rho  \ottsym{)} \,  \mathbin{\cup}  \, \ottsym{\{}  \ell  \ottsym{\}} \\
   \end{array}\]
\end{defn}

\begin{defn}[Row concatenation]
 Row concatenation $\rho_{{\mathrm{1}}}  \odot  \rho_{{\mathrm{2}}}$ is defined as follows:
 \[\begin{array}{lll}
   \cdot   \odot  \rho_{{\mathrm{2}}} &\defeq& \rho_{{\mathrm{2}}} \\
  \ottsym{(}    \ell  \mathbin{:}  \ottnt{A}   ;  \rho_{{\mathrm{1}}}   \ottsym{)}  \odot  \rho_{{\mathrm{2}}} &\defeq&   \ell  \mathbin{:}  \ottnt{A}   ;  \ottsym{(}  \rho_{{\mathrm{1}}}  \odot  \rho_{{\mathrm{2}}}  \ottsym{)}  \\
   \end{array}\]
\end{defn}

\begin{defn}[Rows ending with $ \star $]
 Row type $\rho$ ends with $ \star $ if and only if
 $\rho \,  =  \, \rho'  \odot  \star$ for some $\rho'$.
\end{defn}

\begin{defn}[Consistency]
 Consistency $\ottnt{A}  \sim  \ottnt{B}$ is the smallest relation satisfying the rules given by
 \reffig{sup-grad-consistency}.
\end{defn}

\begin{figure}[h]
 \input{defns/grad_lang/consistent_equiv}
 \caption{Consistent equivalence.}
 \label{fig:sup-grad-consistent-equiv}
\end{figure}

\begin{defn}[Row splitting]
 Row splitting $\rho_{{\mathrm{1}}} \,  \triangleright _{ \ell }  \, \ottnt{A}  \ottsym{,}  \rho_{{\mathrm{2}}}$ is defined as follows.
 \ifpaper
 \[\begin{array}{ll}
  \star \,  \triangleright _{ \ell }  \, \star  \ottsym{,}  \star \qquad
    \ell  \mathbin{:}  \ottnt{A}   ;  \rho  \,  \triangleright _{ \ell }  \, \ottnt{A}  \ottsym{,}  \rho \qquad
    \ell'  \mathbin{:}  \ottnt{B}   ;  \rho_{{\mathrm{1}}}  \,  \triangleright _{ \ell }  \, \ottnt{A}  \ottsym{,}  \ottsym{(}    \ell'  \mathbin{:}  \ottnt{B}   ;  \rho_{{\mathrm{2}}}   \ottsym{)} \quad
   \text{(if $\ell \,  \not=  \, \ell'$ and $\rho_{{\mathrm{1}}} \,  \triangleright _{ \ell }  \, \ottnt{A}  \ottsym{,}  \rho_{{\mathrm{2}}}$)}
   \end{array}\]
 \else
 \[\begin{array}{lcll}
    \ell  \mathbin{:}  \ottnt{A}   ;  \rho    &  \triangleright _{ \ell }  & \ottnt{A}  \ottsym{,}  \rho \\
    \ell'  \mathbin{:}  \ottnt{B}   ;  \rho_{{\mathrm{1}}}  &  \triangleright _{ \ell }  & \ottnt{A}  \ottsym{,}  \ottsym{(}    \ell'  \mathbin{:}  \ottnt{B}   ;  \rho_{{\mathrm{2}}}   \ottsym{)} &
   \text{(if $\ell \,  \not=  \, \ell'$ and $\rho_{{\mathrm{1}}} \,  \triangleright _{ \ell }  \, \ottnt{A}  \ottsym{,}  \rho_{{\mathrm{2}}}$)} \\
   \star        &  \triangleright _{ \ell }  & \star  \ottsym{,}  \star
   \end{array}\]
 \fi
\end{defn}

\begin{defn}[Consistent equivalence]
 Consistency equivalence $\ottnt{A}  \simeq  \ottnt{B}$ is the smallest relation satisfying the
 rules given by \reffig{sup-grad-consistent-equiv}.
\end{defn}

\begin{figure}[h]
 \input{defns/grad_lang/type_match}
 \caption{Type matching.}
 \label{fig:sup-grad-type-match}
\end{figure}

\begin{defn}[Type matching]
 Type matching $\ottnt{A}  \triangleright  \ottnt{B}$ is the smallest relation satisfying the rules given
 by \reffig{sup-grad-type-match}.
\end{defn}

\begin{figure}[h]
 \input{defns/grad_lang/wf_tctx}
 \input{defns/grad_lang/wf_type}
 \input{defns/grad_lang/typing}
 \caption{Typing of {\surfacelang}.}
 \label{fig:sup-grad-typing}
\end{figure}

\begin{defn}[Typing]
 The well-formedness judgments $\vdash  \Gamma$ and $\Gamma  \vdash  \ottnt{A}  \ottsym{:}  \ottnt{K}$, and the typing
 judgment $\Gamma  \vdash  \ottnt{M}  \ottsym{:}  \ottnt{A}$ of {\surfacelang} are the smallest relations
 satisfying the rules given by \reffig{sup-grad-typing}.
\end{defn}

\clearpage

\subsection{Blame calculus {\interlang}}

\subsubsection{Syntax}

{\def\arraystretch{1.2}
\begin{figure}[h]
 \input{defns/blame/syntax}
 \caption{Syntax of {\interlang}.}
 \label{fig:sup-inter-syntax}
\end{figure}
}

\begin{defn}[Comparison between name stores]
 We write $\Sigma \,  \subseteq  \, \Sigma'$ if and only if, for any $\alpha$, $\ottnt{K}$, and $\ottnt{A}$,
 if $ \alpha  \mathord{:}  \ottnt{K}   \ottsym{:=}  \ottnt{A} \,  \in  \, \Sigma$, then $ \alpha  \mathord{:}  \ottnt{K}   \ottsym{:=}  \ottnt{A} \,  \in  \, \Sigma'$.
\end{defn}

\begin{defn}[Substitution]
 Type substitution $ \ottnt{e}    [  \ottnt{A}  /  \mathit{X}  ]  $ of $\ottnt{A}$ for $\mathit{X}$ in $\ottnt{e}$ is
 defined in a capture-avoiding manner as usual.
 Value substitution $ \ottnt{e}    [  \ottnt{v}  \ottsym{/}  \mathit{x}  ]  $ is also defined similarly.
\end{defn}

\subsubsection{Semantics}

\begin{defn}[Record splitting]
 $\ottnt{v} \,  \triangleright _{ \ell }  \, \ottnt{v_{{\mathrm{1}}}}  \ottsym{,}  \ottnt{v_{{\mathrm{2}}}}$ is defined as follows:
 \[\begin{array}{lll}
  \ottsym{\{}  \ell  \ottsym{=}  \ottnt{v_{{\mathrm{1}}}}  \ottsym{;}  \ottnt{v_{{\mathrm{2}}}}  \ottsym{\}} \,  \triangleright _{ \ell }  \, \ottnt{v_{{\mathrm{1}}}}  \ottsym{,}  \ottnt{v_{{\mathrm{2}}}} \\
  \ottsym{\{}  \ell'  \ottsym{=}  \ottnt{v_{{\mathrm{1}}}}  \ottsym{;}  \ottnt{v_{{\mathrm{2}}}}  \ottsym{\}} \,  \triangleright _{ \ell }  \, \ottnt{v_{{\mathrm{21}}}}  \ottsym{,}  \ottsym{\{}  \ell'  \ottsym{=}  \ottnt{v_{{\mathrm{1}}}}  \ottsym{;}  \ottnt{v_{{\mathrm{22}}}}  \ottsym{\}} &
   \text{(where $\ell \,  \not=  \, \ell'$ and $\ottnt{v_{{\mathrm{2}}}} \,  \triangleright _{ \ell }  \, \ottnt{v_{{\mathrm{21}}}}  \ottsym{,}  \ottnt{v_{{\mathrm{22}}}}$)}
   \end{array}\]
\end{defn}

\begin{defn}[Field postpending]
 Field postpending $ \rho  \mathrel{@}   \ell  \mathbin{:}  \ottnt{A}  $ is defined as follows:
 \ifpaper
 \[
  \ottsym{(}    \ell'  \mathbin{:}  \ottnt{B}   ;  \rho   \ottsym{)}  \mathrel{@}   \ell  \mathbin{:}  \ottnt{A}   \defeq   \ell'  \mathbin{:}  \ottnt{B}   ;  \ottsym{(}   \rho  \mathrel{@}   \ell  \mathbin{:}  \ottnt{A}    \ottsym{)}  \qquad
  \star  \mathrel{@}   \ell  \mathbin{:}  \ottnt{A}   \defeq   \ell  \mathbin{:}  \ottnt{A}   ;  \star 
 \]
 \else
 \[\begin{array}{lll}
   \ottsym{(}    \ell'  \mathbin{:}  \ottnt{B}   ;  \rho   \ottsym{)}  \mathrel{@}   \ell  \mathbin{:}  \ottnt{A}   &\defeq&   \ell'  \mathbin{:}  \ottnt{B}   ;  \ottsym{(}   \rho  \mathrel{@}   \ell  \mathbin{:}  \ottnt{A}    \ottsym{)}  \\
     \star  \mathrel{@}   \ell  \mathbin{:}  \ottnt{A}   &\defeq&   \ell  \mathbin{:}  \ottnt{A}   ;  \star 
   \end{array}\]
 \fi
\end{defn}

\begin{defn}[Ground row types of rows]
 \[\begin{array}{lll}
   \mathit{grow}  (   \cdot   )  &\defeq&  \cdot  \\
     \mathit{grow}  (  \alpha  )  &\defeq& \alpha \\
     \mathit{grow}  (    \ell  \mathbin{:}  \ottnt{A}   ;  \rho   )  &\defeq&   \ell  \mathbin{:}  \star   ;  \star  \\
   \end{array}\]
\end{defn}

\begin{defn}[Row embedding]
 Row embedding $ \variantliftrow{ \rho }{ \ottnt{e} } $ is defined as follows:
 \ifpaper
 \[
   \variantliftrow{ \ottsym{(}    \ell  \mathbin{:}  \ottnt{A}   ;  \rho   \ottsym{)} }{ \ottnt{e} }  \ \defeq \  \variantlift{ \ell }{ \ottnt{A} }{ \ottsym{(}   \variantliftrow{ \rho }{ \ottnt{e} }   \ottsym{)} }  \qquad
   \variantliftrow{ \rho }{ \ottnt{e} }          \ \defeq \ \ottnt{e} \quad \text{(if $\rho$ is not a row extension)}
 \]
 \else
 \[\begin{array}{lll}
   \variantliftrow{ \ottsym{(}    \ell  \mathbin{:}  \ottnt{A}   ;  \rho   \ottsym{)} }{ \ottnt{e} }   &\defeq&  \variantlift{ \ell }{ \ottnt{A} }{ \ottsym{(}   \variantliftrow{ \rho }{ \ottnt{e} }   \ottsym{)} }  \\
   \variantliftrow{ \rho }{ \ottnt{e} }           &\defeq& \ottnt{e} \quad \text{(if $\rho \,  \not=  \, \ottsym{(}    \ell  \mathbin{:}  \ottnt{A}   ;  \rho'   \ottsym{)}$)}
   \end{array}\]
 \fi
\end{defn}

\begin{defn}[Field insertion]
 Function $ \variantliftdown{ \rho }{ \ell }{ \ottnt{A} }{ \ottnt{e} } $ embeds a term $\ottnt{e}$ of type $ \langle    \rho  \odot  \rho'    \rangle $ into $ \langle    \rho  \odot  \ottsym{(}    \ell  \mathbin{:}  \ottnt{A}   ;   \cdot    \ottsym{)}  \odot  \rho'    \rangle $.  Formally, it is
 defined as follows:
 \[\begin{array}{lll}
   \variantliftdown{ \ottsym{(}    \ell'  \mathbin{:}  \ottnt{B'}   ;  \rho   \ottsym{)} }{ \ell }{ \ottnt{A} }{ \ottnt{e} }   &\defeq&  \mathsf{case} \,  \ottnt{e}  \,\mathsf{with}\, \langle  \ell' \,  \mathit{x}   \rightarrow   \ell' \, \mathit{x}   \ottsym{;}   \mathit{y}   \rightarrow    \variantlift{ \ell' }{ \ottnt{B'} }{ \ottsym{(}   \variantliftdown{ \rho }{ \ell }{ \ottnt{A} }{ \mathit{y} }   \ottsym{)} }   \rangle  \\
   \variantliftdown{ \rho }{ \ell }{ \ottnt{A} }{ \ottnt{e} }             &\defeq&  \variantlift{ \ell }{ \ottnt{A} }{ \ottnt{e} }  \quad
   \text{(if \ifpaper $\rho$ is not a row extension\else $\rho \,  \not=  \, \ottsym{(}    \ell'  \mathbin{:}  \ottnt{B'}   ;  \rho'   \ottsym{)}$ for any $\ell'$, $\ottnt{B'}$, and $\rho'$\fi)}
   \end{array}\]
\end{defn}

\begin{defn}[Name in conversion label]
 We define $ \mathit{name} (  \ottsym{+}  \alpha  ) $ and $ \mathit{name} (  \ottsym{-}  \alpha  ) $ to be $\alpha$.
\end{defn}

\begin{figure}[h]
 \input{defns/blame/red} %
 \caption{Reduction rules of {\interlang}.}
 \label{fig:sup-inter-red}
\end{figure}

\begin{figure}[h]
 \input{defns/blame/red_record}
 \caption{Cast and conversion reduction rules for record types.}
 \label{fig:sup-inter-red-record}
\end{figure}

\begin{figure}[h]
 \input{defns/blame/red_variant}
 \caption{Cast and conversion reduction rules for variant types.}
 \label{fig:sup-inter-red-variant}
\end{figure}

\begin{figure}[h]
 \input{defns/blame/eval}
 \caption{Evaluation rules of {\interlang}.}
 \label{fig:sup-inter-eval}
\end{figure}

\begin{defn}
 Relations $ \longrightarrow $ and $ \rightsquigarrow $ are the smallest relations satisfying the
 rules in Figures~\ref{fig:sup-inter-red}, \ref{fig:sup-inter-red-record},
 \ref{fig:sup-inter-red-variant}, and \ref{fig:sup-inter-eval}.
\end{defn}

\begin{defn}[Multi-step evaluation]
 Binary relation $ \longrightarrow^{*} $ over terms is the reflexive and transitive closure
 of $ \longrightarrow $.
\end{defn}

\subsection{Typing}

{\def\arraystretch{1.5}
\begin{figure}[h]
 \input{defns/blame/convert}
 \caption{Type convertibility.}
\end{figure}

\begin{figure}[h]
 \input{defns/blame/wf_tctx}
 \input{defns/blame/wf_type}
 \caption{Well-formedness rules of {\interlang}.}
 \label{fig:sup-inter-well-formedness}
\end{figure}

\begin{figure}[h]
 \input{defns/blame/typing}
 \caption{Typing rules of {\interlang}.}
 \label{fig:sup-inter-typing}
\end{figure}
}

\begin{defn}
 Judgments $\Sigma  \vdash  \Gamma$, $\Sigma  \ottsym{;}  \Gamma  \vdash  \ottnt{A}  \ottsym{:}  \ottnt{K}$, and $\Sigma  \ottsym{;}  \Gamma  \vdash  \ottnt{e}  \ottsym{:}  \ottnt{A}$ are the
 smallest relations satisfying the rules in
 Figures~\ref{fig:sup-inter-well-formedness} and \ref{fig:sup-inter-typing}.
\end{defn}

\subsection{Translation}

\begin{figure}[h]
 \input{defns/blame/trans}
 \caption{Translation rules.}
 \label{fig:sup-trans}
\end{figure}

\begin{defn}
 Relation $\Gamma  \vdash  \ottnt{M}  \ottsym{:}  \ottnt{A}  \hookrightarrow  \ottnt{e}$ is the smallest relation satisfying
 the rules in \reffig{sup-trans}.
\end{defn}

\clearpage
\section{Proofs}

\subsection{Consistency}

\begin{lemma}{equiv-qpoly}
 Suppose $\ottnt{A}  \equiv  \ottnt{B}$.
 $\mathbf{QPoly} \, \ottsym{(}  \ottnt{A}  \ottsym{)}$ if and only if $\mathbf{QPoly} \, \ottsym{(}  \ottnt{B}  \ottsym{)}$.
\end{lemma}
\begin{proof}
 Straightforward by induction on the derivation of $\ottnt{A}  \equiv  \ottnt{B}$.
\end{proof}

\begin{lemma}{equiv-free-tyvar}
 If $\ottnt{A}  \equiv  \ottnt{B}$, then $ \mathit{ftv}  (  \ottnt{A}  )  \,  =  \,  \mathit{ftv}  (  \ottnt{B}  ) $.
\end{lemma}
\begin{proof}
 Straightforward by induction on the derivation of $\ottnt{A}  \equiv  \ottnt{B}$.
\end{proof}

\begin{lemma}{equiv-ends-with-dyn}
 Suppose that $\rho_{{\mathrm{1}}}  \equiv  \rho_{{\mathrm{2}}}$.
 $\rho_{{\mathrm{1}}}$ ends with $ \star $ if and only if so does $\rho_{{\mathrm{2}}}$.
\end{lemma}
\begin{proof}
 Straightforward by induction on the derivation of $\ottnt{A}  \equiv  \ottnt{B}$.
\end{proof}

\begin{lemma}{equiv-labels}
 If $\rho_{{\mathrm{1}}}  \equiv  \rho_{{\mathrm{2}}}$, then $\mathit{dom} \, \ottsym{(}  \rho_{{\mathrm{1}}}  \ottsym{)} \,  =  \, \mathit{dom} \, \ottsym{(}  \rho_{{\mathrm{2}}}  \ottsym{)}$.
\end{lemma}
\begin{proof}
 Straightforward by induction on the derivation of $\rho_{{\mathrm{1}}}  \equiv  \rho_{{\mathrm{2}}}$.
\end{proof}

\begin{lemma}{equiv-inversion}
 Suppose $\ottnt{A}  \equiv  \ottnt{B}$.
 \begin{enumerate}
  \item \label{lem:equiv-inversion:dyn}
        $\ottnt{A} \,  =  \, \star$ if and only if $\ottnt{B} \,  =  \, \star$.
  \item \label{lem:equiv-inversion:fun}
        $\ottnt{A} \,  =  \, \ottnt{A_{{\mathrm{1}}}}  \rightarrow  \ottnt{A_{{\mathrm{2}}}}$ if and only if $\ottnt{B} \,  =  \, \ottnt{B_{{\mathrm{1}}}}  \rightarrow  \ottnt{B_{{\mathrm{2}}}}$, and
        $\ottnt{A_{{\mathrm{1}}}}  \equiv  \ottnt{B_{{\mathrm{1}}}}$ and $\ottnt{A_{{\mathrm{2}}}}  \equiv  \ottnt{B_{{\mathrm{2}}}}$.
  \item \label{lem:equiv-inversion:forall}
        $\ottnt{A} \,  =  \,  \text{\unboldmath$\forall\!$}  \,  \mathit{X}  \mathord{:}  \ottnt{K}   \ottsym{.} \, \ottnt{A'}$ if and only if $\ottnt{B} \,  =  \,  \text{\unboldmath$\forall\!$}  \,  \mathit{X}  \mathord{:}  \ottnt{K}   \ottsym{.} \, \ottnt{B'}$, and
        $\ottnt{A'}  \equiv  \ottnt{B'}$.
  \item \label{lem:equiv-inversion:record}
        $\ottnt{A} \,  =  \,  [  \rho_{{\mathrm{1}}}  ] $ if and only if $\ottnt{B} \,  =  \,  [  \rho_{{\mathrm{2}}}  ] $, and
        $\rho_{{\mathrm{1}}}  \equiv  \rho_{{\mathrm{2}}}$.
  \item \label{lem:equiv-inversion:variant}
        $\ottnt{A} \,  =  \,  \langle  \rho_{{\mathrm{1}}}  \rangle $ if and only if $\ottnt{B} \,  =  \,  \langle  \rho_{{\mathrm{2}}}  \rangle $, and
        $\rho_{{\mathrm{1}}}  \equiv  \rho_{{\mathrm{2}}}$.
  \item \label{lem:equiv-inversion:row-label}
        $\ottnt{A} \,  =  \, \rho_{{\mathrm{11}}}  \odot  \ottsym{(}    \ell  \mathbin{:}  \ottnt{A'}   ;   \cdot    \ottsym{)}  \odot  \rho_{{\mathrm{12}}}$ and $\ell \,  \not\in  \, \mathit{dom} \, \ottsym{(}  \rho_{{\mathrm{11}}}  \ottsym{)}$ if and only if
        $\ottnt{B} \,  =  \, \rho_{{\mathrm{21}}}  \odot  \ottsym{(}    \ell  \mathbin{:}  \ottnt{B'}   ;   \cdot    \ottsym{)}  \odot  \rho_{{\mathrm{22}}}$ and $\ell \,  \not\in  \, \mathit{dom} \, \ottsym{(}  \rho_{{\mathrm{21}}}  \ottsym{)}$, and
        $\ottnt{A'}  \equiv  \ottnt{B'}$ and $\rho_{{\mathrm{11}}}  \odot  \rho_{{\mathrm{12}}}  \equiv  \rho_{{\mathrm{21}}}  \odot  \rho_{{\mathrm{22}}}$.
 \end{enumerate}
\end{lemma}
\begin{proof}
 Straightforward by induction on the derivation of $\ottnt{A}  \equiv  \ottnt{B}$.
\end{proof}


\begin{lemma}{consistent-symm}
 If $\ottnt{A}  \simeq  \ottnt{B}$, then $\ottnt{B}  \simeq  \ottnt{A}$.
\end{lemma}
\begin{proof}
 Straightforward by induction on the derivation of $\ottnt{A}  \simeq  \ottnt{B}$.
\end{proof}

\begin{lemma}{consistent-inv-tyname}
 If $\alpha  \simeq  \rho$, then $\rho \,  =  \, \alpha$ or $\rho \,  =  \, \star$.
\end{lemma}
\begin{proof}
 Straightforward by case analysis on the derivation of $\alpha  \simeq  \rho$.
\end{proof}

\begin{lemma}{consistent-inv-remp}
 If $ \cdot   \simeq  \rho$, then $\rho \,  =  \,  \cdot $ or $\rho \,  =  \, \star$.
\end{lemma}
\begin{proof}
 Straightforward by case analysis on the derivation of $ \cdot   \simeq  \rho$.
\end{proof}

\begin{lemma}{consistent-inv-cons}
 If $  \ell  \mathbin{:}  \ottnt{A}   ;  \rho_{{\mathrm{1}}}   \simeq  \rho_{{\mathrm{2}}}$,
 then $\rho_{{\mathrm{2}}} \,  \triangleright _{ \ell }  \, \ottnt{B}  \ottsym{,}  \rho'_{{\mathrm{2}}}$ and $\ottnt{A}  \simeq  \ottnt{B}$ and $\rho_{{\mathrm{1}}}  \simeq  \rho'_{{\mathrm{2}}}$.
\end{lemma}
\begin{proof}
 By induction on $  \ell  \mathbin{:}  \ottnt{A}   ;  \rho_{{\mathrm{1}}}   \simeq  \rho_{{\mathrm{2}}}$.
 \begin{caseanalysis}
  \case \CE{Refl}: Obvious
  since $\rho_{{\mathrm{2}}} \,  =  \,   \ell  \mathbin{:}  \ottnt{A}   ;  \rho_{{\mathrm{1}}} $ and $  \ell  \mathbin{:}  \ottnt{A}   ;  \rho_{{\mathrm{1}}}  \,  \triangleright _{ \ell }  \, \ottnt{A}  \ottsym{,}  \rho_{{\mathrm{1}}}$.

  \case \CE{ConsL}: Obvious by inversion.
  \case \CE{ConsR}:
   We have $\rho_{{\mathrm{2}}} \,  =  \,   \ell'  \mathbin{:}  \ottnt{B}   ;  \rho'_{{\mathrm{2}}} $ for some $\ell'$, $\ottnt{B}$, and $\rho'_{{\mathrm{2}}}$.

   If $\ell \,  =  \, \ell'$, then, since $  \ell  \mathbin{:}  \ottnt{A}   ;  \rho_{{\mathrm{1}}}  \,  \triangleright _{ \ell' }  \, \ottnt{A}  \ottsym{,}  \rho_{{\mathrm{1}}}$,
   we have $\ottnt{A}  \simeq  \ottnt{B}$ and $\rho_{{\mathrm{1}}}  \simeq  \rho'_{{\mathrm{2}}}$ by inversion.
   Since $  \ell'  \mathbin{:}  \ottnt{B}   ;  \rho'_{{\mathrm{2}}}  \,  \triangleright _{ \ell }  \, \ottnt{B}  \ottsym{,}  \rho'_{{\mathrm{2}}}$, we finish.

   Otherwise, suppose $\ell \,  \not=  \, \ell'$.
   Then, by inversion and definition of type matching,
   \begin{itemize}
    \item $  \ell  \mathbin{:}  \ottnt{A}   ;  \rho_{{\mathrm{1}}}  \,  \triangleright _{ \ell' }  \, \ottnt{A'}  \ottsym{,}    \ell  \mathbin{:}  \ottnt{A}   ;  \rho'_{{\mathrm{1}}} $,
    \item $\rho_{{\mathrm{1}}} \,  \triangleright _{ \ell' }  \, \ottnt{A'}  \ottsym{,}  \rho'_{{\mathrm{1}}}$,
    \item $\ottnt{A'}  \simeq  \ottnt{B}$, and
    \item $  \ell  \mathbin{:}  \ottnt{A}   ;  \rho'_{{\mathrm{1}}}   \simeq  \rho'_{{\mathrm{2}}}$
   \end{itemize}
   for some $\ottnt{A'}$ and $\rho'_{{\mathrm{1}}}$.
   By the IH, $\rho'_{{\mathrm{2}}} \,  \triangleright _{ \ell }  \, \ottnt{B'}  \ottsym{,}  \rho''_{{\mathrm{2}}}$ and $\ottnt{A}  \simeq  \ottnt{B'}$ and $\rho'_{{\mathrm{1}}}  \simeq  \rho''_{{\mathrm{2}}}$
   for some $\ottnt{B'}$ and $\rho''_{{\mathrm{2}}}$.
   Since $\ell \,  \not=  \, \ell'$, we have $\rho_{{\mathrm{2}}} =   \ell'  \mathbin{:}  \ottnt{B}   ;  \rho'_{{\mathrm{2}}}  \,  \triangleright _{ \ell }  \, \ottnt{B'}  \ottsym{,}    \ell'  \mathbin{:}  \ottnt{B}   ;  \rho''_{{\mathrm{2}}} $.
   Since $\ottnt{A}  \simeq  \ottnt{B'}$, it suffices to show that $\rho_{{\mathrm{1}}}  \simeq    \ell'  \mathbin{:}  \ottnt{B}   ;  \rho''_{{\mathrm{2}}} $.
   Here, $\rho_{{\mathrm{1}}} \,  \triangleright _{ \ell' }  \, \ottnt{A'}  \ottsym{,}  \rho'_{{\mathrm{1}}}$ and $\ottnt{A'}  \simeq  \ottnt{B}$ and $\rho'_{{\mathrm{1}}}  \simeq  \rho''_{{\mathrm{2}}}$ (obtained
   above).
   Thus, by \CE{ConsR}, $\rho_{{\mathrm{1}}}  \simeq    \ell'  \mathbin{:}  \ottnt{B}   ;  \rho''_{{\mathrm{2}}} $.

  \case the others: Contradictory.
 \end{caseanalysis}
\end{proof}

\begin{lemma}{consistent-inv-fun}
 If $\ottnt{A_{{\mathrm{1}}}}  \rightarrow  \ottnt{A_{{\mathrm{2}}}}  \simeq  \ottnt{B_{{\mathrm{1}}}}  \rightarrow  \ottnt{B_{{\mathrm{2}}}}$, then $\ottnt{A_{{\mathrm{1}}}}  \simeq  \ottnt{B_{{\mathrm{1}}}}$ and $\ottnt{A_{{\mathrm{2}}}}  \simeq  \ottnt{B_{{\mathrm{2}}}}$.
\end{lemma}
\begin{proof}
 Straightforward by case analysis on the derivation of $\ottnt{A_{{\mathrm{1}}}}  \rightarrow  \ottnt{A_{{\mathrm{2}}}}  \simeq  \ottnt{B_{{\mathrm{1}}}}  \rightarrow  \ottnt{B_{{\mathrm{2}}}}$.
\end{proof}

\begin{lemma}{consistent-inv-forall}
 If $ \text{\unboldmath$\forall\!$}  \,  \mathit{X}  \mathord{:}  \ottnt{K}   \ottsym{.} \, \ottnt{A}  \simeq   \text{\unboldmath$\forall\!$}  \,  \mathit{X}  \mathord{:}  \ottnt{K}   \ottsym{.} \, \ottnt{B}$, then $\ottnt{A}  \simeq  \ottnt{B}$.
\end{lemma}
\begin{proof}
 Straightforward by case analysis on the derivation of $\ottnt{A_{{\mathrm{1}}}}  \rightarrow  \ottnt{A_{{\mathrm{2}}}}  \simeq  \ottnt{B_{{\mathrm{1}}}}  \rightarrow  \ottnt{B_{{\mathrm{2}}}}$.
\end{proof}

\begin{lemma}{consistent-inv-forall-qpoly}
 If $ \text{\unboldmath$\forall\!$}  \,  \mathit{X}  \mathord{:}  \ottnt{K}   \ottsym{.} \, \ottnt{A}  \simeq  \ottnt{B}$ and $\mathbf{QPoly} \, \ottsym{(}  \ottnt{B}  \ottsym{)}$,
 then $\mathit{X} \,  \not\in  \,  \mathit{ftv}  (  \ottnt{B}  ) $ and $\ottnt{A}  \simeq  \ottnt{B}$.
\end{lemma}
\begin{proof}
 Straightforward by case analysis on the derivation of $ \text{\unboldmath$\forall\!$}  \,  \mathit{X}  \mathord{:}  \ottnt{K}   \ottsym{.} \, \ottnt{A}  \simeq  \ottnt{B}$.
\end{proof}

\begin{lemma}{consistent-inv-record}
 If $ [  \rho_{{\mathrm{1}}}  ]   \simeq   [  \rho_{{\mathrm{2}}}  ] $, then $\rho_{{\mathrm{1}}}  \simeq  \rho_{{\mathrm{2}}}$.
\end{lemma}
\begin{proof}
 Straightforward by case analysis on the derivation of $ [  \rho_{{\mathrm{1}}}  ]   \simeq   [  \rho_{{\mathrm{2}}}  ] $.
\end{proof}

\begin{lemma}{consistent-inv-variant}
 If $ \langle  \rho_{{\mathrm{1}}}  \rangle   \simeq   \langle  \rho_{{\mathrm{2}}}  \rangle $, then $\rho_{{\mathrm{1}}}  \simeq  \rho_{{\mathrm{2}}}$.
\end{lemma}
\begin{proof}
 Straightforward by case analysis on the derivation of $ \langle  \rho_{{\mathrm{1}}}  \rangle   \simeq   \langle  \rho_{{\mathrm{2}}}  \rangle $.
\end{proof}

\begin{lemmap}{consistent-decomp-aux}
 Suppose that $\ottnt{A}  \sim  \ottnt{B}$.
 If $\rho_{{\mathrm{1}}}  \sim  \rho_{{\mathrm{21}}}  \odot  \rho_{{\mathrm{22}}}$ and $\ell \,  \not\in  \, \mathit{dom} \, \ottsym{(}  \rho_{{\mathrm{21}}}  \ottsym{)}$,
 then there exist some $\rho_{{\mathrm{11}}}$ and $\rho_{{\mathrm{12}}}$ such that
 \begin{itemize}
  \item $\rho_{{\mathrm{1}}}  \equiv  \rho_{{\mathrm{11}}}  \odot  \rho_{{\mathrm{12}}}$,
  \item $\rho_{{\mathrm{11}}}  \odot  \ottsym{(}    \ell  \mathbin{:}  \ottnt{A}   ;   \cdot    \ottsym{)}  \odot  \rho_{{\mathrm{12}}}  \sim  \rho_{{\mathrm{21}}}  \odot  \ottsym{(}    \ell  \mathbin{:}  \ottnt{B}   ;   \cdot    \ottsym{)}  \odot  \rho_{{\mathrm{22}}}$,
  \item $\rho_{{\mathrm{11}}}  \odot  \ottsym{(}    \ell  \mathbin{:}  \ottnt{A}   ;   \cdot    \ottsym{)}  \odot  \rho_{{\mathrm{3}}}  \odot  \rho_{{\mathrm{12}}}  \sim  \rho_{{\mathrm{21}}}  \odot  \ottsym{(}    \ell  \mathbin{:}  \ottnt{B}   ;   \cdot    \ottsym{)}  \odot  \rho_{{\mathrm{22}}}$
        for any $\rho_{{\mathrm{3}}}$ such that $\mathit{dom} \, \ottsym{(}  \rho_{{\mathrm{3}}}  \ottsym{)} \,  \mathbin{\cap}  \, \mathit{dom} \, \ottsym{(}  \rho_{{\mathrm{21}}}  \odot  \rho_{{\mathrm{22}}}  \ottsym{)} \,  =  \,  \emptyset $ if
        $\rho_{{\mathrm{21}}}  \odot  \rho_{{\mathrm{22}}}$ ends with $ \star $, and
  \item $\ell \,  \not\in  \, \mathit{dom} \, \ottsym{(}  \rho_{{\mathrm{11}}}  \ottsym{)}$.
 \end{itemize}
\end{lemmap}
\begin{proof}
 By induction on the derivation of $\rho_{{\mathrm{1}}}  \sim  \rho_{{\mathrm{21}}}  \odot  \rho_{{\mathrm{22}}}$.
  Since $\rho_{{\mathrm{21}}}  \odot  \rho_{{\mathrm{22}}}$ is defined, there are only two cases on $\rho_{{\mathrm{21}}}$ to be considered.
  \begin{caseanalysis}
  \case $\rho_{{\mathrm{21}}} \,  =  \,  \cdot $:
   Let $\rho_{{\mathrm{11}}} \,  =  \,  \cdot $ and $\rho_{{\mathrm{12}}} \,  =  \, \rho_{{\mathrm{1}}}$.  Then, it suffices to show the followings.
   \begin{itemize}
    \item $\rho_{{\mathrm{1}}}  \equiv  \rho_{{\mathrm{1}}}$.  By \Eq{Refl}.
    \item $  \ell  \mathbin{:}  \ottnt{A}   ;  \rho_{{\mathrm{1}}}   \sim    \ell  \mathbin{:}  \ottnt{B}   ;  \rho_{{\mathrm{22}}} $.
          Since $\rho_{{\mathrm{1}}}  \sim  \rho_{{\mathrm{21}}}  \odot  \rho_{{\mathrm{22}}} = \rho_{{\mathrm{22}}}$ and $\ottnt{A}  \sim  \ottnt{B}$,
          we prove this by \Cns{Cons}.
    \item Supposing $\rho_{{\mathrm{22}}}$ ends with $ \star $, we have to show
          $\ottsym{(}    \ell  \mathbin{:}  \ottnt{A}   ;   \cdot    \ottsym{)}  \odot  \rho_{{\mathrm{3}}}  \odot  \rho_{{\mathrm{1}}}  \sim  \ottsym{(}    \ell  \mathbin{:}  \ottnt{B}   ;   \cdot    \ottsym{)}  \odot  \rho_{{\mathrm{22}}}$ for $\rho_{{\mathrm{3}}}$ such that
          $\mathit{dom} \, \ottsym{(}  \rho_{{\mathrm{3}}}  \ottsym{)} \,  \mathbin{\cap}  \, \mathit{dom} \, \ottsym{(}  \rho_{{\mathrm{22}}}  \ottsym{)} \,  =  \,  \emptyset $.
          Since $\rho_{{\mathrm{1}}}  \sim  \rho_{{\mathrm{22}}}$ and $\mathit{dom} \, \ottsym{(}  \rho_{{\mathrm{3}}}  \ottsym{)} \,  \mathbin{\cap}  \, \mathit{dom} \, \ottsym{(}  \rho_{{\mathrm{22}}}  \ottsym{)} \,  =  \,  \emptyset $ and $\rho_{{\mathrm{22}}}$ ends with $ \star $,
          we have $\rho_{{\mathrm{3}}}  \odot  \rho_{{\mathrm{1}}}  \sim  \rho_{{\mathrm{22}}}$ by \Cns{ConsL}.
          Since $\ottnt{A}  \sim  \ottnt{B}$, we have that by \Cns{Cons}.
    \item $\ell \,  \not\in  \, \mathit{dom} \, \ottsym{(}   \cdot   \ottsym{)}$.  Trivial.
   \end{itemize}

  \case $\rho_{{\mathrm{21}}} \,  =  \,   \ell'  \mathbin{:}  \ottnt{C}   ;  \rho'_{{\mathrm{21}}} $:
   We have:
   \begin{eqnarray}
    \rho_{{\mathrm{21}}}  \odot  \rho_{{\mathrm{22}}} \,  =  \,   \ell'  \mathbin{:}  \ottnt{C}   ;  \rho'_{{\mathrm{21}}}  \odot  \rho_{{\mathrm{22}}}  \\
    \ell \,  \not\in  \, \mathit{dom} \, \ottsym{(}    \ell'  \mathbin{:}  \ottnt{C}   ;  \rho'_{{\mathrm{21}}}   \ottsym{)} \label{lem:consistent-decomp-aux:eq:ldom}
   \end{eqnarray}
   By case analysis on the rule applied last to derive $\rho_{{\mathrm{1}}}  \sim    \ell'  \mathbin{:}  \ottnt{C}   ;  \rho'_{{\mathrm{21}}}  \odot  \rho_{{\mathrm{22}}} $.
   \begin{caseanalysis}
    \case \Cns{Refl}:
     We have $\rho_{{\mathrm{1}}} \,  =  \,   \ell'  \mathbin{:}  \ottnt{C}   ;  \rho'_{{\mathrm{21}}}  \odot  \rho_{{\mathrm{22}}} $.
     Let $\rho_{{\mathrm{11}}} \,  =  \,   \ell'  \mathbin{:}  \ottnt{C}   ;  \rho'_{{\mathrm{21}}} $ and $\rho_{{\mathrm{12}}} \,  =  \, \rho_{{\mathrm{22}}}$.
     Then, it suffices to show the followings.
     \begin{itemize}
      \item $\rho_{{\mathrm{1}}}  \equiv  \ottsym{(}    \ell'  \mathbin{:}  \ottnt{C}   ;  \rho'_{{\mathrm{21}}}   \ottsym{)}  \odot  \rho_{{\mathrm{22}}}$.  By \Eq{Refl}.
      \item $  \ell'  \mathbin{:}  \ottnt{C}   ;  \rho'_{{\mathrm{21}}}   \odot  \ottsym{(}    \ell  \mathbin{:}  \ottnt{A}   ;   \cdot    \ottsym{)}  \odot  \rho_{{\mathrm{22}}}  \sim    \ell'  \mathbin{:}  \ottnt{C}   ;  \rho'_{{\mathrm{21}}}   \odot  \ottsym{(}    \ell  \mathbin{:}  \ottnt{B}   ;   \cdot    \ottsym{)}  \odot  \rho_{{\mathrm{22}}}$.
            By \Cns{Refl} and \Cns{Cons}.
      \item Supposing $\ottsym{(}    \ell'  \mathbin{:}  \ottnt{C}   ;  \rho'_{{\mathrm{21}}}   \ottsym{)}  \odot  \rho_{{\mathrm{22}}}$ ends with $ \star $, we have to show
            \[
               \ell'  \mathbin{:}  \ottnt{C}   ;  \rho'_{{\mathrm{21}}}   \odot  \ottsym{(}    \ell  \mathbin{:}  \ottnt{A}   ;   \cdot    \ottsym{)}  \odot  \rho_{{\mathrm{3}}}  \odot  \rho_{{\mathrm{22}}}  \sim    \ell'  \mathbin{:}  \ottnt{C}   ;  \rho'_{{\mathrm{21}}}   \odot  \ottsym{(}    \ell  \mathbin{:}  \ottnt{B}   ;   \cdot    \ottsym{)}  \odot  \rho_{{\mathrm{22}}}
            \]
            for any $\rho_{{\mathrm{3}}}$ such that $\mathit{dom} \, \ottsym{(}  \rho_{{\mathrm{3}}}  \ottsym{)} \,  \mathbin{\cap}  \, \mathit{dom} \, \ottsym{(}    \ell'  \mathbin{:}  \ottnt{C}   ;  \rho'_{{\mathrm{21}}}   \odot  \rho_{{\mathrm{22}}}  \ottsym{)} \,  =  \,  \emptyset $.
            By \Cns{Refl}, \Cns{ConsL}, and \Cns{Cons}.
      \item $\ell \,  \not\in  \, \mathit{dom} \, \ottsym{(}    \ell'  \mathbin{:}  \ottnt{C}   ;  \rho'_{{\mathrm{21}}}   \ottsym{)}$.  By (\ref{lem:consistent-decomp-aux:eq:ldom}).
     \end{itemize}

    \case \Cns{DynL}:
     We have $\rho_{{\mathrm{1}}} \,  =  \, \star$.
     Let $\rho_{{\mathrm{11}}} \,  =  \,  \cdot $ and $\rho_{{\mathrm{12}}} \,  =  \, \star$.
     Then, it suffices to show the followings.
     \begin{itemize}
      \item $\star  \equiv  \star$. By \Eq{Refl}.
      \item $  \ell  \mathbin{:}  \ottnt{A}   ;  \star   \sim    \ell'  \mathbin{:}  \ottnt{C}   ;  \rho'_{{\mathrm{21}}}   \odot  \ottsym{(}    \ell  \mathbin{:}  \ottnt{B}   ;   \cdot    \ottsym{)}  \odot  \rho_{{\mathrm{22}}}$.
            By \Cns{ConsR} and \Cns{Cons} with (\ref{lem:consistent-decomp-aux:eq:ldom}).
      \item Supposing $\ottsym{(}    \ell'  \mathbin{:}  \ottnt{C}   ;  \rho'_{{\mathrm{21}}}   \ottsym{)}  \odot  \rho_{{\mathrm{22}}}$ ends with $ \star $, we have to show
            \[
            \ottsym{(}    \ell  \mathbin{:}  \ottnt{A}   ;   \cdot    \ottsym{)}  \odot  \rho_{{\mathrm{3}}}  \odot  \star  \sim    \ell'  \mathbin{:}  \ottnt{C}   ;  \rho'_{{\mathrm{21}}}   \odot  \ottsym{(}    \ell  \mathbin{:}  \ottnt{B}   ;   \cdot    \ottsym{)}  \odot  \rho_{{\mathrm{22}}}
            \]
            for any $\rho_{{\mathrm{3}}}$ such that $\mathit{dom} \, \ottsym{(}  \rho_{{\mathrm{3}}}  \ottsym{)} \,  \mathbin{\cap}  \, \mathit{dom} \, \ottsym{(}    \ell'  \mathbin{:}  \ottnt{C}   ;  \rho'_{{\mathrm{21}}}   \odot  \rho_{{\mathrm{22}}}  \ottsym{)} \,  =  \,  \emptyset $.
            By \Cns{ConsR}, \Cns{ConsL}, and \Cns{Cons}.
      \item $\ell \,  \not\in  \, \mathit{dom} \, \ottsym{(}   \cdot   \ottsym{)}$. Trivial.
     \end{itemize}

    \case \Cns{Cons}:
     We have $\rho_{{\mathrm{1}}} \,  =  \,   \ell'  \mathbin{:}  \ottnt{D}   ;  \rho'_{{\mathrm{1}}} $ and, by inversion,
     $\ottnt{D}  \sim  \ottnt{C}$ and $\rho'_{{\mathrm{1}}}  \sim  \rho'_{{\mathrm{21}}}  \odot  \rho_{{\mathrm{22}}}$
     for some $\ottnt{D}$ and $\rho'_{{\mathrm{1}}}$.
     By the IH, there exist some $\rho'_{{\mathrm{11}}}$ and $\rho_{{\mathrm{12}}}$ such that
     \begin{itemize}
      \item[(a)] $\rho'_{{\mathrm{1}}}  \equiv  \rho'_{{\mathrm{11}}}  \odot  \rho_{{\mathrm{12}}}$,
      \item[(b)] $\rho'_{{\mathrm{11}}}  \odot  \ottsym{(}    \ell  \mathbin{:}  \ottnt{A}   ;   \cdot    \ottsym{)}  \odot  \rho_{{\mathrm{12}}}  \sim  \rho'_{{\mathrm{21}}}  \odot  \ottsym{(}    \ell  \mathbin{:}  \ottnt{B}   ;   \cdot    \ottsym{)}  \odot  \rho_{{\mathrm{22}}}$,
      \item[(c)] $\rho'_{{\mathrm{11}}}  \odot  \ottsym{(}    \ell  \mathbin{:}  \ottnt{A}   ;   \cdot    \ottsym{)}  \odot  \rho_{{\mathrm{3}}}  \odot  \rho_{{\mathrm{12}}}  \sim  \rho'_{{\mathrm{21}}}  \odot  \ottsym{(}    \ell  \mathbin{:}  \ottnt{B}   ;   \cdot    \ottsym{)}  \odot  \rho_{{\mathrm{22}}}$
                 for any $\rho_{{\mathrm{3}}}$ such that $\mathit{dom} \, \ottsym{(}  \rho_{{\mathrm{3}}}  \ottsym{)} \,  \mathbin{\cap}  \, \mathit{dom} \, \ottsym{(}  \rho'_{{\mathrm{21}}}  \odot  \rho_{{\mathrm{22}}}  \ottsym{)} \,  =  \,  \emptyset $ if
                 $\rho'_{{\mathrm{21}}}  \odot  \rho_{{\mathrm{22}}}$ ends with $ \star $ for some $\rho_{{\mathrm{2}}}$, and
      \item[(d)] $\ell \,  \not\in  \, \mathit{dom} \, \ottsym{(}  \rho'_{{\mathrm{11}}}  \ottsym{)}$
     \end{itemize}
     for some $\rho'_{{\mathrm{11}}}$ and $\rho_{{\mathrm{12}}}$.

     Let $\rho_{{\mathrm{11}}} \,  =  \,   \ell'  \mathbin{:}  \ottnt{D}   ;  \rho'_{{\mathrm{11}}} $.
     Then, it suffices to show the followings.
     \begin{itemize}
      \item $\rho_{{\mathrm{1}}} =   \ell'  \mathbin{:}  \ottnt{D}   ;  \rho'_{{\mathrm{1}}}   \equiv    \ell'  \mathbin{:}  \ottnt{D}   ;  \rho'_{{\mathrm{11}}}   \odot  \rho_{{\mathrm{12}}}$. By (a) and \Eq{Cons}.
      \item $  \ell'  \mathbin{:}  \ottnt{D}   ;  \rho'_{{\mathrm{11}}}   \odot  \ottsym{(}    \ell  \mathbin{:}  \ottnt{A}   ;   \cdot    \ottsym{)}  \odot  \rho_{{\mathrm{12}}}  \sim    \ell'  \mathbin{:}  \ottnt{C}   ;  \rho'_{{\mathrm{21}}}   \odot  \ottsym{(}    \ell  \mathbin{:}  \ottnt{B}   ;   \cdot    \ottsym{)}  \odot  \rho_{{\mathrm{22}}}$.
            By (b) and \Cns{Cons} with $\ottnt{D}  \sim  \ottnt{C}$.
      \item Supposing $\ottsym{(}    \ell'  \mathbin{:}  \ottnt{C}   ;  \rho'_{{\mathrm{21}}}   \ottsym{)}  \odot  \rho_{{\mathrm{22}}}$ ends with $ \star $ for some $\rho_{{\mathrm{2}}}$, we have to show
            \[
              \ell'  \mathbin{:}  \ottnt{D}   ;  \rho'_{{\mathrm{11}}}   \odot  \ottsym{(}    \ell  \mathbin{:}  \ottnt{A}   ;   \cdot    \ottsym{)}  \odot  \rho_{{\mathrm{3}}}  \odot  \rho_{{\mathrm{12}}}  \sim    \ell'  \mathbin{:}  \ottnt{C}   ;  \rho'_{{\mathrm{21}}}   \odot  \ottsym{(}    \ell  \mathbin{:}  \ottnt{B}   ;   \cdot    \ottsym{)}  \odot  \rho_{{\mathrm{22}}}
            \]
            for any $\rho_{{\mathrm{3}}}$ such that $\mathit{dom} \, \ottsym{(}  \rho_{{\mathrm{3}}}  \ottsym{)} \,  \mathbin{\cap}  \, \mathit{dom} \, \ottsym{(}    \ell'  \mathbin{:}  \ottnt{C}   ;  \rho'_{{\mathrm{21}}}   \odot  \rho_{{\mathrm{22}}}  \ottsym{)} \,  =  \,  \emptyset $.
            By (c) and \Cns{Cons} with $\ottnt{D}  \sim  \ottnt{C}$.
      \item $\ell \,  \not\in  \, \mathit{dom} \, \ottsym{(}    \ell'  \mathbin{:}  \ottnt{D}   ;  \rho'_{{\mathrm{11}}}   \ottsym{)}$. By (d) and (\ref{lem:consistent-decomp-aux:eq:ldom}).
     \end{itemize}

    \case \Cns{ConsL}:
     We have $\rho_{{\mathrm{1}}} \,  =  \,   \ell''  \mathbin{:}  \ottnt{D}   ;  \rho'_{{\mathrm{1}}} $ and, by inversion,
     \begin{itemize}
      \item $\ell'' \,  \not\in  \, \mathit{dom} \, \ottsym{(}    \ell'  \mathbin{:}  \ottnt{C}   ;  \rho'_{{\mathrm{21}}}  \odot  \rho_{{\mathrm{22}}}   \ottsym{)}$,
      \item $  \ell'  \mathbin{:}  \ottnt{C}   ;  \rho'_{{\mathrm{21}}}  \odot  \rho_{{\mathrm{22}}} $ ends with $ \star $, and
      \item $\rho'_{{\mathrm{1}}}  \sim    \ell'  \mathbin{:}  \ottnt{C}   ;  \rho'_{{\mathrm{21}}}  \odot  \rho_{{\mathrm{22}}} $
     \end{itemize}
     for some $\ell''$, $\ottnt{D}$, $\rho'_{{\mathrm{1}}}$, and $\rho_{{\mathrm{2}}}$.

     By the IH, there exist some $\rho'_{{\mathrm{11}}}$ and $\rho'_{{\mathrm{12}}}$ such that
     \begin{itemize}
      \item[(a)] $\rho'_{{\mathrm{1}}}  \equiv  \rho'_{{\mathrm{11}}}  \odot  \rho'_{{\mathrm{12}}}$,
      \item[(b)] $\rho'_{{\mathrm{11}}}  \odot  \ottsym{(}    \ell  \mathbin{:}  \ottnt{A}   ;   \cdot    \ottsym{)}  \odot  \rho'_{{\mathrm{12}}}  \sim    \ell'  \mathbin{:}  \ottnt{C}   ;  \rho'_{{\mathrm{21}}}   \odot  \ottsym{(}    \ell  \mathbin{:}  \ottnt{B}   ;   \cdot    \ottsym{)}  \odot  \rho_{{\mathrm{22}}}$, and
      \item[(c)] $\rho'_{{\mathrm{11}}}  \odot  \ottsym{(}    \ell  \mathbin{:}  \ottnt{A}   ;   \cdot    \ottsym{)}  \odot  \rho_{{\mathrm{3}}}  \odot  \rho'_{{\mathrm{12}}}  \sim    \ell'  \mathbin{:}  \ottnt{C}   ;  \rho'_{{\mathrm{21}}}   \odot  \ottsym{(}    \ell  \mathbin{:}  \ottnt{B}   ;   \cdot    \ottsym{)}  \odot  \rho_{{\mathrm{22}}}$
                 for any $\rho_{{\mathrm{3}}}$ such that $\mathit{dom} \, \ottsym{(}  \rho_{{\mathrm{3}}}  \ottsym{)} \,  \mathbin{\cap}  \, \mathit{dom} \, \ottsym{(}    \ell'  \mathbin{:}  \ottnt{C}   ;  \rho'_{{\mathrm{21}}}  \odot  \rho_{{\mathrm{22}}}   \ottsym{)} \,  =  \,  \emptyset $ if
                 $  \ell'  \mathbin{:}  \ottnt{C}   ;  \rho'_{{\mathrm{21}}}  \odot  \rho_{{\mathrm{22}}} $ ends with $ \star $ for some $\rho'_{{\mathrm{2}}}$, and
      \item[(d)] $\ell \,  \not\in  \, \mathit{dom} \, \ottsym{(}  \rho'_{{\mathrm{11}}}  \ottsym{)}$.
     \end{itemize}

     Suppose that $\ell'' \,  =  \, \ell$.  By (d), $\ell'' \,  \not\in  \, \mathit{dom} \, \ottsym{(}  \rho'_{{\mathrm{11}}}  \ottsym{)}$.
     Let $\rho_{{\mathrm{11}}} \,  =  \, \rho'_{{\mathrm{11}}}$ and $\rho_{{\mathrm{12}}} \,  =  \,   \ell''  \mathbin{:}  \ottnt{D}   ;  \rho'_{{\mathrm{12}}} $.
     Then, it suffices to show the followings.
     \begin{itemize}
      \item $  \ell''  \mathbin{:}  \ottnt{D}   ;  \rho'_{{\mathrm{1}}}   \equiv  \rho'_{{\mathrm{11}}}  \odot    \ell''  \mathbin{:}  \ottnt{D}   ;  \rho'_{{\mathrm{12}}} $.
            By (a) and \Eq{Cons}, we have
            \[
               \ell''  \mathbin{:}  \ottnt{D}   ;  \rho'_{{\mathrm{1}}}   \equiv    \ell''  \mathbin{:}  \ottnt{D}   ;   \cdot    \odot  \rho'_{{\mathrm{11}}}  \odot  \rho'_{{\mathrm{12}}}.
            \]
            Since $\ell'' \,  \not\in  \, \mathit{dom} \, \ottsym{(}  \rho'_{{\mathrm{11}}}  \ottsym{)}$, we have
            \[
               \ell''  \mathbin{:}  \ottnt{D}   ;  \rho'_{{\mathrm{1}}}   \equiv  \rho'_{{\mathrm{11}}}  \odot    \ell''  \mathbin{:}  \ottnt{D}   ;  \rho'_{{\mathrm{12}}} .
            \]

      \item $\rho'_{{\mathrm{11}}}  \odot  \ottsym{(}    \ell  \mathbin{:}  \ottnt{A}   ;   \cdot    \ottsym{)}  \odot    \ell''  \mathbin{:}  \ottnt{D}   ;  \rho'_{{\mathrm{12}}}   \sim    \ell'  \mathbin{:}  \ottnt{C}   ;  \rho'_{{\mathrm{21}}}   \odot  \ottsym{(}    \ell  \mathbin{:}  \ottnt{B}   ;   \cdot    \ottsym{)}  \odot  \rho_{{\mathrm{22}}}$.
            Since $\ell'' \,  \not\in  \, \mathit{dom} \, \ottsym{(}    \ell'  \mathbin{:}  \ottnt{C}   ;  \rho'_{{\mathrm{21}}}  \odot  \rho_{{\mathrm{22}}}   \ottsym{)}$ and $\rho_{{\mathrm{22}}}$ ends with $ \star $,
            we have
            \[
             \rho'_{{\mathrm{11}}}  \odot  \ottsym{(}    \ell  \mathbin{:}  \ottnt{A}   ;   \cdot    \ottsym{)}  \odot  \ottsym{(}    \ell''  \mathbin{:}  \ottnt{D}   ;   \cdot    \ottsym{)}  \odot  \rho'_{{\mathrm{12}}}  \sim    \ell'  \mathbin{:}  \ottnt{C}   ;  \rho'_{{\mathrm{21}}}   \odot  \ottsym{(}    \ell  \mathbin{:}  \ottnt{B}   ;   \cdot    \ottsym{)}  \odot  \rho_{{\mathrm{22}}}
            \]
            by (c).

      \item Supposing $\ottsym{(}    \ell'  \mathbin{:}  \ottnt{C}   ;  \rho'_{{\mathrm{21}}}   \ottsym{)}  \odot  \rho_{{\mathrm{22}}}$ ends with $ \star $, we have to show
            \[
            \rho'_{{\mathrm{11}}}  \odot  \ottsym{(}    \ell  \mathbin{:}  \ottnt{A}   ;   \cdot    \ottsym{)}  \odot  \rho_{{\mathrm{3}}}  \odot    \ell''  \mathbin{:}  \ottnt{D}   ;  \rho'_{{\mathrm{12}}}   \sim    \ell'  \mathbin{:}  \ottnt{C}   ;  \rho'_{{\mathrm{21}}}   \odot  \ottsym{(}    \ell  \mathbin{:}  \ottnt{B}   ;   \cdot    \ottsym{)}  \odot  \rho_{{\mathrm{22}}}
            \]
            for any $\rho_{{\mathrm{3}}}$ such that $\mathit{dom} \, \ottsym{(}  \rho_{{\mathrm{3}}}  \ottsym{)} \,  \mathbin{\cap}  \, \mathit{dom} \, \ottsym{(}    \ell'  \mathbin{:}  \ottnt{C}   ;  \rho'_{{\mathrm{21}}}   \odot  \rho_{{\mathrm{22}}}  \ottsym{)} \,  =  \,  \emptyset $.
            Since $\ell'' \,  \not\in  \, \mathit{dom} \, \ottsym{(}    \ell'  \mathbin{:}  \ottnt{C}   ;  \rho'_{{\mathrm{21}}}  \odot  \rho_{{\mathrm{22}}}   \ottsym{)}$, we have 
            \[
             \rho'_{{\mathrm{11}}}  \odot  \ottsym{(}    \ell  \mathbin{:}  \ottnt{A}   ;   \cdot    \ottsym{)}  \odot  \rho_{{\mathrm{3}}}  \odot  \ottsym{(}    \ell''  \mathbin{:}  \ottnt{D}   ;   \cdot    \ottsym{)}  \odot  \rho'_{{\mathrm{12}}}  \sim    \ell'  \mathbin{:}  \ottnt{C}   ;  \rho'_{{\mathrm{21}}}   \odot  \ottsym{(}    \ell  \mathbin{:}  \ottnt{B}   ;   \cdot    \ottsym{)}  \odot  \rho_{{\mathrm{22}}}.
            \]
            by (c).

      \item $\ell \,  \not\in  \, \mathit{dom} \, \ottsym{(}  \rho'_{{\mathrm{11}}}  \ottsym{)}$. By (d).
     \end{itemize}

     Otherwise, suppose that $\ell'' \,  \not=  \, \ell$.
     Let $\rho_{{\mathrm{11}}} \,  =  \,   \ell''  \mathbin{:}  \ottnt{D}   ;  \rho'_{{\mathrm{11}}} $ and $\rho_{{\mathrm{12}}} \,  =  \, \rho'_{{\mathrm{12}}}$.
     Then, it suffices to show the followings.
     \begin{itemize}
      \item $  \ell''  \mathbin{:}  \ottnt{D}   ;  \rho'_{{\mathrm{1}}}   \equiv    \ell''  \mathbin{:}  \ottnt{D}   ;  \rho'_{{\mathrm{11}}}   \odot  \rho'_{{\mathrm{12}}}$.
            By (a) and \Eq{Cons}.
      \item $  \ell''  \mathbin{:}  \ottnt{D}   ;  \rho'_{{\mathrm{11}}}   \odot  \ottsym{(}    \ell  \mathbin{:}  \ottnt{A}   ;   \cdot    \ottsym{)}  \odot  \rho'_{{\mathrm{12}}}  \sim    \ell'  \mathbin{:}  \ottnt{C}   ;  \rho'_{{\mathrm{21}}}   \odot  \ottsym{(}    \ell  \mathbin{:}  \ottnt{B}   ;   \cdot    \ottsym{)}  \odot  \rho_{{\mathrm{22}}}$.
            Since $\ell'' \,  \not\in  \, \mathit{dom} \, \ottsym{(}    \ell'  \mathbin{:}  \ottnt{C}   ;  \rho'_{{\mathrm{21}}}  \odot  \rho_{{\mathrm{22}}}   \ottsym{)}$ and $\ell'' \,  \not=  \, \ell$,
            we have $\ell'' \,  \not\in  \, \mathit{dom} \, \ottsym{(}    \ell'  \mathbin{:}  \ottnt{C}   ;  \rho'_{{\mathrm{21}}}   \odot  \ottsym{(}    \ell  \mathbin{:}  \ottnt{B}   ;   \cdot    \ottsym{)}  \odot  \rho_{{\mathrm{22}}}  \ottsym{)}$.
            Since $\rho_{{\mathrm{22}}}$ ends with $ \star $, we finish by (b) and \Cns{ConsL}.
      \item Supposing $\ottsym{(}    \ell'  \mathbin{:}  \ottnt{C}   ;  \rho'_{{\mathrm{21}}}   \ottsym{)}  \odot  \rho_{{\mathrm{22}}}$ ends with $ \star $, we have to show
            \[
              \ell''  \mathbin{:}  \ottnt{D}   ;  \rho'_{{\mathrm{11}}}   \odot  \ottsym{(}    \ell  \mathbin{:}  \ottnt{A}   ;   \cdot    \ottsym{)}  \odot  \rho_{{\mathrm{3}}}  \odot  \rho'_{{\mathrm{12}}}  \sim    \ell'  \mathbin{:}  \ottnt{C}   ;  \rho'_{{\mathrm{21}}}   \odot  \ottsym{(}    \ell  \mathbin{:}  \ottnt{B}   ;   \cdot    \ottsym{)}  \odot  \rho_{{\mathrm{22}}}
            \]
            for any $\rho_{{\mathrm{3}}}$ such that $\mathit{dom} \, \ottsym{(}  \rho_{{\mathrm{3}}}  \ottsym{)} \,  \mathbin{\cap}  \, \mathit{dom} \, \ottsym{(}    \ell'  \mathbin{:}  \ottnt{C}   ;  \rho'_{{\mathrm{21}}}   \odot  \rho_{{\mathrm{22}}}  \ottsym{)} \,  =  \,  \emptyset $.
            By (c), we have
            \[
             \rho'_{{\mathrm{11}}}  \odot  \ottsym{(}    \ell  \mathbin{:}  \ottnt{A}   ;   \cdot    \ottsym{)}  \odot  \rho_{{\mathrm{3}}}  \odot  \rho'_{{\mathrm{12}}}  \sim    \ell'  \mathbin{:}  \ottnt{C}   ;  \rho'_{{\mathrm{21}}}   \odot  \ottsym{(}    \ell  \mathbin{:}  \ottnt{B}   ;   \cdot    \ottsym{)}  \odot  \rho_{{\mathrm{22}}}.
            \]
            Since $\ell'' \,  \not\in  \, \mathit{dom} \, \ottsym{(}    \ell'  \mathbin{:}  \ottnt{C}   ;  \rho'_{{\mathrm{21}}}  \odot  \rho_{{\mathrm{22}}}   \ottsym{)}$ and $\ell'' \,  \not=  \, \ell$,
            we have $\ell'' \,  \not\in  \, \mathit{dom} \, \ottsym{(}    \ell'  \mathbin{:}  \ottnt{C}   ;  \rho'_{{\mathrm{21}}}   \odot  \ottsym{(}    \ell  \mathbin{:}  \ottnt{B}   ;   \cdot    \ottsym{)}  \odot  \rho_{{\mathrm{22}}}  \ottsym{)}$.
            Since $\rho_{{\mathrm{22}}}$ ends with $ \star $, we finish by \Cns{ConsL}.
      \item $\ell \,  \not\in  \, \mathit{dom} \, \ottsym{(}    \ell''  \mathbin{:}  \ottnt{D}   ;  \rho'_{{\mathrm{11}}}   \ottsym{)}$. By (d) and $\ell'' \,  \not=  \, \ell$.
     \end{itemize}

    \case \Cns{ConsR}:
     By inversion, we have
     \begin{itemize}
      \item $\ell' \,  \not\in  \, \mathit{dom} \, \ottsym{(}  \rho_{{\mathrm{1}}}  \ottsym{)}$,
      \item $\rho_{{\mathrm{1}}}$ ends with $ \star $, and
      \item $\rho_{{\mathrm{1}}}  \sim  \rho'_{{\mathrm{21}}}  \odot  \rho_{{\mathrm{22}}}$.
     \end{itemize}
     By the IH, there exist some $\rho_{{\mathrm{11}}}$ and $\rho_{{\mathrm{12}}}$ such that
     \begin{itemize}
      \item[(a)] $\rho_{{\mathrm{1}}}  \equiv  \rho_{{\mathrm{11}}}  \odot  \rho_{{\mathrm{12}}}$,
      \item[(b)] $\rho_{{\mathrm{11}}}  \odot  \ottsym{(}    \ell  \mathbin{:}  \ottnt{A}   ;   \cdot    \ottsym{)}  \odot  \rho_{{\mathrm{12}}}  \sim  \rho'_{{\mathrm{21}}}  \odot  \ottsym{(}    \ell  \mathbin{:}  \ottnt{B}   ;   \cdot    \ottsym{)}  \odot  \rho_{{\mathrm{22}}}$,
      \item[(c)] $\rho_{{\mathrm{11}}}  \odot  \ottsym{(}    \ell  \mathbin{:}  \ottnt{A}   ;   \cdot    \ottsym{)}  \odot  \rho_{{\mathrm{3}}}  \odot  \rho_{{\mathrm{12}}}  \sim  \rho'_{{\mathrm{21}}}  \odot  \ottsym{(}    \ell  \mathbin{:}  \ottnt{B}   ;   \cdot    \ottsym{)}  \odot  \rho_{{\mathrm{22}}}$
                 for any $\rho_{{\mathrm{3}}}$ such that $\mathit{dom} \, \ottsym{(}  \rho_{{\mathrm{3}}}  \ottsym{)} \,  \mathbin{\cap}  \, \mathit{dom} \, \ottsym{(}  \rho'_{{\mathrm{21}}}  \odot  \rho_{{\mathrm{22}}}  \ottsym{)} \,  =  \,  \emptyset $ if
                 $\rho'_{{\mathrm{21}}}  \odot  \rho_{{\mathrm{22}}}$ ends with $ \star $ for some $\rho_{{\mathrm{2}}}$, and
      \item[(d)] $\ell \,  \not\in  \, \mathit{dom} \, \ottsym{(}  \rho_{{\mathrm{11}}}  \ottsym{)}$.
     \end{itemize}

     First, we show
     \begin{equation}
      \ell' \,  \not\in  \, \mathit{dom} \, \ottsym{(}  \rho_{{\mathrm{11}}}  \odot  \ottsym{(}    \ell  \mathbin{:}  \ottnt{A}   ;   \cdot    \ottsym{)}  \odot  \rho_{{\mathrm{12}}}  \ottsym{)}.
       \label{lem:consistent-decomp-aux:eq:l'notin}
     \end{equation}
     Since $\ell \,  \not\in  \, \mathit{dom} \, \ottsym{(}  \rho_{{\mathrm{21}}}  \ottsym{)}$ from the assumption and $\rho_{{\mathrm{21}}} \,  =  \,   \ell'  \mathbin{:}  \ottnt{C}   ;  \rho'_{{\mathrm{21}}} $, $\ell \,  \not=  \, \ell'$.
     Since $\ell' \,  \not\in  \, \mathit{dom} \, \ottsym{(}  \rho_{{\mathrm{1}}}  \ottsym{)}$ and $\rho_{{\mathrm{1}}}  \equiv  \rho_{{\mathrm{11}}}  \odot  \rho_{{\mathrm{12}}}$,
     we have $\ell' \,  \not\in  \, \mathit{dom} \, \ottsym{(}  \rho_{{\mathrm{11}}}  \odot  \ottsym{(}    \ell  \mathbin{:}  \ottnt{A}   ;   \cdot    \ottsym{)}  \odot  \rho_{{\mathrm{12}}}  \ottsym{)}$.

     It suffices to show the followings.
     \begin{itemize}
      \item $\rho_{{\mathrm{1}}}  \equiv  \rho_{{\mathrm{11}}}  \odot  \rho_{{\mathrm{12}}}$.  By (a).
      \item $\rho_{{\mathrm{11}}}  \odot  \ottsym{(}    \ell  \mathbin{:}  \ottnt{A}   ;   \cdot    \ottsym{)}  \odot  \rho_{{\mathrm{12}}}  \sim    \ell'  \mathbin{:}  \ottnt{C}   ;  \rho'_{{\mathrm{21}}}   \odot  \ottsym{(}    \ell  \mathbin{:}  \ottnt{B}   ;   \cdot    \ottsym{)}  \odot  \rho_{{\mathrm{22}}}$.
            Since $\rho_{{\mathrm{1}}}$ ends with $ \star $ and $\rho_{{\mathrm{1}}}  \equiv  \rho_{{\mathrm{11}}}  \odot  \rho_{{\mathrm{12}}}$,
            $\rho_{{\mathrm{12}}}$ ends with $ \star $.
            Thus, we have
            \[
             \rho_{{\mathrm{11}}}  \odot  \ottsym{(}    \ell  \mathbin{:}  \ottnt{A}   ;   \cdot    \ottsym{)}  \odot  \rho_{{\mathrm{12}}}  \sim    \ell'  \mathbin{:}  \ottnt{C}   ;  \rho'_{{\mathrm{21}}}   \odot  \ottsym{(}    \ell  \mathbin{:}  \ottnt{B}   ;   \cdot    \ottsym{)}  \odot  \rho_{{\mathrm{22}}}
            \]
            by (b), (\ref{lem:consistent-decomp-aux:eq:l'notin}), and \Cns{ConsR}.

      \item Supposing $\ottsym{(}    \ell'  \mathbin{:}  \ottnt{C}   ;  \rho'_{{\mathrm{21}}}   \ottsym{)}  \odot  \rho_{{\mathrm{22}}}$ ends with $ \star $ for some $\rho_{{\mathrm{2}}}$, we have to show
            \[
             \rho_{{\mathrm{11}}}  \odot  \ottsym{(}    \ell  \mathbin{:}  \ottnt{A}   ;   \cdot    \ottsym{)}  \odot  \rho_{{\mathrm{3}}}  \odot  \rho_{{\mathrm{12}}}  \sim    \ell'  \mathbin{:}  \ottnt{C}   ;  \rho'_{{\mathrm{21}}}   \odot  \ottsym{(}    \ell  \mathbin{:}  \ottnt{B}   ;   \cdot    \ottsym{)}  \odot  \rho_{{\mathrm{22}}}
            \]
            for any $\rho_{{\mathrm{3}}}$ such that $\mathit{dom} \, \ottsym{(}  \rho_{{\mathrm{3}}}  \ottsym{)} \,  \mathbin{\cap}  \, \mathit{dom} \, \ottsym{(}    \ell'  \mathbin{:}  \ottnt{C}   ;  \rho'_{{\mathrm{21}}}   \odot  \rho_{{\mathrm{22}}}  \ottsym{)} \,  =  \,  \emptyset $.
            By (c), we have
            \[
             \rho_{{\mathrm{11}}}  \odot  \ottsym{(}    \ell  \mathbin{:}  \ottnt{A}   ;   \cdot    \ottsym{)}  \odot  \rho_{{\mathrm{3}}}  \odot  \rho_{{\mathrm{12}}}  \sim  \rho'_{{\mathrm{21}}}  \odot  \ottsym{(}    \ell  \mathbin{:}  \ottnt{B}   ;   \cdot    \ottsym{)}  \odot  \rho_{{\mathrm{22}}}.
            \]
            By (\ref{lem:consistent-decomp-aux:eq:l'notin}),
            $\ell' \,  \not\in  \, \mathit{dom} \, \ottsym{(}  \rho_{{\mathrm{11}}}  \odot  \ottsym{(}    \ell  \mathbin{:}  \ottnt{A}   ;   \cdot    \ottsym{)}  \odot  \rho_{{\mathrm{3}}}  \odot  \rho_{{\mathrm{12}}}  \ottsym{)}$.
            Since $\rho_{{\mathrm{12}}}$ ends with $ \star $,
            we have
            \[
             \rho_{{\mathrm{11}}}  \odot  \ottsym{(}    \ell  \mathbin{:}  \ottnt{A}   ;   \cdot    \ottsym{)}  \odot  \rho_{{\mathrm{3}}}  \odot  \rho_{{\mathrm{12}}}  \sim    \ell'  \mathbin{:}  \ottnt{C}   ;  \rho'_{{\mathrm{21}}}   \odot  \ottsym{(}    \ell  \mathbin{:}  \ottnt{B}   ;   \cdot    \ottsym{)}  \odot  \rho_{{\mathrm{22}}}
            \]
            by \Cns{ConsR}.
     \end{itemize}

    \case \Cns{DynR}, \Cns{Fun}, \Cns{Poly}, \Cns{PolyL}, \Cns{PolyR}, \Cns{Record}, and \Cns{Variant}:
     Note that the contradiction in the case of \Cns{PolyL} is proven by the definition of $ \mathbf{QPoly} $.
   \end{caseanalysis}
  \end{caseanalysis}
\end{proof}

\begin{lemma}{consistent-decomp}
 If $\ottnt{A}  \simeq  \ottnt{B}$, then $\ottnt{A}  \equiv  \ottnt{C}$ and $\ottnt{C}  \sim  \ottnt{B}$ for some $\ottnt{C}$.
\end{lemma}
\begin{proof}
 By induction on the derivation of $\ottnt{A}  \simeq  \ottnt{B}$.
 \begin{caseanalysis}
  \case \CE{Refl}: Obvious because $ \equiv $ and $ \sim $ are reflexive.
  \case \CE{DynL}: By $\star  \equiv  \star$ \Eq{Refl} and $\star  \sim  \ottnt{B}$ \Cns{DynL}.
  \case \CE{DynR}: By $\ottnt{A}  \equiv  \ottnt{A}$ \Eq{Refl} and $\ottnt{A}  \sim  \star$ \Cns{DynR}.
  \case \CE{Fun}:
   We have $\ottnt{A_{{\mathrm{1}}}}  \rightarrow  \ottnt{A_{{\mathrm{2}}}}  \simeq  \ottnt{B_{{\mathrm{1}}}}  \rightarrow  \ottnt{B_{{\mathrm{2}}}}$ and, by inversion,
   $\ottnt{A_{{\mathrm{1}}}}  \simeq  \ottnt{B_{{\mathrm{1}}}}$ and $\ottnt{A_{{\mathrm{2}}}}  \simeq  \ottnt{B_{{\mathrm{2}}}}$.
   By the IHs,
   \begin{itemize}
    \item $\ottnt{A_{{\mathrm{1}}}}  \equiv  \ottnt{C_{{\mathrm{1}}}}$,
    \item $\ottnt{C_{{\mathrm{1}}}}  \sim  \ottnt{B_{{\mathrm{1}}}}$,
    \item $\ottnt{A_{{\mathrm{2}}}}  \equiv  \ottnt{C_{{\mathrm{2}}}}$, and
    \item $\ottnt{C_{{\mathrm{2}}}}  \sim  \ottnt{B_{{\mathrm{2}}}}$
   \end{itemize}
   for some $\ottnt{C_{{\mathrm{1}}}}$ and $\ottnt{C_{{\mathrm{2}}}}$.
   By \Eq{Fun}, $\ottnt{A_{{\mathrm{1}}}}  \rightarrow  \ottnt{A_{{\mathrm{2}}}}  \equiv  \ottnt{C_{{\mathrm{1}}}}  \rightarrow  \ottnt{C_{{\mathrm{2}}}}$.
   By \Cns{Fun}, $\ottnt{C_{{\mathrm{1}}}}  \rightarrow  \ottnt{C_{{\mathrm{2}}}}  \sim  \ottnt{B_{{\mathrm{1}}}}  \rightarrow  \ottnt{B_{{\mathrm{2}}}}$.

  \case \CE{Poly}:
   We have $ \text{\unboldmath$\forall\!$}  \,  \mathit{X}  \mathord{:}  \ottnt{K}   \ottsym{.} \, \ottnt{A'}  \simeq   \text{\unboldmath$\forall\!$}  \,  \mathit{X}  \mathord{:}  \ottnt{K}   \ottsym{.} \, \ottnt{B'}$ and, by inversion, $\ottnt{A'}  \simeq  \ottnt{B'}$.
   By the IH,
   $\ottnt{A'}  \equiv  \ottnt{C'}$ and $\ottnt{C'}  \sim  \ottnt{B'}$ for some $\ottnt{C'}$.
   By \Eq{Poly}, $ \text{\unboldmath$\forall\!$}  \,  \mathit{X}  \mathord{:}  \ottnt{K}   \ottsym{.} \, \ottnt{A'}  \equiv   \text{\unboldmath$\forall\!$}  \,  \mathit{X}  \mathord{:}  \ottnt{K}   \ottsym{.} \, \ottnt{C'}$.
   By \Cns{Poly}, $ \text{\unboldmath$\forall\!$}  \,  \mathit{X}  \mathord{:}  \ottnt{K}   \ottsym{.} \, \ottnt{C'}  \sim   \text{\unboldmath$\forall\!$}  \,  \mathit{X}  \mathord{:}  \ottnt{K}   \ottsym{.} \, \ottnt{B'}$.

  \case \CE{PolyL}:
   We have $ \text{\unboldmath$\forall\!$}  \,  \mathit{X}  \mathord{:}  \ottnt{K}   \ottsym{.} \, \ottnt{A'}  \simeq  \ottnt{B}$ and, by inversion,
   $\mathbf{QPoly} \, \ottsym{(}  \ottnt{B}  \ottsym{)}$ and $\mathit{X} \,  \not\in  \,  \mathit{ftv}  (  \ottnt{B}  ) $ and $\ottnt{A'}  \simeq  \ottnt{B}$.
   By the IH, $\ottnt{A'}  \equiv  \ottnt{C}$ and $\ottnt{C}  \sim  \ottnt{B}$ for some $\ottnt{C}$.
   By \Eq{Poly}, $ \text{\unboldmath$\forall\!$}  \,  \mathit{X}  \mathord{:}  \ottnt{K}   \ottsym{.} \, \ottnt{A'}  \equiv   \text{\unboldmath$\forall\!$}  \,  \mathit{X}  \mathord{:}  \ottnt{K}   \ottsym{.} \, \ottnt{C}$.
   By \Cns{PolyL}, $ \text{\unboldmath$\forall\!$}  \,  \mathit{X}  \mathord{:}  \ottnt{K}   \ottsym{.} \, \ottnt{C}  \sim  \ottnt{B}$.

  \case \CE{PolyR}:
   We have $\ottnt{A}  \simeq   \text{\unboldmath$\forall\!$}  \,  \mathit{X}  \mathord{:}  \ottnt{K}   \ottsym{.} \, \ottnt{B'}$ and, by inversion,
   $\mathbf{QPoly} \, \ottsym{(}  \ottnt{A}  \ottsym{)}$ and $\mathit{X} \,  \not\in  \,  \mathit{ftv}  (  \ottnt{A}  ) $ and $\ottnt{A}  \simeq  \ottnt{B'}$.
   By the IH, $\ottnt{A}  \equiv  \ottnt{C}$ and $\ottnt{C}  \sim  \ottnt{B'}$ for some $\ottnt{C}$.
   Since $\ottnt{A}  \equiv  \ottnt{C}$,
   we can find $\mathbf{QPoly} \, \ottsym{(}  \ottnt{C}  \ottsym{)}$ by \reflem{equiv-qpoly} and $\mathbf{QPoly} \, \ottsym{(}  \ottnt{A}  \ottsym{)}$, and
   $\mathit{X} \,  \not\in  \,  \mathit{ftv}  (  \ottnt{C}  ) $ by \reflem{equiv-free-tyvar} and $\mathit{X} \,  \not\in  \,  \mathit{ftv}  (  \ottnt{A}  ) $.
   Thus, by \Cns{PolyR}, $\ottnt{C}  \sim   \text{\unboldmath$\forall\!$}  \,  \mathit{X}  \mathord{:}  \ottnt{K}   \ottsym{.} \, \ottnt{B'}$.

  \case \CE{Record}: By the IH, \Eq{Record}, and \Cns{Record}.
  \case \CE{Variant}: By the IH, \Eq{Variant}, and \Cns{Variant}.

  \case \CE{ConsL}:
   We have $  \ell  \mathbin{:}  \ottnt{A'}   ;  \rho_{{\mathrm{1}}}   \simeq  \ottnt{B}$ and, by inversion,
   $\ottnt{B} \,  \triangleright _{ \ell }  \, \ottnt{B'}  \ottsym{,}  \rho_{{\mathrm{2}}}$ and $\ottnt{A'}  \simeq  \ottnt{B'}$ and $\rho_{{\mathrm{1}}}  \simeq  \rho_{{\mathrm{2}}}$.
   By the IHs,
   \begin{itemize}
    \item $\ottnt{A'}  \equiv  \ottnt{C'}$,
    \item $\ottnt{C'}  \sim  \ottnt{B'}$,
    \item $\rho_{{\mathrm{1}}}  \equiv  \rho$, and
    \item $\rho  \sim  \rho_{{\mathrm{2}}}$
   \end{itemize}
   for some $\ottnt{C'}$ and $\rho$.

   If $\ell \,  \in  \, \mathit{dom} \, \ottsym{(}  \ottnt{B}  \ottsym{)}$, then $\ottnt{B} \,  =  \, \rho_{{\mathrm{21}}}  \odot  \ottsym{(}    \ell  \mathbin{:}  \ottnt{B'}   ;   \cdot    \ottsym{)}  \odot  \rho_{{\mathrm{22}}}$
   for some $\rho_{{\mathrm{21}}}$ and $\rho_{{\mathrm{22}}}$ such that $\rho_{{\mathrm{2}}} \,  =  \, \rho_{{\mathrm{21}}}  \odot  \rho_{{\mathrm{22}}}$
   and $\ell \,  \not\in  \, \mathit{dom} \, \ottsym{(}  \rho_{{\mathrm{21}}}  \ottsym{)}$.
   Since $\rho  \sim  \rho_{{\mathrm{21}}}  \odot  \rho_{{\mathrm{22}}}$ and $\ottnt{C'}  \sim  \ottnt{B'}$, there exist some $\rho_{{\mathrm{11}}}$ and $\rho_{{\mathrm{12}}}$ such that
   \begin{itemize}
    \item $\rho  \equiv  \rho_{{\mathrm{11}}}  \odot  \rho_{{\mathrm{12}}}$,
    \item $\rho_{{\mathrm{11}}}  \odot  \ottsym{(}    \ell  \mathbin{:}  \ottnt{C'}   ;   \cdot    \ottsym{)}  \odot  \rho_{{\mathrm{12}}}  \sim  \rho_{{\mathrm{21}}}  \odot  \ottsym{(}    \ell  \mathbin{:}  \ottnt{B'}   ;   \cdot    \ottsym{)}  \odot  \rho_{{\mathrm{22}}}$, and
    \item $\ell \,  \not\in  \, \mathit{dom} \, \ottsym{(}  \rho_{{\mathrm{11}}}  \ottsym{)}$.
   \end{itemize}
   Here, we have
   \[\begin{array}{lll} &
      \ell  \mathbin{:}  \ottnt{A'}   ;  \rho_{{\mathrm{1}}}  \\  \equiv &
      \ell  \mathbin{:}  \ottnt{C'}   ;  \rho                       & \text{since $\ottnt{A'}  \equiv  \ottnt{C'}$ and $\rho_{{\mathrm{1}}}  \equiv  \rho$} \\  \equiv &
      \ell  \mathbin{:}  \ottnt{C'}   ;  \rho_{{\mathrm{11}}}  \odot  \rho_{{\mathrm{12}}}             & \text{since $\rho  \equiv  \rho_{{\mathrm{11}}}  \odot  \rho_{{\mathrm{12}}}$} \\  \equiv &
    \rho_{{\mathrm{11}}}  \odot  \ottsym{(}    \ell  \mathbin{:}  \ottnt{C'}   ;   \cdot    \ottsym{)}  \odot  \rho_{{\mathrm{12}}} & \text{since $\ell \,  \not\in  \, \mathit{dom} \, \ottsym{(}  \rho_{{\mathrm{11}}}  \ottsym{)}$} \\  \sim &
    \rho_{{\mathrm{21}}}  \odot  \ottsym{(}    \ell  \mathbin{:}  \ottnt{B'}   ;   \cdot    \ottsym{)}  \odot  \rho_{{\mathrm{22}}} \\ =&
    \ottnt{B}.
     \end{array}\]

   Otherwise, if $\ell \,  \not\in  \, \mathit{dom} \, \ottsym{(}  \ottnt{B}  \ottsym{)}$,
   it is found from $\ottnt{B} \,  \triangleright _{ \ell }  \, \ottnt{B'}  \ottsym{,}  \rho_{{\mathrm{2}}}$ that $\ottnt{B} \,  =  \, \rho_{{\mathrm{2}}}$ and $\ottnt{B}$ ends with $ \star $.
   Since $\rho  \sim  \rho_{{\mathrm{2}}}$, we have $\rho  \sim  \ottnt{B}$.
   By \Cns{ConsL}, $  \ell  \mathbin{:}  \ottnt{A'}   ;  \rho   \sim  \ottnt{B}$.
   Here, we have
   \[
      \ell  \mathbin{:}  \ottnt{A'}   ;  \rho_{{\mathrm{1}}}   \equiv    \ell  \mathbin{:}  \ottnt{A'}   ;  \rho   \sim  \ottnt{B}.
   \]

  \case \CE{ConsR}:
   We have $\ottnt{A}  \simeq    \ell  \mathbin{:}  \ottnt{B'}   ;  \rho_{{\mathrm{2}}} $ and, by inversion,
   $\ottnt{A} \,  \triangleright _{ \ell }  \, \ottnt{A'}  \ottsym{,}  \rho_{{\mathrm{1}}}$ and $\ottnt{A'}  \simeq  \ottnt{B'}$ and $\rho_{{\mathrm{1}}}  \simeq  \rho_{{\mathrm{2}}}$.
   By the IHs,
   \begin{itemize}
    \item $\ottnt{A'}  \equiv  \ottnt{C'}$,
    \item $\ottnt{C'}  \sim  \ottnt{B'}$,
    \item $\rho_{{\mathrm{1}}}  \equiv  \rho$, and
    \item $\rho  \sim  \rho_{{\mathrm{2}}}$
   \end{itemize}
   for some $\ottnt{C'}$ and $\rho_{{\mathrm{2}}}$.

   If $\ell \,  \in  \, \mathit{dom} \, \ottsym{(}  \ottnt{A}  \ottsym{)}$, then $\ottnt{A} \,  =  \, \rho_{{\mathrm{11}}}  \odot  \ottsym{(}    \ell  \mathbin{:}  \ottnt{A'}   ;   \cdot    \ottsym{)}  \odot  \rho_{{\mathrm{12}}}$
   for some $\rho_{{\mathrm{11}}}$ and $\rho_{{\mathrm{12}}}$ such that
   $\rho_{{\mathrm{1}}} \,  =  \, \rho_{{\mathrm{11}}}  \odot  \rho_{{\mathrm{12}}}$ and $\ell \,  \not\in  \, \mathit{dom} \, \ottsym{(}  \rho_{{\mathrm{11}}}  \ottsym{)}$.
   Here, we have
   \[\begin{array}{lll} &
    \ottnt{A} \\ =&
    \rho_{{\mathrm{11}}}  \odot    \ell  \mathbin{:}  \ottnt{A'}   ;   \cdot    \odot  \rho_{{\mathrm{12}}} \\  \equiv &
    \rho_{{\mathrm{11}}}  \odot    \ell  \mathbin{:}  \ottnt{C'}   ;   \cdot    \odot  \rho_{{\mathrm{12}}} & \text{since $\ottnt{A'}  \equiv  \ottnt{C'}$} \\  \equiv &
      \ell  \mathbin{:}  \ottnt{C'}   ;  \rho_{{\mathrm{11}}}  \odot  \rho_{{\mathrm{12}}}           & \text{since $\ell \,  \not\in  \, \mathit{dom} \, \ottsym{(}  \rho_{{\mathrm{11}}}  \ottsym{)}$} \\ =&
      \ell  \mathbin{:}  \ottnt{C'}   ;  \rho_{{\mathrm{1}}}  \\  \equiv &
      \ell  \mathbin{:}  \ottnt{C'}   ;  \rho                     & \text{since $\rho_{{\mathrm{1}}}  \equiv  \rho$} \\  \sim &
      \ell  \mathbin{:}  \ottnt{B'}   ;  \rho_{{\mathrm{2}}}                    & \text{by \Cns{Cons} since $\ottnt{C'}  \sim  \ottnt{B'}$ and $\rho  \sim  \rho_{{\mathrm{2}}}$} \\ =&
    \ottnt{B}.
     \end{array}\]
 \end{caseanalysis}
\end{proof}

\begin{lemma}{consistent-inv-row-label}
 If $\rho_{{\mathrm{11}}}  \odot  \ottsym{(}    \ell  \mathbin{:}  \ottnt{A}   ;   \cdot    \ottsym{)}  \odot  \rho_{{\mathrm{12}}}  \simeq  \rho_{{\mathrm{21}}}  \odot  \ottsym{(}    \ell  \mathbin{:}  \ottnt{B}   ;   \cdot    \ottsym{)}  \odot  \rho_{{\mathrm{22}}}$
 and $\ell \,  \not\in  \, \mathit{dom} \, \ottsym{(}  \rho_{{\mathrm{11}}}  \ottsym{)} \,  \mathbin{\cup}  \, \mathit{dom} \, \ottsym{(}  \rho_{{\mathrm{21}}}  \ottsym{)}$,
 then $\ottnt{A}  \simeq  \ottnt{B}$ and $\rho_{{\mathrm{11}}}  \odot  \rho_{{\mathrm{12}}}  \simeq  \rho_{{\mathrm{21}}}  \odot  \rho_{{\mathrm{22}}}$.
\end{lemma}
\begin{proof}
 By induction on the derivation of
 $\rho_{{\mathrm{11}}}  \odot  \ottsym{(}    \ell  \mathbin{:}  \ottnt{A}   ;   \cdot    \ottsym{)}  \odot  \rho_{{\mathrm{12}}}  \simeq  \rho_{{\mathrm{21}}}  \odot  \ottsym{(}    \ell  \mathbin{:}  \ottnt{B}   ;   \cdot    \ottsym{)}  \odot  \rho_{{\mathrm{22}}}$.
 \begin{caseanalysis}
  \case \CE{Refl}: Obvious by \CE{Refl}.
  \case \CE{ConsL}: By case analysis on $\rho_{{\mathrm{11}}}$.
   \begin{caseanalysis}
    \case $\rho_{{\mathrm{11}}} \,  =  \,  \cdot $:  We have $\ottnt{A}  \simeq  \ottnt{B}$ and $\rho_{{\mathrm{12}}}  \simeq  \rho_{{\mathrm{21}}}  \odot  \rho_{{\mathrm{22}}}$ by inversion, and therefore we finish.
    \case $\rho_{{\mathrm{11}}} \,  \not=  \,  \cdot $:  We have $\rho_{{\mathrm{11}}} \,  =  \,   \ell'  \mathbin{:}  \ottnt{A'}   ;  \rho'_{{\mathrm{11}}} $.
     Since $\ell \,  \not\in  \, \mathit{dom} \, \ottsym{(}  \rho_{{\mathrm{11}}}  \ottsym{)}$, it is found that $\ell \,  \not=  \, \ell'$.
     \begin{caseanalysis}
      \case $\ell' \,  \in  \, \mathit{dom} \, \ottsym{(}  \rho_{{\mathrm{21}}}  \ottsym{)}$:
       There exist some $\rho_{{\mathrm{211}}}$, $\rho_{{\mathrm{212}}}$, and $\ottnt{B'}$ such that
       \begin{itemize}
        \item $\rho_{{\mathrm{21}}} \,  =  \, \rho_{{\mathrm{211}}}  \odot  \ottsym{(}    \ell'  \mathbin{:}  \ottnt{B'}   ;   \cdot    \ottsym{)}  \odot  \rho_{{\mathrm{212}}}$,
        \item $\ell' \,  \not\in  \, \mathit{dom} \, \ottsym{(}  \rho_{{\mathrm{211}}}  \ottsym{)}$,
        \item $\ottnt{A'}  \simeq  \ottnt{B'}$, and
        \item $\rho'_{{\mathrm{11}}}  \odot  \ottsym{(}    \ell  \mathbin{:}  \ottnt{A}   ;   \cdot    \ottsym{)}  \odot  \rho_{{\mathrm{12}}}  \simeq  \rho_{{\mathrm{211}}}  \odot  \rho_{{\mathrm{212}}}  \odot  \ottsym{(}    \ell  \mathbin{:}  \ottnt{B}   ;   \cdot    \ottsym{)}  \odot  \rho_{{\mathrm{22}}}$
       \end{itemize}
       by inversion.
       By the IH, $\ottnt{A}  \simeq  \ottnt{B}$ and $\rho'_{{\mathrm{11}}}  \odot  \rho_{{\mathrm{12}}}  \simeq  \rho_{{\mathrm{211}}}  \odot  \rho_{{\mathrm{212}}}  \odot  \rho_{{\mathrm{22}}}$.
       By \CE{ConsL},
       \[
        \rho_{{\mathrm{11}}}  \odot  \rho_{{\mathrm{12}}} =   \ell'  \mathbin{:}  \ottnt{A'}   ;  \rho'_{{\mathrm{11}}}   \odot  \rho_{{\mathrm{12}}}  \simeq  \rho_{{\mathrm{211}}}  \odot  \ottsym{(}    \ell'  \mathbin{:}  \ottnt{B'}   ;   \cdot    \ottsym{)}  \odot  \rho_{{\mathrm{212}}}  \odot  \rho_{{\mathrm{22}}} = \rho_{{\mathrm{21}}}  \odot  \rho_{{\mathrm{22}}}.
       \]

      \case $\ell' \,  \not\in  \, \mathit{dom} \, \ottsym{(}  \rho_{{\mathrm{21}}}  \ottsym{)}$ and $\ell' \,  \in  \, \mathit{dom} \, \ottsym{(}  \rho_{{\mathrm{22}}}  \ottsym{)}$:
       There exist some $\rho_{{\mathrm{221}}}$, $\rho_{{\mathrm{222}}}$, and $\ottnt{B'}$ such that
       \begin{itemize}
        \item $\rho_{{\mathrm{22}}} \,  =  \, \rho_{{\mathrm{221}}}  \odot  \ottsym{(}    \ell'  \mathbin{:}  \ottnt{B'}   ;   \cdot    \ottsym{)}  \odot  \rho_{{\mathrm{222}}}$,
        \item $\ell' \,  \not\in  \, \mathit{dom} \, \ottsym{(}  \rho_{{\mathrm{221}}}  \ottsym{)}$,
        \item $\ottnt{A'}  \simeq  \ottnt{B'}$, and
        \item $\rho'_{{\mathrm{11}}}  \odot  \ottsym{(}    \ell  \mathbin{:}  \ottnt{A}   ;   \cdot    \ottsym{)}  \odot  \rho_{{\mathrm{12}}}  \simeq  \rho_{{\mathrm{21}}}  \odot  \ottsym{(}    \ell  \mathbin{:}  \ottnt{B}   ;   \cdot    \ottsym{)}  \odot  \rho_{{\mathrm{221}}}  \odot  \rho_{{\mathrm{222}}}$
       \end{itemize}
       by inversion.
       By the IH, $\ottnt{A}  \simeq  \ottnt{B}$ and $\rho'_{{\mathrm{11}}}  \odot  \rho_{{\mathrm{12}}}  \simeq  \rho_{{\mathrm{21}}}  \odot  \rho_{{\mathrm{221}}}  \odot  \rho_{{\mathrm{222}}}$.
       By \CE{ConsL},
       \[
        \rho_{{\mathrm{11}}}  \odot  \rho_{{\mathrm{12}}} =   \ell'  \mathbin{:}  \ottnt{A'}   ;  \rho'_{{\mathrm{11}}}   \odot  \rho_{{\mathrm{12}}}  \simeq  \rho_{{\mathrm{21}}}  \odot  \rho_{{\mathrm{221}}}  \odot  \ottsym{(}    \ell'  \mathbin{:}  \ottnt{B'}   ;   \cdot    \ottsym{)}  \odot  \rho_{{\mathrm{22}}} = \rho_{{\mathrm{21}}}  \odot  \rho_{{\mathrm{22}}}.
       \]

      \case $\ell' \,  \not\in  \, \mathit{dom} \, \ottsym{(}  \rho_{{\mathrm{21}}}  \odot  \rho_{{\mathrm{22}}}  \ottsym{)}$:
       It is found that
       \begin{itemize}
        \item $\rho_{{\mathrm{21}}}  \odot  \rho_{{\mathrm{22}}}$ ends with $ \star $ and
        \item $\rho'_{{\mathrm{11}}}  \odot  \ottsym{(}    \ell  \mathbin{:}  \ottnt{A}   ;   \cdot    \ottsym{)}  \odot  \rho_{{\mathrm{12}}}  \simeq  \rho_{{\mathrm{21}}}  \odot  \ottsym{(}    \ell  \mathbin{:}  \ottnt{B}   ;   \cdot    \ottsym{)}  \odot  \rho_{{\mathrm{22}}}$
       \end{itemize}
       by inversion.
       By the IH, $\ottnt{A}  \simeq  \ottnt{B}$ and $\rho'_{{\mathrm{11}}}  \odot  \rho_{{\mathrm{12}}}  \simeq  \rho_{{\mathrm{21}}}  \odot  \rho_{{\mathrm{22}}}$.
       By \CE{ConsL}, $  \ell'  \mathbin{:}  \ottnt{A'}   ;  \rho'_{{\mathrm{11}}}   \odot  \rho_{{\mathrm{12}}}  \simeq  \rho_{{\mathrm{21}}}  \odot  \rho_{{\mathrm{22}}}$.
     \end{caseanalysis}
   \end{caseanalysis}

  \case \CE{ConsR}:  Similar to the case for \CE{ConsL}.
  \case the others: Contradictory.
 \end{caseanalysis}
\end{proof}

\begin{lemma}{consistent-inj-row-label}
 If $\ottnt{A}  \simeq  \ottnt{B}$ and $\rho_{{\mathrm{11}}}  \odot  \rho_{{\mathrm{12}}}  \simeq  \rho_{{\mathrm{21}}}  \odot  \rho_{{\mathrm{22}}}$
 and $\ell \,  \not\in  \, \mathit{dom} \, \ottsym{(}  \rho_{{\mathrm{11}}}  \ottsym{)} \,  \mathbin{\cup}  \, \mathit{dom} \, \ottsym{(}  \rho_{{\mathrm{21}}}  \ottsym{)}$,
 then $\rho_{{\mathrm{11}}}  \odot  \ottsym{(}    \ell  \mathbin{:}  \ottnt{A}   ;   \cdot    \ottsym{)}  \odot  \rho_{{\mathrm{12}}}  \simeq  \rho_{{\mathrm{21}}}  \odot  \ottsym{(}    \ell  \mathbin{:}  \ottnt{B}   ;   \cdot    \ottsym{)}  \odot  \rho_{{\mathrm{22}}}$.
\end{lemma}
\begin{proof}
 By induction on the sum of the sizes of $\rho_{{\mathrm{11}}}  \odot  \rho_{{\mathrm{12}}}$ and $\rho_{{\mathrm{21}}}  \odot  \rho_{{\mathrm{22}}}$.
 Since $\rho_{{\mathrm{11}}}  \odot  \rho_{{\mathrm{12}}}$ is defined, there are only two cases on $\rho_{{\mathrm{11}}}$ to be considered.
 \begin{caseanalysis}
  \case $\rho_{{\mathrm{11}}} \,  =  \,  \cdot $: By \CE{ConsL}.
  \case $\rho_{{\mathrm{11}}} \,  =  \,   \ell'  \mathbin{:}  \ottnt{A'}   ;  \rho'_{{\mathrm{11}}} $:
   If $\rho_{{\mathrm{21}}} \,  =  \,  \cdot $, then $\rho_{{\mathrm{11}}}  \odot  \rho_{{\mathrm{21}}} \,  =  \, \rho_{{\mathrm{22}}}$.
   By \CE{ConsR},
   \[
    \rho_{{\mathrm{11}}}  \odot  \ottsym{(}    \ell  \mathbin{:}  \ottnt{A}   ;   \cdot    \ottsym{)}  \odot  \rho_{{\mathrm{12}}}  \simeq    \ell  \mathbin{:}  \ottnt{B}   ;  \rho_{{\mathrm{22}}}  = \rho_{{\mathrm{21}}}  \odot  \ottsym{(}    \ell  \mathbin{:}  \ottnt{B}   ;   \cdot    \ottsym{)}  \odot  \rho_{{\mathrm{22}}},
   \]
   and so we finish.

   In what follows, we suppose $\rho_{{\mathrm{21}}} \,  \not=  \,  \cdot $.
   By case analysis on the rule applied last to derive $  \ell'  \mathbin{:}  \ottnt{A'}   ;  \rho'_{{\mathrm{11}}}   \odot  \rho_{{\mathrm{12}}}  \simeq  \rho_{{\mathrm{21}}}  \odot  \rho_{{\mathrm{22}}}$.
   \begin{caseanalysis}
    \case \CE{Refl}:
     Since $\rho_{{\mathrm{21}}} \,  \not=  \,  \cdot $, we can suppose that $\rho_{{\mathrm{21}}} \,  =  \,   \ell'  \mathbin{:}  \ottnt{A'}   ;  \rho'_{{\mathrm{21}}} $.
     Thus, $\rho'_{{\mathrm{11}}}  \odot  \rho_{{\mathrm{21}}} \,  =  \, \rho'_{{\mathrm{21}}}  \odot  \rho_{{\mathrm{22}}}$, and
     $\rho'_{{\mathrm{11}}}  \odot  \rho_{{\mathrm{21}}}  \simeq  \rho'_{{\mathrm{21}}}  \odot  \rho_{{\mathrm{22}}}$ by \CE{Refl}.
     By the IH, $\rho'_{{\mathrm{11}}}  \odot  \ottsym{(}    \ell  \mathbin{:}  \ottnt{A}   ;   \cdot    \ottsym{)}  \odot  \rho_{{\mathrm{12}}}  \simeq  \rho'_{{\mathrm{21}}}  \odot  \ottsym{(}    \ell  \mathbin{:}  \ottnt{B}   ;   \cdot    \ottsym{)}  \odot  \rho_{{\mathrm{22}}}$.
     By \CE{Refl} and \CE{ConsL},
     \[
      \rho_{{\mathrm{11}}}  \odot  \ottsym{(}    \ell  \mathbin{:}  \ottnt{A}   ;   \cdot    \ottsym{)}  \odot  \rho_{{\mathrm{12}}} =
        \ell'  \mathbin{:}  \ottnt{A'}   ;  \rho'_{{\mathrm{11}}}   \odot  \ottsym{(}    \ell  \mathbin{:}  \ottnt{A}   ;   \cdot    \ottsym{)}  \odot  \rho_{{\mathrm{12}}}  \simeq    \ell  \mathbin{:}  \ottnt{A'}   ;  \rho'_{{\mathrm{21}}}   \odot  \ottsym{(}    \ell  \mathbin{:}  \ottnt{B}   ;   \cdot    \ottsym{)}  \odot  \rho_{{\mathrm{22}}} =
      \rho_{{\mathrm{21}}}  \odot  \ottsym{(}    \ell  \mathbin{:}  \ottnt{B}   ;   \cdot    \ottsym{)}  \odot  \rho_{{\mathrm{22}}}.
     \]

    \case \CE{DynR}: We have $\rho_{{\mathrm{21}}}  \odot  \rho_{{\mathrm{22}}} \,  =  \, \star$.  By \CE{ConsR}.
    \case \CE{ConsL}:
     By inversion, $\rho_{{\mathrm{21}}}  \odot  \rho_{{\mathrm{22}}} \,  \triangleright _{ \ell' }  \, \ottnt{B'}  \ottsym{,}  \rho_{{\mathrm{2}}}$ and
     $\ottnt{A'}  \simeq  \ottnt{B'}$ and $\rho'_{{\mathrm{11}}}  \odot  \rho_{{\mathrm{12}}}  \simeq  \rho_{{\mathrm{2}}}$
     for some $\ottnt{B'}$, and $\rho_{{\mathrm{2}}}$.

     \begin{caseanalysis}
      \case $\ell' \,  \in  \, \mathit{dom} \, \ottsym{(}  \rho_{{\mathrm{21}}}  \ottsym{)}$:
       There exist some $\rho_{{\mathrm{211}}}$ and $\rho_{{\mathrm{212}}}$ such that
       \begin{itemize}
        \item $\rho_{{\mathrm{21}}} \,  =  \, \rho_{{\mathrm{211}}}  \odot  \ottsym{(}    \ell'  \mathbin{:}  \ottnt{B'}   ;   \cdot    \ottsym{)}  \odot  \rho_{{\mathrm{212}}}$,
        \item $\rho_{{\mathrm{2}}} \,  =  \, \rho_{{\mathrm{211}}}  \odot  \rho_{{\mathrm{212}}}  \odot  \rho_{{\mathrm{22}}}$, and
        \item $\ell' \,  \not\in  \, \mathit{dom} \, \ottsym{(}  \rho_{{\mathrm{211}}}  \ottsym{)}$.
       \end{itemize}
       Since $\rho'_{{\mathrm{11}}}  \odot  \rho_{{\mathrm{12}}}  \simeq  \rho_{{\mathrm{2}}}$, we have $\rho'_{{\mathrm{11}}}  \odot  \rho_{{\mathrm{12}}}  \simeq  \rho_{{\mathrm{211}}}  \odot  \rho_{{\mathrm{212}}}  \odot  \rho_{{\mathrm{22}}}$.
       By the IH, $\rho'_{{\mathrm{11}}}  \odot  \ottsym{(}    \ell  \mathbin{:}  \ottnt{A}   ;   \cdot    \ottsym{)}  \odot  \rho_{{\mathrm{12}}}  \simeq  \rho_{{\mathrm{211}}}  \odot  \rho_{{\mathrm{212}}}  \odot  \ottsym{(}    \ell  \mathbin{:}  \ottnt{B}   ;   \cdot    \ottsym{)}  \odot  \rho_{{\mathrm{22}}}$.
       By \CE{ConsL},
       \[
         \ell'  \mathbin{:}  \ottnt{A'}   ;  \rho'_{{\mathrm{11}}}   \odot  \ottsym{(}    \ell  \mathbin{:}  \ottnt{A}   ;   \cdot    \ottsym{)}  \odot  \rho_{{\mathrm{12}}}  \simeq  \rho_{{\mathrm{211}}}  \odot  \ottsym{(}    \ell'  \mathbin{:}  \ottnt{B'}   ;   \cdot    \ottsym{)}  \odot  \rho_{{\mathrm{212}}}  \odot  \ottsym{(}    \ell  \mathbin{:}  \ottnt{B}   ;   \cdot    \ottsym{)}  \odot  \rho_{{\mathrm{22}}}.
       \]
       Thus,
       \[
        \rho_{{\mathrm{11}}}  \odot  \ottsym{(}    \ell  \mathbin{:}  \ottnt{A}   ;   \cdot    \ottsym{)}  \odot  \rho_{{\mathrm{12}}} =
          \ell'  \mathbin{:}  \ottnt{A'}   ;  \rho'_{{\mathrm{11}}}   \odot  \ottsym{(}    \ell  \mathbin{:}  \ottnt{A}   ;   \cdot    \ottsym{)}  \odot  \rho_{{\mathrm{12}}}  \simeq  \rho_{{\mathrm{211}}}  \odot  \ottsym{(}    \ell'  \mathbin{:}  \ottnt{B'}   ;   \cdot    \ottsym{)}  \odot  \rho_{{\mathrm{212}}}  \odot  \ottsym{(}    \ell  \mathbin{:}  \ottnt{B}   ;   \cdot    \ottsym{)}  \odot  \rho_{{\mathrm{22}}} =
        \rho_{{\mathrm{21}}}  \odot  \ottsym{(}    \ell  \mathbin{:}  \ottnt{B}   ;   \cdot    \ottsym{)}  \odot  \rho_{{\mathrm{22}}}.
       \]

      \case $\ell' \,  \not\in  \, \mathit{dom} \, \ottsym{(}  \rho_{{\mathrm{21}}}  \ottsym{)}$ and $\ell' \,  \in  \, \mathit{dom} \, \ottsym{(}  \rho_{{\mathrm{22}}}  \ottsym{)}$:
       There exist some $\rho_{{\mathrm{221}}}$ and $\rho_{{\mathrm{222}}}$ such that
       \begin{itemize}
        \item $\rho_{{\mathrm{22}}} \,  =  \, \rho_{{\mathrm{221}}}  \odot  \ottsym{(}    \ell'  \mathbin{:}  \ottnt{B'}   ;   \cdot    \ottsym{)}  \odot  \rho_{{\mathrm{222}}}$,
        \item $\rho_{{\mathrm{2}}} \,  =  \, \rho_{{\mathrm{21}}}  \odot  \rho_{{\mathrm{221}}}  \odot  \rho_{{\mathrm{222}}}$, and
        \item $\ell' \,  \not\in  \, \mathit{dom} \, \ottsym{(}  \rho_{{\mathrm{221}}}  \ottsym{)}$.
       \end{itemize}
       Since $\rho'_{{\mathrm{11}}}  \odot  \rho_{{\mathrm{12}}}  \simeq  \rho_{{\mathrm{2}}}$, we have $\rho'_{{\mathrm{11}}}  \odot  \rho_{{\mathrm{12}}}  \simeq  \rho_{{\mathrm{21}}}  \odot  \rho_{{\mathrm{221}}}  \odot  \rho_{{\mathrm{222}}}$.
       By the IH, $\rho'_{{\mathrm{11}}}  \odot  \ottsym{(}    \ell  \mathbin{:}  \ottnt{A}   ;   \cdot    \ottsym{)}  \odot  \rho_{{\mathrm{12}}}  \simeq  \rho_{{\mathrm{21}}}  \odot  \ottsym{(}    \ell  \mathbin{:}  \ottnt{B}   ;   \cdot    \ottsym{)}  \odot  \rho_{{\mathrm{221}}}  \odot  \rho_{{\mathrm{222}}}$.
       By \CE{ConsL},
       \[
         \ell'  \mathbin{:}  \ottnt{A'}   ;  \rho'_{{\mathrm{11}}}   \odot  \ottsym{(}    \ell  \mathbin{:}  \ottnt{A}   ;   \cdot    \ottsym{)}  \odot  \rho_{{\mathrm{12}}}  \simeq  \rho_{{\mathrm{21}}}  \odot  \ottsym{(}    \ell  \mathbin{:}  \ottnt{B}   ;   \cdot    \ottsym{)}  \odot  \rho_{{\mathrm{221}}}  \odot  \ottsym{(}    \ell'  \mathbin{:}  \ottnt{B'}   ;   \cdot    \ottsym{)}  \odot  \rho_{{\mathrm{222}}}.
       \]
       Thus,
       \[
        \rho_{{\mathrm{11}}}  \odot  \ottsym{(}    \ell  \mathbin{:}  \ottnt{A}   ;   \cdot    \ottsym{)}  \odot  \rho_{{\mathrm{12}}} =
          \ell'  \mathbin{:}  \ottnt{A'}   ;  \rho'_{{\mathrm{11}}}   \odot  \ottsym{(}    \ell  \mathbin{:}  \ottnt{A}   ;   \cdot    \ottsym{)}  \odot  \rho_{{\mathrm{12}}}  \simeq  \rho_{{\mathrm{21}}}  \odot  \ottsym{(}    \ell  \mathbin{:}  \ottnt{B}   ;   \cdot    \ottsym{)}  \odot  \rho_{{\mathrm{221}}}  \odot  \ottsym{(}    \ell'  \mathbin{:}  \ottnt{B'}   ;   \cdot    \ottsym{)}  \odot  \rho_{{\mathrm{222}}} =
        \rho_{{\mathrm{21}}}  \odot  \ottsym{(}    \ell  \mathbin{:}  \ottnt{B}   ;   \cdot    \ottsym{)}  \odot  \rho_{{\mathrm{22}}}.
      \]

      \case $\ell' \,  \not\in  \, \mathit{dom} \, \ottsym{(}  \rho_{{\mathrm{21}}}  \ottsym{)}$ and $\ell' \,  \not\in  \, \mathit{dom} \, \ottsym{(}  \rho_{{\mathrm{22}}}  \ottsym{)}$:
       It is found that
       \begin{itemize}
        \item $\rho_{{\mathrm{21}}}  \odot  \rho_{{\mathrm{22}}}$ ends with $ \star $ and
        \item $\rho_{{\mathrm{2}}} \,  =  \, \rho_{{\mathrm{21}}}  \odot  \rho_{{\mathrm{22}}}$.
       \end{itemize}
       Since $\rho'_{{\mathrm{11}}}  \odot  \rho_{{\mathrm{12}}}  \simeq  \rho_{{\mathrm{2}}}$, we have $\rho'_{{\mathrm{11}}}  \odot  \rho_{{\mathrm{12}}}  \simeq  \rho_{{\mathrm{21}}}  \odot  \rho_{{\mathrm{22}}}$.
       By the IH, $\rho'_{{\mathrm{11}}}  \odot  \ottsym{(}    \ell  \mathbin{:}  \ottnt{A}   ;   \cdot    \ottsym{)}  \odot  \rho_{{\mathrm{12}}}  \simeq  \rho_{{\mathrm{21}}}  \odot  \ottsym{(}    \ell  \mathbin{:}  \ottnt{B}   ;   \cdot    \ottsym{)}  \odot  \rho_{{\mathrm{22}}}$.
       By \CE{ConsL}, we finish.
     \end{caseanalysis}

    \case \CE{ConsR}: Similar to the case for \CE{ConsL}.
    \case \CE{DynL}, \CE{Fun}, \CE{Poly}, \CE{PolyL}, \CE{PolyR}, \CE{Record}, and \CE{Variant}: Contradictory.
     Note that the contradiction in the case of \Cns{PolyR} is proven by the definition of $ \mathbf{QPoly} $.
   \end{caseanalysis}
 \end{caseanalysis}
\end{proof}

\begin{lemma}{consistent-inj-row-label-dyn}
 If $\rho_{{\mathrm{1}}}  \simeq  \rho_{{\mathrm{21}}}  \odot  \rho_{{\mathrm{22}}}$ and $\rho_{{\mathrm{1}}}$ ends with $ \star $ and $\ell \,  \not\in  \, \mathit{dom} \, \ottsym{(}  \rho_{{\mathrm{1}}}  \ottsym{)}$,
 then $\rho_{{\mathrm{1}}}  \simeq  \rho_{{\mathrm{21}}}  \odot  \ottsym{(}    \ell  \mathbin{:}  \ottnt{A}   ;   \cdot    \ottsym{)}  \odot  \rho_{{\mathrm{22}}}$ for any $\ottnt{A}$.
\end{lemma}
\begin{proof}
 By induction on the sizes of $\rho_{{\mathrm{1}}}$ and $\rho_{{\mathrm{21}}}$.
 If $\rho_{{\mathrm{21}}} \,  =  \,  \cdot $, then we finish by \CE{ConsR}.

 In what follows, since $\rho_{{\mathrm{21}}}  \odot  \rho_{{\mathrm{22}}}$ is defined, we can suppose that
 $\rho_{{\mathrm{21}}} \,  =  \,   \ell'  \mathbin{:}  \ottnt{B}   ;  \rho'_{{\mathrm{21}}} $ for some $\ell'$, $\ottnt{B}$, and $\rho'_{{\mathrm{21}}}$.
 By case analysis on the rule applied last to derive
 $\rho_{{\mathrm{1}}}  \simeq  \rho_{{\mathrm{21}}}  \odot  \rho_{{\mathrm{22}}}$.
 \begin{caseanalysis}
  \case \CE{Refl}:
   We have $\rho_{{\mathrm{1}}} \,  =  \,   \ell'  \mathbin{:}  \ottnt{B}   ;  \rho'_{{\mathrm{1}}} $ for some $\rho'_{{\mathrm{1}}}$ such that
   $\rho'_{{\mathrm{1}}} \,  =  \, \rho'_{{\mathrm{21}}}  \odot  \rho_{{\mathrm{22}}}$.
   Since $\rho'_{{\mathrm{1}}}  \simeq  \rho'_{{\mathrm{21}}}  \odot  \rho_{{\mathrm{22}}}$ by \CE{Refl}, we have $\rho'  \simeq  \rho'_{{\mathrm{21}}}  \odot  \ottsym{(}    \ell  \mathbin{:}  \ottnt{A}   ;   \cdot    \ottsym{)}  \odot  \rho_{{\mathrm{22}}}$
   by the IH.  By \CE{ConsL}, we finish.

  \case \CE{DynL}: By \CE{DynL}.

  \case \CE{ConsL}:
   We have $\rho_{{\mathrm{1}}} \,  =  \,   \ell''  \mathbin{:}  \ottnt{C}   ;  \rho'_{{\mathrm{1}}} $ and, by inversion,
   $\rho_{{\mathrm{21}}}  \odot  \rho_{{\mathrm{22}}} \,  \triangleright _{ \ell'' }  \, \ottnt{B'}  \ottsym{,}  \rho'_{{\mathrm{2}}}$ and $\ottnt{C}  \simeq  \ottnt{B'}$ and $\rho'_{{\mathrm{1}}}  \simeq  \rho'_{{\mathrm{2}}}$
   for some $\ell''$, $\ottnt{B'}$, $\ottnt{C}$, $\rho'_{{\mathrm{1}}}$, and $\rho'_{{\mathrm{2}}}$.
   Note that $\ell \,  \not=  \, \ell''$ since $\ell \,  \not\in  \, \mathit{dom} \, \ottsym{(}  \rho_{{\mathrm{1}}}  \ottsym{)}$.

   \begin{caseanalysis}
    \case $\ell'' \,  \in  \, \mathit{dom} \, \ottsym{(}  \rho_{{\mathrm{21}}}  \ottsym{)}$:
     There exist some $\rho_{{\mathrm{211}}}$ and $\rho_{{\mathrm{212}}}$ such that
     \begin{itemize}
      \item $\rho_{{\mathrm{21}}} \,  =  \, \rho_{{\mathrm{211}}}  \odot  \ottsym{(}    \ell''  \mathbin{:}  \ottnt{B'}   ;   \cdot    \ottsym{)}  \odot  \rho_{{\mathrm{212}}}$,
      \item $\rho'_{{\mathrm{2}}} \,  =  \, \rho_{{\mathrm{211}}}  \odot  \rho_{{\mathrm{212}}}  \odot  \rho_{{\mathrm{22}}}$, and
      \item $\ell'' \,  \not\in  \, \mathit{dom} \, \ottsym{(}  \rho_{{\mathrm{211}}}  \ottsym{)}$.
     \end{itemize}
     Since $\rho'_{{\mathrm{1}}}  \simeq  \rho'_{{\mathrm{2}}}$, we have $\rho'_{{\mathrm{1}}}  \simeq  \rho_{{\mathrm{211}}}  \odot  \rho_{{\mathrm{212}}}  \odot  \rho_{{\mathrm{22}}}$.
     By the IH, $\rho'_{{\mathrm{1}}}  \simeq  \rho_{{\mathrm{211}}}  \odot  \rho_{{\mathrm{212}}}  \odot  \ottsym{(}    \ell  \mathbin{:}  \ottnt{A}   ;   \cdot    \ottsym{)}  \odot  \rho_{{\mathrm{22}}}$.
     Since $\ottsym{(}  \rho_{{\mathrm{21}}}  \odot  \ottsym{(}    \ell  \mathbin{:}  \ottnt{A}   ;   \cdot    \ottsym{)}  \odot  \rho_{{\mathrm{22}}}  \ottsym{)} \,  \triangleright _{ \ell'' }  \, \ottnt{B'}  \ottsym{,}  \rho_{{\mathrm{211}}}  \odot  \rho_{{\mathrm{212}}}  \odot  \ottsym{(}    \ell  \mathbin{:}  \ottnt{A}   ;   \cdot    \ottsym{)}  \odot  \rho_{{\mathrm{22}}}$,
     we finish by \CE{ConsL}.

    \case $\ell'' \,  \not\in  \, \mathit{dom} \, \ottsym{(}  \rho_{{\mathrm{21}}}  \ottsym{)}$ and $\ell'' \,  \in  \, \mathit{dom} \, \ottsym{(}  \rho_{{\mathrm{22}}}  \ottsym{)}$:
     There exist some $\rho_{{\mathrm{221}}}$ and $\rho_{{\mathrm{222}}}$ such that
     \begin{itemize}
      \item $\rho_{{\mathrm{22}}} \,  =  \, \rho_{{\mathrm{221}}}  \odot  \ottsym{(}    \ell''  \mathbin{:}  \ottnt{B'}   ;   \cdot    \ottsym{)}  \odot  \rho_{{\mathrm{222}}}$,
      \item $\rho'_{{\mathrm{2}}} \,  =  \, \rho_{{\mathrm{21}}}  \odot  \rho_{{\mathrm{221}}}  \odot  \rho_{{\mathrm{222}}}$, and
      \item $\ell'' \,  \not\in  \, \mathit{dom} \, \ottsym{(}  \rho_{{\mathrm{221}}}  \ottsym{)}$.
     \end{itemize}
     Since $\rho'_{{\mathrm{1}}}  \simeq  \rho'_{{\mathrm{2}}}$, we have $\rho'_{{\mathrm{1}}}  \simeq  \rho_{{\mathrm{21}}}  \odot  \rho_{{\mathrm{221}}}  \odot  \rho_{{\mathrm{222}}}$.
     By the IH, $\rho'_{{\mathrm{1}}}  \simeq  \rho_{{\mathrm{21}}}  \odot  \ottsym{(}    \ell  \mathbin{:}  \ottnt{A}   ;   \cdot    \ottsym{)}  \odot  \rho_{{\mathrm{221}}}  \odot  \rho_{{\mathrm{222}}}$.
     Since $\ottsym{(}  \rho_{{\mathrm{21}}}  \odot  \ottsym{(}    \ell  \mathbin{:}  \ottnt{A}   ;   \cdot    \ottsym{)}  \odot  \rho_{{\mathrm{22}}}  \ottsym{)} \,  \triangleright _{ \ell'' }  \, \ottnt{B'}  \ottsym{,}  \rho_{{\mathrm{21}}}  \odot  \ottsym{(}    \ell  \mathbin{:}  \ottnt{A}   ;   \cdot    \ottsym{)}  \odot  \rho_{{\mathrm{221}}}  \odot  \rho_{{\mathrm{222}}}$,
     we finish by \CE{ConsL}.

    \case $\ell'' \,  \not\in  \, \mathit{dom} \, \ottsym{(}  \rho_{{\mathrm{21}}}  \ottsym{)}$ and $\ell'' \,  \not\in  \, \mathit{dom} \, \ottsym{(}  \rho_{{\mathrm{22}}}  \ottsym{)}$:
     We have $\ottnt{B'} \,  =  \, \star$ and $\rho'_{{\mathrm{2}}} \,  =  \, \rho_{{\mathrm{21}}}  \odot  \rho_{{\mathrm{22}}}$ and $\rho_{{\mathrm{21}}}  \odot  \rho_{{\mathrm{22}}}$ ends with $ \star $.
     Since $\rho'_{{\mathrm{1}}}  \simeq  \rho'_{{\mathrm{2}}}$, we have $\rho'_{{\mathrm{1}}}  \simeq  \rho_{{\mathrm{21}}}  \odot  \rho_{{\mathrm{22}}}$.
     By the IH, $\rho'_{{\mathrm{1}}}  \simeq  \rho_{{\mathrm{21}}}  \odot  \ottsym{(}    \ell  \mathbin{:}  \ottnt{A}   ;   \cdot    \ottsym{)}  \odot  \rho_{{\mathrm{22}}}$.
     By \CE{ConsL}, we finish.
   \end{caseanalysis}

  \case \CE{ConsR}:
   Since $\rho_{{\mathrm{21}}} \,  =  \,   \ell'  \mathbin{:}  \ottnt{B}   ;  \rho'_{{\mathrm{21}}} $, by inversion
   $\rho_{{\mathrm{1}}} \,  \triangleright _{ \ell' }  \, \ottnt{C}  \ottsym{,}  \rho'_{{\mathrm{1}}}$ and $\ottnt{C}  \simeq  \ottnt{B}$ and $\rho'_{{\mathrm{1}}}  \simeq  \rho'_{{\mathrm{21}}}  \odot  \rho_{{\mathrm{22}}}$
   for some $\ottnt{C}$ and $\rho'_{{\mathrm{1}}}$.
   By the IH, $\rho'_{{\mathrm{1}}}  \simeq  \rho'_{{\mathrm{21}}}  \odot  \ottsym{(}    \ell  \mathbin{:}  \ottnt{A}   ;   \cdot    \ottsym{)}  \odot  \rho_{{\mathrm{22}}}$.
   By \CE{ConsR}, we finish.

  \case the others: Contradictory.
     Note that the contradiction in the case of \Cns{PolyL} is proven by the definition of $ \mathbf{QPoly} $.
 \end{caseanalysis}
\end{proof}

\begin{lemma}{consistent-subsume-equiv-trans}
 If $\ottnt{A}  \equiv  \ottnt{C}$ and $\ottnt{C}  \equiv  \ottnt{B}$ and $\ottnt{A}  \simeq  \ottnt{C}$ and $\ottnt{C}  \simeq  \ottnt{B}$,
 then $\ottnt{A}  \simeq  \ottnt{B}$.
\end{lemma}
\begin{proof}
 By induction on $\ottnt{A}  \simeq  \ottnt{C}$.
 \begin{caseanalysis}
  \case \CE{Refl}: Obvious.
  \case \CE{DynL}: By \CE{DynL}.

  \case \CE{DynR}:
   We have $\ottnt{C} \,  =  \, \star$.
   By \reflem{equiv-inversion} (\ref{lem:equiv-inversion:dyn}), $\ottnt{A} \,  =  \, \star$.
   Thus, we finish by \CE{DynL}.

  \case \CE{Fun}:
   We have $\ottnt{A} \,  =  \, \ottnt{A_{{\mathrm{1}}}}  \rightarrow  \ottnt{A_{{\mathrm{2}}}}$ and $\ottnt{C} \,  =  \, \ottnt{C_{{\mathrm{1}}}}  \rightarrow  \ottnt{C_{{\mathrm{2}}}}$ and, by inversion,
   $\ottnt{A_{{\mathrm{1}}}}  \simeq  \ottnt{C_{{\mathrm{1}}}}$ and $\ottnt{A_{{\mathrm{2}}}}  \simeq  \ottnt{C_{{\mathrm{2}}}}$
   for some $\ottnt{A_{{\mathrm{1}}}}$, $\ottnt{A_{{\mathrm{2}}}}$, $\ottnt{C_{{\mathrm{1}}}}$, and $\ottnt{C_{{\mathrm{2}}}}$.
   Since $\ottnt{A}  \equiv  \ottnt{C}$, we have $\ottnt{A_{{\mathrm{1}}}}  \equiv  \ottnt{C_{{\mathrm{1}}}}$ and $\ottnt{A_{{\mathrm{2}}}}  \equiv  \ottnt{C_{{\mathrm{2}}}}$
   by \reflem{equiv-inversion} (\ref{lem:equiv-inversion:fun}).
   Again, by \reflem{equiv-inversion} (\ref{lem:equiv-inversion:fun}),
   since $\ottnt{C}  \equiv  \ottnt{B}$, there exist some $\ottnt{B_{{\mathrm{1}}}}$ and $\ottnt{B_{{\mathrm{2}}}}$ such that
   $\ottnt{B} \,  =  \, \ottnt{B_{{\mathrm{1}}}}  \rightarrow  \ottnt{B_{{\mathrm{2}}}}$ and $\ottnt{C_{{\mathrm{1}}}}  \equiv  \ottnt{B_{{\mathrm{1}}}}$ and $\ottnt{C_{{\mathrm{2}}}}  \equiv  \ottnt{B_{{\mathrm{2}}}}$.
   Since $\ottnt{C}  \simeq  \ottnt{B}$, we have $\ottnt{C_{{\mathrm{1}}}}  \simeq  \ottnt{B_{{\mathrm{1}}}}$ and $\ottnt{C_{{\mathrm{2}}}}  \simeq  \ottnt{B_{{\mathrm{2}}}}$
   by \reflem{consistent-inv-fun}.
   Thus, by the IHs, $\ottnt{A_{{\mathrm{1}}}}  \simeq  \ottnt{B_{{\mathrm{1}}}}$ and $\ottnt{A_{{\mathrm{2}}}}  \simeq  \ottnt{B_{{\mathrm{2}}}}$.
   By \CE{Fun}, $\ottnt{A_{{\mathrm{1}}}}  \rightarrow  \ottnt{A_{{\mathrm{2}}}}  \simeq  \ottnt{B_{{\mathrm{1}}}}  \rightarrow  \ottnt{B_{{\mathrm{2}}}}$.

  \case \CE{Poly}:
   We have $\ottnt{A} \,  =  \,  \text{\unboldmath$\forall\!$}  \,  \mathit{X}  \mathord{:}  \ottnt{K}   \ottsym{.} \, \ottnt{A'}$ and $\ottnt{C} \,  =  \,  \text{\unboldmath$\forall\!$}  \,  \mathit{X}  \mathord{:}  \ottnt{K}   \ottsym{.} \, \ottnt{C'}$ and, by inversion,
   $\ottnt{A'}  \simeq  \ottnt{C'}$ for some $\mathit{X}$, $\ottnt{K}$, $\ottnt{A'}$, and $\ottnt{C'}$.
   Since $\ottnt{A}  \equiv  \ottnt{C}$, we have $\ottnt{A'}  \equiv  \ottnt{C'}$
   by \reflem{equiv-inversion} (\ref{lem:equiv-inversion:forall}).
   Again, by \reflem{equiv-inversion} (\ref{lem:equiv-inversion:forall}),
   since $\ottnt{C}  \equiv  \ottnt{B}$, there exist some $\ottnt{B'}$ such that
   $\ottnt{B} \,  =  \,  \text{\unboldmath$\forall\!$}  \,  \mathit{X}  \mathord{:}  \ottnt{K}   \ottsym{.} \, \ottnt{B'}$ and $\ottnt{C'}  \equiv  \ottnt{B'}$.
   Since $\ottnt{C}  \simeq  \ottnt{B}$, we have $\ottnt{C'}  \simeq  \ottnt{B'}$
   by \reflem{consistent-inv-forall}.
   Thus, by the IH, $\ottnt{A'}  \simeq  \ottnt{B'}$.
   By \CE{Poly}, $ \text{\unboldmath$\forall\!$}  \,  \mathit{X}  \mathord{:}  \ottnt{K}   \ottsym{.} \, \ottnt{A'}  \simeq   \text{\unboldmath$\forall\!$}  \,  \mathit{X}  \mathord{:}  \ottnt{K}   \ottsym{.} \, \ottnt{B'}$.

  \case \CE{PolyL}:
   We have $\ottnt{A} \,  =  \,  \text{\unboldmath$\forall\!$}  \,  \mathit{X}  \mathord{:}  \ottnt{K}   \ottsym{.} \, \ottnt{A'}$ and, by inversion,
   $\mathbf{QPoly} \, \ottsym{(}  \ottnt{C}  \ottsym{)}$ and $\mathit{X} \,  \not\in  \,  \mathit{ftv}  (  \ottnt{C}  ) $, for some $\mathit{X}$, $\ottnt{K}$, and $\ottnt{A'}$.
   $\mathbf{QPoly} \, \ottsym{(}  \ottnt{C}  \ottsym{)}$ is contradictory with the fact that $\ottnt{C} \,  =  \,  \text{\unboldmath$\forall\!$}  \,  \mathit{X}  \mathord{:}  \ottnt{K}   \ottsym{.} \, \ottnt{C'}$ for some $\ottnt{C'}$,
   which is implied by \reflem{equiv-inversion} (\ref{lem:equiv-inversion:forall}) with $\ottnt{A}  \equiv  \ottnt{C}$ and $\ottnt{A} \,  =  \,  \text{\unboldmath$\forall\!$}  \,  \mathit{X}  \mathord{:}  \ottnt{K}   \ottsym{.} \, \ottnt{A'}$.

  \case \CE{PolyR}:
   We have $\ottnt{C} \,  =  \,  \text{\unboldmath$\forall\!$}  \,  \mathit{X}  \mathord{:}  \ottnt{K}   \ottsym{.} \, \ottnt{C'}$ and, by inversion,
   $\mathbf{QPoly} \, \ottsym{(}  \ottnt{A}  \ottsym{)}$ and $\mathit{X} \,  \not\in  \,  \mathit{ftv}  (  \ottnt{A}  ) $, for some $\mathit{X}$, $\ottnt{K}$, and $\ottnt{C'}$.
   $\mathbf{QPoly} \, \ottsym{(}  \ottnt{A}  \ottsym{)}$ is contradictory with the fact that $\ottnt{A} \,  =  \,  \text{\unboldmath$\forall\!$}  \,  \mathit{X}  \mathord{:}  \ottnt{K}   \ottsym{.} \, \ottnt{A'}$ for some $\ottnt{A'}$,
   which is implied by \reflem{equiv-inversion} (\ref{lem:equiv-inversion:forall}) with $\ottnt{A}  \equiv  \ottnt{C}$ and $\ottnt{C} \,  =  \,  \text{\unboldmath$\forall\!$}  \,  \mathit{X}  \mathord{:}  \ottnt{K}   \ottsym{.} \, \ottnt{C'}$.

  \case \CE{Record}:
   We have $\ottnt{A} \,  =  \,  [  \rho_{{\mathrm{1}}}  ] $ and $\ottnt{C} \,  =  \,  [  \rho_{{\mathrm{3}}}  ] $ and, by inversion,
   $\rho_{{\mathrm{1}}}  \simeq  \rho_{{\mathrm{3}}}$ for some $\rho_{{\mathrm{1}}}$ and $\rho_{{\mathrm{3}}}$.
   Since $\ottnt{A}  \equiv  \ottnt{C}$, we have $\rho_{{\mathrm{1}}}  \equiv  \rho_{{\mathrm{3}}}$
   by \reflem{equiv-inversion} (\ref{lem:equiv-inversion:record}).
   Again, by \reflem{equiv-inversion} (\ref{lem:equiv-inversion:record}),
   since $\ottnt{C}  \equiv  \ottnt{B}$, there exists some $\rho_{{\mathrm{2}}}$ such that
   $\ottnt{B} \,  =  \,  [  \rho_{{\mathrm{2}}}  ] $ and $\rho_{{\mathrm{3}}}  \equiv  \rho_{{\mathrm{2}}}$.
   Since $\ottnt{C}  \simeq  \ottnt{B}$, we have $\rho_{{\mathrm{3}}}  \simeq  \rho_{{\mathrm{2}}}$
   by \reflem{consistent-inv-record}.
   By the IH, $\rho_{{\mathrm{1}}}  \simeq  \rho_{{\mathrm{2}}}$.
   By \CE{Record}, $ [  \rho_{{\mathrm{1}}}  ]   \simeq   [  \rho_{{\mathrm{2}}}  ] $.

  \case \CE{Variant}:
   We have $\ottnt{A} \,  =  \,  \langle  \rho_{{\mathrm{1}}}  \rangle $ and $\ottnt{C} \,  =  \,  \langle  \rho_{{\mathrm{3}}}  \rangle $ and, by inversion,
   $\rho_{{\mathrm{1}}}  \simeq  \rho_{{\mathrm{3}}}$ for some $\rho_{{\mathrm{1}}}$ and $\rho_{{\mathrm{3}}}$.
   Since $\ottnt{A}  \equiv  \ottnt{C}$, we have $\rho_{{\mathrm{1}}}  \equiv  \rho_{{\mathrm{3}}}$
   by \reflem{equiv-inversion} (\ref{lem:equiv-inversion:variant}).
   Again, by \reflem{equiv-inversion} (\ref{lem:equiv-inversion:variant}),
   since $\ottnt{C}  \equiv  \ottnt{B}$, there exists some $\rho_{{\mathrm{2}}}$ such that
   $\ottnt{B} \,  =  \,  \langle  \rho_{{\mathrm{2}}}  \rangle $ and $\rho_{{\mathrm{3}}}  \equiv  \rho_{{\mathrm{2}}}$.
   Since $\ottnt{C}  \simeq  \ottnt{B}$, we have $\rho_{{\mathrm{3}}}  \simeq  \rho_{{\mathrm{2}}}$
   by \reflem{consistent-inv-variant}.
   By the IH, $\rho_{{\mathrm{1}}}  \simeq  \rho_{{\mathrm{2}}}$.
   By \CE{Variant}, $ \langle  \rho_{{\mathrm{1}}}  \rangle   \simeq   \langle  \rho_{{\mathrm{2}}}  \rangle $.

  \case \CE{ConsL}:
   We have $\ottnt{A} \,  =  \,   \ell  \mathbin{:}  \ottnt{A'}   ;  \rho_{{\mathrm{1}}} $ and, by inversion,
   $\ottnt{C} \,  \triangleright _{ \ell }  \, \ottnt{C'}  \ottsym{,}  \rho_{{\mathrm{3}}}$ and $\ottnt{A'}  \simeq  \ottnt{C'}$ and $\rho_{{\mathrm{1}}}  \simeq  \rho_{{\mathrm{3}}}$
   for some $\ell$, $\ottnt{A'}$, $\ottnt{C'}$, $\rho_{{\mathrm{1}}}$, and $\rho_{{\mathrm{3}}}$.
   Since $\ottnt{A}  \equiv  \ottnt{C}$, there exist $\rho_{{\mathrm{31}}}$ and $\rho_{{\mathrm{32}}}$ such that
   \begin{itemize}
    \item $\ottnt{C} \,  =  \, \rho_{{\mathrm{31}}}  \odot  \ottsym{(}    \ell  \mathbin{:}  \ottnt{C'}   ;   \cdot    \ottsym{)}  \odot  \rho_{{\mathrm{32}}}$,
    \item $\ottnt{A'}  \equiv  \ottnt{C'}$,
    \item $\rho_{{\mathrm{1}}}  \equiv  \rho_{{\mathrm{31}}}  \odot  \rho_{{\mathrm{32}}}$, and
    \item $\ell \,  \not\in  \, \mathit{dom} \, \ottsym{(}  \rho_{{\mathrm{31}}}  \ottsym{)}$
   \end{itemize}
   by \reflem{equiv-inversion} (\ref{lem:equiv-inversion:row-label}).
   Again, by \reflem{equiv-inversion} (\ref{lem:equiv-inversion:row-label}),
   since $\ottnt{C}  \equiv  \ottnt{B}$, there exists some $\ottnt{B'}$, $\rho_{{\mathrm{21}}}$, and $\rho_{{\mathrm{22}}}$ such that
   \begin{itemize}
    \item $\ottnt{B} \,  =  \, \rho_{{\mathrm{21}}}  \odot  \ottsym{(}    \ell  \mathbin{:}  \ottnt{B'}   ;   \cdot    \ottsym{)}  \odot  \rho_{{\mathrm{22}}}$,
    \item $\ottnt{C'}  \equiv  \ottnt{B'}$,
    \item $\rho_{{\mathrm{31}}}  \odot  \rho_{{\mathrm{32}}}  \equiv  \rho_{{\mathrm{21}}}  \odot  \rho_{{\mathrm{22}}}$, and
    \item $\ell \,  \not\in  \, \mathit{dom} \, \ottsym{(}  \rho_{{\mathrm{21}}}  \ottsym{)}$.
   \end{itemize}
   Since $\rho_{{\mathrm{31}}}  \odot  \ottsym{(}    \ell  \mathbin{:}  \ottnt{C'}   ;   \cdot    \ottsym{)}  \odot  \rho_{{\mathrm{32}}} = \ottnt{C}  \simeq  \ottnt{B} = \rho_{{\mathrm{21}}}  \odot  \ottsym{(}    \ell  \mathbin{:}  \ottnt{B'}   ;   \cdot    \ottsym{)}  \odot  \rho_{{\mathrm{22}}}$
   and $\ell \,  \not\in  \, \mathit{dom} \, \ottsym{(}  \rho_{{\mathrm{31}}}  \ottsym{)} \,  \mathbin{\cup}  \, \mathit{dom} \, \ottsym{(}  \rho_{{\mathrm{21}}}  \ottsym{)}$,
   we have $\ottnt{C'}  \simeq  \ottnt{B'}$ and $\rho_{{\mathrm{31}}}  \odot  \rho_{{\mathrm{32}}}  \simeq  \rho_{{\mathrm{21}}}  \odot  \rho_{{\mathrm{22}}}$
   by \reflem{consistent-inv-row-label}.
   Since $\ottnt{C} \,  \triangleright _{ \ell }  \, \ottnt{C'}  \ottsym{,}  \rho_{{\mathrm{3}}}$, we have $\rho_{{\mathrm{3}}} \,  =  \, \rho_{{\mathrm{31}}}  \odot  \rho_{{\mathrm{32}}}$, so $\rho_{{\mathrm{1}}}  \simeq  \rho_{{\mathrm{31}}}  \odot  \rho_{{\mathrm{32}}}$.
   By the IHs, $\ottnt{A'}  \simeq  \ottnt{B'}$ and $\rho_{{\mathrm{1}}}  \simeq  \rho_{{\mathrm{21}}}  \odot  \rho_{{\mathrm{22}}}$.
   Since $\ottsym{(}  \rho_{{\mathrm{21}}}  \odot  \ottsym{(}    \ell  \mathbin{:}  \ottnt{B'}   ;   \cdot    \ottsym{)}  \odot  \rho_{{\mathrm{22}}}  \ottsym{)} \,  \triangleright _{ \ell }  \, \ottnt{B'}  \ottsym{,}  \rho_{{\mathrm{21}}}  \odot  \rho_{{\mathrm{22}}}$, we have
   $  \ell  \mathbin{:}  \ottnt{A'}   ;  \rho_{{\mathrm{1}}}   \simeq  \rho_{{\mathrm{21}}}  \odot  \ottsym{(}    \ell  \mathbin{:}  \ottnt{B'}   ;   \cdot    \ottsym{)}  \odot  \rho_{{\mathrm{22}}} = \ottnt{B}$ by \CE{ConsL}.

  \CE{ConsR}:
   We have $\ottnt{C} \,  =  \,   \ell  \mathbin{:}  \ottnt{C'}   ;  \rho_{{\mathrm{3}}} $ and, by inversion,
   $\ottnt{A} \,  \triangleright _{ \ell }  \, \ottnt{A'}  \ottsym{,}  \rho_{{\mathrm{1}}}$ and $\ottnt{A'}  \simeq  \ottnt{C'}$ and $\rho_{{\mathrm{1}}}  \simeq  \rho_{{\mathrm{3}}}$
   for some $\ell$, $\ottnt{A'}$, $\ottnt{C'}$, $\rho_{{\mathrm{1}}}$, and $\rho_{{\mathrm{3}}}$.
   Since $\ottnt{A}  \equiv  \ottnt{C}$, there exist $\rho_{{\mathrm{11}}}$ and $\rho_{{\mathrm{12}}}$ such that
   \begin{itemize}
    \item $\ottnt{A} \,  =  \, \rho_{{\mathrm{11}}}  \odot  \ottsym{(}    \ell  \mathbin{:}  \ottnt{A'}   ;   \cdot    \ottsym{)}  \odot  \rho_{{\mathrm{12}}}$,
    \item $\ottnt{A'}  \equiv  \ottnt{C'}$,
    \item $\rho_{{\mathrm{11}}}  \odot  \rho_{{\mathrm{12}}}  \equiv  \rho_{{\mathrm{3}}}$, and
    \item $\ell \,  \not\in  \, \mathit{dom} \, \ottsym{(}  \rho_{{\mathrm{11}}}  \ottsym{)}$
   \end{itemize}
   by \reflem{equiv-inversion} (\ref{lem:equiv-inversion:row-label}).
   Again, by \reflem{equiv-inversion} (\ref{lem:equiv-inversion:row-label}),
   since $\ottnt{C}  \equiv  \ottnt{B}$, there exists some $\ottnt{B'}$, $\rho_{{\mathrm{21}}}$, and $\rho_{{\mathrm{22}}}$ such that
   \begin{itemize}
    \item $\ottnt{B} \,  =  \, \rho_{{\mathrm{21}}}  \odot  \ottsym{(}    \ell  \mathbin{:}  \ottnt{B'}   ;   \cdot    \ottsym{)}  \odot  \rho_{{\mathrm{22}}}$,
    \item $\ottnt{C'}  \equiv  \ottnt{B'}$,
    \item $\rho_{{\mathrm{3}}}  \equiv  \rho_{{\mathrm{21}}}  \odot  \rho_{{\mathrm{22}}}$, and
    \item $\ell \,  \not\in  \, \mathit{dom} \, \ottsym{(}  \rho_{{\mathrm{21}}}  \ottsym{)}$.
   \end{itemize}
   Since $  \ell  \mathbin{:}  \ottnt{C'}   ;  \rho_{{\mathrm{3}}}  = \ottnt{C}  \simeq  \ottnt{B} = \rho_{{\mathrm{21}}}  \odot  \ottsym{(}    \ell  \mathbin{:}  \ottnt{B'}   ;   \cdot    \ottsym{)}  \odot  \rho_{{\mathrm{22}}}$
   and $\ell \,  \not\in  \, \mathit{dom} \, \ottsym{(}  \rho_{{\mathrm{21}}}  \ottsym{)}$,
   we have $\ottnt{C'}  \simeq  \ottnt{B'}$ and $\rho_{{\mathrm{3}}}  \simeq  \rho_{{\mathrm{21}}}  \odot  \rho_{{\mathrm{22}}}$
   by \reflem{consistent-inv-row-label}.
   Since $\ottnt{A} \,  \triangleright _{ \ell }  \, \ottnt{A'}  \ottsym{,}  \rho_{{\mathrm{1}}}$, we have $\rho_{{\mathrm{1}}} \,  =  \, \rho_{{\mathrm{11}}}  \odot  \rho_{{\mathrm{12}}}$, so $\rho_{{\mathrm{11}}}  \odot  \rho_{{\mathrm{12}}}  \simeq  \rho_{{\mathrm{3}}}$.
   By the IHs, $\ottnt{A'}  \simeq  \ottnt{B'}$ and $\rho_{{\mathrm{11}}}  \odot  \rho_{{\mathrm{12}}}  \simeq  \rho_{{\mathrm{21}}}  \odot  \rho_{{\mathrm{22}}}$.
   Since $\ell \,  \not\in  \, \mathit{dom} \, \ottsym{(}  \rho_{{\mathrm{11}}}  \ottsym{)} \,  \mathbin{\cup}  \, \mathit{dom} \, \ottsym{(}  \rho_{{\mathrm{21}}}  \ottsym{)}$ and $\ottnt{A'}  \simeq  \ottnt{B'}$,
   we have $\rho_{{\mathrm{11}}}  \odot  \ottsym{(}    \ell  \mathbin{:}  \ottnt{A'}   ;   \cdot    \ottsym{)}  \odot  \rho_{{\mathrm{12}}}  \simeq  \rho_{{\mathrm{21}}}  \odot  \ottsym{(}    \ell  \mathbin{:}  \ottnt{B'}   ;   \cdot    \ottsym{)}  \odot  \rho_{{\mathrm{22}}}$.
 \end{caseanalysis}
\end{proof}

\begin{lemma}{consistent-subsume-equiv}
 If $\ottnt{A}  \equiv  \ottnt{B}$, then $\ottnt{A}  \simeq  \ottnt{B}$.
\end{lemma}
\begin{proof}
 By induction on the derivation of $\ottnt{A}  \equiv  \ottnt{B}$.
 \begin{caseanalysis}
  \case \Eq{Refl}: By \CE{Refl}.
  \case \Eq{Trans}:
   By inversion, $\ottnt{A}  \equiv  \ottnt{C}$ and $\ottnt{C}  \equiv  \ottnt{B}$ for some $\ottnt{C}$.
   By the IHs, $\ottnt{A}  \simeq  \ottnt{C}$ and $\ottnt{C}  \simeq  \ottnt{B}$.
   We have $\ottnt{A}  \simeq  \ottnt{B}$ by \reflem{consistent-subsume-equiv-trans}.

  \case \Eq{Sym}:
   By inversion, $\ottnt{B}  \equiv  \ottnt{A}$.  By the IH, $\ottnt{B}  \simeq  \ottnt{A}$.
   By \reflem{consistent-symm}, $\ottnt{A}  \simeq  \ottnt{B}$.

  \case \Eq{Fun}: By the IHs.
  \case \Eq{Poly}: By the IH.
  \case \Eq{Record}: By the IH.
  \case \Eq{Variant}: By the IH.
  \case \Eq{Cons}: By the IH, \CE{Refl}, and \CE{Cons}.
  \case \Eq{Swap}: By \CE{Refl} and \CE{ConsL}.
 \end{caseanalysis}
\end{proof}

\begin{lemma}{consistent-comp-equiv-consistent}
 If $\ottnt{A}  \equiv  \ottnt{C}$ and $\ottnt{C}  \sim  \ottnt{B}$, then $\ottnt{A}  \simeq  \ottnt{B}$.
\end{lemma}
\begin{proof}
 By induction on $\ottnt{C}  \sim  \ottnt{B}$.
 \begin{caseanalysis}
  \case \Cns{Refl}: By \reflem{consistent-subsume-equiv}.

  \case \Cns{DynL}:
   We have $\ottnt{C} \,  =  \, \star$.
   By \reflem{equiv-inversion} (\ref{lem:equiv-inversion:dyn}), $\ottnt{A} \,  =  \, \star$.
   Thus, we finish by \CE{DynL}.

  \case \Cns{DynR}:
   We have $\ottnt{B} \,  =  \, \star$.  By \CE{DynR}.

  \case \Cns{Fun}:
   We have $\ottnt{C} \,  =  \, \ottnt{C_{{\mathrm{1}}}}  \rightarrow  \ottnt{C_{{\mathrm{2}}}}$ and $\ottnt{B} \,  =  \, \ottnt{B_{{\mathrm{1}}}}  \rightarrow  \ottnt{B_{{\mathrm{2}}}}$ and, by inversion,
   $\ottnt{C_{{\mathrm{1}}}}  \sim  \ottnt{B_{{\mathrm{1}}}}$ and $\ottnt{C_{{\mathrm{2}}}}  \sim  \ottnt{B_{{\mathrm{2}}}}$ for some $\ottnt{C_{{\mathrm{1}}}}$, $\ottnt{C_{{\mathrm{2}}}}$, $\ottnt{B_{{\mathrm{1}}}}$, and $\ottnt{B_{{\mathrm{2}}}}$.
   Since $\ottnt{A}  \equiv  \ottnt{C_{{\mathrm{1}}}}  \rightarrow  \ottnt{C_{{\mathrm{2}}}}$, there exist some $\ottnt{A_{{\mathrm{11}}}}$ and $\ottnt{A_{{\mathrm{12}}}}$
   such that $\ottnt{A} \,  =  \, \ottnt{A_{{\mathrm{1}}}}  \rightarrow  \ottnt{A_{{\mathrm{2}}}}$ and $\ottnt{A_{{\mathrm{1}}}}  \equiv  \ottnt{C_{{\mathrm{1}}}}$ and $\ottnt{A_{{\mathrm{2}}}}  \equiv  \ottnt{C_{{\mathrm{2}}}}$,
   by \reflem{equiv-inversion} (\ref{lem:equiv-inversion:fun}).
   By the IHs,
   $\ottnt{A_{{\mathrm{1}}}}  \simeq  \ottnt{B_{{\mathrm{1}}}}$ and $\ottnt{A_{{\mathrm{2}}}}  \simeq  \ottnt{B_{{\mathrm{2}}}}$.
   By \CE{Fun}, $\ottnt{A_{{\mathrm{1}}}}  \rightarrow  \ottnt{A_{{\mathrm{2}}}}  \simeq  \ottnt{B_{{\mathrm{1}}}}  \rightarrow  \ottnt{B_{{\mathrm{2}}}}$.

  \case \Cns{Poly}:
   We have $\ottnt{C} \,  =  \,  \text{\unboldmath$\forall\!$}  \,  \mathit{X}  \mathord{:}  \ottnt{K}   \ottsym{.} \, \ottnt{C'}$ and $\ottnt{B} \,  =  \,  \text{\unboldmath$\forall\!$}  \,  \mathit{X}  \mathord{:}  \ottnt{K}   \ottsym{.} \, \ottnt{B'}$ and, by inversion,
   $\ottnt{C'}  \sim  \ottnt{B'}$ for some $\mathit{X}$, $\ottnt{K}$, $\ottnt{C'}$, and $\ottnt{B'}$.
   Since $\ottnt{A}  \equiv   \text{\unboldmath$\forall\!$}  \,  \mathit{X}  \mathord{:}  \ottnt{K}   \ottsym{.} \, \ottnt{C'}$, there exists some $\ottnt{A'}$ such that
   $\ottnt{A} \,  =  \,  \text{\unboldmath$\forall\!$}  \,  \mathit{X}  \mathord{:}  \ottnt{K}   \ottsym{.} \, \ottnt{A'}$ and $\ottnt{A'}  \equiv  \ottnt{C'}$,
   by \reflem{equiv-inversion} (\ref{lem:equiv-inversion:forall}).
   By the IH, $\ottnt{A'}  \simeq  \ottnt{B'}$.
   By \CE{Poly}, we finish.

  \case \Cns{PolyL}:
   We have $\ottnt{C} \,  =  \,  \text{\unboldmath$\forall\!$}  \,  \mathit{X}  \mathord{:}  \ottnt{K}   \ottsym{.} \, \ottnt{C'}$ and, by inversion,
   $\mathbf{QPoly} \, \ottsym{(}  \ottnt{B}  \ottsym{)}$ and $\mathit{X} \,  \not\in  \,  \mathit{ftv}  (  \ottnt{B}  ) $ and $\ottnt{C'}  \sim  \ottnt{B}$.
   Since $\ottnt{A} \,  =  \,  \text{\unboldmath$\forall\!$}  \,  \mathit{X}  \mathord{:}  \ottnt{K}   \ottsym{.} \, \ottnt{C'}$, there exists some $\ottnt{A'}$ such that
   $\ottnt{A} \,  =  \,  \text{\unboldmath$\forall\!$}  \,  \mathit{X}  \mathord{:}  \ottnt{K}   \ottsym{.} \, \ottnt{A'}$ and $\ottnt{A'}  \equiv  \ottnt{C'}$,
   by \reflem{equiv-inversion} (\ref{lem:equiv-inversion:forall}).
   By the IH, $\ottnt{A'}  \simeq  \ottnt{B}$.
   By \CE{PolyL}, we finish.

  \case \Cns{PolyR}:
   We have $\ottnt{B} \,  =  \,  \text{\unboldmath$\forall\!$}  \,  \mathit{X}  \mathord{:}  \ottnt{K}   \ottsym{.} \, \ottnt{B'}$ and, by inversion,
   $\mathbf{QPoly} \, \ottsym{(}  \ottnt{C}  \ottsym{)}$ and $\mathit{X} \,  \not\in  \,  \mathit{ftv}  (  \ottnt{C}  ) $ and $\ottnt{C}  \sim  \ottnt{B'}$.
   By the IH, $\ottnt{A}  \simeq  \ottnt{B'}$.
   Since $\ottnt{A}  \equiv  \ottnt{C}$ and $\mathbf{QPoly} \, \ottsym{(}  \ottnt{C}  \ottsym{)}$ and $\mathit{X} \,  \not\in  \,  \mathit{ftv}  (  \ottnt{C}  ) $,
   we have $\mathbf{QPoly} \, \ottsym{(}  \ottnt{A}  \ottsym{)}$ and $\mathit{X} \,  \not\in  \,  \mathit{ftv}  (  \ottnt{A}  ) $ by
   Lemmas~\ref{lem:equiv-qpoly} and \ref{lem:equiv-free-tyvar}.
   Thus, by \CE{PolyR}, we have $\ottnt{A}  \simeq   \text{\unboldmath$\forall\!$}  \,  \mathit{X}  \mathord{:}  \ottnt{K}   \ottsym{.} \, \ottnt{B'}$.

  \case \Cns{Record}:
   We have $\ottnt{C} \,  =  \,  [  \rho_{{\mathrm{3}}}  ] $ and $\ottnt{B} \,  =  \,  [  \rho_{{\mathrm{2}}}  ] $ and, by inversion,
   $\rho_{{\mathrm{3}}}  \sim  \rho_{{\mathrm{2}}}$ for some $\rho_{{\mathrm{3}}}$ and $\rho_{{\mathrm{2}}}$.
   Since $\ottnt{A}  \equiv  \ottnt{C}$, there exists some $\rho_{{\mathrm{1}}}$ such that
   $\ottnt{A} \,  =  \,  [  \rho_{{\mathrm{1}}}  ] $ and $\rho_{{\mathrm{1}}}  \equiv  \rho_{{\mathrm{3}}}$,
   by \reflem{equiv-inversion} (\ref{lem:equiv-inversion:record}).
   By the IH, $\rho_{{\mathrm{1}}}  \simeq  \rho_{{\mathrm{2}}}$.
   Thus, by \CE{Record}, we have $ [  \rho_{{\mathrm{1}}}  ]   \simeq   [  \rho_{{\mathrm{2}}}  ] $.

  \case \Cns{Variant}:
   We have $\ottnt{C} \,  =  \,  \langle  \rho_{{\mathrm{3}}}  \rangle $ and $\ottnt{B} \,  =  \,  \langle  \rho_{{\mathrm{2}}}  \rangle $ and, by inversion,
   $\rho_{{\mathrm{3}}}  \sim  \rho_{{\mathrm{2}}}$ for some $\rho_{{\mathrm{3}}}$ and $\rho_{{\mathrm{2}}}$.
   Since $\ottnt{A}  \equiv  \ottnt{C}$, there exists some $\rho_{{\mathrm{1}}}$ such that
   $\ottnt{A} \,  =  \,  \langle  \rho_{{\mathrm{1}}}  \rangle $ and $\rho_{{\mathrm{1}}}  \equiv  \rho_{{\mathrm{3}}}$,
   by \reflem{equiv-inversion} (\ref{lem:equiv-inversion:variant}).
   By the IH, $\rho_{{\mathrm{1}}}  \simeq  \rho_{{\mathrm{2}}}$.
   Thus, by \CE{Variant}, we have $ \langle  \rho_{{\mathrm{1}}}  \rangle   \simeq   \langle  \rho_{{\mathrm{2}}}  \rangle $.

  \case \Cns{Cons}:
   We have $\ottnt{C} \,  =  \,   \ell  \mathbin{:}  \ottnt{C'}   ;  \rho_{{\mathrm{3}}} $ and $\ottnt{B} \,  =  \,   \ell  \mathbin{:}  \ottnt{B'}   ;  \rho_{{\mathrm{2}}} $ and, by inversion,
   $\ottnt{C'}  \sim  \ottnt{B'}$ and $\rho_{{\mathrm{3}}}  \sim  \rho_{{\mathrm{2}}}$
   for some $\ell$, $\ottnt{C'}$, $\ottnt{B'}$, $\rho_{{\mathrm{3}}}$, and $\rho_{{\mathrm{2}}}$.
   Since $\ottnt{A}  \equiv  \ottnt{C}$, there exist some $\ottnt{A'}$, $\rho_{{\mathrm{11}}}$, $\rho_{{\mathrm{12}}}$ such that
   \begin{itemize}
    \item $\ottnt{A} \,  =  \, \rho_{{\mathrm{11}}}  \odot  \ottsym{(}    \ell  \mathbin{:}  \ottnt{A'}   ;   \cdot    \ottsym{)}  \odot  \rho_{{\mathrm{12}}}$,
    \item $\ottnt{A'}  \equiv  \ottnt{C'}$,
    \item $\rho_{{\mathrm{11}}}  \odot  \rho_{{\mathrm{12}}}  \equiv  \rho_{{\mathrm{3}}}$, and
    \item $\ell \,  \not\in  \, \mathit{dom} \, \ottsym{(}  \rho_{{\mathrm{11}}}  \ottsym{)}$
   \end{itemize}
   by \reflem{equiv-inversion} (\ref{lem:equiv-inversion:row-label}).
   By the IHs, $\ottnt{A'}  \simeq  \ottnt{B'}$ and $\rho_{{\mathrm{11}}}  \odot  \rho_{{\mathrm{12}}}  \simeq  \rho_{{\mathrm{2}}}$.
   We have $\ottnt{A} \,  \triangleright _{ \ell }  \, \ottnt{A'}  \ottsym{,}  \rho_{{\mathrm{11}}}  \odot  \rho_{{\mathrm{12}}}$.
   Thus, by \CE{ConsR}, $\ottnt{A}  \simeq    \ell  \mathbin{:}  \ottnt{B'}   ;  \rho_{{\mathrm{2}}} $.

  \case \Cns{ConsL}:
   We have $\ottnt{C} \,  =  \,   \ell  \mathbin{:}  \ottnt{C'}   ;  \rho_{{\mathrm{3}}} $ and, by inversion,
   $\ell \,  \not\in  \, \mathit{dom} \, \ottsym{(}  \ottnt{B}  \ottsym{)}$ and $\ottnt{B}$ ends with $ \star $ and $\rho_{{\mathrm{3}}}  \sim  \ottnt{B}$
   for some $\ell$, $\ottnt{C'}$, and $\rho_{{\mathrm{3}}}$.
   Since $\ottnt{A}  \equiv  \ottnt{C}$, there exist some $\ottnt{A'}$, $\rho_{{\mathrm{11}}}$, $\rho_{{\mathrm{12}}}$ such that
   \begin{itemize}
   \item $\ottnt{A} \,  =  \, \rho_{{\mathrm{11}}}  \odot  \ottsym{(}    \ell  \mathbin{:}  \ottnt{A'}   ;   \cdot    \ottsym{)}  \odot  \rho_{{\mathrm{12}}}$,
   \item $\ottnt{A'}  \equiv  \ottnt{C'}$,
   \item $\rho_{{\mathrm{11}}}  \odot  \rho_{{\mathrm{12}}}  \equiv  \rho_{{\mathrm{3}}}$, and
   \item $\ell \,  \not\in  \, \mathit{dom} \, \ottsym{(}  \rho_{{\mathrm{11}}}  \ottsym{)}$
   \end{itemize}
   by \reflem{equiv-inversion} (\ref{lem:equiv-inversion:row-label}).
   By the IH, $\rho_{{\mathrm{11}}}  \odot  \rho_{{\mathrm{12}}}  \simeq  \ottnt{B}$.
   By Lemmas~\ref{lem:consistent-symm} and \ref{lem:consistent-inj-row-label-dyn},
   $\rho_{{\mathrm{11}}}  \odot  \ottsym{(}    \ell  \mathbin{:}  \ottnt{A'}   ;   \cdot    \ottsym{)}  \odot  \rho_{{\mathrm{12}}}  \simeq  \ottnt{B}$.

  \case \Cns{ConsR}:
   we have $\ottnt{B} \,  =  \,   \ell  \mathbin{:}  \ottnt{B'}   ;  \rho_{{\mathrm{2}}} $ and, by inversion,
   $\ell \,  \not\in  \, \mathit{dom} \, \ottsym{(}  \ottnt{C}  \ottsym{)}$ and $\ottnt{C}$ ends with $ \star $ and $\ottnt{C}  \sim  \rho_{{\mathrm{2}}}$
   for some $\ell$, $\ottnt{B'}$, and $\rho_{{\mathrm{2}}}$.
   By the IH, $\ottnt{A}  \simeq  \rho_{{\mathrm{2}}}$.
   Since $\ottnt{A}  \equiv  \ottnt{C}$ and $\ell \,  \not\in  \, \mathit{dom} \, \ottsym{(}  \ottnt{C}  \ottsym{)}$ and $\ottnt{C}$ ends with $ \star $,
   we have $\ell \,  \not\in  \, \mathit{dom} \, \ottsym{(}  \ottnt{A}  \ottsym{)}$ and $\ottnt{A}$ ends with $ \star $
   by Lemmas~\ref{lem:equiv-free-tyvar} and \ref{lem:equiv-ends-with-dyn}.
   Thus, by \CE{ConsR}, we have $\ottnt{A}  \simeq    \ell  \mathbin{:}  \ottnt{B'}   ;  \rho_{{\mathrm{2}}} $.
 \end{caseanalysis}
\end{proof}

\begin{lemma}{consistent-comp-consistent-equiv}
 If $\ottnt{A}  \simeq  \ottnt{B}$ and $\ottnt{B}  \equiv  \ottnt{C}$, then $\ottnt{A}  \simeq  \ottnt{C}$.
\end{lemma}
\begin{proof}
 By induction on the derivation of $\ottnt{A}  \simeq  \ottnt{B}$.
 \begin{caseanalysis}
  \case \CE{Refl}: By \reflem{consistent-subsume-equiv}.
  \case \CE{DynL}: By \CE{DynL}.
  \case \CE{DynR}: We have $\ottnt{B} \,  =  \, \star$.
   Since $\ottnt{B}  \equiv  \ottnt{C}$, we have $\ottnt{C} \,  =  \, \star$
   by \reflem{equiv-inversion} (\ref{lem:equiv-inversion:dyn}).
   By \CE{DynR}.
  \case \CE{Fun}:
   We have $\ottnt{A} \,  =  \, \ottnt{A_{{\mathrm{1}}}}  \rightarrow  \ottnt{A_{{\mathrm{2}}}}$ and $\ottnt{B} \,  =  \, \ottnt{B_{{\mathrm{1}}}}  \rightarrow  \ottnt{B_{{\mathrm{2}}}}$ and, by inversion,
   $\ottnt{A_{{\mathrm{1}}}}  \simeq  \ottnt{B_{{\mathrm{1}}}}$ and $\ottnt{A_{{\mathrm{2}}}}  \simeq  \ottnt{B_{{\mathrm{2}}}}$ for some $\ottnt{A_{{\mathrm{1}}}}$, $\ottnt{A_{{\mathrm{2}}}}$, $\ottnt{B_{{\mathrm{1}}}}$, and $\ottnt{B_{{\mathrm{2}}}}$.
   Since $\ottnt{B}  \equiv  \ottnt{C}$, there exist some $\ottnt{C_{{\mathrm{1}}}}$ and $\ottnt{C_{{\mathrm{2}}}}$ such that
   $\ottnt{C} \,  =  \, \ottnt{C_{{\mathrm{1}}}}  \rightarrow  \ottnt{C_{{\mathrm{2}}}}$ and
   $\ottnt{B_{{\mathrm{1}}}}  \equiv  \ottnt{C_{{\mathrm{1}}}}$ and $\ottnt{B_{{\mathrm{2}}}}  \equiv  \ottnt{C_{{\mathrm{2}}}}$
   by \reflem{equiv-inversion} (\ref{lem:equiv-inversion:fun}).
   By the IHs, $\ottnt{A_{{\mathrm{1}}}}  \simeq  \ottnt{C_{{\mathrm{1}}}}$ and $\ottnt{A_{{\mathrm{2}}}}  \simeq  \ottnt{C_{{\mathrm{2}}}}$.
   Thus, $\ottnt{A_{{\mathrm{1}}}}  \rightarrow  \ottnt{A_{{\mathrm{2}}}}  \simeq  \ottnt{C_{{\mathrm{1}}}}  \rightarrow  \ottnt{C_{{\mathrm{2}}}}$ by \CE{Fun}.

  \case \CE{Poly}:
   We have $\ottnt{A} \,  =  \,  \text{\unboldmath$\forall\!$}  \,  \mathit{X}  \mathord{:}  \ottnt{K}   \ottsym{.} \, \ottnt{A'}$ and $\ottnt{B} \,  =  \,  \text{\unboldmath$\forall\!$}  \,  \mathit{X}  \mathord{:}  \ottnt{K}   \ottsym{.} \, \ottnt{B'}$ and, by inversion,
   $\ottnt{A'}  \simeq  \ottnt{B'}$ for some $\mathit{X}$, $\ottnt{K}$, $\ottnt{A'}$, and $\ottnt{B'}$.
   Since $\ottnt{B}  \equiv  \ottnt{C}$, there exist some $\ottnt{C'}$ such that
   $\ottnt{C} \,  =  \,  \text{\unboldmath$\forall\!$}  \,  \mathit{X}  \mathord{:}  \ottnt{K}   \ottsym{.} \, \ottnt{C'}$ and $\ottnt{B'}  \equiv  \ottnt{C'}$
   by \reflem{equiv-inversion} (\ref{lem:equiv-inversion:forall}).
   By the IH, $\ottnt{A'}  \simeq  \ottnt{C'}$.
   Thus, $ \text{\unboldmath$\forall\!$}  \,  \mathit{X}  \mathord{:}  \ottnt{K}   \ottsym{.} \, \ottnt{A'}  \simeq   \text{\unboldmath$\forall\!$}  \,  \mathit{X}  \mathord{:}  \ottnt{K}   \ottsym{.} \, \ottnt{C'}$ by \CE{Poly}.

  \case \CE{PolyL}:
   We have $\ottnt{A} \,  =  \,  \text{\unboldmath$\forall\!$}  \,  \mathit{X}  \mathord{:}  \ottnt{K}   \ottsym{.} \, \ottnt{A'}$ and, by inversion,
   $\mathbf{QPoly} \, \ottsym{(}  \ottnt{B}  \ottsym{)}$ and $\mathit{X} \,  \not\in  \,  \mathit{ftv}  (  \ottnt{B}  ) $ and $\ottnt{A'}  \simeq  \ottnt{B}$
   for some $\mathit{X}$, $\ottnt{K}$, and $\ottnt{A'}$.
   By the IH, $\ottnt{A'}  \simeq  \ottnt{C}$.
   Since $\ottnt{B}  \equiv  \ottnt{C}$ and $\mathbf{QPoly} \, \ottsym{(}  \ottnt{B}  \ottsym{)}$ and $\mathit{X} \,  \not\in  \,  \mathit{ftv}  (  \ottnt{B}  ) $,
   we have $\mathbf{QPoly} \, \ottsym{(}  \ottnt{C}  \ottsym{)}$ and $\mathit{X} \,  \not\in  \,  \mathit{ftv}  (  \ottnt{C}  ) $
   by Lemmas~\ref{lem:equiv-qpoly} and \ref{lem:equiv-free-tyvar}.
   Thus, $ \text{\unboldmath$\forall\!$}  \,  \mathit{X}  \mathord{:}  \ottnt{K}   \ottsym{.} \, \ottnt{A'}  \simeq  \ottnt{C}$ by \CE{PolyL}.

  \case \CE{PolyR}:
   We have $\ottnt{B} \,  =  \,  \text{\unboldmath$\forall\!$}  \,  \mathit{X}  \mathord{:}  \ottnt{K}   \ottsym{.} \, \ottnt{B'}$ and, by inversion,
   $\mathbf{QPoly} \, \ottsym{(}  \ottnt{A}  \ottsym{)}$ and $\mathit{X} \,  \not\in  \,  \mathit{ftv}  (  \ottnt{A}  ) $ and $\ottnt{A}  \simeq  \ottnt{B'}$
   for some $\mathit{X}$, $\ottnt{K}$, and $\ottnt{B'}$.
   By \reflem{equiv-inversion} (\ref{lem:equiv-inversion:forall}),
   since $\ottnt{B}  \equiv  \ottnt{C}$, there exists some $\ottnt{C'}$ such that
   $\ottnt{C} \,  =  \,  \text{\unboldmath$\forall\!$}  \,  \mathit{X}  \mathord{:}  \ottnt{K}   \ottsym{.} \, \ottnt{C'}$ and $\ottnt{B'}  \equiv  \ottnt{C'}$.
   By the IH, $\ottnt{A}  \simeq  \ottnt{C'}$.
   By \CE{PolyR}, $\ottnt{A}  \simeq   \text{\unboldmath$\forall\!$}  \,  \mathit{X}  \mathord{:}  \ottnt{K}   \ottsym{.} \, \ottnt{C'}$.

  \case \CE{Record}:
   We have $\ottnt{A} \,  =  \,  [  \rho_{{\mathrm{1}}}  ] $ and $\ottnt{B} \,  =  \,  [  \rho_{{\mathrm{2}}}  ] $ and, by inversion,
   $\rho_{{\mathrm{1}}}  \simeq  \rho_{{\mathrm{2}}}$ for some $\rho_{{\mathrm{1}}}$ and $\rho_{{\mathrm{2}}}$.
   By \reflem{equiv-inversion} (\ref{lem:equiv-inversion:record}),
   since $\ottnt{B}  \equiv  \ottnt{C}$, there exists some $\rho_{{\mathrm{3}}}$ such that
   $\ottnt{C} \,  =  \,  [  \rho_{{\mathrm{3}}}  ] $ and $\rho_{{\mathrm{2}}}  \equiv  \rho_{{\mathrm{3}}}$.
   By the IH, $\rho_{{\mathrm{1}}}  \simeq  \rho_{{\mathrm{3}}}$.
   By \CE{Record}, $ [  \rho_{{\mathrm{1}}}  ]   \simeq   [  \rho_{{\mathrm{3}}}  ] $.

  \case \CE{Variant}:
   We have $\ottnt{A} \,  =  \,  \langle  \rho_{{\mathrm{1}}}  \rangle $ and $\ottnt{B} \,  =  \,  \langle  \rho_{{\mathrm{2}}}  \rangle $ and, by inversion,
   $\rho_{{\mathrm{1}}}  \simeq  \rho_{{\mathrm{2}}}$ for some $\rho_{{\mathrm{1}}}$ and $\rho_{{\mathrm{2}}}$.
   By \reflem{equiv-inversion} (\ref{lem:equiv-inversion:variant}),
   since $\ottnt{B}  \equiv  \ottnt{C}$, there exists some $\rho_{{\mathrm{3}}}$ such that
   $\ottnt{C} \,  =  \,  \langle  \rho_{{\mathrm{3}}}  \rangle $ and $\rho_{{\mathrm{2}}}  \equiv  \rho_{{\mathrm{3}}}$.
   By the IH, $\rho_{{\mathrm{1}}}  \simeq  \rho_{{\mathrm{3}}}$.
   By \CE{Variant}, $ \langle  \rho_{{\mathrm{1}}}  \rangle   \simeq   \langle  \rho_{{\mathrm{3}}}  \rangle $.

  \case \CE{ConsL}:
   We have $\ottnt{A} \,  =  \,   \ell  \mathbin{:}  \ottnt{A'}   ;  \rho_{{\mathrm{1}}} $ and, by inversion,
   $\ottnt{B} \,  \triangleright _{ \ell }  \, \ottnt{B'}  \ottsym{,}  \rho_{{\mathrm{2}}}$ and $\ottnt{A'}  \simeq  \ottnt{B'}$ and $\rho_{{\mathrm{1}}}  \simeq  \rho_{{\mathrm{2}}}$
   for some $\ell$, $\ottnt{A'}$, $\ottnt{B'}$, $\rho_{{\mathrm{1}}}$, and $\rho_{{\mathrm{2}}}$.

   If $\ell \,  \in  \, \mathit{dom} \, \ottsym{(}  \ottnt{B}  \ottsym{)}$, then there exist some $\rho_{{\mathrm{21}}}$ and $\rho_{{\mathrm{22}}}$ such that
   \begin{itemize}
    \item $\ottnt{B} \,  =  \, \rho_{{\mathrm{21}}}  \odot  \ottsym{(}    \ell  \mathbin{:}  \ottnt{B'}   ;   \cdot    \ottsym{)}  \odot  \rho_{{\mathrm{22}}}$,
    \item $\rho_{{\mathrm{2}}} \,  =  \, \rho_{{\mathrm{21}}}  \odot  \rho_{{\mathrm{22}}}$, and 
    \item $\ell \,  \not\in  \, \mathit{dom} \, \ottsym{(}  \rho_{{\mathrm{21}}}  \ottsym{)}$.
   \end{itemize}
   Since $\ottnt{B}  \equiv  \ottnt{C}$, there exist some $\ottnt{C'}$, $\rho_{{\mathrm{31}}}$, and $\rho_{{\mathrm{32}}}$ such that
   \begin{itemize}
    \item $\ottnt{C} \,  =  \, \rho_{{\mathrm{31}}}  \odot  \ottsym{(}    \ell  \mathbin{:}  \ottnt{C'}   ;   \cdot    \ottsym{)}  \odot  \rho_{{\mathrm{32}}}$,
    \item $\ottnt{B'}  \equiv  \ottnt{C'}$,
    \item $\ell \,  \not\in  \, \mathit{dom} \, \ottsym{(}  \rho_{{\mathrm{31}}}  \ottsym{)}$, and
    \item $\rho_{{\mathrm{21}}}  \odot  \rho_{{\mathrm{22}}}  \equiv  \rho_{{\mathrm{31}}}  \odot  \rho_{{\mathrm{32}}}$
   \end{itemize}
   by \reflem{equiv-inversion} (\ref{lem:equiv-inversion:row-label}).
   Since $\rho_{{\mathrm{1}}}  \simeq  \rho_{{\mathrm{2}}}$ and $\rho_{{\mathrm{2}}} = \rho_{{\mathrm{21}}}  \odot  \rho_{{\mathrm{22}}}  \equiv  \rho_{{\mathrm{31}}}  \odot  \rho_{{\mathrm{32}}}$,
   we have $\rho_{{\mathrm{1}}}  \simeq  \rho_{{\mathrm{31}}}  \odot  \rho_{{\mathrm{32}}}$ by the IH.
   Besides, $\ottnt{A'}  \simeq  \ottnt{B'}$ and $\ottnt{B'}  \equiv  \ottnt{C'}$, we have $\ottnt{A'}  \simeq  \ottnt{C'}$ by the IH.
   Since $\ottnt{C} \,  \triangleright _{ \ell }  \, \ottnt{C'}  \ottsym{,}  \rho_{{\mathrm{31}}}  \odot  \rho_{{\mathrm{32}}}$, we have $  \ell  \mathbin{:}  \ottnt{A'}   ;  \rho_{{\mathrm{1}}}   \simeq  \ottnt{C}$ by \CE{ConsL}.

   Otherwise, if $\ell \,  \not\in  \, \mathit{dom} \, \ottsym{(}  \ottnt{B}  \ottsym{)}$, then $\ottnt{B'} \,  =  \, \star$ and $\rho_{{\mathrm{2}}} \,  =  \, \ottnt{B}$ and
   $\ottnt{B}$ ends with $ \star $.
   Since $\rho_{{\mathrm{1}}}  \simeq  \rho_{{\mathrm{2}}}$ and $\rho_{{\mathrm{2}}} = \ottnt{B}  \equiv  \ottnt{C}$,
   we have $\rho_{{\mathrm{1}}}  \simeq  \ottnt{C}$ by the IH.
   Since $\ottnt{B}  \equiv  \ottnt{C}$, we can find $\ottnt{C} \,  \triangleright _{ \ell }  \, \star  \ottsym{,}  \ottnt{C}$ by
   Lemmas~\ref{lem:equiv-labels} and \ref{lem:equiv-ends-with-dyn}.
   Since $\ottnt{A'}  \simeq  \star$ by \CE{DynR} and $\rho_{{\mathrm{1}}}  \simeq  \ottnt{C}$,
   we have $  \ell  \mathbin{:}  \ottnt{A'}   ;  \rho_{{\mathrm{1}}}   \simeq  \ottnt{C}$ by \CE{ConsL}.

  \case \CE{ConsR}:
   We have $\ottnt{B} \,  =  \,   \ell  \mathbin{:}  \ottnt{B'}   ;  \rho_{{\mathrm{2}}} $ and, by inversion,
   $\ottnt{A} \,  \triangleright _{ \ell }  \, \ottnt{A'}  \ottsym{,}  \rho_{{\mathrm{1}}}$ and $\ottnt{A'}  \simeq  \ottnt{B'}$ and $\rho_{{\mathrm{1}}}  \simeq  \rho_{{\mathrm{2}}}$
   for some $\ell$, $\ottnt{A'}$, $\ottnt{B'}$, $\rho_{{\mathrm{1}}}$, and $\rho_{{\mathrm{2}}}$.
   Since $\ottnt{B}  \equiv  \ottnt{C}$, there exist some $\ottnt{C'}$, $\rho_{{\mathrm{31}}}$, and $\rho_{{\mathrm{32}}}$ such that
   \begin{itemize}
    \item $\ottnt{C} \,  =  \, \rho_{{\mathrm{31}}}  \odot  \ottsym{(}    \ell  \mathbin{:}  \ottnt{C'}   ;   \cdot    \ottsym{)}  \odot  \rho_{{\mathrm{32}}}$,
    \item $\ottnt{B'}  \equiv  \ottnt{C'}$,
    \item $\ell \,  \not\in  \, \mathit{dom} \, \ottsym{(}  \rho_{{\mathrm{31}}}  \ottsym{)}$, and
    \item $\rho_{{\mathrm{2}}}  \equiv  \rho_{{\mathrm{31}}}  \odot  \rho_{{\mathrm{32}}}$
   \end{itemize}
   by \reflem{equiv-inversion} (\ref{lem:equiv-inversion:row-label}).

   If $\ell \,  \in  \, \mathit{dom} \, \ottsym{(}  \ottnt{A}  \ottsym{)}$, then there exist some $\rho_{{\mathrm{11}}}$ and $\rho_{{\mathrm{12}}}$ such that
   \begin{itemize}
    \item $\ottnt{A} \,  =  \, \rho_{{\mathrm{11}}}  \odot  \ottsym{(}    \ell  \mathbin{:}  \ottnt{A'}   ;   \cdot    \ottsym{)}  \odot  \rho_{{\mathrm{12}}}$,
    \item $\rho_{{\mathrm{1}}} \,  =  \, \rho_{{\mathrm{11}}}  \odot  \rho_{{\mathrm{12}}}$, and
    \item $\ell \,  \not\in  \, \mathit{dom} \, \ottsym{(}  \rho_{{\mathrm{11}}}  \ottsym{)}$.
   \end{itemize}
   Since $\rho_{{\mathrm{1}}}  \simeq  \rho_{{\mathrm{2}}}$ and $\rho_{{\mathrm{2}}}  \equiv  \rho_{{\mathrm{31}}}  \odot  \rho_{{\mathrm{32}}}$,
   we have $\rho_{{\mathrm{1}}}  \simeq  \rho_{{\mathrm{31}}}  \odot  \rho_{{\mathrm{32}}}$ by the IH.
   Besides, $\ottnt{A'}  \simeq  \ottnt{B'}$ and $\ottnt{B'}  \equiv  \ottnt{C'}$, we have $\ottnt{A'}  \simeq  \ottnt{C'}$ by the IH.
   By \reflem{consistent-inj-row-label},
   \[
    \ottnt{A} = \rho_{{\mathrm{11}}}  \odot  \ottsym{(}    \ell  \mathbin{:}  \ottnt{A'}   ;   \cdot    \ottsym{)}  \odot  \rho_{{\mathrm{12}}}  \simeq  \rho_{{\mathrm{31}}}  \odot  \ottsym{(}    \ell  \mathbin{:}  \ottnt{C'}   ;   \cdot    \ottsym{)}  \odot  \rho_{{\mathrm{32}}} = \ottnt{C}.
   \]

   Otherwise, if $\ell \,  \not\in  \, \mathit{dom} \, \ottsym{(}  \ottnt{A}  \ottsym{)}$, then $\ottnt{A'} \,  =  \, \star$ and $\rho_{{\mathrm{1}}} \,  =  \, \ottnt{A}$ and
   $\ottnt{A}$ ends with $ \star $.
   Since $\ottnt{A} = \rho_{{\mathrm{1}}}  \simeq  \rho_{{\mathrm{2}}}$ and $\rho_{{\mathrm{2}}}  \equiv  \rho_{{\mathrm{31}}}  \odot  \rho_{{\mathrm{32}}}$,
   we have $\ottnt{A}  \simeq  \rho_{{\mathrm{31}}}  \odot  \rho_{{\mathrm{32}}}$ by the IH.
   By \reflem{consistent-inj-row-label-dyn},
   $\ottnt{A}  \simeq  \rho_{{\mathrm{31}}}  \odot  \ottsym{(}    \ell  \mathbin{:}  \ottnt{C'}   ;   \cdot    \ottsym{)}  \odot  \rho_{{\mathrm{32}}} = \ottnt{C}$.
 \end{caseanalysis}
\end{proof}

\begin{thm} \label{thm:ce-consistency-equiv}
 $\ottnt{A}  \simeq  \ottnt{B}$ if and only if $\ottnt{A}  \equiv  \ottnt{A'}$ and $\ottnt{A'}  \sim  \ottnt{B'}$ and $\ottnt{B'}  \equiv  \ottnt{B}$
 for some $\ottnt{A'}$ and $\ottnt{B'}$.
\end{thm}
\begin{proof}
 First, we show the left-to-right direction.
 Suppose $\ottnt{A}  \simeq  \ottnt{B}$.  By \reflem{consistent-decomp}, there exists some $\ottnt{A'}$ such that
 $\ottnt{A}  \equiv  \ottnt{A'}$ and $\ottnt{A'}  \sim  \ottnt{B}$.  Since ${[B == B]}$ by \Eq{Refl}, we finish

 Next, we show the right-to-left.
 Suppose that $\ottnt{A}  \equiv  \ottnt{A'}$ and $\ottnt{A'}  \sim  \ottnt{B'}$ and $\ottnt{B'}  \equiv  \ottnt{B}$.
 By \reflem{consistent-comp-equiv-consistent}, $\ottnt{A}  \simeq  \ottnt{B'}$.
 By \reflem{consistent-comp-consistent-equiv}, $\ottnt{A}  \simeq  \ottnt{B}$.
\end{proof}

\subsection{Type Soundness}

\begin{lemmap}{Weakening}{weakening}
 Suppose that $\Sigma  \vdash  \Gamma_{{\mathrm{1}}}  \ottsym{,}  \Gamma_{{\mathrm{2}}}$.
 Let $\Gamma_{{\mathrm{3}}}$ be a typing context such that
 $\mathit{dom} \, \ottsym{(}  \Gamma_{{\mathrm{2}}}  \ottsym{)} \,  \mathbin{\cap}  \, \mathit{dom} \, \ottsym{(}  \Gamma_{{\mathrm{3}}}  \ottsym{)} \,  =  \, \emptyset$.
 \begin{enumerate}
  \item \label{lem:weakening:typctx}
        If $\Sigma  \vdash  \Gamma_{{\mathrm{1}}}  \ottsym{,}  \Gamma_{{\mathrm{3}}}$,
        then $\Sigma  \vdash  \Gamma_{{\mathrm{1}}}  \ottsym{,}  \Gamma_{{\mathrm{2}}}  \ottsym{,}  \Gamma_{{\mathrm{3}}}$.
  \item \label{lem:weakening:type}
        If $\Sigma  \ottsym{;}  \Gamma_{{\mathrm{1}}}  \ottsym{,}  \Gamma_{{\mathrm{3}}}  \vdash  \ottnt{A}  \ottsym{:}  \ottnt{K}$,
        then $\Sigma  \ottsym{;}  \Gamma_{{\mathrm{1}}}  \ottsym{,}  \Gamma_{{\mathrm{2}}}  \ottsym{,}  \Gamma_{{\mathrm{3}}}  \vdash  \ottnt{A}  \ottsym{:}  \ottnt{K}$.
  \item \label{lem:weakening:term}
        If $\Sigma  \ottsym{;}  \Gamma_{{\mathrm{1}}}  \ottsym{,}  \Gamma_{{\mathrm{3}}}  \vdash  \ottnt{e}  \ottsym{:}  \ottnt{A}$,
        then $\Sigma  \ottsym{;}  \Gamma_{{\mathrm{1}}}  \ottsym{,}  \Gamma_{{\mathrm{2}}}  \ottsym{,}  \Gamma_{{\mathrm{3}}}  \vdash  \ottnt{e}  \ottsym{:}  \ottnt{A}$.
 \end{enumerate}
\end{lemmap}
\begin{proof}
 Straightforward by mutual induction on the derivations.
\end{proof}

\begin{lemmap}{Weakening type names}{weakening-tyname}
 Suppose that $\Sigma \,  \subseteq  \, \Sigma'$.
 \begin{enumerate}
  \item If $ \Sigma   \vdash   \ottnt{B}  \prec^{ \Phi }  \ottnt{C} $, then $ \Sigma'   \vdash   \ottnt{B}  \prec^{ \Phi }  \ottnt{C} $.
  \item If $\Sigma  \vdash  \Gamma$, then $\Sigma'  \vdash  \Gamma$.
  \item If $\Sigma  \ottsym{;}  \Gamma  \vdash  \ottnt{B}  \ottsym{:}  \ottnt{K}$, then $\Sigma'  \ottsym{;}  \Gamma  \vdash  \ottnt{B}  \ottsym{:}  \ottnt{K}$.
  \item If $\Sigma  \ottsym{;}  \Gamma  \vdash  \ottnt{e}  \ottsym{:}  \ottnt{B}$, then $\Sigma'  \ottsym{;}  \Gamma  \vdash  \ottnt{e}  \ottsym{:}  \ottnt{B}$.
 \end{enumerate}
\end{lemmap}
\begin{proof}
 Straightforward by mutual induction on the derivations.
\end{proof}

\begin{lemma}{type-subst-qpoly}
 If $\mathbf{QPoly} \, \ottsym{(}  \ottnt{A}  \ottsym{)}$, then $\mathbf{QPoly} \, \ottsym{(}   \ottnt{A}    [  \ottnt{B}  /  \mathit{X}  ]    \ottsym{)}$.
\end{lemma}
\begin{proof}
 First, we show $ \ottnt{A}    [  \ottnt{B}  /  \mathit{X}  ]  $ is not a polymorphic type
 by case analysis on $\ottnt{A}$.
 \begin{caseanalysis}
  \case $\ottnt{A} \,  =  \, \star$, $\mathit{Y}$ (where $\mathit{X} \,  \not=  \, \mathit{Y}$), $\alpha$, $\iota$,
   $\ottnt{A'}  \rightarrow  \ottnt{B'}$, $ [  \rho  ] $, $ \langle  \rho  \rangle $, $ \cdot $, and
   $  \ell  \mathbin{:}  \ottnt{C}   ;  \rho $:
   Obvious.
  \case $\ottnt{A} \,  =  \, \mathit{X}$: Since $\mathbf{QPoly} \, \ottsym{(}  \ottnt{A}  \ottsym{)}$, $\ottnt{A}$ must contain the dynamic
   type; thus, contradictory.
  \case $\ottnt{A} \,  =  \,  \text{\unboldmath$\forall\!$}  \,  \mathit{Y}  \mathord{:}  \ottnt{K}   \ottsym{.} \, \ottnt{C}$: Contradictory with $\mathbf{QPoly} \, \ottsym{(}  \ottnt{A}  \ottsym{)}$.
 \end{caseanalysis}

 Thus, it suffices to show that $ \ottnt{A}    [  \ottnt{B}  /  \mathit{X}  ]  $ contains the dynamic type, which is
 obvious since $\ottnt{A}$ contains the dynamic type (from $\mathbf{QPoly} \, \ottsym{(}  \ottnt{A}  \ottsym{)}$) and type
 substitution preserves that property.
\end{proof}

\begin{lemma}{type-subst-row-match}
 If $\rho_{{\mathrm{1}}} \,  \triangleright _{ \ell }  \, \ottnt{A}  \ottsym{,}  \rho_{{\mathrm{2}}}$, then $ \rho_{{\mathrm{1}}}    [  \ottnt{B}  /  \mathit{X}  ]   \,  \triangleright _{ \ell }  \,  \ottnt{A}    [  \ottnt{B}  /  \mathit{X}  ]    \ottsym{,}   \rho_{{\mathrm{2}}}    [  \ottnt{B}  /  \mathit{X}  ]  $.
\end{lemma}
\begin{proof}
 By induction on $\rho_{{\mathrm{1}}}$.
 \begin{caseanalysis}
  \case $\rho_{{\mathrm{1}}} \,  =  \,   \ell'  \mathbin{:}  \ottnt{C}   ;  \rho'_{{\mathrm{1}}} $:
   If $\ell' \,  =  \, \ell$, then $\ottnt{A} \,  =  \, \ottnt{C}$ and $\rho_{{\mathrm{2}}} \,  =  \, \rho'_{{\mathrm{1}}}$, and, therefore,
   the statement holds obviously.

   Otherwise, if $\ell' \,  \not=  \, \ell$, then
   we have $\rho'_{{\mathrm{1}}} \,  \triangleright _{ \ell }  \, \ottnt{A}  \ottsym{,}  \rho'_{{\mathrm{2}}}$ and
   $\rho_{{\mathrm{2}}} \,  =  \,   \ell'  \mathbin{:}  \ottnt{C}   ;  \rho'_{{\mathrm{2}}} $.
   By the IH, $ \rho'_{{\mathrm{1}}}    [  \ottnt{B}  /  \mathit{X}  ]   \,  \triangleright _{ \ell }  \,  \ottnt{A}    [  \ottnt{B}  /  \mathit{X}  ]    \ottsym{,}   \rho'_{{\mathrm{2}}}    [  \ottnt{B}  /  \mathit{X}  ]  $.
   Thus, $  \ell'  \mathbin{:}   \ottnt{C}    [  \ottnt{B}  /  \mathit{X}  ]     ;   \rho'_{{\mathrm{1}}}    [  \ottnt{B}  /  \mathit{X}  ]    \,  \triangleright _{ \ell }  \,  \ottnt{A}    [  \ottnt{B}  /  \mathit{X}  ]    \ottsym{,}    \ell'  \mathbin{:}   \ottnt{C}    [  \ottnt{B}  /  \mathit{X}  ]     ;   \rho'_{{\mathrm{2}}}    [  \ottnt{B}  /  \mathit{X}  ]   $,
   which is what we have to prove.

  \case $\rho_{{\mathrm{1}}} \,  =  \, \star$: Obvious.
 \end{caseanalysis}
\end{proof}

\begin{lemmap}{Type substitution preserves consistency}{type-subst-consistency}
 If $\ottnt{A}  \simeq  \ottnt{B}$, then $ \ottnt{A}    [  \ottnt{C}  /  \mathit{X}  ]    \simeq   \ottnt{B}    [  \ottnt{C}  /  \mathit{X}  ]  $.
\end{lemmap}
\begin{proof}
 By induction on the derivation of $\ottnt{A}  \simeq  \ottnt{B}$.
 We mention only the interesting cases below.
 \begin{caseanalysis}
  \case \CE{PolyL}:
   We have $ \text{\unboldmath$\forall\!$}  \,  \mathit{Y}  \mathord{:}  \ottnt{K}   \ottsym{.} \, \ottnt{A'}  \simeq  \ottnt{B}$ and, by inversion,
   $\mathbf{QPoly} \, \ottsym{(}  \ottnt{B}  \ottsym{)}$ and $\mathit{Y} \,  \not\in  \,  \mathit{ftv}  (  \ottnt{B}  ) $ and $\ottnt{A'}  \simeq  \ottnt{B}$.
   Without loss of generality, we can suppose that
   $\mathit{Y} \,  \not\in  \,  \mathit{ftv}  (  \ottnt{C}  ) $.
   Thus, $\mathit{Y} \,  \not\in  \,  \mathit{ftv}  (   \ottnt{B}    [  \ottnt{C}  /  \mathit{X}  ]    ) $.
   By the IH, $ \ottnt{A'}    [  \ottnt{C}  /  \mathit{X}  ]    \simeq   \ottnt{B}    [  \ottnt{C}  /  \mathit{X}  ]  $.
   By \reflem{type-subst-qpoly}, $\mathbf{QPoly} \, \ottsym{(}   \ottnt{B}    [  \ottnt{C}  /  \mathit{X}  ]    \ottsym{)}$.
   Thus, by \CE{PolyL}, $  \text{\unboldmath$\forall\!$}  \,  \mathit{Y}  \mathord{:}  \ottnt{K}   \ottsym{.} \, \ottnt{A'}    [  \ottnt{C}  /  \mathit{X}  ]    \simeq   \ottnt{B}    [  \ottnt{C}  /  \mathit{X}  ]  $
  \case \CE{PolyR}: Similar to the case for \CE{PolyL}.
  \case \CE{ConsL}:
   We have $  \ell  \mathbin{:}  \ottnt{A'}   ;  \rho_{{\mathrm{1}}}   \simeq  \ottnt{B}$ and, by inversion,
   $\ottnt{B} \,  \triangleright _{ \ell }  \, \ottnt{B'}  \ottsym{,}  \rho_{{\mathrm{2}}}$ and $\ottnt{A'}  \simeq  \ottnt{B'}$ and $\rho_{{\mathrm{1}}}  \simeq  \rho_{{\mathrm{2}}}$.
   By the IHs, $ \ottnt{A'}    [  \ottnt{C}  /  \mathit{X}  ]    \simeq   \ottnt{B'}    [  \ottnt{C}  /  \mathit{X}  ]  $ and $ \rho_{{\mathrm{1}}}    [  \ottnt{C}  /  \mathit{X}  ]    \simeq   \rho_{{\mathrm{2}}}    [  \ottnt{C}  /  \mathit{X}  ]  $.
   By \reflem{type-subst-row-match},
   $ \ottnt{B}    [  \ottnt{C}  /  \mathit{X}  ]   \,  \triangleright _{ \ell }  \,  \ottnt{B'}    [  \ottnt{C}  /  \mathit{X}  ]    \ottsym{,}   \rho_{{\mathrm{2}}}    [  \ottnt{C}  /  \mathit{X}  ]  $.
   Thus, by \CE{ConsL}, $  \ell  \mathbin{:}   \ottnt{A'}    [  \ottnt{C}  /  \mathit{X}  ]     ;   \rho_{{\mathrm{1}}}    [  \ottnt{C}  /  \mathit{X}  ]     \simeq   \ottnt{B}    [  \ottnt{C}  /  \mathit{X}  ]  $.

  \case \CE{ConsR}: Similar to the case for \CE{ConsL}.
 \end{caseanalysis}
\end{proof}

\begin{lemmap}{Type substitution}{type-subst}
 Suppose that $\Sigma  \ottsym{;}  \Gamma_{{\mathrm{1}}}  \vdash  \ottnt{A}  \ottsym{:}  \ottnt{K}$.
 \begin{enumerate}
  \item \label{lem:type-subst:typctx}
        If $\Sigma  \vdash  \Gamma_{{\mathrm{1}}}  \ottsym{,}   \mathit{X}  \mathord{:}  \ottnt{K}   \ottsym{,}  \Gamma_{{\mathrm{2}}}$,
        then $\Sigma  \vdash  \Gamma_{{\mathrm{1}}}  \ottsym{,}  \Gamma_{{\mathrm{2}}} \,  [  \ottnt{A}  /  \mathit{X}  ] $.
  \item \label{lem:type-subst:type}
        If $\Sigma  \ottsym{;}  \Gamma_{{\mathrm{1}}}  \ottsym{,}   \mathit{X}  \mathord{:}  \ottnt{K}   \ottsym{,}  \Gamma_{{\mathrm{2}}}  \vdash  \ottnt{B}  \ottsym{:}  \ottnt{K'}$,
        then $\Sigma  \ottsym{;}  \Gamma_{{\mathrm{1}}}  \ottsym{,}  \Gamma_{{\mathrm{2}}} \,  [  \ottnt{A}  /  \mathit{X}  ]   \vdash   \ottnt{B}    [  \ottnt{A}  /  \mathit{X}  ]    \ottsym{:}  \ottnt{K'}$.
  \item \label{lem:type-subst:term}
        If $\Sigma  \ottsym{;}  \Gamma_{{\mathrm{1}}}  \ottsym{,}   \mathit{X}  \mathord{:}  \ottnt{K}   \ottsym{,}  \Gamma_{{\mathrm{2}}}  \vdash  \ottnt{e}  \ottsym{:}  \ottnt{B}$,
        then $\Sigma  \ottsym{;}  \Gamma_{{\mathrm{1}}}  \ottsym{,}  \Gamma_{{\mathrm{2}}} \,  [  \ottnt{A}  /  \mathit{X}  ]   \vdash   \ottnt{e}    [  \ottnt{A}  /  \mathit{X}  ]    \ottsym{:}   \ottnt{B}    [  \ottnt{A}  /  \mathit{X}  ]  $.
 \end{enumerate}
\end{lemmap}
\begin{proof}
 Straightforward by mutual induction on the derivations.
 Only the interesting case is \WF{TyVar}.
 Suppose we have $\Sigma  \ottsym{;}  \Gamma_{{\mathrm{1}}}  \ottsym{,}   \mathit{X}  \mathord{:}  \ottnt{K}   \ottsym{,}  \Gamma_{{\mathrm{2}}}  \vdash  \mathit{Y}  \ottsym{:}  \ottnt{K'}$.
 By inversion, $\Sigma  \vdash  \Gamma_{{\mathrm{1}}}  \ottsym{,}   \mathit{X}  \mathord{:}  \ottnt{K}   \ottsym{,}  \Gamma_{{\mathrm{2}}}$ and $ \mathit{Y}  \mathord{:}  \ottnt{K'}  \,  \in  \, \Gamma_{{\mathrm{1}}}  \ottsym{,}   \mathit{X}  \mathord{:}  \ottnt{K}   \ottsym{,}  \Gamma_{{\mathrm{2}}}$.
 By the IH, $\Sigma  \vdash  \Gamma_{{\mathrm{1}}}  \ottsym{,}  \Gamma_{{\mathrm{2}}} \,  [  \ottnt{A}  /  \mathit{X}  ] $.
 If $\mathit{X} \,  \not=  \, \mathit{Y}$, then $ \mathit{Y}  \mathord{:}  \ottnt{K'}  \,  \in  \, \Gamma_{{\mathrm{1}}}  \ottsym{,}  \Gamma_{{\mathrm{2}}} \,  [  \ottnt{A}  /  \mathit{X}  ] $ and, therefore, by \WF{TyVar},
 $\Sigma  \ottsym{;}  \Gamma_{{\mathrm{1}}}  \ottsym{,}  \Gamma_{{\mathrm{2}}} \,  [  \ottnt{A}  /  \mathit{X}  ]   \vdash  \mathit{Y}  \ottsym{:}  \ottnt{K'}$.
 Otherwise, if $\mathit{X} \,  =  \, \mathit{Y}$, then we have to show
 $\Sigma  \ottsym{;}  \Gamma_{{\mathrm{1}}}  \ottsym{,}  \Gamma_{{\mathrm{2}}} \,  [  \ottnt{A}  /  \mathit{X}  ]   \vdash  \ottnt{A}  \ottsym{:}  \ottnt{K}$.
 Since $\Sigma  \ottsym{;}  \Gamma_{{\mathrm{1}}}  \vdash  \ottnt{A}  \ottsym{:}  \ottnt{K}$ and $\Sigma  \vdash  \Gamma_{{\mathrm{1}}}  \ottsym{,}  \Gamma_{{\mathrm{2}}} \,  [  \ottnt{A}  /  \mathit{X}  ] $,
 we have $\Sigma  \ottsym{;}  \Gamma_{{\mathrm{1}}}  \ottsym{,}  \Gamma_{{\mathrm{2}}} \,  [  \ottnt{A}  /  \mathit{X}  ]   \vdash  \ottnt{A}  \ottsym{:}  \ottnt{K}$
 by \reflem{weakening} (\ref{lem:weakening:type}).

 Note that the case for \T{Cast} uses \reflem{type-subst-consistency} and that
 the case for \T{Conv} depends on the fact that $\ottnt{e}$ and $\ottnt{B}$ are closed.
\end{proof}

\begin{lemmap}{Type substitution on convertibility}{convert-type-subst}
 Suppose that $\alpha$ does not occur in $\ottnt{A}$.
 \begin{enumerate}
  \item $ \Sigma  \ottsym{,}   \alpha  \mathord{:}  \ottnt{K}   \ottsym{:=}  \ottnt{B}   \vdash    \ottnt{A}    [  \alpha  /  \mathit{X}  ]    \prec^{ \ottsym{+}  \alpha }   \ottnt{A}    [  \ottnt{B}  /  \mathit{X}  ]   $.
  \item $ \Sigma  \ottsym{,}   \alpha  \mathord{:}  \ottnt{K}   \ottsym{:=}  \ottnt{B}   \vdash    \ottnt{A}    [  \ottnt{B}  /  \mathit{X}  ]    \prec^{ \ottsym{-}  \alpha }   \ottnt{A}    [  \alpha  /  \mathit{X}  ]   $.
 \end{enumerate}
\end{lemmap}
\begin{proof}
 Let $\Sigma' \,  =  \, \Sigma  \ottsym{,}   \alpha  \mathord{:}  \ottnt{K}   \ottsym{:=}  \ottnt{B}$.
 By induction on $\ottnt{A}$.
 \begin{caseanalysis}
  \case $\ottnt{A} \,  =  \, \mathit{X}$:
   We have $ \ottnt{A}    [  \alpha  /  \mathit{X}  ]   \,  =  \, \alpha$ and $ \ottnt{A}    [  \ottnt{B}  /  \mathit{X}  ]   \,  =  \, \ottnt{B}$.

   First, we have to show $ \Sigma'   \vdash   \alpha  \prec^{ \ottsym{+}  \alpha }  \ottnt{B} $,
   which is shwon by \Cv{Reveal}.

   Next, we have to show $ \Sigma'   \vdash   \ottnt{B}  \prec^{ \ottsym{-}  \alpha }  \alpha $,
   which is shown by \Cv{Conceal}.

  \case $\ottnt{A} \,  =  \, \mathit{Y}$ where $\mathit{X} \,  \not=  \, \mathit{Y}$: By \Cv{TyVar}.
  \case $\ottnt{A} \,  =  \, \alpha$:
   Contradictory with the assumption that $\alpha$ does not occur in $\ottnt{A}$.
  \case $\ottnt{A} \,  =  \, \alpha'$ where $\alpha \,  \not=  \, \alpha'$: By \Cv{TyName}.
  \case $\ottnt{A} \,  =  \, \star$: By \Cv{Dyn}.
  \case $\ottnt{A} \,  =  \, \iota$: By \Cv{Base}.
  \case $\ottnt{A} \,  =  \, \ottnt{A_{{\mathrm{1}}}}  \rightarrow  \ottnt{A_{{\mathrm{2}}}}$:
   By the IHs, we have
   \begin{itemize}
    \item $ \Sigma'   \vdash    \ottnt{A_{{\mathrm{1}}}}    [  \alpha  /  \mathit{X}  ]    \prec^{ \ottsym{+}  \alpha }   \ottnt{A_{{\mathrm{1}}}}    [  \ottnt{B}  /  \mathit{X}  ]   $,
    \item $ \Sigma'   \vdash    \ottnt{A_{{\mathrm{2}}}}    [  \alpha  /  \mathit{X}  ]    \prec^{ \ottsym{+}  \alpha }   \ottnt{A_{{\mathrm{2}}}}    [  \ottnt{B}  /  \mathit{X}  ]   $,
    \item $ \Sigma'   \vdash    \ottnt{A_{{\mathrm{1}}}}    [  \ottnt{B}  /  \mathit{X}  ]    \prec^{ \ottsym{-}  \alpha }   \ottnt{A_{{\mathrm{1}}}}    [  \alpha  /  \mathit{X}  ]   $, and
    \item $ \Sigma'   \vdash    \ottnt{A_{{\mathrm{2}}}}    [  \ottnt{B}  /  \mathit{X}  ]    \prec^{ \ottsym{-}  \alpha }   \ottnt{A_{{\mathrm{2}}}}    [  \alpha  /  \mathit{X}  ]   $.
   \end{itemize}
   By \Cv{Fun},
   $ \Sigma'   \vdash     \ottnt{A_{{\mathrm{1}}}}    [  \alpha  /  \mathit{X}  ]    \rightarrow  \ottnt{A_{{\mathrm{2}}}}    [  \alpha  /  \mathit{X}  ]    \prec^{ \ottsym{+}  \alpha }    \ottnt{A_{{\mathrm{1}}}}    [  \ottnt{B}  /  \mathit{X}  ]    \rightarrow  \ottnt{A_{{\mathrm{2}}}}    [  \ottnt{B}  /  \mathit{X}  ]   $ and
   $ \Sigma'   \vdash     \ottnt{A_{{\mathrm{1}}}}    [  \ottnt{B}  /  \mathit{X}  ]    \rightarrow  \ottnt{A_{{\mathrm{2}}}}    [  \ottnt{B}  /  \mathit{X}  ]    \prec^{ \ottsym{-}  \alpha }    \ottnt{A_{{\mathrm{1}}}}    [  \alpha  /  \mathit{X}  ]    \rightarrow  \ottnt{A_{{\mathrm{2}}}}    [  \alpha  /  \mathit{X}  ]   $.

  \case $\ottnt{A} \,  =  \,  \text{\unboldmath$\forall\!$}  \,  \mathit{X'}  \mathord{:}  \ottnt{K}   \ottsym{.} \, \ottnt{A'}$: By the IH and \Cv{Poly}.
  \case $\ottnt{A} \,  =  \,  [  \rho  ] $: By the IH and \Cv{Record}.
  \case $\ottnt{A} \,  =  \,  \langle  \rho  \rangle $: By the IH and \Cv{Variant}.
  \case $\ottnt{A} \,  =  \,  \cdot $: By \Cv{REmp}.
  \case $\ottnt{A} \,  =  \,   \ell  \mathbin{:}  \ottnt{A'}   ;  \rho $: By the IHs and \Cv{Cons}.
 \end{caseanalysis}
\end{proof}

\begin{lemma}{value-subst-wf}
 \begin{enumerate}
  \item \label{lem:value-subst-wf:typctx}
        If $\Sigma  \vdash  \Gamma_{{\mathrm{1}}}  \ottsym{,}   \mathit{x}  \mathord{:}  \ottnt{A}   \ottsym{,}  \Gamma_{{\mathrm{2}}}$, then $\Sigma  \vdash  \Gamma_{{\mathrm{1}}}  \ottsym{,}  \Gamma_{{\mathrm{2}}}$.
  \item \label{lem:typeue-subst-wf:type}
        If $\Sigma  \ottsym{;}  \Gamma_{{\mathrm{1}}}  \ottsym{,}   \mathit{x}  \mathord{:}  \ottnt{A}   \ottsym{,}  \Gamma_{{\mathrm{2}}}  \vdash  \ottnt{B}  \ottsym{:}  \ottnt{K}$, then $\Sigma  \ottsym{;}  \Gamma_{{\mathrm{1}}}  \ottsym{,}  \Gamma_{{\mathrm{2}}}  \vdash  \ottnt{B}  \ottsym{:}  \ottnt{K}$.
 \end{enumerate}
\end{lemma}
\begin{proof}
 Straightforward by mutual induction on the derivations.
\end{proof}

\begin{lemmap}{Value substitution}{value-subst}
 If $\Sigma  \ottsym{;}  \Gamma_{{\mathrm{1}}}  \vdash  \ottnt{v}  \ottsym{:}  \ottnt{A}$ and $\Sigma  \ottsym{;}  \Gamma_{{\mathrm{1}}}  \ottsym{,}   \mathit{x}  \mathord{:}  \ottnt{A}   \ottsym{,}  \Gamma_{{\mathrm{2}}}  \vdash  \ottnt{e}  \ottsym{:}  \ottnt{B}$,
 then $\Sigma  \ottsym{;}  \Gamma_{{\mathrm{1}}}  \ottsym{,}  \Gamma_{{\mathrm{2}}}  \vdash   \ottnt{e}    [  \ottnt{v}  \ottsym{/}  \mathit{x}  ]    \ottsym{:}  \ottnt{B}$.
\end{lemmap}
\begin{proof}
 By mutual induction on the derivations.
 The only interesting case is \T{Var}.

 Suppose that $\Sigma  \ottsym{;}  \Gamma_{{\mathrm{1}}}  \ottsym{,}   \mathit{x}  \mathord{:}  \ottnt{A}   \ottsym{,}  \Gamma_{{\mathrm{2}}}  \vdash  \mathit{y}  \ottsym{:}  \ottnt{B}$.
 By inversion, $\Sigma  \vdash  \Gamma_{{\mathrm{1}}}  \ottsym{,}   \mathit{x}  \mathord{:}  \ottnt{A}   \ottsym{,}  \Gamma_{{\mathrm{2}}}$ and $ \mathit{y}  \mathord{:}  \ottnt{B}  \,  \in  \, \Gamma_{{\mathrm{1}}}  \ottsym{,}   \mathit{x}  \mathord{:}  \ottnt{A}   \ottsym{,}  \Gamma_{{\mathrm{2}}}$.
 By \reflem{value-subst-wf}, $\Sigma  \vdash  \Gamma_{{\mathrm{1}}}  \ottsym{,}  \Gamma_{{\mathrm{2}}}$.
 If $\mathit{x} \,  \not=  \, \mathit{y}$, then $ \mathit{y}  \mathord{:}  \ottnt{B}  \,  \in  \, \Gamma_{{\mathrm{1}}}  \ottsym{,}  \Gamma_{{\mathrm{2}}}$.
 Thus, by \T{Var}, $\Sigma  \ottsym{;}  \Gamma_{{\mathrm{1}}}  \ottsym{,}  \Gamma_{{\mathrm{2}}}  \vdash  \mathit{y}  \ottsym{:}  \ottnt{B}$.
 Since $ \mathit{y}    [  \ottnt{v}  \ottsym{/}  \mathit{x}  ]   \,  =  \, \mathit{y}$, we finish.
 Otherwise, if $\mathit{x} \,  =  \, \mathit{y}$, then we have to show that
 $\Sigma  \ottsym{;}  \Gamma_{{\mathrm{1}}}  \ottsym{,}  \Gamma_{{\mathrm{2}}}  \vdash  \ottnt{v}  \ottsym{:}  \ottnt{A}$
 (note that $ \mathit{y}    [  \ottnt{v}  \ottsym{/}  \mathit{x}  ]   \,  =  \, \ottnt{v}$ and that $\ottnt{A} \,  =  \, \ottnt{B}$ since $ \mathit{y}  \mathord{:}  \ottnt{B}  \,  \in  \, \Gamma_{{\mathrm{1}}}  \ottsym{,}   \mathit{x}  \mathord{:}  \ottnt{A}   \ottsym{,}  \Gamma_{{\mathrm{2}}}$).
 Since $\Sigma  \ottsym{;}  \Gamma_{{\mathrm{1}}}  \vdash  \ottnt{v}  \ottsym{:}  \ottnt{A}$ and $\Sigma  \vdash  \Gamma_{{\mathrm{1}}}  \ottsym{,}  \Gamma_{{\mathrm{2}}}$,
 we have $\Sigma  \ottsym{;}  \Gamma_{{\mathrm{1}}}  \ottsym{,}  \Gamma_{{\mathrm{2}}}  \vdash  \ottnt{v}  \ottsym{:}  \ottnt{A}$ by \reflem{weakening} (\ref{lem:weakening:term}).

 The cases for \T{Const}, \T{TApp}, \T{REmp}, \T{VInj}, \T{VLift}, \T{Cast}, and
 \T{Conv} also use \reflem{value-subst-wf}.
\end{proof}

\begin{lemmap}{Canonical forms}{canonical-forms}
 Suppose that $\Sigma  \ottsym{;}   \emptyset   \vdash  \ottnt{v}  \ottsym{:}  \ottnt{A}$.
 \begin{enumerate}
  \item If $\ottnt{A} \,  =  \, \iota$, then $\ottnt{v} \,  =  \, \kappa$ for some $\kappa$.
  \item If $\ottnt{A} \,  =  \, \ottnt{B}  \rightarrow  \ottnt{C}$, then $\ottnt{v} \,  =  \,  \lambda\!  \,  \mathit{x}  \mathord{:}  \ottnt{B}   \ottsym{.}  \ottnt{e}$ for some $\mathit{x}$ and $\ottnt{e}$,
        or $\ottnt{v} \,  =  \, \kappa$ for some $\kappa$ such that $ \mathit{ty}  (  \kappa  )  \,  =  \, \ottnt{B}  \rightarrow  \ottnt{C}$.
  \item If $\ottnt{A} \,  =  \,  \text{\unboldmath$\forall\!$}  \,  \mathit{X}  \mathord{:}  \ottnt{K}   \ottsym{.} \, \ottnt{B}$, then $\ottnt{v} \,  =  \,   \Lambda\!  \,  \mathit{X}  \mathord{:}  \ottnt{K}   \ottsym{.}   \ottnt{e}  ::  \ottnt{B} $ for some $\ottnt{e}$.
  \item If $\ottnt{A} \,  =  \,  [   \cdot   ] $, then $\ottnt{v} \,  =  \, \ottsym{\{}  \ottsym{\}}$.
  \item If $\ottnt{A} \,  =  \,  [    \ell  \mathbin{:}  \ottnt{B}   ;  \rho   ] $,
        then $\ottnt{v} \,  =  \, \ottsym{\{}  \ell  \ottsym{=}  \ottnt{v_{{\mathrm{1}}}}  \ottsym{;}  \ottnt{v_{{\mathrm{2}}}}  \ottsym{\}}$ for some $\ottnt{v_{{\mathrm{1}}}}$ and $\ottnt{v_{{\mathrm{2}}}}$.
  \item If $\ottnt{A} \,  =  \,  \langle    \ell  \mathbin{:}  \ottnt{B}   ;  \rho   \rangle $,
        then $\ottnt{v} \,  =  \, \ell \, \ottnt{v'}$ or $\ottnt{v} \,  =  \,  \variantlift{ \ell }{ \ottnt{B} }{ \ottnt{v'} } $ for some $\ottnt{v'}$.

  \item If $\ottnt{A} \,  =  \, \star$,
        then $\ottnt{v} \,  =  \, \ottnt{v'}  \ottsym{:}  \mathit{G} \,  \stackrel{ \ottnt{p} }{\Rightarrow}  \star $ for some $\ottnt{v'}$, $\mathit{G}$, and $\ottnt{p}$.
  \item If $\ottnt{A} \,  =  \,  [  \star  ] $,
        then $\ottnt{v} \,  =  \, \ottnt{v'}  \ottsym{:}   [  \gamma  ]  \,  \stackrel{ \ottnt{p} }{\Rightarrow}   [  \star  ]  $ for some
        $\ottnt{v'}$, $\gamma$, and $\ottnt{p}$.
  \item If $\ottnt{A} \,  =  \,  \langle  \star  \rangle $,
        then $\ottnt{v} \,  =  \, \ottnt{v'}  \ottsym{:}   \langle  \gamma  \rangle  \,  \stackrel{ \ottnt{p} }{\Rightarrow}   \langle  \star  \rangle  $ for some
        $\ottnt{v'}$, $\gamma$, and $\ottnt{p}$.

  \item If $\ottnt{A} \,  =  \, \alpha$,
        then $\ottnt{v} \,  =  \, \ottnt{v'}  \ottsym{:}  \ottnt{B} \,  \stackrel{ \ottsym{-}  \alpha }{\Rightarrow}  \alpha $ for some $\ottnt{v'}$ and $\ottnt{B}$.
  \item If $\ottnt{A} \,  =  \,  [  \alpha  ] $,
        then $\ottnt{v} \,  =  \, \ottnt{v'}  \ottsym{:}   [  \rho  ]  \,  \stackrel{ \ottsym{-}  \alpha }{\Rightarrow}   [  \alpha  ]  $ for some $\ottnt{v'}$ and $\rho$.
  \item If $\ottnt{A} \,  =  \,  \langle  \alpha  \rangle $,
        then $\ottnt{v} \,  =  \, \ottnt{v'}  \ottsym{:}   \langle  \rho  \rangle  \,  \stackrel{ \ottsym{-}  \alpha }{\Rightarrow}   \langle  \alpha  \rangle  $ for some $\ottnt{v'}$ and $\rho$.
 \end{enumerate}
\end{lemmap}
\begin{proof}
 By case analysis on the typing rule applied to derive $\Sigma  \ottsym{;}   \emptyset   \vdash  \ottnt{v}  \ottsym{:}  \ottnt{A}$.
 \begin{caseanalysis}
  \case \T{Var}, \T{App}, \T{TApp}, \T{RLet}, \T{VCase}, and \T{Blame}: Contradictory.
  \case \T{Const}, \T{Lam}, \T{TLam}, \T{REmp}, \T{RExt}, \T{VInj}, \T{VLift}: Obvious.
  \case \T{Cast}:
   We have $\Sigma  \ottsym{;}   \emptyset   \vdash  \ottnt{e}  \ottsym{:}  \ottnt{B} \,  \stackrel{ \ottnt{p} }{\Rightarrow}  \ottnt{A}   \ottsym{:}  \ottnt{A}$
   for some $\ottnt{e}$, $\ottnt{B}$, and $\ottnt{p}$.
   By inversion, $\Sigma  \ottsym{;}   \emptyset   \vdash  \ottnt{A}  \ottsym{:}   \mathsf{T} $.
   We do case analysis on the rule applied last to derive $\Sigma  \ottsym{;}   \emptyset   \vdash  \ottnt{A}  \ottsym{:}   \mathsf{T} $.
   \begin{caseanalysis}
    \case \WF{TyVar}, \WF{REmp}, and \WF{Cons}: Contradictory.
    \case \WF{TyName}, \WF{Base}, \WF{Fun}, and \WF{Poly}:
     Contradictory because there are no values of the form $\ottnt{e}  \ottsym{:}  \ottnt{B} \,  \stackrel{ \ottnt{p} }{\Rightarrow}  \ottnt{A} $
     in these cases.
    \case \WF{Dyn}, \WF{Record}, and \WF{Variant}:
     Obvious because of the definition of values.
   \end{caseanalysis}

  \case \T{Conv}:
   We have $\Sigma  \ottsym{;}   \emptyset   \vdash  \ottnt{e}  \ottsym{:}  \ottnt{B} \,  \stackrel{ \Phi }{\Rightarrow}  \ottnt{A}   \ottsym{:}  \ottnt{A}$
   for some $\ottnt{e}$, $\ottnt{B}$, and $\Phi$.
   By inversion, $\Sigma  \ottsym{;}   \emptyset   \vdash  \ottnt{A}  \ottsym{:}   \mathsf{T} $.
   We do case analysis on the rule applied last to derive $\Sigma  \ottsym{;}   \emptyset   \vdash  \ottnt{A}  \ottsym{:}   \mathsf{T} $.
   \begin{caseanalysis}
    \case \WF{TyVar}, \WF{REmp}, and \WF{Cons}: Contradictory.
    \case \WF{Dyn}, \WF{Base}, \WF{Fun}, and \WF{Poly}:
     Contradictory because there are no values of the form $\ottnt{e}  \ottsym{:}  \ottnt{B} \,  \stackrel{ \Phi }{\Rightarrow}  \ottnt{A} $
     in these cases.
    \case \WF{TyName}, \WF{Record}, and \WF{Variant}:
     Obvious because of the definition of values.
   \end{caseanalysis}
 \end{caseanalysis}
\end{proof}

\begin{lemma}{canonical-forms-variant-remp}
 If $\Sigma  \ottsym{;}   \emptyset   \vdash  \ottnt{v}  \ottsym{:}   \langle   \cdot   \rangle $, contradictory.
\end{lemma}
\begin{proof}
 Straightforward by case analysis on the rule applied last to derive
 $\Sigma  \ottsym{;}   \emptyset   \vdash  \ottnt{v}  \ottsym{:}   \langle   \cdot   \rangle $.
\end{proof}

\begin{lemmap}{Value inversion: constants}{value-inversion-constant}
 If $\Sigma  \ottsym{;}   \emptyset   \vdash  \kappa  \ottsym{:}  \ottnt{A}$, then $\ottnt{A} \,  =  \,  \mathit{ty}  (  \kappa  ) $.
\end{lemmap}
\begin{proof}
 Straightforward by case analysis on the derivation of $\Sigma  \ottsym{;}   \emptyset   \vdash  \kappa  \ottsym{:}  \ottnt{A}$.
\end{proof}

\begin{lemmap}{Value inversion: constants}{value-inversion-lambda}
 If $\Sigma  \ottsym{;}   \emptyset   \vdash   \lambda\!  \,  \mathit{x}  \mathord{:}  \ottnt{A}   \ottsym{.}  \ottnt{e}  \ottsym{:}  \ottnt{A'}  \rightarrow  \ottnt{B}$, then $\ottnt{A} \,  =  \, \ottnt{A'}$ and $\Sigma  \ottsym{;}   \mathit{x}  \mathord{:}  \ottnt{A}   \vdash  \ottnt{e}  \ottsym{:}  \ottnt{B}$.
\end{lemmap}
\begin{proof}
 Straightforward by case analysis on the derivation of $\Sigma  \ottsym{;}   \emptyset   \vdash   \lambda\!  \,  \mathit{x}  \mathord{:}  \ottnt{A}   \ottsym{.}  \ottnt{e}  \ottsym{:}  \ottnt{A'}  \rightarrow  \ottnt{B}$.
\end{proof}

\begin{lemmap}{Value inversion: constants}{value-inversion-typelambda}
 If $\Sigma  \ottsym{;}   \emptyset   \vdash    \Lambda\!  \,  \mathit{X}  \mathord{:}  \ottnt{K}   \ottsym{.}   \ottnt{e}  ::  \ottnt{A}   \ottsym{:}   \text{\unboldmath$\forall\!$}  \,  \mathit{X'}  \mathord{:}  \ottnt{K'}   \ottsym{.} \, \ottnt{A'}$, then
 $\mathit{X} \,  =  \, \mathit{X'}$ and $\ottnt{K} \,  =  \, \ottnt{K'}$ and $\ottnt{A} \,  =  \, \ottnt{A'}$ and $\Sigma  \ottsym{;}   \mathit{X}  \mathord{:}  \ottnt{K}   \vdash  \ottnt{e}  \ottsym{:}  \ottnt{A}$.
\end{lemmap}
\begin{proof}
 Straightforward by case analysis on the derivation of $\Sigma  \ottsym{;}   \emptyset   \vdash    \Lambda\!  \,  \mathit{X}  \mathord{:}  \ottnt{K}   \ottsym{.}   \ottnt{e}  ::  \ottnt{A}   \ottsym{:}   \text{\unboldmath$\forall\!$}  \,  \mathit{X'}  \mathord{:}  \ottnt{K'}   \ottsym{.} \, \ottnt{A'}$.
\end{proof}

\begin{lemmap}{Value inversion: record extensions}{value-inversion-record-cons}
 If $\Sigma  \ottsym{;}   \emptyset   \vdash  \ottsym{\{}  \ell  \ottsym{=}  \ottnt{v_{{\mathrm{1}}}}  \ottsym{;}  \ottnt{v_{{\mathrm{2}}}}  \ottsym{\}}  \ottsym{:}   [  \rho  ] $,
 there exist some $\ottnt{A}$ and $\rho'$ such that
 $\rho \,  =  \,  [    \ell  \mathbin{:}  \ottnt{A}   ;  \rho'   ] $ and
 $\Sigma  \ottsym{;}   \emptyset   \vdash  \ottnt{v_{{\mathrm{1}}}}  \ottsym{:}  \ottnt{A}$ and
 $\Sigma  \ottsym{;}   \emptyset   \vdash  \ottnt{v_{{\mathrm{2}}}}  \ottsym{:}   [  \rho'  ] $.
\end{lemmap}
\begin{proof}
 Straightforward by case analysis on the derivation of $\Sigma  \ottsym{;}   \emptyset   \vdash  \ottsym{\{}  \ell  \ottsym{=}  \ottnt{v_{{\mathrm{1}}}}  \ottsym{;}  \ottnt{v_{{\mathrm{2}}}}  \ottsym{\}}  \ottsym{:}   [  \rho  ] $.
\end{proof}

\begin{lemmap}{Value inversion: variant injections}{value-inversion-variant-inj}
 If $\Sigma  \ottsym{;}   \emptyset   \vdash  \ell \, \ottnt{v}  \ottsym{:}   \langle    \ell  \mathbin{:}  \ottnt{A}   ;  \rho   \rangle $, then $\Sigma  \ottsym{;}   \emptyset   \vdash  \ottnt{v}  \ottsym{:}  \ottnt{A}$.
\end{lemmap}
\begin{proof}
 Straightforward by case analysis on the derivation of $\Sigma  \ottsym{;}   \emptyset   \vdash  \ell \, \ottnt{v}  \ottsym{:}   \langle    \ell  \mathbin{:}  \ottnt{A}   ;  \rho   \rangle $.
\end{proof}

\begin{lemmap}{Value inversion: variant lifts}{value-inversion-variant-lift}
 If $\Sigma  \ottsym{;}   \emptyset   \vdash   \variantlift{ \ell }{ \ottnt{A} }{ \ottnt{v} }   \ottsym{:}   \langle    \ell  \mathbin{:}  \ottnt{B}   ;  \rho   \rangle $, then $\Sigma  \ottsym{;}   \emptyset   \vdash  \ottnt{v}  \ottsym{:}   \langle  \rho  \rangle $ and $\ottnt{A} \,  =  \, \ottnt{B}$.
\end{lemmap}
\begin{proof}
 Straightforward by case analysis on the derivation of $\Sigma  \ottsym{;}   \emptyset   \vdash   \variantlift{ \ell }{ \ottnt{A} }{ \ottnt{v} }   \ottsym{:}   \langle    \ell  \mathbin{:}  \ottnt{A}   ;  \rho   \rangle $.
\end{proof}

\begin{lemmap}{Value inversion: casts}{value-inversion-cast}
 If $\Sigma  \ottsym{;}   \emptyset   \vdash  \ottnt{v}  \ottsym{:}  \ottnt{A} \,  \stackrel{ \ottnt{p} }{\Rightarrow}  \ottnt{B}   \ottsym{:}  \ottnt{B}$,
 then $\Sigma  \ottsym{;}   \emptyset   \vdash  \ottnt{v}  \ottsym{:}  \ottnt{A}$ and $\ottnt{A}  \simeq  \ottnt{B}$.
\end{lemmap}
\begin{proof}
 Straightforward by case analysis on the derivation of $\Sigma  \ottsym{;}   \emptyset   \vdash  \ottnt{v}  \ottsym{:}  \ottnt{A} \,  \stackrel{ \ottnt{p} }{\Rightarrow}  \ottnt{B}   \ottsym{:}  \ottnt{B}$.
\end{proof}

\begin{lemmap}{Value inversion: conversions}{value-inversion-conv}
 If $\Sigma  \ottsym{;}   \emptyset   \vdash  \ottnt{v}  \ottsym{:}  \ottnt{A} \,  \stackrel{ \ottsym{-}  \alpha }{\Rightarrow}  \alpha   \ottsym{:}  \alpha$, then $\Sigma  \ottsym{;}   \emptyset   \vdash  \ottnt{v}  \ottsym{:}  \ottnt{A}$ and $\Sigma  \ottsym{(}  \alpha  \ottsym{)} \,  =  \, \ottnt{A}$.
\end{lemmap}
\begin{proof}
 Straightforward by case analysis on the derivation of $\Sigma  \ottsym{;}   \emptyset   \vdash  \ottnt{v}  \ottsym{:}  \ottnt{A} \,  \stackrel{ \ottsym{-}  \alpha }{\Rightarrow}  \alpha   \ottsym{:}  \alpha$.
\end{proof}

\begin{lemmap}{Value inversion: conversions with records}{value-inversion-conv-record}
 If $\Sigma  \ottsym{;}   \emptyset   \vdash  \ottnt{v}  \ottsym{:}   [  \rho  ]  \,  \stackrel{ \ottsym{-}  \alpha }{\Rightarrow}   [  \alpha  ]    \ottsym{:}  \alpha$,
 then $\Sigma  \ottsym{;}   \emptyset   \vdash  \ottnt{v}  \ottsym{:}   [  \rho  ] $ and $\Sigma  \ottsym{(}  \alpha  \ottsym{)} \,  =  \, \rho$.
\end{lemmap}
\begin{proof}
 Straightforward by case analysis on the derivation of $\Sigma  \ottsym{;}   \emptyset   \vdash  \ottnt{v}  \ottsym{:}   [  \rho  ]  \,  \stackrel{ \ottsym{-}  \alpha }{\Rightarrow}   [  \alpha  ]    \ottsym{:}  \alpha$.
\end{proof}

\begin{lemmap}{Value inversion: conversions with variants}{value-inversion-conv-variant}
 If $\Sigma  \ottsym{;}   \emptyset   \vdash  \ottnt{v}  \ottsym{:}   \langle  \rho  \rangle  \,  \stackrel{ \ottsym{-}  \alpha }{\Rightarrow}   \langle  \alpha  \rangle    \ottsym{:}  \alpha$,
 then $\Sigma  \ottsym{;}   \emptyset   \vdash  \ottnt{v}  \ottsym{:}   \langle  \rho  \rangle $ and $\Sigma  \ottsym{(}  \alpha  \ottsym{)} \,  =  \, \rho$.
\end{lemmap}
\begin{proof}
 Straightforward by case analysis on the derivation of $\Sigma  \ottsym{;}   \emptyset   \vdash  \ottnt{v}  \ottsym{:}   \langle  \rho  \rangle  \,  \stackrel{ \ottsym{-}  \alpha }{\Rightarrow}   \langle  \alpha  \rangle    \ottsym{:}  \alpha$.
\end{proof}

\begin{lemma}{progress-aux}
 If $ \Sigma  \mid  \ottnt{e}   \longrightarrow   \Sigma'  \mid  \ottnt{e'} $ or $\ottnt{e} \,  =  \, \mathsf{blame} \, \ottnt{p}$,
 then $ \Sigma  \mid   \ottnt{E}  [  \ottnt{e}  ]    \longrightarrow   \Sigma'  \mid  \ottnt{e''} $ for some $\ottnt{e''}$.
\end{lemma}
\begin{proof}
 If $\ottnt{e} \,  =  \, \mathsf{blame} \, \ottnt{p}$, then we finish by \E{Blame}.
 If $ \Sigma  \mid  \ottnt{e}   \longrightarrow   \Sigma'  \mid  \ottnt{e'} $, we can prove the statement straightforwardly by
 case analysis on the evaluation rule applied to derive
 $ \Sigma  \mid  \ottnt{e}   \longrightarrow   \Sigma'  \mid  \ottnt{e'} $.
\end{proof}

\begin{lemmap}{Unique ground type}{ground-type-unique}
 If $\Sigma  \ottsym{;}   \emptyset   \vdash  \ottnt{A}  \ottsym{:}   \mathsf{T} $ and $\ottnt{A} \,  \not=  \, \star$ and $\ottnt{A}$ is not an universal
 type, then there exists an unique ground type $\mathit{G}$ such that
 $\ottnt{A}  \simeq  \mathit{G}$.
\end{lemmap}
\begin{proof}
 By case analysis on $\ottnt{A}$.
 \begin{caseanalysis}
  \case $\ottnt{A} \,  =  \, \mathit{X}$: Contradictory with $\Sigma  \ottsym{;}   \emptyset   \vdash  \ottnt{A}  \ottsym{:}   \mathsf{T} $
  \case $\ottnt{A} \,  =  \, \alpha$: Only ground type $\alpha$ is consistent with $\alpha$.
  \case $\ottnt{A} \,  =  \, \star$: Contradictory with $\ottnt{A} \,  \not=  \, \star$.
  \case $\ottnt{A} \,  =  \, \iota$: Only ground type $\iota$ is consistent with $\iota$.
  \case $\ottnt{A} \,  =  \, \ottnt{B}  \rightarrow  \ottnt{C}$: Only ground type $\star  \rightarrow  \star$ is consistent with $\ottnt{B}  \rightarrow  \ottnt{C}$.
  \case $\ottnt{A} \,  =  \,  \text{\unboldmath$\forall\!$}  \,  \mathit{X}  \mathord{:}  \ottnt{K}   \ottsym{.} \, \ottnt{B}$: Contradictory.
  \case $\ottnt{A} \,  =  \,  [  \rho  ] $: Only ground type $ [  \star  ] $ is consistent with $ [  \rho  ] $.
  \case $\ottnt{A} \,  =  \,  \langle  \rho  \rangle $: Only ground type $ \langle  \star  \rangle $ is consistent with $ \langle  \rho  \rangle $.
  \case $\ottnt{A} \,  =  \,  \cdot $: Contradictory with $\Sigma  \ottsym{;}   \emptyset   \vdash  \ottnt{A}  \ottsym{:}   \mathsf{T} $.
  \case $\ottnt{A} \,  =  \,   \ell  \mathbin{:}  \ottnt{B}   ;  \rho $: Contradictory with $\Sigma  \ottsym{;}   \emptyset   \vdash  \ottnt{A}  \ottsym{:}   \mathsf{T} $.
 \end{caseanalysis}
\end{proof}

\begin{lemma}{grow-is-defined}
 If $\Sigma  \ottsym{;}   \emptyset   \vdash  \rho  \ottsym{:}   \mathsf{R} $ and $\rho \,  \not=  \, \star$, then $ \mathit{grow}  (  \rho  ) $ is defined
 and $ \mathit{grow}  (  \rho  ) $ is a ground row type.
\end{lemma}
\begin{proof}
 Straightforward by case analysis on the derivation of $\Sigma  \ottsym{;}   \emptyset   \vdash  \rho  \ottsym{:}   \mathsf{R} $.
\end{proof}

\begin{lemma}{consistent-grow}
 If $ \mathit{grow}  (  \rho  ) $ is defined, then $\rho  \simeq   \mathit{grow}  (  \rho  ) $.
\end{lemma}
\begin{proof}
 Obvious by definition of $ \mathit{grow} $.
\end{proof}

\begin{lemma}{wf-grow}
 If $ \mathit{grow}  (  \rho  ) $ is defined and $\Sigma  \ottsym{;}  \Gamma  \vdash  \rho  \ottsym{:}  \ottnt{K}$,
 then $\Sigma  \ottsym{;}  \Gamma  \vdash   \mathit{grow}  (  \rho  )   \ottsym{:}  \ottnt{K}$.
\end{lemma}
\begin{proof}
 Obvious by definition of $ \mathit{grow} $.
\end{proof}

\begin{lemma}{typctx-type-wf}
 \begin{enumerate}
  \item If $\Sigma  \ottsym{;}  \Gamma  \vdash  \ottnt{A}  \ottsym{:}  \ottnt{K}$, then $\Sigma  \vdash  \Gamma$.
  \item If $\Sigma  \ottsym{;}  \Gamma  \vdash  \ottnt{e}  \ottsym{:}  \ottnt{A}$, then $\Sigma  \vdash  \Gamma$ and $\Sigma  \ottsym{;}  \Gamma  \vdash  \ottnt{A}  \ottsym{:}   \mathsf{T} $.
 \end{enumerate}
\end{lemma}
\begin{proof}
 Straightforward by induction on the typing derivations with
 Lemmas~\ref{lem:weakening} and \ref{lem:type-subst}.
\end{proof}

\begin{lemmap}{Progress}{progress}
 If $\Sigma  \ottsym{;}   \emptyset   \vdash  \ottnt{e}  \ottsym{:}  \ottnt{A}$, then one of the followings holds:
 \begin{itemize}
  \item $\ottnt{e}$ is a value;
  \item $\ottnt{e} \,  =  \, \mathsf{blame} \, \ottnt{p}$ for some $\ell$; or
  \item $ \Sigma  \mid  \ottnt{e}   \longrightarrow   \Sigma'  \mid  \ottnt{e'} $ for some $\Sigma'$ and $\ottnt{e'}$.
 \end{itemize}

\end{lemmap}
\begin{proof}
 By induction on the derivation of $\Sigma  \ottsym{;}   \emptyset   \vdash  \ottnt{e}  \ottsym{:}  \ottnt{A}$.
 \begin{caseanalysis}
  \case \T{Var}: Contradictory.
  \case \T{Const}, \T{Lam}, \T{TLam}, \T{REmp}, and \T{Blame}: Obvious.
  \case \T{App}:
   We have $\Sigma  \ottsym{;}   \emptyset   \vdash  \ottnt{e_{{\mathrm{1}}}} \, \ottnt{e_{{\mathrm{2}}}}  \ottsym{:}  \ottnt{A}$ and, by inversion,
   $\Sigma  \ottsym{;}   \emptyset   \vdash  \ottnt{e_{{\mathrm{1}}}}  \ottsym{:}  \ottnt{B}  \rightarrow  \ottnt{A}$ and $\Sigma  \ottsym{;}   \emptyset   \vdash  \ottnt{e_{{\mathrm{2}}}}  \ottsym{:}  \ottnt{B}$.
   If $\ottnt{e_{{\mathrm{1}}}} \,  =  \, \mathsf{blame} \, \ottnt{p}$ for some $\ottnt{p}$, or $ \Sigma  \mid  \ottnt{e_{{\mathrm{1}}}}   \longrightarrow   \Sigma'  \mid  \ottnt{e'_{{\mathrm{1}}}} $ for some $\Sigma'$ and $\ottnt{e'_{{\mathrm{1}}}}$,
   then we finish by \reflem{progress-aux}.

   In what follows, we can suppose that $\ottnt{e_{{\mathrm{1}}}} \,  =  \, \ottnt{v_{{\mathrm{1}}}}$ for some $\ottnt{v_{{\mathrm{1}}}}$
   by the IH.
   If $\ottnt{e_{{\mathrm{2}}}} \,  =  \, \mathsf{blame} \, \ottnt{p}$ for some $\ottnt{p}$, or $ \Sigma  \mid  \ottnt{e_{{\mathrm{2}}}}   \longrightarrow   \Sigma'  \mid  \ottnt{e'_{{\mathrm{2}}}} $ for some $\Sigma'$ and $\ottnt{e'_{{\mathrm{2}}}}$,
   then we finish by \reflem{progress-aux}.

   In what follows, we can suppose that $\ottnt{e_{{\mathrm{2}}}} \,  =  \, \ottnt{v_{{\mathrm{2}}}}$ for some $\ottnt{v_{{\mathrm{2}}}}$
   by the IH.
   We have $\Sigma  \ottsym{;}   \emptyset   \vdash  \ottnt{v_{{\mathrm{1}}}}  \ottsym{:}  \ottnt{B}  \rightarrow  \ottnt{A}$.
   Thus, by \reflem{canonical-forms}, there are two cases on
   $\ottnt{v_{{\mathrm{1}}}}$ to be considered.
   \begin{caseanalysis}
    \case $\ottnt{v_{{\mathrm{1}}}} \,  =  \,  \lambda\!  \,  \mathit{x}  \mathord{:}  \ottnt{B}   \ottsym{.}  \ottnt{e'_{{\mathrm{1}}}}$ for some $\mathit{x}$ and $\ottnt{e'_{{\mathrm{1}}}}$:
     By \R{Beta}/\E{Red}.
    \case $\ottnt{v_{{\mathrm{1}}}} \,  =  \, \kappa_{{\mathrm{1}}}$ and $ \mathit{ty}  (  \kappa_{{\mathrm{1}}}  )  \,  =  \, \ottnt{B}  \rightarrow  \ottnt{A}$ for some $\kappa_{{\mathrm{1}}}$:
     By the assumption on constants, $\ottnt{B} \,  =  \, \iota$ for some $\iota$.
     Since $\Sigma  \ottsym{;}   \emptyset   \vdash  \ottnt{v_{{\mathrm{2}}}}  \ottsym{:}  \iota$, we have $\ottnt{v_{{\mathrm{2}}}} \,  =  \, \kappa_{{\mathrm{2}}}$ for some $\kappa_{{\mathrm{2}}}$.
     By the assumption on constants, $ \zeta  (  \kappa_{{\mathrm{1}}}  ,  \kappa_{{\mathrm{2}}}  ) $ is defined.
     Thus, we finish by \E{Const}/\R{Red}.
   \end{caseanalysis}

  \case \T{TApp}:
   We have $\Sigma  \ottsym{;}   \emptyset   \vdash  \ottnt{e_{{\mathrm{1}}}} \, \ottnt{B}  \ottsym{:}   \ottnt{C}    [  \ottnt{B}  /  \mathit{X}  ]  $ and, by inversion,
   $\Sigma  \ottsym{;}   \emptyset   \vdash  \ottnt{e_{{\mathrm{1}}}}  \ottsym{:}   \text{\unboldmath$\forall\!$}  \,  \mathit{X}  \mathord{:}  \ottnt{K}   \ottsym{.} \, \ottnt{C}$ and $\Sigma  \ottsym{;}   \emptyset   \vdash  \ottnt{B}  \ottsym{:}  \ottnt{K}$.
   If $\ottnt{e_{{\mathrm{1}}}} \,  =  \, \mathsf{blame} \, \ottnt{p}$ for some $[p]$, or $ \Sigma  \mid  \ottnt{e_{{\mathrm{1}}}}   \longrightarrow   \Sigma'  \mid  \ottnt{e'_{{\mathrm{1}}}} $ for some $\Sigma'$ and $\ottnt{e'_{{\mathrm{1}}}}$,
   then we finish by \reflem{progress-aux}.

   In what follows, we can suppose that $\ottnt{e_{{\mathrm{1}}}} \,  =  \, \ottnt{v_{{\mathrm{1}}}}$ for some $\ottnt{v_{{\mathrm{1}}}}$
   by the IH.
   We have $\Sigma  \ottsym{;}   \emptyset   \vdash  \ottnt{v_{{\mathrm{1}}}}  \ottsym{:}   \text{\unboldmath$\forall\!$}  \,  \mathit{X}  \mathord{:}  \ottnt{K}   \ottsym{.} \, \ottnt{C}$.
   Thus, by \reflem{canonical-forms},
   $\ottnt{v_{{\mathrm{1}}}} \,  =  \,   \Lambda\!  \,  \mathit{X}  \mathord{:}  \ottnt{K}   \ottsym{.}   \ottnt{e'_{{\mathrm{1}}}}  ::  \ottnt{C} $ for some $\ottnt{e'_{{\mathrm{1}}}}$.
   By \E{TyBeta}, we finish.

  \case \T{RExt}:
   We have $\Sigma  \ottsym{;}   \emptyset   \vdash  \ottsym{\{}  \ell  \ottsym{=}  \ottnt{e_{{\mathrm{1}}}}  \ottsym{;}  \ottnt{e_{{\mathrm{2}}}}  \ottsym{\}}  \ottsym{:}   [    \ell  \mathbin{:}  \ottnt{B}   ;  \rho   ] $ and, by inversion,
   $\Sigma  \ottsym{;}   \emptyset   \vdash  \ottnt{e_{{\mathrm{1}}}}  \ottsym{:}  \ottnt{B}$ and
   $\Sigma  \ottsym{;}   \emptyset   \vdash  \ottnt{e_{{\mathrm{2}}}}  \ottsym{:}   [  \rho  ] $.
   If $\ottnt{e_{{\mathrm{1}}}} \,  =  \, \mathsf{blame} \, \ottnt{p}$ for some $\ottnt{p}$, or $ \Sigma  \mid  \ottnt{e_{{\mathrm{1}}}}   \longrightarrow   \Sigma'  \mid  \ottnt{e'_{{\mathrm{1}}}} $ for some $\Sigma'$ and $\ottnt{e'_{{\mathrm{1}}}}$,
   then we finish by \reflem{progress-aux}.

   In what follows, we can suppose that $\ottnt{e_{{\mathrm{1}}}} \,  =  \, \ottnt{v_{{\mathrm{1}}}}$ for some $\ottnt{v_{{\mathrm{1}}}}$
   by the IH.
   If $\ottnt{e_{{\mathrm{2}}}} \,  =  \, \mathsf{blame} \, \ottnt{p}$ for some $\ottnt{p}$, or $ \Sigma  \mid  \ottnt{e_{{\mathrm{2}}}}   \longrightarrow   \Sigma'  \mid  \ottnt{e'_{{\mathrm{2}}}} $ for some $\Sigma'$ and $\ottnt{e'_{{\mathrm{2}}}}$,
   then we finish by \reflem{progress-aux}.

   In what follows, we can suppose that $\ottnt{e_{{\mathrm{2}}}} \,  =  \, \ottnt{v_{{\mathrm{2}}}}$ for some $\ottnt{v_{{\mathrm{2}}}}$
   by the IH.
   Then, we finish because $\ottnt{e} \,  =  \, \ottsym{\{}  \ell  \ottsym{=}  \ottnt{v_{{\mathrm{1}}}}  \ottsym{;}  \ottnt{v_{{\mathrm{2}}}}  \ottsym{\}}$ is a value.

  \case \T{RLet}:
   We have $\Sigma  \ottsym{;}   \emptyset   \vdash  \mathsf{let} \, \ottsym{\{}  \ell  \ottsym{=}  \mathit{x}  \ottsym{;}  \mathit{y}  \ottsym{\}}  \ottsym{=}  \ottnt{e_{{\mathrm{1}}}} \, \mathsf{in} \, \ottnt{e_{{\mathrm{2}}}}  \ottsym{:}  \ottnt{A}$ and, by inversion,
   $\Sigma  \ottsym{;}   \emptyset   \vdash  \ottnt{e_{{\mathrm{1}}}}  \ottsym{:}   [    \ell  \mathbin{:}  \ottnt{B}   ;  \rho   ] $ and
   $\Sigma  \ottsym{;}   \mathit{x}  \mathord{:}  \ottnt{B}   \ottsym{,}   \mathit{y}  \mathord{:}   [  \rho  ]    \vdash  \ottnt{e_{{\mathrm{2}}}}  \ottsym{:}  \ottnt{A}$.
   If $\ottnt{e_{{\mathrm{1}}}} \,  =  \, \mathsf{blame} \, \ottnt{p}$ for some $\ottnt{p}$, or $ \Sigma  \mid  \ottnt{e_{{\mathrm{1}}}}   \longrightarrow   \Sigma'  \mid  \ottnt{e'_{{\mathrm{1}}}} $ for some $\Sigma'$ and $\ottnt{e'_{{\mathrm{1}}}}$,
   then we finish by \reflem{progress-aux}.

   In what follows, we can suppose that $\ottnt{e_{{\mathrm{1}}}} \,  =  \, \ottnt{v_{{\mathrm{1}}}}$ for some $\ottnt{v_{{\mathrm{1}}}}$
   by the IH.
   Since $\Sigma  \ottsym{;}   \emptyset   \vdash  \ottnt{v_{{\mathrm{1}}}}  \ottsym{:}   [    \ell  \mathbin{:}  \ottnt{B}   ;  \rho   ] $,
   we have $\ottnt{v_{{\mathrm{1}}}} \,  =  \, \ottsym{\{}  \ell  \ottsym{=}  \ottnt{v'_{{\mathrm{1}}}}  \ottsym{;}  \ottnt{v'_{{\mathrm{2}}}}  \ottsym{\}}$ for some $\ottnt{v'_{{\mathrm{1}}}}$ and $\ottnt{v'_{{\mathrm{2}}}}$
   by \reflem{canonical-forms}.
   Thus, we finish by \R{Record}/\E{Red}.

  \case \T{VInj}:
   We have $\Sigma  \ottsym{;}   \emptyset   \vdash  \ell \, \ottnt{e'}  \ottsym{:}   \langle    \ell  \mathbin{:}  \ottnt{B}   ;  \rho   \rangle $ and, by inversion,
   $\Sigma  \ottsym{;}   \emptyset   \vdash  \ottnt{e'}  \ottsym{:}  \ottnt{B}$ and $\Sigma  \ottsym{;}   \emptyset   \vdash  \rho  \ottsym{:}   \mathsf{R} $.
   If $\ottnt{e'} \,  =  \, \mathsf{blame} \, \ottnt{p}$ for some $\ottnt{p}$, or $ \Sigma  \mid  \ottnt{e'}   \longrightarrow   \Sigma'  \mid  \ottnt{e''} $ for some $\Sigma'$ and $\ottnt{e''}$,
   then we finish by \reflem{progress-aux}.

   In what follows, we can suppose that $\ottnt{e'} \,  =  \, \ottnt{v}$ for some $\ottnt{v}$
   by the IH.
   Then, we finish because $\ottnt{e} \,  =  \, \ell \, \ottnt{v}$ is a value.

  \case \T{VLift}:
   We have $\Sigma  \ottsym{;}   \emptyset   \vdash   \variantlift{ \ell }{ \ottnt{B} }{ \ottnt{e'} }   \ottsym{:}   \langle    \ell  \mathbin{:}  \ottnt{B}   ;  \rho   \rangle $ and, by inversion,
   $\Sigma  \ottsym{;}   \emptyset   \vdash  \ottnt{e'}  \ottsym{:}   \langle  \rho  \rangle $ and $\Sigma  \ottsym{;}   \emptyset   \vdash  \ottnt{B}  \ottsym{:}   \mathsf{T} $.
   If $\ottnt{e'} \,  =  \, \mathsf{blame} \, \ottnt{p}$ for some $\ottnt{p}$, or $ \Sigma  \mid  \ottnt{e'}   \longrightarrow   \Sigma'  \mid  \ottnt{e''} $ for some $\Sigma'$ and $\ottnt{e''}$,
   then we finish by \reflem{progress-aux}.

   In what follows, we can suppose that $\ottnt{e'} \,  =  \, \ottnt{v}$ for some $\ottnt{v}$
   by the IH.
   Then, we finish because $\ottnt{e} \,  =  \,  \variantlift{ \ell }{ \ottnt{B} }{ \ottnt{v} } $ is a value.

  \case \T{VCase}:
   We have $\Sigma  \ottsym{;}   \emptyset   \vdash   \mathsf{case} \,  \ottnt{e'}  \,\mathsf{with}\, \langle  \ell \,  \mathit{x}   \rightarrow   \ottnt{e_{{\mathrm{1}}}}   \ottsym{;}   \mathit{y}   \rightarrow   \ottnt{e_{{\mathrm{2}}}}  \rangle   \ottsym{:}  \ottnt{A}$ and, by inversion,
   $\Sigma  \ottsym{;}   \emptyset   \vdash  \ottnt{e'}  \ottsym{:}   \langle    \ell  \mathbin{:}  \ottnt{B}   ;  \rho   \rangle $.
   If $\ottnt{e'} \,  =  \, \mathsf{blame} \, \ottnt{p}$ for some $\ottnt{p}$, or $ \Sigma  \mid  \ottnt{e'}   \longrightarrow   \Sigma'  \mid  \ottnt{e''} $ for some $\Sigma'$ and $\ottnt{e''}$,
   then we finish by \reflem{progress-aux}.

   In what follows, we can suppose that $\ottnt{e'} \,  =  \, \ottnt{v}$ for some $\ottnt{v}$
   by the IH.
   We have $\Sigma  \ottsym{;}   \emptyset   \vdash  \ottnt{v}  \ottsym{:}   \langle    \ell  \mathbin{:}  \ottnt{B}   ;  \rho   \rangle $.
   Thus, by \reflem{canonical-forms}, there are two cases on $\ottnt{v}$ to be considered.
   \begin{caseanalysis}
    \case $\ottnt{v} \,  =  \, \ell \, \ottnt{v'}$ for some $\ottnt{v'}$: By \R{CaseL}\E{Red}.
    \case $\ottnt{v} \,  =  \,  \variantlift{ \ell }{ \ottnt{B} }{ \ottnt{v'} } $ for some $\ottnt{v'}$: By \R{CaseR}\E{Red}.
   \end{caseanalysis}

  \case \T{Cast}:
   We have $\Sigma  \ottsym{;}   \emptyset   \vdash  \ottnt{e'}  \ottsym{:}  \ottnt{B} \,  \stackrel{ \ottnt{p} }{\Rightarrow}  \ottnt{A}   \ottsym{:}  \ottnt{A}$ and, by inversion,
   $\Sigma  \ottsym{;}   \emptyset   \vdash  \ottnt{e'}  \ottsym{:}  \ottnt{B}$ and $\ottnt{B}  \simeq  \ottnt{A}$ and $\Sigma  \ottsym{;}   \emptyset   \vdash  \ottnt{A}  \ottsym{:}   \mathsf{T} $.
   If $\ottnt{e'} \,  =  \, \mathsf{blame} \, \ottnt{q}$ for some $\ottnt{q}$, or $ \Sigma  \mid  \ottnt{e'}   \longrightarrow   \Sigma'  \mid  \ottnt{e''} $ for some $\Sigma'$ and $\ottnt{e''}$,
   then we finish by \reflem{progress-aux}.

   In what follows, we can suppose that $\ottnt{e'} \,  =  \, \ottnt{v}$ for some $\ottnt{v}$
   by the IH.
   By case analysis on $\ottnt{B}  \simeq  \ottnt{A}$.
   \begin{caseanalysis}
    \case \CE{Refl}:
     We have $\ottnt{B} \,  =  \, \ottnt{A}$.
     By case analysis on $\ottnt{A}$.
     \begin{caseanalysis}
      \case $\ottnt{A} \,  =  \, \mathit{X}$:  Contradictory with $\Sigma  \ottsym{;}   \emptyset   \vdash  \ottnt{A}  \ottsym{:}   \mathsf{T} $.
      \case $\ottnt{A} \,  =  \, \alpha$: By \R{IdName}/\E{Red}.
      \case $\ottnt{A} \,  =  \, \star$: By \R{IdDyn}/\E{Red}.
      \case $\ottnt{A} \,  =  \, \iota$: By \R{IdBase}/\E{Red}.
      \case $\ottnt{A} \,  =  \, \ottnt{A_{{\mathrm{1}}}}  \rightarrow  \ottnt{A_{{\mathrm{2}}}}$: By \R{Wrap}/\E{Red}.
      \case $\ottnt{A} \,  =  \,  \text{\unboldmath$\forall\!$}  \,  \mathit{X}  \mathord{:}  \ottnt{K}   \ottsym{.} \, \ottnt{A'}$: By \R{Content}/\E{Red}.

      \case $\ottnt{A} \,  =  \,  [  \rho  ] $: By case analysis on $\rho$.
       Note that $\Sigma  \ottsym{;}   \emptyset   \vdash  \rho  \ottsym{:}   \mathsf{R} $ since $\Sigma  \ottsym{;}   \emptyset   \vdash   [  \rho  ]   \ottsym{:}   \mathsf{T} $.
       \begin{caseanalysis}
        \case $\rho \,  =  \, \mathit{X}$, $\iota$, $\ottnt{A'}  \rightarrow  \ottnt{B'}$, $ \text{\unboldmath$\forall\!$}  \,  \mathit{X}  \mathord{:}  \ottnt{K}   \ottsym{.} \, \ottnt{A'}$, $ [  \rho'  ] $, and $ \langle  \rho'  \rangle $:
         Contradictory with $\Sigma  \ottsym{;}   \emptyset   \vdash  \rho  \ottsym{:}   \mathsf{R} $.
        \case $\rho \,  =  \, \alpha$: By \E{RIdName}/\E{Red}.
        \case $\rho \,  =  \, \star$: \sloppy{By \reflem{canonical-forms},
         $\ottnt{v} \,  =  \, \ottnt{v'}  \ottsym{:}   [  \gamma'  ]  \,  \stackrel{ \ottnt{q} }{\Rightarrow}   [  \star  ]  $ for some $\ottnt{v'}$, $\gamma'$, and $\ottnt{q}$.
         We have $ [  \gamma'  ]   \simeq   [  \star  ] $ by \CE{DynR}/\CE{Record}.
         Thus, we finish by \R{RFromDyn}/\E{Red}.}
        \case $\rho \,  =  \,  \cdot $: By \R{Remp}/\E{Red}.
        \case $\rho \,  =  \,   \ell  \mathbin{:}  \ottnt{C}   ;  \rho' $: By \reflem{canonical-forms},
         $\ottnt{v} \,  =  \, \ottsym{\{}  \ell  \ottsym{=}  \ottnt{v_{{\mathrm{1}}}}  \ottsym{;}  \ottnt{v_{{\mathrm{2}}}}  \ottsym{\}}$ for some $\ottnt{v_{{\mathrm{1}}}}$ and $\ottnt{v_{{\mathrm{2}}}}$.
         Thus, $\ottnt{v} \,  \triangleright _{ \ell }  \, \ottnt{v_{{\mathrm{1}}}}  \ottsym{,}  \ottnt{v_{{\mathrm{2}}}}$.
         Since $  \ell  \mathbin{:}  \ottnt{C}   ;  \rho'  \,  \triangleright _{ \ell }  \, \ottnt{C}  \ottsym{,}  \rho'$, we finish by \R{Rev}/\E{Red}.
       \end{caseanalysis}

      \case $\ottnt{A} \,  =  \,  \langle  \rho  \rangle $: By case analysis on $\rho$.
       Note that $\Sigma  \ottsym{;}   \emptyset   \vdash  \rho  \ottsym{:}   \mathsf{R} $ since $\Sigma  \ottsym{;}   \emptyset   \vdash   \langle  \rho  \rangle   \ottsym{:}   \mathsf{T} $.
       \begin{caseanalysis}
        \case $\rho \,  =  \, \mathit{X}$, $\iota$, $\ottnt{A'}  \rightarrow  \ottnt{B'}$, $ \text{\unboldmath$\forall\!$}  \,  \mathit{X}  \mathord{:}  \ottnt{K}   \ottsym{.} \, \ottnt{A'}$, $ [  \rho'  ] $, and $ \langle  \rho'  \rangle $:
         Contradictory with $\Sigma  \ottsym{;}   \emptyset   \vdash  \rho  \ottsym{:}   \mathsf{R} $.
        \case $\rho \,  =  \, \alpha$: By \E{VIdName}/\E{Red}.
        \case $\rho \,  =  \, \star$: By \reflem{canonical-forms},
         $\ottnt{v} \,  =  \, \ottnt{v'}  \ottsym{:}   \langle  \gamma'  \rangle  \,  \stackrel{ \ottnt{q} }{\Rightarrow}   \langle  \star  \rangle  $ for some $\ottnt{v'}$, $\gamma'$, and $\ottnt{q}$.
         We have $ \langle  \gamma'  \rangle   \simeq   \langle  \star  \rangle $ by \CE{DynR}/\CE{Variant}.
         Thus, we finish by \R{VFromDyn}/\E{Red}.
        \case $\rho \,  =  \,  \cdot $: By \reflem{canonical-forms-variant-remp}.
        \case $\rho \,  =  \,   \ell  \mathbin{:}  \ottnt{C}   ;  \rho' $: By \reflem{canonical-forms}, there are two cases to be considered.

         If $\ottnt{v} \,  =  \, \ell \, \ottnt{v'}$ for some $\ottnt{v'}$, then we finish by \R{VRevInj}/\E{Red}.

         Otherwise, if $\ottnt{v} \,  =  \,  \variantlift{ \ell }{ \ottnt{C} }{ \ottnt{v'} } $ for some $\ottnt{v'}$,
         then we finish by \R{VRevLift}\E{Red}.
       \end{caseanalysis}

      \case $\ottnt{A} \,  =  \,  \cdot $ and $  \ell  \mathbin{:}  \ottnt{B}   ;  \rho $: Contradictory with $\Sigma  \ottsym{;}   \emptyset   \vdash  \ottnt{A}  \ottsym{:}   \mathsf{T} $.
     \end{caseanalysis}

    \case \CE{DynL}:  We have $\ottnt{B} \,  =  \, \star$.
     By \reflem{canonical-forms},
     $\ottnt{v} \,  =  \, \ottnt{v'}  \ottsym{:}  \mathit{G} \,  \stackrel{ \ottnt{q} }{\Rightarrow}  \star $ for some $\ottnt{v'}$, $\mathit{G}$, and $\ottnt{q}$.
     By case analysis on $\ottnt{A}$.
     \begin{caseanalysis}
      \case $\ottnt{A} \,  =  \, \mathit{H}$: By \R{Ground}/\E{Red} or \R{Blame}/\E{Red}.
      \case $\ottnt{A} \,  =  \, \mathit{X}$: Contradictory with $\Sigma  \ottsym{;}   \emptyset   \vdash  \ottnt{A}  \ottsym{:}   \mathsf{T} $.
      \case $\ottnt{A} \,  =  \, \star$: By \R{IdDyn}/\E{Red}.
      \case $\ottnt{A} \,  =  \, \ottnt{A_{{\mathrm{1}}}}  \rightarrow  \ottnt{A_{{\mathrm{2}}}}$ ($\ottnt{A_{{\mathrm{1}}}}  \rightarrow  \ottnt{A_{{\mathrm{2}}}} \,  \not=  \, \star  \rightarrow  \star$):
       Since $\ottnt{A_{{\mathrm{1}}}}  \rightarrow  \ottnt{A_{{\mathrm{2}}}}  \simeq  \star  \rightarrow  \star$, we finish by \R{FromDyn}/\E{Red}.
      \case $\ottnt{A} \,  =  \,  \text{\unboldmath$\forall\!$}  \,  \mathit{X}  \mathord{:}  \ottnt{K}   \ottsym{.} \, \ottnt{A'}$: By \R{Gen}/\E{Red}.
      \case $\ottnt{A} \,  =  \,  [  \rho  ] $ ($\rho \,  \not=  \, \star$):
       Since $ [  \rho  ]   \simeq   [  \star  ] $, we finish by \R{FromDyn}/\E{Red}.
      \case $\ottnt{A} \,  =  \,  \langle  \rho  \rangle $ ($\rho \,  \not=  \, \star$):
       Since $ \langle  \rho  \rangle   \simeq   \langle  \star  \rangle $, we finish by \R{FromDyn}/\E{Red}.
      \case $\ottnt{A} \,  =  \,  \cdot $: Contradictory with $\Sigma  \ottsym{;}   \emptyset   \vdash  \ottnt{A}  \ottsym{:}   \mathsf{T} $.
      \case $\ottnt{A} \,  =  \,   \ell  \mathbin{:}  \ottnt{C}   ;  \rho $: Contradictory with $\Sigma  \ottsym{;}   \emptyset   \vdash  \ottnt{A}  \ottsym{:}   \mathsf{T} $.
     \end{caseanalysis}

    \case \CE{DynR}:  We have $\ottnt{A} \,  =  \, \star$.

     If $\ottnt{B} \,  =  \, \star$, then we finish by \R{IdDyn}/\E{Red}.

     If $\ottnt{B} \,  =  \,  \text{\unboldmath$\forall\!$}  \,  \mathit{X}  \mathord{:}  \ottnt{K}   \ottsym{.} \, \ottnt{B'}$, then we finish by \R{Inst}/\E{Red}.

     Otherwise, by \reflem{ground-type-unique}, there exists some $\mathit{G}$ such that
     $\ottnt{B}  \simeq  \mathit{G}$.
     If $\ottnt{B} \,  =  \, \mathit{G}$, then $\ottnt{e} \,  =  \, \ottnt{v}  \ottsym{:}  \mathit{G} \,  \stackrel{ \ottnt{p} }{\Rightarrow}  \star $ is a value.
     Otherwise, we finish by \R{ToDyn}/\E{Red}.

    \case \CE{Fun}: By \R{Wrap}/\E{Red}.
    \case \CE{Poly}: By \R{Content}/\E{Red}.
    \case \CE{PolyL}: By \R{Inst}/\E{Red}.
    \case \CE{PolyR}: By \R{Gen}/\E{Red}.
    \case \CE{Record}: We have $\ottnt{A} \,  =  \,  [  \rho_{{\mathrm{1}}}  ] $ and $\ottnt{B} \,  =  \,  [  \rho_{{\mathrm{2}}}  ] $ and $\rho_{{\mathrm{2}}}  \simeq  \rho_{{\mathrm{1}}}$
     for some $\rho_{{\mathrm{1}}}$ and $\rho_{{\mathrm{2}}}$.
     Since $\Sigma  \ottsym{;}   \emptyset   \vdash   [  \rho_{{\mathrm{1}}}  ]   \ottsym{:}   \mathsf{T} $, we have $\Sigma  \ottsym{;}   \emptyset   \vdash  \rho_{{\mathrm{1}}}  \ottsym{:}   \mathsf{R} $.
     By \reflem{typctx-type-wf}, $\Sigma  \ottsym{;}   \emptyset   \vdash   [  \rho_{{\mathrm{2}}}  ]   \ottsym{:}   \mathsf{T} $, so $\Sigma  \ottsym{;}   \emptyset   \vdash  \rho_{{\mathrm{2}}}  \ottsym{:}   \mathsf{R} $.

     If $\rho_{{\mathrm{2}}} \,  =  \, \star$, then we finish by \reflem{canonical-forms}, and \R{RFromDyn}/\E{Red} or \R{RBlame}/\E{Red}.

     In what follows, we suppose $\rho_{{\mathrm{2}}} \,  \not=  \, \star$.
     By case analysis on $\rho_{{\mathrm{1}}}$.
     \begin{caseanalysis}
      \case $\rho_{{\mathrm{1}}} \,  =  \, \star$:
       Since $\rho_{{\mathrm{2}}} \,  \not=  \, \star$ and $\Sigma  \ottsym{;}   \emptyset   \vdash  \rho_{{\mathrm{2}}}  \ottsym{:}   \mathsf{R} $,
       $ \mathit{grow}  (  \rho_{{\mathrm{2}}}  ) $ is defined and is a ground row type by \reflem{grow-is-defined}.

       If $ \mathit{grow}  (  \rho_{{\mathrm{2}}}  )  \,  =  \, \rho_{{\mathrm{2}}}$, then $\ottnt{v}  \ottsym{:}   [  \rho_{{\mathrm{2}}}  ]  \,  \stackrel{ \ottnt{p} }{\Rightarrow}   [  \star  ]  $ is a value.

       Otherwise, if $ \mathit{grow}  (  \rho_{{\mathrm{2}}}  )  \,  \not=  \, \rho_{{\mathrm{2}}}$, we finish by \R{RToDyn}/\E{Red}.

      \case $\rho_{{\mathrm{1}}} \,  =  \, \alpha$: \sloppy{
       Since $\rho_{{\mathrm{2}}}  \simeq  \alpha$ and $\rho_{{\mathrm{2}}} \,  \not=  \, \star$, we have $\rho_{{\mathrm{2}}} \,  =  \, \alpha$
       by Lemmas~\ref{lem:consistent-symm} and \ref{lem:consistent-inv-tyname}.
       We finish by \R{RIdName}/\E{Red}.
      }

      \case $\rho_{{\mathrm{1}}} \,  =  \,  \cdot $:
       Since $\rho_{{\mathrm{2}}}  \simeq   \cdot $ and $\rho_{{\mathrm{2}}} \,  \not=  \, \star$, we have $\rho_{{\mathrm{2}}} \,  =  \,  \cdot $
       by Lemmas~\ref{lem:consistent-symm} and \ref{lem:consistent-inv-remp}.
       We finish by \R{REmp}/\E{Red}.

      \case $\rho_{{\mathrm{1}}} \,  =  \,   \ell  \mathbin{:}  \ottnt{C_{{\mathrm{1}}}}   ;  \rho'_{{\mathrm{1}}} $:
       By Lemmas~\ref{lem:consistent-symm} and \ref{lem:consistent-inv-cons},
       $\rho_{{\mathrm{2}}} \,  \triangleright _{ \ell }  \, \ottnt{C_{{\mathrm{2}}}}  \ottsym{,}  \rho'_{{\mathrm{2}}}$ and $\ottnt{C_{{\mathrm{2}}}}  \simeq  \ottnt{C_{{\mathrm{1}}}}$ and $\rho'_{{\mathrm{2}}}  \simeq  \rho'_{{\mathrm{1}}}$
       for some $\ottnt{C_{{\mathrm{2}}}}$ and $\rho'_{{\mathrm{2}}}$.

       If $\ell \,  \in  \, \mathit{dom} \, \ottsym{(}  \rho_{{\mathrm{2}}}  \ottsym{)}$, then there exist some $\rho_{{\mathrm{21}}}$ and $\rho_{{\mathrm{22}}}$ such that
       \begin{itemize}
        \item $\rho_{{\mathrm{2}}} \,  =  \, \rho_{{\mathrm{21}}}  \odot  \ottsym{(}    \ell  \mathbin{:}  \ottnt{C_{{\mathrm{2}}}}   ;   \cdot    \ottsym{)}  \odot  \rho_{{\mathrm{22}}}$,
        \item $\rho'_{{\mathrm{2}}} \,  =  \, \rho_{{\mathrm{21}}}  \odot  \rho_{{\mathrm{22}}}$, and
        \item $\ell \,  \not\in  \, \mathit{dom} \, \ottsym{(}  \rho_{{\mathrm{21}}}  \ottsym{)}$.
       \end{itemize}
       Since $\Sigma  \ottsym{;}   \emptyset   \vdash  \ottnt{v}  \ottsym{:}   [  \rho_{{\mathrm{2}}}  ] $, there exist some $\ottnt{v_{{\mathrm{1}}}}$ and $\ottnt{v_{{\mathrm{2}}}}$
       such that $\ottnt{v} \,  \triangleright _{ \ell }  \, \ottnt{v_{{\mathrm{1}}}}  \ottsym{,}  \ottnt{v_{{\mathrm{2}}}}$
       by Lemmas~\ref{lem:canonical-forms} and \ref{lem:value-inversion-record-cons}.
       Thus, we finish by \R{RRev}/\E{Red}.

       If $\ell \,  \not\in  \, \mathit{dom} \, \ottsym{(}  \rho_{{\mathrm{2}}}  \ottsym{)}$, then we finish by \R{RCon}/\E{Red}.

      \case $\rho_{{\mathrm{1}}} \,  =  \, \mathit{X}$, $\iota$, $\ottnt{C}  \rightarrow  \ottnt{D}$, $ \text{\unboldmath$\forall\!$}  \,  \mathit{X}  \mathord{:}  \ottnt{K}   \ottsym{.} \, \ottnt{C}$, $ [  \rho'  ] $, and $ \langle  \rho'  \rangle $:
       Contradictory with $\Sigma  \ottsym{;}   \emptyset   \vdash  \rho_{{\mathrm{1}}}  \ottsym{:}   \mathsf{R} $.
     \end{caseanalysis}

    \case \CE{Variant}: We have $\ottnt{A} \,  =  \,  \langle  \rho_{{\mathrm{1}}}  \rangle $ and $\ottnt{B} \,  =  \,  \langle  \rho_{{\mathrm{2}}}  \rangle $ and $\rho_{{\mathrm{2}}}  \simeq  \rho_{{\mathrm{1}}}$
     for some $\rho_{{\mathrm{1}}}$ and $\rho_{{\mathrm{2}}}$.
     Since $\Sigma  \ottsym{;}   \emptyset   \vdash   \langle  \rho_{{\mathrm{1}}}  \rangle   \ottsym{:}   \mathsf{T} $, we have $\Sigma  \ottsym{;}   \emptyset   \vdash  \rho_{{\mathrm{1}}}  \ottsym{:}   \mathsf{R} $.
     By \reflem{typctx-type-wf}, $\Sigma  \ottsym{;}   \emptyset   \vdash   \langle  \rho_{{\mathrm{2}}}  \rangle   \ottsym{:}   \mathsf{T} $, so $\Sigma  \ottsym{;}   \emptyset   \vdash  \rho_{{\mathrm{2}}}  \ottsym{:}   \mathsf{R} $.

     By case analysis on $\rho_{{\mathrm{2}}}$.
     \begin{caseanalysis}
      \case $\rho_{{\mathrm{2}}} \,  =  \, \star$:
       We finish by \reflem{canonical-forms}, and \R{VFromDyn}/\E{Red} or \R{VBlame}\E{Red}.

      \case $\rho_{{\mathrm{2}}} \,  =  \, \alpha$:
       Since $\rho_{{\mathrm{2}}}  \simeq  \rho_{{\mathrm{1}}}$, we have $\rho_{{\mathrm{1}}} \,  =  \, \alpha$ or $\rho_{{\mathrm{1}}} \,  =  \, \star$ by \reflem{consistent-inv-tyname}.

       If $\rho_{{\mathrm{1}}} \,  =  \, \star$, then $\ottnt{v}  \ottsym{:}   [  \alpha  ]  \,  \stackrel{ \ottnt{p} }{\Rightarrow}   [  \star  ]  $ is a value.

       Otherwise, if $\rho_{{\mathrm{1}}} \,  =  \, \alpha$, then we finish by \R{VIdName}/\E{Red}.

      \case $\rho_{{\mathrm{2}}} \,  =  \,  \cdot $:
       Contradictory by \reflem{canonical-forms-variant-remp}.

      \case $\rho_{{\mathrm{2}}} \,  =  \,   \ell  \mathbin{:}  \ottnt{C_{{\mathrm{2}}}}   ;  \rho'_{{\mathrm{2}}} $: \sloppy{
       If $\ell \,  \in  \, \mathit{dom} \, \ottsym{(}  \rho_{{\mathrm{1}}}  \ottsym{)}$,
       then we finish by \reflem{canonical-forms}, and \R{VRevInj}/\E{Red} or \R{VRevLift}/\E{Red}.
      }

       Otherwise, suppose $\ell \,  \not\in  \, \mathit{dom} \, \ottsym{(}  \rho_{{\mathrm{1}}}  \ottsym{)}$.
       Since $\Sigma  \ottsym{;}   \emptyset   \vdash  \rho_{{\mathrm{2}}}  \ottsym{:}   \mathsf{R} $ and $\rho_{{\mathrm{2}}} \,  \not=  \, \star$,
       it is found that $ \mathit{grow}  (  \rho_{{\mathrm{2}}}  ) $ is defined.
       If $\rho_{{\mathrm{1}}} \,  =  \, \star$ and $ \mathit{grow}  (  \rho_{{\mathrm{2}}}  )  \,  =  \, \rho_{{\mathrm{2}}}$, then $\ottnt{v}  \ottsym{:}   [  \rho_{{\mathrm{2}}}  ]  \,  \stackrel{ \ottnt{p} }{\Rightarrow}   [  \star  ]  $ is a value.
       If $\rho_{{\mathrm{1}}} \,  =  \, \star$ and $ \mathit{grow}  (  \rho_{{\mathrm{2}}}  )  \,  \not=  \, \rho_{{\mathrm{2}}}$, then we finish by \R{VToDyn}/\E{Red}.

      \sloppy{
       Otherwise, suppose $\rho_{{\mathrm{1}}} \,  \not=  \, \star$.
       Then, we finish by \reflem{canonical-forms}, and \R{VConInj}/\E{Red} or \R{VConLift}/\E{Red}.
      }

      \case $\rho_{{\mathrm{2}}} \,  =  \, \mathit{X}$, $\iota$, $\ottnt{C}  \rightarrow  \ottnt{D}$, $ \text{\unboldmath$\forall\!$}  \,  \mathit{X}  \mathord{:}  \ottnt{K}   \ottsym{.} \, \ottnt{C}$, $ [  \rho'  ] $, and $ \langle  \rho'  \rangle $:
       Contradictory with $\Sigma  \ottsym{;}   \emptyset   \vdash  \rho_{{\mathrm{2}}}  \ottsym{:}   \mathsf{R} $.
     \end{caseanalysis}

    \case \CE{ConsL}:
     We have $\ottnt{B} \,  =  \,   \ell  \mathbin{:}  \ottnt{C_{{\mathrm{2}}}}   ;  \rho_{{\mathrm{2}}} $ for some $\ell$, $\ottnt{C_{{\mathrm{2}}}}$, and $\rho_{{\mathrm{2}}}$.
     Since $\Sigma  \ottsym{;}   \emptyset   \vdash  \ottnt{e'}  \ottsym{:}  \ottnt{B}$, we have $\Sigma  \ottsym{;}   \emptyset   \vdash  \ottnt{B}  \ottsym{:}   \mathsf{T} $ by \reflem{typctx-type-wf}.
     However, there is a contradiction that $\Sigma  \ottsym{;}   \emptyset   \vdash    \ell  \mathbin{:}  \ottnt{C_{{\mathrm{2}}}}   ;  \rho_{{\mathrm{2}}}   \ottsym{:}   \mathsf{T} $ does not hold.

    \case \CE{ConsR}:
     We have $\ottnt{A} \,  =  \,   \ell  \mathbin{:}  \ottnt{C_{{\mathrm{1}}}}   ;  \rho_{{\mathrm{1}}} $ for some $\ell$, $\ottnt{C_{{\mathrm{1}}}}$, and $\rho_{{\mathrm{1}}}$.
     However, it is contradictory with $\Sigma  \ottsym{;}   \emptyset   \vdash  \ottnt{A}  \ottsym{:}   \mathsf{T} $.
   \end{caseanalysis}

  \case \T{Conv}:
   We have $\Sigma  \ottsym{;}   \emptyset   \vdash  \ottnt{e'}  \ottsym{:}  \ottnt{B} \,  \stackrel{ \Phi }{\Rightarrow}  \ottnt{A}   \ottsym{:}  \ottnt{A}$ and, by inversion,
   $\Sigma  \ottsym{;}   \emptyset   \vdash  \ottnt{e'}  \ottsym{:}  \ottnt{B}$ and $ \Sigma   \vdash   \ottnt{B}  \prec^{ \Phi }  \ottnt{A} $ and $\Sigma  \ottsym{;}   \emptyset   \vdash  \ottnt{A}  \ottsym{:}   \mathsf{T} $.
   If $\ottnt{e'} \,  =  \, \mathsf{blame} \, \ottnt{q}$ for some $\ottnt{q}$, or $ \Sigma  \mid  \ottnt{e'}   \longrightarrow   \Sigma'  \mid  \ottnt{e''} $ for some $\Sigma'$ and $\ottnt{e''}$,
   then we finish by \reflem{progress-aux}.

   In what follows, we can suppose that $\ottnt{e'} \,  =  \, \ottnt{v}$ for some $\ottnt{v}$
   by the IH.
   By case analysis on $ \Sigma   \vdash   \ottnt{B}  \prec^{ \Phi }  \ottnt{A} $.
   \begin{caseanalysis}
    \case \Cv{Dyn}: By \R{CIdDyn}/\E{Red}.
    \case \Cv{TyVar}: Contradictory with $\Sigma  \ottsym{;}   \emptyset   \vdash  \ottnt{A}  \ottsym{:}   \mathsf{T} $.
    \case \Cv{TyName}: By \R{CIdName}/\E{Red}.
    \case \Cv{Reveal}:
     We have $\ottnt{B} \,  =  \, \alpha$ and $\Phi \,  =  \, \ottsym{+}  \alpha$ and $\Sigma  \ottsym{(}  \alpha  \ottsym{)} \,  =  \, \ottnt{A}$ for some $\alpha$.
     By \reflem{canonical-forms}, $\ottnt{v} \,  =  \, \ottnt{v'}  \ottsym{:}  \ottnt{C} \,  \stackrel{ \ottsym{-}  \alpha }{\Rightarrow}  \alpha $ for some $\ottnt{C}$.
     Since $\Sigma  \ottsym{;}   \emptyset   \vdash  \ottnt{v}  \ottsym{:}  \ottnt{B}$, we have $\Sigma  \ottsym{(}  \alpha  \ottsym{)} \,  =  \, \ottnt{C}$ by \reflem{value-inversion-conv}, so $\ottnt{A} \,  =  \, \ottnt{C}$.
     We finish by \R{CName}/\E{Red}.

    \case \Cv{Conceal}: $\ottnt{v}  \ottsym{:}  \ottnt{B} \,  \stackrel{ \Phi }{\Rightarrow}  \ottnt{A} $ is a value.

    \case \Cv{Base}: By \R{CIdBase}/\E{Red}.
    \case \Cv{Fun}: By \R{CFun}/\E{Red}.
    \case \Cv{Poly}: By \R{CForall}/\E{Red}.
    \case \Cv{Record}: We have $\ottnt{B} \,  =  \,  [  \rho_{{\mathrm{2}}}  ] $ and $\ottnt{A} \,  =  \,  [  \rho_{{\mathrm{1}}}  ] $ and $ \Sigma   \vdash   \rho_{{\mathrm{2}}}  \prec^{ \Phi }  \rho_{{\mathrm{1}}} $
     for some $\rho_{{\mathrm{1}}}$ and $\rho_{{\mathrm{2}}}$.
     Since $\Sigma  \ottsym{;}   \emptyset   \vdash  \ottnt{A}  \ottsym{:}   \mathsf{T} $, we have $\Sigma  \ottsym{;}   \emptyset   \vdash  \rho_{{\mathrm{1}}}  \ottsym{:}   \mathsf{R} $.
     By case analysis on $ \Sigma   \vdash   \rho_{{\mathrm{2}}}  \prec^{ \Phi }  \rho_{{\mathrm{1}}} $.
     \begin{caseanalysis}
      \case \Cv{Dyn}: By \R{CRIdDyn}/\E{Red}.
      \case \Cv{TyName}: By \R{CRIdName}/\E{Red}.
      \case \Cv{Reveal}:
       We have $\rho_{{\mathrm{2}}} \,  =  \, \alpha$ and $\Phi \,  =  \, \ottsym{+}  \alpha$ and $\Sigma  \ottsym{(}  \alpha  \ottsym{)} \,  =  \, \rho_{{\mathrm{1}}}$ for some $\alpha$.
       By \reflem{canonical-forms}, $\ottnt{v} \,  =  \, \ottnt{v'}  \ottsym{:}   [  \rho'  ]  \,  \stackrel{ \ottsym{-}  \alpha }{\Rightarrow}   [  \alpha  ]  $ for some $\ottnt{v'}$ and $\rho'$.
       By \reflem{value-inversion-conv-record}, $\Sigma  \ottsym{(}  \alpha  \ottsym{)} \,  =  \, \rho'$, so $\rho' \,  =  \, \rho$.
       We finish by \R{CRName}/\E{Red}.

      \case \Cv{Conceal}: $\ottnt{v}  \ottsym{:}  \ottnt{B} \,  \stackrel{ \Phi }{\Rightarrow}  \ottnt{A} $ is a value.

      \case \Cv{REmp}: By \R{CREmp}/\E{Red}.
      \case \Cv{Cons}: By \R{CRExt}/\E{Red}.
      \case \Cv{TyVar}, \Cv{Base}, \Cv{Fun}, \Cv{Poly}, \Cv{Record}, and \Cv{Variant}:
       Contradictory with $\Sigma  \ottsym{;}   \emptyset   \vdash  \rho_{{\mathrm{1}}}  \ottsym{:}   \mathsf{R} $.
     \end{caseanalysis}

    \case \Cv{Variant}: We have $\ottnt{B} \,  =  \,  \langle  \rho_{{\mathrm{2}}}  \rangle $ and $\ottnt{A} \,  =  \,  \langle  \rho_{{\mathrm{1}}}  \rangle $ and $ \Sigma   \vdash   \rho_{{\mathrm{2}}}  \prec^{ \Phi }  \rho_{{\mathrm{1}}} $
     for some $\rho_{{\mathrm{1}}}$ and $\rho_{{\mathrm{2}}}$.
     Since $\Sigma  \ottsym{;}   \emptyset   \vdash  \ottnt{A}  \ottsym{:}   \mathsf{T} $, we have $\Sigma  \ottsym{;}   \emptyset   \vdash  \rho_{{\mathrm{1}}}  \ottsym{:}   \mathsf{R} $.
     By case analysis on $ \Sigma   \vdash   \rho_{{\mathrm{2}}}  \prec^{ \Phi }  \rho_{{\mathrm{1}}} $.
     \begin{caseanalysis}
      \case \Cv{Dyn}: By \R{CVIdDyn}/\E{Red}.
      \case \Cv{TyName}: By \R{CVIdName}/\E{Red}.
      \case \Cv{Reveal}:
       We have $\rho_{{\mathrm{2}}} \,  =  \, \alpha$ and $\Phi \,  =  \, \ottsym{+}  \alpha$ and $\Sigma  \ottsym{(}  \alpha  \ottsym{)} \,  =  \, \rho_{{\mathrm{1}}}$ for some $\alpha$.
       By \reflem{canonical-forms}, $\ottnt{v} \,  =  \, \ottnt{v'}  \ottsym{:}   \langle  \rho'  \rangle  \,  \stackrel{ \ottsym{-}  \alpha }{\Rightarrow}   \langle  \alpha  \rangle  $ for some $\ottnt{v'}$ and $\rho'$.
       By \reflem{value-inversion-conv-variant}, $\Sigma  \ottsym{(}  \alpha  \ottsym{)} \,  =  \, \rho'$, so $\rho' \,  =  \, \rho_{{\mathrm{1}}}$.
       We finish by \R{CVName}/\E{Red}.

      \case \Cv{Conceal}: $\ottnt{v}  \ottsym{:}  \ottnt{B} \,  \stackrel{ \Phi }{\Rightarrow}  \ottnt{A} $ is a value.
      \case \Cv{REmp}: We have $\Sigma  \ottsym{;}   \emptyset   \vdash  \ottnt{v}  \ottsym{:}   [   \cdot   ] $, which is contradictory by \reflem{canonical-forms-variant-remp}.
      \case \Cv{Cons}: By \R{CVar}/\E{Red}.
      \case \Cv{TyVar}, \Cv{Base}, \Cv{Fun}, \Cv{Poly}, \Cv{Record}, and \Cv{Variant}:
       Contradictory with $\Sigma  \ottsym{;}   \emptyset   \vdash  \rho_{{\mathrm{1}}}  \ottsym{:}   \mathsf{R} $.
     \end{caseanalysis}

    \case \Cv{REmp} and \Cv{Cons}: Contradictory with $\Sigma  \ottsym{;}   \emptyset   \vdash  \ottnt{A}  \ottsym{:}   \mathsf{T} $.
   \end{caseanalysis}
 \end{caseanalysis}
\end{proof}


\begin{lemma}{consistent-ground-type-wf}
 If $\Sigma  \ottsym{;}  \Gamma  \vdash  \ottnt{A}  \ottsym{:}   \mathsf{T} $ and $\ottnt{A}  \simeq  \mathit{G}$,
 then $\Sigma  \ottsym{;}  \Gamma  \vdash  \mathit{G}  \ottsym{:}   \mathsf{T} $.
\end{lemma}
\begin{proof}
 By case analysis on $\mathit{G}$.
 \begin{caseanalysis}
  \case $\mathit{G} \,  =  \, \iota$, $\star  \rightarrow  \star$, $ [  \star  ] $, and $ \langle  \star  \rangle $:
   Obvious.
  \case $\mathit{G} \,  =  \, \alpha$:
   Since $\ottnt{A}  \simeq  \alpha$, we have $\ottnt{A} \,  =  \, \alpha$ or $\ottnt{A} \,  =  \, \star$
   by Lemmas~\ref{lem:consistent-symm} and \ref{lem:consistent-inv-tyname}.
   In either case, $\Sigma  \ottsym{;}  \Gamma  \vdash  \ottnt{A}  \ottsym{:}   \mathsf{T} $.
 \end{caseanalysis}
\end{proof}

\begin{lemma}{typing-record-ext-decomp}
 If $\Sigma  \ottsym{;}   \emptyset   \vdash  \ottnt{v}  \ottsym{:}   [  \rho  ] $ and $\ottnt{v} \,  \triangleright _{ \ell }  \, \ottnt{v_{{\mathrm{1}}}}  \ottsym{,}  \ottnt{v_{{\mathrm{2}}}}$,
 then there exist some $\rho_{{\mathrm{1}}}$, $\rho_{{\mathrm{2}}}$, and $\ottnt{A}$ such that
 $\rho \,  =  \, \rho_{{\mathrm{1}}}  \odot  \ottsym{(}    \ell  \mathbin{:}  \ottnt{A}   ;   \cdot    \ottsym{)}  \odot  \rho_{{\mathrm{2}}}$ and
 $\ell \,  \not\in  \, \mathit{dom} \, \ottsym{(}  \rho_{{\mathrm{1}}}  \ottsym{)}$ and
 $\Sigma  \ottsym{;}   \emptyset   \vdash  \ottnt{v_{{\mathrm{1}}}}  \ottsym{:}  \ottnt{A}$ and
 $\Sigma  \ottsym{;}   \emptyset   \vdash  \ottnt{v_{{\mathrm{2}}}}  \ottsym{:}   [    \rho_{{\mathrm{1}}}  \odot  \rho_{{\mathrm{2}}}    ] $.
\end{lemma}
\begin{proof}
 By induction on the derivation of $\ottnt{v} \,  \triangleright _{ \ell }  \, \ottnt{v_{{\mathrm{1}}}}  \ottsym{,}  \ottnt{v_{{\mathrm{2}}}}$.
 \begin{caseanalysis}
  \case $\ottsym{\{}  \ell  \ottsym{=}  \ottnt{v_{{\mathrm{1}}}}  \ottsym{;}  \ottnt{v_{{\mathrm{2}}}}  \ottsym{\}} \,  \triangleright _{ \ell }  \, \ottnt{v_{{\mathrm{1}}}}  \ottsym{,}  \ottnt{v_{{\mathrm{2}}}}$:
   We have $\ottnt{v} \,  =  \, \ottsym{\{}  \ell  \ottsym{=}  \ottnt{v_{{\mathrm{1}}}}  \ottsym{;}  \ottnt{v_{{\mathrm{2}}}}  \ottsym{\}}$.
   Since $\Sigma  \ottsym{;}   \emptyset   \vdash  \ottnt{v}  \ottsym{:}   [  \rho  ] $,
   there exist $\ottnt{A}$ and $\rho'$ such that
   $\rho \,  =  \,   \ell  \mathbin{:}  \ottnt{A}   ;   [  \rho'  ]  $ and
   $\Sigma  \ottsym{;}   \emptyset   \vdash  \ottnt{v_{{\mathrm{1}}}}  \ottsym{:}  \ottnt{A}$ and
   $\Sigma  \ottsym{;}   \emptyset   \vdash  \ottnt{v_{{\mathrm{2}}}}  \ottsym{:}   [  \rho'  ] $.

  \case $\ottsym{\{}  \ell'  \ottsym{=}  \ottnt{v'_{{\mathrm{1}}}}  \ottsym{;}  \ottnt{v'_{{\mathrm{2}}}}  \ottsym{\}} \,  \triangleright _{ \ell }  \, \ottnt{v_{{\mathrm{1}}}}  \ottsym{,}  \ottsym{\{}  \ell'  \ottsym{=}  \ottnt{v'_{{\mathrm{1}}}}  \ottsym{;}  \ottnt{v''_{{\mathrm{2}}}}  \ottsym{\}}$
        where $\ell \,  \not=  \, \ell'$ and $\ottnt{v'_{{\mathrm{2}}}} \,  \triangleright _{ \ell }  \, \ottnt{v_{{\mathrm{1}}}}  \ottsym{,}  \ottnt{v''_{{\mathrm{2}}}}$:
   We have $\ottnt{v} \,  =  \, \ottsym{\{}  \ell'  \ottsym{=}  \ottnt{v'_{{\mathrm{1}}}}  \ottsym{;}  \ottnt{v'_{{\mathrm{2}}}}  \ottsym{\}}$ and $\ottnt{v_{{\mathrm{2}}}} \,  =  \, \ottsym{\{}  \ell'  \ottsym{=}  \ottnt{v'_{{\mathrm{1}}}}  \ottsym{;}  \ottnt{v''_{{\mathrm{2}}}}  \ottsym{\}}$.
   Since $\Sigma  \ottsym{;}   \emptyset   \vdash  \ottnt{v}  \ottsym{:}   [  \rho  ] $,
   there exist some $\ottnt{B}$ and $\rho'$ such that
   $\Sigma  \ottsym{;}   \emptyset   \vdash  \ottnt{v'_{{\mathrm{1}}}}  \ottsym{:}  \ottnt{B}$ and
   $\Sigma  \ottsym{;}   \emptyset   \vdash  \ottnt{v'_{{\mathrm{2}}}}  \ottsym{:}   [  \rho'  ] $ and 
   $\rho \,  =  \,   \ell  \mathbin{:}  \ottnt{B}   ;  \rho' $.
   Since $\Sigma  \ottsym{;}   \emptyset   \vdash  \ottnt{v'_{{\mathrm{2}}}}  \ottsym{:}   [  \rho'  ] $ and $\ottnt{v'_{{\mathrm{2}}}} \,  \triangleright _{ \ell }  \, \ottnt{v_{{\mathrm{1}}}}  \ottsym{,}  \ottnt{v''_{{\mathrm{2}}}}$,
   there exist some $\rho'_{{\mathrm{1}}}$, $\rho'_{{\mathrm{2}}}$, and $\ottnt{A}$ such that
   \begin{itemize}
    \item $\rho' \,  =  \, \rho'_{{\mathrm{1}}}  \odot  \ottsym{(}    \ell  \mathbin{:}  \ottnt{A}   ;   \cdot    \ottsym{)}  \odot  \rho'_{{\mathrm{2}}}$,
    \item $\ell \,  \not\in  \, \mathit{dom} \, \ottsym{(}  \rho'_{{\mathrm{1}}}  \ottsym{)}$,
    \item $\Sigma  \ottsym{;}   \emptyset   \vdash  \ottnt{v_{{\mathrm{1}}}}  \ottsym{:}  \ottnt{A}$, and
    \item $\Sigma  \ottsym{;}   \emptyset   \vdash  \ottnt{v''_{{\mathrm{2}}}}  \ottsym{:}   [    \rho'_{{\mathrm{1}}}  \odot  \rho'_{{\mathrm{2}}}    ] $
   \end{itemize}
   by the IH.
   Since $\Sigma  \ottsym{;}   \emptyset   \vdash  \ottnt{v'_{{\mathrm{1}}}}  \ottsym{:}  \ottnt{B}$ and $\Sigma  \ottsym{;}   \emptyset   \vdash  \ottnt{v''_{{\mathrm{2}}}}  \ottsym{:}   [    \rho'_{{\mathrm{1}}}  \odot  \rho'_{{\mathrm{2}}}    ] $,
   we have $\Sigma  \ottsym{;}   \emptyset   \vdash  \ottsym{\{}  \ell'  \ottsym{=}  \ottnt{v'_{{\mathrm{1}}}}  \ottsym{;}  \ottnt{v''_{{\mathrm{2}}}}  \ottsym{\}}  \ottsym{:}   [    \ell'  \mathbin{:}  \ottnt{B}   ;  \ottsym{(}  \rho'_{{\mathrm{1}}}  \odot  \rho'_{{\mathrm{2}}}  \ottsym{)}   ] $
   by \T{RExt}.
 \end{caseanalysis}
\end{proof}

\begin{lemma}{typing-lift-rows}
 If $\Sigma  \ottsym{;}  \Gamma  \vdash  \ottnt{e}  \ottsym{:}   \langle  \rho  \rangle $ and $\Sigma  \ottsym{;}  \Gamma  \vdash  \rho'  \ottsym{:}   \mathsf{R} $ and $\rho'  \odot  \rho$ is defined,
 then $\Sigma  \ottsym{;}  \Gamma  \vdash   \variantliftrow{ \rho' }{ \ottnt{e} }   \ottsym{:}   \langle    \rho'  \odot  \rho    \rangle $.
\end{lemma}
\begin{proof}
 By induction on $\rho'$.
 \begin{caseanalysis}
  \case $\rho' \,  =  \,  \cdot $: Trivial since $ \variantliftrow{  \cdot  }{ \ottnt{e} }  \,  =  \, \ottnt{e}$.
  \case $\rho' \,  =  \,   \ell  \mathbin{:}  \ottnt{A}   ;  \rho'' $:
   We have $ \variantliftrow{ \rho' }{ \ottnt{e} }  \,  =  \,  \variantlift{ \ell }{ \ottnt{A} }{ \ottsym{(}   \variantliftrow{ \rho'' }{ \ottnt{e} }   \ottsym{)} } $.
   Since $\Sigma  \ottsym{;}  \Gamma  \vdash  \rho'  \ottsym{:}   \mathsf{R} $, we have $\Sigma  \ottsym{;}  \Gamma  \vdash  \ottnt{A}  \ottsym{:}   \mathsf{T} $ and $\Sigma  \ottsym{;}  \Gamma  \vdash  \rho''  \ottsym{:}   \mathsf{R} $.
   By the IH, $\Sigma  \ottsym{;}  \Gamma  \vdash   \variantliftrow{ \rho'' }{ \ottnt{e} }   \ottsym{:}   \langle    \rho''  \odot  \rho    \rangle $.
   By \T{VLift}, $\Sigma  \ottsym{;}  \Gamma  \vdash   \variantlift{ \ell }{ \ottnt{A} }{ \ottsym{(}   \variantliftrow{ \rho'' }{ \ottnt{e} }   \ottsym{)} }   \ottsym{:}   \langle    \ell  \mathbin{:}  \ottnt{A}   ;  \rho''  \odot  \rho   \rangle $.
  \case otherwise: Contradictory with $\rho'  \odot  \rho$ is defined.
 \end{caseanalysis}
\end{proof}

\begin{lemma}{typing-lift-rows-after-down}
 If $\Sigma  \ottsym{;}  \Gamma  \vdash  \ottnt{e}  \ottsym{:}   \langle    \rho_{{\mathrm{1}}}  \odot  \rho_{{\mathrm{2}}}    \rangle $ and $\Sigma  \ottsym{;}  \Gamma  \vdash  \ottnt{A}  \ottsym{:}   \mathsf{T} $,
 then $\Sigma  \ottsym{;}  \Gamma  \vdash   \variantliftdown{ \rho_{{\mathrm{1}}} }{ \ell }{ \ottnt{A} }{ \ottnt{e} }   \ottsym{:}   \langle    \rho_{{\mathrm{1}}}  \odot  \ottsym{(}    \ell  \mathbin{:}  \ottnt{A}   ;   \cdot    \ottsym{)}  \odot  \rho_{{\mathrm{2}}}    \rangle $.
\end{lemma}
\begin{proof}
 By induction on $\rho_{{\mathrm{1}}}$.
 \begin{caseanalysis}
  \case $\rho_{{\mathrm{1}}} \,  =  \,   \ell'  \mathbin{:}  \ottnt{B}   ;  \rho'_{{\mathrm{1}}} $:
   We have $ \variantliftdown{ \rho_{{\mathrm{1}}} }{ \ell }{ \ottnt{A} }{ \ottnt{e} }  \,  =  \,  \mathsf{case} \,  \ottnt{e}  \,\mathsf{with}\, \langle  \ell' \,  \mathit{x}   \rightarrow   \ell' \, \mathit{x}   \ottsym{;}   \mathit{y}   \rightarrow    \variantlift{ \ell' }{ \ottnt{B} }{ \ottsym{(}   \variantliftdown{ \rho'_{{\mathrm{1}}} }{ \ell }{ \ottnt{A} }{ \mathit{y} }   \ottsym{)} }   \rangle $.
   It suffices to show that
   \[
    \Sigma  \ottsym{;}  \Gamma  \vdash   \mathsf{case} \,  \ottnt{e}  \,\mathsf{with}\, \langle  \ell' \,  \mathit{x}   \rightarrow   \ell' \, \mathit{x}   \ottsym{;}   \mathit{y}   \rightarrow    \variantlift{ \ell' }{ \ottnt{B} }{ \ottsym{(}   \variantliftdown{ \rho'_{{\mathrm{1}}} }{ \ell }{ \ottnt{A} }{ \mathit{y} }   \ottsym{)} }   \rangle   \ottsym{:}   \langle      \ell'  \mathbin{:}  \ottnt{B}   ;  \rho'_{{\mathrm{1}}}   \odot  \ottsym{(}    \ell  \mathbin{:}  \ottnt{A}   ;   \cdot    \ottsym{)}  \odot  \rho_{{\mathrm{2}}}    \rangle 
   \]

   Since $\Sigma  \ottsym{;}  \Gamma  \ottsym{,}   \mathit{y}  \mathord{:}   \langle    \rho'_{{\mathrm{1}}}  \odot  \rho_{{\mathrm{2}}}    \rangle    \vdash  \ottnt{A}  \ottsym{:}   \mathsf{T} $ by Lemmas~\ref{lem:typctx-type-wf} and \ref{lem:weakening},
   we have
   \[
    \Sigma  \ottsym{;}  \Gamma  \ottsym{,}   \mathit{y}  \mathord{:}   \langle    \rho'_{{\mathrm{1}}}  \odot  \rho_{{\mathrm{2}}}    \rangle    \vdash   \variantliftdown{ \rho'_{{\mathrm{1}}} }{ \ell }{ \ottnt{A} }{ \mathit{y} }   \ottsym{:}   \langle    \rho'_{{\mathrm{1}}}  \odot  \ottsym{(}    \ell  \mathbin{:}  \ottnt{A}   ;   \cdot    \ottsym{)}  \odot  \rho_{{\mathrm{2}}}    \rangle 
   \]
   by the IH.  Thus, by \T{VLift},
   \[
    \Sigma  \ottsym{;}  \Gamma  \ottsym{,}   \mathit{y}  \mathord{:}   \langle    \rho'_{{\mathrm{1}}}  \odot  \rho_{{\mathrm{2}}}    \rangle    \vdash   \variantlift{ \ell' }{ \ottnt{B} }{ \ottsym{(}   \variantliftdown{ \rho'_{{\mathrm{1}}} }{ \ell }{ \ottnt{A} }{ \mathit{y} }   \ottsym{)} }   \ottsym{:}   \langle      \ell'  \mathbin{:}  \ottnt{B}   ;  \rho'_{{\mathrm{1}}}   \odot  \ottsym{(}    \ell  \mathbin{:}  \ottnt{A}   ;   \cdot    \ottsym{)}  \odot  \rho_{{\mathrm{2}}}    \rangle 
   \]
   (note that $\Sigma  \ottsym{;}  \Gamma  \ottsym{,}   \mathit{y}  \mathord{:}   \langle    \rho'_{{\mathrm{1}}}  \odot  \rho_{{\mathrm{2}}}    \rangle    \vdash  \ottnt{B}  \ottsym{:}   \mathsf{T} $ by Lemmas~\ref{lem:typctx-type-wf} and \ref{lem:weakening}).
   Since $\Sigma  \ottsym{;}  \Gamma  \ottsym{,}   \mathit{x}  \mathord{:}  \ottnt{B}   \vdash  \ell' \, \mathit{x}  \ottsym{:}   \langle      \ell'  \mathbin{:}  \ottnt{B}   ;  \rho'_{{\mathrm{1}}}   \odot  \ottsym{(}    \ell  \mathbin{:}  \ottnt{A}   ;   \cdot    \ottsym{)}  \odot  \rho_{{\mathrm{2}}}    \rangle $
   by \T{VInj} (note that $\Sigma  \ottsym{;}  \Gamma  \ottsym{,}   \mathit{x}  \mathord{:}  \ottnt{B}   \vdash  \rho'_{{\mathrm{1}}}  \odot  \ottsym{(}    \ell  \mathbin{:}  \ottnt{A}   ;   \cdot    \ottsym{)}  \odot  \rho_{{\mathrm{2}}}  \ottsym{:}   \mathsf{R} $ by \reflem{weakening}),
   and $\Sigma  \ottsym{;}  \Gamma  \vdash  \ottnt{e}  \ottsym{:}   \langle      \ell'  \mathbin{:}  \ottnt{B}   ;  \rho'_{{\mathrm{1}}}   \odot  \rho_{{\mathrm{2}}}    \rangle $,
   we have
   \[
    \Sigma  \ottsym{;}  \Gamma  \vdash   \mathsf{case} \,  \ottnt{e}  \,\mathsf{with}\, \langle  \ell' \,  \mathit{x}   \rightarrow   \ell' \, \mathit{x}   \ottsym{;}   \mathit{y}   \rightarrow    \variantlift{ \ell' }{ \ottnt{B} }{ \ottsym{(}   \variantliftdown{ \rho'_{{\mathrm{1}}} }{ \ell }{ \ottnt{A} }{ \mathit{y} }   \ottsym{)} }   \rangle   \ottsym{:}   \langle      \ell'  \mathbin{:}  \ottnt{B}   ;  \rho'_{{\mathrm{1}}}   \odot  \ottsym{(}    \ell  \mathbin{:}  \ottnt{A}   ;   \cdot    \ottsym{)}  \odot  \rho_{{\mathrm{2}}}    \rangle 
   \]
   by \T{VCase}.

  \case $\rho_{{\mathrm{1}}} \,  =  \,  \cdot $:
   We have $ \variantliftdown{ \rho_{{\mathrm{1}}} }{ \ell }{ \ottnt{A} }{ \ottnt{e} }  \,  =  \,  \variantlift{ \ell }{ \ottnt{A} }{ \ottnt{e} } $.
   It suffices to show that
   $\Sigma  \ottsym{;}  \Gamma  \vdash   \variantlift{ \ell }{ \ottnt{A} }{ \ottnt{e} }   \ottsym{:}   \langle    \ell  \mathbin{:}  \ottnt{A}   ;  \rho_{{\mathrm{2}}}   \rangle $, which is shown by \T{VLift}.

  \case otherwise: Contradictory with the fact that $\rho_{{\mathrm{1}}}  \odot  \rho_{{\mathrm{2}}}$ is defined.
 \end{caseanalysis}
\end{proof}

\begin{lemmap}{Convertibility inversion: function types}{convert-inversion-fun}
 If $ \Sigma   \vdash   \ottnt{A_{{\mathrm{1}}}}  \rightarrow  \ottnt{B_{{\mathrm{1}}}}  \prec^{ \Phi }  \ottnt{A_{{\mathrm{2}}}}  \rightarrow  \ottnt{B_{{\mathrm{2}}}} $, then $ \Sigma   \vdash   \ottnt{A_{{\mathrm{2}}}}  \prec^{  \overline{ \Phi }  }  \ottnt{A_{{\mathrm{1}}}} $ and $ \Sigma   \vdash   \ottnt{B_{{\mathrm{1}}}}  \prec^{ \Phi }  \ottnt{B_{{\mathrm{2}}}} $.
\end{lemmap}
\begin{proof}
 Straightforward by case analysis on $ \Sigma   \vdash   \ottnt{A_{{\mathrm{1}}}}  \rightarrow  \ottnt{B_{{\mathrm{1}}}}  \prec^{ \Phi }  \ottnt{A_{{\mathrm{2}}}}  \rightarrow  \ottnt{B_{{\mathrm{2}}}} $.
\end{proof}

\begin{lemmap}{Convertibility inversion: universal types}{convert-inversion-forall}
 If $ \Sigma   \vdash    \text{\unboldmath$\forall\!$}  \,  \mathit{X}  \mathord{:}  \ottnt{K}   \ottsym{.} \, \ottnt{A}  \prec^{ \Phi }   \text{\unboldmath$\forall\!$}  \,  \mathit{X}  \mathord{:}  \ottnt{K}   \ottsym{.} \, \ottnt{B} $, then $ \Sigma   \vdash   \ottnt{A}  \prec^{ \Phi }  \ottnt{B} $.
\end{lemmap}
\begin{proof}
 Straightforward by case analysis on $ \Sigma   \vdash    \text{\unboldmath$\forall\!$}  \,  \mathit{X}  \mathord{:}  \ottnt{K}   \ottsym{.} \, \ottnt{A}  \prec^{ \Phi }   \text{\unboldmath$\forall\!$}  \,  \mathit{X}  \mathord{:}  \ottnt{K}   \ottsym{.} \, \ottnt{B} $.
\end{proof}

\begin{lemmap}{Convertibility inversion: record types}{convert-inversion-record}
 If $ \Sigma   \vdash    [  \rho_{{\mathrm{1}}}  ]   \prec^{ \Phi }   [  \rho_{{\mathrm{2}}}  ]  $, then $ \Sigma   \vdash   \rho_{{\mathrm{1}}}  \prec^{ \Phi }  \rho_{{\mathrm{2}}} $.
\end{lemmap}
\begin{proof}
 Straightforward by case analysis on $ \Sigma   \vdash    [  \rho_{{\mathrm{1}}}  ]   \prec^{ \Phi }   [  \rho_{{\mathrm{2}}}  ]  $.
\end{proof}

\begin{lemmap}{Convertibility inversion: variant types}{convert-inversion-variant}
 If $ \Sigma   \vdash    \langle  \rho_{{\mathrm{1}}}  \rangle   \prec^{ \Phi }   \langle  \rho_{{\mathrm{2}}}  \rangle  $, then $ \Sigma   \vdash   \rho_{{\mathrm{1}}}  \prec^{ \Phi }  \rho_{{\mathrm{2}}} $.
\end{lemmap}
\begin{proof}
 Straightforward by case analysis on $ \Sigma   \vdash    \langle  \rho_{{\mathrm{1}}}  \rangle   \prec^{ \Phi }   \langle  \rho_{{\mathrm{2}}}  \rangle  $.
\end{proof}

\begin{lemmap}{Convertibility inversion: row cons}{convert-inversion-row-cons}
 If $ \Sigma   \vdash     \ell  \mathbin{:}  \ottnt{A}   ;  \rho_{{\mathrm{1}}}   \prec^{ \Phi }    \ell  \mathbin{:}  \ottnt{B}   ;  \rho_{{\mathrm{2}}}  $, then $ \Sigma   \vdash   \ottnt{A}  \prec^{ \Phi }  \ottnt{B} $ and $ \Sigma   \vdash   \rho_{{\mathrm{1}}}  \prec^{ \Phi }  \rho_{{\mathrm{2}}} $.
\end{lemmap}
\begin{proof}
 Straightforward by case analysis on $ \Sigma   \vdash     \ell  \mathbin{:}  \ottnt{A}   ;  \rho_{{\mathrm{1}}}   \prec^{ \Phi }    \ell  \mathbin{:}  \ottnt{B}   ;  \rho_{{\mathrm{2}}}  $.
\end{proof}

\begin{lemmap}{Subject reduction on reduction}{subject-red-red}
 If $\Sigma  \ottsym{;}   \emptyset   \vdash  \ottnt{e}  \ottsym{:}  \ottnt{A}$ and $\ottnt{e}  \rightsquigarrow  \ottnt{e'}$,
 then $\Sigma  \ottsym{;}   \emptyset   \vdash  \ottnt{e'}  \ottsym{:}  \ottnt{A}$.
\end{lemmap}
\begin{proof}
 By case analysis on the derivation of $\Sigma  \ottsym{;}   \emptyset   \vdash  \ottnt{e}  \ottsym{:}  \ottnt{A}$.
 \begin{caseanalysis}
  \case \T{Var}, \T{Const}, \T{Lam}, \T{TLam}, \T{REmp}, \T{Blame}:
   Contradictory; there are no reduction rules to apply.
  \case \T{TApp}, \T{RExt}, \T{VInj}, \T{VLift}:
   Contradictory; there are no applicable reduction rules.

  \case \T{App}:
   We have $\ottnt{e} \,  =  \, \ottnt{e_{{\mathrm{1}}}} \, \ottnt{e_{{\mathrm{2}}}}$ and, by inversion,
   $\Sigma  \ottsym{;}   \emptyset   \vdash  \ottnt{e_{{\mathrm{1}}}}  \ottsym{:}  \ottnt{B}  \rightarrow  \ottnt{A}$ and $\Sigma  \ottsym{;}   \emptyset   \vdash  \ottnt{e_{{\mathrm{2}}}}  \ottsym{:}  \ottnt{B}$
   for some $\ottnt{e_{{\mathrm{1}}}}$, $\ottnt{e_{{\mathrm{2}}}}$, and $\ottnt{B}$.
   By case analysis on the reduction rules applicable to $\ottnt{e_{{\mathrm{1}}}} \, \ottnt{e_{{\mathrm{2}}}}$.
   \begin{caseanalysis}
    \case \R{Cons}:
     We have $\ottnt{e_{{\mathrm{1}}}} \,  =  \, \kappa_{{\mathrm{1}}}$ and $\ottnt{e_{{\mathrm{2}}}} \,  =  \, \kappa_{{\mathrm{2}}}$ and $\ottnt{e'} \,  =  \,  \zeta  (  \kappa_{{\mathrm{1}}}  ,  \kappa_{{\mathrm{2}}}  ) $
     for some $\kappa_{{\mathrm{1}}}$ and $\kappa_{{\mathrm{2}}}$.
     By \reflem{value-inversion-constant}, $ \mathit{ty}  (  \kappa_{{\mathrm{1}}}  )  \,  =  \, \ottnt{B}  \rightarrow  \ottnt{A}$.
     By the assumptions about constants, $ \mathit{ty}  (   \zeta  (  \kappa_{{\mathrm{1}}}  ,  \kappa_{{\mathrm{2}}}  )   )  \,  =  \, \ottnt{A}$.
     Thus, $\Sigma  \ottsym{;}   \emptyset   \vdash   \zeta  (  \kappa_{{\mathrm{1}}}  ,  \kappa_{{\mathrm{2}}}  )   \ottsym{:}  \ottnt{A}$ by \T{Const}.

    \case \R{Beta}:
     By \reflem{canonical-forms},
     $\ottnt{e_{{\mathrm{1}}}} \,  =  \,  \lambda\!  \,  \mathit{x}  \mathord{:}  \ottnt{B}   \ottsym{.}  \ottnt{e'_{{\mathrm{1}}}}$ and $\ottnt{e_{{\mathrm{2}}}} \,  =  \, \ottnt{v_{{\mathrm{2}}}}$ and $\ottnt{e'} \,  =  \,  \ottnt{e'_{{\mathrm{1}}}}    [  \ottnt{v_{{\mathrm{2}}}}  \ottsym{/}  \mathit{x}  ]  $
     for some $\mathit{x}$, $\ottnt{e'_{{\mathrm{1}}}}$, and $\ottnt{v_{{\mathrm{2}}}}$.
     By \reflem{value-inversion-lambda},
     $\Sigma  \ottsym{;}   \mathit{x}  \mathord{:}  \ottnt{B}   \vdash  \ottnt{e'_{{\mathrm{1}}}}  \ottsym{:}  \ottnt{A}$.
     Since $\Sigma  \ottsym{;}   \emptyset   \vdash  \ottnt{v_{{\mathrm{2}}}}  \ottsym{:}  \ottnt{B}$, we have
     $\Sigma  \ottsym{;}   \emptyset   \vdash   \ottnt{e'_{{\mathrm{1}}}}    [  \ottnt{v_{{\mathrm{2}}}}  \ottsym{/}  \mathit{x}  ]    \ottsym{:}  \ottnt{A}$ by \reflem{value-subst}.
   \end{caseanalysis}

  \case \T{RLet}:
   We have $\ottnt{e} \,  =  \, \mathsf{let} \, \ottsym{\{}  \ell  \ottsym{=}  \mathit{x}  \ottsym{;}  \mathit{y}  \ottsym{\}}  \ottsym{=}  \ottnt{e_{{\mathrm{1}}}} \, \mathsf{in} \, \ottnt{e_{{\mathrm{2}}}}$ and, by inversion,
   $\Sigma  \ottsym{;}   \emptyset   \vdash  \ottnt{e_{{\mathrm{1}}}}  \ottsym{:}   [    \ell  \mathbin{:}  \ottnt{B}   ;  \rho   ] $ and
   $\Sigma  \ottsym{;}   \mathit{x}  \mathord{:}  \ottnt{B}   \ottsym{,}   \mathit{y}  \mathord{:}   [  \rho  ]    \vdash  \ottnt{e_{{\mathrm{2}}}}  \ottsym{:}  \ottnt{A}$.
   The reduction rules applicable to $\ottnt{e}$ is only \R{Record}.
   We can suppose that $\ottnt{e_{{\mathrm{1}}}} \,  =  \, \ottsym{\{}  \ell  \ottsym{=}  \ottnt{v_{{\mathrm{1}}}}  \ottsym{;}  \ottnt{v_{{\mathrm{2}}}}  \ottsym{\}}$ and $\ottnt{e'} \,  =  \,  \ottnt{e_{{\mathrm{2}}}}    [  \ottnt{v_{{\mathrm{1}}}}  \ottsym{/}  \mathit{x}  \ottsym{,}  \ottnt{v_{{\mathrm{2}}}}  \ottsym{/}  \mathit{y}  ]  $.
   By \reflem{value-inversion-record-cons},
   $\Sigma  \ottsym{;}   \emptyset   \vdash  \ottnt{v_{{\mathrm{1}}}}  \ottsym{:}  \ottnt{B}$ and $\Sigma  \ottsym{;}   \emptyset   \vdash  \ottnt{v_{{\mathrm{2}}}}  \ottsym{:}   [  \rho  ] $.
   Since $\Sigma  \ottsym{;}   \mathit{x}  \mathord{:}  \ottnt{B}   \ottsym{,}   \mathit{y}  \mathord{:}   [  \rho  ]    \vdash  \ottnt{e_{{\mathrm{2}}}}  \ottsym{:}  \ottnt{A}$,
   we have $\Sigma  \ottsym{;}   \emptyset   \vdash   \ottnt{e_{{\mathrm{2}}}}    [  \ottnt{v_{{\mathrm{1}}}}  \ottsym{/}  \mathit{x}  \ottsym{,}  \ottnt{v_{{\mathrm{2}}}}  \ottsym{/}  \mathit{y}  ]    \ottsym{:}  \ottnt{A}$ by \reflem{value-subst}.

  \case \T{VCase}:
   We have $\ottnt{e} \,  =  \,  \mathsf{case} \,  \ottnt{e_{{\mathrm{0}}}}  \,\mathsf{with}\, \langle  \ell \,  \mathit{x}   \rightarrow   \ottnt{e_{{\mathrm{1}}}}   \ottsym{;}   \mathit{y}   \rightarrow   \ottnt{e_{{\mathrm{2}}}}  \rangle $ and, by inversion,
   $\Sigma  \ottsym{;}   \emptyset   \vdash  \ottnt{e_{{\mathrm{0}}}}  \ottsym{:}   \langle    \ell  \mathbin{:}  \ottnt{B}   ;  \rho   \rangle $ and
   $\Sigma  \ottsym{;}   \mathit{x}  \mathord{:}  \ottnt{B}   \vdash  \ottnt{e_{{\mathrm{1}}}}  \ottsym{:}  \ottnt{A}$ and
   $\Sigma  \ottsym{;}   \mathit{y}  \mathord{:}   \langle  \rho  \rangle    \vdash  \ottnt{e_{{\mathrm{2}}}}  \ottsym{:}  \ottnt{A}$
   for some $\ottnt{e_{{\mathrm{0}}}}$, $\ottnt{e_{{\mathrm{1}}}}$, $\ottnt{e_{{\mathrm{2}}}}$, $\ell$, $\mathit{x}$, $\mathit{y}$, $\ottnt{B}$, and $\rho$.
   By case analysis on the reduction rules applicable to $\ottnt{e}$.
   \begin{caseanalysis}
    \case \R{CaseL}:
     We can suppose that $\ottnt{e_{{\mathrm{0}}}} \,  =  \, \ell \, \ottnt{v}$ and $\ottnt{e'} \,  =  \,  \ottnt{e_{{\mathrm{1}}}}    [  \ottnt{v}  \ottsym{/}  \mathit{x}  ]  $
     for some $\ottnt{v}$.
     By \reflem{value-inversion-variant-inj}, $\Sigma  \ottsym{;}   \emptyset   \vdash  \ottnt{v}  \ottsym{:}  \ottnt{B}$.
     Since $\Sigma  \ottsym{;}   \mathit{x}  \mathord{:}  \ottnt{B}   \vdash  \ottnt{e_{{\mathrm{1}}}}  \ottsym{:}  \ottnt{A}$,
     we have $\Sigma  \ottsym{;}   \emptyset   \vdash   \ottnt{e_{{\mathrm{1}}}}    [  \ottnt{v}  \ottsym{/}  \mathit{x}  ]    \ottsym{:}  \ottnt{A}$ by \reflem{value-subst}.

    \case \R{CaseR}:
     We can suppose that $\ottnt{e_{{\mathrm{0}}}} \,  =  \,  \variantlift{ \ell }{ \ottnt{C} }{ \ottnt{v} } $ and $\ottnt{e'} \,  =  \,  \ottnt{e_{{\mathrm{2}}}}    [  \ottnt{v}  \ottsym{/}  \mathit{y}  ]  $
     for some $\ottnt{C}$ and $\ottnt{v}$.
     By \reflem{value-inversion-variant-lift}, $\Sigma  \ottsym{;}   \emptyset   \vdash  \ottnt{v}  \ottsym{:}   \langle  \rho  \rangle $.
     Since $\Sigma  \ottsym{;}   \mathit{y}  \mathord{:}   \langle  \rho  \rangle    \vdash  \ottnt{e_{{\mathrm{2}}}}  \ottsym{:}  \ottnt{A}$,
     we have $\Sigma  \ottsym{;}   \emptyset   \vdash   \ottnt{e_{{\mathrm{2}}}}    [  \ottnt{v}  \ottsym{/}  \mathit{y}  ]    \ottsym{:}  \ottnt{A}$ by \reflem{value-subst}.
   \end{caseanalysis}

  \case \T{Cast}:
   We have $\ottnt{e} \,  =  \, \ottnt{e_{{\mathrm{0}}}}  \ottsym{:}  \ottnt{B} \,  \stackrel{ \ottnt{p} }{\Rightarrow}  \ottnt{A} $ and, by inversion,
   $\Sigma  \ottsym{;}   \emptyset   \vdash  \ottnt{e_{{\mathrm{0}}}}  \ottsym{:}  \ottnt{B}$ and $\ottnt{B}  \simeq  \ottnt{A}$ and $\Sigma  \ottsym{;}   \emptyset   \vdash  \ottnt{A}  \ottsym{:}   \mathsf{T} $
   for some $\ottnt{e_{{\mathrm{0}}}}$, $\ottnt{B}$, and $\ottnt{p}$.
   Besides, we have $\Sigma  \ottsym{;}   \emptyset   \vdash  \ottnt{B}  \ottsym{:}   \mathsf{T} $ by \reflem{typctx-type-wf}.
   By case analysis on the reduction rules applicable to $\ottnt{e}$.
   \begin{caseanalysis}
    \case \R{IdDyn}, \R{IdBase}, \R{IdName}, \R{REmp}, \R{RIdName}, and \R{VIdName}:
     We have $\ottnt{B} \,  =  \, \ottnt{A}$ and $\ottnt{e_{{\mathrm{0}}}} \,  =  \, \ottnt{v}$ and $\ottnt{e'} \,  =  \, \ottnt{v}$ for some $\ottnt{v}$.
     Since $\Sigma  \ottsym{;}   \emptyset   \vdash  \ottnt{e_{{\mathrm{0}}}}  \ottsym{:}  \ottnt{B}$, we have $\Sigma  \ottsym{;}   \emptyset   \vdash  \ottnt{v}  \ottsym{:}  \ottnt{A}$, which
     is what we have to show.

    \case \R{Blame}, \R{RBlame}, \R{VBlame}:
     Obvious by \T{Blame} since $\ottnt{e'} \,  =  \, \mathsf{blame} \, \ottnt{q}$ for some $\ottnt{q}$.

    \case \R{ToDyn}:
     We have $\ottnt{e_{{\mathrm{0}}}} \,  =  \, \ottnt{v}$ and $\ottnt{A} \,  =  \, \star$ and $\ottnt{e'} \,  =  \, \ottnt{v}  \ottsym{:}  \ottnt{B} \,  \stackrel{ \ottnt{p} }{\Rightarrow}  \mathit{G}  \,  \stackrel{ \ottnt{p} }{\Rightarrow}  \star $
     for some $\ottnt{v}$ and $\mathit{G}$ such that $\ottnt{B}  \simeq  \mathit{G}$.
     Since $\Sigma  \ottsym{;}   \emptyset   \vdash  \ottnt{v}  \ottsym{:}  \ottnt{B}$ and $\ottnt{B}  \simeq  \mathit{G}$,
     we have $\Sigma  \ottsym{;}   \emptyset   \vdash  \mathit{G}  \ottsym{:}   \mathsf{T} $ by \reflem{consistent-ground-type-wf}.
     Thus, $\Sigma  \ottsym{;}   \emptyset   \vdash  \ottnt{v}  \ottsym{:}  \ottnt{B} \,  \stackrel{ \ottnt{p} }{\Rightarrow}  \mathit{G}  \,  \stackrel{ \ottnt{p} }{\Rightarrow}  \star   \ottsym{:}  \star$
     by \T{Cast}.

    \case \R{FromDyn}:
     We have $\ottnt{e_{{\mathrm{0}}}} \,  =  \, \ottnt{v}$ and $\ottnt{B} \,  =  \, \star$ and
     $\ottnt{e'} \,  =  \, \ottnt{v}  \ottsym{:}  \star \,  \stackrel{ \ottnt{p} }{\Rightarrow}  \mathit{G}  \,  \stackrel{ \ottnt{p} }{\Rightarrow}  \ottnt{A} $ for some $\ottnt{v}$ and $\mathit{G}$ such that
     $\ottnt{A}  \simeq  \mathit{G}$.
     Since $\Sigma  \ottsym{;}   \emptyset   \vdash  \ottnt{A}  \ottsym{:}   \mathsf{T} $ and $\ottnt{A}  \simeq  \mathit{G}$,
     we have $\Sigma  \ottsym{;}   \emptyset   \vdash  \mathit{G}  \ottsym{:}   \mathsf{T} $ by \reflem{consistent-ground-type-wf}.
     Since $\Sigma  \ottsym{;}   \emptyset   \vdash  \ottnt{v}  \ottsym{:}  \star$, we have
     $\Sigma  \ottsym{;}   \emptyset   \vdash  \ottnt{v}  \ottsym{:}  \star \,  \stackrel{ \ottnt{p} }{\Rightarrow}  \mathit{G}  \,  \stackrel{ \ottnt{p} }{\Rightarrow}  \ottnt{A}   \ottsym{:}  \ottnt{A}$ by \T{Cast}
     (note that $\mathit{G}  \simeq  \ottnt{A}$ by \reflem{consistent-symm}).

    \case \R{Ground}:
     We have $\ottnt{e_{{\mathrm{0}}}} \,  =  \, \ottnt{v}  \ottsym{:}  \mathit{G} \,  \stackrel{ \ottnt{p} }{\Rightarrow}  \star $ and $\ottnt{B} \,  =  \, \star$ and $\ottnt{A} \,  =  \, \mathit{G}$ and $\ottnt{e'} \,  =  \, \ottnt{v}$
     for some $\ottnt{v}$ and $\mathit{G}$.
     Since $\Sigma  \ottsym{;}   \emptyset   \vdash  \ottnt{e_{{\mathrm{0}}}}  \ottsym{:}  \ottnt{B}$, i.e., $\Sigma  \ottsym{;}   \emptyset   \vdash  \ottnt{v}  \ottsym{:}  \mathit{G} \,  \stackrel{ \ottnt{p} }{\Rightarrow}  \star   \ottsym{:}  \star$,
     we have $\Sigma  \ottsym{;}   \emptyset   \vdash  \ottnt{v}  \ottsym{:}  \mathit{G}$ by \reflem{value-inversion-cast}.
     Thus, we have $\Sigma  \ottsym{;}   \emptyset   \vdash  \ottnt{e'}  \ottsym{:}  \ottnt{A}$.

    \case \R{Wrap}:
     We have $\ottnt{e_{{\mathrm{0}}}} \,  =  \, \ottnt{v}$ and $\ottnt{B} \,  =  \, \ottnt{B_{{\mathrm{1}}}}  \rightarrow  \ottnt{B_{{\mathrm{2}}}}$ and $\ottnt{A} \,  =  \, \ottnt{A_{{\mathrm{1}}}}  \rightarrow  \ottnt{A_{{\mathrm{2}}}}$ and
     $\ottnt{e'} \,  =  \,  \lambda\!  \,  \mathit{x}  \mathord{:}  \ottnt{A_{{\mathrm{1}}}}   \ottsym{.}  \ottnt{v} \, \ottsym{(}  \mathit{x}  \ottsym{:}  \ottnt{A_{{\mathrm{1}}}} \,  \stackrel{  \overline{ \ottnt{p} }  }{\Rightarrow}  \ottnt{B_{{\mathrm{1}}}}   \ottsym{)}  \ottsym{:}  \ottnt{B_{{\mathrm{2}}}} \,  \stackrel{ \ottnt{p} }{\Rightarrow}  \ottnt{A_{{\mathrm{2}}}} $.
     Since $\ottnt{B_{{\mathrm{1}}}}  \rightarrow  \ottnt{B_{{\mathrm{2}}}}  \simeq  \ottnt{A_{{\mathrm{1}}}}  \rightarrow  \ottnt{A_{{\mathrm{2}}}}$,
     we have $\ottnt{A_{{\mathrm{1}}}}  \simeq  \ottnt{B_{{\mathrm{1}}}}$ and $\ottnt{B_{{\mathrm{2}}}}  \simeq  \ottnt{A_{{\mathrm{2}}}}$
     by Lemmas~\ref{lem:consistent-inv-fun} and \ref{lem:consistent-symm}.
     Besides, we have $\Sigma  \ottsym{;}   \emptyset   \vdash  \ottnt{A_{{\mathrm{1}}}}  \ottsym{:}   \mathsf{T} $, $\Sigma  \ottsym{;}   \emptyset   \vdash  \ottnt{A_{{\mathrm{2}}}}  \ottsym{:}   \mathsf{T} $,
     $\Sigma  \ottsym{;}   \emptyset   \vdash  \ottnt{B_{{\mathrm{1}}}}  \ottsym{:}   \mathsf{T} $, and $\Sigma  \ottsym{;}   \emptyset   \vdash  \ottnt{B_{{\mathrm{2}}}}  \ottsym{:}   \mathsf{T} $
     since $\Sigma  \ottsym{;}   \emptyset   \vdash  \ottnt{A_{{\mathrm{1}}}}  \rightarrow  \ottnt{A_{{\mathrm{2}}}}  \ottsym{:}   \mathsf{T} $ and $\Sigma  \ottsym{;}   \emptyset   \vdash  \ottnt{B_{{\mathrm{1}}}}  \rightarrow  \ottnt{B_{{\mathrm{2}}}}  \ottsym{:}   \mathsf{T} $.
     Thus, since $\Sigma  \ottsym{;}   \mathit{x}  \mathord{:}  \ottnt{A_{{\mathrm{1}}}}   \vdash  \ottnt{v}  \ottsym{:}  \ottnt{B_{{\mathrm{1}}}}  \rightarrow  \ottnt{B_{{\mathrm{2}}}}$ by \reflem{weakening},
     we have $\Sigma  \ottsym{;}   \emptyset   \vdash   \lambda\!  \,  \mathit{x}  \mathord{:}  \ottnt{A_{{\mathrm{1}}}}   \ottsym{.}  \ottnt{v} \, \ottsym{(}  \mathit{x}  \ottsym{:}  \ottnt{A_{{\mathrm{1}}}} \,  \stackrel{  \overline{ \ottnt{p} }  }{\Rightarrow}  \ottnt{B_{{\mathrm{1}}}}   \ottsym{)}  \ottsym{:}  \ottnt{B_{{\mathrm{2}}}} \,  \stackrel{ \ottnt{p} }{\Rightarrow}  \ottnt{A_{{\mathrm{2}}}}   \ottsym{:}  \ottnt{A_{{\mathrm{1}}}}  \rightarrow  \ottnt{A_{{\mathrm{2}}}}$.

    \case \R{Content}:
     We have $\ottnt{e_{{\mathrm{0}}}} \,  =  \, \ottnt{v}$ and $\ottnt{B} \,  =  \,  \text{\unboldmath$\forall\!$}  \,  \mathit{X}  \mathord{:}  \ottnt{K}   \ottsym{.} \, \ottnt{B'}$ and $\ottnt{A} \,  =  \,  \text{\unboldmath$\forall\!$}  \,  \mathit{X}  \mathord{:}  \ottnt{K}   \ottsym{.} \, \ottnt{A'}$ and
     $\ottnt{e'} \,  =  \,   \Lambda\!  \,  \mathit{X}  \mathord{:}  \ottnt{K}   \ottsym{.}   \ottsym{(}  \ottnt{v} \, \mathit{X}  \ottsym{:}  \ottnt{B'} \,  \stackrel{ \ottnt{p} }{\Rightarrow}  \ottnt{A'}   \ottsym{)}  ::  \ottnt{A'} $
     for some $\ottnt{v}$, $\mathit{X}$, $\ottnt{K}$, $\ottnt{A'}$, and $\ottnt{B'}$.
     Since $ \text{\unboldmath$\forall\!$}  \,  \mathit{X}  \mathord{:}  \ottnt{K}   \ottsym{.} \, \ottnt{B'}  \simeq   \text{\unboldmath$\forall\!$}  \,  \mathit{X}  \mathord{:}  \ottnt{K}   \ottsym{.} \, \ottnt{A'}$,
     we have $\ottnt{B'}  \simeq  \ottnt{A'}$ by \reflem{consistent-inv-forall}.
     Since $\Sigma  \ottsym{;}   \emptyset   \vdash   \text{\unboldmath$\forall\!$}  \,  \mathit{X}  \mathord{:}  \ottnt{K}   \ottsym{.} \, \ottnt{A'}  \ottsym{:}   \mathsf{T} $,
     we have $\Sigma  \ottsym{;}   \mathit{X}  \mathord{:}  \ottnt{K}   \vdash  \ottnt{A'}  \ottsym{:}   \mathsf{T} $.
     Thus, since $\Sigma  \ottsym{;}   \mathit{X}  \mathord{:}  \ottnt{K}   \vdash  \ottnt{v}  \ottsym{:}   \text{\unboldmath$\forall\!$}  \,  \mathit{X}  \mathord{:}  \ottnt{K}   \ottsym{.} \, \ottnt{B'}$ by \reflem{weakening},
     we have $\Sigma  \ottsym{;}   \emptyset   \vdash    \Lambda\!  \,  \mathit{X}  \mathord{:}  \ottnt{K}   \ottsym{.}   \ottsym{(}  \ottnt{v} \, \mathit{X}  \ottsym{:}  \ottnt{B'} \,  \stackrel{ \ottnt{p} }{\Rightarrow}  \ottnt{A'}   \ottsym{)}  ::  \ottnt{A'}   \ottsym{:}   \text{\unboldmath$\forall\!$}  \,  \mathit{X}  \mathord{:}  \ottnt{K}   \ottsym{.} \, \ottnt{A'}$.

    \case \R{Inst}:
     We have $\ottnt{e_{{\mathrm{0}}}} \,  =  \, \ottnt{v}$ and $\ottnt{B} \,  =  \,  \text{\unboldmath$\forall\!$}  \,  \mathit{X}  \mathord{:}  \ottnt{K}   \ottsym{.} \, \ottnt{B'}$ and $\ottnt{e'} \,  =  \, \ottsym{(}  \ottnt{v} \, \star  \ottsym{)}  \ottsym{:}   \ottnt{B'}    [  \star  /  \mathit{X}  ]   \,  \stackrel{ \ottnt{p} }{\Rightarrow}  \ottnt{A} $
     for some $\ottnt{v}$, $\mathit{X}$, $\ottnt{K}$, and $\ottnt{B'}$.
     Besides, $\mathbf{QPoly} \, \ottsym{(}  \ottnt{A}  \ottsym{)}$.

     Since $\Sigma  \ottsym{;}   \emptyset   \vdash  \ottnt{v}  \ottsym{:}   \text{\unboldmath$\forall\!$}  \,  \mathit{X}  \mathord{:}  \ottnt{K}   \ottsym{.} \, \ottnt{B'}$, we have
     \[
     \Sigma  \ottsym{;}   \emptyset   \vdash  \ottnt{v} \, \star  \ottsym{:}   \ottnt{B'}    [  \star  /  \mathit{X}  ]  .
     \]

     Since $\mathbf{QPoly} \, \ottsym{(}  \ottnt{A}  \ottsym{)}$ and $\ottnt{B}  \simeq  \ottnt{A}$, i.e., $ \text{\unboldmath$\forall\!$}  \,  \mathit{X}  \mathord{:}  \ottnt{K}   \ottsym{.} \, \ottnt{B'}  \simeq  \ottnt{A}$, we have
     $\ottnt{B'}  \simeq  \ottnt{A}$ and $[[X notin ftv(A)]]$.
     Thus, by \reflem{type-subst-consistency}, $ \ottnt{B'}    [  \star  /  \mathit{X}  ]    \simeq  \ottnt{A}$.
     By \T{Cast},
     \[
      \Sigma  \ottsym{;}   \emptyset   \vdash  \ottsym{(}  \ottnt{v} \, \star  \ottsym{)}  \ottsym{:}   \ottnt{B'}    [  \star  /  \mathit{X}  ]   \,  \stackrel{ \ottnt{p} }{\Rightarrow}  \ottnt{A}   \ottsym{:}  \ottnt{A}.
     \]

    \case \R{Gen}:
     We have $\ottnt{e_{{\mathrm{0}}}} \,  =  \, \ottnt{v}$ and $\ottnt{A} \,  =  \,  \text{\unboldmath$\forall\!$}  \,  \mathit{X}  \mathord{:}  \ottnt{K}   \ottsym{.} \, \ottnt{A'}$ and $\ottnt{e'} \,  =  \,   \Lambda\!  \,  \mathit{X}  \mathord{:}  \ottnt{K}   \ottsym{.}   \ottsym{(}  \ottnt{v}  \ottsym{:}  \ottnt{B} \,  \stackrel{ \ottnt{p} }{\Rightarrow}  \ottnt{A'}   \ottsym{)}  ::  \ottnt{A'} $
     for some $\ottnt{v}$, $\mathit{X}$, $\ottnt{K}$, and $\ottnt{A'}$.
     Besides, $\mathbf{QPoly} \, \ottsym{(}  \ottnt{B}  \ottsym{)}$.

     Since $\Sigma  \ottsym{;}   \emptyset   \vdash  \ottnt{v}  \ottsym{:}  \ottnt{B}$, we have
     \[
     \Sigma  \ottsym{;}   \mathit{X}  \mathord{:}  \ottnt{K}   \vdash  \ottnt{v}  \ottsym{:}  \ottnt{B}
     \]
     by \reflem{weakening}.

     Since $\mathbf{QPoly} \, \ottsym{(}  \ottnt{B}  \ottsym{)}$ and $\ottnt{B}  \simeq  \ottnt{A}$, i.e., $\ottnt{B}  \simeq   \text{\unboldmath$\forall\!$}  \,  \mathit{X}  \mathord{:}  \ottnt{K}   \ottsym{.} \, \ottnt{A'}$, we have
     $\ottnt{B}  \simeq  \ottnt{A'}$ and $[[X notin ftv(B)]]$
     by Lemmas~\ref{lem:consistent-symm} and \ref{lem:consistent-inv-forall-qpoly}.
     Furthermore, $\Sigma  \ottsym{;}   \emptyset   \vdash   \text{\unboldmath$\forall\!$}  \,  \mathit{X}  \mathord{:}  \ottnt{K}   \ottsym{.} \, \ottnt{A'}  \ottsym{:}   \mathsf{T} $, we have $\Sigma  \ottsym{;}   \mathit{X}  \mathord{:}  \ottnt{K}   \vdash  \ottnt{A'}  \ottsym{:}   \mathsf{T} $.
     Thus, we have $\Sigma  \ottsym{;}   \emptyset   \vdash    \Lambda\!  \,  \mathit{X}  \mathord{:}  \ottnt{K}   \ottsym{.}   \ottsym{(}  \ottnt{v}  \ottsym{:}  \ottnt{B} \,  \stackrel{ \ottnt{p} }{\Rightarrow}  \ottnt{A'}   \ottsym{)}  ::  \ottnt{A'}   \ottsym{:}   \text{\unboldmath$\forall\!$}  \,  \mathit{X}  \mathord{:}  \ottnt{K}   \ottsym{.} \, \ottnt{A'}$.

    \case \R{RToDyn}:
     We have $\ottnt{e_{{\mathrm{0}}}} \,  =  \, \ottnt{v}$ and $\ottnt{A} \,  =  \,  [  \star  ] $ and $\ottnt{B} \,  =  \,  [  \rho  ] $ and
     $\ottnt{e'} \,  =  \, \ottnt{v}  \ottsym{:}   [  \rho  ]  \,  \stackrel{ \ottnt{p} }{\Rightarrow}   [   \mathit{grow}  (  \rho  )   ]   \,  \stackrel{ \ottnt{p} }{\Rightarrow}   [  \star  ]  $
     for some $\ottnt{v}$ and $\rho$ such that $\rho \,  \not=  \,  \mathit{grow}  (  \rho  ) $.
     By \reflem{consistent-grow}, $\rho  \simeq   \mathit{grow}  (  \rho  ) $, and therefore
     $ [  \rho  ]   \simeq   [   \mathit{grow}  (  \rho  )   ] $ by \CE{Record}.
     Since $\Sigma  \ottsym{;}   \emptyset   \vdash  \ottnt{v}  \ottsym{:}   [  \rho  ] $ and $\Sigma  \ottsym{;}   \emptyset   \vdash   [   \mathit{grow}  (  \rho  )   ]   \ottsym{:}   \mathsf{T} $ by \reflem{wf-grow}, we have
     $\Sigma  \ottsym{;}   \emptyset   \vdash  \ottnt{v}  \ottsym{:}   [  \rho  ]  \,  \stackrel{ \ottnt{p} }{\Rightarrow}   [   \mathit{grow}  (  \rho  )   ]   \,  \stackrel{ \ottnt{p} }{\Rightarrow}   [  \star  ]    \ottsym{:}   [  \star  ] $.

    \case \R{RFromDyn}:
     We have
     $\ottnt{e_{{\mathrm{0}}}} \,  =  \, \ottnt{v}  \ottsym{:}   [  \gamma  ]  \,  \stackrel{ \ottnt{q} }{\Rightarrow}   [  \star  ]  $ and
     $\ottnt{B} \,  =  \,  [  \star  ] $ and
     $\ottnt{A} \,  =  \,  [  \rho_{{\mathrm{1}}}  ] $ and
     $\ottnt{e'} \,  =  \, \ottnt{v}  \ottsym{:}   [  \gamma  ]  \,  \stackrel{ \ottnt{q} }{\Rightarrow}   [  \rho_{{\mathrm{1}}}  ]  $
     for some $\ottnt{v}$, $\gamma$, $\rho_{{\mathrm{1}}}$, and $\ottnt{q}$ such that
     $\gamma  \simeq  \rho_{{\mathrm{1}}}$.
     Since $\gamma  \simeq  \rho_{{\mathrm{1}}}$, we have $ [  \gamma  ]   \simeq   [  \rho_{{\mathrm{1}}}  ] $ by \CE{Record}.
     Since $\Sigma  \ottsym{;}   \emptyset   \vdash  \ottnt{e_{{\mathrm{0}}}}  \ottsym{:}  \ottnt{B}$, i.e., $\Sigma  \ottsym{;}   \emptyset   \vdash  \ottnt{v}  \ottsym{:}   [  \gamma  ]  \,  \stackrel{ \ottnt{q} }{\Rightarrow}   [  \star  ]    \ottsym{:}   [  \star  ] $,
     we have $\Sigma  \ottsym{;}   \emptyset   \vdash  \ottnt{v}  \ottsym{:}   [  \gamma  ] $ by \reflem{value-inversion-cast}.
     Thus, we have $\Sigma  \ottsym{;}   \emptyset   \vdash  \ottnt{v}  \ottsym{:}   [  \gamma  ]  \,  \stackrel{ \ottnt{q} }{\Rightarrow}   [  \rho_{{\mathrm{1}}}  ]    \ottsym{:}   [  \rho_{{\mathrm{1}}}  ] $ by \T{Cast}.

    \case \R{RRev}:
     We have
     \begin{itemize}
      \item $\ottnt{e_{{\mathrm{0}}}} \,  =  \, \ottnt{v}$,
      \item $\ottnt{A} \,  =  \,  [    \ell  \mathbin{:}  \ottnt{A'}   ;  \rho_{{\mathrm{1}}}   ] $,
      \item $\ottnt{B} \,  =  \,  [  \rho_{{\mathrm{2}}}  ] $, and
      \item $\ottnt{e'} \,  =  \, \ottsym{\{}  \ell  \ottsym{=}  \ottsym{(}  \ottnt{v_{{\mathrm{1}}}}  \ottsym{:}  \ottnt{B'} \,  \stackrel{ \ottnt{p} }{\Rightarrow}  \ottnt{A'}   \ottsym{)}  \ottsym{;}  \ottnt{v_{{\mathrm{2}}}}  \ottsym{:}   [  \rho'_{{\mathrm{2}}}  ]  \,  \stackrel{ \ottnt{p} }{\Rightarrow}   [  \rho_{{\mathrm{1}}}  ]    \ottsym{\}}$
     \end{itemize}
     for some $\ottnt{v}$, $\ell$, $\ottnt{A'}$, $\ottnt{B'}$, $\rho_{{\mathrm{1}}}$, $\rho_{{\mathrm{2}}}$, and $\rho'_{{\mathrm{2}}}$
     such that $\ottnt{v} \,  \triangleright _{ \ell }  \, \ottnt{v_{{\mathrm{1}}}}  \ottsym{,}  \ottnt{v_{{\mathrm{2}}}}$ and $\rho_{{\mathrm{2}}} \,  \triangleright _{ \ell }  \, \ottnt{B'}  \ottsym{,}  \rho'_{{\mathrm{2}}}$.
     Since $\Sigma  \ottsym{;}   \emptyset   \vdash  \ottnt{v}  \ottsym{:}  \ottnt{B}$ $\ottnt{B} \,  =  \,  [  \rho_{{\mathrm{2}}}  ] $ and $\ottnt{v} \,  \triangleright _{ \ell }  \, \ottnt{v_{{\mathrm{1}}}}  \ottsym{,}  \ottnt{v_{{\mathrm{2}}}}$,
     there exist some $\rho_{{\mathrm{21}}}$, $\rho_{{\mathrm{22}}}$, and $\ottnt{B'}$ such that
     \begin{itemize}
      \item $\rho_{{\mathrm{2}}} \,  =  \, \rho_{{\mathrm{21}}}  \odot  \ottsym{(}    \ell  \mathbin{:}  \ottnt{B'}   ;   \cdot    \ottsym{)}  \odot  \rho_{{\mathrm{22}}}$,
      \item $\rho'_{{\mathrm{2}}} \,  =  \, \rho_{{\mathrm{21}}}  \odot  \rho_{{\mathrm{22}}}$,
      \item $\ell \,  \not\in  \, \mathit{dom} \, \ottsym{(}  \rho_{{\mathrm{21}}}  \ottsym{)}$,
      \item $\Sigma  \ottsym{;}   \emptyset   \vdash  \ottnt{v_{{\mathrm{1}}}}  \ottsym{:}  \ottnt{B'}$, and
      \item $\Sigma  \ottsym{;}   \emptyset   \vdash  \ottnt{v_{{\mathrm{2}}}}  \ottsym{:}   [    \rho_{{\mathrm{21}}}  \odot  \rho_{{\mathrm{22}}}    ] $.
     \end{itemize}
     Since $\ottnt{B}  \simeq  \ottnt{A}$, i.e., $ [    \rho_{{\mathrm{21}}}  \odot  \ottsym{(}    \ell  \mathbin{:}  \ottnt{B'}   ;   \cdot    \ottsym{)}  \odot  \rho_{{\mathrm{22}}}    ]   \simeq   [    \ell  \mathbin{:}  \ottnt{A'}   ;  \rho_{{\mathrm{1}}}   ] $,
     we have $\ottnt{B'}  \simeq  \ottnt{A'}$ and $\rho_{{\mathrm{21}}}  \odot  \rho_{{\mathrm{22}}}  \simeq  \rho_{{\mathrm{1}}}$
     by Lemmas~\ref{lem:consistent-symm}, \ref{lem:consistent-inv-record}, and \ref{lem:consistent-inv-cons}.
     Since $\Sigma  \ottsym{;}   \emptyset   \vdash  \ottnt{A}  \ottsym{:}   \mathsf{T} $, i.e., $\Sigma  \ottsym{;}   \emptyset   \vdash   [    \ell  \mathbin{:}  \ottnt{A'}   ;  \rho_{{\mathrm{1}}}   ]   \ottsym{:}   \mathsf{T} $,
     we have $\Sigma  \ottsym{;}   \emptyset   \vdash  \ottnt{A'}  \ottsym{:}   \mathsf{T} $.
     Thus,
     \[
      \Sigma  \ottsym{;}   \emptyset   \vdash  \ottnt{v_{{\mathrm{1}}}}  \ottsym{:}  \ottnt{B'} \,  \stackrel{ \ottnt{p} }{\Rightarrow}  \ottnt{A'}   \ottsym{:}  \ottnt{A'}
     \]
     by \T{Cast}.

     Since $\rho_{{\mathrm{21}}}  \odot  \rho_{{\mathrm{22}}}  \simeq  \rho_{{\mathrm{1}}}$, i.e., $\rho'_{{\mathrm{2}}}  \simeq  \rho_{{\mathrm{1}}}$, we have $ [  \rho'_{{\mathrm{2}}}  ]   \simeq   [  \rho_{{\mathrm{1}}}  ] $ by \CE{Record}.
     Since $\Sigma  \ottsym{;}   \emptyset   \vdash  \ottnt{v_{{\mathrm{2}}}}  \ottsym{:}   [  \rho'_{{\mathrm{2}}}  ] $ (note that $\rho'_{{\mathrm{2}}} \,  =  \, \rho_{{\mathrm{21}}}  \odot  \rho_{{\mathrm{22}}}$)
     and $\Sigma  \ottsym{;}   \emptyset   \vdash   [  \rho_{{\mathrm{1}}}  ]   \ottsym{:}   \mathsf{T} $ (from $\Sigma  \ottsym{;}   \emptyset   \vdash  \ottnt{A}  \ottsym{:}   \mathsf{T} $),
     we have
     \[
      \Sigma  \ottsym{;}   \emptyset   \vdash  \ottnt{v_{{\mathrm{2}}}}  \ottsym{:}   [  \rho'_{{\mathrm{2}}}  ]  \,  \stackrel{ \ottnt{p} }{\Rightarrow}   [  \rho_{{\mathrm{1}}}  ]    \ottsym{:}   [  \rho_{{\mathrm{1}}}  ] 
     \]
     by \T{Cast}.

     Thus, by \T{RExt},
     \[
      \Sigma  \ottsym{;}   \emptyset   \vdash  \ottsym{\{}  \ell  \ottsym{=}  \ottsym{(}  \ottnt{v_{{\mathrm{1}}}}  \ottsym{:}  \ottnt{B'} \,  \stackrel{ \ottnt{p} }{\Rightarrow}  \ottnt{A'}   \ottsym{)}  \ottsym{;}  \ottnt{v_{{\mathrm{2}}}}  \ottsym{:}   [  \rho'_{{\mathrm{2}}}  ]  \,  \stackrel{ \ottnt{p} }{\Rightarrow}   [  \rho_{{\mathrm{1}}}  ]    \ottsym{\}}  \ottsym{:}   [    \ell  \mathbin{:}  \ottnt{A'}   ;  \rho_{{\mathrm{1}}}   ] .
     \]

    \case \R{RCon}:
     We have
     \begin{itemize}
      \item $\ottnt{e_{{\mathrm{0}}}} \,  =  \, \ottnt{v}$,
      \item $\ottnt{A} \,  =  \,  [    \ell  \mathbin{:}  \ottnt{A'}   ;  \rho_{{\mathrm{1}}}   ] $,
      \item $\ottnt{B} \,  =  \,  [  \rho_{{\mathrm{2}}}  ] $, and
      \item $\ottnt{e'} \,  =  \, \ottnt{v}  \ottsym{:}   [  \rho_{{\mathrm{2}}}  ]  \,  \stackrel{ \ottnt{p} }{\Rightarrow}   [     \rho_{{\mathrm{2}}}  \mathrel{@}   \ell  \mathbin{:}  \ottnt{A'}      ]   \,  \stackrel{ \ottnt{p} }{\Rightarrow}   [    \ell  \mathbin{:}  \ottnt{A'}   ;  \rho_{{\mathrm{1}}}   ]  $
     \end{itemize}
     for some $\ottnt{v}$, $\ell$, $\ottnt{A'}$, $\rho_{{\mathrm{1}}}$, $\rho_{{\mathrm{2}}}$
     such that $\ell \,  \not\in  \, \mathit{dom} \, \ottsym{(}  \rho_{{\mathrm{2}}}  \ottsym{)}$ and $\rho_{{\mathrm{2}}} \,  \not=  \, \star$.

     Since $\ottnt{B}  \simeq  \ottnt{A}$, there exist some $\ottnt{B'}$ and $\rho'_{{\mathrm{2}}}$ such that
     \begin{itemize}
      \item $\rho_{{\mathrm{2}}} \,  \triangleright _{ \ell }  \, \ottnt{B'}  \ottsym{,}  \rho'_{{\mathrm{2}}}$,
      \item $\ottnt{B'}  \simeq  \ottnt{A'}$, and
      \item $\rho'_{{\mathrm{2}}}  \simeq  \rho_{{\mathrm{1}}}$
     \end{itemize}
     by Lemmas~\ref{lem:consistent-inv-record}, \ref{lem:consistent-symm}, and \ref{lem:consistent-inv-cons}.
     Since $\ell \,  \not\in  \, \mathit{dom} \, \ottsym{(}  \rho_{{\mathrm{2}}}  \ottsym{)}$ and $\rho_{{\mathrm{2}}} \,  \triangleright _{ \ell }  \, \ottnt{B'}  \ottsym{,}  \rho'_{{\mathrm{2}}}$,
     it is found that $\rho_{{\mathrm{2}}}$ ends with $ \star $ and $\ottnt{B'} \,  =  \, \star$ and $\rho'_{{\mathrm{2}}} \,  =  \, \rho_{{\mathrm{2}}}$.
     Thus, by \reflem{consistent-inj-row-label-dyn},
     $\rho_{{\mathrm{2}}}  \simeq   \rho_{{\mathrm{2}}}  \mathrel{@}   \ell  \mathbin{:}  \ottnt{A'}  $.
     Since $\Sigma  \ottsym{;}   \emptyset   \vdash  \rho_{{\mathrm{2}}}  \ottsym{:}   \mathsf{R} $ and $\Sigma  \ottsym{;}   \emptyset   \vdash  \ottnt{A}  \ottsym{:}   \mathsf{T} $,
     we have $\Sigma  \ottsym{;}   \emptyset   \vdash   [     \rho_{{\mathrm{2}}}  \mathrel{@}   \ell  \mathbin{:}  \ottnt{A'}      ]   \ottsym{:}   \mathsf{R} $.
     Thus,
     \[
      \Sigma  \ottsym{;}   \emptyset   \vdash  \ottnt{v}  \ottsym{:}   [  \rho_{{\mathrm{2}}}  ]  \,  \stackrel{ \ottnt{p} }{\Rightarrow}   [     \rho_{{\mathrm{2}}}  \mathrel{@}   \ell  \mathbin{:}  \ottnt{A'}      ]    \ottsym{:}   [     \rho_{{\mathrm{2}}}  \mathrel{@}   \ell  \mathbin{:}  \ottnt{A'}      ] 
     \]
     by \T{Cast}.

     Since $\rho'_{{\mathrm{2}}}  \simeq  \rho_{{\mathrm{1}}}$ and $\ottnt{A'}  \simeq  \ottnt{A'}$ \CE{Refl} and $\ell \,  \not\in  \, \mathit{dom} \, \ottsym{(}  \rho'_{{\mathrm{2}}}  \ottsym{)}$ (since $\ell \,  \not\in  \, \mathit{dom} \, \ottsym{(}  \rho_{{\mathrm{2}}}  \ottsym{)}$ and $\rho_{{\mathrm{2}}} \,  =  \, \rho'_{{\mathrm{2}}}$),
     we have $ \rho'_{{\mathrm{2}}}  \mathrel{@}   \ell  \mathbin{:}  \ottnt{A'}    \simeq    \ell  \mathbin{:}  \ottnt{A'}   ;  \rho_{{\mathrm{1}}} $ by \CE{ConsR}.
     Thus,
     \[
      \Sigma  \ottsym{;}   \emptyset   \vdash  \ottnt{v}  \ottsym{:}   [  \rho_{{\mathrm{2}}}  ]  \,  \stackrel{ \ottnt{p} }{\Rightarrow}   [     \rho_{{\mathrm{2}}}  \mathrel{@}   \ell  \mathbin{:}  \ottnt{A'}      ]   \,  \stackrel{ \ottnt{p} }{\Rightarrow}   [    \ell  \mathbin{:}  \ottnt{A'}   ;  \rho_{{\mathrm{1}}}   ]    \ottsym{:}   [    \ell  \mathbin{:}  \ottnt{A'}   ;  \rho_{{\mathrm{1}}}   ] 
     \]
     by \T{Cast}.

    \case \R{VToDyn}:
     We have
     $\ottnt{e_{{\mathrm{0}}}} \,  =  \, \ottnt{v}$ and
     $\ottnt{A} \,  =  \,  \langle  \star  \rangle $ and
     $\ottnt{B} \,  =  \,  \langle  \rho  \rangle $ and
     $\ottnt{e'} \,  =  \, \ottnt{v}  \ottsym{:}   \langle  \rho  \rangle  \,  \stackrel{ \ottnt{p} }{\Rightarrow}   \langle   \mathit{grow}  (  \rho  )   \rangle   \,  \stackrel{ \ottnt{p} }{\Rightarrow}   \langle  \star  \rangle  $
     for some $\ottnt{v}$ and $\rho$
     $\rho \,  \not=  \,  \mathit{grow}  (  \rho  ) $.

     By \reflem{consistent-grow}, $\rho  \simeq   \mathit{grow}  (  \rho  ) $, and therefore
     $ \langle  \rho  \rangle   \simeq   \langle   \mathit{grow}  (  \rho  )   \rangle $ by \CE{Variant}.
     Since $\Sigma  \ottsym{;}   \emptyset   \vdash  \ottnt{v}  \ottsym{:}   \langle  \rho  \rangle $ and $\Sigma  \ottsym{;}   \emptyset   \vdash   \langle   \mathit{grow}  (  \rho  )   \rangle   \ottsym{:}   \mathsf{T} $ by \reflem{wf-grow}, we have
     $\Sigma  \ottsym{;}   \emptyset   \vdash  \ottnt{v}  \ottsym{:}   \langle  \rho  \rangle  \,  \stackrel{ \ottnt{p} }{\Rightarrow}   \langle   \mathit{grow}  (  \rho  )   \rangle   \,  \stackrel{ \ottnt{p} }{\Rightarrow}   \langle  \star  \rangle    \ottsym{:}   \langle  \star  \rangle $.

    \case \R{VFromDyn}:
     We have
     $\ottnt{e_{{\mathrm{0}}}} \,  =  \, \ottnt{v}  \ottsym{:}   \langle  \gamma  \rangle  \,  \stackrel{ \ottnt{q} }{\Rightarrow}   \langle  \star  \rangle  $ and
     $\ottnt{B} \,  =  \,  \langle  \star  \rangle $ and
     $\ottnt{A} \,  =  \,  \langle  \rho_{{\mathrm{1}}}  \rangle $ and
     $\ottnt{e'} \,  =  \, \ottnt{v}  \ottsym{:}   \langle  \gamma  \rangle  \,  \stackrel{ \ottnt{q} }{\Rightarrow}   \langle  \rho_{{\mathrm{1}}}  \rangle  $
     for some $\ottnt{v}$, $\gamma$, $\rho_{{\mathrm{1}}}$, and $\ottnt{q}$ such that
     $\gamma  \simeq  \rho_{{\mathrm{1}}}$.
     Since $\gamma  \simeq  \rho_{{\mathrm{1}}}$, we have $ \langle  \gamma  \rangle   \simeq   \langle  \rho_{{\mathrm{1}}}  \rangle $ by \CE{Variant}.
     Since $\Sigma  \ottsym{;}   \emptyset   \vdash  \ottnt{e_{{\mathrm{0}}}}  \ottsym{:}  \ottnt{B}$, i.e., $\Sigma  \ottsym{;}   \emptyset   \vdash  \ottnt{v}  \ottsym{:}   \langle  \gamma  \rangle  \,  \stackrel{ \ottnt{q} }{\Rightarrow}   \langle  \star  \rangle    \ottsym{:}   \langle  \star  \rangle $,
     we have $\Sigma  \ottsym{;}   \emptyset   \vdash  \ottnt{v}  \ottsym{:}   \langle  \gamma  \rangle $ by \reflem{value-inversion-cast}.
     Thus, we have $\Sigma  \ottsym{;}   \emptyset   \vdash  \ottnt{v}  \ottsym{:}   [  \gamma  ]  \,  \stackrel{ \ottnt{q} }{\Rightarrow}   [  \rho_{{\mathrm{1}}}  ]    \ottsym{:}   [  \rho_{{\mathrm{1}}}  ] $
     by \T{Cast}.

    \case \R{VRevInj}:
     We have
     $\ottnt{e_{{\mathrm{0}}}} \,  =  \, \ell \, \ottnt{v}$ and
     $\ottnt{A} \,  =  \,  \langle    \rho_{{\mathrm{11}}}  \odot  \ottsym{(}    \ell  \mathbin{:}  \ottnt{A'}   ;   \cdot    \ottsym{)}  \odot  \rho_{{\mathrm{12}}}    \rangle $ and
     $\ottnt{B} \,  =  \,  \langle    \ell  \mathbin{:}  \ottnt{B'}   ;  \rho_{{\mathrm{2}}}   \rangle $ and
     $\ottnt{e'} \,  =  \,  \variantliftrow{ \rho_{{\mathrm{11}}} }{ \ottsym{(}  \ell \, \ottsym{(}  \ottnt{v}  \ottsym{:}  \ottnt{B'} \,  \stackrel{ \ottnt{p} }{\Rightarrow}  \ottnt{A'}   \ottsym{)}  \ottsym{)} } $
     for some $\ell$, $\ottnt{v}$, $\rho_{{\mathrm{2}}}$, $\rho_{{\mathrm{11}}}$, $\ottnt{A'}$, and $\ottnt{B'}$ such that
     $\ell \,  \not\in  \, \mathit{dom} \, \ottsym{(}  \rho_{{\mathrm{11}}}  \ottsym{)}$.

     Since $\Sigma  \ottsym{;}   \emptyset   \vdash  \ell \, \ottnt{v}  \ottsym{:}   \langle    \ell  \mathbin{:}  \ottnt{B'}   ;  \rho_{{\mathrm{2}}}   \rangle $,
     we have $\Sigma  \ottsym{;}   \emptyset   \vdash  \ottnt{v}  \ottsym{:}  \ottnt{B'}$ by \reflem{value-inversion-variant-inj}.
     Since $ \langle    \ell  \mathbin{:}  \ottnt{B'}   ;  \rho_{{\mathrm{2}}}   \rangle   \simeq   \langle    \rho_{{\mathrm{11}}}  \odot  \ottsym{(}    \ell  \mathbin{:}  \ottnt{A'}   ;   \cdot    \ottsym{)}  \odot  \rho_{{\mathrm{12}}}    \rangle $,
     we have $\ottnt{B'}  \simeq  \ottnt{A'}$ by Lemmas~\ref{lem:consistent-inv-variant} and \ref{lem:consistent-inv-cons}.
     Since $\Sigma  \ottsym{;}   \emptyset   \vdash   \langle    \rho_{{\mathrm{11}}}  \odot  \ottsym{(}    \ell  \mathbin{:}  \ottnt{A'}   ;   \cdot    \ottsym{)}  \odot  \rho_{{\mathrm{12}}}    \rangle   \ottsym{:}   \mathsf{T} $,
     we have $\Sigma  \ottsym{;}   \emptyset   \vdash  \ottnt{A'}  \ottsym{:}   \mathsf{T} $ and $\Sigma  \ottsym{;}   \emptyset   \vdash  \rho_{{\mathrm{12}}}  \ottsym{:}   \mathsf{R} $.
     \[
      \Sigma  \ottsym{;}   \emptyset   \vdash  \ell \, \ottsym{(}  \ottnt{v}  \ottsym{:}  \ottnt{B'} \,  \stackrel{ \ottnt{p} }{\Rightarrow}  \ottnt{A'}   \ottsym{)}  \ottsym{:}   \langle    \ell  \mathbin{:}  \ottnt{A'}   ;  \rho_{{\mathrm{12}}}   \rangle 
     \]
     by \T{Cast} and \T{VInj}.

     Since $\Sigma  \ottsym{;}   \emptyset   \vdash   \langle    \rho_{{\mathrm{11}}}  \odot  \ottsym{(}    \ell  \mathbin{:}  \ottnt{A'}   ;   \cdot    \ottsym{)}  \odot  \rho_{{\mathrm{12}}}    \rangle   \ottsym{:}   \mathsf{T} $, we have $\Sigma  \ottsym{;}   \emptyset   \vdash  \rho_{{\mathrm{11}}}  \ottsym{:}   \mathsf{R} $.
     Thus, by \reflem{typing-lift-rows},
     \[
      \Sigma  \ottsym{;}   \emptyset   \vdash   \variantliftrow{ \rho_{{\mathrm{11}}} }{ \ottsym{(}  \ell \, \ottsym{(}  \ottnt{v}  \ottsym{:}  \ottnt{B'} \,  \stackrel{ \ottnt{p} }{\Rightarrow}  \ottnt{A'}   \ottsym{)}  \ottsym{)} }   \ottsym{:}   \langle    \rho_{{\mathrm{11}}}  \odot  \ottsym{(}    \ell  \mathbin{:}  \ottnt{A'}   ;  \rho_{{\mathrm{12}}}   \ottsym{)}    \rangle .
     \]

    \case \R{VRevLift}: \sloppy{
     We have
     $\ottnt{e_{{\mathrm{0}}}} \,  =  \,  \variantlift{ \ell }{ \ottnt{C} }{ \ottnt{v} } $ and
     $\ottnt{B} \,  =  \,  \langle    \ell  \mathbin{:}  \ottnt{C}   ;  \rho_{{\mathrm{2}}}   \rangle $ and
     $\ottnt{A} \,  =  \,  \langle  \rho_{{\mathrm{1}}}  \rangle $ and
     $\ottnt{e'} \,  =  \,  \variantliftdown{ \rho_{{\mathrm{11}}} }{ \ell }{ \ottnt{C} }{ \ottsym{(}  \ottnt{v}  \ottsym{:}   \langle  \rho_{{\mathrm{2}}}  \rangle  \,  \stackrel{ \ottnt{p} }{\Rightarrow}   \langle    \rho_{{\mathrm{11}}}  \odot  \rho_{{\mathrm{12}}}    \rangle    \ottsym{)} } $
     for some $\ell$, $\ottnt{C}$, $\ottnt{v}$, $\rho_{{\mathrm{1}}}$, $\rho_{{\mathrm{2}}}$, $\rho_{{\mathrm{11}}}$, and $\rho_{{\mathrm{12}}}$ sch such that
     $\rho_{{\mathrm{1}}} \,  =  \, \rho_{{\mathrm{11}}}  \odot  \ottsym{(}    \ell  \mathbin{:}  \ottnt{C}   ;   \cdot    \ottsym{)}  \odot  \rho_{{\mathrm{12}}}$ and
     $\ell \,  \not\in  \, \mathit{dom} \, \ottsym{(}  \rho_{{\mathrm{11}}}  \ottsym{)}$.
    }

     Since $\Sigma  \ottsym{;}   \emptyset   \vdash  \ottnt{e_{{\mathrm{0}}}}  \ottsym{:}  \ottnt{B}$, i.e., $\Sigma  \ottsym{;}   \emptyset   \vdash   \variantlift{ \ell }{ \ottnt{C} }{ \ottnt{v} }   \ottsym{:}   \langle    \ell  \mathbin{:}  \ottnt{C}   ;  \rho_{{\mathrm{2}}}   \rangle $,
     we have $\Sigma  \ottsym{;}   \emptyset   \vdash  \ottnt{v}  \ottsym{:}   \langle  \rho_{{\mathrm{2}}}  \rangle $ by \reflem{value-inversion-variant-lift}.
     Since $\ottnt{B}  \simeq  \ottnt{A}$, i.e., $ \langle    \ell  \mathbin{:}  \ottnt{C}   ;  \rho_{{\mathrm{2}}}   \rangle   \simeq   \langle  \rho_{{\mathrm{1}}}  \rangle $, and $\rho_{{\mathrm{1}}} \,  =  \, \rho_{{\mathrm{11}}}  \odot  \ottsym{(}    \ell  \mathbin{:}  \ottnt{C}   ;   \cdot    \ottsym{)}  \odot  \rho_{{\mathrm{12}}}$ and $\ell \,  \not\in  \, \mathit{dom} \, \ottsym{(}  \rho_{{\mathrm{11}}}  \ottsym{)}$,
     we have $\rho_{{\mathrm{2}}}  \simeq  \rho_{{\mathrm{11}}}  \odot  \rho_{{\mathrm{12}}}$ by Lemmas~\ref{lem:consistent-inv-variant} and \ref{lem:consistent-inv-cons}.
     Thus, \CE{Variant}, $ \langle  \rho_{{\mathrm{2}}}  \rangle   \simeq   \langle    \rho_{{\mathrm{11}}}  \odot  \rho_{{\mathrm{12}}}    \rangle $.
     Since $\Sigma  \ottsym{;}   \emptyset   \vdash  \ottnt{A}  \ottsym{:}   \mathsf{T} $, i.e., $\Sigma  \ottsym{;}   \emptyset   \vdash   \langle  \rho_{{\mathrm{1}}}  \rangle   \ottsym{:}   \mathsf{T} $,
     we have $\Sigma  \ottsym{;}   \emptyset   \vdash   \langle    \rho_{{\mathrm{11}}}  \odot  \rho_{{\mathrm{12}}}    \rangle   \ottsym{:}   \mathsf{T} $.
     Thus, by \T{Cast},
     \[
      \Sigma  \ottsym{;}   \emptyset   \vdash  \ottnt{v}  \ottsym{:}   \langle  \rho_{{\mathrm{2}}}  \rangle  \,  \stackrel{ \ottnt{p} }{\Rightarrow}   \langle    \rho_{{\mathrm{11}}}  \odot  \rho_{{\mathrm{12}}}    \rangle    \ottsym{:}   \langle    \rho_{{\mathrm{11}}}  \odot  \rho_{{\mathrm{12}}}    \rangle .
     \]
     Since $\Sigma  \ottsym{;}   \emptyset   \vdash  \ottnt{B}  \ottsym{:}   \mathsf{T} $, i.e., $\Sigma  \ottsym{;}   \emptyset   \vdash   \langle    \ell  \mathbin{:}  \ottnt{C}   ;  \rho_{{\mathrm{2}}}   \rangle   \ottsym{:}   \mathsf{T} $,
     we have $\Sigma  \ottsym{;}   \emptyset   \vdash  \ottnt{C}  \ottsym{:}   \mathsf{T} $.
     Thus, by \reflem{typing-lift-rows-after-down},
     \[
      \Sigma  \ottsym{;}   \emptyset   \vdash   \variantliftdown{ \rho_{{\mathrm{11}}} }{ \ell }{ \ottnt{C} }{ \ottsym{(}  \ottnt{v}  \ottsym{:}   \langle  \rho_{{\mathrm{2}}}  \rangle  \,  \stackrel{ \ottnt{p} }{\Rightarrow}   \langle    \rho_{{\mathrm{11}}}  \odot  \rho_{{\mathrm{12}}}    \rangle    \ottsym{)} }   \ottsym{:}   \langle    \rho_{{\mathrm{11}}}  \odot  \ottsym{(}    \ell  \mathbin{:}  \ottnt{C}   ;   \cdot    \ottsym{)}  \odot  \rho_{{\mathrm{12}}}    \rangle ,
     \]
     which is what we have to show.

    \case \R{VConInj}:
     We have
     $\ottnt{e_{{\mathrm{0}}}} \,  =  \, \ell \, \ottnt{v}$ and
     $\ottnt{B} \,  =  \,  \langle    \ell  \mathbin{:}  \ottnt{B'}   ;  \rho_{{\mathrm{2}}}   \rangle $ and
     $\ottnt{A} \,  =  \,  \langle  \rho_{{\mathrm{1}}}  \rangle $ and
     $\ottnt{e'} \,  =  \,  \variantliftrow{ \rho_{{\mathrm{1}}} }{ \ottsym{(}  \ell \, \ottnt{v}  \ottsym{:}   \langle    \ell  \mathbin{:}  \ottnt{B'}   ;  \star   \rangle  \,  \stackrel{ \ottnt{p} }{\Rightarrow}   \langle  \star  \rangle    \ottsym{)} } $
     for some $\ell$, $\ottnt{v}$, $\ottnt{B'}$, $\rho_{{\mathrm{1}}}$, and $\rho_{{\mathrm{2}}}$
     such that $\ell \,  \not\in  \, \mathit{dom} \, \ottsym{(}  \rho_{{\mathrm{1}}}  \ottsym{)}$ and $\rho_{{\mathrm{1}}} \,  \not=  \, \star$.

     Since $\ottnt{B}  \simeq  \ottnt{A}$, i.e., $ \langle    \ell  \mathbin{:}  \ottnt{B'}   ;  \rho_{{\mathrm{2}}}   \rangle   \simeq   \langle  \rho_{{\mathrm{1}}}  \rangle $.
     By \reflem{consistent-inv-variant}, $  \ell  \mathbin{:}  \ottnt{B'}   ;  \rho_{{\mathrm{2}}}   \simeq  \rho_{{\mathrm{1}}}$.
     By \ref{lem:consistent-inv-cons}, $\rho_{{\mathrm{1}}} \,  \triangleright _{ \ell }  \, \ottnt{A'}  \ottsym{,}  \rho'_{{\mathrm{1}}}$ for some $\ottnt{A'}$ and $\rho'_{{\mathrm{1}}}$.
     Since $\ell \,  \not\in  \, \mathit{dom} \, \ottsym{(}  \rho_{{\mathrm{1}}}  \ottsym{)}$, $\rho_{{\mathrm{1}}}$ ends with $ \star $, that is,
     there exists some $\rho''_{{\mathrm{1}}}$ such that $\rho''_{{\mathrm{1}}}  \odot  \star \,  =  \, \rho_{{\mathrm{1}}}$.
     Since $ \variantliftrow{ \rho_{{\mathrm{1}}} }{ \ottnt{e''} }  \,  =  \,  \variantliftrow{ \rho''_{{\mathrm{1}}} }{ \ottnt{e''} } $ for any $\ottnt{e''}$, it suffices to show that
     \[
      \Sigma  \ottsym{;}   \emptyset   \vdash   \variantliftrow{ \rho''_{{\mathrm{1}}} }{ \ottsym{(}  \ell \, \ottnt{v}  \ottsym{:}   \langle    \ell  \mathbin{:}  \ottnt{B'}   ;  \star   \rangle  \,  \stackrel{ \ottnt{p} }{\Rightarrow}   \langle  \star  \rangle    \ottsym{)} }   \ottsym{:}   \langle    \rho''_{{\mathrm{1}}}  \odot  \star    \rangle .
     \]

     Since $\Sigma  \ottsym{;}   \emptyset   \vdash  \ottnt{e_{{\mathrm{0}}}}  \ottsym{:}  \ottnt{B}$, i.e., $\Sigma  \ottsym{;}   \emptyset   \vdash  \ell \, \ottnt{v}  \ottsym{:}   \langle    \ell  \mathbin{:}  \ottnt{B'}   ;  \rho_{{\mathrm{2}}}   \rangle $,
     we have $\Sigma  \ottsym{;}   \emptyset   \vdash  \ottnt{v}  \ottsym{:}  \ottnt{B'}$ by \reflem{value-inversion-variant-inj}.
     Thus, by \T{VInj}, $\Sigma  \ottsym{;}   \emptyset   \vdash  \ell \, \ottnt{v}  \ottsym{:}   \langle    \ell  \mathbin{:}  \ottnt{B'}   ;  \star   \rangle $.
     We have $ \langle    \ell  \mathbin{:}  \ottnt{B'}   ;  \star   \rangle   \simeq   \langle  \star  \rangle $ by \CE{Refl}, \CE{ConsL}, and \CE{Variant}.
     Since $\Sigma  \ottsym{;}   \emptyset   \vdash   \langle  \star  \rangle   \ottsym{:}   \mathsf{R} $,
     we have
     \[
      \Sigma  \ottsym{;}   \emptyset   \vdash  \ell \, \ottnt{v}  \ottsym{:}   \langle    \ell  \mathbin{:}  \ottnt{B'}   ;  \star   \rangle  \,  \stackrel{ \ottnt{p} }{\Rightarrow}   \langle  \star  \rangle    \ottsym{:}   \langle  \star  \rangle 
     \]
     by \T{Cast}.
     Since $\Sigma  \ottsym{;}   \emptyset   \vdash  \ottnt{A}  \ottsym{:}   \mathsf{T} $, i.e., $\Sigma  \ottsym{;}   \emptyset   \vdash   [  \rho_{{\mathrm{1}}}  ]   \ottsym{:}   \mathsf{T} $,
     we have $\Sigma  \ottsym{;}   \emptyset   \vdash  \rho''_{{\mathrm{1}}}  \ottsym{:}   \mathsf{R} $.
     Thus, by \reflem{typing-lift-rows},
     \[
      \Sigma  \ottsym{;}   \emptyset   \vdash   \variantliftrow{ \rho''_{{\mathrm{1}}} }{ \ottsym{(}  \ell \, \ottnt{v}  \ottsym{:}   \langle    \ell  \mathbin{:}  \ottnt{B'}   ;  \star   \rangle  \,  \stackrel{ \ottnt{p} }{\Rightarrow}   \langle  \star  \rangle    \ottsym{)} }   \ottsym{:}   \langle    \rho''_{{\mathrm{1}}}  \odot  \star    \rangle .
     \]

    \case \R{VConLift}: \sloppy{
     We have
     $\ottnt{e_{{\mathrm{0}}}} \,  =  \,  \variantlift{ \ell }{ \ottnt{B'} }{ \ottnt{v} } $ and
     $\ottnt{B} \,  =  \,  \langle    \ell  \mathbin{:}  \ottnt{B'}   ;  \rho_{{\mathrm{2}}}   \rangle $ and
     $\ottnt{A} \,  =  \,  \langle  \rho_{{\mathrm{1}}}  \rangle $ and
     $\ottnt{e'} \,  =  \, \ottsym{(}   \variantliftdown{ \rho_{{\mathrm{1}}} }{ \ell }{ \ottnt{B'} }{ \ottsym{(}  \ottnt{v}  \ottsym{:}   \langle  \rho_{{\mathrm{2}}}  \rangle  \,  \stackrel{ \ottnt{p} }{\Rightarrow}   \langle  \rho_{{\mathrm{1}}}  \rangle    \ottsym{)} }   \ottsym{)}  \ottsym{:}   \langle     \rho_{{\mathrm{1}}}  \mathrel{@}   \ell  \mathbin{:}  \ottnt{B'}      \rangle  \,  \stackrel{ \ottnt{p} }{\Rightarrow}   \langle  \rho_{{\mathrm{1}}}  \rangle  $
     for some $\ell$, $\ottnt{v}$, $\ottnt{B'}$, $\rho_{{\mathrm{1}}}$, and $\rho_{{\mathrm{2}}}$ such that
     $\ell \,  \not\in  \, \mathit{dom} \, \ottsym{(}  \rho_{{\mathrm{1}}}  \ottsym{)}$ and $\rho_{{\mathrm{1}}} \,  \not=  \, \star$.
    }

     Since $[NS; gemp |- e0 : B]$, i.e., $\Sigma  \ottsym{;}   \emptyset   \vdash   \variantlift{ \ell }{ \ottnt{B'} }{ \ottnt{v} }   \ottsym{:}   \langle    \ell  \mathbin{:}  \ottnt{B'}   ;  \rho_{{\mathrm{2}}}   \rangle $,
     we have $\Sigma  \ottsym{;}   \emptyset   \vdash  \ottnt{v}  \ottsym{:}   \langle  \rho_{{\mathrm{2}}}  \rangle $ by \reflem{value-inversion-variant-lift}.
     Since $\ottnt{B}  \simeq  \ottnt{A}$, i.e., $ \langle    \ell  \mathbin{:}  \ottnt{B'}   ;  \rho_{{\mathrm{2}}}   \rangle   \simeq   \langle  \rho_{{\mathrm{1}}}  \rangle $,
     there exist some $\ottnt{A'}$ and $\rho'_{{\mathrm{1}}}$ such that
     \begin{itemize}
      \item $\rho_{{\mathrm{1}}} \,  \triangleright _{ \ell }  \, \ottnt{A'}  \ottsym{,}  \rho'_{{\mathrm{1}}}$,
      \item $\ottnt{B'}  \simeq  \ottnt{A'}$, and
      \item $\rho_{{\mathrm{2}}}  \simeq  \rho'_{{\mathrm{1}}}$
     \end{itemize}
     by Lemmas~\ref{lem:consistent-inv-variant} and \ref{lem:consistent-inv-cons}.
     Since $[l notin dom(r1)]$, it is found that
     \begin{itemize}
      \item $\rho_{{\mathrm{1}}}$ ends with $ \star $, i.e., $\rho_{{\mathrm{1}}} \,  =  \, \rho''_{{\mathrm{1}}}  \odot  \star$ for some $\rho''_{{\mathrm{1}}}$,
      \item $\ottnt{A'} \,  =  \, \star$, and
      \item $\rho'_{{\mathrm{1}}} \,  =  \, \rho_{{\mathrm{1}}}$.
     \end{itemize}
     Thus, $\rho_{{\mathrm{2}}}  \simeq  \rho_{{\mathrm{1}}}$, and therefore $ \langle  \rho_{{\mathrm{2}}}  \rangle   \simeq   \langle  \rho_{{\mathrm{1}}}  \rangle $ by \CE{Variant}.
     Since $\Sigma  \ottsym{;}   \emptyset   \vdash  \ottnt{A}  \ottsym{:}   \mathsf{T} $, i.e., $\Sigma  \ottsym{;}   \emptyset   \vdash   \langle  \rho_{{\mathrm{1}}}  \rangle   \ottsym{:}   \mathsf{T} $,
     we have
     \[
      \Sigma  \ottsym{;}   \emptyset   \vdash  \ottnt{v}  \ottsym{:}   \langle  \rho_{{\mathrm{2}}}  \rangle  \,  \stackrel{ \ottnt{p} }{\Rightarrow}   \langle  \rho_{{\mathrm{1}}}  \rangle    \ottsym{:}   \langle  \rho_{{\mathrm{1}}}  \rangle 
     \]
     by \T{Cast}.
     Since $ \variantliftdown{ \rho_{{\mathrm{1}}} }{ \ell }{ \ottnt{B'} }{ \ottnt{e''} }  \,  =  \,  \variantliftdown{ \rho''_{{\mathrm{1}}} }{ \ell }{ \ottnt{B'} }{ \ottnt{e''} } $ for any $\ottnt{e''}$,
     and $\Sigma  \ottsym{;}  \Gamma  \vdash  \ottnt{B'}  \ottsym{:}   \mathsf{T} $ from $\Sigma  \ottsym{;}   \emptyset   \vdash  \ottnt{B}  \ottsym{:}   \mathsf{T} $, i.e., $\Sigma  \ottsym{;}   \emptyset   \vdash   \langle    \ell  \mathbin{:}  \ottnt{B'}   ;  \rho_{{\mathrm{2}}}   \rangle   \ottsym{:}   \mathsf{T} $,
     we have
     \[
      \Sigma  \ottsym{;}   \emptyset   \vdash   \variantliftdown{ \rho_{{\mathrm{1}}} }{ \ell }{ \ottnt{B'} }{ \ottsym{(}  \ottnt{v}  \ottsym{:}   \langle  \rho_{{\mathrm{2}}}  \rangle  \,  \stackrel{ \ottnt{p} }{\Rightarrow}   \langle  \rho_{{\mathrm{1}}}  \rangle    \ottsym{)} }   \ottsym{:}   \langle    \rho''_{{\mathrm{1}}}  \odot  \ottsym{(}    \ell  \mathbin{:}  \ottnt{B'}   ;   \cdot    \ottsym{)}  \odot  \star    \rangle .
     \]
     Since $\rho''_{{\mathrm{1}}}  \odot  \ottsym{(}    \ell  \mathbin{:}  \ottnt{B'}   ;   \cdot    \ottsym{)}  \odot  \star \,  =  \,  \rho_{{\mathrm{1}}}  \mathrel{@}   \ell  \mathbin{:}  \ottnt{B'}  $,
     we have
     \[
      \Sigma  \ottsym{;}   \emptyset   \vdash   \variantliftdown{ \rho_{{\mathrm{1}}} }{ \ell }{ \ottnt{B'} }{ \ottsym{(}  \ottnt{v}  \ottsym{:}   \langle  \rho_{{\mathrm{2}}}  \rangle  \,  \stackrel{ \ottnt{p} }{\Rightarrow}   \langle  \rho_{{\mathrm{1}}}  \rangle    \ottsym{)} }   \ottsym{:}   \langle     \rho_{{\mathrm{1}}}  \mathrel{@}   \ell  \mathbin{:}  \ottnt{B'}      \rangle .
     \]
     Since $\rho_{{\mathrm{1}}}  \simeq  \rho_{{\mathrm{1}}}$ by \CE{Refl}, and $\rho_{{\mathrm{1}}}$ ends with $ \star $ and $\ell \,  \not\in  \, \mathit{dom} \, \ottsym{(}  \rho_{{\mathrm{1}}}  \ottsym{)}$,
     we have $\rho''_{{\mathrm{1}}}  \odot  \ottsym{(}    \ell  \mathbin{:}  \ottnt{B'}   ;   \cdot    \ottsym{)}  \odot  \star  \simeq  \rho_{{\mathrm{1}}}$, i.e.,
     $ \rho_{{\mathrm{1}}}  \mathrel{@}   \ell  \mathbin{:}  \ottnt{B'}    \simeq  \rho_{{\mathrm{1}}}$ by \reflem{consistent-inj-row-label-dyn}.
     Thus, $ \langle     \rho_{{\mathrm{1}}}  \mathrel{@}   \ell  \mathbin{:}  \ottnt{B'}      \rangle   \simeq   \langle  \rho_{{\mathrm{1}}}  \rangle $ by \CE{Variant}.
     Since $\Sigma  \ottsym{;}   \emptyset   \vdash  \ottnt{A}  \ottsym{:}   \mathsf{T} $, i.e., $\Sigma  \ottsym{;}   \emptyset   \vdash   \langle  \rho_{{\mathrm{1}}}  \rangle   \ottsym{:}   \mathsf{T} $,
     we have
     \[
      \Sigma  \ottsym{;}   \emptyset   \vdash  \ottsym{(}   \variantliftdown{ \rho_{{\mathrm{1}}} }{ \ell }{ \ottnt{B'} }{ \ottsym{(}  \ottnt{v}  \ottsym{:}   \langle  \rho_{{\mathrm{2}}}  \rangle  \,  \stackrel{ \ottnt{p} }{\Rightarrow}   \langle  \rho_{{\mathrm{1}}}  \rangle    \ottsym{)} }   \ottsym{)}  \ottsym{:}   \langle     \rho_{{\mathrm{1}}}  \mathrel{@}   \ell  \mathbin{:}  \ottnt{B'}      \rangle  \,  \stackrel{ \ottnt{p} }{\Rightarrow}   \langle  \rho_{{\mathrm{1}}}  \rangle    \ottsym{:}   \langle  \rho_{{\mathrm{1}}}  \rangle 
     \]
     by \T{Cast}, which is what we have to show.
   \end{caseanalysis}

  \case \T{Conv}:
   We have $\ottnt{e} \,  =  \, \ottnt{e_{{\mathrm{0}}}}  \ottsym{:}  \ottnt{B} \,  \stackrel{ \Phi }{\Rightarrow}  \ottnt{A} $ and, by inversion,
   $\Sigma  \ottsym{;}   \emptyset   \vdash  \ottnt{e_{{\mathrm{0}}}}  \ottsym{:}  \ottnt{B}$ and $\Sigma  \ottsym{;}   \emptyset   \vdash  \ottnt{A}  \ottsym{:}   \mathsf{T} $ and $ \Sigma   \vdash   \ottnt{B}  \prec^{ \Phi }  \ottnt{A} $
   for some $\ottnt{e_{{\mathrm{0}}}}$, $\ottnt{B}$, and $\Phi$.
   Besides, we have $\Sigma  \ottsym{;}   \emptyset   \vdash  \ottnt{B}  \ottsym{:}   \mathsf{T} $ by \reflem{typctx-type-wf}.
   By case analysis on the reduction rules applicable to $\ottnt{e}$.
   \begin{caseanalysis}
    \case \R{CName}, \R{CRName}, and \R{CVName}:
     We have $\ottnt{e_{{\mathrm{0}}}} \,  =  \, \ottnt{v}  \ottsym{:}  \ottnt{A} \,  \stackrel{ \ottsym{-}  \alpha }{\Rightarrow}  \ottnt{B} $ and $\Phi \,  =  \, \ottsym{+}  \alpha$ and $\ottnt{e'} \,  =  \, \ottnt{v}$
     for some $\ottnt{v}$ and $\alpha$.
     Since $\Sigma  \ottsym{;}   \emptyset   \vdash  \ottnt{e_{{\mathrm{0}}}}  \ottsym{:}  \ottnt{B}$, i.e., $\Sigma  \ottsym{;}   \emptyset   \vdash  \ottnt{v}  \ottsym{:}  \ottnt{A} \,  \stackrel{ \ottsym{-}  \alpha }{\Rightarrow}  \ottnt{B}   \ottsym{:}  \ottnt{B}$,
     we have $\Sigma  \ottsym{;}   \emptyset   \vdash  \ottnt{v}  \ottsym{:}  \ottnt{A}$
     by Lemma~\ref{lem:value-inversion-conv}, \ref{lem:value-inversion-conv-record}, or \ref{lem:value-inversion-conv-variant}.
     This is what we have to show.

    \case \R{CIdDyn}: \sloppy{
     We have $\ottnt{e_{{\mathrm{0}}}} \,  =  \, \ottnt{v}$ and $\ottnt{e'} \,  =  \, \ottnt{v}$ and $\ottnt{A} \,  =  \, \ottnt{B}$ for some $\ottnt{v}$.
     Since $\Sigma  \ottsym{;}   \emptyset   \vdash  \ottnt{e_{{\mathrm{0}}}}  \ottsym{:}  \ottnt{B}$, we finish.
     The cases for \R{CIdName}, \R{CIdBase}, \R{CREmp}, \R{CRIdDyn}, \R{CRIdName}, \R{CVIdDyn}, and \R{CVIdName} can be shown similarly.
    }

    \case \R{CFun}:
     We have $\ottnt{e_{{\mathrm{0}}}} \,  =  \, \ottnt{v}$ and $\ottnt{B} \,  =  \, \ottnt{B_{{\mathrm{1}}}}  \rightarrow  \ottnt{B_{{\mathrm{2}}}}$ and $\ottnt{A} \,  =  \, \ottnt{A_{{\mathrm{1}}}}  \rightarrow  \ottnt{A_{{\mathrm{2}}}}$ and
     $\ottnt{e'} \,  =  \,  \lambda\!  \,  \mathit{x}  \mathord{:}  \ottnt{A_{{\mathrm{1}}}}   \ottsym{.}  \ottnt{v} \, \ottsym{(}  \mathit{x}  \ottsym{:}  \ottnt{A_{{\mathrm{1}}}} \,  \stackrel{  \overline{ \Phi }  }{\Rightarrow}  \ottnt{B_{{\mathrm{1}}}}   \ottsym{)}  \ottsym{:}  \ottnt{B_{{\mathrm{2}}}} \,  \stackrel{ \Phi }{\Rightarrow}  \ottnt{A_{{\mathrm{2}}}} $
     for some $\ottnt{v}$, $\ottnt{A_{{\mathrm{1}}}}$, $\ottnt{A_{{\mathrm{2}}}}$, $\ottnt{B_{{\mathrm{1}}}}$, $\ottnt{B_{{\mathrm{2}}}}$, and $\mathit{x}$.
     Since $ \Sigma   \vdash   \ottnt{B}  \prec^{ \Phi }  \ottnt{A} $, i.e., $ \Sigma   \vdash   \ottnt{B_{{\mathrm{1}}}}  \rightarrow  \ottnt{B_{{\mathrm{2}}}}  \prec^{ \Phi }  \ottnt{A_{{\mathrm{1}}}}  \rightarrow  \ottnt{A_{{\mathrm{2}}}} $,
     we have $ \Sigma   \vdash   \ottnt{A_{{\mathrm{1}}}}  \prec^{  \overline{ \Phi }  }  \ottnt{B_{{\mathrm{1}}}} $ and $ \Sigma   \vdash   \ottnt{B_{{\mathrm{2}}}}  \prec^{ \Phi }  \ottnt{A_{{\mathrm{2}}}} $
     by \reflem{convert-inversion-fun}.
     Since $\Sigma  \ottsym{;}   \emptyset   \vdash  \ottnt{A_{{\mathrm{1}}}}  \rightarrow  \ottnt{A_{{\mathrm{2}}}}  \ottsym{:}   \mathsf{T} $ and $\Sigma  \ottsym{;}   \emptyset   \vdash  \ottnt{B_{{\mathrm{1}}}}  \rightarrow  \ottnt{B_{{\mathrm{2}}}}  \ottsym{:}   \mathsf{T} $,
     we have
     $\Sigma  \ottsym{;}   \emptyset   \vdash  \ottnt{A_{{\mathrm{1}}}}  \ottsym{:}   \mathsf{T} $ and
     $\Sigma  \ottsym{;}   \emptyset   \vdash  \ottnt{A_{{\mathrm{2}}}}  \ottsym{:}   \mathsf{T} $ and
     $\Sigma  \ottsym{;}   \emptyset   \vdash  \ottnt{B_{{\mathrm{1}}}}  \ottsym{:}   \mathsf{T} $ and
     $\Sigma  \ottsym{;}   \emptyset   \vdash  \ottnt{B_{{\mathrm{2}}}}  \ottsym{:}   \mathsf{T} $.
     Thus, we have $\Sigma  \ottsym{;}   \mathit{x}  \mathord{:}  \ottnt{A_{{\mathrm{1}}}}   \vdash  \mathit{x}  \ottsym{:}  \ottnt{A_{{\mathrm{1}}}} \,  \stackrel{  \overline{ \Phi }  }{\Rightarrow}  \ottnt{B_{{\mathrm{1}}}}   \ottsym{:}  \ottnt{B_{{\mathrm{1}}}}$ by \T{Conv}.
     Since $\Sigma  \ottsym{;}   \mathit{x}  \mathord{:}  \ottnt{A_{{\mathrm{1}}}}   \vdash  \ottnt{v}  \ottsym{:}  \ottnt{B_{{\mathrm{1}}}}  \rightarrow  \ottnt{B_{{\mathrm{2}}}}$ by \reflem{weakening},
     we have
     \[
      \Sigma  \ottsym{;}   \mathit{x}  \mathord{:}  \ottnt{A_{{\mathrm{1}}}}   \vdash  \ottnt{v} \, \ottsym{(}  \mathit{x}  \ottsym{:}  \ottnt{A_{{\mathrm{1}}}} \,  \stackrel{  \overline{ \Phi }  }{\Rightarrow}  \ottnt{B_{{\mathrm{1}}}}   \ottsym{)}  \ottsym{:}  \ottnt{B_{{\mathrm{2}}}} \,  \stackrel{ \Phi }{\Rightarrow}  \ottnt{A_{{\mathrm{2}}}}   \ottsym{:}  \ottnt{A_{{\mathrm{2}}}}
     \]
     by \T{App} and \T{Conv}.
     Thus,
     \[
      \Sigma  \ottsym{;}   \emptyset   \vdash   \lambda\!  \,  \mathit{x}  \mathord{:}  \ottnt{A_{{\mathrm{1}}}}   \ottsym{.}  \ottnt{v} \, \ottsym{(}  \mathit{x}  \ottsym{:}  \ottnt{A_{{\mathrm{1}}}} \,  \stackrel{  \overline{ \Phi }  }{\Rightarrow}  \ottnt{B_{{\mathrm{1}}}}   \ottsym{)}  \ottsym{:}  \ottnt{B_{{\mathrm{2}}}} \,  \stackrel{ \Phi }{\Rightarrow}  \ottnt{A_{{\mathrm{2}}}}   \ottsym{:}  \ottnt{A_{{\mathrm{1}}}}  \rightarrow  \ottnt{A_{{\mathrm{2}}}}
     \]
     by \T{Lam}.

    \case \R{CForall}:
     We have $\ottnt{e_{{\mathrm{0}}}} \,  =  \, \ottnt{v}$ and $\ottnt{B} \,  =  \,  \text{\unboldmath$\forall\!$}  \,  \mathit{X}  \mathord{:}  \ottnt{K}   \ottsym{.} \, \ottnt{B'}$ and $\ottnt{A} \,  =  \,  \text{\unboldmath$\forall\!$}  \,  \mathit{X}  \mathord{:}  \ottnt{K}   \ottsym{.} \, \ottnt{A'}$ and
     $\ottnt{e'} \,  =  \,   \Lambda\!  \,  \mathit{X}  \mathord{:}  \ottnt{K}   \ottsym{.}   \ottsym{(}  \ottnt{v} \, \mathit{x}  \ottsym{:}  \ottnt{B'} \,  \stackrel{ \Phi }{\Rightarrow}  \ottnt{A'}   \ottsym{)}  ::  \ottnt{A'} $
     for some $\ottnt{v}$, $\mathit{X}$, $\ottnt{K}$, $\ottnt{A'}$, $\ottnt{B'}$, and $\mathit{x}$.
     Since $ \Sigma   \vdash   \ottnt{B}  \prec^{ \Phi }  \ottnt{A} $, i.e., $ \Sigma   \vdash    \text{\unboldmath$\forall\!$}  \,  \mathit{X}  \mathord{:}  \ottnt{K}   \ottsym{.} \, \ottnt{B'}  \prec^{ \Phi }   \text{\unboldmath$\forall\!$}  \,  \mathit{X}  \mathord{:}  \ottnt{K}   \ottsym{.} \, \ottnt{A'} $,
     we have $ \Sigma   \vdash   \ottnt{B'}  \prec^{ \Phi }  \ottnt{A'} $
     by \reflem{convert-inversion-forall}.
     Since $\Sigma  \ottsym{;}   \emptyset   \vdash   \text{\unboldmath$\forall\!$}  \,  \mathit{X}  \mathord{:}  \ottnt{K}   \ottsym{.} \, \ottnt{A'}  \ottsym{:}   \mathsf{T} $ and $\Sigma  \ottsym{;}   \emptyset   \vdash   \text{\unboldmath$\forall\!$}  \,  \mathit{X}  \mathord{:}  \ottnt{K}   \ottsym{.} \, \ottnt{B'}  \ottsym{:}   \mathsf{T} $,
     we have
     $\Sigma  \ottsym{;}   \mathit{X}  \mathord{:}  \ottnt{K}   \vdash  \ottnt{A'}  \ottsym{:}   \mathsf{T} $ and
     $\Sigma  \ottsym{;}   \mathit{X}  \mathord{:}  \ottnt{K}   \vdash  \ottnt{B'}  \ottsym{:}   \mathsf{T} $.
     Since $\Sigma  \ottsym{;}   \mathit{X}  \mathord{:}  \ottnt{K}   \vdash  \ottnt{v}  \ottsym{:}   \text{\unboldmath$\forall\!$}  \,  \mathit{X}  \mathord{:}  \ottnt{K}   \ottsym{.} \, \ottnt{B'}$ by \reflem{weakening},
     we have
     \[
      \Sigma  \ottsym{;}   \mathit{X}  \mathord{:}  \ottnt{K}   \vdash  \ottnt{v} \, \mathit{x}  \ottsym{:}  \ottnt{B'} \,  \stackrel{ \Phi }{\Rightarrow}  \ottnt{A'}   \ottsym{:}  \ottnt{A'}
     \]
     by \T{Var}, \T{App}, and \T{Conv}.
     Thus,
     \[
      \Sigma  \ottsym{;}   \emptyset   \vdash    \Lambda\!  \,  \mathit{X}  \mathord{:}  \ottnt{K}   \ottsym{.}   \ottsym{(}  \ottnt{v} \, \mathit{x}  \ottsym{:}  \ottnt{B'} \,  \stackrel{ \Phi }{\Rightarrow}  \ottnt{A'}   \ottsym{)}  ::  \ottnt{A'}   \ottsym{:}   \text{\unboldmath$\forall\!$}  \,  \mathit{X}  \mathord{:}  \ottnt{K}   \ottsym{.} \, \ottnt{A'}
     \]
     by \T{TLam}.

    \case \R{CRExt}:
     We have
     $\ottnt{e_{{\mathrm{0}}}} \,  =  \, \ottnt{v}$ and
     $\ottnt{B} \,  =  \,  [    \ell  \mathbin{:}  \ottnt{B'}   ;  \rho_{{\mathrm{2}}}   ] $ and
     $\ottnt{A} \,  =  \,  [    \ell  \mathbin{:}  \ottnt{A'}   ;  \rho_{{\mathrm{1}}}   ] $ and
     $\ottnt{e'} \,  =  \, \mathsf{let} \, \ottsym{\{}  \ell  \ottsym{=}  \mathit{x}  \ottsym{;}  \mathit{y}  \ottsym{\}}  \ottsym{=}  \ottnt{v} \, \mathsf{in} \, \ottsym{\{}  \ell  \ottsym{=}  \mathit{x}  \ottsym{:}  \ottnt{B'} \,  \stackrel{ \Phi }{\Rightarrow}  \ottnt{A'}   \ottsym{;}  \mathit{y}  \ottsym{:}   [  \rho_{{\mathrm{2}}}  ]  \,  \stackrel{ \Phi }{\Rightarrow}   [  \rho_{{\mathrm{1}}}  ]    \ottsym{\}}$
     for some $\ottnt{v}$, $\ell$, $\ottnt{A'}$, $\ottnt{B'}$, $\rho_{{\mathrm{1}}}$, $\rho_{{\mathrm{2}}}$, $\mathit{x}$, and $\mathit{y}$.
     Since $ \Sigma   \vdash   \ottnt{B}  \prec^{ \Phi }  \ottnt{A} $, i.e., $ \Sigma   \vdash    [    \ell  \mathbin{:}  \ottnt{B'}   ;  \rho_{{\mathrm{2}}}   ]   \prec^{ \Phi }   [    \ell  \mathbin{:}  \ottnt{A'}   ;  \rho_{{\mathrm{1}}}   ]  $,
     we have $ \Sigma   \vdash   \ottnt{B'}  \prec^{ \Phi }  \ottnt{A'} $ and $ \Sigma   \vdash   \rho_{{\mathrm{2}}}  \prec^{ \Phi }  \rho_{{\mathrm{1}}} $
     by Lemmas~\ref{lem:convert-inversion-record} and \ref{lem:convert-inversion-row-cons},
     and $ \Sigma   \vdash    [  \rho_{{\mathrm{2}}}  ]   \prec^{ \Phi }   [  \rho_{{\mathrm{1}}}  ]  $ by \Cv{Record}.
     Since $\Sigma  \ottsym{;}   \emptyset   \vdash  \ottnt{A}  \ottsym{:}   \mathsf{T} $, i.e., $\Sigma  \ottsym{;}   \emptyset   \vdash   [    \ell  \mathbin{:}  \ottnt{A'}   ;  \rho_{{\mathrm{1}}}   ]   \ottsym{:}   \mathsf{T} $,
     we have $\Sigma  \ottsym{;}   \emptyset   \vdash  \ottnt{A'}  \ottsym{:}   \mathsf{T} $ and $\Sigma  \ottsym{;}   \emptyset   \vdash  \rho_{{\mathrm{1}}}  \ottsym{:}   \mathsf{R} $, and therefore $\Sigma  \ottsym{;}   \emptyset   \vdash   [  \rho_{{\mathrm{1}}}  ]   \ottsym{:}   \mathsf{T} $.
     Thus,
     \[
      \Sigma  \ottsym{;}   \mathit{x}  \mathord{:}  \ottnt{B'}   \ottsym{,}   \mathit{y}  \mathord{:}   [  \rho_{{\mathrm{2}}}  ]    \vdash  \ottsym{\{}  \ell  \ottsym{=}  \mathit{x}  \ottsym{:}  \ottnt{B'} \,  \stackrel{ \Phi }{\Rightarrow}  \ottnt{A'}   \ottsym{;}  \mathit{y}  \ottsym{:}   [  \rho_{{\mathrm{2}}}  ]  \,  \stackrel{ \Phi }{\Rightarrow}   [  \rho_{{\mathrm{1}}}  ]    \ottsym{\}}  \ottsym{:}   [    \ell  \mathbin{:}  \ottnt{A'}   ;  \rho_{{\mathrm{1}}}   ] 
     \]
     by \T{Conv} and \T{RExt}.
     Since $\Sigma  \ottsym{;}   \emptyset   \vdash  \ottnt{e_{{\mathrm{0}}}}  \ottsym{:}  \ottnt{B}$, i.e., $\Sigma  \ottsym{;}   \emptyset   \vdash  \ottnt{v}  \ottsym{:}   [    \ell  \mathbin{:}  \ottnt{B'}   ;  \rho_{{\mathrm{2}}}   ] $,
     we have
     \[
      \Sigma  \ottsym{;}   \emptyset   \vdash  \mathsf{let} \, \ottsym{\{}  \ell  \ottsym{=}  \mathit{x}  \ottsym{;}  \mathit{y}  \ottsym{\}}  \ottsym{=}  \ottnt{v} \, \mathsf{in} \, \ottsym{\{}  \ell  \ottsym{=}  \mathit{x}  \ottsym{:}  \ottnt{B'} \,  \stackrel{ \Phi }{\Rightarrow}  \ottnt{A'}   \ottsym{;}  \mathit{y}  \ottsym{:}   [  \rho_{{\mathrm{2}}}  ]  \,  \stackrel{ \Phi }{\Rightarrow}   [  \rho_{{\mathrm{1}}}  ]    \ottsym{\}}  \ottsym{:}   [    \ell  \mathbin{:}  \ottnt{A'}   ;  \rho_{{\mathrm{1}}}   ] ,
     \]
     which is what we have to prove.

    \case \R{CVar}:
     We have
     $\ottnt{e_{{\mathrm{0}}}} \,  =  \, \ottnt{v}$ and $\ottnt{B} \,  =  \,  \langle    \ell  \mathbin{:}  \ottnt{B'}   ;  \rho_{{\mathrm{2}}}   \rangle $ and $\ottnt{A} \,  =  \,  \langle    \ell  \mathbin{:}  \ottnt{A'}   ;  \rho_{{\mathrm{1}}}   \rangle $ and
     $\ottnt{e'} \,  =  \,  \mathsf{case} \,  \ottnt{v}  \,\mathsf{with}\, \langle  \ell \,  \mathit{x}   \rightarrow   \ell \, \ottsym{(}  \mathit{x}  \ottsym{:}  \ottnt{B'} \,  \stackrel{ \Phi }{\Rightarrow}  \ottnt{A'}   \ottsym{)}   \ottsym{;}   \mathit{y}   \rightarrow    \variantlift{ \ell }{ \ottnt{A'} }{ \ottsym{(}  \mathit{y}  \ottsym{:}   \langle  \rho_{{\mathrm{2}}}  \rangle  \,  \stackrel{ \Phi }{\Rightarrow}   \langle  \rho_{{\mathrm{1}}}  \rangle    \ottsym{)} }   \rangle $
     for some $\ottnt{v}$, $\ell$, $\ottnt{A'}$, $\ottnt{B'}$, $\rho_{{\mathrm{1}}}$, ${[r2]}$, $\mathit{x}$, and $\mathit{y}$.
     Since $ \Sigma   \vdash   \ottnt{B}  \prec^{ \Phi }  \ottnt{A} $, i.e., $ \Sigma   \vdash    \langle    \ell  \mathbin{:}  \ottnt{B'}   ;  \rho_{{\mathrm{2}}}   \rangle   \prec^{ \Phi }   \langle    \ell  \mathbin{:}  \ottnt{A'}   ;  \rho_{{\mathrm{1}}}   \rangle  $,
     we have $ \Sigma   \vdash   \ottnt{B'}  \prec^{ \Phi }  \ottnt{A'} $ and $ \Sigma   \vdash   \rho_{{\mathrm{2}}}  \prec^{ \Phi }  \rho_{{\mathrm{1}}} $
     by Lemmas~\ref{lem:convert-inversion-variant} and \ref{lem:convert-inversion-row-cons},
     and $ \Sigma   \vdash    \langle  \rho_{{\mathrm{2}}}  \rangle   \prec^{ \Phi }   \langle  \rho_{{\mathrm{1}}}  \rangle  $ by \Cv{Variant}.
     Since $\Sigma  \ottsym{;}   \emptyset   \vdash  \ottnt{A}  \ottsym{:}   \mathsf{T} $, i.e., $\Sigma  \ottsym{;}   \emptyset   \vdash   \langle    \ell  \mathbin{:}  \ottnt{A'}   ;  \rho_{{\mathrm{1}}}   \rangle   \ottsym{:}   \mathsf{T} $,
     we have $\Sigma  \ottsym{;}   \emptyset   \vdash  \ottnt{A'}  \ottsym{:}   \mathsf{T} $ and $\Sigma  \ottsym{;}   \emptyset   \vdash  \rho_{{\mathrm{1}}}  \ottsym{:}   \mathsf{R} $, and therefore $\Sigma  \ottsym{;}   \emptyset   \vdash   \langle  \rho_{{\mathrm{1}}}  \rangle   \ottsym{:}   \mathsf{T} $.
     Thus,
     \[
      \Sigma  \ottsym{;}   \mathit{x}  \mathord{:}  \ottnt{B'}   \vdash  \ell \, \ottsym{(}  \mathit{x}  \ottsym{:}  \ottnt{B'} \,  \stackrel{ \Phi }{\Rightarrow}  \ottnt{A'}   \ottsym{)}  \ottsym{:}   \langle    \ell  \mathbin{:}  \ottnt{A'}   ;  \rho_{{\mathrm{1}}}   \rangle 
     \]
     and
     \[
      \Sigma  \ottsym{;}   \mathit{y}  \mathord{:}   \langle  \rho_{{\mathrm{2}}}  \rangle    \vdash   \variantlift{ \ell }{ \ottnt{A'} }{ \ottsym{(}  \mathit{y}  \ottsym{:}   \langle  \rho_{{\mathrm{2}}}  \rangle  \,  \stackrel{ \Phi }{\Rightarrow}   \langle  \rho_{{\mathrm{1}}}  \rangle    \ottsym{)} }   \ottsym{:}   \langle    \ell  \mathbin{:}  \ottnt{A'}   ;  \rho_{{\mathrm{1}}}   \rangle 
     \]
     by \T{Conv} and \T{VInj}.
     Since $\Sigma  \ottsym{;}   \emptyset   \vdash  \ottnt{e_{{\mathrm{0}}}}  \ottsym{:}  \ottnt{B}$, i.e., $\Sigma  \ottsym{;}   \emptyset   \vdash  \ottnt{v}  \ottsym{:}   \langle    \ell  \mathbin{:}  \ottnt{B'}   ;  \rho_{{\mathrm{2}}}   \rangle $,
     we have
     \[
      \Sigma  \ottsym{;}   \emptyset   \vdash   \mathsf{case} \,  \ottnt{v}  \,\mathsf{with}\, \langle  \ell \,  \mathit{x}   \rightarrow   \ell \, \ottsym{(}  \mathit{x}  \ottsym{:}  \ottnt{B'} \,  \stackrel{ \Phi }{\Rightarrow}  \ottnt{A'}   \ottsym{)}   \ottsym{;}   \mathit{y}   \rightarrow    \variantlift{ \ell }{ \ottnt{A'} }{ \ottsym{(}  \mathit{y}  \ottsym{:}   \langle  \rho_{{\mathrm{2}}}  \rangle  \,  \stackrel{ \Phi }{\Rightarrow}   \langle  \rho_{{\mathrm{1}}}  \rangle    \ottsym{)} }   \rangle   \ottsym{:}   \langle    \ell  \mathbin{:}  \ottnt{A'}   ;  \rho_{{\mathrm{1}}}   \rangle ,
     \]
     which is what we have to show.

   \end{caseanalysis}

 \end{caseanalysis}
\end{proof}

\begin{lemma}{eval-growing-tyname}
 If $ \Sigma  \mid  \ottnt{e}   \longrightarrow   \Sigma'  \mid  \ottnt{e'} $, then $\Sigma \,  \subseteq  \, \Sigma'$.
\end{lemma}
\begin{proof}
 Obvious by case analysis on the evaluation rule applied to derive $ \Sigma  \mid  \ottnt{e}   \longrightarrow   \Sigma'  \mid  \ottnt{e'} $.
\end{proof}

\begin{lemmap}{Subject reduction}{subject-red}
 If $\Sigma  \ottsym{;}   \emptyset   \vdash  \ottnt{e}  \ottsym{:}  \ottnt{A}$ and $ \Sigma  \mid  \ottnt{e}   \longrightarrow   \Sigma'  \mid  \ottnt{e'} $,
 then $\Sigma'  \ottsym{;}   \emptyset   \vdash  \ottnt{e'}  \ottsym{:}  \ottnt{A}$.
\end{lemmap}
\begin{proof}
 By induction on the derivation of $\Sigma  \ottsym{;}   \emptyset   \vdash  \ottnt{e}  \ottsym{:}  \ottnt{A}$.
 \begin{caseanalysis}
  \case \T{Var}, \T{Const}, \T{Lam}, \T{TLam}, \T{REmp}, \T{Blame}:
   Contradictory; there are no reduction rules to apply.

  \case \T{App}:
   We have $\ottnt{e} \,  =  \, \ottnt{e_{{\mathrm{1}}}} \, \ottnt{e_{{\mathrm{2}}}}$ and, by inversion,
   $\Sigma  \ottsym{;}   \emptyset   \vdash  \ottnt{e_{{\mathrm{1}}}}  \ottsym{:}  \ottnt{B}  \rightarrow  \ottnt{A}$ and $\Sigma  \ottsym{;}   \emptyset   \vdash  \ottnt{e_{{\mathrm{2}}}}  \ottsym{:}  \ottnt{B}$
   for some $\ottnt{e_{{\mathrm{1}}}}$, $\ottnt{e_{{\mathrm{2}}}}$, and $\ottnt{B}$.

   If $ \Sigma  \mid  \ottnt{e_{{\mathrm{1}}}}   \longrightarrow   \Sigma'  \mid  \ottnt{e'_{{\mathrm{1}}}} $ for some $\ottnt{e'_{{\mathrm{1}}}}$, then
   we have $\Sigma'  \ottsym{;}   \emptyset   \vdash  \ottnt{e'_{{\mathrm{1}}}}  \ottsym{:}  \ottnt{B}  \rightarrow  \ottnt{A}$ by the IH, and therefore
   $\Sigma'  \ottsym{;}   \emptyset   \vdash  \ottnt{e'_{{\mathrm{1}}}} \, \ottnt{e_{{\mathrm{2}}}}  \ottsym{:}  \ottnt{A}$
   by Lemmas~\ref{lem:eval-growing-tyname} and \ref{lem:weakening-tyname}, and \T{App}.

   If $ \Sigma  \mid  \ottnt{e_{{\mathrm{2}}}}   \longrightarrow   \Sigma'  \mid  \ottnt{e'_{{\mathrm{2}}}} $ for some $\ottnt{e'_{{\mathrm{2}}}}$, then
   we have $\Sigma'  \ottsym{;}   \emptyset   \vdash  \ottnt{e'_{{\mathrm{2}}}}  \ottsym{:}  \ottnt{B}$ by the IH, and therefore
   we have $\Sigma'  \ottsym{;}   \emptyset   \vdash  \ottnt{e_{{\mathrm{1}}}} \, \ottnt{e'_{{\mathrm{2}}}}  \ottsym{:}  \ottnt{A}$
   by Lemmas~\ref{lem:eval-growing-tyname} and \ref{lem:weakening-tyname}, and \T{App}.

   In what follows, we suppose that neither $\ottnt{e_{{\mathrm{1}}}}$ nor $\ottnt{e_{{\mathrm{2}}}}$ cannot be evaluated under $\Sigma$.
   By case analysis on the reduction rule applied to $\ottnt{e}$.
   \begin{caseanalysis}
    \case \E{Red}: We have $\ottnt{e_{{\mathrm{1}}}} \, \ottnt{e_{{\mathrm{2}}}} \,  =  \,  \ottnt{E}  [  \ottnt{e'_{{\mathrm{1}}}}  ] $ and $\ottnt{e'} \,  =  \,  \ottnt{E}  [  \ottnt{e'_{{\mathrm{2}}}}  ] $
     for some $\ottnt{E}$, $\ottnt{e'_{{\mathrm{1}}}}$, and $\ottnt{e'_{{\mathrm{2}}}}$ such that $\ottnt{e'_{{\mathrm{1}}}}  \rightsquigarrow  \ottnt{e'_{{\mathrm{2}}}}$.
     Besides, $\Sigma' \,  =  \, \Sigma$.
     By case analysis on $\ottnt{E}$.
     \begin{caseanalysis}
      \case $\ottnt{E} \,  =  \,  [\,] $: By \reflem{subject-red-red}.
      \case $\ottnt{E} \,  =  \, \ottnt{E'} \, \ottnt{e_{{\mathrm{2}}}}$: Contradictory with the assumption that
       $\ottnt{e_{{\mathrm{1}}}} \,  =  \,  \ottnt{E'}  [  \ottnt{e'_{{\mathrm{1}}}}  ] $ cannot be evaluated under $\Sigma$.
      \case $\ottnt{E} \,  =  \, \ottnt{v_{{\mathrm{1}}}} \, \ottnt{E'}$: Contradictory with the assumption that
       $\ottnt{e_{{\mathrm{2}}}} \,  =  \,  \ottnt{E'}  [  \ottnt{e'_{{\mathrm{1}}}}  ] $ cannot be evaluated under $\Sigma$.
      \case otherwise: Contradictory with the assumption that $\ottnt{e_{{\mathrm{1}}}} \, \ottnt{e_{{\mathrm{2}}}} \,  =  \,  \ottnt{E}  [  \ottnt{e'_{{\mathrm{1}}}}  ] $.
     \end{caseanalysis}

    \case \E{Blame}: By \T{Blame}.
    \case \E{TyBeta}: Contradictory with the assumption that
     neither $\ottnt{e_{{\mathrm{1}}}}$ nor $[e2]$ cannot be evaluated under $\Sigma$.
   \end{caseanalysis}

  \case \T{TApp}:
   We have $\ottnt{e} \,  =  \, \ottnt{e_{{\mathrm{1}}}} \, \ottnt{B}$ and, by inversion,
   $\Sigma  \ottsym{;}   \emptyset   \vdash  \ottnt{e_{{\mathrm{1}}}}  \ottsym{:}   \text{\unboldmath$\forall\!$}  \,  \mathit{X}  \mathord{:}  \ottnt{K}   \ottsym{.} \, \ottnt{C}$ and $\Sigma  \ottsym{;}   \emptyset   \vdash  \ottnt{B}  \ottsym{:}  \ottnt{K}$ and $\ottnt{A} \,  =  \,  \ottnt{C}    [  \ottnt{B}  /  \mathit{X}  ]  $
   for some $\ottnt{e_{{\mathrm{1}}}}$, $\mathit{X}$, $\ottnt{K}$, $\ottnt{B}$, and $\ottnt{C}$.

   If $ \Sigma  \mid  \ottnt{e_{{\mathrm{1}}}}   \longrightarrow   \Sigma'  \mid  \ottnt{e'_{{\mathrm{1}}}} $ for some $\ottnt{e'_{{\mathrm{1}}}}$, then
   we have $\Sigma'  \ottsym{;}   \emptyset   \vdash  \ottnt{e'_{{\mathrm{1}}}}  \ottsym{:}   \text{\unboldmath$\forall\!$}  \,  \mathit{X}  \mathord{:}  \ottnt{K}   \ottsym{.} \, \ottnt{C}$ by the IH, and therefore
   $\Sigma'  \ottsym{;}   \emptyset   \vdash  \ottnt{e'_{{\mathrm{1}}}} \, \ottnt{B}  \ottsym{:}   \ottnt{C}    [  \ottnt{B}  /  \mathit{X}  ]  $
   by Lemmas~\ref{lem:eval-growing-tyname} and \ref{lem:weakening-tyname}, and \T{TApp}.

   In what follows, we suppose that $\ottnt{e_{{\mathrm{1}}}}$ cannot be evaluated under $\Sigma$.
   By case analysis on the reduction rule applied to $\ottnt{e}$.
   \begin{caseanalysis}
    \case \E{Red}: We have $\ottnt{e_{{\mathrm{1}}}} \, \ottnt{B} \,  =  \,  \ottnt{E}  [  \ottnt{e'_{{\mathrm{1}}}}  ] $ and $\ottnt{e'} \,  =  \,  \ottnt{E}  [  \ottnt{e'_{{\mathrm{2}}}}  ] $
     for some $\ottnt{E}$, $\ottnt{e'_{{\mathrm{1}}}}$, and $\ottnt{e'_{{\mathrm{2}}}}$ such that $\ottnt{e'_{{\mathrm{1}}}}  \rightsquigarrow  \ottnt{e'_{{\mathrm{2}}}}$.
     Besides, $\Sigma' \,  =  \, \Sigma$.
     By case analysis on $\ottnt{E}$.
     \begin{caseanalysis}
      \case $\ottnt{E} \,  =  \,  [\,] $: By \reflem{subject-red-red}.
      \case $\ottnt{E} \,  =  \, \ottnt{E'} \, \ottnt{B}$: Contradictory with the assumption that
       $\ottnt{e_{{\mathrm{1}}}} \,  =  \,  \ottnt{E'}  [  \ottnt{e'_{{\mathrm{1}}}}  ] $ cannot be evaluated under $\Sigma$.
      \case otherwise: Contradictory with the assumption that $\ottnt{e_{{\mathrm{1}}}} \, \ottnt{B} \,  =  \,  \ottnt{E}  [  \ottnt{e'_{{\mathrm{1}}}}  ] $.
     \end{caseanalysis}

    \case \E{Blame}: By \T{Blame}.
    \case \E{TyBeta}:
     We have $\ottnt{e_{{\mathrm{1}}}} \, \ottnt{B} \,  =  \,  \ottnt{E}  [  \ottsym{(}    \Lambda\!  \,  \mathit{X'}  \mathord{:}  \ottnt{K'}   \ottsym{.}   \ottnt{e'_{{\mathrm{0}}}}  ::  \ottnt{C'}   \ottsym{)} \, \ottnt{B'}  ] $ and $\ottnt{e'} \,  =  \,  \ottnt{E}  [    \ottnt{e'_{{\mathrm{0}}}}    [  \alpha  /  \mathit{X'}  ]    \ottsym{:}   \ottnt{C'}    [  \alpha  /  \mathit{X'}  ]   \,  \stackrel{ \ottsym{+}  \alpha }{\Rightarrow}  \ottnt{C'}     [  \ottnt{B'}  /  \mathit{X}  ]    ] $
     and $\Sigma' \,  =  \, \Sigma  \ottsym{,}   \alpha  \mathord{:}  \ottnt{K'}   \ottsym{:=}  \ottnt{B'}$
     for some $\ottnt{E}$, $\mathit{X'}$, $\ottnt{K'}$, $\ottnt{e'_{{\mathrm{0}}}}$, $\ottnt{B'}$, $\ottnt{C'}$, and $\alpha$.
     By case analysis on $\ottnt{E}$.
     \begin{caseanalysis}
      \case $\ottnt{E} \,  =  \,  [\,] $:
       We have $\ottnt{e_{{\mathrm{1}}}} \,  =  \,   \Lambda\!  \,  \mathit{X}  \mathord{:}  \ottnt{K}   \ottsym{.}   \ottnt{e'_{{\mathrm{0}}}}  ::  \ottnt{C} $ by \reflem{canonical-forms} (note that
       $\mathit{X} \,  =  \, \mathit{X'}$ and $\ottnt{K} \,  =  \, \ottnt{K'}$ and $\ottnt{C} \,  =  \, \ottnt{C'}$) and $\ottnt{B'} \,  =  \, \ottnt{B}$.

       It suffices to show that
       \[
        \Sigma  \ottsym{,}   \alpha  \mathord{:}  \ottnt{K}   \ottsym{:=}  \ottnt{B}  \ottsym{;}   \emptyset   \vdash    \ottnt{e'_{{\mathrm{0}}}}    [  \alpha  /  \mathit{X}  ]    \ottsym{:}   \ottnt{C}    [  \alpha  /  \mathit{X}  ]   \,  \stackrel{ \ottsym{+}  \alpha }{\Rightarrow}  \ottnt{C}     [  \ottnt{B}  /  \mathit{X}  ]    \ottsym{:}   \ottnt{C}    [  \ottnt{B}  /  \mathit{X}  ]  .
       \]

       Since $\Sigma  \ottsym{;}   \emptyset   \vdash  \ottnt{e_{{\mathrm{1}}}}  \ottsym{:}   \text{\unboldmath$\forall\!$}  \,  \mathit{X}  \mathord{:}  \ottnt{K}   \ottsym{.} \, \ottnt{C}$, i.e., $\Sigma  \ottsym{;}   \emptyset   \vdash    \Lambda\!  \,  \mathit{X}  \mathord{:}  \ottnt{K}   \ottsym{.}   \ottnt{e'_{{\mathrm{0}}}}  ::  \ottnt{C}   \ottsym{:}   \text{\unboldmath$\forall\!$}  \,  \mathit{X}  \mathord{:}  \ottnt{K}   \ottsym{.} \, \ottnt{C}$,
       we have $\Sigma  \ottsym{;}   \mathit{X}  \mathord{:}  \ottnt{K}   \vdash  \ottnt{e'_{{\mathrm{0}}}}  \ottsym{:}  \ottnt{C}$ by \reflem{value-inversion-typelambda}.
       Thus, $\Sigma  \ottsym{,}   \alpha  \mathord{:}  \ottnt{K}   \ottsym{:=}  \ottnt{B}  \ottsym{;}   \mathit{X}  \mathord{:}  \ottnt{K}   \vdash  \ottnt{e'_{{\mathrm{0}}}}  \ottsym{:}  \ottnt{C}$ by \reflem{weakening-tyname}.
       Since $\Sigma  \ottsym{,}   \alpha  \mathord{:}  \ottnt{K}   \ottsym{:=}  \ottnt{B}  \ottsym{;}   \emptyset   \vdash  \alpha  \ottsym{:}  \ottnt{K}$ by \WF{TyName}, we have
       \[
        \Sigma  \ottsym{,}   \alpha  \mathord{:}  \ottnt{K}   \ottsym{:=}  \ottnt{B}  \ottsym{;}   \emptyset   \vdash   \ottnt{e'_{{\mathrm{0}}}}    [  \alpha  /  \mathit{X}  ]    \ottsym{:}   \ottnt{C}    [  \alpha  /  \mathit{X}  ]  
       \]
       by \reflem{type-subst}.

       Since $\Sigma  \ottsym{;}   \emptyset   \vdash  \ottnt{e_{{\mathrm{1}}}}  \ottsym{:}   \text{\unboldmath$\forall\!$}  \,  \mathit{X}  \mathord{:}  \ottnt{K}   \ottsym{.} \, \ottnt{C}$, we have $\Sigma  \ottsym{;}   \emptyset   \vdash   \text{\unboldmath$\forall\!$}  \,  \mathit{X}  \mathord{:}  \ottnt{K}   \ottsym{.} \, \ottnt{C}  \ottsym{:}   \mathsf{T} $ by \reflem{typctx-type-wf}.
       Thus, since $\alpha$ is a fresh type name for $\Sigma$,
       $\alpha$ does not occur in $\ottnt{C}$.
       Therefore, we have
       \[
         \Sigma  \ottsym{,}   \alpha  \mathord{:}  \ottnt{K}   \ottsym{:=}  \ottnt{B}   \vdash    \ottnt{C}    [  \alpha  /  \mathit{X}  ]    \prec^{ \ottsym{+}  \alpha }   \ottnt{C}    [  \ottnt{B}  /  \mathit{X}  ]   
       \]
       by \reflem{convert-type-subst}.
       Since $\Sigma  \ottsym{;}   \emptyset   \vdash  \ottnt{e}  \ottsym{:}  \ottnt{A}$, we have $\Sigma  \ottsym{;}   \emptyset   \vdash  \ottnt{A}  \ottsym{:}   \mathsf{T} $ by \reflem{typctx-type-wf}, and therefore
       $\Sigma  \ottsym{;}   \emptyset   \vdash   \ottnt{C}    [  \ottnt{B}  /  \mathit{X}  ]    \ottsym{:}   \mathsf{T} $.
       Thus, by \T{Conv},
       \[
        \Sigma  \ottsym{,}   \alpha  \mathord{:}  \ottnt{K}   \ottsym{:=}  \ottnt{B}  \ottsym{;}   \emptyset   \vdash    \ottnt{e'_{{\mathrm{0}}}}    [  \alpha  /  \mathit{X}  ]    \ottsym{:}   \ottnt{C}    [  \alpha  /  \mathit{X}  ]   \,  \stackrel{ \ottsym{+}  \alpha }{\Rightarrow}  \ottnt{C}     [  \ottnt{B}  /  \mathit{X}  ]    \ottsym{:}   \ottnt{C}    [  \ottnt{B}  /  \mathit{X}  ]  .
       \]

      \case $\ottnt{E} \,  =  \, \ottnt{E'} \, \ottnt{B}$: Contradictory with the assumption that
       $\ottnt{e_{{\mathrm{1}}}} \,  =  \,  \ottnt{E'}  [  \ottsym{(}    \Lambda\!  \,  \mathit{X'}  \mathord{:}  \ottnt{K'}   \ottsym{.}   \ottnt{e'_{{\mathrm{0}}}}  ::  \ottnt{C'}   \ottsym{)} \, \ottnt{B'}  ] $ cannot be evaluated under $\Sigma$.
      \case otherwise: Contradictory with the assumption that $\ottnt{e_{{\mathrm{1}}}} \, \ottnt{B} \,  =  \,  \ottnt{E}  [  \ottnt{e'_{{\mathrm{1}}}}  ] $.
     \end{caseanalysis}
   \end{caseanalysis}

  \case \T{RExt}:
   We have $\ottnt{e} \,  =  \, \ottsym{\{}  \ell  \ottsym{=}  \ottnt{e_{{\mathrm{1}}}}  \ottsym{;}  \ottnt{e_{{\mathrm{2}}}}  \ottsym{\}}$ and, by inversion,
   $\Sigma  \ottsym{;}   \emptyset   \vdash  \ottnt{e_{{\mathrm{1}}}}  \ottsym{:}  \ottnt{B}$ and $\Sigma  \ottsym{;}   \emptyset   \vdash  \ottnt{e_{{\mathrm{2}}}}  \ottsym{:}   [  \rho  ] $ and
   $\ottnt{A} \,  =  \,  [    \ell  \mathbin{:}  \ottnt{B}   ;  \rho   ] $
   for some $\ell$, $\ottnt{e_{{\mathrm{1}}}}$, $\ottnt{e_{{\mathrm{2}}}}$, $\ottnt{B}$, and $\rho$.

   If $ \Sigma  \mid  \ottnt{e_{{\mathrm{1}}}}   \longrightarrow   \Sigma'  \mid  \ottnt{e'_{{\mathrm{1}}}} $ for some $\ottnt{e'_{{\mathrm{1}}}}$, then
   we have $\Sigma'  \ottsym{;}   \emptyset   \vdash  \ottnt{e'_{{\mathrm{1}}}}  \ottsym{:}  \ottnt{B}$ by the IH, and therefore
   $\Sigma'  \ottsym{;}   \emptyset   \vdash  \ottsym{\{}  \ell  \ottsym{=}  \ottnt{e'_{{\mathrm{1}}}}  \ottsym{;}  \ottnt{e_{{\mathrm{2}}}}  \ottsym{\}}  \ottsym{:}   [    \ell  \mathbin{:}  \ottnt{B}   ;  \rho   ] $
   by Lemmas~\ref{lem:eval-growing-tyname} and \ref{lem:weakening-tyname}, and \T{RExt}.

   If $ \Sigma  \mid  \ottnt{e_{{\mathrm{2}}}}   \longrightarrow   \Sigma'  \mid  \ottnt{e'_{{\mathrm{2}}}} $ for some $\ottnt{e'_{{\mathrm{2}}}}$, then
   we have $\Sigma'  \ottsym{;}   \emptyset   \vdash  \ottnt{e'_{{\mathrm{2}}}}  \ottsym{:}   [  \rho  ] $ by the IH, and therefore
   we have $\Sigma'  \ottsym{;}   \emptyset   \vdash  \ottsym{\{}  \ell  \ottsym{=}  \ottnt{e_{{\mathrm{1}}}}  \ottsym{;}  \ottnt{e'_{{\mathrm{2}}}}  \ottsym{\}}  \ottsym{:}   [    \ell  \mathbin{:}  \ottnt{B}   ;  \rho   ] $.
   by Lemmas~\ref{lem:eval-growing-tyname} and \ref{lem:weakening-tyname}, and \T{RExt}.

   In what follows, we suppose that neither $\ottnt{e_{{\mathrm{1}}}}$ nor $\ottnt{e_{{\mathrm{2}}}}$ cannot be evaluated under $\Sigma$.
   By case analysis on the reduction rule applied to $\ottnt{e}$.
   \begin{caseanalysis}
    \case \E{Red}: We have $\ottsym{\{}  \ell  \ottsym{=}  \ottnt{e_{{\mathrm{1}}}}  \ottsym{;}  \ottnt{e_{{\mathrm{2}}}}  \ottsym{\}} \,  =  \,  \ottnt{E}  [  \ottnt{e'_{{\mathrm{1}}}}  ] $ and $\ottnt{e'} \,  =  \,  \ottnt{E}  [  \ottnt{e'_{{\mathrm{2}}}}  ] $
     for some $\ottnt{E}$, $\ottnt{e'_{{\mathrm{1}}}}$, and $\ottnt{e'_{{\mathrm{2}}}}$ such that $\ottnt{e'_{{\mathrm{1}}}}  \rightsquigarrow  \ottnt{e'_{{\mathrm{2}}}}$.
     Besides, $\Sigma' \,  =  \, \Sigma$.
     By case analysis on $\ottnt{E}$.
     \begin{caseanalysis}
      \case $\ottnt{E} \,  =  \,  [\,] $: By \reflem{subject-red-red}.
      \case $\ottnt{E} \,  =  \, \ottsym{\{}  \ell  \ottsym{=}  \ottnt{E'}  \ottsym{;}  \ottnt{e_{{\mathrm{2}}}}  \ottsym{\}}$: Contradictory with the assumption that
       $\ottnt{e_{{\mathrm{1}}}} \,  =  \,  \ottnt{E'}  [  \ottnt{e'_{{\mathrm{1}}}}  ] $ cannot be evaluated under $\Sigma$.
      \case $\ottnt{E} \,  =  \, \ottsym{\{}  \ell  \ottsym{=}  \ottnt{v_{{\mathrm{1}}}}  \ottsym{;}  \ottnt{E'}  \ottsym{\}}$: Contradictory with the assumption that
       $\ottnt{e_{{\mathrm{2}}}} \,  =  \,  \ottnt{E'}  [  \ottnt{e'_{{\mathrm{1}}}}  ] $ cannot be evaluated under $\Sigma$.
      \case otherwise: Contradictory with the assumption that $\ottsym{\{}  \ell  \ottsym{=}  \ottnt{e_{{\mathrm{1}}}}  \ottsym{;}  \ottnt{e_{{\mathrm{2}}}}  \ottsym{\}} \,  =  \,  \ottnt{E}  [  \ottnt{e'_{{\mathrm{1}}}}  ] $.
     \end{caseanalysis}

    \case \E{Blame}: By \T{Blame}.
    \case \E{TyBeta}: Contradictory with the assumption that
     neither $\ottnt{e_{{\mathrm{1}}}}$ nor $[e2]$ cannot be evaluated under $\Sigma$.
   \end{caseanalysis}

  \case \T{RLet}:
   We have $\ottnt{e} \,  =  \, \mathsf{let} \, \ottsym{\{}  \ell  \ottsym{=}  \mathit{x}  \ottsym{;}  \mathit{y}  \ottsym{\}}  \ottsym{=}  \ottnt{e_{{\mathrm{1}}}} \, \mathsf{in} \, \ottnt{e_{{\mathrm{2}}}}$ and, by inversion,
   $\Sigma  \ottsym{;}   \emptyset   \vdash  \ottnt{e_{{\mathrm{1}}}}  \ottsym{:}   [    \ell  \mathbin{:}  \ottnt{B}   ;  \rho   ] $ and $\Sigma  \ottsym{;}   \mathit{x}  \mathord{:}  \ottnt{B}   \ottsym{,}   \mathit{y}  \mathord{:}   [  \rho  ]    \vdash  \ottnt{e_{{\mathrm{2}}}}  \ottsym{:}  \ottnt{A}$
   for some $\ell$, $\mathit{x}$, $\mathit{y}$, $\ottnt{e_{{\mathrm{1}}}}$, $\ottnt{e_{{\mathrm{2}}}}$, $\ottnt{B}$, and $\rho$.

   If $ \Sigma  \mid  \ottnt{e_{{\mathrm{1}}}}   \longrightarrow   \Sigma'  \mid  \ottnt{e'_{{\mathrm{1}}}} $ for some $\ottnt{e'_{{\mathrm{1}}}}$, then
   we have $\Sigma'  \ottsym{;}   \emptyset   \vdash  \ottnt{e'_{{\mathrm{1}}}}  \ottsym{:}   [    \ell  \mathbin{:}  \ottnt{B}   ;  \rho   ] $ by the IH, and therefore
   $\Sigma'  \ottsym{;}   \emptyset   \vdash  \mathsf{let} \, \ottsym{\{}  \ell  \ottsym{=}  \mathit{x}  \ottsym{;}  \mathit{y}  \ottsym{\}}  \ottsym{=}  \ottnt{e'_{{\mathrm{1}}}} \, \mathsf{in} \, \ottnt{e_{{\mathrm{2}}}}  \ottsym{:}  \ottnt{A}$
   by Lemmas~\ref{lem:eval-growing-tyname} and \ref{lem:weakening-tyname}, and \T{RLet}.

   In what follows, we suppose that $\ottnt{e_{{\mathrm{1}}}}$ cannot be evaluated under $\Sigma$.
   By case analysis on the reduction rule applied to $\ottnt{e}$.
   \begin{caseanalysis}
    \case \E{Red}: We have $\mathsf{let} \, \ottsym{\{}  \ell  \ottsym{=}  \mathit{x}  \ottsym{;}  \mathit{y}  \ottsym{\}}  \ottsym{=}  \ottnt{e_{{\mathrm{1}}}} \, \mathsf{in} \, \ottnt{e_{{\mathrm{2}}}} \,  =  \,  \ottnt{E}  [  \ottnt{e'_{{\mathrm{1}}}}  ] $ and $\ottnt{e'} \,  =  \,  \ottnt{E}  [  \ottnt{e'_{{\mathrm{2}}}}  ] $
     for some $\ottnt{E}$, $\ottnt{e'_{{\mathrm{1}}}}$, and $\ottnt{e'_{{\mathrm{2}}}}$ such that $\ottnt{e'_{{\mathrm{1}}}}  \rightsquigarrow  \ottnt{e'_{{\mathrm{2}}}}$.
     Besides, $\Sigma' \,  =  \, \Sigma$.
     By case analysis on $\ottnt{E}$.
     \begin{caseanalysis}
      \case $\ottnt{E} \,  =  \,  [\,] $: By \reflem{subject-red-red}.
      \case $\ottnt{E} \,  =  \, \mathsf{let} \, \ottsym{\{}  \ell  \ottsym{=}  \mathit{x}  \ottsym{;}  \mathit{y}  \ottsym{\}}  \ottsym{=}  \ottnt{E'} \, \mathsf{in} \, \ottnt{e_{{\mathrm{2}}}}$: Contradictory with the assumption that
       $\ottnt{e_{{\mathrm{1}}}} \,  =  \,  \ottnt{E'}  [  \ottnt{e'_{{\mathrm{1}}}}  ] $ cannot be evaluated under $\Sigma$.
      \case otherwise: Contradictory with the assumption that $\mathsf{let} \, \ottsym{\{}  \ell  \ottsym{=}  \mathit{x}  \ottsym{;}  \mathit{y}  \ottsym{\}}  \ottsym{=}  \ottnt{e_{{\mathrm{1}}}} \, \mathsf{in} \, \ottnt{e_{{\mathrm{2}}}} \,  =  \,  \ottnt{E}  [  \ottnt{e'_{{\mathrm{1}}}}  ] $.
     \end{caseanalysis}

    \case \E{Blame}: By \T{Blame}.
    \case \E{TyBeta}: Contradictory with the assumption that
     $\ottnt{e_{{\mathrm{1}}}}$ cannot be evaluated under $\Sigma$.
   \end{caseanalysis}

  \case \T{VInj}:
   We have $\ottnt{e} \,  =  \, \ell \, \ottnt{e_{{\mathrm{0}}}}$ and, by inversion,
   $\Sigma  \ottsym{;}   \emptyset   \vdash  \ottnt{e_{{\mathrm{0}}}}  \ottsym{:}  \ottnt{B}$ and $\ottnt{A} \,  =  \,  \langle    \ell  \mathbin{:}  \ottnt{B}   ;  \rho   \rangle $
   for some $\ell$, $\ottnt{e_{{\mathrm{0}}}}$, $\ottnt{B}$, and $\rho$ such that
   $\Sigma  \ottsym{;}   \emptyset   \vdash  \rho  \ottsym{:}   \mathsf{R} $.

   If $ \Sigma  \mid  \ottnt{e_{{\mathrm{0}}}}   \longrightarrow   \Sigma'  \mid  \ottnt{e'_{{\mathrm{0}}}} $ for some $\ottnt{e'_{{\mathrm{0}}}}$, then
   we have $\Sigma'  \ottsym{;}   \emptyset   \vdash  \ottnt{e'_{{\mathrm{0}}}}  \ottsym{:}  \ottnt{B}$ by the IH, and therefore
   $\Sigma'  \ottsym{;}   \emptyset   \vdash  \ell \, \ottnt{e'_{{\mathrm{0}}}}  \ottsym{:}   \langle    \ell  \mathbin{:}  \ottnt{B}   ;  \rho   \rangle $
   by Lemmas~\ref{lem:eval-growing-tyname} and \ref{lem:weakening-tyname}, and \T{VInj}.

   In what follows, we suppose that $\ottnt{e_{{\mathrm{0}}}}$ cannot be evaluated under $\Sigma$.
   By case analysis on the reduction rule applied to $\ottnt{e}$.
   \begin{caseanalysis}
    \case \E{Red}: We have $\ell \, \ottnt{e_{{\mathrm{0}}}} \,  =  \,  \ottnt{E}  [  \ottnt{e'_{{\mathrm{1}}}}  ] $ and $\ottnt{e'} \,  =  \,  \ottnt{E}  [  \ottnt{e'_{{\mathrm{2}}}}  ] $
     for some $\ottnt{E}$, $\ottnt{e'_{{\mathrm{1}}}}$, and $\ottnt{e'_{{\mathrm{2}}}}$ such that $\ottnt{e'_{{\mathrm{1}}}}  \rightsquigarrow  \ottnt{e'_{{\mathrm{2}}}}$.
     Besides, $\Sigma' \,  =  \, \Sigma$.
     By case analysis on $\ottnt{E}$.
     \begin{caseanalysis}
      \case $\ottnt{E} \,  =  \,  [\,] $: By \reflem{subject-red-red}.
      \case $\ottnt{E} \,  =  \, \ell \, \ottnt{E'}$: Contradictory with the assumption that
       $\ottnt{e_{{\mathrm{0}}}} \,  =  \,  \ottnt{E'}  [  \ottnt{e'_{{\mathrm{1}}}}  ] $ cannot be evaluated under $\Sigma$.
      \case otherwise: Contradictory with the assumption that $\ell \, \ottnt{e_{{\mathrm{0}}}} \,  =  \,  \ottnt{E}  [  \ottnt{e'_{{\mathrm{1}}}}  ] $.
     \end{caseanalysis}

    \case \E{Blame}: By \T{Blame}.
    \case \E{TyBeta}: Contradictory with the assumption that
     $\ottnt{e_{{\mathrm{0}}}}$ cannot be evaluated under $\Sigma$.
   \end{caseanalysis}

  \case \T{VLift}:
   We have $\ottnt{e} \,  =  \,  \variantlift{ \ell }{ \ottnt{B} }{ \ottnt{e_{{\mathrm{0}}}} } $ and, by inversion,
   $\Sigma  \ottsym{;}   \emptyset   \vdash  \ottnt{e_{{\mathrm{0}}}}  \ottsym{:}   \langle  \rho  \rangle $ and $\ottnt{A} \,  =  \,  \langle    \ell  \mathbin{:}  \ottnt{B}   ;  \rho   \rangle $
   for some $\ell$, $\ottnt{e_{{\mathrm{0}}}}$, $\ottnt{B}$, and $\rho$ such that
   $\Sigma  \ottsym{;}   \emptyset   \vdash  \ottnt{B}  \ottsym{:}   \mathsf{T} $.

   If $ \Sigma  \mid  \ottnt{e_{{\mathrm{0}}}}   \longrightarrow   \Sigma'  \mid  \ottnt{e'_{{\mathrm{0}}}} $ for some $\ottnt{e'_{{\mathrm{0}}}}$, then
   we have $\Sigma'  \ottsym{;}   \emptyset   \vdash  \ottnt{e'_{{\mathrm{0}}}}  \ottsym{:}   \langle  \rho  \rangle $ by the IH, and therefore
   $\Sigma'  \ottsym{;}   \emptyset   \vdash   \variantlift{ \ell }{ \ottnt{B} }{ \ottnt{e'_{{\mathrm{0}}}} }   \ottsym{:}   \langle    \ell  \mathbin{:}  \ottnt{B}   ;  \rho   \rangle $
   by Lemmas~\ref{lem:eval-growing-tyname} and \ref{lem:weakening-tyname}, and \T{VLift}.

   In what follows, we suppose that $\ottnt{e_{{\mathrm{0}}}}$ cannot be evaluated under $\Sigma$.
   By case analysis on the reduction rule applied to $\ottnt{e}$.
   \begin{caseanalysis}
    \case \E{Red}: We have $ \variantlift{ \ell }{ \ottnt{B} }{ \ottnt{e_{{\mathrm{0}}}} }  \,  =  \,  \ottnt{E}  [  \ottnt{e'_{{\mathrm{1}}}}  ] $ and $\ottnt{e'} \,  =  \,  \ottnt{E}  [  \ottnt{e'_{{\mathrm{2}}}}  ] $
     for some $\ottnt{E}$, $\ottnt{e'_{{\mathrm{1}}}}$, and $\ottnt{e'_{{\mathrm{2}}}}$ such that $\ottnt{e'_{{\mathrm{1}}}}  \rightsquigarrow  \ottnt{e'_{{\mathrm{2}}}}$.
     Besides, $\Sigma' \,  =  \, \Sigma$.
     By case analysis on $\ottnt{E}$.
     \begin{caseanalysis}
      \case $\ottnt{E} \,  =  \,  [\,] $: By \reflem{subject-red-red}.
      \case $\ottnt{E} \,  =  \,  \variantlift{ \ell }{ \ottnt{B} }{ \ottnt{E'} } $: Contradictory with the assumption that
       $\ottnt{e_{{\mathrm{0}}}} \,  =  \,  \ottnt{E'}  [  \ottnt{e'_{{\mathrm{1}}}}  ] $ cannot be evaluated under $\Sigma$.
      \case otherwise: Contradictory with the assumption that $ \variantlift{ \ell }{ \ottnt{B} }{ \ottnt{e_{{\mathrm{0}}}} }  \,  =  \,  \ottnt{E}  [  \ottnt{e'_{{\mathrm{1}}}}  ] $.
     \end{caseanalysis}

    \case \E{Blame}: By \T{Blame}.
    \case \E{TyBeta}: Contradictory with the assumption that
     $\ottnt{e_{{\mathrm{0}}}}$ cannot be evaluated under $\Sigma$.
   \end{caseanalysis}

  \case \T{VCase}:
   We have $\ottnt{e} \,  =  \,  \mathsf{case} \,  \ottnt{e_{{\mathrm{0}}}}  \,\mathsf{with}\, \langle  \ell \,  \mathit{x}   \rightarrow   \ottnt{e_{{\mathrm{1}}}}   \ottsym{;}   \mathit{y}   \rightarrow   \ottnt{e_{{\mathrm{2}}}}  \rangle $ and, by inversion,
   $\Sigma  \ottsym{;}   \emptyset   \vdash  \ottnt{e_{{\mathrm{0}}}}  \ottsym{:}   \langle    \ell  \mathbin{:}  \ottnt{B}   ;  \rho   \rangle $ and
   $\Sigma  \ottsym{;}   \mathit{x}  \mathord{:}  \ottnt{B}   \vdash  \ottnt{e_{{\mathrm{1}}}}  \ottsym{:}  \ottnt{A}$ and
   $\Sigma  \ottsym{;}   \mathit{y}  \mathord{:}   \langle  \rho  \rangle    \vdash  \ottnt{e_{{\mathrm{2}}}}  \ottsym{:}  \ottnt{A}$
   for some $\ell$, $\ottnt{e_{{\mathrm{0}}}}$, $\ottnt{e_{{\mathrm{1}}}}$, $\ottnt{e_{{\mathrm{2}}}}$, $\ottnt{B}$, $\rho$, $\mathit{x}$, and $\mathit{y}$.

   If $ \Sigma  \mid  \ottnt{e_{{\mathrm{0}}}}   \longrightarrow   \Sigma'  \mid  \ottnt{e'_{{\mathrm{0}}}} $ for some $\ottnt{e'_{{\mathrm{0}}}}$, then
   we have $\Sigma'  \ottsym{;}   \emptyset   \vdash  \ottnt{e'_{{\mathrm{0}}}}  \ottsym{:}   \langle    \ell  \mathbin{:}  \ottnt{B}   ;  \rho   \rangle $ by the IH, and therefore
   $\Sigma'  \ottsym{;}   \emptyset   \vdash   \mathsf{case} \,  \ottnt{e'_{{\mathrm{0}}}}  \,\mathsf{with}\, \langle  \ell \,  \mathit{x}   \rightarrow   \ottnt{e_{{\mathrm{1}}}}   \ottsym{;}   \mathit{y}   \rightarrow   \ottnt{e_{{\mathrm{2}}}}  \rangle   \ottsym{:}  \ottnt{A}$
   by Lemmas~\ref{lem:eval-growing-tyname} and \ref{lem:weakening-tyname}, and \T{VCase}.

   In what follows, we suppose that $\ottnt{e_{{\mathrm{0}}}}$ cannot be evaluated under $\Sigma$.
   By case analysis on the reduction rule applied to $\ottnt{e}$.
   \begin{caseanalysis}
    \case \E{Red}: We have $ \mathsf{case} \,  \ottnt{e_{{\mathrm{0}}}}  \,\mathsf{with}\, \langle  \ell \,  \mathit{x}   \rightarrow   \ottnt{e_{{\mathrm{1}}}}   \ottsym{;}   \mathit{y}   \rightarrow   \ottnt{e_{{\mathrm{2}}}}  \rangle  \,  =  \,  \ottnt{E}  [  \ottnt{e'_{{\mathrm{1}}}}  ] $ and $\ottnt{e'} \,  =  \,  \ottnt{E}  [  \ottnt{e'_{{\mathrm{2}}}}  ] $
     for some $\ottnt{E}$, $\ottnt{e'_{{\mathrm{1}}}}$, and $\ottnt{e'_{{\mathrm{2}}}}$ such that $\ottnt{e'_{{\mathrm{1}}}}  \rightsquigarrow  \ottnt{e'_{{\mathrm{2}}}}$.
     Besides, $\Sigma' \,  =  \, \Sigma$.
     By case analysis on $\ottnt{E}$.
     \begin{caseanalysis}
      \case $\ottnt{E} \,  =  \,  [\,] $: By \reflem{subject-red-red}.
      \case $\ottnt{E} \,  =  \,  \mathsf{case} \,  \ottnt{E'}  \,\mathsf{with}\, \langle  \ell \,  \mathit{x}   \rightarrow   \ottnt{e_{{\mathrm{1}}}}   \ottsym{;}   \mathit{y}   \rightarrow   \ottnt{e_{{\mathrm{2}}}}  \rangle $: Contradictory with the assumption that
       $\ottnt{e_{{\mathrm{0}}}} \,  =  \,  \ottnt{E'}  [  \ottnt{e'_{{\mathrm{1}}}}  ] $ cannot be evaluated under $\Sigma$.
      \case otherwise: Contradictory with the assumption that $ \mathsf{case} \,  \ottnt{e_{{\mathrm{0}}}}  \,\mathsf{with}\, \langle  \ell \,  \mathit{x}   \rightarrow   \mathit{x}   \ottsym{;}   \mathit{y}   \rightarrow   \ottnt{e_{{\mathrm{2}}}}  \rangle  \,  =  \,  \ottnt{E}  [  \ottnt{e'_{{\mathrm{1}}}}  ] $.
     \end{caseanalysis}

    \case \E{Blame}: By \T{Blame}.
    \case \E{TyBeta}: Contradictory with the assumption that
     $\ottnt{e_{{\mathrm{0}}}}$ cannot be evaluated under $\Sigma$.
   \end{caseanalysis}

  \case \T{Cast}:
   We have $\ottnt{e} \,  =  \, \ottnt{e_{{\mathrm{0}}}}  \ottsym{:}  \ottnt{B} \,  \stackrel{ \ottnt{p} }{\Rightarrow}  \ottnt{A} $ and, by inversion,
   $\Sigma  \ottsym{;}   \emptyset   \vdash  \ottnt{e_{{\mathrm{0}}}}  \ottsym{:}  \ottnt{B}$ and $\ottnt{B}  \simeq  \ottnt{A}$ and $\Sigma  \ottsym{;}   \emptyset   \vdash  \ottnt{A}  \ottsym{:}   \mathsf{T} $
   for some $\ottnt{e_{{\mathrm{0}}}}$, $\ottnt{A}$, $\ottnt{B}$, and $\ottnt{p}$.

   If $ \Sigma  \mid  \ottnt{e_{{\mathrm{0}}}}   \longrightarrow   \Sigma'  \mid  \ottnt{e'_{{\mathrm{0}}}} $ for some $\ottnt{e'_{{\mathrm{0}}}}$, then
   we have $\Sigma'  \ottsym{;}   \emptyset   \vdash  \ottnt{e'_{{\mathrm{0}}}}  \ottsym{:}  \ottnt{B}$ by the IH, and therefore
   $\Sigma'  \ottsym{;}   \emptyset   \vdash  \ottnt{e'_{{\mathrm{0}}}}  \ottsym{:}  \ottnt{B} \,  \stackrel{ \ottnt{p} }{\Rightarrow}  \ottnt{A}   \ottsym{:}  \ottnt{A}$
   by Lemmas~\ref{lem:eval-growing-tyname} and \ref{lem:weakening-tyname}, and \T{Cast}.

   In what follows, we suppose that $\ottnt{e_{{\mathrm{0}}}}$ cannot be evaluated under $\Sigma$.
   By case analysis on the reduction rule applied to $\ottnt{e}$.
   \begin{caseanalysis}
    \case \E{Red}: We have $\ottnt{e_{{\mathrm{0}}}}  \ottsym{:}  \ottnt{B} \,  \stackrel{ \ottnt{p} }{\Rightarrow}  \ottnt{A}  \,  =  \,  \ottnt{E}  [  \ottnt{e'_{{\mathrm{1}}}}  ] $ and $\ottnt{e'} \,  =  \,  \ottnt{E}  [  \ottnt{e'_{{\mathrm{2}}}}  ] $
     for some $\ottnt{E}$, $\ottnt{e'_{{\mathrm{1}}}}$, and $\ottnt{e'_{{\mathrm{2}}}}$ such that $\ottnt{e'_{{\mathrm{1}}}}  \rightsquigarrow  \ottnt{e'_{{\mathrm{2}}}}$.
     Besides, $\Sigma' \,  =  \, \Sigma$.
     By case analysis on $\ottnt{E}$.
     \begin{caseanalysis}
      \case $\ottnt{E} \,  =  \,  [\,] $: By \reflem{subject-red-red}.
      \case $\ottnt{E} \,  =  \, \ottnt{E'}  \ottsym{:}  \ottnt{B} \,  \stackrel{ \ottnt{p} }{\Rightarrow}  \ottnt{A} $: Contradictory with the assumption that
       $\ottnt{e_{{\mathrm{0}}}} \,  =  \,  \ottnt{E'}  [  \ottnt{e'_{{\mathrm{1}}}}  ] $ cannot be evaluated under $\Sigma$.
      \case otherwise: Contradictory with the assumption that $\ottnt{e_{{\mathrm{0}}}}  \ottsym{:}  \ottnt{B} \,  \stackrel{ \ottnt{p} }{\Rightarrow}  \ottnt{A}  \,  =  \,  \ottnt{E}  [  \ottnt{e'_{{\mathrm{1}}}}  ] $.
     \end{caseanalysis}

    \case \E{Blame}: By \T{Blame}.
    \case \E{TyBeta}: Contradictory with the assumption that
     $\ottnt{e_{{\mathrm{0}}}}$ cannot be evaluated under $\Sigma$.
   \end{caseanalysis}

  \case \T{Conv}:
   We have $\ottnt{e} \,  =  \, \ottnt{e_{{\mathrm{0}}}}  \ottsym{:}  \ottnt{B} \,  \stackrel{ \Phi }{\Rightarrow}  \ottnt{A} $ and, by inversion,
   $\Sigma  \ottsym{;}   \emptyset   \vdash  \ottnt{e_{{\mathrm{0}}}}  \ottsym{:}  \ottnt{B}$ and $ \Sigma   \vdash   \ottnt{B}  \prec^{ \Phi }  \ottnt{A} $ and $\Sigma  \ottsym{;}   \emptyset   \vdash  \ottnt{A}  \ottsym{:}   \mathsf{T} $
   for some $\ottnt{e_{{\mathrm{0}}}}$, $\ottnt{A}$, $\ottnt{B}$, and $\Phi$.

   If $ \Sigma  \mid  \ottnt{e_{{\mathrm{0}}}}   \longrightarrow   \Sigma'  \mid  \ottnt{e'_{{\mathrm{0}}}} $ for some $\ottnt{e'_{{\mathrm{0}}}}$, then
   we have $\Sigma'  \ottsym{;}   \emptyset   \vdash  \ottnt{e'_{{\mathrm{0}}}}  \ottsym{:}  \ottnt{B}$ by the IH, and therefore
   $\Sigma'  \ottsym{;}   \emptyset   \vdash  \ottnt{e'_{{\mathrm{0}}}}  \ottsym{:}  \ottnt{B} \,  \stackrel{ \Phi }{\Rightarrow}  \ottnt{A}   \ottsym{:}  \ottnt{A}$
   by Lemmas~\ref{lem:eval-growing-tyname} and \ref{lem:weakening-tyname}, and \T{Conv}.

   In what follows, we suppose that $\ottnt{e_{{\mathrm{0}}}}$ cannot be evaluated under $\Sigma$.
   By case analysis on the reduction rule applied to $\ottnt{e}$.
   \begin{caseanalysis}
    \case \E{Red}: We have $\ottnt{e_{{\mathrm{0}}}}  \ottsym{:}  \ottnt{B} \,  \stackrel{ \Phi }{\Rightarrow}  \ottnt{A}  \,  =  \,  \ottnt{E}  [  \ottnt{e'_{{\mathrm{1}}}}  ] $ and $\ottnt{e'} \,  =  \,  \ottnt{E}  [  \ottnt{e'_{{\mathrm{2}}}}  ] $
     for some $\ottnt{E}$, $\ottnt{e'_{{\mathrm{1}}}}$, and $\ottnt{e'_{{\mathrm{2}}}}$ such that $\ottnt{e'_{{\mathrm{1}}}}  \rightsquigarrow  \ottnt{e'_{{\mathrm{2}}}}$.
     Besides, $\Sigma' \,  =  \, \Sigma$.
     By case analysis on $\ottnt{E}$.
     \begin{caseanalysis}
      \case $\ottnt{E} \,  =  \,  [\,] $: By \reflem{subject-red-red}.
      \case $\ottnt{E} \,  =  \, \ottnt{E'}  \ottsym{:}  \ottnt{B} \,  \stackrel{ \Phi }{\Rightarrow}  \ottnt{A} $: Contradictory with the assumption that
       $\ottnt{e_{{\mathrm{0}}}} \,  =  \,  \ottnt{E'}  [  \ottnt{e'_{{\mathrm{1}}}}  ] $ cannot be evaluated under $\Sigma$.
      \case otherwise: Contradictory with the assumption that $\ottnt{e_{{\mathrm{0}}}}  \ottsym{:}  \ottnt{B} \,  \stackrel{ \Phi }{\Rightarrow}  \ottnt{A}  \,  =  \,  \ottnt{E}  [  \ottnt{e'_{{\mathrm{1}}}}  ] $.
     \end{caseanalysis}

    \case \E{Blame}: By \T{Blame}.
    \case \E{TyBeta}: Contradictory with the assumption that
     $\ottnt{e_{{\mathrm{0}}}}$ cannot be evaluated under $\Sigma$.
   \end{caseanalysis}
 \end{caseanalysis}
\end{proof}

\begin{thm}[Type soundness]
 If $ \emptyset   \ottsym{;}   \emptyset   \vdash  \ottnt{e}  \ottsym{:}  \ottnt{A}$ and $  \emptyset   \mid  \ottnt{e}   \longrightarrow^{*}   \Sigma'  \mid  \ottnt{e'} $ and
 $\ottnt{e'}$ cannot be evaluated under $\Sigma'$, then
 either $\ottnt{e'}$ is a value or $\ottnt{e'} \,  =  \, \mathsf{blame} \, \ottnt{p}$ for some $\ottnt{p}$.
\end{thm}
\begin{proof}
 By Lemmas~\ref{lem:subject-red} and \ref{lem:progress}.
\end{proof}

\subsection{Type-preserving translation}
\label{sec:proof:trans}

\begin{assum}
 We assume that $\ottnt{A}  \simeq  \ottnt{A}  \oplus  \ottnt{B}$ and $\ottnt{B}  \simeq  \ottnt{A}  \oplus  \ottnt{B}$ and that
 if $\Gamma  \vdash  \ottnt{A}  \ottsym{:}   \mathsf{T} $ and $\Gamma  \vdash  \ottnt{B}  \ottsym{:}   \mathsf{T} $, then $\Gamma  \vdash  \ottnt{A}  \oplus  \ottnt{B}  \ottsym{:}   \mathsf{T} $.
\end{assum}

\begin{lemma}{ce-type-matching}
 \begin{enumerate}
  \item If $\ottnt{A}  \triangleright  \ottnt{B}$, then $\ottnt{A}  \simeq  \ottnt{B}$.
        Furthermore, if $\Sigma  \ottsym{;}  \Gamma  \vdash  \ottnt{A}  \ottsym{:}  \ottnt{K}$, then $\Sigma  \ottsym{;}  \Gamma  \vdash  \ottnt{B}  \ottsym{:}  \ottnt{K}$.
  \item If $\ottnt{A}  \triangleright   [  \rho  ] $ and $\rho \,  \triangleright _{ \ell }  \, \ottnt{B}  \ottsym{,}  \rho'$,
        then $\ottnt{A}  \simeq   [    \ell  \mathbin{:}  \ottnt{B}   ;  \rho'   ] $.
        Furthermore, if $\Sigma  \ottsym{;}  \Gamma  \vdash  \ottnt{A}  \ottsym{:}   \mathsf{T} $,
        then $\Sigma  \ottsym{;}  \Gamma  \vdash   [    \ell  \mathbin{:}  \ottnt{B}   ;  \rho'   ]   \ottsym{:}   \mathsf{T} $.
  \item If $\ottnt{A}  \triangleright   \langle  \rho  \rangle $ and $\rho \,  \triangleright _{ \ell }  \, \ottnt{B}  \ottsym{,}  \rho'$,
        then $\ottnt{A}  \simeq   \langle    \ell  \mathbin{:}  \ottnt{B}   ;  \rho'   \rangle $.
        Furthermore, if $\Sigma  \ottsym{;}  \Gamma  \vdash  \ottnt{A}  \ottsym{:}   \mathsf{T} $,
        then $\Sigma  \ottsym{;}  \Gamma  \vdash   \langle    \ell  \mathbin{:}  \ottnt{B}   ;  \rho'   \rangle   \ottsym{:}   \mathsf{T} $.
 \end{enumerate}
\end{lemma}
\begin{proof}
 \begin{enumerate}
  \item Obvious by the definition of type matching.
  \item If $\ottnt{A}$ is $ \star $, it is trivial to show.
        Otherwise, $\ottnt{A} \,  =  \,  [  \rho  ] $.
        If $\ell \,  \in  \, \mathit{dom} \, \ottsym{(}  \rho  \ottsym{)}$, then $\rho  \equiv    \ell  \mathbin{:}  \ottnt{B}   ;  \rho' $.
        Thus, $\rho  \simeq    \ell  \mathbin{:}  \ottnt{B}   ;  \rho' $ by \reflem{consistent-subsume-equiv}.
        Thus, by \CE{Record}, $ [  \rho  ]   \simeq   [    \ell  \mathbin{:}  \ottnt{B}   ;  \rho'   ] $
        Since $\Sigma  \ottsym{;}  \Gamma  \vdash  \rho  \ottsym{:}   \mathsf{R} $, we find that
        $\Sigma  \ottsym{;}  \Gamma  \vdash  \ottnt{B}  \ottsym{:}   \mathsf{T} $ and $\Sigma  \ottsym{;}  \Gamma  \vdash  \rho'  \ottsym{:}   \mathsf{R} $.
        Thus, $\Sigma  \ottsym{;}  \Gamma  \vdash   [    \ell  \mathbin{:}  \ottnt{B}   ;  \rho'   ]   \ottsym{:}   \mathsf{T} $ by \WF{Cons} and \WF{Record}.
  \item Similarly to the case for record types.
 \end{enumerate}
\end{proof}

\begin{lemma}{typctx-type-wf-surface2inter}
 \begin{enumerate}
  \item If $\vdash  \Gamma$, then $ \emptyset   \vdash  \Gamma$.
  \item If $\Gamma  \vdash  \ottnt{A}  \ottsym{:}  \ottnt{K}$, then $ \emptyset   \ottsym{;}  \Gamma  \vdash  \ottnt{A}  \ottsym{:}  \ottnt{K}$.
 \end{enumerate}
\end{lemma}
\begin{proof}
 Straightforward by mutual induction on the derivations.
\end{proof}

\begin{lemma}{trans-well-typed}
 If $\Gamma  \vdash  \ottnt{M}  \ottsym{:}  \ottnt{A}  \hookrightarrow  \ottnt{e}$, then $ \emptyset   \ottsym{;}  \Gamma  \vdash  \ottnt{e}  \ottsym{:}  \ottnt{A}$.
\end{lemma}
\begin{proof}
 By induction on the derivation of $\Gamma  \vdash  \ottnt{M}  \ottsym{:}  \ottnt{A}  \hookrightarrow  \ottnt{e}$.
 The proof is straightforward by using the assumption about $ \oplus $ stated in
 this section and Lemmas~\ref{lem:ce-type-matching},
 \ref{lem:typctx-type-wf-surface2inter}, and \ref{lem:typctx-type-wf}.
\end{proof}

\begin{lemma}{trans-succeed}
 If $\Gamma  \vdash  \ottnt{M}  \ottsym{:}  \ottnt{A}$, then $\Gamma  \vdash  \ottnt{M}  \ottsym{:}  \ottnt{A}  \hookrightarrow  \ottnt{e}$ for some $\ottnt{e}$.
\end{lemma}
\begin{proof}
 Straightforward by induction on the typing derivation.
\end{proof}

\begin{thm}
 If $\Gamma  \vdash  \ottnt{M}  \ottsym{:}  \ottnt{A}$, then there exists some $\ottnt{e}$ such that
 $\Gamma  \vdash  \ottnt{M}  \ottsym{:}  \ottnt{A}  \hookrightarrow  \ottnt{e}$ and $ \emptyset   \ottsym{;}  \Gamma  \vdash  \ottnt{e}  \ottsym{:}  \ottnt{A}$.
\end{thm}
\begin{proof}
 By Lemmas~\ref{lem:trans-succeed} and \ref{lem:trans-well-typed}.
\end{proof}

\subsection{Conservativity over typing}
\label{sec:proof:conservativity:typing}

In this section, we write $\makestatic{\Gamma}$, $\makestatic{\mathit{A} }$, $\makestatic{\rho}$, $\makestatic{\mathit{M} }$ for typing
contexts, types, rows, and terms where $ \star $ and any type name do not appear.

\begin{defn}
 We write $\Gamma_{{\mathrm{1}}}  \equiv  \Gamma_{{\mathrm{2}}}$ if and only if
 (1) $\Gamma_{{\mathrm{1}}} \,  =  \,  \emptyset $ and $\Gamma_{{\mathrm{2}}} \,  =  \,  \emptyset $;
 (2) $\Gamma_{{\mathrm{1}}} \,  =  \, \Gamma'_{{\mathrm{1}}}  \ottsym{,}   \mathit{x}  \mathord{:}  \ottnt{A} $ and $\Gamma_{{\mathrm{2}}} \,  =  \, \Gamma'_{{\mathrm{2}}}  \ottsym{,}   \mathit{x}  \mathord{:}  \ottnt{B} $ and $\Gamma'_{{\mathrm{1}}}  \equiv  \Gamma'_{{\mathrm{2}}}$ and $\ottnt{A}  \equiv  \ottnt{B}$; or
 (3) $\Gamma_{{\mathrm{1}}} \,  =  \, \Gamma'_{{\mathrm{1}}}  \ottsym{,}   \mathit{X}  \mathord{:}  \ottnt{K} $ and $\Gamma_{{\mathrm{2}}} \,  =  \, \Gamma'_{{\mathrm{2}}}  \ottsym{,}   \mathit{X}  \mathord{:}  \ottnt{K} $ and $\Gamma'_{{\mathrm{1}}}  \equiv  \Gamma'_{{\mathrm{2}}}$.
\end{defn}

\begin{assum}
 We assume that $\makestatic{\mathit{A} }  \oplus  \makestatic{\mathit{B} }$ is defined if and only if $\makestatic{\mathit{A} }  \equiv  \makestatic{\mathit{B} }$, and
 if $\makestatic{\mathit{A} }  \equiv  \makestatic{\mathit{B} }$, then $\makestatic{\mathit{A} }  \oplus  \makestatic{\mathit{B} }  \equiv  \makestatic{\mathit{A} }$.
\end{assum}

\begin{assum}
 We assume that, if $\ottnt{A_{{\mathrm{1}}}}  \equiv  \ottnt{A_{{\mathrm{2}}}}$ and $\ottnt{B_{{\mathrm{1}}}}  \equiv  \ottnt{B_{{\mathrm{2}}}}$, then
 $\ottnt{A_{{\mathrm{1}}}}  \oplus  \ottnt{B_{{\mathrm{1}}}}  \equiv  \ottnt{A_{{\mathrm{2}}}}  \oplus  \ottnt{B_{{\mathrm{2}}}}$.
\end{assum}

\begin{lemma}{typing-tctx-reorder}
 Suppose that $\Gamma  \equiv  \Gamma'$.
 \begin{enumerate}
  \item If $\vdash  \Gamma$, then $\vdash  \Gamma'$.
  \item If $\Gamma  \vdash  \ottnt{A}  \ottsym{:}  \ottnt{K}$, then $\Gamma'  \vdash  \ottnt{A'}  \ottsym{:}  \ottnt{K}$ for any $\ottnt{A'}$ such that
        $\ottnt{A}  \equiv  \ottnt{A'}$.
  \item If $\Gamma  \vdash  \ottnt{M}  \ottsym{:}  \ottnt{A}$,
        then $\Gamma'  \vdash  \ottnt{M}  \ottsym{:}  \ottnt{A'}$ for some $\ottnt{A'}$ such that $\ottnt{A}  \equiv  \ottnt{A'}$.
 \end{enumerate}
 We mention only the interesting cases.
 \begin{caseanalysis}
  \case \WFg{Cons}:
   We are given $\Gamma  \vdash    \ell  \mathbin{:}  \ottnt{B}   ;  \rho   \ottsym{:}   \mathsf{R} $ and, by inversion,
   $\Gamma  \vdash  \ottnt{B}  \ottsym{:}   \mathsf{T} $ and $\Gamma  \vdash  \rho  \ottsym{:}   \mathsf{R} $.

   We suppose that some $\rho'$ such that $  \ell  \mathbin{:}  \ottnt{B}   ;  \rho   \equiv  \rho'$ is given.
   Since $  \ell  \mathbin{:}  \ottnt{B}   ;  \rho   \equiv  \rho'$, there exists some $\ottnt{B''}$ and $\rho''$ such that
   $\rho' \,  \triangleright _{ \ell }  \, \ottnt{B''}  \ottsym{,}  \rho''$ and $\ottnt{B}  \equiv  \ottnt{B''}$ and $\rho  \equiv  \rho''$.
   By the IHs, $\Gamma'  \vdash  \ottnt{B''}  \ottsym{:}   \mathsf{T} $ and $\Gamma'  \vdash  \rho''  \ottsym{:}   \mathsf{R} $.
   Thus, $\Gamma'  \vdash    \ell  \mathbin{:}  \ottnt{B''}   ;  \rho''   \ottsym{:}   \mathsf{R} $ by \WFg{Cons}.
   We can show that $\Gamma'  \vdash  \rho'  \ottsym{:}   \mathsf{R} $ by the fact that $\rho' \,  \triangleright _{ \ell }  \, \ottnt{B''}  \ottsym{,}  \rho''$.

  \case \Tg{App}: We are given $\Gamma  \vdash  \ottnt{M_{{\mathrm{1}}}} \, \ottnt{M_{{\mathrm{2}}}}  \ottsym{:}  \ottnt{A}$ and, by inversion,
  $\Gamma  \vdash  \ottnt{M_{{\mathrm{1}}}}  \ottsym{:}  \ottnt{A_{{\mathrm{1}}}}$ and $\Gamma  \vdash  \ottnt{M_{{\mathrm{2}}}}  \ottsym{:}  \ottnt{A_{{\mathrm{2}}}}$ and
  $\ottnt{A_{{\mathrm{1}}}}  \triangleright  \ottnt{A_{{\mathrm{11}}}}  \rightarrow  \ottnt{A}$ and $\ottnt{A_{{\mathrm{2}}}}  \simeq  \ottnt{A_{{\mathrm{11}}}}$.

  If $\ottnt{A} \,  =  \, \star$, it is easy to show.

  Otherwise, we can suppose that $\ottnt{A} \,  =  \, \ottnt{A_{{\mathrm{11}}}}  \rightarrow  \ottnt{A}$.
  By the IHs with \reflem{equiv-inversion},
  $\Gamma'  \vdash  \ottnt{M_{{\mathrm{1}}}}  \ottsym{:}  \ottnt{A'_{{\mathrm{11}}}}  \rightarrow  \ottnt{A'}$ and $\Gamma'  \vdash  \ottnt{M_{{\mathrm{2}}}}  \ottsym{:}  \ottnt{A'_{{\mathrm{2}}}}$
  for some $\ottnt{A'}$, $\ottnt{A'_{{\mathrm{11}}}}$, $\ottnt{A'_{{\mathrm{2}}}}$ such that
  $\ottnt{A}  \equiv  \ottnt{A'}$, $\ottnt{A_{{\mathrm{11}}}}  \equiv  \ottnt{A'_{{\mathrm{11}}}}$, and $\ottnt{A_{{\mathrm{2}}}}  \equiv  \ottnt{A'_{{\mathrm{2}}}}$.
  By Theorem~\ref{thm:ce-consistency-equiv},
  $\ottnt{A'_{{\mathrm{2}}}}  \simeq  \ottnt{A'_{{\mathrm{11}}}}$.  Thus, we finish by \Tg{App}.
  
  \case \Tg{TApp}: This case uses the fact that $\ottnt{A}  \equiv  \ottnt{B}$, then $ \ottnt{A}    [  \ottnt{C}  /  \mathit{X}  ]    \equiv   \ottnt{B}    [  \ottnt{C}  /  \mathit{X}  ]  $.

  \case \Tg{VCase}: This cases uses the second assumption about $ \oplus $ stated
  in this section.
 \end{caseanalysis}
\end{lemma}

\begin{lemma}{cnsv-static-consistent-equiv}
 If $\makestatic{\mathit{A} }  \simeq  \makestatic{\mathit{B} }$, then $\makestatic{\mathit{A} }  \equiv  \makestatic{\mathit{B} }$.
\end{lemma}
\begin{proof}
 By \reflem{consistent-decomp}, there exists some $\makestatic{\mathit{C} }$ such that
 $\makestatic{\mathit{A} }  \equiv  \makestatic{\mathit{C} }$ and $\makestatic{\mathit{C} }  \sim  \makestatic{\mathit{B} }$.
 Then, it is easy to show that $\makestatic{\mathit{C} } \,  =  \, \makestatic{\mathit{B} }$ by induction on the derivation of $\makestatic{\mathit{C} }  \sim  \makestatic{\mathit{B} }$.
\end{proof}

\begin{lemma}{cnsv-typing-gs}
 \begin{enumerate}
  \item If $\vdash  \makestatic{\Gamma}$, then $ \mathrel{ \makestatic{\vdash} }  \makestatic{\Gamma} $.
  \item If $\makestatic{\Gamma}  \vdash  \makestatic{\mathit{A} }  \ottsym{:}  \ottnt{K}$, then $ \makestatic{\Gamma}  \mathrel{ \makestatic{\vdash} }  \makestatic{\mathit{A} }  :  \ottnt{K} $.
  \item If $\makestatic{\Gamma}  \vdash  \makestatic{\mathit{M} }  \ottsym{:}  \ottnt{A}$, then $ \makestatic{\Gamma}  \mathrel{ \makestatic{\vdash} }  \makestatic{\mathit{M} }  :  \ottnt{A} $.
 \end{enumerate}
\end{lemma}
xo\begin{proof}
 By mutual induction on the derivations.

 Below are important facts to show this lemma.
 \begin{enumerate}
  \item If $ \makestatic{\Gamma}  \mathrel{ \makestatic{\vdash} }  \makestatic{\mathit{M} }  :  \ottnt{A} $, then $ \star $ and any type name do not appear in $\ottnt{A}$.
  \item If $\makestatic{\mathit{A} }  \triangleright  \makestatic{\mathit{B} }$, then $\makestatic{\mathit{A} } \,  =  \, \makestatic{\mathit{B} }$.
  \item If $\makestatic{\rho}_{{\mathrm{1}}} \,  \triangleright _{ \ell }  \, \makestatic{\mathit{A} }  \ottsym{,}  \makestatic{\rho}_{{\mathrm{2}}}$, then $\makestatic{\rho}_{{\mathrm{1}}}  \equiv    \ell  \mathbin{:}  \makestatic{\mathit{A} }   ;  \makestatic{\rho}_{{\mathrm{2}}} $.
 \end{enumerate}

 The case for \Tg{App} is interesting, so we mention only that case.
 We are given $\makestatic{\Gamma}  \vdash  \makestatic{\mathit{M} }_{{\mathrm{1}}} \, \makestatic{\mathit{M} }_{{\mathrm{2}}}  \ottsym{:}  \ottnt{A}$ and, by inversion,
 $\makestatic{\Gamma}  \vdash  \makestatic{\mathit{M} }_{{\mathrm{1}}}  \ottsym{:}  \ottnt{B}$ and $\makestatic{\Gamma}  \vdash  \makestatic{\mathit{M} }_{{\mathrm{2}}}  \ottsym{:}  \ottnt{C}$ and
 $\ottnt{B}  \triangleright  \ottnt{B_{{\mathrm{1}}}}  \rightarrow  \ottnt{A}$ and $\ottnt{C}  \simeq  \ottnt{B_{{\mathrm{1}}}}$.
 By the IHs, $ \makestatic{\Gamma}  \mathrel{ \makestatic{\vdash} }  \makestatic{\mathit{M} }_{{\mathrm{1}}}  :  \ottnt{B} $ and $ \makestatic{\Gamma}  \mathrel{ \makestatic{\vdash} }  \makestatic{\mathit{M} }_{{\mathrm{2}}}  :  \ottnt{C} $.
 Thus, we can find $ \star $ and any type name do not appear in
 $\ottnt{B}$ nor $\ottnt{C}$.
 Thus, $\ottnt{B} \,  =  \, \ottnt{B_{{\mathrm{1}}}}  \rightarrow  \ottnt{A}$.
 Since $\ottnt{C}  \simeq  \ottnt{B_{{\mathrm{1}}}}$, we find $\ottnt{C}  \equiv  \ottnt{B_{{\mathrm{1}}}}$
 by \reflem{cnsv-static-consistent-equiv}.
 Thus, by \Ts{Equiv}, $ \makestatic{\Gamma}  \mathrel{ \makestatic{\vdash} }  \makestatic{\mathit{M} }_{{\mathrm{2}}}  :  \ottnt{B_{{\mathrm{1}}}} $.
 By \Ts{App}, we have $ \makestatic{\Gamma}  \mathrel{ \makestatic{\vdash} }  \makestatic{\mathit{M} }_{{\mathrm{1}}} \, \makestatic{\mathit{M} }_{{\mathrm{2}}}  :  \ottnt{A} $.

 The first assumption about $ \oplus $ stated in this section
 is used in the case for \Tg{VCase}.
\end{proof}

\begin{lemma}{cnsv-typing-sg}
 \begin{enumerate}
  \item If $ \mathrel{ \makestatic{\vdash} }  \makestatic{\Gamma} $, then $\vdash  \makestatic{\Gamma}$.
  \item If $ \makestatic{\Gamma}  \mathrel{ \makestatic{\vdash} }  \makestatic{\mathit{A} }  :  \ottnt{K} $, then $\makestatic{\Gamma}  \vdash  \makestatic{\mathit{A} }  \ottsym{:}  \ottnt{K}$.
  \item If $ \makestatic{\Gamma}  \mathrel{ \makestatic{\vdash} }  \makestatic{\mathit{M} }  :  \makestatic{\mathit{A} } $, then $\makestatic{\Gamma}  \vdash  \makestatic{\mathit{M} }  \ottsym{:}  \makestatic{\mathit{B} }$
        for some $\makestatic{\mathit{B} }$ such that $\makestatic{\mathit{A} }  \equiv  \makestatic{\mathit{B} }$.
 \end{enumerate}
\end{lemma}
\begin{proof}
 By mutual induction on the derivations.
 We mention only the interesting cases.
 \begin{caseanalysis}
  \case \Ts{App}:
   We are given $ \makestatic{\Gamma}  \mathrel{ \makestatic{\vdash} }  \makestatic{\mathit{M} }_{{\mathrm{1}}} \, \makestatic{\mathit{M} }_{{\mathrm{2}}}  :  \makestatic{\mathit{A} } $ and, by inversion,
   $ \makestatic{\Gamma}  \mathrel{ \makestatic{\vdash} }  \makestatic{\mathit{M} }_{{\mathrm{1}}}  :  \makestatic{\mathit{B} }  \rightarrow  \makestatic{\mathit{A} } $ and $ \makestatic{\Gamma}  \mathrel{ \makestatic{\vdash} }  \makestatic{\mathit{M} }_{{\mathrm{2}}}  :  \makestatic{\mathit{B} } $.
   By the IHs, $\makestatic{\Gamma}  \vdash  \makestatic{\mathit{M} }_{{\mathrm{1}}}  \ottsym{:}  \makestatic{\mathit{B} }_{{\mathrm{1}}}  \rightarrow  \makestatic{\mathit{A} }_{{\mathrm{1}}}$ and $\makestatic{\Gamma}  \vdash  \makestatic{\mathit{M} }_{{\mathrm{2}}}  \ottsym{:}  \makestatic{\mathit{B} }_{{\mathrm{2}}}$ and
   $\makestatic{\mathit{B} }  \rightarrow  \makestatic{\mathit{A} }  \equiv  \makestatic{\mathit{B} }_{{\mathrm{1}}}  \rightarrow  \makestatic{\mathit{A} }_{{\mathrm{1}}}$ and $\makestatic{\mathit{B} }  \equiv  \makestatic{\mathit{B} }_{{\mathrm{2}}}$
   for some $\makestatic{\mathit{B} }_{{\mathrm{1}}}$, $\makestatic{\mathit{B} }_{{\mathrm{2}}}$, and $\makestatic{\mathit{A} }_{{\mathrm{1}}}$.

   We have $\makestatic{\mathit{B} }_{{\mathrm{1}}}  \rightarrow  \makestatic{\mathit{A} }_{{\mathrm{1}}}  \triangleright  \makestatic{\mathit{B} }_{{\mathrm{1}}}  \rightarrow  \makestatic{\mathit{A} }_{{\mathrm{1}}}$.
   By \reflem{equiv-inversion} (\ref{lem:equiv-inversion:fun}),
   we have
   $\makestatic{\mathit{B} }  \equiv  \makestatic{\mathit{B} }_{{\mathrm{1}}}$ and $\makestatic{\mathit{A} }  \equiv  \makestatic{\mathit{A} }_{{\mathrm{1}}}$.
   Thus, $\makestatic{\mathit{B} }_{{\mathrm{2}}}  \equiv  \makestatic{\mathit{B} }_{{\mathrm{1}}}$.
   By \reflem{consistent-comp-equiv-consistent}, $\makestatic{\mathit{B} }_{{\mathrm{2}}}  \simeq  \makestatic{\mathit{B} }_{{\mathrm{1}}}$.
   Thus, by \Tg{App}, $\makestatic{\Gamma}  \vdash  \makestatic{\mathit{M} }_{{\mathrm{1}}} \, \makestatic{\mathit{M} }_{{\mathrm{2}}}  \ottsym{:}  \makestatic{\mathit{A} }_{{\mathrm{1}}}$.

  \case \Ts{TApp}:
   Similar to the case of \Ts{App}; we use the fact that,
   if $\ottnt{A}  \equiv  \ottnt{B}$, then $ \ottnt{A}    [  \ottnt{C}  /  \mathit{X}  ]    \equiv   \ottnt{B}    [  \ottnt{C}  /  \mathit{X}  ]  $.

  \case \Ts{RLet}:
   We are give $ \makestatic{\Gamma}  \mathrel{ \makestatic{\vdash} }  \mathsf{let} \, \ottsym{\{}  \ell  \ottsym{=}  \mathit{x}  \ottsym{;}  \mathit{y}  \ottsym{\}}  \ottsym{=}  \makestatic{\mathit{M} }_{{\mathrm{1}}} \, \mathsf{in} \, \makestatic{\mathit{M} }_{{\mathrm{2}}}  :  \makestatic{\mathit{A} } $ and,
   by inversion, $ \makestatic{\Gamma}  \mathrel{ \makestatic{\vdash} }  \makestatic{\mathit{M} }_{{\mathrm{1}}}  :   [    \ell  \mathbin{:}  \makestatic{\mathit{B} }   ;  \makestatic{\rho}   ]  $ and
   $ \makestatic{\Gamma}  \ottsym{,}   \mathit{x}  \mathord{:}  \makestatic{\mathit{B} }   \ottsym{,}   \mathit{y}  \mathord{:}   [  \makestatic{\rho}  ]    \mathrel{ \makestatic{\vdash} }  \makestatic{\mathit{M} }_{{\mathrm{2}}}  :  \makestatic{\mathit{A} } $.

   By the IHs with \reflem{equiv-inversion},
   $\makestatic{\Gamma}  \vdash  \makestatic{\mathit{M} }_{{\mathrm{1}}}  \ottsym{:}   [    \ell  \mathbin{:}  \makestatic{\mathit{B} }_{{\mathrm{0}}}   ;  \makestatic{\rho}_{{\mathrm{0}}}   ] $ and
   $\makestatic{\Gamma}  \ottsym{,}   \mathit{x}  \mathord{:}  \makestatic{\mathit{B} }   \ottsym{,}   \mathit{y}  \mathord{:}   [  \makestatic{\rho}  ]    \vdash  \makestatic{\mathit{M} }_{{\mathrm{2}}}  \ottsym{:}  \makestatic{\mathit{A} }_{{\mathrm{0}}}$
   for some $\makestatic{\rho}_{{\mathrm{0}}}$, $\makestatic{\mathit{A} }_{{\mathrm{0}}}$, and $\makestatic{\mathit{B} }_{{\mathrm{0}}}$
   such that $\makestatic{\rho}  \equiv  \makestatic{\rho}_{{\mathrm{0}}}$ and $\makestatic{\mathit{A} }  \equiv  \makestatic{\mathit{A} }_{{\mathrm{0}}}$ and $\makestatic{\mathit{B} }  \equiv  \makestatic{\mathit{B} }_{{\mathrm{0}}}$.

   Since $\makestatic{\Gamma}  \ottsym{,}   \mathit{x}  \mathord{:}  \makestatic{\mathit{B} }   \ottsym{,}   \mathit{y}  \mathord{:}   [  \makestatic{\rho}  ]    \equiv  \makestatic{\Gamma}  \ottsym{,}   \mathit{x}  \mathord{:}  \makestatic{\mathit{B} }_{{\mathrm{0}}}   \ottsym{,}   \mathit{y}  \mathord{:}   [  \makestatic{\rho}_{{\mathrm{0}}}  ]  $,
   we have $\makestatic{\Gamma}  \ottsym{,}   \mathit{x}  \mathord{:}  \makestatic{\mathit{B} }_{{\mathrm{0}}}   \ottsym{,}   \mathit{y}  \mathord{:}   [  \makestatic{\rho}_{{\mathrm{0}}}  ]    \vdash  \makestatic{\mathit{M} }_{{\mathrm{2}}}  \ottsym{:}  \makestatic{\mathit{A} }_{{\mathrm{1}}}$
   for some $\makestatic{\mathit{A} }_{{\mathrm{1}}}$ such that $\makestatic{\mathit{A} }_{{\mathrm{0}}}  \equiv  \makestatic{\mathit{A} }_{{\mathrm{1}}}$ by \reflem{typing-tctx-reorder}.
   Since $\makestatic{\mathit{A} }  \equiv  \makestatic{\mathit{A} }_{{\mathrm{1}}}$, we finish by \T{RLet}.

  \case \Ts{VCase}: Similar to the case of \Ts{RLet}.  This case also uses the
  first assumption about $ \oplus $ stated in this section.
 \end{caseanalysis}
\end{proof}

\begin{thm}
 \begin{enumerate}
  \item If $\makestatic{\Gamma}  \vdash  \makestatic{\mathit{M} }  \ottsym{:}  \makestatic{\mathit{A} }$, then $ \makestatic{\Gamma}  \mathrel{ \makestatic{\vdash} }  \makestatic{\mathit{M} }  :  \makestatic{\mathit{A} } $.
  \item If $ \makestatic{\Gamma}  \mathrel{ \makestatic{\vdash} }  \makestatic{\mathit{M} }  :  \makestatic{\mathit{A} } $, then $\makestatic{\Gamma}  \vdash  \makestatic{\mathit{M} }  \ottsym{:}  \makestatic{\mathit{B} }$
        for some $\makestatic{\mathit{B} }$ such that $\makestatic{\mathit{A} }  \equiv  \makestatic{\mathit{B} }$.
 \end{enumerate}
\end{thm}
\begin{proof}
 By Lemmas~\ref{lem:cnsv-typing-gs} and \ref{lem:cnsv-typing-sg}.
\end{proof}